\documentclass[3p]{elsarticle}
\usepackage{amssymb}
\usepackage{amsmath}

\usepackage{graphicx}
\usepackage{subfigure}
\usepackage{color}
\graphicspath{{./img/}{./mathematicaimg/}{./}}

\usepackage{booktabs}
\usepackage{longtable}
\usepackage{multirow}

\usepackage[bookmarks]{hyperref}

\newcommand{\td}{\mathrm{d}}
\newcommand{\te}{\mathrm{e}}
\newcommand{\ti}{\mathrm{i}}
\newcommand{\ft}{\mathfrak{t}}

\newcommand{\tL}{\mathrm{L}}
\newcommand{\tT}{\mathrm{T}}
\newcommand{\tK}{\kappa}
\newcommand{\tR}{\mathrm{R}}

\newcommand{\corr}[1]{\langle #1 \rangle}
\newcommand{\abs}[1]{\left\vert#1\right\vert}
\newcommand{\set}[1]{\{#1\}}
\newcommand{\im}{i}
\newcommand{\doo}{\partial}
\newcommand{\de}{\mathrm{d}}
\newcommand{\integers}{\mathbb{Z}}
\newcommand{\real}{\mathbb{R}}

\journal{Physics Reports}

\begin{document}
 
 \begin{frontmatter}
 
 \title {Dual gauge field theory of quantum liquid crystals in two dimensions}

\author[keio,nims,riken,cor1]{Aron~J.~Beekman}
\ead{aron@phys-h.keio.ac.jp}

\author[leiden,cor2]{Jaakko Nissinen}

\author[stanford]{Kai Wu}

\author[leiden]{Ke Liu}

\author[leiden]{Robert-Jan Slager}

\author[stlouis]{Zohar Nussinov}

\author[leiden]{Vladimir Cvetkovic}

\author[leiden]{Jan Zaanen}

\cortext[cor1]{Corresponding author}
\cortext[cor2]{Correspondence to: Low Temperature Laboratory, Aalto University, P.O. Box 15100, FI-00076 Aalto, Finland}

\address[keio]{Department of Physics, and Research and Education Center for Natural Sciences,
Keio University, Hiyoshi 4-1-1, Yokohama, Kanagawa 223-8521, Japan}
\address[nims]{Computational Materials Science Unit, National Institute for Materials 
Science, Namiki 1-1, Tsukuba, Ibaraki 305-0044, Japan}
\address[riken]{RIKEN Center for Emergent Matter Science (CEMS), Wako, Saitama 351-0198, Japan}
\address[leiden]{Institute-Lorentz for Theoretical Physics, Leiden University, PO Box 9506, NL-2300 RA Leiden, The Netherlands}
\address[stanford]{Stanford Institute for Materials and Energy Sciences, SLAC National Accelerator Laboratory and Stanford University, Menlo Park, CA 94025, USA}
\address[stlouis]{Department of Physics, Washington University, St. Louis, MO 63160, USA}

 \begin{abstract}
We present a self-contained review of the theory of dislocation-mediated quantum melting at zero temperature in two spatial dimensions. The theory describes the liquid-crystalline phases with spatial symmetries in between a quantum crystalline solid and an isotropic superfluid: quantum nematics and smectics. It is based on an Abelian-Higgs-type duality mapping of phonons onto gauge bosons (``stress photons''), which encode for the capacity of the crystal to propagate stresses. Dislocations and disclinations, the topological defects of the crystal, are sources for the gauge fields and the melting of the crystal can be understood as the proliferation (condensation) of these defects, giving rise to the Anderson-Higgs mechanism on the dual side. For the liquid crystal phases, the shear sector of the gauge bosons becomes massive signaling that shear rigidity is lost. After providing the necessary background knowledge, including the order parameter theory of two-dimensional quantum liquid crystals and the dual theory of stress gauge bosons in bosonic crystals, the theory of melting is developed step-by-step via the disorder theory of dislocation-mediated melting. Resting on symmetry principles, we derive the phenomenological imaginary time actions of quantum nematics and smectics and analyze the full spectrum of collective modes. The quantum nematic is a superfluid having a true rotational Goldstone mode due to rotational symmetry breaking, and the origin of this `deconfined' mode is traced back to the crystalline phase. The two-dimensional quantum smectic turns out to be a dizzyingly anisotropic phase with the collective modes interpolating between the solid and nematic in a non-trivial way. We also consider electrically charged bosonic crystals and liquid crystals, and carefully analyze the electromagnetic response of the quantum liquid crystal phases. In particular, the quantum nematic is a real superconductor and shows the Meissner effect. Their special properties inherited from spatial symmetry breaking show up mostly at finite momentum, and should be accessible by momentum-sensitive spectroscopy.
 \end{abstract}
 
\begin{keyword}
 quantum liquid crystals; quantum phase transitions; Abelian-Higgs duality; superconductivity
\end{keyword}

\end{frontmatter}

\tableofcontents

 \begin{longtable}{l l} 
 \caption{List of symbols. Field quantities will be denoted with their spacetime argument $(x) = (t,\mathbf{x})$.}\\
  \toprule 
 Symbol & Description \\
 \midrule
\endfirsthead
 \toprule 
 Symbol & Description \\
 \midrule
 \endhead
 $a_{ij}$ & lattice gauge field\\
 $\widetilde{a}_{\widetilde{i}\widetilde{j}}$ & dual lattice gauge field\\
 $A (\omega,q)$ & spectral function \\
 $A_\mu (x)$ & electromagnetic (photon) gauge field \\
 $B^a$ & Burgers vector \\
 $B(x)$ & magnetic field \\
 $b^a_\mu (x) $ & stress gauge field \\
 $c_\tL$ & longitudinal phonon velocity \\
 $c_\tT$ & transverse phonon velocity \\
 $c_\tK$ & compression (sound) velocity \\
 $c_\td$ & dislocation velocity \\
 $c_\tR$ & rotational velocity \\
 $c_l$ & velocity of light\\
 $C_{\mu\nu ab}$ &  elastic constants \\
 $D$ & number of dimensions\\
 $e^*$ & electric charge \\
 $E_m (x)$ & electric field \\
 $h_\mu (x)$ & torque stress gauge field \\
 $G (\omega,q)$ & stress propagator \\
 $G_\tL (\omega,q)$ & longitudinal propagator \\
 $G_\tT (\omega,q)$ & transverse propagator \\
 $\mathcal{G}$ & electromagnetic propagator \\
 $\mathcal{H}$ & Hamiltonian \\
 $\mathcal{H} (x)$ & Hamiltonian density \\
 $\mathcal{K} (x)$ & compressional source \\
 $\mathcal{J}(x)$ & external source / rotation source \\
 $\mathcal{J}^a(x)$ & displacement source \\
 $J^a_\mu (x)$ & dislocation current \\
 $j_m(x)$ & electric current density \\
 $j_\mu (x)$ & spacetime electric current density  \\
 $\mathcal{L}(x)$ & Lagrangian density \\
 $\mathcal{L}_\mathrm{E} (x)$ & Euclidean Lagrangian density \\
 $\ell$ & rotational stiffness length \\
 $\ell'$ & higher-order compressional length \\
 $m^*$ & mass \\
 $n$ & electric particle density \\
 $n^a$ & Burgers vector \\
 $q$ & momentum absolute value \\
 $q_m$ & momentum \\
 $p$ & spacetime momentum \\
 $P$, $\bar{P}$ & discrete point group \\
 $\mathbf{R}$ & particle coordinate \\
 $\mathcal{S}$ & action \\
 $\mathcal{S}_\mathrm{E}$ & Euclidean action \\
 $t$ & time \\
 $\tau$ & imaginary time \\
 $\ft$ & imaginary time $\times$ a velocity \\
 $x$ & space coordinates \\
 $u^a(x)$ & displacement field \\
 $\mathcal{Z}$ & partition function\\
 $\delta$ & imaginary time convergence factor \\
 $\hat{\varepsilon}_{ab} (\omega,q)$ & dielectric function \\
 $\varepsilon_0$ & dielectric constant \\
 $\varepsilon(x)$, $\varepsilon^a(x)$ & gauge transformation field \\
 $\eta$ & smectic interrogation angle \\
 $\theta(x), \theta_i$ & nematic order parameter field\\
 $\Theta_\mu(x)$, $\Theta^a_\mu(x)$ & disclination current \\
 $\kappa$ & compression modulus \\
 $\lambda(x)$ & Lagrange multiplier field \\
 $\lambda$, $\lambda(\omega,q)$ & screening length / penetration depth \\
 $\lambda_\mathrm{L}$ & London penetration depth \\
 $\lambda_\mathrm{d}$ & dislocation penetration depth \\
 $\lambda_\mathrm{s}$ & shear penetration depth \\
 $\mu$ & shear modulus \\
 $\mu_0$ & magnetic constant \\
 $\nu$ & Poisson ratio \\
 $\rho$ & mass density \\
 $\rho_Q(x)$ & electric charge density \\
 $\sigma^a_\mu (x)$ & relativistic stress tensor \\
 $\hat{\sigma}_{ab} (\omega,q)$ & conductivity tensor \\
 $\tau_\mu (x)$  & torque stress  \\
 $\tau^\ell_\mu$  & second-gradient torque stress  \\
 $\Phi (x)$ & condensate field \\
 $\Phi^a (x)$ & dislocation condensate field \\
 $\omega (x)$, $\omega^{ab}(x)$ & rotational strain field \\
 $\omega$ & frequency \\
 $\omega_n$ & Matsubara frequency \\
 $\omega_\mathrm{p}$ & plasma frequency\\
 $\Omega$ & Higgs mass \\
 $\Omega^c$ & Frank vector \\
 $\Omega^{ab}$ & deficient rotation  \\
  \bottomrule
 \end{longtable}

\section{Introduction}\label{sec:Introduction}
How do crystals melt at zero temperature into quantum liquids? This would seem to be a question that was answered a long time ago. The $^4$He superfluid solidifies under pressure, through a first-order transition that is regarded as well understood.  Similarly, it is widely believed that at low density the Fermi liquid formed from electrons will turn into a Wigner crystal also involving a first-order transition. However, dealing with microscopic constituents which are less simple than helium atoms, in principle zero-temperature phases can be formed which are in between the crystal and the isotropic superfluid: the `vestigial' quantum liquid-crystalline phases. 

It appears that such phases are realized in the strongly-interacting electron systems found in iron and copper superconductors, and in recent years this has grown into a sizable research field~\cite{AndoEtAl02, Vojta09,OganesyanKivelsonFradkin01, BorziEtAl07, FradkinEtAl10,Fradkin12, ChuEtAl12,FernandesChubukovSchmalian14}. This development was started a long time ago, by a seminal paper due to  Kivelson, Fradkin and Emery~\cite{KivelsonFradkinEmery98} that explained the potential for such vestigial phases to exist in the electron systems. Inspired by this work, one of the authors of this review (J.Z.)  in 1997 asked himself the question ``what can be learned in general about such quantum liquid crystals?'' He decided to focus first on circumstances that simplify the life of a theorist: bosonic matter living in the maximally symmetric Galilean space, and two space and one time dimensions (2+1D). For physical reasons explained below, the interest was particularly focused on the case that the (crystalline) correlations in the quantum liquid states are as pronounced as possible. 

As a lucky circumstance it turned out that a rather powerful mathematical methodology was already lying in wait, originating in the field of `mathematical elasticity theory' that was rewritten in a systematic field-theoretical language by Hagen Kleinert in the 1980s~\cite{Kleinert89b,Kleinert08}. It dealt with the classical statistical physics of crystal melting in three space dimensions, but can be extended to handle the quantum problem in 2+1D.  This revolves around a weak--strong duality: it can be viewed as an extension of the well-known {\em Abelian-Higgs} or {\em vortex--boson} duality used in quantum melting of superfluid order in 2+1D, which is exquisitely able to deal with the strong-correlation aspect given its non-perturbative powers.  However, this generalization is far from trivial.  Deeply rooted in the intricacies of the symmetries of spacetime itself, the duality reveals that the physics of crystal quantum melting is much richer than in the superfluid case. This is the story that we wish to expose in this review. 

The groundwork was laid down in the period 1998--2003 and these results were published in comprehensive form~\cite{ZaanenNussinovMukhin04}. This description was however in a number of regards rather crude and contained several flaws. It turned out that ambitious junior scientists that passed by in Leiden since then were allured by the subject, perfecting gradually the 2003 case and discovering new depths in the problem. Crucial parts were never published~\cite{Cvetkovic06} while the remainder got scattered over the literature~\cite{KleinertZaanen04,CvetkovicNussinovZaanen06,CvetkovicZaanen06a,CvetkovicZaanen06b,CvetkovicNussinovMukhinZaanen08,ZaanenBeekman12,BeekmanWuCvetkovicZaanen13,LiuEtAl15}. With the advent of the present third generation of students it appears that the last missing pieces have fallen into place and we believe that we have now an essentially complete theory in our hands at least for 2+1 dimensions. All that remains are some quantitative details such as the effects of crystalline anisotropy that are easy but tedious. We decided to endeavor to present the whole story in a comprehensive and coherent fashion, making it accessible for the larger community beyond the `Leiden school'. 

More recently we have been concentrating increasingly on the 3+1D case: this is technically considerably harder but rewarding a much richer physical landscape. Several research directions are presently under investigation but a thorough understanding of the easier 2+1D case is a necessity to appreciate fully the vestigial marvels that one discovers in the most physical of all dimensions. This formed an extra motivation for us to write this extensive review.

Before we turn to the mathematical substances described in the bulk of this review let us first present a gross overview of the context, and the main features of this {\em quantum field theory of liquid crystalline matter}.

\subsection{The prehistory: fluctuating stripes and other high-$T_\mathrm{c}$ empiricisms}

It all started in the late 1990s during the heydays of the subject of {\em fluctuating order} in the copper-oxide high-$T_\mathrm{c}$ superconductors.  Before the discovery of high-$T_\mathrm{c}$ superconductivity in 1986 it was taken for granted that electron systems realized in solids were ruled by the principles of the highly-itinerant electron gas. The essence of such systems is that they are quite featureless: the Fermi liquid is a simple, homogeneous state of matter, `equalized' by the highly delocalized nature of its quasiparticles.  Structure can emerge in the form of spontaneous symmetry breaking but this should be of the Bardeen--Cooper--Schrieffer (BCS) kind where it becomes discernible only at the long time and length scales associated with the weak-coupling gap. It came then as a big surprise when inelastic neutron scattering measurements seemed to reveal that at least in the underdoped ``pseudogap'' regime of the cuprates, the electron quantum liquid is much more textured. Spin fluctuations were observed at energies associated with highly collective physics ($\sim$0-80 meV) that reveal a high degree of spatial organization, although there is no sign of static translational symmetry breaking (for recent experimental results, see Refs.~\cite{KivelsonEtAl03, VigEtAl15}). 

In the 1990s it was discovered that in the family of doped La$_2$CuO$_4$  (``214'') superconductors, under     
specific circumstances, static order can occur in the form of {\em stripes}~\cite{TranquadaEtAl95, KivelsonEtAl03}.  These stripes are a ubiquitous ordering phenomenon found generically in doped Mott insulators other than cuprates (nickelates, cobaltates, manganites \ldots). These are best understood as a lattice of electronic discommensurations (``rivers of charge'') that are formed when the charge-commensurate Mott insulator is doped. For the present purposes these might be viewed as `crystals', likely formed from (preformed) electronic Cooper pairs that break the rotational symmetry (tetragonal, $C_4$) of the underlying square lattice of ions  in a unidirectional (orthorhombic, $C_2$) way.  This in turn goes hand-in-hand with an incommensurate antiferromagnetic order.  This view was initially received with quite some skepticism.  It was argued that this could well be a specialty of the 214-family, being also in other regards atypical (e.g. relatively low superconducting
$T_\mathrm{c}$). This changed with the discovery of charge order in the other underdoped cuprates, at first on the surface by scanning tunneling spectroscopy \cite{HanaguriEtAl04},  followed by a barrage of other experimental observations~\cite{CominEtAl15, DaSilvaNetoEtAl14, CominDamascelli16}. This has turned in recent years into a mainstream research subject in the community: see the review Ref.~\cite{KeimerEtAl15}. There are still fierce debates over the question whether this charge order should be understood as a weak-coupling `Peierls-like' charge density wave (CDW) instability, or as a strongly-coupled affair arising in a doped Mott insulator~\cite{ZaanenGunnarsson89}. The latest experimental results appear to largely support the strong coupling view~\cite{MesarosEtAl16}. Similarly, impressive progress
has been made studying doped Mott insulators using several numerical methods.  Although the various methods are characterized by multiple a-priori uncontrolled assumptions, it was very recently demonstrated that these invariably predict the ground state of the doped Hubbard model to be of the stripe-ordered kind~\cite{ZhengEtAl16}.

Dealing with a weak-coupling CDW associated with a Fermi surface instability the `solid-like' correlations should rapidly disappear when the charge order melts, be it as function of temperature or by quantum fluctuations in the zero-temperature state. The physical ramification of ``strongly coupled'' in this context is that such correlations should remain quite strong even in the liquid state. The energy scale associated with the formation of the charge order on the microscope scale is by definition  assumed to be large, and the melting process is driven by {\em highly collective} degrees of freedom --- the main theme of this review. Observation of the consequences of such ``fluctuating order'' in electron systems is not easy~\cite{ZaanenEtAl01, KivelsonEtAl03}. One has to have experimental access to the {\em dynamical} 
responses of the electron system in a large window of relevant energies and momenta. Until very recently only spin fluctuations could be measured, and it was early on pointed out that these should be able to reveal information about such fluctuations in the case that the charge order is accompanied by stripe antiferromagnetism~\cite{ZaanenHorbachSaarloos96}. Elastic neutron scattering then revealed the surprise  that at somewhat higher energies the spin fluctuations in superconducting, underdoped cuprates, which lack any sign of static stripe order, look very similar to those of the {\em striped} cuprates~\cite{TranquadaEtAl04}. The difference is that in the former a gap is opening up at small energies in the spin-wave spectrum of the latter.  On basis of these observations the idea of {\em dynamical} or {\em fluctuating} stripes was born: at mesoscopic distances ($\sim$ nanometers) and energies ($\sim 10$ meV)  the electron liquid approaches closely a striped state but eventually quantum fluctuations take over, turning it into a featureless superconducting state at macroscopic distances. 

It proved very difficult to make this notion more precise, a main obstacle being the absence of experimental means of directly observing fluctuating charge order. The spin fluctuations represent an
inherently indirect measure and  one would like  to measure instead the {\em charge} fluctuations. Roughly twenty years after the idea of ``fluctuating stripes'' emerged, this appears to be now on the verge of happening due to the arrival of the high resolution RIXS beam lines and of a novel EELS spectrometer~\cite{VigEtAl15}. Also in this context the computational progress is adding urgency to this affair by the very recent Quantum Monte Carlo results for a three-band model signaling strong stripe fluctuations at elevated temperatures~\cite{HuangEtAl16}.   

However, at first sight it sounds like a tall marching order to {\em interpret} such results. This is about the quantum physics of strongly-interacting forms of matter and without the help of powerful mathematics 
it may well be that no sense can be made of the observations. The theory presented in this review is precisely aiming at making a difference in this regard. It is very useful in physics to know what happens
 in the {\em limits}. The established paradigm dealing with order in electron systems is heavily resting on the weak-coupling limit: one starts from a free Fermi gas, to find out how this is modified by interactions.
 But this is inherently perturbative and when the interactions become strong one loses mathematical control.  By mobilizing some big guns of quantum field theory (gauge theory, weak--strong duality) we will 
 demonstrate here that exactly the {\em opposite} limit describing in a mathematically precise way the `maximal solid-like' quantum fluid becomes also easy to compute, at least once one has mastered the 
 use of the field-theoretical toolbox. The only restriction is that this works solely for bosons, preformed Cooper pairs in the present empirical context.  This is powerful mathematics and it predicts a `universe' of novel phenomena, that we will outline in the remainder of this introduction.  

\begin{figure*}
 \subfigure[ solid -- $\mathbb{Z}^2 \rtimes \bar{P}$]{
  \includegraphics[width=3.7cm]{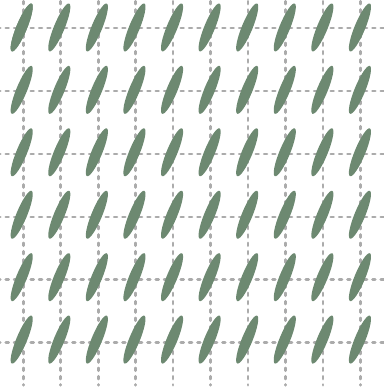}\label{fig:classical melting solid}
 }
 \subfigure[ smectic -- $(\mathbb{R} \times \mathbb{Z}) \rtimes \bar{P}$]{
  \includegraphics[width=3.7cm]{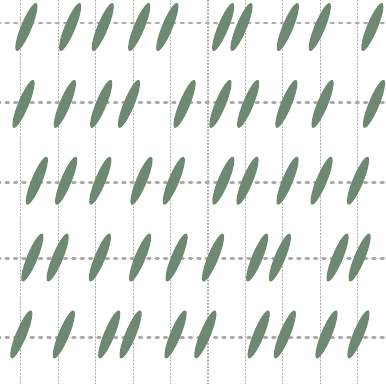}\label{fig:classical melting smectic}
 }
 \subfigure[ nematic -- $\mathbb{R}^2 \rtimes \bar{P}$]{
  \includegraphics[width=3.7cm]{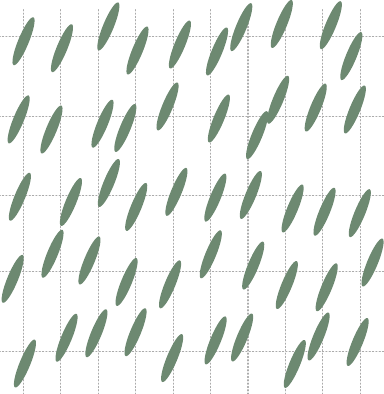}\label{fig:classical melting nematic}
 }
 \subfigure[ liquid -- $\mathbb{R}^2 \rtimes O(2)$]{
  \includegraphics[width=3.7cm]{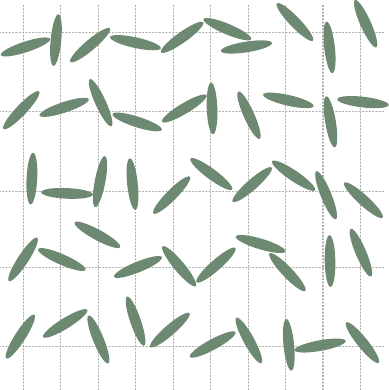}\label{fig:classical melting liquid}
 }
 \caption{Sketch of classical melting of a 2D crystalline solid. The thick dashed lines denote crystal axes along which the `particles' are ordered while the thin dashed lines are for reference with the previous situation only. Denoted in the figure labels is the symmetry group as a subgroup of the 2D Euclidean group $E(2) \simeq \mathbb{R}^2 \rtimes O(2)$, where $\bar{P}$ denotes the discrete point group of the lattice. \subref{fig:classical melting solid} Solid: regularly ordered breaking both translations and rotations completely to a discrete space group. \subref{fig:classical melting nematic} Smectic: translational symmetry is restored in the horizontal direction, but these liquid layers are still periodically stacked in the vertical direction. Rotations are broken. \subref{fig:classical melting nematic} Nematic: all translational symmetry is now restored like a liquid, but the constituents maintain orientational order. 
\subref{fig:classical melting liquid}. Liquid: completely disordered and all symmetry is restored. For the complementary dual picture, see Fig.~\ref{fig:phase diagram sketch}}\label{fig:melting sketch}
\end{figure*}

Returning to the historical development, the main source of inspiration for this work has been all along the seminal 1998 paper by Kivelson, Fradkin and Emery~\cite{KivelsonFradkinEmery98}. These authors argued that the fluctuating stripe physics forms a natural stage for the formation of new zero-temperature phases of matter: the {\em quantum liquid crystals}. In most general terms it follows a wisdom  which is well tested in the realms of the physics of classical, finite-temperature matter \cite{DeGennesProst95}. Typically the system forms a fully symmetric liquid at high temperatures, while it breaks the translations and rotations of Euclidean space at low temperatures, forming a solid. However, given particular microscopic conditions (e.g. `rod-like molecules') one finds the {\em partially ordered} or {\em vestigial} phases. One manifestation corresponds to the nematic-type liquid crystalline order where translational symmetry is restored --- the {\em liquid} aspect --- while rotational symmetry is still broken (``the rods are lined up''). There are also smectic-type phases which break translations in one direction while the system remains fluid in the other directions (``stack of liquid layers''), see Fig.~\ref{fig:melting sketch}.  A priori, the same hierarchy 
of symmetry breakings can occur at zero temperature, with the difference that the liquids are now identified as quantum liquids. Crudely speaking, one can now envisage that the `stripiness' takes the role of the rod-like molecules on the microscopic scale. Subsequently one can picture that a {\em quantum 
smectic} is formed which behaves like a zero-temperature metal or superconductor in one spatial direction, while it insulates in other directions. Similarly, metallic or superconducting 
zero-temperature states can be imagined which are anisotropic because of the spontaneous breaking of spatial rotations: the {\em quantum nematics}. The notion of quantum liquid crystals  
appeared to be a fruitful idea. Not long thereafter evidences were found for the occurrence of such quantum nematic order in part of the underdoped regime of 
YBa$_2$Cu$_3$O$_{6+x}$ and Bi$_2$Sr$_2$CaCu$_2$O$_{8+x}$ cuprate superconductors~\cite{AndoEtAl02, HinkovEtAl08, DaouEtAl10, HowaldEtAl03, KohsakaEtAl07}. 

However,  Kivelson {\em et al.}~\cite{KivelsonFradkinEmery98} took it a step further by conceptualizing it in the language of the celebrated Kosterlitz--Thouless--Nelson--Halperin--Young (KTNHY) theory of topological 
melting in two `classical' dimensions~\cite{KosterlitzThouless72,KosterlitzThouless73, HalperinNelson78,NelsonHalperin79, Young79}. In this framework the liquid is not understood as the state where the constituents of the solid are liberated, freely moving around in a gaseous state. Instead it is asserted that the solid stays locally fully intact, and instead the `isolated' topological excitations associated with the restoration of translational invariance (dislocations, Sec.~\ref{sec:Topological defects in solids}) proliferate. Such a liquid still breaks the rotational symmetry since rotational-symmetry restoration
requires different defects: disclinations. Therefore states of matter where the dislocations are `condensed' while the disclinations are still `massive' are symmetry-wise identical to the smectics and nematics formed
from the rods of Fig.~\ref{fig:melting sketch}. In fact, the theory we will present here starts out from this basic notion: is just the generalization of the KTNHY theory to the zero-temperature quantum realms in 2+1D, showing that in this quantum setting there is a lot more going on. 

Also in other areas the concept of quantum liquid crystals flourished. It became clear that the stripe phases formed in high-Landau-level 
quantum Hall systems turn into quantum nematic phases.  The theme became particular prominent in the iron-pnictide superconductors~\cite{ChuangEtAl10} where such nematic order appears to be very pronounced although the debate about its precise microscopic origin as well as its relation to the superconductivity is still raging~\cite{FernandesChubukovSchmalian14}. For completeness we will shortly review these matters in Sec.~\ref{subsec:Quantum liquid crystals: the full landscape}. They are interesting 
subjects by themselves, revealing physics of a different kind than we are addressing. It is questionable whether the material in this paper is of any consequence in these realms. The quantum Hall nematics may be `sufficiently orderly', but the dynamical information which is our main output cannot possibly be measured in two-dimensional electron gases. The pnictides are almost surely situated on the weak-coupling side: no evidence of any kind emerged for strong charge-order correlations in their electron systems. 

With regard to the potential empirical relevance of the theory the only obvious theater of which we are presently aware are the underdoped cuprates. Even in this context it remains to be seen whether 
any of the phenomena that the theory predicts will occur in a literal fashion in nature.  It hinges after all on an extreme limit, and it depends on whether the microscopic conditions in real electron 
systems permit getting near enough to this ``maximal solid behavior'' such that the remnants of its physics are discernible in experiment. At present this work is therefore in first instance of a general 
{\em theoretical} interest. However, anybody who will take the effort to master this affair will be rewarded by the striking elegance and beauty of the physics of the maximally-correlated quantum fluid, 
making one wonder whether nature can ignore this opportunity.

\subsection{Platonic perfection and the big guns of quantum field theory}  

Quantum field theory as it comes alive in condensed matter physics is precisely tied to the universal long-wavelength physics associated with zero-temperature matter. Inspired by the empirical developments described above we became aware that actually the general description of the quantum liquid crystals is among the remaining open problems that can be tackled at least in principle by the established machinery of quantum field theory. More generally, this is about \emph{quantum} many-body systems that spontaneously break spatial symmetries. This is what we set out to explore some 15 years ago. This program is not quite completed yet. Dimensionality is a particularly important factor and quite serious complications arise in 3+1 and higher dimensions. However, in two space dimensions the theory is brought under complete control, which this review is intended to present in a comprehensive and coherent fashion. 

In order to get anywhere we consider matter formed from {\em bosons}: there are surely some very deep questions related to {\em fermionic} quantum liquid crystals but there is just no controlled mathematical technology available that can tackle the fermion sign problem (see also Sec.~\ref{subsec:Quantum liquid crystals: the full landscape}). As related to the empirical context of the previous paragraphs, at zero temperature one is invariably dealing with nematic (or smectic) {\em superconductors} formed  from Cooper pairs which are bosons. Therefore, insofar as any of our findings can be of direct relevance in this empirical context, it is natural to explore what the bosonic theory has to tell. 

The next crucial assumption is that we start out with a system living in Galilean-invariant space. There is no ionic background lattice and our bosonic system has to break the spatial symmetries all by itself. This assumption detaches our theoretical work from a literal application to the empirical electron systems.  This is however the natural stage for the elegant physics associated with the field theory and it is just useful to know what happens in this limit, as we hope to demonstrate. After all, there are signs that the strength of the `anisotropy' coming from the lattice might not be at all that large: the case in point is that the scanning tunneling spectroscopy (STS) images of cuprate stripes are littered with rather smooth dislocation textures of a type that would not occur when the effective lattice potential would be dominant \cite{MesarosEtAl11}.  We will later present several results following from the continuum theory that might be still of relevance to the lattice incarnation when the pinning energy of the lattice is sufficiently weak.  Of course, the experimentalists should take up the challenge to engineer such  a continuum bosonic quantum liquid crystal, for instance by exploiting cold atoms etc.

The experienced condensed matter physicist might now be tempted to stop reading: what new is to be learned about a system of bosons in the Galilean continuum? This realm of physics is supposed to be completely charted: dealing with bosonic particles like $^4$He atoms these are well known to form either close packed (in 2+1D, triangular) crystals or superfluids at zero temperature. Dealing with `rod-like bosons' there is surely room to have an intermediate quantum nematic phase corresponding to a superfluid breaking space rotations in addition. Resting on generic wisdoms of order parameter theory it is obvious that a Goldstone boson will be present in this phase associated with the rotational symmetry breaking that can be sorted out in a couple of lines of algebra. What is the big deal? 

This industry standard paradigm is based on a weak-interaction, `gaseous' perspective. To describe the superfluid one takes the free boson Bose--Einstein condensate perspective dressed by weak interactions (Bogoliubov theory). In helium one typically finds a strong first-order transition to the crystal phase, which can be well understood as a classical crystal dressed by mild zero-point motions. 
The reason, however, for this review to be quite long is that we will mobilize the `big gun' machinery of quantum field theory. This is actually geared to deal with a physics regime that might be described as a `maximally strongly-interacting' regime of the microscopic bosons. In fact, the reader will find out that these bosons have disappeared altogether from the mathematical description that is entirely concerned with the emergent collective degrees of freedom that are formed from a near infinity of microscopic degrees of freedom. 

We will find out that the  zero-temperature liquids are invariably superfluids or superconductors. However, these are now characterized by transient crystalline correlations extending on length scales that are large compared to the lattice constant. From this non-perturbative starting point, it is rather natural for the field theory to describe the kind of  physics that is envisioned by the {\em fluctuating stripes} hypothesis, where the superconductor is locally, at the smallest length scales, still behaving as a crystal. 

The big gun machinery that we will employ is weak--strong or Kramers--Wannier duality~\cite{KramersWannier41}.  The immediate predecessor of the present pursuit is the intense activity in the 1970s revolving around the Berezinskii--Kosterlitz--Thouless (BKT) topological melting theory~\cite{Berezinskii70,KosterlitzThouless72,KosterlitzThouless73}, and the particular implementation in the form of the Kosterlitz--Thouless--Nelson--Halperin--Young (KTNHY) theory  of finite-temperature melting of  a crystal in two dimensions, involving the {\em hexatic} vestigial phase~\cite{KosterlitzThouless72,KosterlitzThouless73, HalperinNelson78,NelsonHalperin79, Young79}. The central notion is that the destruction of the ordered state can be best understood in terms of the unbinding (proliferation) of the topological defects associated with the broken symmetry. The topological defects of the superfluid are vortices. The vortex is thereby the unique agent associated with the destruction of the order: in a strict sense a single delocalized vortex suffices to destroy the order parameter of the whole system. In the ordered state the excitations can therefore be divided in {\em smooth configurations} corresponding to the Goldstone bosons, whose existence is tied to the presence of order, and the {\em singular} or {\em multivalued configurations} characterized by topological quantum numbers. As long as the latter occur only as neutral combinations (e.g.  bound vortex--antivortex pairs) the order parameter cannot be destroyed.  Conversely, when the topological excitations unbind and proliferate, the system turns automatically into the disordered state, which now can be seen as a condensate, a `dually ordered state' formed out of topological defects. This is in essence the basic principle of field-theoretical weak--strong or Kramers--Wannier dualities. Many weak--strong mappings have been developed since, ranging all the way to the fanciful dualities discovered in string theory such as the AdS/CFT correspondence \cite{ZaanenEtAl15}.

If this principle applies universally (which is not at all clear) it will lead to the staggering consequence that, away from the critical state, all field-theoretical systems are always to be regarded as ordered states. It is just pending the access of the observer to {\em order operators} or {\em disorder operators} whether he/she perceives the disordered state as ordered or the other way around. The benefit for theorists is that the mathematical description of the {\em weakly-coupled} ordered/symmetry broken state is very well controlled (Goldstone bosons and so forth) while {\em strongly-coupled} disordered states are typically much harder to describe. Now in the dual description the latter are yet again of the tranquil, ordered kind. This review explores in detail the workings of the weak--strong duality as applied to zero-temperature quantum crystals and its duals in 2+1 spacetime dimensions. 

Turning to crystals, the symmetry that is broken is the Euclidean group associated with space itself which is a much richer affair than the internal $U(1)$ global symmetry of $XY$-spin systems/superfluids. In crystals this can be broken to the smallest possible subgroups as classified in terms of the space groups.  Solids are of course overly familiar from daily life but to a degree this familiarity is deceptive.  The symmetry principles which are involved are much more intricate than the usual internal symmetries. The Euclidean group $E(D)$ in $D$ space dimensions involves $D$ independent translations, $\mathbb{R}^D$, forming an infinite group, in semidirect relation with the orthogonal group $O(D)$  including rotations and reflections. Semidirect here means that rotating, translating and rotating back is in general not the same as simply translating (the rotation group {\em acts} on the translation group). This is denoted as $E(D) = \mathbb{R}^D \rtimes O(D)$. Crystals are described by space groups $S \subset E(D)$ which are comprised of lattice translations $\mathbb{Z}^D$, again in semidirect relation to discrete point group symmetries $\bar{P} \subset O(D)$, augmented by non-symmorphic symmetries such as glide reflections.  This will be the underlying theme throughout this review: the surprising richness of the physics is an expression of this intricate symmetry structure. 

The point of departure will be the theory describing the nature of the maximally ordered phase (the solid): this is the 19th century theory of elasticity, promoted to the Lagrangian of the quantum theory by adding a kinetic term. Although it has been around for years, as a field theory it is quite involved given its tensor structure. The topological content is also remarkably rich, despite the fact that the basics have been identified a long time ago, and much of it has been exported to engineering departments.  The topological defect associated with the restoration of  translational invariance is the {\em dislocation} identified by Burgers in the 1930s~\cite{Burgers39}.  Its topological invariant is the {\em Burgers vector} associated with the discrete lattice translation symmetry of the crystal; this translational defect does not affect the rotational symmetry and its charge has therefore to keep track of this information. The rotational $O(D)$ symmetry is restored by a separate {\em disclination} defect with the Frank vector (or {\em Frank scalar} in 2D) as topological charge~\cite{Frank58}. The dislocation can be in turn viewed as a bound disclination--antidisclination pair while the disclination also corresponds to a bound state of an infinite number of dislocations with parallel Burgers vector, see Fig.~\ref{fig:disclination dislocation interdependence} below. 

In a solid, dislocations and disclinations are topologically distinct defects with an explicit hierarchy: the {\em deconfined} (in the solid) dislocations are intrinsically easier to produce than the confined disclinations, although {\em a priori} one cannot exclude the possibility that the disclinations will proliferate together with the dislocations giving rise to the first-order transition directly from the solid to the isotropic liquid. This depends on the details of the `UV'  (the `chemistry' on the molecular scale). 
 
We now insist that the disclinations stay massive and thereby the breaking of the rotational symmetry of the solid is maintained. However, when the dislocations proliferate, translational symmetry is restored and the system turns into a fluid. Given the vectorial nature of the Burgers vectors this can be accomplished in different ways. When free dislocations occur with precisely equal probability the translational symmetry is restored in all possible directions while the rotational symmetry is still broken as characterized by the point group of the `parent' crystal. These form the family of `nematic-like' liquid crystals. There is actually an ambiguity in the vocabulary that is not settled: in the soft-matter community it is convention to reserve {\em nematic} for the uniaxial, $D_{h\infty}$-symmetric variety (ordered states of 'rod-like' molecules), while for instance the nomenclature {\em $p$-atics} has been suggested for 2D nematics characterized by a $p$-fold axis~\cite{ParkLubensky96}. In full generality, these substances are classified by their point group symmetries. Because we are mainly interested in long-distance hydrodynamic properties which do not really differ between the different point groups, and by lack of a generally accepted convention, we will call all these substances {\em nematics}, with a point group prefix when needed. See  Sec.~\ref{sec:Order parameters for 2+1-dimensional nematics} for a more nuanced view.

The topology allows for yet another possibility~\cite{OstlundHalperin81}, which is sometimes overlooked. It is a topological requirement for the nematic order that the Burgers vectors in the dislocation condensate are locally anti-parallel since a net `Burgers uniform magnetization' corresponds to a finite disclination density, which we excluded from the start. However, there is no requirement to populate all Burgers directions equally as happens in the nematics. Instead, one can just preferentially populate the Burgers vectors in one direction. The effect is that in this direction the system turns into a fluid while translations are still broken in the orthogonal direction: this is the topological description of the smectic as the state that can occur in between the crystal and the nematic. Obviously, when the disclinations proliferate one will eventually end up in an isotropic liquid although still other phases are possible with a higher point group symmetry associated with a preferential population of certain Frank vectors. 

In the present context of crystal quantum melting in two spatial dimensions, the crucial ingredient is that the dislocations are `quite like' vortices with regard to their status in the duality, as was already realized by KTNHY in the 2+0D case.  It was then asserted that an unbinding BKT transition can take place involving only the dislocations (keeping the disclinations massive) into a {\em hexatic} state (in our terminology:  $C_6$-nematic). Our pursuit is in essence just the next logical generalization of this affair: how does this {\em topological melting} work out in the 2+1D bosonic quantum theory realized at zero temperature? 

The quantum version is however richer in a number of regards, and in a way more closely approaching a platonic perfected incarnation of liquid crystalline order. As we will see, the fluids associated with the quantum disordered crystal can also be viewed as  `dual superconductors' but now the ``gauge bosons that acquire a mass'' are associated with shear forces while the `dual matter' corresponds to the Bose condensate formed out of dislocations. The similarity between dislocations and vortices is rooted in the fact that, like the vortices, the dislocations restore an {\em Abelian} symmetry: the translational symmetry of space. For this reason, the crystal duality is still governed by the general rules of Abelian dualities. However, there are also fundamental differences: in constructing the duality the richness associated with field-theoretical elasticity comes to life.  For instance, in the dual {\em dislocation superconductors} the information associated with the rotational symmetry breaking in the liquid crystals is carried by the Burgers vectors of the dislocations.  In the gauge-theoretical description these take the role of additional `flavors' which in turn determine the couplings to the `stress photons' mediating the interactions between the  dislocations. 

As we will explain at length in Secs.~\ref{sec:Dynamics of disorder fields}--\ref{sec:Quantum smectic}, this has the net effect that {\em shear stress} has a similar fate in the liquid as the magnetic field in a normal superconductor: the capacity to propagate shear forces is expelled from the {\em dual stress superconductor} at distances larger than the {\em shear penetration depth}. By just populating the Burgers charges equally or preferentially in a particular orientational direction these dislocation condensates describe equally well nematic- and smectic-type phases where now the solid-like behavior of the latter is captured naturally by the incapacity of the dislocations to `Higgs' shear stress in a direction perpendicular to their Burgers vector.  

In fact, one could nonchalantly anticipate that this general recipe applies {\em a priori} equally well to the classical, finite-temperature liquid crystals in three space dimensions as to the 2+1D quantum case in the usual guise of thermal field theory.  In equilibrium, one can compute matters first in a spacetime with Euclidean signature and Wick rotate to Minkowski time afterwards. Where is, then, the difference between 3D classical and 2+1D quantum `elastic matter'? The quantum matter is formed from conserved constituents (like electrons, atoms) at finite density, and we are interested in phenomena occurring at energies which are small compared to the thermodynamic potential. Under these conditions Lorentz invariance is badly broken:  the `crystal' formed in spacetime is made from worldlines and although these do displace in space directions they are incompressible in the time direction. Compared to 3D crystals this `spacetime crystal' is singularly anisotropic; as realized by Nelson and coworkers its only sibling in the classical world is the Abrikosov lattice formed from flux lines in superconductors \cite{MarchettiNelson90}. 

Nevertheless, one can take the bold step to postulate the existence  of a Lorentz-invariant `world crystal' corresponding to a spacetime as an isotropic elastic medium. This is characterized by stress tensors that are symmetric in spacetime labels and it is very easy to demonstrate that the nematic-type quantum liquid crystal which is dual to this medium is the vacuum of strictly {\em linearized} gravity where the disclinations are the exclusive sources of curvature~\cite{KleinertZaanen04,ZaanenBeekman12}. As an intriguing consequence, since gravity is incompressible in 3D there are no massless propagating modes in 2+1D while in 3+1D one just finds the two spin-2 gravitons.  As we will see, this is very distinct from the mode spectrum of the real life non-relativistic quantum nematics. 

However, a significant simplification is associated with the fact that the symmetry breaking only affects the 2D space in the non-relativistic case.  In the soft-matter tradition it is well understood that the classification of nematic-type orders is in terms of the point groups, and in 2D this is a rather simple affair given that the 2D rotational groups are all Abelian.  We will discuss the precise nature of these orientational order parameters in Sec.~\ref{sec:Order parameters for 2+1-dimensional nematics}, actually making the case that these are most conveniently approached in the language of discrete $O(2)/\integers_N$-gauge theory.  In three space dimensions hell breaks loose since the point groups turn non-Abelian with the effect that the order parameters acquire a highly non-trivial tensor structure. This can be also brought under control employing discrete non-Abelian gauge theory; this will be subject of a separate publication.  In fact, in the duality construction we will close our eyes for the intricacies associated with particular nematic symmetries and concentrate instead of the maximally symmetric  `spherical cow' cases descending from isotropic elasticity; the lower symmetry cases just invoke adding details like anisotropic velocities which do not play any interesting role in the duality {\em per se}. 

There is yet another aspect that is special to the non-relativistic {\em quantum} liquid crystals, which in turn plays a crucial role for their physics.  In the classical setting, dynamics does not affect the thermodynamics, but this is different in the quantum incarnation since quantum physics `entangles' space and time. From the study of the motions of classical dislocations in solids it is well known that these are subjected to a special principle rooted in topology: dislocations can move {\em ballistically} in the direction of their Burgers vector (called {\em glide motion}) while in the absence of interstitial and/or substitutional defects {\em climb motion} perpendicular to the Burgers vector is completely impeded.  In addition, the inertial mass associated with the climb motion is identical to the mass of the constituents of the solid, but dislocations do not fall in the gravitational field of the earth, the reason being that the dislocation ``does not carry volume''. It can only accelerate by applying shear forces to the medium.   This {\em glide principle} will play a remarkable role in the quantum problem. As we will see that it is responsible for the capacity of the zero-temperature quantum fluid, that is ultimately a dislocation condensate, to propagate sound. 

\subsection{The warped dual view on quantum liquid crystals}\label{subsec:The warped dual view on quantum liquid crystals}

This completes the exposition of the basic ingredients for the {\em dual quantum theory of elasticity} for bosonic matter in 2+1D.  These in turn form the building blocks for a quantum field theory with remarkable mathematical qualities. By just blind computation one obtains results that shed a different, often surprising light on a seemingly very classical physics topic. This will be the substance of the remainder of this review but to whet the appetite of the reader let us present a list of these surprises, roughly in the same order as they appear below.
\begin{enumerate}
\item  
{\em Phonons are gauge bosons.}
The theory of quantum elasticity is just the 19th century theory of elasticity with a kinetic energy term added to its Lagrangian. This is nothing else than the long-wavelength theory of acoustic phonons. Using Kleinert's way of employing the stress--strain duality~\cite{Kleinert89b,Kleinert08} we show how to rewrite this in terms of  $U(1)$-gauge fields. In contrast to textbook wisdoms, phonons can be regarded as `photons' when the question is asked how the medium propagates forces. The elastic medium is richer than the vacuum of electromagnetism in the regard that the crystal directions enter as `flavors' in the gauge theory. These {\em stress photons} are sourced by external shear and compressional stresses but also by the dislocations, which in turn have the same status as charged particles in electromagnetism, with the same complication that they carry the {\em Burgers vector charge} as a 'flavor' (Sec.~\ref{sec:Dual elasticity}). 
\item
{\em The disordered solid is a stress superconductor.}
Since individual dislocations are `charged particles' interacting via `stress photons', when the dislocations proliferate and condense the resulting quantum fluid can be viewed as a {\em stress superconductor}.  Shear stress is the rigidity exclusively associated with translational symmetry breaking, and it is this form of stress that falls prey to the analogue of the Meissner effect. Shear stress is ``expelled from the liquid'' in the same way that magnetic fields cannot enter the superconductor.  One can identify a {\em shear penetration depth} having the meaning that at short distances the medium remembers its elastic nature with the effect that shear forces propagate. At a length scale larger than the average distance between (anti-)dislocations, shear stress becomes perfectly `screened' by the response of the Bose-condensed dislocations  (Sec.~\ref{sec:Quantum nematic}).
\item
{\em The disordered solid is a real superfluid.}
The fact that dislocations ``do not occupy volume'' is in the duality encapsulated by the glide principle. After incorporating this glide constraint, one finds that the dislocation condensate decouples from the purely compressional stress photons: the quantum liquid carries massless sound, which in turn can be viewed as the longitudinal phonon of the disordered crystal that ``lost its shear contributions'' (Sec.~\ref{sec:Quantum nematic}). The mechanism involves a mode coupling between the longitudinal phonon and a condensate mode having surprising ramifications for experiment.  Besides the specialties associated with the orientational symmetry breaking, we find a bosonic quantum fluid that just carries sound. By studying the response of its EM charged version to external magnetic fields  (item 9.) we prove that this liquid is actually a superfluid! At first sight this might sound alarming since we have constructed it from ingredients (phonons, dislocations) that have no knowledge of the constituent bosonic particles forming the crystal. It seems to violate the principle that superfluidity is governed by the off-diagonal long range order (ODLRO) of the (constituent) bosonic fields. This is less dramatic than it appears at first sight: the braiding of the dislocations will give rise to the braiding of the worldlines of the constituent bosons.  In fact, it amounts to a reformulation of the usual ODLRO principle to the limit of maximal ‘crystalline correlations’ in the fluid: ``a bosonic crystal that has lost its shear rigidity is a superfluid''. 
\item
{\em The rotational Goldstone mode deconfines in the quantum nematic.}
By insisting that the disclinations stay massive while the dislocations proliferate, the quantum liquids we describe are automatically quantum liquid crystals, where we just learned that these are actually superfluids in so far their `liquid' aspect is concerned. Given that the isotropy of space is still spontaneously broken there should be a rigidity present, including the associated Goldstone boson: the `rotational phonon' and the reactive response to {\em torque stress}.  But now we face a next conundrum: the phonons of the crystal are purely translational modes, how does this rotational sector ``appear out of thin air'' when the shear rigidity is destroyed? Why is there no separate mode in the crystal associated with the rotational symmetry breaking? The reason is that translations and rotations are in semidirect relation in the space groups describing the crystal: one cannot break translations without breaking rotations. To do full justice to this symmetry principle, in Sec.~\ref{subsec:Torque stress gauge field} we introduce  a more fanciful `dynamical Ehrenfest constraint' duality construction, where the role of rotations and the associated disclination sources is made explicit already in the crystal. We find, elegantly, that in this description the {\em torque photon} as well as the associated disclination sources are literally {\em confined} in the crystal, in the same physical (although not mathematical) sense as quarks are confined in QCD. When the dislocations condense the torque stresses and the disclinations {\em deconfine} becoming the objects encapsulating the `rotational physics' of the quantum nematics~(Sec.~\ref{subsec:torque stress nematic}). As compared to the classical (finite-temperature) nematics there is one striking difference. It is well known that the rotational Goldstone mode couples to the dissipative circulation in the normal hydrodynamical fluid, and is overdamped. The superfluid is however irrotational, thereby protecting the rotational Goldstone mode as a propagating excitation.
\item
{\em Partial translational melting is a quantum smectic.}
The dislocation condensate consists in essence of $D$ $U(1)$-fields in $D$ space dimensions. By preferentially condensing one Burgers orientation in the dislocation condensate we construct the {\em quantum smectic} in Sec.~\ref{sec:Quantum smectic}. Although one anticipates the conventional picture of ``stacks of liquid layers'', the zero temperature quantum case defies such intuitions. It cannot be viewed as simply a `solid $\times$ liquid', and the transverse and longitudinal characters mix in the mode spectrum, depending on the interrogation angle of the linear response (see Fig.~\ref{fig:smectic spectral functions special angles} below). Surprisingly, the `most isotropic' response is found with field propagating at 45$^\circ$ to the layers. Nevertheless, for fields propagating along the layers (in the `liquid direction'), the duality construction flawlessly reproduces the transverse ``undulation mode'' in the solid direction with a quadratic ($\omega \propto q^2$) dispersion, that is known in classical smectics~\cite{DeGennesProst95,ChaikinLubensky00}.
\item
{\em Order parameter theory of 2+1-dimensional nematics.}
In Sec.~\ref{sec:Order parameters for 2+1-dimensional nematics} we develop a completely general theory of order parameters arising due to broken rotational symmetry in 2+1 dimensions. The conventional `uniaxial' nematic is just one example of a host of possible $p$-atic orders, which derive directly from the space group of the crystal that undergoes dislocation-mediated melting. This is extended to a gauge theory formulation, where the topological defects in the nematic phase (disclinations) are represented by $\mathbb{Z}_N$-fluxes. At zero temperature, this leads to the prediction of a new phase, the $\mathbb{Z}_N$-deconfined phase, where these gauge fluxes are frozen but not condensed.

\item
{\em Transverse phonons become massive shear modes in the quantum liquid crystal.}
Up to this point we have highlighted universal features of the long-wavelength limit. These are in fact not depending on the assumption of maximal crystalline correlations intrinsic to the duality description. By adiabatic continuation they are smoothly connected to the outcomes of the weakly-interacting, gaseous description. The difference between the two descriptions becomes manifest considering the spectrum of {\em finite-energy} excitations. The dual stress superconductor description yields a plethora of propagating {\em massive} modes in the quantum liquid crystals which depend critically on the assumption that the crystalline correlation length/shear penetration depth is large compared to the lattice constant.  Their origin is easily understood in terms of the dual relativistic superconductor description. In the Higgs phase of a real superconductor the photon becomes massive; in the stress superconductor the stress photons (the phonons) become massive, and  they propagate shear forces only over a short range.  In other words, these new modes in the liquid phases correspond to {\em massive shear photons}.  A simple example is the transverse phonons of the crystal that just acquire a `Higgs mass' in the liquid. However,  the stress superconductor is more intricate than just the Abelian-Higgs condensate. The case in point is the way that the longitudinal phonon of the crystal, sensitive to both the compression and the shear modulus, turns into the sound mode of the superfluid, rendering the sound mode of the quantum nematic to be of a purely compressional nature. The existence of these massive  modes is critically dependent on the assumption that {\em interstitials} (the constituent bosons) are absent. When the crystalline correlation length shrinks towards the lattice constant these modes will get damped to eventually disappear in the gaseous limit where they are completely absent at small momenta. We believe that the roton of e.g. $^4$He can be viewed as a remnant of such a shear photon in the regime where the crystalline correlation length has become a few lattice constants. 
\item
{\em Elasticity and the charged bosonic Wigner crystal.}
Up to here we have dealt with electromagnetically neutral systems, but as we will show in Secs.~\ref{sec:Dual elasticity of charged media},\ref{sec:Electromagnetic observables} it is straightforward to extend the description to electrically charged systems. The first step is to derive the elastic theory of the charged {\em bosonic Wigner crystal}. We shall obtain the spectrum of (coupled) stress and electromagnetic (EM) photons, considering the coupling of the 2+1D matter to 2+1D electrodynamics. The unpinned Wigner crystal behaves as a perfect conductor, where the plasmon now propagates with the longitudinal phonon velocity. In the  transverse optical response however, next to the expected plasmon there is also a weak massless mode, with quadratic dispersion at low energies, extrapolating to the transverse phonon at high energies.
\item
{\em The charged quantum nematic shows the Meissner effect.}
Since a dislocation does not carry volume it does not carry charge either. Accordingly, in the dislocation condensate the longitudinal EM response is characterized by a `true' plasmon, a sound wave that has acquired a plasmon energy. The surprise is in the transverse EM response: the EM photon acquires a mass and the system expels magnetic fields  according to the  Meissner effect of a superconductor. This proves the earlier assertion that the bosonic quantum nematic, described in terms of a dual stress superconductor, is indeed a genuine superfluid that turns into a superconductor when it is gauged with electromagnetic fields. The mechanism is fascinating: the Meissner effect is in a way hiding in the Wigner crystal where it is killed by a term arising from  the massless shear photons. When the latter acquire a mass this compensation is no longer complete with the outcome that EM photons are expelled (Sec.~\ref{subsec:Superconductivity and the electromagnetism of the quantum nematic}).
\item
{\em The charged quantum smectic shows strongly anisotropic properties.}
The quantum nematic is just an isotropic superconductor but the charged quantum smectic is equally intriguing as the neutral counterpart: in the `liquid' direction it is characterized by a finite superfluid density and the capacity to expel EM fields, but at an angle of 45$^{\circ}$ with respect to the `solid' direction its EM response is identical to that of the Wigner crystal. For momenta along the `solid' direction, there is a massive transverse plasma polariton and massive mode from arising from mode coupling with the shear photon (i.e. phonon). At intermediate angles, the plasma polariton persists but the spectral weights of the coupled modes interpolate between the `magic' angles with massless and massive modes at finite momenta. For transverse fields propagating at finite momenta near but not exactly in the liquid direction, magnetic screening at \emph{finite} frequencies (the skin effect) is enhanced with respect to the Wigner crystal (Sec.~\ref{subsec:Electromagnetism of the charged quantum smectic}).
\item
{\em The massive shear mode is detectable by finite-momentum spectroscopy.}
The charged case is the one of greatest potential relevance to the physics of the electron systems. Given the very small mass of electrons the only way to exert external forces on the system probing the physical properties of the liquid crystalline states is by its electromagnetic responses. A key prediction is that the massive shear photons are in principle observable by electromagnetic means, albeit with the practical difficulty that these carry finite optical weight only at finite momenta. This runs into the usual difficulty that because of the large mismatch between the `material velocity' and the speed of light the experimentalists can only easily interrogate the zero-momentum limit. However, using modern techniques, such as electron energy-loss spectroscopy or inelastic X-ray scattering, this regime becomes accessible and in Sec.~\ref{subsec:Superconductivity and the electromagnetism of the quantum nematic} we will present a very precise prediction regarding the massive shear photon that should become visible in the longitudinal EM channel. 
\item
{\em Directly probing the liquid crystalline order parameter.}
Last but not least, is it possible to measure the order parameters of the quantum-liquid crystals directly using electromagnetic means? Given the presence of the pinning energy in the real solids, an even more pressing issue is whether there is any way to couple into the rotational Goldstone boson of the nematic that is expected to be characterized in any case by a finite `anisotropy gap' caused by the pinning. In Sec.~\ref{subsec:Superconductivity and the electromagnetism of the quantum nematic} we shall see that the rotational Goldstone mode leaves its signature in the transverse conductivity at low but non-zero momenta.
\end{enumerate}

\subsection{Quantum liquid crystals: the full landscape}\label{subsec:Quantum liquid crystals: the full landscape}

We will now shortly discuss the relation of our `maximally-correlated' quantum liquid crystals to other manifestations of quantum liquid-crystalline order.

Electrons in solids are a natural theater to look for quantum fluids such as our present quantum liquid crystals. The greatly complicating circumstance is that 
electrons are fermions. One is facing a monumental obstruction attempting to describe systems of interacting fermions at any finite density in general mathematical terms: the {\em fermion sign} problem.  In constructing the theory we employ the general methodology of quantum field theory: 
 mapping the quantum problem onto an equivalent statistical physics problem in Euclidean spacetime, to then mobilize the powerful probabilistic machinery of statistical physics to solve the problem~\cite{Sachdev99}. As we emphasized before, we just generalize the classical KTNHY story to 2+1 Euclidean dimensions and after 
Wick rotation a quantum story unfolds. But this does not work for fermions since the fermionic path integral does not lead to valid statistical physics, since fermion signs correspond to ``negative probabilities''.  It is just not known how to generalize deeply non-perturbative operations like the weak--strong
duality behind the present theory to a fermionic setting at any finite density.

To circumvent this trouble we assume that the electrons are subjected to very strong interactions that first bind them in `local pairs' which subsequently form a tightly-bound crystal that can only melt by topological means. Given the sign problem, the only other option is to start from the opposite end: depart from the non-interacting Fermi gas to find out what happens when interactions are switched on. There is surely interesting physics to be found here which is however rather tangential to the theme of this review. For completeness let us present here a short sketch of these other approaches (see also Ref.~\cite{FradkinEtAl10}).

How to describe nematic order departing from a free Fermi gas? The object to work with is the Fermi surface, an isotropic sphere in momentum space (when working in the Galilean continuum).  When the Fermi surface turns into an ellipsoid, the isotropy is lost and symmetry-wise it corresponds to a uniaxial nematic deformation. This can be accomplished by switching on an electron--electron interaction of a quadrupolar nature (associated with the Landau Fermi liquid parameter $F_2$)~\cite{OganesyanKivelsonFradkin01}; the surprise is that this is perturbatively unstable! Upon inspecting the leading order perturbative corrections one discovers an extremely bad IR divergence. As it turns out, the rotational Goldstone mode does not decouple from the quasiparticle excitations in the deep infrared and just as in the classical nematic it is overdamped.  The quasiparticles pick up an IR divergence as well. This is perhaps the most profound problem in this field: although the interactions are weak the nematic Fermi fluid cannot be a Fermi liquid, but the fermion sign problem is in the way of finding out what it is instead!

In real electron systems the anisotropy of the underlying lattice will render the rotational symmetry to become discrete: in pnictides and cuprates one is typically dealing with a square lattice with a fourfold ($C_4$) symmetry axis that turns into a twofold ($C_2$) axis in the nematic state. This anisotropy gap of this ``Ising nematic''  protects the physical systems from this divergence. It is still debated whether the nematic order found in pnictides~\cite{FernandesChubukovSchmalian14} is of this `near Fermi-liquid' kind or rather of a strongly-coupled ``spin nematic'' nature, where also the complications of orbital degeneracy~\cite{KruegerEtAl09} may play a crucial role. 

One can subsequently ask the question what happens in such a metallic nematic when the order disappears at a zero-temperature quantum phase transition.  This is typically approached from the Hertz--Millis perspective~\cite{Hertz76, Millis93}. One assumes that the order is governed by a bosonic order-parameter theory such that the quantum phase transition is equivalent to a thermal phase transition in Euclidean space time~\cite{Sachdev99}. However, the critical order-parameter fluctuations are perturbatively coupled to the electron--hole excitations around the Fermi surface of a free fermion metal.  The latter can in turn give rise to IR singularities changing the nature of the universality class: the (Ising) nematic transition in 2+1D is a case in point~\cite{MetlitskiSachdev10}.

The most recent results are associated with a special model characterized by sign cancellations, making it possible to unleash the powers of Monte Carlo simulations; these indicate that the perturbative assumptions wired in the Hertz--Millis approach break down, instead showing a non-Fermi-liquid behavior in the metallic state and a strong tendency towards superconductivity at the quantum critical point~\cite{SchattnerEtAl16, LedererEtAl16}. 

There is yet another series of ideas that are more closely related to the strong-coupling bosonic perspective of this review. One can arrive at a notion of a quantum smectic which is quite different from the quantum smectics we will highlight, identified by Emery \emph{et al.}~\cite{EmeryFradkinKivelsonLubensky00}. One starts out from static stripes assuming that metallic 1+1D Luttinger liquids are formed on every stripe, which are subsequently coupled into a 2+1D system. One can now demonstrate that for particular forward-scattering-dominated interactions the inter-stripe interactions become {\em irrelevant} with the effect that the system continues to behave like a Luttinger liquid in the direction parallel to the stripes, while becoming, in the scaling limit, insulating in the perpendicular direction. These ideas were taken up and further elaborated on in the context of quantum Hall (QH) physics. Upon going to high filling fractions in the QH two-dimensional electron gases, it is easy to show that at some point one will form the QH ``stripes''~\cite{FoglerKoulakovShklovskii96, MoessnerChalker96}. These are microscopically very different from the doped-Mott-insulator stripes: they consist of linear arrays of filling fraction $n$ and $n+1$ incompressible QH fluids, while chiral edge states that propagate on the boundaries are quite literal incarnations of the Luttinger liquids of Kivelson \emph{et al.}~\cite{KivelsonFradkinEmery98}. Besides this `edge mode' quantum smectic, one can also contemplate it be subjected to dislocation melting~\cite{FradkinKivelson99} and there is experimental evidence for the formation of nematic states as well, observed in terms of anisotropic QH transport~\cite{LillyEtAl99, XiaEtAl11}.  
The transport properties in such quantum nematics are tied to the chiral QH edge states and, for the topologically-ordered quantum liquid crystal QH phases, these are governed by Chern--Simons topological field theory. This adds an extra layer of physics since the associated topological order communicates with the nematic order parameter~\cite{Balents96, MulliganNayakKachru10, MulliganNayakKachru11, MaciejkoEtAl13, YouChoFradkin14}. The general hydrodynamics of QH liquid crystals is in turn deeply rooted in the effectively non-commutative geometry associated with 2+1D matter in magnetic fields~\cite{RadzihovskyDorsey02}.

Finally, there is a set of new topological-order phenomena associated with the effects of topological crystal melting dealing with more complicated ``intertwined" orders~\cite{ChoEtAl15}. Perhaps the simplest way to understand this is to see how topological order associated with the deconfining states of discrete gauge theories can arise as an emergent phenomenon related to stripes. This was born in the context of the magnetic stripes, where it was named ``stripe fractionalization''~\cite{ZaanenEtAl01, ZaanenNussinov02, ZhangDemlerSachdev02}. 
The charge stripes are at the same time domain walls in the antiferromagnet. Imagine now that the charge order is subjected to dislocation quantum melting. Insisting that the antiferromagnetic order persists, the magnet domain walls should stay intact and as a consequence only {\em double} dislocations proliferate. Since the dislocations are Bose condensed, the staggered antiferromagnetic order parameter becomes identified with its opposite: the system turns into a {\em spin nematic} which breaks magnetic rotation symmetry but has vanishing staggered magnetization (``headless arrows''). One can now unbind a double-charge dislocation pair into an isolated dislocation but this implies a frustration in the spin background that is identified as the {\em $\pi$-disclination of the spin nematic}. When these proliferate as well, one enters a quantum-disordered magnetic phase. Apart from this it is also possible that the antiferromagnet quantum disorders by itself but since locally the antiferromagnet correlations are still strong, the double dislocations continue to be bound, and this is somehow a different state than the fully disordered one. As it turns out, this is precisely described by ``$O(3)/\mathbb{Z}_2$" lattice gauge theory, where the spin nematic and fully disordered phase correspond to the Higgs and the confining phase of the $\mathbb{Z}_2$ gauge theory. The topologically ordered deconfining phase of the Ising gauge theory has now the simple interpretation of the 
otherwise featureless condensate of the double charge dislocations! Putting this on a torus, this phase would boast an {\em either even or odd} number of charge stripes when traversing the circumferences of the torus, although one cannot hope to detect the stripes since they form a nematic dislocation condensate.  

Recently evidence emerged that the {\em superconductivity} in static stripe systems may behave similar to  the antiferromagnetism, with the phase of the order  parameter reversing from stripe to stripe: the ``pair density  waves''~\cite{FradkinKivelsonTranquada15, HamidianEtAl16}. 
 One can imagine that similar ``stripe fractionalization'' topological orders may occur, now involving the superconducting order parameter~\cite{BergFradkinKivelson09}. The jury is still out on whether this is of relevance to cuprates~\cite{MrossSenthil12, MrossSenthil15}. Another context where this could be relevant are the `unbalanced condensates' formed by cold atoms. Conventionally one expects here the so-called FFLO states~\cite{FuldeFerrel64, LarkinOvchinnokov65} which are symmetry-wise of the same kind as pair density waves. One can envisage in this context similarly partially-melted phases~\cite{RadzihovskyViswanath09}.

\subsection{Organization of this paper}

This is a review paper and we have therefore striven to present a reasonably self-contained exposition. Understanding of basic condensed matter physics (phase transitions, Green's functions) as well as field theory and path integrals, is assumed. This article supersedes our earlier works~\cite{ZaanenNussinovMukhin04,KleinertZaanen04,Cvetkovic06,CvetkovicNussinovZaanen06,CvetkovicZaanen06a,CvetkovicZaanen06b,CvetkovicNussinovMukhinZaanen08,ZaanenBeekman12,BeekmanWuCvetkovicZaanen13,LiuEtAl15} wherever results are contradictory. 
We warm up in Sec.~\ref{sec:XY-duality} by treating a simpler and well-studied problem: vortex--boson or Abelian-Higgs duality for interacting bosons. Here the principles of the duality construction are laid out, and we will refer to it often in the remainder of the text. Classical elasticity is restated in field theory language in Sec.~\ref{sec:Field-theoretic elasticity}, up to the derivation of the phonon propagators. The topological defects of spatially ordered systems, dislocations and disclinations, are introduced in Sec.~\ref{sec:Topological defects in solids}, including field-theoretic {\em defect currents} leading to a compact form for the glide constraint for dislocations. Up to this point, nothing essentially new has been offered. In Sec.~\ref{sec:Order parameters for 2+1-dimensional nematics} an independent, gauge-theoretic treatment of the order parameters of the nematic--isotropic liquid is included, which also contains a justification for our use of the word {\em nematic} for any state with complete translational symmetry but broken rotational symmetry. The remainder of the review concerns the main topic: dislocation-mediated quantum melting of two-dimensional crystals. In Sec.~\ref{sec:Dual elasticity} we perform the duality transformation by going from the strain variables of Sec.~\ref{sec:Field-theoretic elasticity} to stress variables expressed as gauge fields. Here we rederive the phonon propagators in this dual language. Higher-order elasticity is considered as well, leading to the introduction of torque stress gauge fields. Sec.~\ref{sec:Dynamics of disorder fields} deals with the condensate of dislocation worldlines, culminating in a Higgs term for the dual stress gauge fields. This is then used to derive the hydrodynamics of the 
quantum nematic in Sec.~\ref{sec:Quantum nematic} and the quantum smectic in Sec.~\ref{sec:Quantum smectic}. These are the main results presented in this review. In the quantum nematic, where translational symmetry is restored, a Goldstone mode related to rotational symmetry breaking emerges. This `deconfinement' is discussed  as the release of a constraint  in the latter half of Sec.~\ref{sec:Quantum nematic}. Finally, it is straightforward to incorporate electrically charged media into the melting program, which is the topic of Sec.~\ref{sec:Dual elasticity of charged media}. In Sec.~\ref{sec:Electromagnetic observables} we derive and compare several experimental signatures such as conductivity, spectral functions and the electron energy-loss function, as well as the Meissner effect, indicating superconductivity. Together with a summary, an outlook for future developments is presented in Sec.~\ref{sec:Conclusions}. \ref{sec:Fourier space coordinate systems} details the Fourier space coordinate systems which are employed throughout this text, while \ref{sec:Euclidean electromagnetism conventions}, \ref{sec:Dual Kubo formula} contain details about derivations in electrically charged elastic media.

\subsection{Conventions and notation}\label{subsec:Conventions and notation}
 \begin{itemize} 
\item 
 For the temporal components we use the following notation. Italic $t$ denotes real time; Greek $\tau$ denotes imaginary time, and fraktur $\ft$ denotes an imaginary temporal component with dimensions of length, rescaled with a factor of (shear) velocity $c$ via $\ft = c \tau$ (cf. Eqs.~\eqref{eq:superfluid Lagrangian}, \eqref{eq:imaginary time shear velocity rescaling}). Most of the calculations are performed in imaginary time $\tau = \ti t$, for which the partition function $Z = \exp(iS) \to \exp(-S_\mathrm{E})$, where $S_\mathrm{E} = \int \td \tau \td^D x\; \mathcal{L}_\mathrm{E}$ is the Euclidean action. Note that the Euclidean Lagrangian $\mathcal{L}_\mathrm{E}$ differs by a sign compared to the real-time Lagrangian in the non-kinetic components due to the choice $\tau = \ti t$. We will suppress the subscript $_\mathrm{E}$ when there is no confusion. Consequently there is no distinction between contravariant and covariant (upper and lower) indices.
\item 
 Greek indices $\mu,\nu,\ldots$ refer to spacetime indices, while Roman indices $m,n,\ldots$ refer to spatial indices only. Spatial Fourier directions (see below) use capital indices $E,F,\ldots$. For quantum elasticity fields in Euclidean time, like the stress tensor $\sigma_\mu^a$, the lower index is the space-time vector index while the upper index is a purely spatial index related to the Burgers charge. This is important since these tensors are not fully symmetric unlike the `classical' stress tensor $\sigma_{mn}$. However, there is no essential difference between upper and lower indices since the metric is Euclidean.
\item
 The fully antisymmetric symbols in 3+1, 3, 2+1 and 2 dimensions respectively obey $\epsilon_{\ft xyz} = \epsilon_{xyz} = \epsilon_{\ft xy} = \epsilon_{xy} = +1$.
 \item The Fourier transforms are:
 \begin{align}
  f(\mathbf{q},\omega) &= \frac{1}{\sqrt{(2\pi)^{D+1}}} \int \td^Dx\, \td t\, f(\mathbf{x},t) \te^{\ti (\mathbf{q}\cdot \mathbf{x} - \omega t)}\nonumber\\
  f(\mathbf{x},t) &= \frac{1}{\sqrt{(2\pi)^{D+1}}} \int \td^D q\, \td \omega \, f(\mathbf{q},\omega) \te^{-\ti (\mathbf{q}\cdot \mathbf{x} - \omega t)}\nonumber
 \end{align}
\item
 In Fourier space we use Matsubara frequencies $\omega_n$. Explicitly $\partial_\tau \to \ti \omega_n$ and $\partial_m \to \ti q_m$. The momentum is $p_\mu = ( \frac{1}{c}\omega_n , \mathbf{q})$ and $p = |p_\mu| = \sqrt{\frac{1}{c^2}\omega_n^2 + q^2}$, where $c$ is an appropriate velocity to be defined in the text. The analytic continuation is $-\ti \omega_n \to \omega + \ti \delta$ where $\delta \ll 1$ is a convergence factor, such that the Fourier transforms are calculated using $\te^{\ti \omega t - \ti \mathbf{q}\cdot \mathbf{x} } = \te^{- \ti \omega_n \tau - \ti \mathbf{q}\cdot \mathbf{x}}$.
 \item
In position space, all fields will be real-valued; in Fourier space we therefore use the notation $A^\dagger(p) = A(-p)$.
\item
Next to the standard $(\ft,x,y)$-coordinate system, we employ two other coordinate systems in Fourier-Matsubara space. The $(\ft,\tL,\tT)$-system has spatial components parallel ($\tL$) and orthogonal ($\tT$) to the spatial momentum $\mathbf{q}$. The $(0,+1,-1)$-system has components parallel ($0$) and orthogonal ($+1$, $-1$) to the spacetime momentum $p_\mu$, where $-1 \parallel \tT$. These systems are detailed in \ref{sec:Fourier space coordinate systems}, which we encourage the reader to study before repeating any calculations. We shall often employ longitudinal and transverse projectors ($P^2 = P$) for spatial coordinates,
\begin{align}
 P^\tL_{mn}(\mathbf{q})&= \frac{q_m q_n}{q^2} \label{eq:longitudinal projector}\\
 P^\tT_{mn}(\mathbf{q}) &= \delta_{mn} - \frac{q_m q_n}{q^2} \label{eq:transverse projector}
\end{align}
Obviously $P^\tL + P^\tT = \mathbb{I}$. 
\item Planck's constant is set $\hbar \equiv 1$, energies are equivalent to frequencies $\omega$ via $E = \hbar \omega \to \omega$.
\end{itemize}

\section{Vortex--boson duality}\label{sec:XY-duality}
There are two reasons to start with a different problem: vortex--boson or Abelian-Higgs or $XY$-duality in the superfluid--Bose-Mott insulator phase transition~\cite{FisherEtAl89,FisherLee89,LeeFisher91,Kleinert89a,NguyenSudbo99,HoveSudbo00,HerbutTessanovic96}. First, it is an easy warm up for the more complicated physics and dualities relevant to elasticity later. And second, it is the simplest duality that yet contains the most important ingredients we need, namely the nature of the condensate of topological defects and the status of dual gauge fields. Because the interpretation will, from time to time, differ from that found in some of the literature, we advise even researchers well-versed in this field to read through this section.

The idea behind vortex--boson duality is the following. An ordered state is eventually destroyed by the proliferation (unbinding) of vortices or topological defects, which in two  spatial  dimensions are point-like, and quantum mechanically it is the Bose--Einstein condensation thereof. The defects are topological singularities in the order parameter field and disturbances in the order parameter field are communicated by the Goldstone modes. This neatly comes together in the dual formulation, where it turns out that the Goldstone modes are in fact dual gauge fields, that precisely mediate the interaction between topological defect sources. Thus we shall adopt the view that the topological defects are ordinary bosonic particles that interact with each other by exchanging gauge fields, like electric charges interact by exchanging photons. Next, the proliferation of topological defects introduces a collective bosonic condensate field to which the gauge fields are minimally coupled. This is exactly as in a superconductor where photons are minimally coupled to the Cooper-pair condensate. We know what will happen: the photons acquire a mass (gap) through the Anderson--Higgs mechanism such that magnetic fields are expelled. Thus the dual gauge fields correspond to massless Goldstone modes in the ordered phase and massive excitations in the disordered phase.

This general framework is virtually literally realized in the superfluid--Bose-Mott insulator phase transition, which we will detail in this section. It will also form the basis of the dislocation-mediated melting transitions in the remainder of the paper. More details can be found in Refs.~\cite{CvetkovicZaanen06a,BeekmanSadriZaanen11}.

\subsection{Bose--Hubbard model}
The Bose-Hubbard model~\cite{FisherEtAl89} describes bosons, created by operators $\hat{b}^\dagger_i$ on lattice sites $i$ with commutation relations $[\hat{b}_i,\hat{b}^\dagger_j] = \delta_{ij}$, with Hamiltonian
\begin{equation}
 \mathcal{H} = -t \sum_{\langle ij \rangle} (\hat{b}^\dagger_i \hat{b}_j + \hat{b}_i \hat{b}^\dagger_j) + U \sum_i (\hat{b}^\dagger_i \hat{b}_i)^2.
\end{equation}
Here the first sum is over nearest-neighbor sites, $t$ is the hopping parameter and $U$ is an on-site repulsion term. A chemical potential was fixed to ensure a large, integer average number $\bar{n}$ of bosons per site such that $\bar{n} +1 \approx \bar{n}$. Now we can write $\hat{b}_i = \sqrt{n_i} \te^{\ti \varphi_i}$ where $n_i = \hat{b}^\dagger_i \hat{b}_i$ is the number operator and $\varphi_i$ is the phase operator. They obey $[ \varphi_i, n_j ] = \ti \delta_{ij}$ and are canonically conjugate. Then the Hamiltonian reads
\begin{equation}\label{eq:Bose-Hubbard phase Hamiltonian}
\mathcal{H} = -t \sum_{\langle ij \rangle} \cos ( \varphi_i - \varphi_j ) + U \sum_i n_i^2.
\end{equation}
Here we have rescaled $t$ with the number of bosons per site $\bar{n}$. The physics of this model is as follows. For $t \gg U$ we have a well-defined phase variable nearly constant in space, and the first term denotes the fluctuations from that phase. This is a superfluid with a spontaneously broken $U(1)$-symmetry and phase rigidity. In this limit Eq.~\eqref{eq:Bose-Hubbard phase Hamiltonian} is also called the $XY$ or quantum rotor model. For $U \gg t$ the repulsion dominates, and the boson number is fixed at each site to the average value, with a large gap to the particle and hole excitations. This is the Mott insulator of bosonic particles (there are no spin degrees of freedom). The parameter $\widetilde{g} \propto U/t$ tunes the phase transition. In the following, we start out deep in the weak-coupling limit $\widetilde{g} \ll 1$, the superfluid phase, where we can neglect the quantized bosonic degrees of freedom in the second term. Furthermore we take the continuum limit, and approximate $\cos(\nabla \varphi) \sim (1-  \tfrac{1}{2} (\nabla \varphi)^2)$ since deviations from the preferred phase value are costly. The continuum limit Hamiltonian of the system is now
\begin{equation}
\mathcal{H} = \frac{t}{2a^{D-2}} \int \td^D x \left[ (\nabla \varphi(x))^2 + \widetilde{g} n(x)^2 \right],
\end{equation}
where $a$ is the lattice constant and a constant term has been dropped. Following from the commutation relation, the number excitations $n(x)\sim n_{i}$ are canonical momenta conjugate to $\varphi(x)$. The quantum partition sum $\mathcal{Z} = \mathrm{Tr }(\te^{-\beta \mathcal{H}})$ of this Hamiltonian becomes the Euclidean imaginary time path integral $\mathcal{Z}=\int \mathcal{D}
\varphi \; \te^{-\mathcal{S}_{\rm E}}$ with the following action~\cite{Sachdev99}, 
\begin{equation}
 \mathcal{S}_{\rm E} = \tfrac{1}{2g}  \int_0^{\beta} \td^D x \td \tau \ \tfrac{1}{c_{\mathrm{ph}}^2} (\partial_\tau \varphi)^2 + (\nabla \varphi)^2, \label{eq: Sphase}
\end{equation}
where $\beta = 1/k_\mathrm{B} T$, $g \sim a^{D-2}/t$ and $c_\mathrm{ph} \sim t\sqrt{\widetilde{g}} a$ is the phase velocity of the superfluid. In this section and the rest of the review, we shall work exclusively in the imaginary time $\tau = \ti t$ path integral formalism as is standard for quantum phase transitions \cite{Sachdev99}. In this convention we have $\mathcal{L}_{\rm E} = \ti \pi_{\varphi}\partial_\tau \varphi + \mathcal{H}$, the canonical momenta being $\pi_{\varphi} \sim n$, in order to have the partition function $\mathcal{Z} = \int \mathcal{D}\varphi \exp( - \mathcal{S})$, where $\mathcal{S}$ is the Euclidean action and we have dropped the subscript $_\mathrm{E}$. While the Euclidean theory of the phase field $\varphi$ in Eq. \eqref{eq: Sphase} is emergently relativistic with `speed of light' $c_\mathrm{ph}$, we must keep in mind that the phase-disordered side in fact corresponds to the Bose-Mott insulator.

\subsection{The superfluid as a Coulomb gas of vortices}
When weakly-interacting bosons condense they form a superfluid, spontaneously breaking global internal $U(1)$-symmetry. The resulting Goldstone mode is the zero-sound mode of the superfluid, and it is a single free massless mode described by a scalar field, as derived from the Bose-Hubbard model above. The partition function for the  relativistic zero-sound mode is
\begin{align}
 \mathcal{Z} &= \int \mathcal{D} \varphi\ \te^{- \mathcal{S}},\\
 \mathcal{S} &= \int_0^\beta \td \tau \int \td^D x \ \mathcal{L},\\
 \mathcal{L} &= \frac{1}{2g} \big( \tfrac{1}{c_\mathrm{ph}^2}(\partial_\tau \varphi)^2 + (\partial_m \varphi)^2\big) \equiv \frac{1}{2g} \big( (\partial_\ft \varphi)^2 + (\partial_m \varphi)^2\big) \nonumber\\
 &\equiv \frac{1}{2g}(\partial_\mu \varphi)^2. \label{eq:superfluid Lagrangian}
\end{align}
Here $\beta = 1/k_\mathrm{B}T$ is the inverse temperature; by $c_\mathrm{ph}$ we denote the phase velocity of the superfluid condensate, and we have defined $\partial_\mu \equiv (\partial_\ft,\partial_m)$. All components have dimension of inverse length via $\partial_\ft \equiv \frac{1}{c_\mathrm{ph}} \partial_\tau$. Furthermore, $g$ is the coupling constant: for small $g$, deviations from the spontaneously chosen value of the superfluid phase $\varphi$ are very costly and thus strongly suppressed, stabilizing the superfluid regime. When $g$ is large the phase can fluctuate wildly. Now we need to remember that $\varphi$ is in fact a compact variable originating from $\cos(\nabla \varphi)$ in Eq.~\eqref{eq:Bose-Hubbard phase Hamiltonian}, and a change by $2\pi$ brings it back to its original value. Such windings of the phase variable correspond to {\em vortices} in the superfluid, and at large $g$ can be created easily. To incorporate the vortices, we must treat the phase $\varphi$ as a multivalued field \cite{Kleinert89a,Kleinert08,zee_book}, having both smooth and singular contributions
\begin{equation}\label{eq:multivalued phase field}
 \varphi(x) = \varphi_\mathrm{smooth}(x) + \varphi_\mathrm{sing}(x).
\end{equation}
Here $\varphi_\mathrm{sing}(x)$ obeys 
\begin{equation}\label{eq:phase field contour integral}
 \oint_{\partial \mathfrak{S}} \td x^\mu \partial_\mu \varphi_\mathrm{sing}(x) = 2\pi N,
\end{equation}
for any contour $\partial \mathfrak{S}$ that encloses the singularity, and $N$ is the winding number of the quantized vorticity. By definition, the same contour integral for the smooth part $\varphi_\mathrm{smooth}$ is zero. We shall restrict ourselves to 2+1 dimensions, the generalization to higher dimensions can be found in Ref. \cite{BeekmanSadriZaanen11}. For a vortex of winding number $N$ we have by Stokes' theorem
\begin{align}
 2\pi N &= \oint_{\partial \mathfrak{S}} \td x^\mu \partial_\mu \varphi = \int_{\mathfrak{S}} \td S^\lambda \epsilon_{\lambda \nu \mu} \partial_\nu \partial_\mu (\varphi_\mathrm{smooth} + \varphi_\mathrm{sing}) 
 = \int_{\mathfrak{S}} \td S^\lambda \epsilon_{\lambda \nu \mu} \partial_\nu \partial_\mu\varphi_\mathrm{sing} \equiv  \int_{\mathfrak{S}} \td S^\lambda J^\mathrm{V}_\lambda.\label{eq:superfluid vortex current}
\end{align}
Here $\mathfrak{S}$ is a surface pierced by the vortex, and $\partial \mathfrak{S}$ is its boundary. Note that derivatives operating on a multivalued field do not commute. We have defined the {\em vortex current}
\begin{equation}\label{eq:vortex current definition}
J^\mathrm{V}_\lambda (x) = \epsilon_{\lambda \nu \mu} \partial_\nu \partial_\mu\varphi_\mathrm{sing} (x).
\end{equation}
We can also conveniently write it as,
\begin{equation}\label{eq:vortex worldline}
J^\mathrm{V}_\lambda (x) = 2\pi \delta_\lambda (L,x).
\end{equation}
Here $\delta_\lambda (L,x)$ is a delta function that is non-zero only on the vortex worldline $L$ pointing in direction $\lambda$~\cite{Kleinert89a,Kleinert08}; $\int_\mathfrak{S} \td S^\lambda \delta_\lambda(L,x) = 1$, if and only if the worldline $L$ pierces the surface $\mathfrak{S}$ (see also Sec.~\ref{subsec:defect densities}). We shall from now on treat topological defects as `ordinary', bosonic particles in 2+1 dimensions encoded by these currents. They obey an integrability condition,
\begin{equation}
 \partial_\lambda J^\mathrm{V}_\lambda =  \partial_\lambda (\epsilon_{\lambda \nu \mu} \partial_\nu \partial_\mu) \varphi_\mathrm{sing} =0, \label{eq:vortex current conservation}
\end{equation}
implying that vortex worldlines cannot begin or end but must either appear as closed spacetime loops, or extend to the boundary of the system.

To see how vortices interact in the superfluid, we perform the duality operation, which is in fact a Legendre transformation to the relativistic canonical momentum associated with the field-theoretic velocity field $\partial_\mu \varphi$. The canonical momentum is
\begin{equation}\label{eq:superfluid canonical momentum}
 \xi_\mu = -\ti \frac{\partial \mathcal{S}}{\partial(\partial_\mu \varphi)} = -\ti\frac{1}{g} \partial_\mu \varphi. 
\end{equation}
The factor $-\ti$ is used in the imaginary time formalism and is consistent with Refs.~\cite{Kleinert89a,ZaanenNussinovMukhin04,CvetkovicZaanen06a,Cvetkovic06}. The field $\xi_\mu$ is related to the supercurrent of the superfluid by analytic continuation to real time. Now we can go from a Lagrangian in terms of the velocity $\partial_\mu \varphi$ to a Hamiltonian density in terms of the momentum $\xi_\mu$,
\begin{equation}
 \mathcal{H} = - \ti \xi_\mu \partial_\mu \varphi + \mathcal{L} = \frac{g}{2} \xi_\mu^2.
\end{equation}
Here we used Eq.~\eqref{eq:superfluid Lagrangian} and Eq.~\eqref{eq:superfluid canonical momentum}. Again, this form of the Hamiltonian density stems from the convention for imaginary time $\tau = \ti t$. The partition function is given by $\mathcal{Z} \propto \exp(- \int \td \tau \td^D x \; \mathcal{H})$. As we are interested in a quantum field-theoretic description, we actually prefer to work with Lagrangians. Therefore we will formally reobtain a Lagrangian, but keep the $\xi_\mu$ as the principal field, thus $\mathcal{L}_\mathrm{dual} = \mathcal{L}_\mathrm{dual}(\xi_\mu)$. Hence
\begin{equation}\label{eq:dual Lagrangian definition}
 \mathcal{L}_\mathrm{dual} = \mathcal{H} + \ti\xi_\mu \partial_\mu \varphi = \frac{g}{2} \xi_\mu^2 + \ti \xi_\mu \partial_\mu \varphi. 
\end{equation}
Note that we could have also directly obtained this form by a Hubbard--Stratonovich transformation applied to Eq. \eqref{eq:superfluid Lagrangian}, but we would not have the interpretation of $\xi_\mu$ as the canonical momentum. Also note that the dual coupling constant $g$ is inversely proportional to the original coupling constant $1/g$. This is therefore a weak--strong duality, in the sense of Kramers and Wannier~\cite{KramersWannier41}.

Now comes the important step in the duality construction. We again separate the phase field in smooth and singular parts as in Eq.~\eqref{eq:multivalued phase field}:
\begin{equation}
 \mathcal{Z} = \int \mathcal{D}\xi_\mu \mathcal{D}\varphi_\mathrm{smooth}\mathcal{D}\varphi_\mathrm{sing}\ \te^{-\int \mathcal{L}_\mathrm{dual}}.
\end{equation}
On the smooth part, one is allowed to perform integration by parts $ \xi_\mu \partial_\mu \varphi_\mathrm{smooth} \to -(\partial_\mu \xi_\mu) \varphi_\mathrm{smooth}$, and to then integrate out $\varphi_\mathrm{smooth}$ in the path integral as a Lagrange multiplier field, producing the constraint
\begin{equation}
 \partial_\mu \xi_\mu = 0.\label{eq:supercurrent constraint}
\end{equation}
It follows that $\xi_\mu$ is the conserved supercurrent of the superfluid, and the conservation law (continuity equation) arises because the phase fluctuations have to be smooth. The constraint Eq. \eqref{eq:supercurrent constraint} can be enforced explicitly by expressing it as the curl of another vector field, the {\em dual gauge potential} or {\em dual gauge field} $b_\lambda(x)$,
\begin{equation}\label{eq:superfluid gauge field definition}
 \xi_\mu(x) = \epsilon_{\mu\nu\lambda} \partial_\nu b_\lambda(x).
\end{equation}
The physical field $\xi_\mu$ is invariant under the addition of the gradient of any smooth scalar field
\begin{equation}\label{eq:superfluid gauge transformation}
 b_\lambda (x) \to b_\lambda (x) + \partial_\lambda \varepsilon(x),
\end{equation}
so $b_\lambda$ is a gauge potential for the `field strength' $\xi_\mu$. We substitute the definition Eq. \eqref{eq:superfluid gauge field definition} into Eq. \eqref{eq:dual Lagrangian definition} to find
\begin{align}
 \mathcal{L}_\mathrm{dual} &= \frac{g}{2} (\epsilon_{\mu\nu\lambda} \partial_\nu b_\lambda)^2 + \ti (\epsilon_{\mu\nu\lambda} \partial_\nu b_\lambda)\partial_\mu \varphi_\mathrm{sing} \nonumber \\
 &= \frac{g}{4} (\partial_\nu b_\lambda -\partial_\lambda b_\nu )^2 + \ti b_\lambda J^\mathrm{V}_\lambda. \label{eq:XY Coulomb action}
\end{align}
Here we performed integration by parts on $b_\lambda$ and substituted the definition of the vortex current Eq. \eqref{eq:superfluid vortex current}. 
This form is identical to the Maxwell Lagrangian for charged particles interacting by exchanging photon fields. Vortices in a 2+1D superfluid are identical to a gas of charged particles, interacting by the electromagnetic-like dual gauge fields $b_\lambda$. Therefore we refer to this as the Coulomb phase for vortex particles. Recall that $b_\lambda$ encodes for the supercurrent $\xi_\mu$ which encodes for the superfluid sound mode $\partial_\mu \varphi$. We conclude that the Goldstone modes that communicate the rigidity of the superfluid medium also mediate interactions between the topological defects of that medium: the vortices. The gauge freedom Eq.~\eqref{eq:superfluid gauge transformation} can in turn be viewed as enforcing vortex conservation Eq. \eqref{eq:vortex current conservation}.

\subsection{The phase-disordered superfluid as a vortex condensate}\label{subsec:The XY-disordered phase as a vortex condensate}
The real power of duality is that it reveals how to take the analogy with electrodynamics one step further. When the free photon encounters a superconductor, which is a condensate of bosons (Cooper pairs), it minimally couples to the condensate and acquires a mass through the Anderson--Higgs mechanism. The interaction mediated by massive gauge bosons is not long- but short-ranged. The rigidity communicated by the dual gauge fields persists over a short distance only, signaling that long-range order has been lost. In other words, a hallmark of the disordered phase is that the dual gauge fields become massive.

In the dual language such a disordered state corresponds to a condensate of vortices. When the coupling constant $g$ increases above a critical value, vortices can be freely created and annihilated and undergo an unbinding transition, which is very similar to the Berezinskii--Kosterlitz--Thouless transition in 2+0 dimensions. This statement can be made very precise on a lattice by considering a grand canonical ensemble of meandering vortex worldlines \cite{Kleinert89a} and has been confirmed by numerical calculations \cite{NguyenSudbo99,HoveSudbo00}, see also Sec.~\ref{sec:Dynamics of disorder fields}. The vortex condensate is described by a collective complex scalar field $\Phi = |\Phi|\te^{\ti\phi}$ to which the dual gauge fields are minimally coupled. 

Since the flow around the vortex core {\em is} superflow, the vortex condensate is governed by the same velocity $c_\mathrm{ph}$ as the superfluid phase fluctuations. And as we are working with a relativistic superfluid, the vortices also have relativistic dynamics. (In other words, we assume there is no `normal component' of the fluid, which is valid near the zero-temperature quantum phase transition. Therefore the vortices do not carry entropy and move relativistically.) It will however be useful to keep track of the influence of the condensate degrees of freedom by assigning to it a hypothetical velocity $c_\mathrm{V}$, which could be different from $c_\mathrm{ph}$. This turns out to be even more useful in the dual elasticity context. The minimal coupling term then becomes
\begin{align}\label{eq:vortex condensate minimal coupling}
 \mathcal{L}_\mathrm{min.coup.} &= \frac{1}{2}| \tfrac{1}{c_\mathrm{V}} (\partial_\tau - \ti b_\tau) \Phi|^2 +  \frac{1}{2}| 
(\partial_l- \ti b_l) \Phi|^2 \nonumber \\
&\equiv \frac{1}{2}|(\tilde{\partial}_\lambda - \ti \tilde{b}_\lambda) \Phi|^2.
\end{align}
The tildes $\tilde{}$ indicates rescaling with the  different velocity $c_\mathrm{V}$. Recall from  Eqs.~\eqref{eq:superfluid canonical momentum}, \eqref{eq:supercurrent constraint} and \eqref{eq:superfluid gauge field definition} that $b_\ft = \frac{1}{c_\mathrm{ph}} b_\tau$, so we have $\tilde{b}_\ft = \frac{c_\mathrm{ph}}{c_\mathrm{V}} b_\ft$. The action for the dual superconductor is
\begin{equation}\label{eq:superfluid Higgs Lagrangian}
 \mathcal{L}_\mathrm{SC} = \frac{g}{2} (\epsilon_{\mu\nu\lambda} \partial_\nu b_\lambda)^2  + \frac{1}{2}|(\tilde{\partial}_\mu - \ti \tilde{b}_\mu)\Phi| + \frac{\alpha}{2} |\Phi|^2 + \frac{\beta}{4} |\Phi|^4. 
\end{equation}
Here $\alpha$ and $\beta$ are dual Ginzburg--Landau parameters. When $\alpha <0$ the condensate field obtains a finite vacuum expectation value $|\Phi| = \sqrt{-\alpha/\beta}= \Phi_0$ and its phase $\phi$ (not to be confused with the superfluid phase $\varphi$) obtains a spontaneously chosen value $\phi_0$. We will always work in the London limit or strong type-II limit, where fluctuations in the condensate amplitude $|\Phi| = \Phi_0$ are neglected. The dual gauge field acquires a Higgs mass $\Omega$ with units of energy, defined by $c_\mathrm{ph}^2 \Phi_0^2/g \equiv \Omega^2$ . The interactions mediated by $b_\lambda$ have turned short-ranged: the dual Meissner effect. In addition to the zero-sound mode, which has turned massive, there is a second mode provided by the condensate itself. By gauge transformation this can also be interpreted as the longitudinal polarization of the dual gauge field. In the physical case where $c_\mathrm{V} = c_\mathrm{ph}$, the two modes are degenerate. In the Bose-Hubbard model, these two degenerate gapped modes correspond to the particle and hole excitations of the Bose-Mott insulator. 

At this point we would like to dispel one common misconception regarding vortex--boson duality, namely that instead of the massive dual gauge field, the dual vortices in the Mott insulator phase (singularities in $\phi$) would correspond to the particle and hole (doublon/holon) excitations. As the insulator is a condensate of vortices which has the action of a type-II superconductor Eq.~\eqref{eq:superfluid Higgs Lagrangian}, there are the dual incarnations of Abrikosov vortices, sometimes called vortex pancakes in two dimensions. In several works, following the ideas laid out in Refs.~\cite{FisherLee89,LeeFisher91}, it is surmised that these dual vortex and antivortex excitations be the particle and hole excitation in the Mott insulator. As we mentioned, that role is played by the dual gauge fields, coding for the Goldstone modes in the superfluid but acquiring a mass through the Anderson--Higgs mechanism in the Mott insulator~\cite{CvetkovicZaanen06a}. There are several ways to see this. First, vortices are topological defects as opposed to localized particle excitations. The energy of a single vortex grows with the system size whereas the particle/hole excitations are local and at rest have a well-defined energy, the Mott gap. Second, the particle and hole have a well-defined boson number. Conversely, a vortex in the Mott insulator is precisely that state where boson number is no longer well-defined, since number and phase are canonically conjugate operators, and according to the duality the core of the Mott vortex recovers the superfluid phase. Similarly, since the Mott insulator is a dual Ginzburg--Landau superconductor, if the charged particle (hole) were a vortex (antivortex), this would imply that an Abrikosov vortex in a real superconductor would carry an elementary charge $2e$, which to the best of our knowledge and observations is not the case. Third, in higher dimensions the vortices become extended objects, for instance vortex lines in three dimensions. But the particle and hole excitations in the Mott insulator are always particle-like in any dimension. Instead, the dual vortices are just that: topological defects that carry a quantized value of the current whose rigidity was lost in the disordered state~\cite{ZaanenBeekman12}. This statement was proven by the derivation in Ref.~\cite{CvetkovicZaanen06a} that the propagating modes on each side of the phase transition are represented by the dual gauge fields. Will we repeat this argument below.

For more details on the duality construction see for instance Refs.~\cite{Kleinert08,BeekmanSadriZaanen11}. One concludes that the disordered superfluid is dual to a superconductor in 2+1D, where vortices no longer have long-range correlations. Such a disordering transition is reflected by the gapping out of the dual gauge field representing the original Goldstone mode, a theme that will feature prominently in quantum elasticity.

\subsection{Propagators and duality}\label{subsec:Correlation functions}
Although we can read off most of the physics directly from the Lagrangian for the superfluid Eq. \eqref{eq:superfluid Lagrangian}, more rigor is acquired by studying the Green's functions or propagators. The phase--phase propagator (two-point correlation function) is easy to compute in momentum space, 
\begin{equation}\label{eq:superfluid phase-phase correlator}
 \langle \varphi(p) \varphi(-p) \rangle = g \frac{1}{p^2},
\end{equation}
as one expects for a massless scalar field. However, the phase variables are not well defined across the phase transition and therefore we cannot compare the correlator Eq. \eqref{eq:superfluid phase-phase correlator} with anything on the disordered side. This was called ``dual censorship'' in Ref.~\cite{CvetkovicZaanen06a}, and conveys the fact that one cannot interrogate the disordered phase with ordered means. In this light, it is better to consider the velocity--velocity correlator, where  $v_\mu = \partial_\mu \varphi$, and this is precisely the point of the duality construction. By definition $\langle v_\mu v_\nu  \rangle= p_\mu p_\nu \langle\varphi \varphi\rangle$, but we can also calculate it directly from the generating functional $\mathcal{Z}$ with external sources $\mathcal{J}_\mu$:
\begin{align}
  \langle v_\mu (p) v_\nu(-p) \rangle &= \frac{1}{\mathcal{Z}[0]} \frac{\delta}{\delta \mathcal{J}_\nu(-p)} \frac{\delta}{\delta \mathcal{J}_\mu (p)} \mathcal{Z}[J] \Big|_{\mathcal{J}=0}, \nonumber\\
 \mathcal{Z}[J] &= \int \mathcal{D}\varphi \exp\big[ -\int \frac{1}{2g} (\partial_\mu \varphi)^2 + (\partial_\mu \varphi) \mathcal{J}_\mu \big].
\end{align}
Integrating out the phase field will leave a Lagrangian $\mathcal{L} \sim \frac{g}{2} \mathcal{J}_\mu \frac{p_\mu p_\nu}{p^2} \mathcal{J}_\nu$, leading to
\begin{equation}\label{eq:superfluid velocity propagator}
 \langle v_\mu v_\nu  \rangle = g\frac{p_\mu p_\nu}{p^2}.
\end{equation}
Obviously, since the duality is a mathematical identity, calculating the same quantity using dual variables $\xi_\mu$ one should obtain the same results. This is a more subtle issue than it seems at face value. With the external source $\mathcal{J}_\mu$ in place, the dual current is redefined through the left-hand side of Eq. \eqref{eq:superfluid canonical momentum}, so that
\begin{equation}
 \xi_\mu = -\ti\frac{1}{g} \partial_\mu \varphi -\ti \mathcal{J}_\mu.
\end{equation}
Thus, the above equation relates the original velocity fields to the dual variables.
The dual Lagrangian becomes
\begin{equation}
 \mathcal{L}_\mathrm{dual} = \frac{g}{2} \xi_\mu^2 - \frac{g}{2}\mathcal{J}_\mu \mathcal{J}_\mu + \ti g \mathcal{J}_\mu \xi_\mu  + \ti \xi_\mu \partial_\mu \varphi. 
\end{equation}
It follows directly from $\mathcal{Z} = \int \exp(- \int\mathcal{L}_\mathrm{dual})$ that
\begin{equation}\label{eq:superfluid Zaanen--Mukhin relation}
 \langle v_\mu v_\nu \rangle = \frac{1}{\mathcal{Z}[0]} \frac{\delta}{\delta \mathcal{J}_\nu} \frac{\delta}{\delta \mathcal{J}_\mu } \mathcal{Z}[J] \Big|_{\mathcal{J}=0} =g\delta_{\mu\nu} - g^2\langle \xi_\mu \xi_\nu \rangle. 
\end{equation}
This relation was first derived in Ref.~\cite{ZaanenNussinovMukhin04}, relating propagators on the original and the dual sides. The propagator $\langle \xi_\mu \xi_\nu \rangle$ can be calculated directly from \eqref{eq:dual Lagrangian definition} but only after implementing the constraint $\partial_\mu \xi_\mu =0$. Another route automatically taking care of this constraint is to substitute the dual gauge field Eq. \eqref{eq:superfluid gauge field definition}, via
\begin{equation}\label{eq:superfluid current gauge field propagator}
 \langle \xi_\mu \xi_\nu \rangle = \epsilon_{\mu\kappa\lambda} \epsilon_{\nu \rho \sigma} (- \ti p_\kappa) (\ti p_\sigma) \langle b_\lambda b_\rho \rangle.
\end{equation}

Since the dual gauge fields couple to the vortex sources, their propagators have a meaningful interpretation by themselves. Since their action is just that of massless free vector fields, transforming to the $(\ft,\tL,\tT)$-system (see Sec.~\ref{subsec:Conventions and notation} and \ref{sec:Fourier space coordinate systems}) and using relation Eq.~\eqref{eq:curl-curl contraction tLT}, we find 
\begin{align}\label{eq:superfluid dual gauge field Lagrangian}
 \mathcal{L}_\mathrm{dual} &= \frac{g}{2} 
 \begin{pmatrix} b^\dagger_\ft \\ b^\dagger_\tL \\ b^\dagger_\tT \end{pmatrix}^\mathrm{T}
 \begin{pmatrix} q^2 & - \frac{\ti}{c_\mathrm{ph}} \omega_n q & 0\\
  \frac{\ti}{c_\mathrm{ph}} \omega_n q & \frac{1}{c_\mathrm{ph}^2} \omega_n^2 & 0\\
 0 & 0& p^2 
 \end{pmatrix}
\begin{pmatrix} b_\ft \\ b_\tL \\ b_\tL \end{pmatrix}
+ \ti b^\dagger_\ft  J^\mathrm{V}_\ft  + \ti b^\dagger_\tL  J^\mathrm{V}_\tL + \ti b^\dagger_\tT J^\mathrm{V}_\tT  .
\end{align}
Recall from Sec.~\ref{subsec:Conventions and notation} that $b_\mu^\dagger(p) = b_\mu(-p)$, and $b^\dagger_\lambda J^\mathrm{V}_\lambda$ is a shorthand for $\frac{1}{2} ( b^\dagger_\lambda J^\mathrm{V}_\lambda + {J^\mathrm{V}_\lambda}^\dagger b_\lambda)$. The propagators can be obtained as $\langle b_\mu b_\nu \rangle = \frac{1}{\mathcal{Z}[0]}\frac{\delta}{\delta J^{\mathrm{V}\dagger}_\nu} \frac{\delta}{\delta J^{\mathrm{V}}_\mu} \mathcal{Z}[J]$ etc.  One should integrate out the fluctuating gauge fields, which we do for $b_\ft$ and then $b_\tT$ to find,
\begin{align}
 \mathcal{L}_\mathrm{dual} 
 &=  \frac{g}{2}(\tfrac{1}{c_\mathrm{ph}^2}\omega_n^2 + q^2) |b_\tT|^2  +  \frac{1}{2}\ti ( b^\dagger_\tT J^\mathrm{V}_\tT - J^\mathrm{V\dagger}_\tT b_\tT) +\frac{1}{2g}  \frac{1}{q^2} |J^\mathrm{V}_\ft|^2  + \tfrac{1}{2}\ti b_\tL^\dagger ( -\frac{\ti \omega_n}{ c_\mathrm{ph}q}J^\mathrm{V}_\ft + J^\mathrm{V}_\tL) + \mathrm{h.c.} \nonumber\\
 &= \frac{1}{2g} \frac{1}{\tfrac{1}{c_\mathrm{ph}^2}\omega_n^2 + q^2}|J^\mathrm{V}_\tT|^2 +\frac{1}{2g}  \frac{1}{q^2} |J^\mathrm{V}_\ft|^2 +\tfrac{1}{2}\ti  b_\tL^\dagger ( -\frac{\ti \omega_n}{ c_\mathrm{ph}q}J^\mathrm{V}_\ft + J^\mathrm{V}_\tL) + \mathrm{h.c.}
\end{align}
In the last line, we see that $b_\tL$ is to be integrated out as a Lagrange multiplier for the constraint $\ti \frac{\omega_n}{c_\mathrm{ph}}J^\mathrm{V}_\ft -q  J^\mathrm{V}_\tL = \partial_\ft J^\mathrm{V}_\tau + \partial_m J^\mathrm{V}_m = 0$, the conservation of vortex current. This is a consequence of the gauge invariance Eq. \eqref{eq:superfluid gauge transformation}. This ascertains that in future calculations, is it allowed to impose the Coulomb gauge $\partial_m b_m = -q b_\tL= 0$, or any other gauge fix at will, right at the Lagrangian level Eq. \eqref{eq:superfluid dual gauge field Lagrangian}. We find two propagators in the Coulomb gauge,
\begin{align}
 \langle b_\ft b_\ft \rangle &= \frac{1}{g} \frac{1}{q^2}, \\
 \langle b_\tT b_\tT \rangle &= \frac{1}{g} \frac{1}{\frac{1}{c_\mathrm{ph}^2}\omega_n^2 + q^2} =\frac{1}{g} \frac{1}{p^2}.
\end{align}
Next to the expected massless propagating mode corresponding to the transverse polarization $b_\tT$, we also discern the static Coulomb force $b_\ft$ between `vortex charges' $J^\mathrm{V}_\ft$, with propagator $\langle b_\ft b_\ft \rangle \sim 1/q^2$. It is interesting that the single scalar field $\varphi$ through the duality transformation also immediately yields these static forces. This feature will become more revealing in the elasticity context. With these results one finds the supercurrent correlators,
\begin{align}
 \langle \xi_\ft \xi_\ft \rangle &=  \frac{q^2}{gp^2}, \\
 \langle \xi_m \xi_n \rangle &=  \frac{1}{g} \Big[ \frac{ \frac{1}{c_\mathrm{ph}^2} \omega_n^2 }{p^2}P^\tL_{mn} + P^\tT_{mn} \Big].
\end{align}
Here we used the longitudinal and transverse projectors Eqs.~\eqref{eq:longitudinal projector} and \eqref{eq:transverse projector}. Given these equations and the generating functional relation Eq.~\eqref{eq:superfluid Zaanen--Mukhin relation}, one can reobtain Eq.~\eqref{eq:superfluid velocity propagator}. Note that the propagating mode shows up in $\langle \xi_\ft \xi_\ft \rangle$ and $\langle \xi_\tL \xi_\tL \rangle$ while the transverse current correlator $\langle \xi_\tT \xi_\tT \rangle$ is constant: a sound mode has no transverse component.

Eq.~\eqref{eq:superfluid Zaanen--Mukhin relation} holds both in the ordered (Coulomb) and the disordered (Higgs) phase. In the disordered phase, we add the Higgs term as in Eq.~\eqref{eq:superfluid Higgs Lagrangian}. We can now choose the unitary gauge $\phi \equiv 0$, simplifying the Higgs term to $\Phi_0^2 b_\lambda^2$. The gauge freedom is absorbed, and we find the Lagrangian
\begin{align}
 \mathcal{L}_\mathrm{H} &=  \frac{g}{2} 
 \begin{pmatrix} b^\dagger_\ft \\ b^\dagger_\tL \\ b^\dagger_\tT \end{pmatrix}^{\!\!\mathrm{T}}\!\!
 \begin{pmatrix} q^2 + \frac{1}{c_\mathrm{V}^2}\Omega^2 & - \frac{\ti}{c_\mathrm{ph}} \omega_n q & 0\\
  \frac{\ti}{c_\mathrm{ph}} \omega_n q & \frac{1}{c_\mathrm{ph}^2} \omega_n^2 + \frac{1}{c_\mathrm{ph}^2} \Omega^2 & 0\\
  0&0 &  p^2 + \frac{1}{c_\mathrm{ph}^2}\Omega^2 
 \end{pmatrix}\!
\begin{pmatrix} b_\ft \\ b_\tL \\ b_\tL \end{pmatrix}
+ \ti b^\dagger_\ft  J^\mathrm{V}_\ft  + \ti b^\dagger_\tL  J^\mathrm{V}_\tL + \ti b^\dagger_\tT J^\mathrm{V}_\tT.\label{eq:superfluid Higgs Lagrangian matrix}
\end{align}
Here $\Omega^2 = c_\mathrm{ph}^2\Phi_0^2/g$ is the Higgs mass. Note the condensate velocity $c_\mathrm{V}$ in the top-left matrix element. Employing external sources to find the propagators, we must integrate out the fluctuating gauge fields. Starting with $b_\ft$, we obtain
\begin{align}
 \mathcal{L}_\mathrm{H} &= \frac{g}{2} b^\dagger_\tL \frac{\frac{1}{c_\mathrm{ph}^2} \Omega^2 (\frac{1}{c_\mathrm{V}^2}\omega_n^2 +q^2 + \frac{1}{c_\mathrm{V}^2} \Omega^2)}{q^2 + \frac{1}{c_\mathrm{V}^2} \Omega^2} b_\tL  + \frac{1}{2g} \frac{|J^\mathrm{V}_\ft|^2  }{q^2 + \frac{1}{c_\mathrm{V}^2}\Omega^2}+ \frac{g}{2}  (p^2 + \frac{1}{c_\mathrm{ph}^2}\Omega^2) b^\dagger_\tT b_\tT +   \ti b^\dagger_\tT J^\mathrm{V}_\tT\nonumber\\
 &\phantom{=}+ \ti\frac{1}{2} b^\dagger_\tL ( J^\mathrm{V}_\tL - \ti\frac{\frac{1}{c_\mathrm{ph}}\omega_n q}{q^2+\frac{1}{c_\mathrm{V}^2}\Omega^2} J^\mathrm{V}_\ft) + \mathrm{h.c.}
\end{align}
From here on, one can integrate out the remaining gauge field components. In the end, we are simply inverting the matrix in Eq.~\eqref{eq:superfluid Higgs Lagrangian matrix}. We can read off the propagators
\begin{align}
 \langle b_\tL b_\tL \rangle &= \frac{1}{g} \frac{q^2 +\frac{1}{c_\mathrm{V}^2}\Omega^2}{ \frac{1}{c_\mathrm{ph}^2}\Omega^2(\frac{1}{c^2_\mathrm{V}}\omega_n^2 + q^2 +\frac{1}{c_\mathrm{V}^2} \Omega^2)}, \label{eq:disordered superfluid condensate mode}\\
 \langle b_\tT b_\tT \rangle &= \frac{1}{g} \frac{1}{\frac{1}{c_\mathrm{ph}^2}\omega_n^2 + q^2 +  \frac{1}{c_\mathrm{ph}^2}\Omega^2 } =\frac{1}{g} \frac{1}{p^2 +  \frac{1}{c_\mathrm{ph}^2}\Omega^2}.\label{eq:disordered superfluid sound mode}
\end{align}
Furthermore, in the static limit $\omega_n \to 0$, we find
\begin{equation}
\langle b_\ft b_\ft \rangle = \frac{1}{g} \frac{1}{q^2 +  \frac{1}{c_\mathrm{ph}^2}\Omega^2}. \label{eq:disordered superfluid Coulomb force}
\end{equation}
From these relations we read off the following physical phenomena: the static Coulomb force Eq. \eqref{eq:disordered superfluid Coulomb force} turns short ranged, with a length scale inversely proportional to the Higgs mass. The superfluid sound mode Eq. \eqref{eq:disordered superfluid sound mode} is Higgsed and decays exponentially indicating the loss of superfluid order. Next to these, there is a second propagating, massive mode Eq. \eqref{eq:disordered superfluid condensate mode}, as the gauge freedom was fixed by removing the condensate phase $\phi$. This is sometimes referred to as that the gauge field $b_\lambda$ ``has eaten the Goldstone boson" (the phase mode of the vortex condensate) to become massive. This is the Anderson--Higgs mechanism for gauge fields to obtain a mass. Noticing that the propagation velocity of this mode is $c_\mathrm{V}$, it is clear that this mode originates from the vortex condensate. In a real-world superfluid, the velocities $c_\mathrm{ph}$ and $c_\mathrm{V}$ are identical. In conclusion, the spectrum of the disordered superfluid contains two massive propagating modes and one short-ranged static force. As mentioned above, in the Bose-Hubbard model the doublet of massive modes corresponds to the doublon and holon excitations of the Bose-Mott insulator. 
 
 \section{Field-theoretic elasticity}\label{sec:Field-theoretic elasticity}
 
In this section, we present the theory of elasticity as a quantum field theory of broken translational symmetries. Crystalline states of matter and their correlations are in some sense the most tangible and everyday case of spontaneous symmetry breaking in macroscopic quantum systems. We now discuss this briefly, since it is the starting point of the framework of this review. 
The path-integral formulation will allow a straightforward introduction of an vortex--boson-type dualization of elasticity in Sec.~\ref{sec:Dual elasticity}. As opposed to the classical theory of elasticity, see e.g. Ref.~\cite{LandauLifshitz86}, the roots of which are centuries old, the field theory of elasticity as we formulate it is a recent development, in great
part by Kleinert~\cite{Kleinert89b}. We will now explain in detail how the theory of elasticity emerges from the many-body path-integral formalism, followed by a discussion on the excitations of the model.

\subsection{Crystalline states and displacement fields}

The quantum Hamiltonian of a non-relativistic many-body system is very generally
\begin{align}
H = \sum_{i=1}^N \tfrac{\mathbf{P}_i^2}{2M_i} + \mathcal{V}(\mathbf{x}_1,\dots, \mathbf{x}_N), \label{eq:Hmanybody}
\end{align}
where $\mathbf{P}_i$ are the momenta, $M_i$ the masses, $\mathbf{x}_i$ coordinates of the $N$ `particles', and $\mathcal{V}$ is the many-body potential. The above form of the Hamiltonian represents the starting point of all condensed matter systems, regardless whether these are gaseous, liquid, or solid. In terms of symmetries, the Hamiltonian Eq.~\eqref{eq:Hmanybody} is invariant under all spatial symmetries of isotropic Euclidean space $E(D)=\mathbb{R}^{D}\rtimes O(D)$. We will be interested in states which spontaneously break the symmetries $E(D)$ to a subgroup; this means that the ground state of \eqref{eq:Hmanybody} breaks them even though the Hamiltonian and the potential $\mathcal{V}(\mathbf{x}_1,\dots,\mathbf{x}_N)$ are invariant under these symmetries.

Our definition of a (crystalline) solid is that both the translations and rotations are broken down to a discrete subgroup, and the groundstate consists of the particles localized around some equilibrium positions ${\mathbf{R}}_i^0$. We define the groundstate as a product state of semiclassical coherent states, where the `one-particle coherent states' localized around $\mathbf{R}_i$ are denoted as $\psi(\mathbf{R}_i)$ (the detailed form of the wave packets is not important in the following). The ground state is
\begin{equation}
\vert \lbrace {\mathbf{R}}_i \rbrace \rangle = \Big( \prod_{i}^N \hat{\psi}^{\dagger} ({\mathbf{R}}_{i}) \Big)  \vert \mathrm{vacuum} \rangle. \label{eq:Slater sum}
\end{equation}
Here $\hat{\psi}^{\dagger}(\mathbf{R}_i)$ is an operator that creates the localized state $\psi(\mathbf{R}_i)$. Since the states are highly localized in position, the center-of-mass momentum {\em a priori} fluctuates and Eq.~\eqref{eq:Slater sum} is not an eigenstate of Eq.~\eqref{eq:Hmanybody}, although the classical state emerges as the many-body quantum state. In the case of bosons, $\hat{\psi}(\mathbf{R}_i)$ are commuting operators, whereas for fermions the Pauli principle has to be obeyed; in this case the $\hat{\psi}(\mathbf{R}_i)$ anticommute and the groundstate wave function is a Slater determinant of the one-particle localized states. In the crystalline state the particles hardly move from their equilibrium positions and the statistics are unimportant. However, we will later focus on states involving quantum melting, which means that the particle positions $\{\mathbf{R}_i\}$ eventually wind around each other in arbitrary ways, and this simplification no longer holds. Therefore we restrict ourselves to bosonic matter only, avoiding any possible fermion-sign problems. 

In the (coherent state) path-integral formulation, the partition function of the quantum system is obtained as a path integral with periodic boundary conditions (trace) over all possible intermediate states $\vert \{\mathbf{R}_i(\tau)\} \rangle$ of the system weighed by the Euclidean action,
\begin{align}
 \mathcal{Z} &= \int \mathcal{D} \{\mathbf{R}_{i}(\tau)\} ~ \left \lbrack \te^{-\mathcal{S}_\mathrm{E}[{\mathbf{R}_i(\tau)}]} \right \rbrack, \label {eq:general path integral}  \\
\mathcal{S}_\mathrm{E} &= \int_0^\beta \td \tau \ {\mathcal{L}} ({\mathbf{R}_i(\tau),\partial_{\tau}\mathbf{R}_i(\tau)}), \label{eq:general action}\\
  \mathcal{L} &= \sum_i^N \ti \mathbf{P}_i\partial_{\tau} \mathbf{R}_i + H (\{ \mathbf{P}_i, \mathbf{R}_i(\tau)\}) . \label{eq:general Lagrangian}
\end{align}
Here we make the assumption that the localized states form a sensible set of states for the problem at hand, as a starting point to study fluctuations. In the crystalline solid, the low-energy excitations are described by small displacements $\mathbf{u}_i(\tau)$ from equilibrium
\begin{equation}
 {\mathbf{R}}_i (\tau) = {\mathbf{R}}^0_i + {\mathbf{u}}_i (\tau), \label{eq:displacement definition}
\end{equation}
which are highly collective excitations in terms of the original particle degrees of freedom. Note that these are the fluctuations around the broken symmetry state described by Eq.~\eqref{eq:Slater sum}. In this representation of the theory of elasticity, crystalline displacements ${\mathbf{u}}_i (\tau)$ (or later $\mathbf{u}(\mathbf{x},\tau)$ in the continuum limit) take the role of the low energy, Goldstone degrees of freedom: the phonons. Their gapless nature is protected by the Goldstone theorem. They will be the principal actors in what follows, analogous to the phase field $\varphi$ in Eq.~\eqref{eq:superfluid Lagrangian} of the superfluid. Compared to the scalar phase field, a very important distinction is that $\mathbf{u}(\tau) = u^a(\tau)$ is a \emph{spatial} vector and displacements in the timelike direction are prohibited, $u^\tau \equiv 0$. Thus, the Euclidean action cannot possibly be subject to emergent Lorentz-invariance, and this is the origin of much of the special properties of the quantum field theory describing quantum solids and liquid crystals (Lorentz-invariant quantum nematics are discussed in Refs.~\cite{KleinertZaanen04,ZaanenBeekman12}).

Let us now consider how the displacements enter the elastic action of the model. The Lagrangian in Eq.~\eqref{eq:general Lagrangian},
obtained from the Hamiltonian Eq.~\eqref{eq:Hmanybody}, contains two contributions:
(i) the inter-particle interaction $\mathcal{V}$, which acts as a potential energy, and (ii) the quantum kinetic term $\mathcal{T}$. First consider the classical limit $M_i \to \infty$ when the quantum fluctuations are completely suppressed. The contribution to the action from the potential energy can be written as an integral over imaginary time
\begin{equation}
  \mathcal{S}_\mathrm{pot} = \int \td \tau ~ {\mathcal{V}} \lbrack \lbrace {\mathbf{R}}_i (\tau) \rbrace \rbrack, \label{eq:equilibrium solid potential}
\end{equation}
with ${\mathcal{V}}\lbrack \lbrace {\mathbf{R}}_i (\tau) \rbrace \rbrack$ being the potential energy of a given configuration of particles at a given imaginary time $\tau$. 

Due to the translational symmetry of the multi-particle Hamiltonian, the potential energy must be a function of the relative displacement between two particles
\begin{align}
  {\mathbf{R}}_{ij} &= {\mathbf{R}}_i - {\mathbf{R}}_j 
  = ({\mathbf{R}}_i^0 + {\mathbf{u}}_i) - ({\mathbf{R}}_j^0 + {\mathbf{u}}_j) 
  \equiv {\mathbf{R}}_{ij}^{(0)} + {\mathbf{\Delta}}_{ij}. \label {Rij}
\end{align}
As a consequence, the expansion of Eq.~\eqref{eq:equilibrium solid potential}  depends only on the relative displacements ${\mathbf{\Delta}}_{ij} \equiv {\mathbf{u}}_i - {\mathbf{u}}_j$ and not on the displacements themselves. We expand the potential energy in
powers of the relative displacements:
\begin{equation}
  {\mathcal{V}} =  {\mathcal{V}}_0 + \frac {\partial {\mathcal{V}}}{\partial {R}_{ij}^a}
  \Delta_{ij}^a + \tfrac 1 2 \frac {\partial^2 {\mathcal{V}}}{\partial {R}_{ij}^a
  \partial {R}_{kl}^b}
  \Delta_{ij}^a \Delta_{kl}^b + O (\Delta^3). \label{eq:solid potential expansion}
\end{equation}
The first term is the binding energy of the solid
and it represents an overall constant which we neglect from now on.  Given the fact that the expansion Eq.~\eqref{eq:solid potential expansion} is performed
around the equilibrium state, the second, linear term in the expansion which represents the restoring force, must vanish.

The next step is a coarse-graining procedure, turning the lattice displacements $\mathbf{u}_i(\tau)$ to continuum displacements $\mathbf{u}(x) \equiv \mathbf{u}(\mathbf{x},\tau)$. The gradient expansion in terms of relative displacements is,
\begin{align}
  \Delta_{ij}^a &= u_i^a - u_j^a 
  = {R}_{ij}^{(0)m} \partial_m u^a + \frac {R_{ij}^{(0)m}
  R_{ij}^{(0)n}}{2} \partial_m \partial_n u^a + \ldots\ . \label{eq:displacements expanded}
\end{align}
Insert this into the second-order term of Eq.\ \eqref {eq:solid potential expansion},
yielding the leading harmonic terms of the elastic potential energy:
\begin{align}
  e_1 ({\mathbf{x}}) &= \tfrac 1 2 \frac {\partial^2 {\mathcal{V}}}{\partial {R}_{ij}^a
  \partial {R}_{kl}^b} R_{ij}^{(0)m} R_{kl}^{(0)n} \partial_m u^a \partial_n u^b
  \equiv \tfrac 1 2 \partial_m u^a C_{mnab} \partial_n u^b, \label{eq:first gradient elastic energy} \\
  e_2 ({\mathbf{x}}) &= \tfrac 1 2 \frac {\partial^2 {\mathcal{V}}}{\partial {R}_{ij}^a
  \partial {R}_{kl}^b} R_{ij}^{(0)m} R_{ij}^{(0)r} R_{kl}^{(0)n} R_{kl}^{(0)s}
  \partial_m \partial_r u^a \partial_n \partial_s u^b
  \equiv \tfrac 1 2 \partial_m \partial_r u^a C^{(2)}_{mrnsab} \partial_n \partial_s u^b.
  \label{eq:second gradient elastic energy}
\end{align}
The first line is the elastic energy governing the physics at large scales, while the second line represents the second-order gradient elastic potential energy, which is important only at shorter length scales. However, in the quantum liquid-crystalline phases central to this work, the second-order gradient terms become relevant again at long distances. Therefore, we will keep track of both the first- and second-order gradient elastic terms in the analysis. Higher order terms in expansion Eq.~\eqref{eq:displacements expanded} contain even higher-order derivatives, and can be safely ignored.

The tensor $C_{mnab}$ is called the elastic tensor and $C^{(2)}_{mnrsab}$ represents second-order contributions. These tensors are derived from the derivatives of the microscopic energy ${\mathcal{V}}$. Regardless of the specific microscopic details of ${\mathcal{V}}$, elasticity exhibits universal behavior at macroscopic scales. This is related to the spontaneous symmetry breaking by the crystalline state, such that Goldstone's theorem governs the physics associated  with the fluctuations $u^a(x)$. First, the case $u^a(x)\sim\rm{const.}$ corresponds to a uniform center-of-mass displacement and cannot change the energy of the system. In other words, when the momentum of a fluctuation $u^a(k)$ goes to zero, the excitation is gapless. Accordingly, it is immediately obvious that only derivatives $\partial_m u^a$ appear in the action. From this perspective the theory of elasticity can be seen as a field theory of suitable tensors constructed from the derivatives of the displacement field $u^a(x)$. Let us now briefly discuss the structure of these tensors. 

{\em A priori}, in general space dimension $D$, the elastic tensor $C_{mnab}$ may have up to $D^4$ 
independent elements but the actual number of independent components is always smaller due to symmetry constraints. The structure of the theory is not as straightforward in terms of the symmetries as perhaps one would expect. This is due to the fact that the symmetry group $E(D)=\mathbb{R}^D \rtimes O(D)$ is the \emph{semidirect} product of translations and rotations and these are not independent. One can see that the \emph{antisymmetric} quantity
\begin{equation}
\omega^{ab} = \tfrac{1}{2}(\partial_a u^b - \partial_b u^a). \label{eq:strain Ehrenfest constraint}
\end{equation}
corresponds to an infinitesimal rotation induced by $u^a(x)$. Since a constant rotation $\omega^{ab} \sim \rm{const.}$ cannot change the energy of the system either, this rotation field has to drop out from Eq.~\eqref{eq:first gradient elastic energy}. Nevertheless, at higher order in the expansion, derivatives $\partial_m\omega^{ab}$ can appear, see Eq.~\eqref{eq:isotropic solid second gradient energy}. Another way of looking at the matter is that to linear order a local rotation is indistinguishable from a local translation \cite{LowManohar02, WatanabeMurayama13}. The absence of local rotations is also responsible for the suppression of rotational Goldstone modes in systems with broken translation symmetry, as discussed in Sec.~\ref{subsec:torque stress nematic}. Summarizing, only the symmetrized combinations of the derivatives of $u^a$ or
{\em strains}
\begin{equation}\label{eq:strain definition}
 u^{ab} = \tfrac{1}{2}(\partial_a u^b + \partial_b u^a)
\end{equation}
can appear in Eq.~\eqref{eq:first gradient elastic energy} and this implies that $C_{mnab}$ is symmetric in the corresponding indices $m,a$ and $n,b$. Throughout this review, we will assume the derivatives $\partial_{a}u^b$ to be small and will not consider non-linear strains $u^{\rm non-linear}_{ab} = \frac{1}{2}\doo_{a}u^{c} \doo_{b} u^c$~\cite{LandauLifshitz86}. We can now count the number of the
independent elements as follows: there are $M = D (D +1) / 2$
strain components, and each can couple with any other one, except
that the off-diagonal couplings are counted twice ($C_{mnab}$
and $C_{nmba}$ must be the same). Thus, the number
of the independent elements in the elasticity tensor can be
at most $M (M+1) /2 = \lbrack D (D+1) (D^2 + D +2) \rbrack / 8$.
In two  dimensions there can be up to 6 independent elements,
while in three dimensions one finds 21 such entries.

\subsection{Isotropic elastic solid}\label{subsec:Isotropic solid}
The number of independent elements of $C_{mnab}$ reduces further due to the point group symmetries of the crystal~\cite{LandauLifshitz86, Kleinert89b}. We will focus on the case of the isotropic
solid as it is fully representative for the theory of the liquid crystals we wish to present. Obviously, as rotational symmetry is broken, the crystalline solid is never truly isotropic but the long-distance physics of several solids is well approximated by the isotropic limit. This holds in particular for the triangular lattice, as well as amorphous (disordered) solids. We would like to focus here on the first case where translational symmetry is broken down to a discrete subgroup, such that the remaining periodicity allows for well-defined lattice momenta of the phonon excitations. Nevertheless, the low-energy physics, in linear response, is completely rotationally invariant even though microscopically it is not. The reader can generalize the theory for an arbitrary elasticity tensor if desired, as this amounts to just more independent components of $C_{mnab}$. 

Here we present the theory of elasticity in arbitrary space dimensions $D$ until we 
specialize to 2+1 dimensions in Sec. \ref{sec:Dual elasticity}. In order for the system to be isotropic, i.e. invariant under 
the group of improper rotations $O (D)$, the most general symmetric elastic tensor takes the form~\cite{Kleinert89b}
\begin{equation}
  C_{mnab} = D \kappa P_{mnab}^{(0)} + 2 \mu P_{mnab}^{(2)}. \label {eq:elasticity tensor}
\end{equation}
Here, the rigidities of the solid are parametrized through
the compression and shear moduli, $\kappa$ and $\mu$ respectively. They are related to the Lam\'e constant $\lambda$ via $\lambda = \kappa - \frac{2}{D} \mu$.
$P_{mnab}^{(s)}$ is the projector on the subspace of
matrices that are irreducible tensors with angular momentum $s$.
There are three such projectors, 
\begin{align}
  P_{mnab}^{(0)} &= \frac 1 D \delta_{ma} \delta_{nb}, \label {eq:spin-0 projector} \\
  P_{mnab}^{(1)} &= \frac 1 2 \lbrack \delta_{mn} \delta_{ab} -
  \delta_{mb} \delta_{na} \rbrack,  \label {eq:spin-1 projector} \\
  P_{mnab}^{(2)} &= \frac 1 2 \lbrack \delta_{mn} \delta_{ab} +
  \delta_{mb} \delta_{na} \rbrack - \frac 1 D \delta_{ma} \delta_{nb}, \label {eq:spin-2 projector}
\end{align}
which, together, satisfy the closure identity (decomposition of unity) $\sum_s P_{mnab}^{(s)} = \delta_{mn} \delta_{ab}$. The absence of the angular momentum $s=1$ projector in Eq.~\eqref{eq:elasticity tensor} reflects the absence of local rotations Eq.~\eqref{eq:strain Ehrenfest constraint}, since $P_{mnab}^{(1)} \partial_n u^b = \omega^{ma}$.

In our analysis, we will substitute one of the elastic rigidities with the Poisson ratio $\nu$,
the ratio of transverse to longitudinal strain when the material is put under external longitudinal stress.
The compression modulus is related to $\nu$ through \cite{Kleinert89b}
\begin{equation}\label{eq:compression modulus definition}
  \kappa = \mu \frac 2 D \Big[ \frac {1 + \nu}{1 - (D - 1) \nu} \Big].
\end{equation}
It is useful to notice that, for stability in  equilibrium, both the compression
and the shear modulus must be positive. This, in
turn, implies that the Poisson ratio can vary
between $\nu = - 1$ (for a compressionless solid)
and $\nu = 1 / (D-1)$ (for a body with no shear rigidity).
The majority of solids have a positive Poisson ratio.
Solids with negative Poisson ratio are also known as
auxetics and they are common in systems where the shape deformation
(i.e., shear) is more costly than compression on the macroscopic level \cite{YangEtAl04}.
Such a property implies that the solid will reduce its cross section upon pressure 
which defies our everyday experience with solids.
Known examples are the Abrikosov lattice in type-II superconductors ($D=2$)~\cite{KleinerRothAutler64}, $\alpha$-quartz ($D=3$)~\cite{KittingerTichyBertagnolli81},
and some foams~\cite{Lakes87}.
Some phases of biopolymers have been shown to exhibit negative Poisson
ratio~\cite{AhsanRudnickBruinsma07}.

The exact form
of the isotropic elasticity tensor in two dimensions is
\begin{equation}
  C_{mnab} = \mu \left \lbrack \delta_{mn} \delta_{ab} + \delta_{mb}
  \delta_{na} + \tfrac {2 \nu}{1 - \nu} \delta_{ma} \delta_{nb} \right \rbrack.
\end{equation}
Two-dimensional solids with the triangular type of the
lattice symmetry group ($C_{3n}$, $D_{3n}$)
share this elasticity tensor. In other words, the notion of the
crystalline axes is lost in the long-distance physics.
Thus, the results derived in this work will also be applicable
not only to an isotropic solid but also to a solid with triangular or
hexagonal lattice symmetry. As one of the consequences,
we will  recover the 
quantum generalization of the KTNHY transition in two dimensions~\cite{KosterlitzThouless72,KosterlitzThouless73, HalperinNelson78,NelsonHalperin79, Young79}.

Let us briefly examine the second-order gradient
term Eq.~\eqref{eq:second gradient elastic energy} for the case of an isotropic solid.
From symmetry, it follows that it can
have only two independent coupling constants
defined via two lengths $\ell '$ and $\ell$~\cite{Kleinert89b}.
The second-order term can be expressed as
\begin{equation}
 e_2 (\mathbf{x}) = \mu \left \lbrack \tfrac {1 - (D-2) \nu}{1 - (D-1) \nu}
 \ell'^2 \partial_m \partial_j u^j \partial_m \partial_k u^k + \ell^2 \partial_m
  \omega^{ab} \partial_m \omega^{ab} \right \rbrack. \label{eq:isotropic solid second gradient energy}
 \end{equation}
Here, $\ell '$ defines the length below which the gradient of
the compression strain $\partial_m \partial_j u^j$ becomes relevant. 
The rotational stiffness length $\ell$ defines the length scale at which  torque/curvature become
physically relevant; remember that local rotations are absent in first-gradient elasticity. Most calculations can be performed in first-gradient elasticity only, but the term $\sim \ell^2$ will be useful in examining the quantum nematic phase of Sec. \ref{sec:Quantum nematic}.

\subsection{Quantum elasticity}\label{sec:Quantum elasticity}

Let us now add the quantum Euclidean time dynamics to the static elasticity obtained so far, following Ref.~\cite{ZaanenNussinovMukhin04}. The kinetic energy for the particle coordinates $\{\mathbf{R}_i(\tau)\}$ is given by $\mathcal{T} = \frac{1}{2}\sum_i^N M_i (\partial_{\tau} \mathbf{R}_i)^2$. In the coherent state path integral for the displacement fields \eqref{eq:displacement definition} in imaginary time this turns into the term in the Lagrangian~\cite{ZaanenNussinovMukhin04}
\begin{equation}
  \mathcal{L}_\mathrm{kin} = \tfrac 1 2 \rho (\partial_\tau \mathbf{u})^2, \label{eq:elastivity kinetic Lagrangian}
\end{equation}
with $\rho$ the mass density of the solid. In the case of most real crystalline solids, the lattice is formed by atoms with very big masses $M_i\to \infty$, quenching the quantum fluctuations. This is the theory of low-energy (acoustic) phonons, which are then just independent harmonic oscillators in the normal-mode decomposition of the lattice. The only quantum effect to quadratic order is the zero-point fluctuations of the harmonic oscillators. One of the central points for the theory of quantum elasticity is therefore the assumption of small masses $M_i$ such that the quantum fluctuations implied by Eq.~\eqref{eq:elastivity kinetic Lagrangian} are important, which is the opposite limit of the usual harmonic lattice story.

We are now ready to define the field theory of quantum elasticity. We combine the kinetic energy contribution Eq.~\eqref{eq:elastivity kinetic Lagrangian} and the potential energy Eq.~\eqref{eq:first gradient elastic energy}
into the total Lagrangian, and recast it in a useful form
as
\begin{equation}
  \mathcal{L}_0[\partial_{\mu}u^a] = \tfrac 1 2 \partial_\mu u^a C_{\mu \nu a b}
  \partial_\nu u^b. \label{eq:elasticity relativistic Lagrangian}
\end{equation}
The sum over the Greek indices includes not only the spatial, but also the temporal direction. To do this, we must define a velocity $c_\tT = \sqrt{\mu / \rho}$, called the {\em shear velocity} or {\em transverse velocity}, see below. We have defined $\partial_\mu \equiv (\partial_\ft , \partial_m) \equiv ( \frac{1}{c_\tT} \partial_\tau , \partial_m)$ and the extended elasticity tensor becomes
\begin{equation}
C_{\mu \nu a b} = \mu \delta_{ab} \delta_{\mu \ft} \delta_{\nu \ft} + C_{mnab}. 
\end{equation}
The Lagrangian Eq.~\eqref{eq:elasticity relativistic Lagrangian}, with the
isotropic elasticity tensor Eq.~\eqref{eq:elasticity tensor}, defines the Euclidean action $\mathcal{S} = \int_0^\beta \td \tau \int \td^D x~ \mathcal{L}_0[\partial_{\mu}u^a]$ of quantum elasticity in two dimensions, with partition sum,
\begin{align}
\mathcal{Z} = \int \mathcal{D} u^a ~ e^{-\mathcal{S}[\partial_{\mu} u^a]}.
\end{align}
This action, which is quadratic in the strains $\partial_{\mu} u^a$ due to the harmonic expansion Eqs.~\eqref{eq:first gradient elastic energy} and \eqref{eq:second gradient elastic energy}, will figure as the starting point for the field theoretical description of a (quantum) solid and its melting. 

We shall now discuss the basic results of the theory of elasticity followed by elastic response functions of the isotropic solid. For more details, see for example Refs.~\cite{LandauLifshitz86, Kleinert89b}. In the solid phase, the displacements $u^a$ are well-defined fields in terms of which the action is Gaussian. It is most convenient to consider the generating functional with external sources $\mathcal{J}_a$,
\begin{equation}
  \mathcal{Z} \lbrack \mathcal{J}_a \rbrack = \int \mathcal{D} u^a ~
  \te^{- \int \td x \td \tau ~ \mathcal{L}\lbrack \mathcal{J}_a \rbrack }.
  \label{eq:external force elasticity partition function}
\end{equation}
where the external sources couple to displacement as
$\mathcal{L} \lbrack \mathcal{J}_a \rbrack = \mathcal{L}_0 + \mathcal{J}^{a} u^a$. The Euler--Lagrange equations of motion are
\begin{align}
\frac{\partial \mathcal{L}[\mathcal{J}_a]}{\partial u^a}-\doo_{\mu}\frac{\partial \mathcal{L}[\mathcal{J}_a]}{\partial (\doo_{\mu} u^a)} = \mathcal{J}_a - \doo_{\mu} C_{\mu\nu a b}\partial_{\nu}u^b =0 .
\end{align}
It is convenient to define the imaginary time stress tensor $\sigma^a_{\mu}$ and rewrite the equations of motion as
\begin{align}
\doo_{\mu} \sigma^a_{\mu}+\ti \mathcal{J}_a = 0, \quad \sigma^a_{\mu} = -\ti C_{\mu\nu a b}\partial_{\nu}u^b,\label{eq:stress}
\end{align}
where the imaginary factors are the same as in Sec.~\ref{sec:XY-duality}. If we temporarily set the temporal part to zero, the Lagrangian becomes the classical elastic energy density $\mathcal{E}_{\rm cl}$ and the equations reduce to those of classical elasticity
\begin{align}
\underline{\sigma}_{m}^a &= C_{mnab}\doo_{n}u^b,&  
\doo_{m}\underline{\sigma}_m^a + \mathcal{J}_a &= 0,
\end{align}
where the $\mathcal{J}_{a}$ have the interpretation as external force densities. Correspondingly, the temporal component $\sigma_{\tau}^a = \rho \ti \doo_{\tau}u^a$ can be interpreted as the momentum density in imaginary time. In first-order elasticity, the elastic tensor $C_{mnab}$ is symmetric and therefore the stress tensor is symmetric as well, and can be alternatively defined in terms of the symmetric strains as \cite{LandauLifshitz86}
\begin{align}\label{eq:classical stress tensor}
\underline{\sigma}_m^a = \frac{\delta \mathcal{E}_{\rm cl}}{\delta u^{ma}}.
\end{align}
We will however define the stress tensor $\sigma^a_{\mu}$ explicitly in terms of the strains $\doo_{\mu}u^a$ without invoking any \emph{a priori} symmetry properties. As already shown by Kleinert~\cite{Kleinert89b}, the classical theory of elasticity can be written in terms of the stresses by a dualization procedure. This duality will be discussed extensively in Sec.~\ref{sec:Dual elasticity} for the quantum case.

\subsection{Propagators of the isotropic elastic solid}\label{subsec:Elastic response functions}

The displacement propagators are given by taking the second derivative of $\mathcal{Z}[\mathcal{J}_a]$  with respect to the
external sources
\begin{equation}
  \langle  u^a  u^b \rangle  = \frac 1 {\mathcal{Z}[0]}
  \left . \frac {\partial^2 \mathcal{Z} \lbrack \mathcal{J} \rbrack}{\partial \mathcal{J}^b \partial \mathcal{J}^a}
  \right |_{\mathcal{J} \to 0}. \label{eq:displacement propagator}
\end{equation}
The absence of topological defects in the crystal phase
means that the path integral in Eq.~\eqref{eq:external force elasticity partition function} incorporates
only smooth contributions of the $u^a$ fields and these
can be resolved by Gaussian integration. For convenience,
in the following we switch to the Matsubara--Fourier transformed
displacement fields. Denoting the bosonic Matsubara frequency
by $-\ti \partial_\tau \mapsto \omega_n$ and the spatial momentum by $-\ti \nabla \mapsto \mathbf{q}$,
the action with the external source terms included turns into
\begin{align}
  \mathcal{L}&= \tfrac 1 2 u^{a\dagger}
  \left \lbrack (\rho \omega_n^2 + \mu q^2) \delta_{ab}
  + \mu \tfrac {1- (D-3) \nu}{1- (D-1) \nu} q_a q_b \right \rbrack
  u^b + \mathcal{J}^{a\dagger} u^a  \nonumber \\
  &= \tfrac 1 2 \rho\; u^{a \dagger} \left \lbrack
  (\omega_n^2 + c_\tL^2 q^2) P_{ab}^\tL +
  (\omega_n^2 + c_\tT^2 q^2) P_{ab}^\tT \right \rbrack
  u^b + \mathcal{J}^{a \dagger} u^a. \label{eq:Matsubara solid Lagrangian}
\end{align}
In the second line,  $u^{a\dagger} = u^a(-\omega_n,-\mathbf{q})$; even though the displacements are real fields, the Fourier transformed components contain complex values, and the Hermitian conjugate must be properly taken into account, see Sec.~\ref{subsec:Conventions and notation}. We will employ this shorthand notation 
throughout this work. Eq.~\eqref{eq:Matsubara solid Lagrangian} introduces both longitudinal and transverse projectors Eqs.~\eqref{eq:longitudinal projector}, \eqref{eq:transverse projector},
\begin{equation}
  P_{ab}^\tL = \frac {q_a q_b}{q^2}, \qquad P_{ab}^\tT = \delta_{ab} - P_{ab}^\tL. \label{eq:longitudinal transverse projectors}
\end{equation}
These separate the displacements into components parallel 
and perpendicular to the wavevector $\mathbf{q}$. The two 
velocities $c_\tL$, $c_\tT$ in Eq.~\eqref{eq:Matsubara solid Lagrangian} correspond
to the longitudinal and the transverse phonon velocity
\begin{align}
  c_\tL^2 &= \frac {2 \mu}{\rho} \frac {1 - (D-2) \nu}{1 - (D-1) \nu}
  = \frac {\kappa +2 \mu \tfrac {D-1}{D}}{\rho}, \label{eq:longitudinal phonon velocity}\\
  c_\tT^2 &= \frac \mu \rho \label{eq:transverse phonon velocity}.
\end{align}
While the longitudinal phonon (sound, compression) exists even in
shearless media, the transverse phonons
owe their existence to the shear rigidity associated with spontaneous translational symmetry breaking and these are absent in gases and liquids.
Notice that the deformations due to longitudinal phonons in a solid
involve both shear and compression as reflected
in the propagation velocity of this mode. From Eq.~\eqref{eq:longitudinal phonon velocity} we see that the longitudinal velocity is always larger than the transverse velocity for $D\ge 2$.

After integration of one longitudinal and
$D-1$ transverse displacement components in  Eq.~\eqref{eq:Matsubara solid Lagrangian}, we find the
generating functional
\begin{align}
  \mathcal{Z} \lbrack \mathcal{J}^a \rbrack &= \prod_{\mathbf{q}, \omega_n} \sqrt
  {\frac {(2 \pi)^D}{\rho^D ( \omega_n^2 + c_\tL^2 q^2 )
  ( \omega_n^2 + c_\tT^2 q^2 )^{D-1} }}  \exp \Big (\tfrac 1 2 \mathcal{J}^{a\dagger} \left \lbrack
  \tfrac {P_{ab}^\tL}{\rho ( \omega_n^2 + c_\tL^2 q^2 )} + \tfrac {P_{ab}^\tT}{\rho
  ( \omega_n^2 + c_\tT^2 q^2 )} \right \rbrack \mathcal{J}^b\Big) \nonumber \\
  &= \mathcal{Z}[0] ~ \exp \Big (\tfrac 1 2 \mathcal{J}^{a\dagger} \left \lbrack \tfrac {P_{ab}^\tL}{\rho ( \omega_n^2 +
  c_\tL^2 q^2 )} + \tfrac {P_{ab}^\tT}{\rho ( \omega_n^2 + c_\tT^2 q^2 )} \right \rbrack \mathcal{J}^b\Big).
\end{align}
The imaginary-time displacement propagator for the elastic ideal
solid Eq.~\eqref{eq:displacement propagator} is found to be
\begin{equation}
  \langle  u^a  u^b  \rangle = \frac 1 \rho
  \left ( \frac {P_{ab}^\tL}{\omega_n^2 + c_\tL^2 q^2} +
  \frac {P_{ab}^\tT}{\omega_n^2 + c_\tT^2 q^2} \right ). \label{eq:Matsubara isotropic displacement propagator}
\end{equation}
The excitation spectrum of an ideal {\em isotropic} crystal
has one massless linear pole
associated with the longitudinal displacements,
and $D-1$ degenerate massless linear
poles associated with the transverse displacements. These are respectively, of course,
the longitudinal and transverse acoustic phonons of the crystal. 

Let us consider a solid with the second-order gradient
terms Eqs.\eqref{eq:second gradient elastic energy} or \eqref{eq:isotropic solid second gradient energy} added. The resulting displacement
propagator is
\begin{align}
  \langle  u^a  u^b  \rangle &= \frac 1 \rho \Big (
  \frac {P_{ab}^\tL}{\omega_n^2 + c_\tL^2 q^2 (1 + \ell'^2 q^2)} 
  + \frac {P_{ab}^\tT}{\omega_n^2 + c_\tT^2 q^2 (1 + \ell^2 q^2)} \Big).
  \label {eq:second gradient displacement propagator}
\end{align}
The choice of the lengths $\ell'$ and $\ell$ in the second order
gradient energy density Eq.~\eqref{eq:isotropic solid second gradient energy} is now
clear: the second-order gradient elasticity effects become visible in the
phonon spectra only at lengths smaller than $\ell'$ in the longitudinal and
$\ell$ in the transverse sector.

The displacement fields are only well defined in the total absence of topological defects, and so are the propagators Eq.~\eqref{eq:displacement propagator}. This situation mimics the phase propagator of the superfluid in Sec. \ref{subsec:Correlation functions}. There, it was shown that the velocity--velocity propagators are also valid in the presence of defects and even in the defect-mediated molten phase. Similarly, we now go over to strain propagators that can be dualized, and which can be used on both sides of the melting phase transition. The general strain propagator (Green's function) is
\begin{equation}
 G_{\mathrm{strain},mnab} = \langle \partial_m u^a \partial_n u^b \rangle,
\end{equation}
of which certain combinations are of greater physical relevance. In particular, we define the longitudinal and transverse strain propagators,
\begin{align}
 G_\tL (\omega_n , q) & \equiv  \langle \partial_a u^a \partial_b u^b \rangle = \frac{1}{\rho} \frac{q^2}{\omega_n^2 + c_\tL^2 q^2(1 + \ell'^2 q^2)}, \label{eq:longitudinal strain propagator}\\
 G_\tT (\omega_n , q)& \equiv 2\langle \omega^{ab} \omega^{ab} \rangle = \frac{1}{\rho} \frac{q^2}{\omega_n^2 + c_\tT^2 q^2(1 + \ell^2 q^2)}.\label{eq:transverse strain propagator}
\end{align}
In more than two spatial dimensions, there is more than one rotation field $\omega^{ab}$, and Eq.~\eqref{eq:transverse strain propagator} is a sum of all rotation contributions. Multiple excitations (phonons) show up as separate poles in this propagator. Interestingly, we see that the local rotation fields acquire
a nontrivial propagator
although they are not present explicitly in the action of Eq.~\eqref{eq:elasticity relativistic Lagrangian}. This is due to the interdependence between translations (shear) and rotations, cf. Sec.~\ref{subsec:Interdependence between dislocations and disclinations}.

One could construct a correlation function between the
compression strain and the local rotation
\begin{equation}
  G_{bc}^\mathrm{chiral} =\langle \partial_a u^a  \omega^{bc} \rangle. \label{eq:chiral strain propagator}
\end{equation}
If a phase of the system possesses a mirror symmetry
such that it is invariant under the change of sign in direction
$b$, the above propagator has to vanish since the
local rotation changes sign under this mirror symmetry.
In isotropic phases the propagator Eq.~\eqref{eq:chiral strain propagator}
acquires a finite value only if the chirality is violated,
meaning that the mirror image of the ground
state represents a different physical state. 
This is the origin of the name of this ``chiral'' 
propagator.

The smectic phase that we construct in Sec. \ref{sec:Quantum smectic} is not isotropic,
and it possesses the mirror symmetry only with respect to axes parallel and
perpendicular to the order parameter. Accordingly,
the chiral propagator will have to vanish for
an excitation propagating in either of these two directions.
When propagating at some different angle with respect to the
layers of the smectic, the mirror symmetry is violated, and the propagator of
Eq.~\eqref{eq:chiral strain propagator} may acquire a finite value.

We now want to perform the dualization of the theory in terms of strains variables into stresses \cite{Kleinert89b, ZaanenNussinovMukhin04}, but in order to do that, we must first become acquainted with the topological defects in solids.
 
 \section{Topological defects in solids}\label{sec:Topological defects in solids}
 
The groundstate of the ordered system is perturbed by excitations. Ordinary excitations like shear forces and propagating phonons are smooth disturbances of the order parameter. There are also excitations that diminish the order in the form of local singularities in the order parameter fields. Because the order parameter field is smooth everywhere outside of the singularity, the influence of the singularity can be noticed throughout the rest of the system. As this influence does not depend on local details, these singularities are called {\em topological defects}. Because the order parameter is an otherwise smooth field, the classification of topological defects employs the mathematics of homotopy theory, which studies the properties of mappings of closed loops and other hyperspheres that are independent of local, smooth deformations~\cite{Mermin79}. 

Since the topological defect perturbs the order, and its influence is noticeable even at long distances from the singularity, the energy of a single defect grows with distance. For instance the energy of a vortex in a superfluid grows logarithmically with distance. Therefore, one hardly finds isolated defects in long-range ordered materials; instead one finds pairs of defects--antidefects that together are topologically neutral. The creation of such a pair costs a finite amount of energy, and can be thermally excited. This excitation energy of closely-bound defect pairs is typically lower than the energy to create an {\em interstitial} which is a missing or additional particle in the crystal, i.e. a non-topological defect. From here on, we will always consider such processes prohibitively expensive, so that we take the number of particles to be conserved.

In elasticity, we are primarily interested in point defects in two, and line defects in three dimensions. They are classified by the homotopy properties of closed loops, collected in the so-called fundamental group $\pi_1(\mathcal{C})$ where $\mathcal{C}$ is the order parameter space or configuration space~\cite{Mermin79}. The order parameter space depends on the symmetry breaking pattern, which for crystals is from the continuous Euclidean group to the discrete space group $\mathbb{R}^D \rtimes O(D) \to \mathbb{Z}^D \rtimes \bar{P}$ where $\bar{P}$ is the point group of the crystal. Since both translational and rotational symmetries are broken, we expect translational and rotational types of topological defects. However the symbol $\rtimes$ denotes a {\em semidirect product} indicating that the operations of performing a translation and of a rotation do not commute. As a consequence, the translational and rotational defects are not independent, as we shall see in Sec.~\ref{subsec:Interdependence between dislocations and disclinations}.

\subsection{Volterra processes}\label{subsec:Volterra processes}
The construction of topological defects in a piece perfectly ordered crystal, can be done via a so-called Volterra process. This entails making an imaginary cut in the material, adding or removing certain particles near the cut and then gluing the altered pieces back together such that locally the crystal order is untainted, except for a point (in 2D) or line (3D) singularity. This singularity is called the defect core. An example of such a Volterra process is given in Fig.~\ref{fig:disclination volterra}. More details on the Volterra construction can be found in textbooks, for instance Refs.~\cite{Friedel64,Kleinert89b, KlemanFriedel08}.

\subsection{Dislocations and disclinations}\label{subsec:Dislocations and disclinations}
\begin{figure*}
\hfill
 \subfigure[ dislocation]{
  \includegraphics[height=4cm]{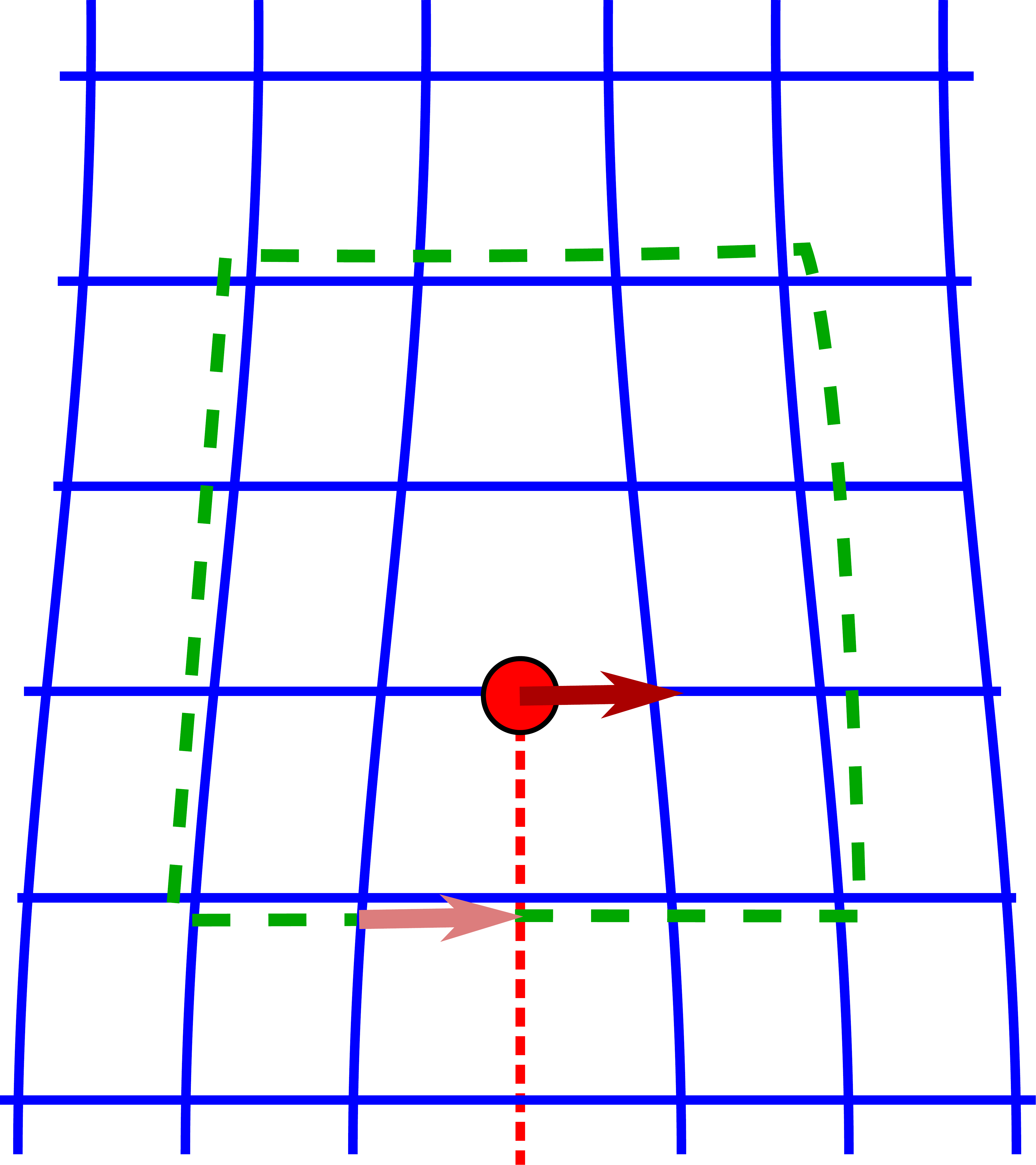}\label{fig:dislocation}
 }
 \hfill
 \subfigure[ disclination]{
  \includegraphics[height=4cm]{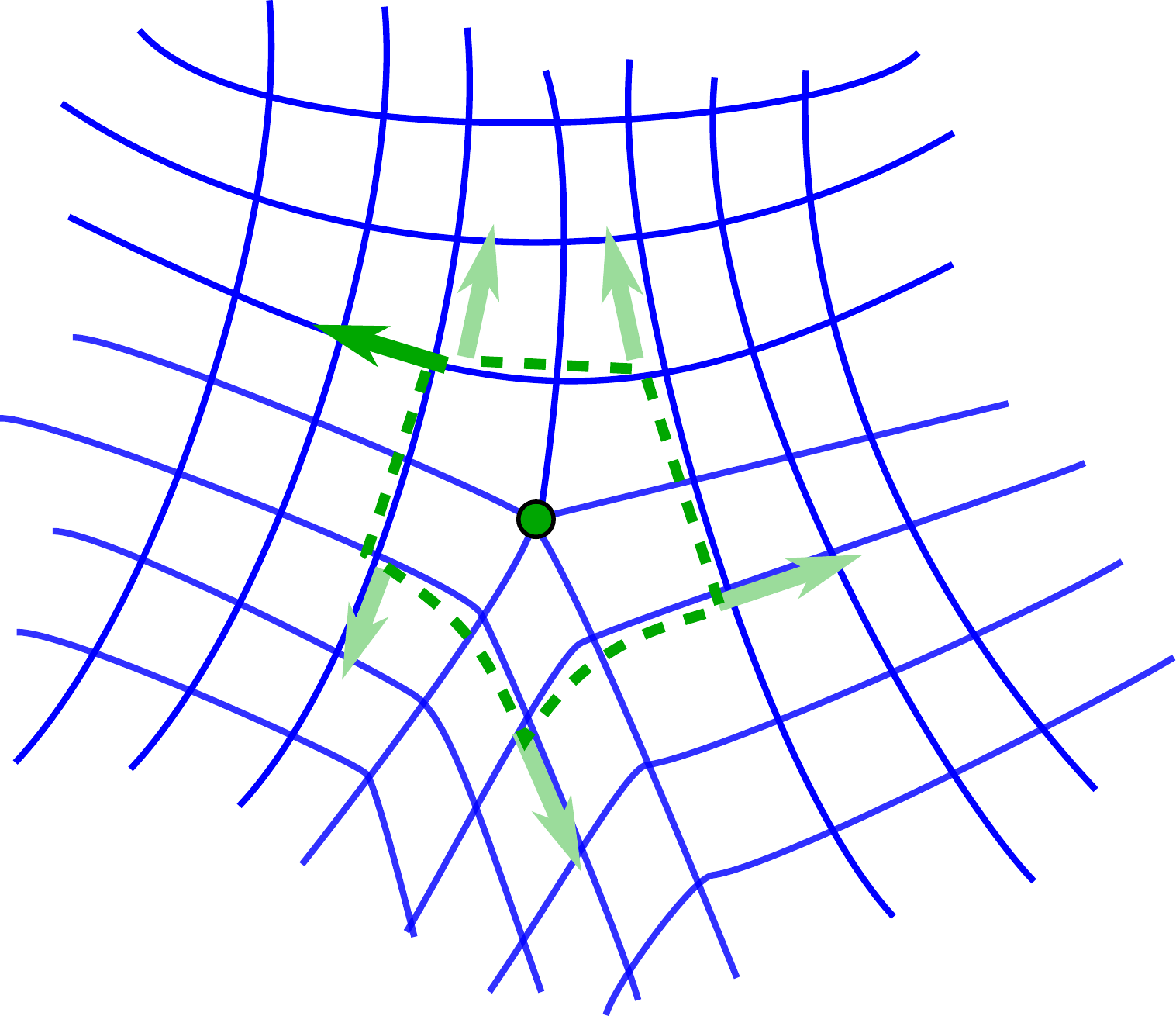}\label{fig:disclination}
 }
 \hfill
 \subfigure[ Volterra construction]{
  \includegraphics[height=4cm]{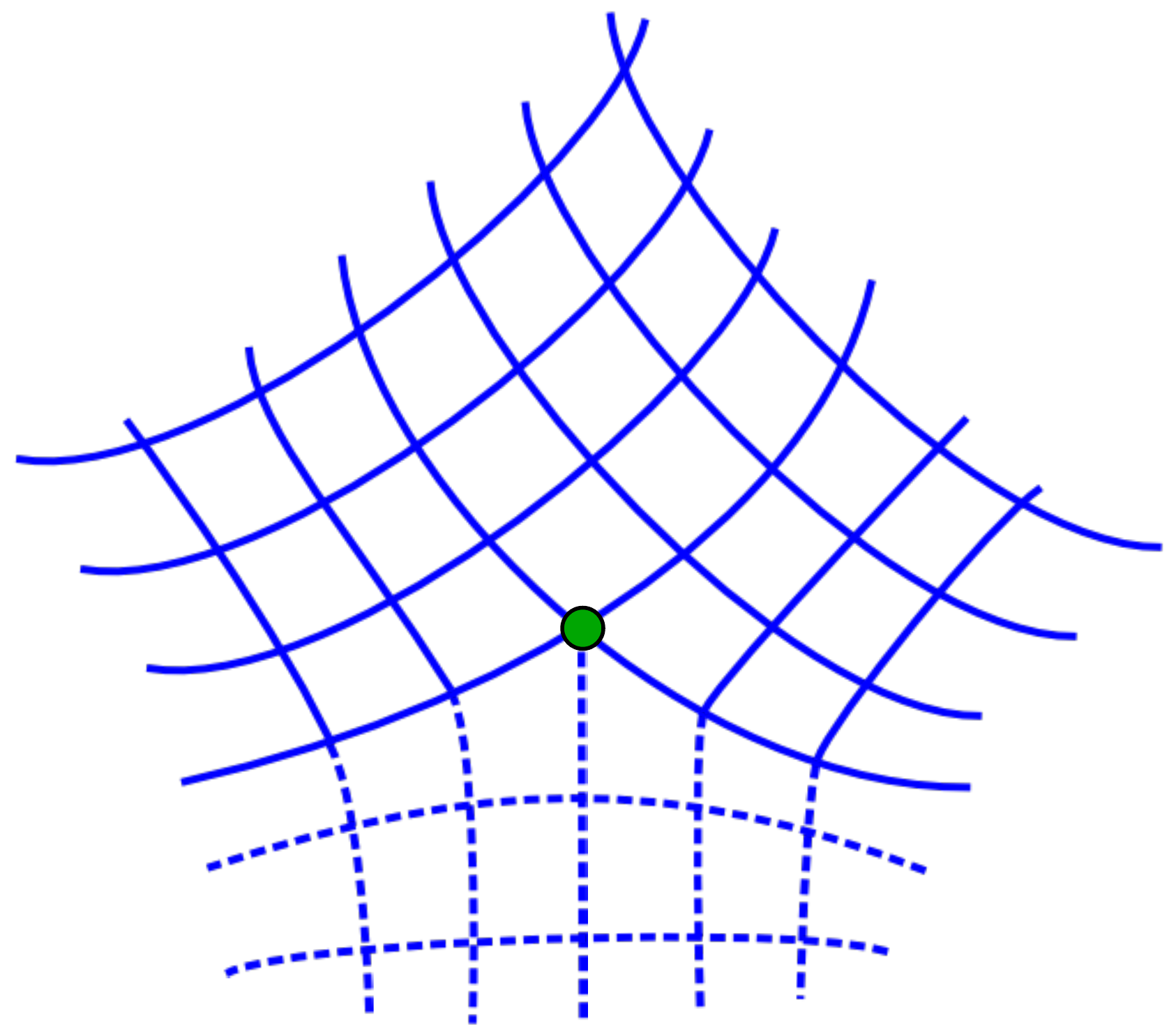}\label{fig:disclination volterra}
 }
 \hfill\null 
 \caption{Topological defects in a 2D square lattice crystal. The core of the defects is indicated by a dot. \subref{fig:dislocation} A dislocation is a half-line insertion, indicated by the dashed red line. Its topological charge is a Burgers vector, indicated by an arrow. This charge can be picked up by traversing a contour around the defect as indicated by the dotted line. \subref{fig:disclination} A disclination is the insertion of a wedge. The site at the defect core has five instead of four neighboring sites, therefore the Frank charge is $+90^\circ$. This can be picked up by parallel transport of a vector along a closed circuit around the disclination core, indicated by the dotted line.  \subref{fig:disclination volterra} The Volterra construction of a disclination. The inserted material is indicated by dashed lines. \label{fig:2D topological defects}}
\end{figure*}

The topological defects associated with translational order are called {\em dislocations}. The Volterra process for creating a dislocation is cutting the material and inserting a half-line that ends at the singularity. If we traverse a circuit around the dislocation core, we notice that there is a deficient displacement. In other words: the lattice distance becomes ill-defined near the dislocation. It is in this sense that dislocations perturb translational order. In the extreme case, dislocations proliferate and completely destroy translational order. This is the defect-mediated melting process that is the main topic of this work. The deficient lattice displacement is a vector quantity called the {\em Burgers vector} $\mathbf{B} = B^a$~\cite{Burgers39}, cf. a vortex in a superfluid Eq.~\eqref{eq:phase field contour integral},
\begin{equation}\label{eq:Burgers circuit dislocation}
\oint_{\partial \mathfrak{S}} \td x^m \partial_m u^a = B^a
\end{equation}
It is the topological charge associated with the dislocation. The circuit $\partial \mathfrak{S}$ around the dislocation core used to measure the displacement is sometimes called Burgers circuit (see Fig.~\ref{fig:dislocation}). In three dimensions there is a dislocation line $L$ with Burgers vector $B^a$. If the Burgers vector is orthogonal to the line, it is called an {\em edge} dislocation, and when the Burgers vector is parallel to the dislocation it is called a {\em screw} dislocation. By viewing a 2D system as a planar cut orthogonal to the defect line, 2D dislocations can be said to be edge dislocations. 

In real materials, degradation of integrity, e.g. metal fatigue, is due to the introduction of dislocations caused by severe strain. Once these dislocations have formed, they are very difficult to remove and have non-trivial dynamics, precisely because of their topological nature. Dislocations can move, but only parallel to their Burgers vector (glide motion). Motion in the perpendicular direction (climb motion) would come down to the addition or removal of particles, and is therefore energetically very costly. This can be turned into a mathematically precise statement, the glide constraint, see Sec.~\ref{subsec:Kinematic constraints}. Dislocations are sources of torsion \cite{Kleinert89b}: a material with dislocations is torsionally strained, and imposing torsion from the outside on a material can induce the formation of dislocations.

The topological defects associated with rotational order are called {\em disclinations}. The Volterra process for creating a disclination is making a cut and inserting or removing a whole wedge of material, see Fig.~\ref{fig:disclination volterra}. In a crystal, the operation must always be an element of the point group (e.g. a multiple of 90$^\circ$ rotation in a square lattice) to ensure the gluing together is locally smooth, but in the continuum limit we may envisage wedges of any angle. The topological charge is obtained by the angle of rotation in parallel transporting of a vector around the disclination core, see Fig.~\ref{fig:disclination}. It is called the {\em Frank angle} $\Omega^{ab}$ or \emph{tensor} $\Omega^{c_1 \cdots c_{D-2}} = \epsilon_{c_1 \cdots c_{D-2} ab} \tfrac{1}{2}\Omega^{ab}$~\cite{Frank58},
\begin{equation}\label{eq:Burgers circuit disclination}
\oint_{\partial \mathfrak{S}} \td x^m \partial_m \epsilon_{c_1 \cdots c_{D-2} ab}\tfrac{1}{2} \omega^{ab} = \Omega^{c_1 \cdots c_{D-2}}.
\end{equation}
In $D$ spatial dimensions, the Frank tensor has $D-2$ components. So in 2D it is a scalar $\Omega$ and in 3D a vector $\Omega^c$. The Frank tensor points perpendicular to the plane of rotation, and its magnitude denotes the deficient angle. In 3D, the straightforward generalization of a 2D disclination is to have the disclination line pointing out of the plane in Fig.~\ref{fig:disclination}, that is, parallel to its Frank vector. This is called a {\em wedge} disclination for obvious reasons. When the Frank vector is orthogonal to the disclination line it is called a {\em twist} disclination.

Since the disclination Volterra process involves much more inserted or removed material than a dislocation, it is energetically much more costly. In practice, in a crystal one never finds isolated disclinations. One may find bound disclination--antidisclination pairs which actually corresponds to dislocations as we will explain in Sec.~\ref{subsec:Interdependence between dislocations and disclinations}. Disclinations are sources of curvature. If one tries to impose a regular lattice on a curved surface, it necessarily involves the introduction of disclinations. Famously, it follows from Euler's characteristic that putting a triangular lattice on a sphere involves twelve $-60^\circ$-disclinations.

\subsection{Defect densities}\label{subsec:defect densities}
The Burgers vector and Frank tensor are topological charges of the defects as a whole. However, the cores of the defects are localized, and we can describe them as local, field-theoretic objects as follows. For clarity, we first introduce static line defects in 3+0D, and then discuss the peculiarities of going to 2+1D with a special time axis.

Let $\delta_m(L,x)$ be the Dirac delta function of unit vectors tangent to the line $L$ of the defect core in 3+0D. That is, if we integrate over a two-dimensional surface with normal $\widehat{\bf{n}}$ that is pierced by the line $L$ with tangent $\bf{t}$, the integral is $\widehat{\bf{n}} \cdot \bf{t}$. To be precise, let the line $L$ have coordinates $x_k^L(s)$ parametrized by $s$, then~\cite{Kleinert89b,Kleinert08},
\begin{equation}\label{eq:line delta function}
 \delta_m (L,x) = \int_L \td s\ \partial_{s} x^L_m(s)  \delta^{(D)}\big(x - x_k^L(s) \big).
\end{equation}
Here the integral is over the whole line $L$. This definition can be generalized to other dimensions, in particular to 3+1D dimensions where the line defects trace out worldsheets in spacetime, which are divided into surface elements with two spacetime indices $\delta_{\mu\nu}(L,x)$, or in $D+1$ dimensions to hypersurfaces with $D-1$ spacetime elements  $\delta_{\mu_1 \cdots \mu_{D-1}}(L,x)$. If the line $L$ is closed (or extends to infinity), this function satisfies $\partial_m \delta_m(L,x) = 0$ everywhere.

We can now define the {\em dislocation density} $J^a_m(x)$ and the {\em disclination density} $\Theta^c_m(x)$ as follows:
\begin{align}
J^a_m(x) &= \delta_m (L,x) B^a, \label{eq:dislocation density}\\
\Theta^c_m(x) &= \delta_m (L,x) \Omega^c. \label{eq:disclination density}
\end{align}
In this text, upper indices always refer to the spatial components of the topological charge, while the lower indices correspond to normal space(time) indices of the worldline. 

Just as in vortex--boson-duality Eq.~\eqref{eq:multivalued phase field}, the displacement field $u^a(x)$ of Eq.~\eqref{eq:displacement definition} can be split into smooth and singular contributions: it is a multivalued field. The defect densities follow directly. Everywhere away from the singularity, the displacement field is smooth and single valued. Then derivatives commute, and $(\partial_m \partial_n - \partial_n \partial_m) u^a(x)$ always vanishes. At the singularity the displacement field is multivalued, and we can define in 3+0D,
\begin{equation}\label{eq:dislocation density multivalued displacement}
 J^a_m (x) = \epsilon_{mkl} \partial_k \partial_l u^a(x). 
\end{equation}
The combination of derivatives $(\partial_m \partial_n - \partial_n \partial_m)$ can be pictured as the tiniest Burgers circuit around the defect core. This agrees with Eq.~\eqref{eq:dislocation density} as the dislocation core is located at the singularity. Similarly we have,
\begin{equation}\label{eq:disclination density multivalued rotations}
 \Theta^c_m (x) = \epsilon_{mkl} \partial_k \partial_l \tfrac{1}{2}\epsilon^{cab} \omega^{ab}(x). 
\end{equation}
Here $\omega^{ab} = \partial_a u^b - \partial_b u^a$ is the local rotation tensor. These definitions generalize straightforwardly to higher dimensions $D$+0 as
\begin{align}
 J^a_{m_1 \cdots m_{D-2}}(x)  &= \epsilon_{m_1 \cdots m_{D-2}kl} \partial_k \partial_l u^a(x), \label{eq:D+0 dislocation density}\\
 \Theta^{c_1 \cdots c_{D-2}}_{m_1 \cdots m_{D-2}}(x)  &= \epsilon_{m_1 \cdots m_{D-2}kl} \partial_k \partial_l \tfrac{1}{2}\epsilon^{c_1 \cdots c_{D-2}ab} \omega^{ab}(x). \label{eq:D+0 disclination density}
\end{align}
Using Stokes' theorem and Eqs.~\eqref{eq:line delta function}, \eqref{eq:dislocation density} and \eqref{eq:disclination density}, Eqs.~\eqref{eq:dislocation density multivalued displacement} and \eqref{eq:disclination density multivalued rotations} lead back to Eqs.~\eqref{eq:Burgers circuit dislocation} and \eqref{eq:Burgers circuit disclination}.

In quantum elasticity, we introduce a time axis. However, the displacements $u^a(x,t)$ are always spatial; there can be no temporal component of the displacement field ($u^t=0)$. This also implies that Burgers vectors $B^a$ are always spatial, as are rotations $\omega^{ab}$, such that one of the indices in the Frank tensor $\Omega^{c_1 \cdots c_{D-2}}$ must be the temporal one. For instance, in 2+1D, there is only one rotational plane, and $\Omega^\ft \equiv \Omega$ is the Frank scalar. Thus in $D+1$ dimensions we have
\begin{align}
 J^a_{\mu_1 \cdots \mu_{D-1}}(x)  &= \epsilon_{\mu_1 \cdots \mu_{D-1}\kappa\lambda} \partial_\kappa \partial_\lambda u^a(x),\label{eq:D+1 dislocation density}\\
 \Theta^{c_1 \cdots c_{D-2}}_{\mu_1 \cdots \mu_{D-1}} (x)  &= \epsilon_{\mu_1 \cdots \mu_{D-1}\kappa\lambda} \partial_\kappa \partial_\lambda \tfrac{1}{2}\epsilon^{t c_1 \cdots c_{D-2}ab} \omega^{ab}(x).\label{eq:D+1 disclination density}
\end{align}
Specializing back to 2+1D, the defect densities we shall use here are
\begin{align}
  J^a_\mu &= \epsilon_{\mu \kappa\lambda} \partial_\kappa \partial_\lambda u^a(x),\label{eq:2D dislocation density}\\
 \Theta_\mu &= \epsilon_{\mu\kappa\lambda} \partial_\kappa \partial_\lambda \tfrac{1}{2}\epsilon^{t ab} \omega^{ab}(x).\label{eq:2D disclination density}
\end{align}

\begin{figure*}
\hfill
 \subfigure[ atoms]{
  \includegraphics[height=4cm]{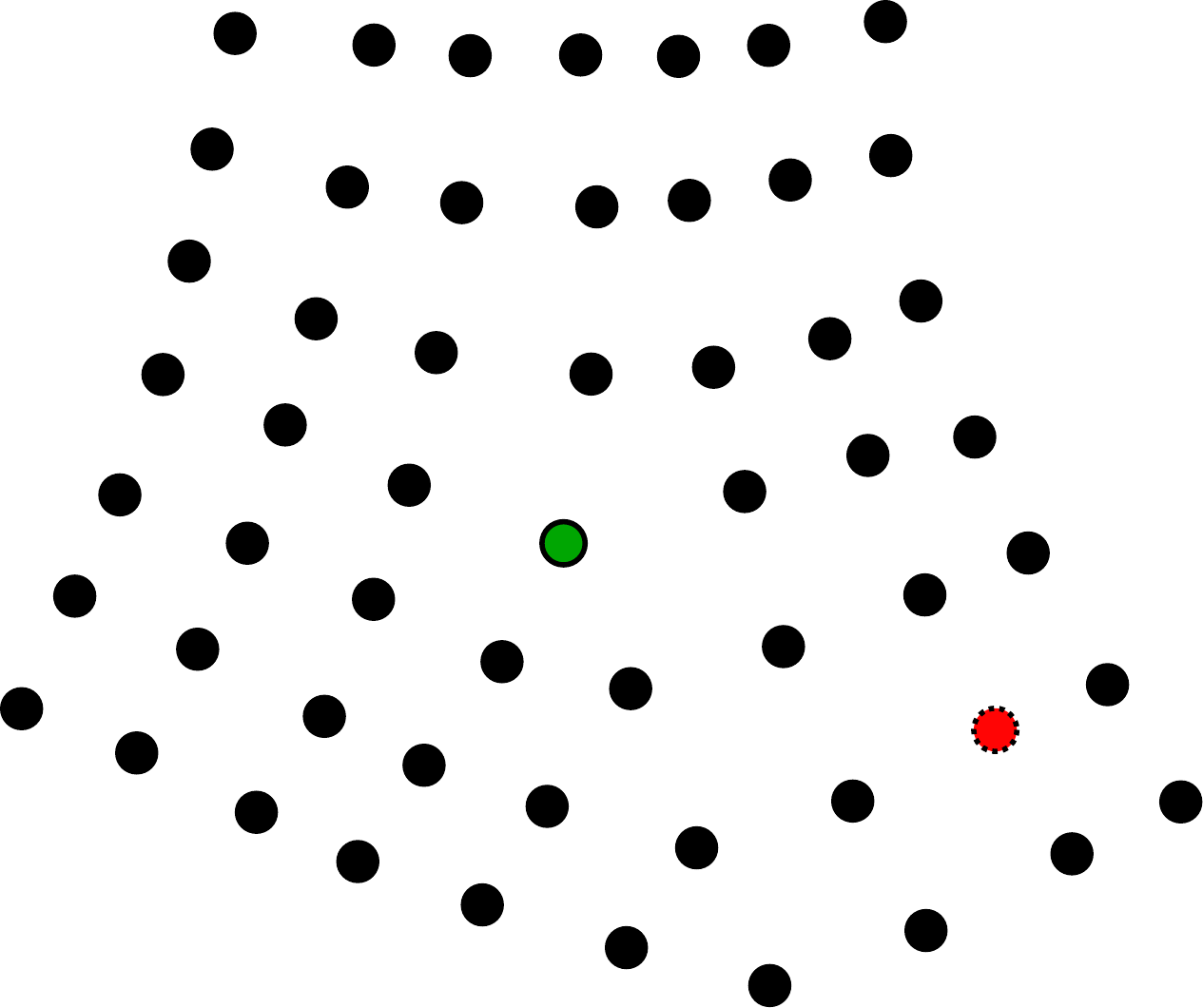}\label{fig:disclination as atoms}
 }
 \hfill
 \subfigure[ disclination]{
  \includegraphics[height=4cm]{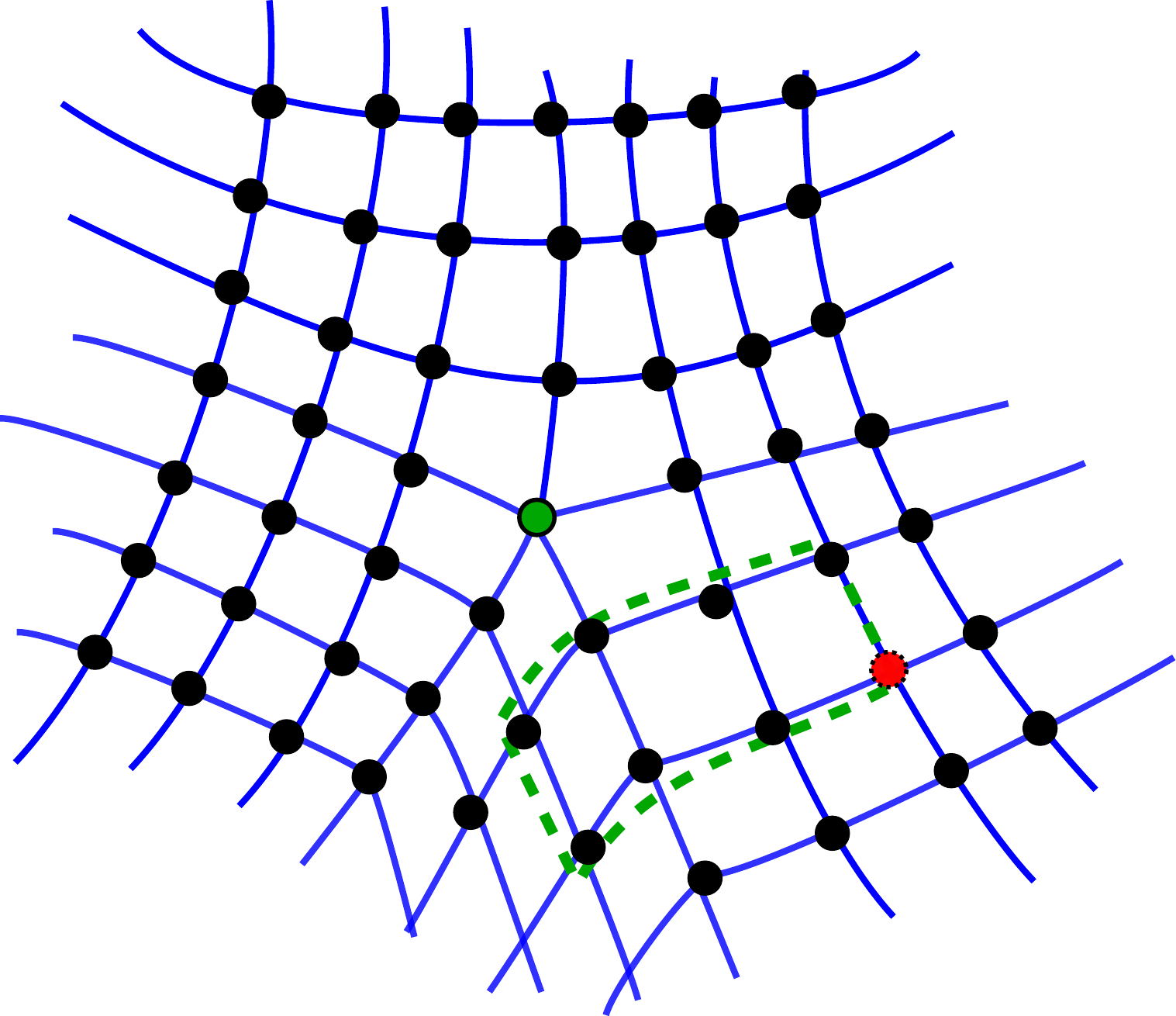}\label{fig:disclination as disclination}
 }
 \hfill
 \subfigure[ stack of dislocations]{
   \includegraphics[height=4cm]{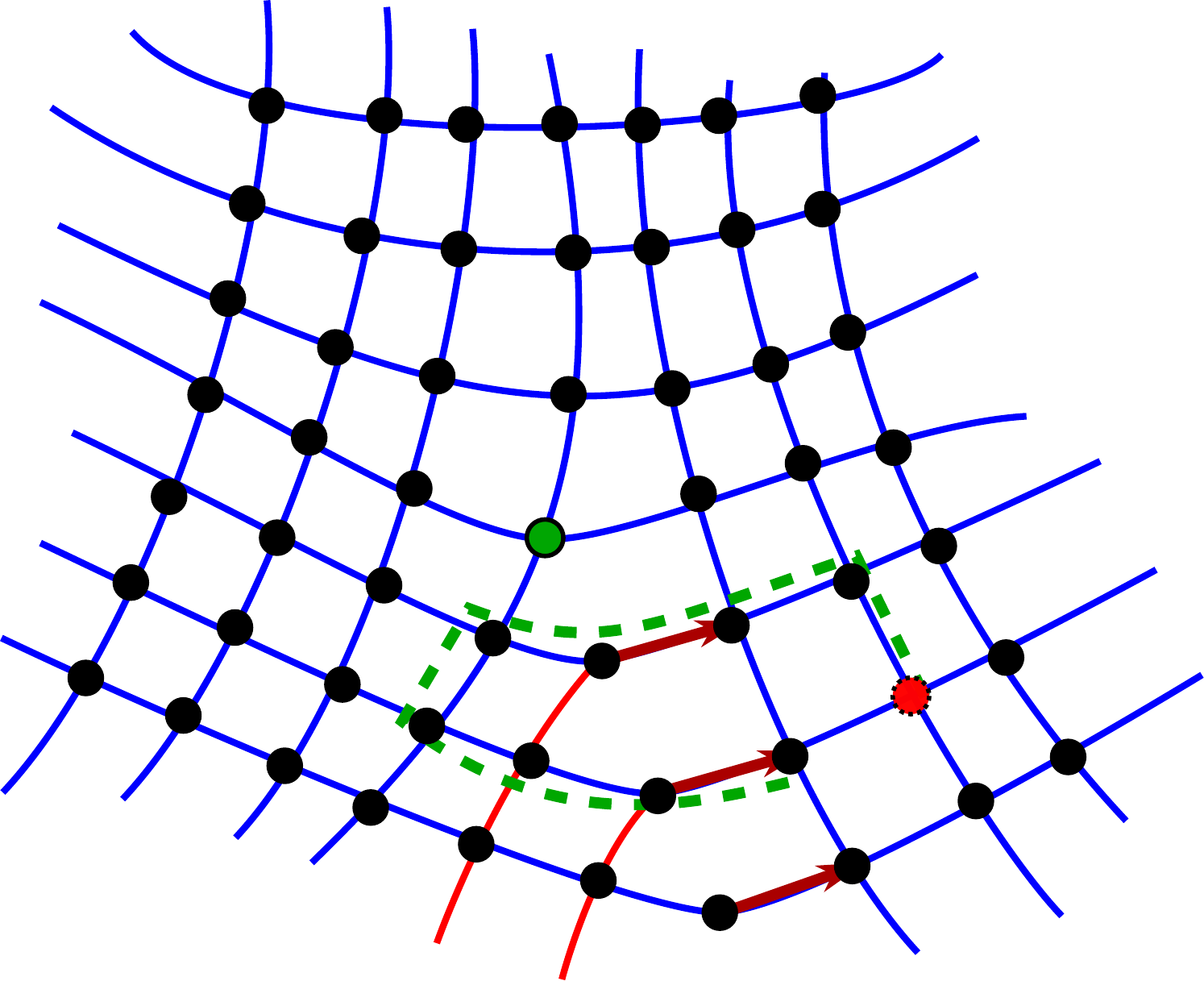}\label{fig:disclination as dislocations}
 }
 \hfill\null
 \caption{Equivalence between a disclination and a stack of dislocations. \subref{fig:disclination as atoms} Lattice sites of a strained piece of crystal. The two colored points are indicated for reference. The assignment of the topological charge is partly arbitrary. \subref{fig:disclination as disclination} The configuration corresponds to a $+90^\circ$-disclination, the same one as in Fig.~\ref{fig:disclination}. Taking a Burgers circuit starting from the red, dotted point shows there is no displacement deficiency, and the Burgers vector is zero. \subref{fig:disclination as dislocations} The same configuration can be assigned as an infinite stack of dislocations, indicated by red lines and arrows. The Frank charge is zero, but a Burgers circuit from the same starting point yields a non-vanishing Burgers vector. \label{fig:disclination dislocation interdependence}}
\end{figure*}

For individual topological defects, the property $\partial_m \delta_m(L,x) =0$ ensures, in 3+0D, 
\begin{align}
 \partial_m J_m^a &= 0,& \partial_m\Theta^c_m &= 0.
\end{align}
These equations state that dislocation and disclination lines cannot end within a piece of material; they must extend to infinity. In other words, there are no monopoles associated with dislocations and disclinations. In 2+1D, these equations become,
\begin{align}
 \partial_\mu J_\mu^a &= 0,& \partial_\mu\Theta^c_\mu &= 0.
\end{align}
These are continuity equations for the particle-like topological defects, stating that the defect density can only increase (decrease) when it flows into (out of) the region in question. These equations hold for individual, isolated defects. In the presence of disclinations, the equation for dislocations change, which is the topic of the next subsection.

\subsection{Interdependence between dislocations and disclinations}\label{subsec:Interdependence between dislocations and disclinations}

For the same reason that translations and rotations are not independent in the symmetry group $\mathbb{R}^D \rtimes O(D)$, dislocations and disclinations are not independent. This is shown pictorially in Fig.~\ref{fig:disclination dislocation interdependence}. We will sharpen this statement in the following.

In the presence of a topological defect at the origin, we want to know the deficient displacement across the cutting surface where some slab of material was inserted in the Volterra process. For a single dislocation, the deficient displacement $\Delta u^a$ is everywhere the same across the inserted half-line, as can be seen in Fig.~\ref{fig:dislocation}. For a disclination however, since a wedge of material is inserted, the deficient displacement increases with distance from the disclination core. Take two reference points $x^1$, $x^2$, then the deficient displacement in the presence of a disclination with deficient rotation $\bar{\Omega}^{ab} \equiv \epsilon^{abc} \Omega^c \equiv \Delta \omega^{ab}$ is
\begin{equation}\label{eq:displacement with rotation}
 \Delta u^a (x^2) = \Delta u^a(x^1) + \bar{\Omega}^{ab} (x^2_b - x^1_b).
\end{equation}
This simply states that the displacement in the $x$-direction increases with separation in the $y$-direction from the core of the rotational defect in the $xy$-plane. A more thorough derivation can, for instance, be found in Ref.~\cite{Kleinert89b}. Thus a deficient displacement can be caused both by dislocations and disclinations. Then a disclination can be interpreted as a collection of dislocations. 

\begin{figure*}
\hfill
 \subfigure[ initial dislocation]{
  \includegraphics[height=4cm]{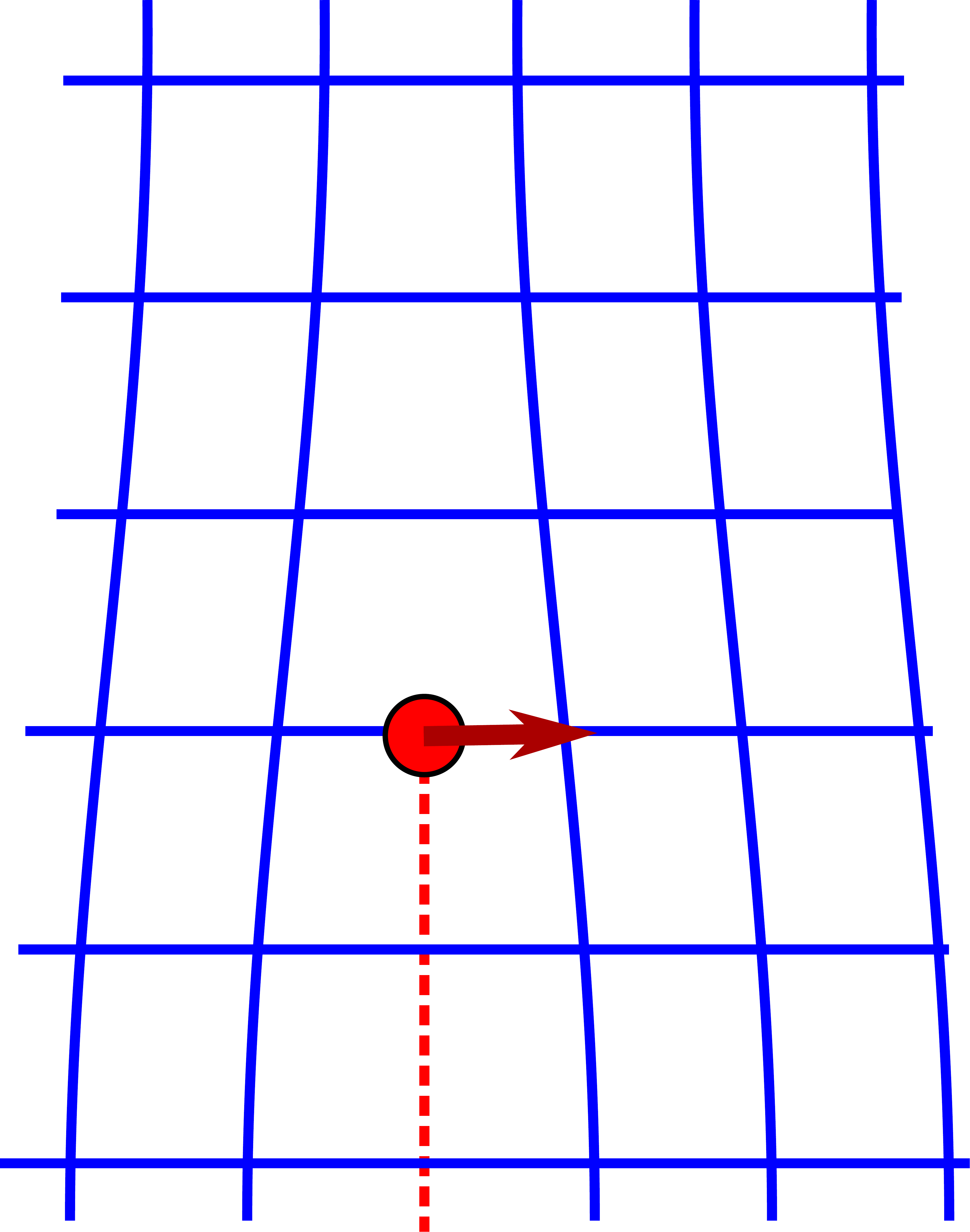}\label{fig:dislocation start}
 }
 \hfill
 \subfigure[ glide motion]{
  \includegraphics[height=4cm]{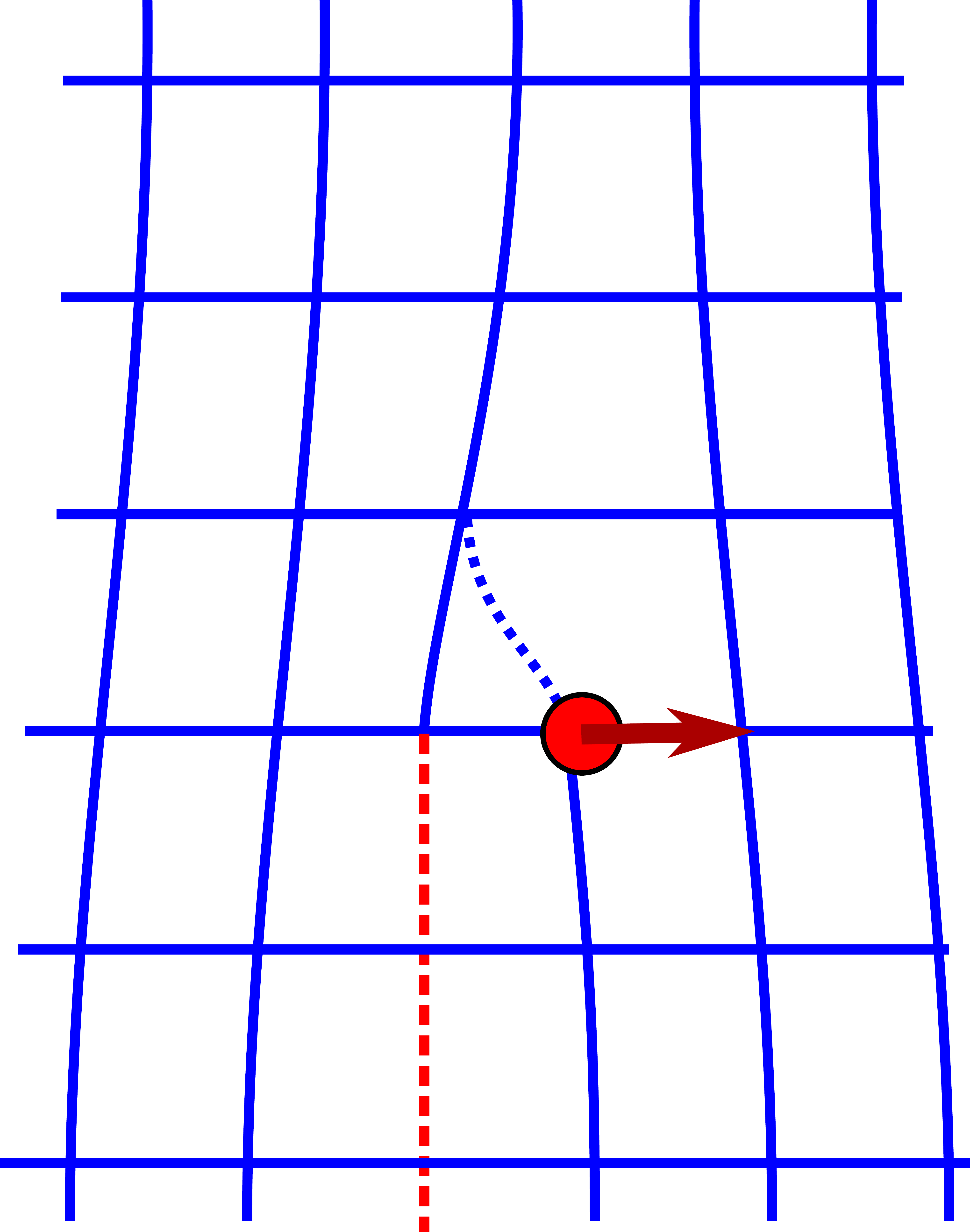}\label{fig:dislocation glide}
 }
 \hfill
 \subfigure[ climb motion]{
  \includegraphics[height=4cm]{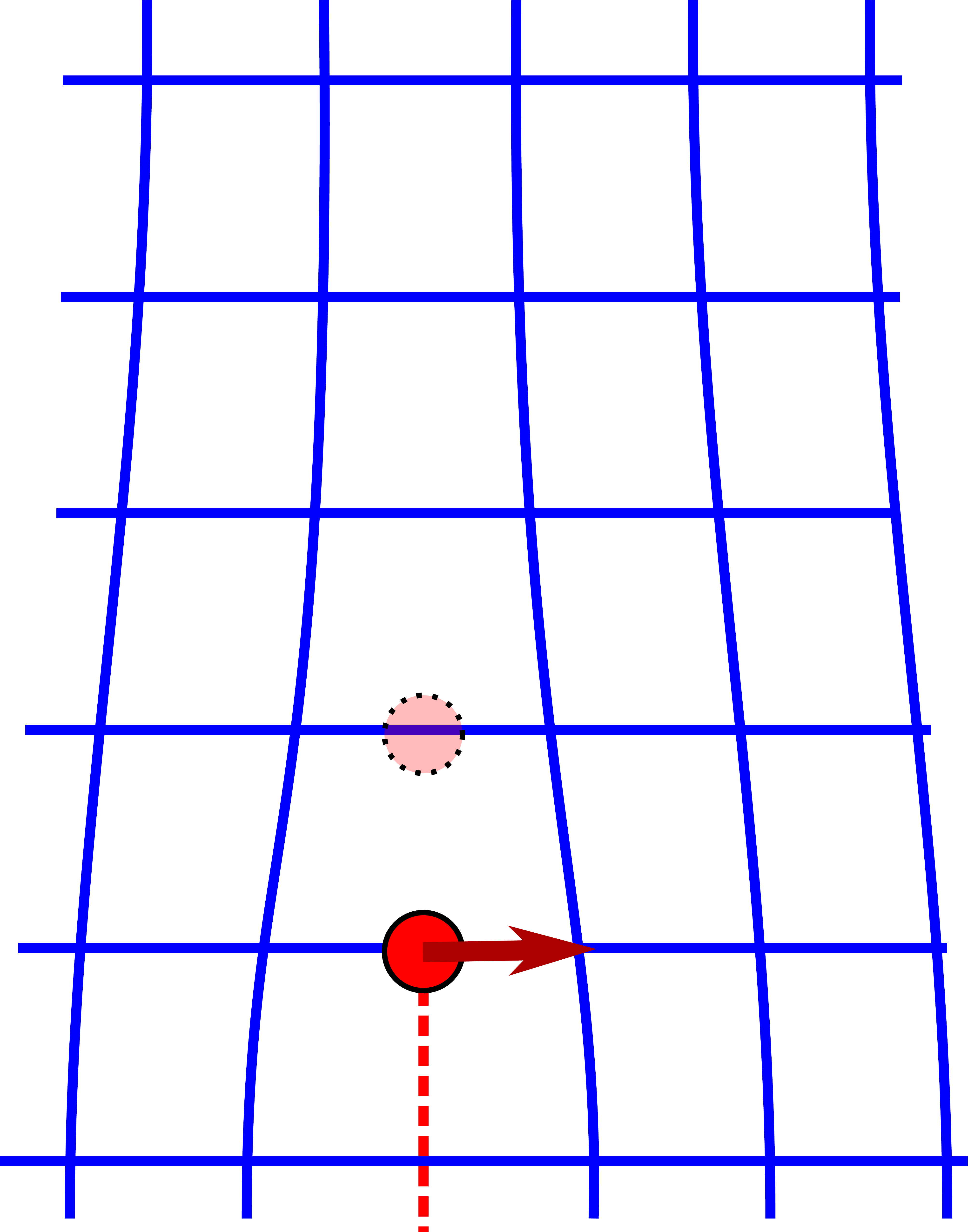}\label{fig:dislocation climb}
 }
 \hfill\null
 \caption{Motion of a 2D dislocation. \ref{fig:dislocation start} Initial configuration with Burgers vector indicated by the arrow. \ref{fig:dislocation glide} Glide motion parallel to the Burger vector. This is only a change of a single `molecular' bond, and the energy cost for such a process is limited and of dynamic nature. \ref{fig:dislocation climb} Climb motion orthogonal to the Burgers vector. The particle with dotted outline does not fit into the resulting lattice. Climb necessarily entails the addition or removal of an interstitial particle and is energetically very costly.}\label{fig:dislocation motion}
\end{figure*}

We would like to translate the above equation into a statement about defect densities. Away from the rotational defect, we could assign the deficient displacement as a local dislocation density. That is, in the presence of a disclination with Frank vector $\bar{\Omega}^c = \tfrac{1}{2}\epsilon^{cab} \bar{\Omega}^{ab}$ at the origin, consider the dislocation density of a line $L$ at a point $\bar{x}$ away from the origin, which itself possibly has a Burgers vector $B^a$. Then from Eq.~\eqref{eq:displacement with rotation} we find
\begin{equation}\label{eq:dislocation density from Frank vector}
 J^a_m (x) = \delta_m(L,x) (B^a + \bar{\Omega}^{ab} \bar{x}^b),
\end{equation}
But since the delta function is non-zero only at $L$, we can identify $\bar{x} = x$. This equation states that a stack of dislocations emanates from the disclination core, as pictured in Fig.~\ref{fig:disclination dislocation interdependence}. Another important consequence of this equation is that a disclination--antidisclination pair is in fact a dislocation with Burgers vector equal to the separation between the disclination and antidisclination cores. Therefore a disclination--antidisclination pair is topologically neutral as far as deficient rotations are concerned, but not so for translations. A completely topologically neutral combination involves at least one other disclination--antidisclination pair that cancels the Burgers vector of the first pair. The other side of the coin is that a net non-zero Burgers vector corresponds to a finite density of disclination--antidisclination pairs. If the presence of disclinations is restricted, for instance dynamically, then dislocations can only appear in topologically neutral combinations, that is, in dislocation--antidislocation pairs. In 2+1D spacetime this corresponds to having closed spacetime loops of such pairs, which can be viewed as the spontaneous creation and annihilation of dislocation pairs.

From the non-local equation Eq.~\eqref{eq:dislocation density from Frank vector}, we can obtain a very important local equation by taking a derivative. Using the Leibniz rule and $\partial_m \delta_m(L,x) =0$ we find
\begin{align}
 \partial_m J^a_m (x) &= \delta_b(L,x) \bar{\Omega}^{ab} = \epsilon^{abc} \delta_b(L,x) \Omega^c
 = \epsilon^{abc} \Theta^c_b (x).
\end{align}
Here we used $\bar{\Omega}^{ab}  = \epsilon^{abc} \Omega^c$ and Eq.~\eqref{eq:disclination density}. It can also be derived from Eqs.~\eqref{eq:dislocation density multivalued displacement} and \eqref{eq:disclination density multivalued rotations}. This equation states that the dislocation density is not conserved in any region where the disclination line is not parallel to its Frank vector. Recall that a twist dislocation has its Frank vector orthogonal to its defect line. Thus twist disclinations necessarily involve dislocation lines ending on the disclination line. The change of the rotational axis from one position to the next necessitates the addition and removal of dislocations. 

In 2+1 dimensions, this becomes even clearer:
\begin{equation}\label{eq:dislocation conservation with disclination}
  \partial_\mu J^a_\mu (x) = \epsilon^{ab} \Theta_b (x).
\end{equation}
Now the left hand side can be considered as an actual dislocation creation process. The right-hand side contains the current or flow $\Theta_b$ of disclination density $\Theta_t$ in the direction $b$. Thus, when the disclination moves, it creates or annihilates dislocations with Burgers vector in the orthogonal direction.  This is also clear from Fig.~\ref{fig:disclination dislocation interdependence}: if we were to move the position of the disclination core up one lattice spacing, we need to insert an additional dislocation line. The same point of view is taken in Ref.~\cite{KlemanFriedel08}.

Let us stress that any disclination line can be pictured as a stack of dislocations. This is a static picture. But the motion of the disclination perpendicular to its Frank vector will create and annihilate dislocations. In 2+1D, any motion of the disclination is perpendicular to its Frank vector, which necessarily lies in the temporal direction.

In the following, we will see how the dislocations and disclinations are, respectively, the sources of torsional stress and curvature stress~\cite{Kleinert89b}. It is important to keep Eq.~\eqref{eq:dislocation conservation with disclination} in mind, as it shows that these sources are not always independent of each other.

\subsection{Kinematic constraints}\label{subsec:Kinematic constraints}
As we mentioned, the motion that dislocations can undergo is restricted to {\em glide} motion, parallel to the Burgers vector, whereas {\em climb} motion orthogonal to its Burgers vector is energetically strongly disfavored. The intuitive reason is simple, see Fig.~\ref{fig:dislocation motion}: for glide motion, it is only necessary that some `molecular' bond be broken and restored somewhere else. The energy of the before- and after-configurations are the same. The reader can already imagine that having many dislocations facilitates this sort of gliding motion, that is, the resistance to shear is lessened. The presence of dislocations leads to the loss of shear rigidity, and the melting to a liquid crystal is the proliferation of dislocations where the restoring forces for shear are completely absent.

Conversely, climb motion requires the addition or removal of a particle from the regular lattice, in other words, an interstitial excitation. Such interstitials are usually very costly. If we are operating at energy scales well below the interstitial creation energy, we can neglect climb altogether, and this would be the whole story. But we can do one better than that, using the dislocation density defined in Eq.~\eqref{eq:dislocation density multivalued displacement}. We summarize here the arguments put forward in Ref.~\cite{CvetkovicNussinovZaanen06}.

Recall that $J^a_t(x)$ is the density of dislocations with Burgers vector $B^a$ at point $x$, and $J^a_m(x)$ is the current of those dislocations in direction $m$. In the absence of any disclinations, they satisfy the continuity equation $\partial_\mu J^a_\mu = 0$ via Eq.~\eqref{eq:dislocation conservation with disclination}. Now the glide constraint is a dynamical constraint, and as such does not pertain to the (static) density $J^a_t$ but only to the current $J^a_m$. Clearly $J^a_a$ (no sum) is the flow in the direction of the Burgers vector, i.e. glide, while $J^a_m,\ a\neq m$ is flow orthogonal to the Burgers vector, i.e. climb. The constraint turns out to be
\begin{equation}\label{eq:2D glide constraint}
\epsilon_{am} J^a_m (x) = 0.
\end{equation}
Here $\epsilon_{am} \equiv \epsilon_{t am}$. 
To better appreciate the physical content of Eq.~\eqref{eq:2D glide constraint}, 
we may examine Fig.~\ref{fig:dislocation start} with $y$ denoting the vertical and $x$ the horizontal direction.
Thus, $J^x_y$ is the current of the dislocation along the $y$ direction associated with the Burgers vector that in Fig. \ref{fig:dislocation} lies along the $x$ axis.
Eq.~\eqref{eq:2D glide constraint} states that $J^y_x= J^x_y$. However, in Fig. \ref{fig:dislocation start}, the current $J^y_x$ identically vanishes as the Burgers vector only has an $x$-component. Thus, if correct,
Eq.~\eqref{eq:2D glide constraint} mandates that $J^y_x=0$. The equality $J^y_x=0$ states that climb motion (current orthogonal to the Burgers vector direction) is impossible.
The derivation of Eq.~\eqref{eq:2D glide constraint} is therefore tantamount to proving that only glide motion is possible.  
This equation actually has a very sound basis in conservation of mass~\cite{CvetkovicNussinovZaanen06}. To see this, substitute Eq. \eqref{eq:dislocation density multivalued displacement} in this equation to find
\begin{equation}\label{eq:glide constraint mass conservation}
 0 = \epsilon_{am} J^a_m (x) = \epsilon_{am} \epsilon_{m \kappa \lambda} \partial_\kappa \partial_\lambda u^a = (\partial_t \partial_a - \partial_a \partial_t) u^a.
\end{equation}
Now let the mass density be $\rho(x)$. To lowest order it is the uniform density $\rho_0$ minus volume distortions $\rho = \rho_0 (1 - \partial_a u^a)$. Meanwhile, the mass density current is $j_a = \rho_0 \partial_t u^a$. Then the conservation of mass, or continuity equation $\partial_t \rho + \partial_a j_a = 0$ coincides precisely with Eq.~\eqref{eq:glide constraint mass conservation}. 

This derivation is valid for higher dimensions as well. For instance, in 3+1D, $J^a_{mn}$ denotes the flow in direction $n$ of the dislocation line along direction $m$ with Burgers vector $B^a$. The associated density is $J^a_{m t}$. Contracting the mass conservation law $(\partial_t \partial_a - \partial_a \partial_t) u^a =0$ with the four-dimensional Levi-Civita symbol $\epsilon_{\mu\nu\kappa\lambda}$ leads to the glide constraint $\epsilon_{tamn} J^a_{mn} = 0$. In general dimensions $D+1$, the statement is~\cite{CvetkovicNussinovZaanen06}:
\begin{equation}\label{eq:glide constraint arbitrary dimension}
 \epsilon_{t a m_1 \cdots m_{D-1}} J^a_{ m_1 \cdots m_{D-1}} = 0.
\end{equation}
Note that only spatial components of the dislocation current are involved, and that all the spacetime indices $m_i$ are orthogonal to the Burgers vector. In dimensions higher than two, one could imagine that the dislocation line could not necessarily be a straight line, and hence an interstitial at one point could be `borrowed' from another point along the line. This is sometimes called {\em restricted climb}. In that case, the constraint Eq.~\eqref{eq:glide constraint arbitrary dimension} is not strictly zero locally, but instead we have $\int_V \td^D x\; \epsilon_{ta m_1 \cdots m_{D-1}} J^a_{ m_1 \cdots m_{D-1}} = 0$. Such an integral constraint can be enforced in a partition function by a Lagrange multiplier. This is exactly what we need in order to implement the glide constraint dealing with an ensemble of dislocation lines that is used in defect-mediated melting, which we will do in Sec.~\ref{subsec:Dislocation disorder field theory}. It turns out that the glide constraint ensures that compression rigidity is maintained even in the liquid crystal. The consequence is that the zero temperature bosonic liquid crystal is at the same time a superfluid with a longitudinal zero sound mode. We will come back to this point in Secs.~\ref{subsec:Mode content of the quantum nematic} and \ref{subsec:Mode content of the quantum smectic}.

\subsection{Interstitials and vacancies}

There is yet another crystalline defect that we have not discussed at length~\cite{ZaanenNussinovMukhin04, Friedel64}. These are the interstitials or vacancies that are formed by extra or missing atoms in the lattice and are \emph{non-topological} point-like defects. In a real crystal, these are finite energy excitations and always present at finite temperatures. In fact, the real-world crystal is a mixture of the crystalline condensate with a gas of interstitials and vacancies, similarly as e.g. a superfluid has both normal and superfluid components. The conservation of mass implies the combined conservation of interstitials and vacancies. Upon the melting of the lattice, these excitations should clearly become effectively liberated as the constituents are free to meander to arbitrary positions. At zero temperature, the crystal is always well defined and for strong potentials $\mathcal{V}$, the interstitials and vacancies are just very heavy virtual excitations dressing the fluctuating zero-temperature crystal. At some critical couplings these bosons can in principle be liberated and form a finite density Bose-condensate living in harmony with the background crystal: this is the disputed and elusive supersolid \cite{KimChan04, NussinovEtAl07, BalatskyEtAl07, Prokofev07, BoninsegniProkofev12}. We feel that this vestigial phase, however, is not a generic situation and requires a fined tuned competition between the kinetic energy (the constituent mass $M_i$), the lattice potential $\mathcal{V}$ and a small energy for the excitations well below the topological defects for the superfluid to form.

In two dimensions in fact, topologically we can interpret an interstitial or a vacancy as a bound pair of a dislocation and an anti-dislocation with a separation of the order of the lattice spacing. Importantly, such dislocation dipoles do not disrupt the lattice at long distances. In 2+1-dimensions these are small dislocation loops that can be in principle be incorporated in perturbation theory dressing the vacuum (the crystal). However, we will actually develop the theory in the opposite limit of dilute and well-separated dislocations that proliferate by unbinding.

In the following we will neglect the interstitials and vacancies altogether focusing on the orderly crystalline lattice with the displacement fields encoding the associated symmetry breaking.

\begin{figure}[t]
\null\hfill
 \subfigure[ triangular]{\includegraphics[height=3.5cm]{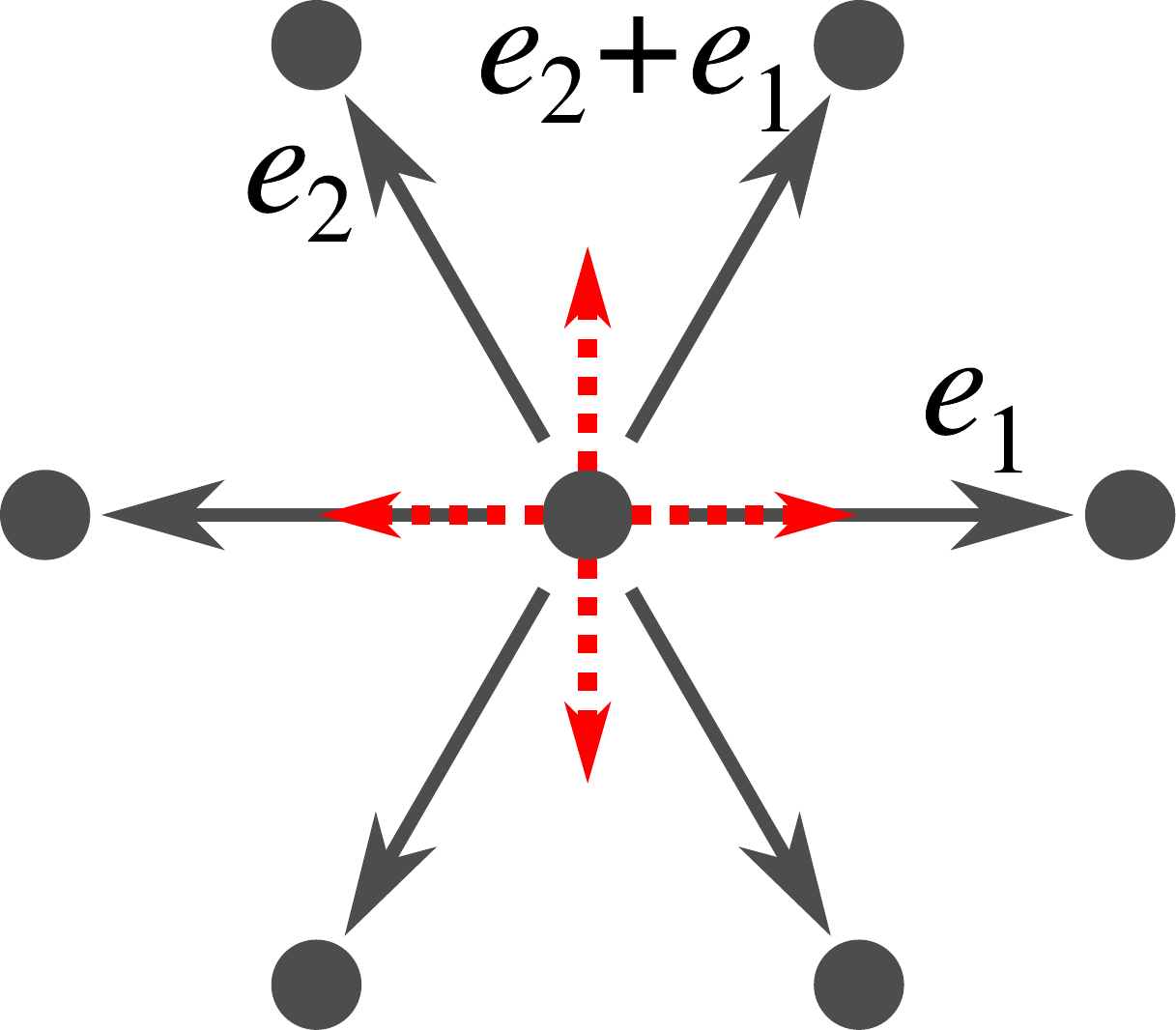}
   \label{fig:triangular lattice vectors} 
 }
 \hfill
 \subfigure[ square]{\includegraphics[height=3.5cm]{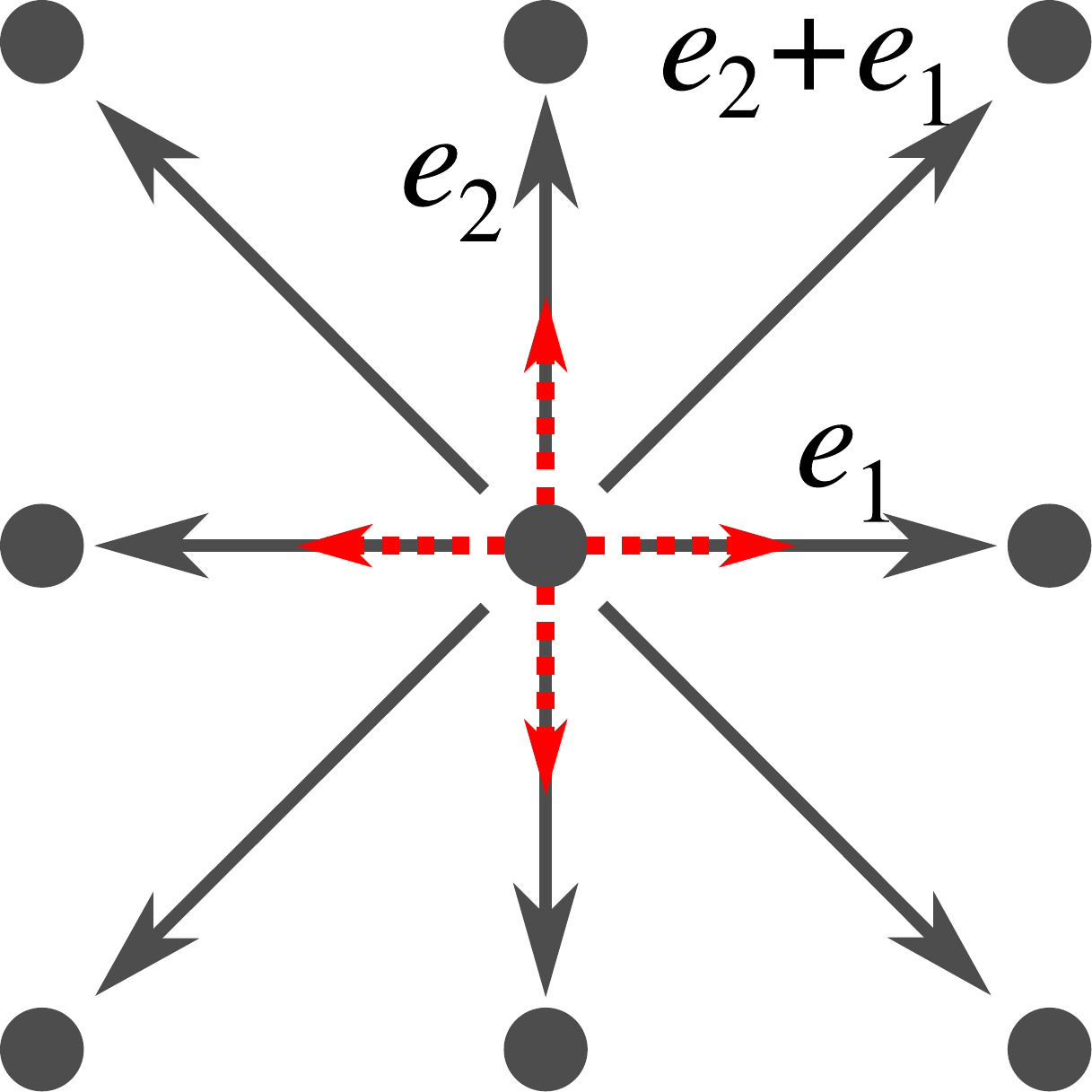}
   \label{fig:square lattice vectors} 
 }
 \hfill\null
 \caption{Lattice vectors for triangular \subref{fig:triangular lattice vectors} and square \subref{fig:square lattice vectors} lattices. In two dimensions, any crystal is described by two elementary lattice vectors $e_1$ and $e_2$. Burgers vectors of dislocations are linear combinations of these lattice vectors. Upon condensing dislocations along one lattice direction, say $e_1$, all lattice points which differ by multiples of $e_1$ become equivalent, such that for instance $e_2$ is equivalent to $e_2 + e_1$. This is the restoration of translation symmetry in the $e_1$-direction. The remaining translational order is always perpendicular to $e_1$, regardless of whether $e_1$ and $e_2$ are orthogonal. Effectively, we can describe dislocation melting by condensates $\Phi^x$ and $\Phi^y$ regardless of the underlying crystal structure, which is indicated in red dashed lines.}\label{fig:lattice vectors}
\end{figure}

\subsection{Preview of defect-mediated melting}\label{subsec:Preview to defect-mediated melting}
Armed with just the classification and properties of the topological defects of the crystalline solid, we can analyze the pattern of defect-mediated quantum melting phenomenologically. This follows by the general wisdom of vortex--boson duality that the topological defects perturb the ordered state and their proliferation restores the symmetry that was spontaneously broken by the order parameter.

Let us first discuss the heuristic idea of melting where the proliferation of defects causes restoration of the associated symmetry. From Fig. \ref{fig:dislocation} one can see that the notion of distance between lattice points becomes ill defined in the presence of a dislocation. Depending on whether one counts distance either including or excluding the inserted half-line will give different results. A condensate of dislocations is {\em that} state where on each point in space there is a superposition of zero, one, or any number of Burgers vectors~\cite{ZaanenBeekman12}. It has acquired gauge freedom in the sense that there is an arbitrariness in choosing which (lattice) distance to assign between two points. Thus, the lattice has disappeared and this state has now become fully translationally symmetric.  

Second, the rotational symmetries are not affected since disclinations are much more energetically costly than dislocations. In fact, in real-life solids, disclinations never appear whereas dislocations are easily identified and are, for instance, the cause of metal fatigue. From the figures \ref{fig:dislocation}--\subref{fig:disclination} one can immediately tell that a disclination is a much stronger disturbance of order than a dislocation (an explicit statement is made in Eq.~\eqref{eq:disclination confinement}). Similarly from Eq.~\eqref{eq:dislocation conservation with disclination} one infers that any disclination may serve as a source of dislocations which are themselves topological defects costing energy growing with the system size. Therefore it is justified to assume the absence of disclinations in any plane where there is still translational order. Topological neutral combinations of disclinations (which on the square lattice involve at least four disclinations) may in principle appear as low-energy fluctuations, that are nevertheless much more costly than dislocation--antidislocation pairs.

Moreover, we argued above that in the absence of disclinations, there can be no net Burgers vector in the system. As such, only dislocation--antidislocation pairs can appear. When these topological defects proliferate in a melting transition restoring the translational symmetry, they must do so with zero net Burgers vector. Therefore, translational symmetry will be restored in an {\em orientation}  along an axis (``the left-right axis''), and not in a {\em direction} (``left'' or ``right''), see Fig. \ref{fig:lattice vectors}. When a condensate forms along a single Burgers orientation, the resulting phase is the quantum smectic and when two linearly independent Burgers directions condense, the quantum nematic phase is obtained.

Just as for its classical sibling, the topological defects of the quantum nematic (liquid crystal) are disclinations, related to the remaining rotational order. These defects are `deconfined' in the nematic phase when the translational dislocations condense to be are part of the ground state. Proliferation of these disclinations lead to the fully symmetric state: the isotropic liquid, or rather the superfluid pertaining to bosonic quantum matter at zero temperature. This is again a second-order phase transition.

Let us now discuss ways to realize the above heuristic picture theoretically. In a rigorous symmetry classification, Mathy and Bais applied the formalism of Hopf symmetry or quantum double symmetry to two-dimensional quantum liquid crystals~\cite{BaisMathy06,MathyBais07}. In their work, the topological defects are treated on equal footing with the particle-like matter excitations both of which are representations of quantum double symmetry groups that describe the topological braiding of arbitrary quantum systems in two spatial dimensions. The topological defects are representations of the Hopf algebra that corresponds to the residual symmetry group in the ordered phase. Just as symmetry breaking occurs when a state vector in a particular representation obtains a vacuum expectation value (forms a condensate), the defect condensate is signaled also by the finite vacuum expectation value of a state vector in the Hopf algebra symmetry representation. In the framework of the continuum field theory of (dual) elasticity used in this review, the important result from this rigorous Hopf algebra derivation is that the condensate state vector can be {\em any superposition of topological defect states}. In particular, we verify the conclusion that the quantum nematic is a condensate of all possible dislocation orientations in Fig.~\ref{fig:lattice vectors}, whereas the quantum smectic condensate is a superposition of dislocations with Burgers vector along one orientation only; this is also the only possible loophole evading full translational symmetry in two dimensions after dislocation condensation.

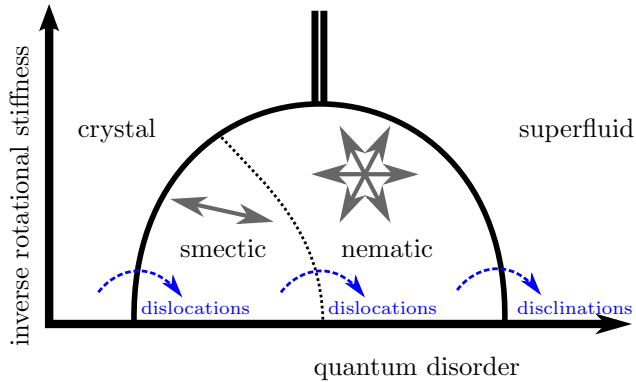
\begin{figure}[t]
\null\hfill
\begingroup%
  \makeatletter%
  \providecommand\rotatebox[2]{#2}%
  \ifx\svgwidth\undefined%
    \setlength{\unitlength}{8.5cm}%
    \ifx\svgscale\undefined%
      \relax%
    \else%
      \setlength{\unitlength}{\unitlength * \real{\svgscale}}%
    \fi%
  \else%
    \setlength{\unitlength}{\svgwidth}%
  \fi%
  \global\let\svgwidth\undefined%
  \global\let\svgscale\undefined%
  \makeatother%
    \begin{picture}(1,0.60519892)%
    \put(0,0){\includegraphics[width=\unitlength]{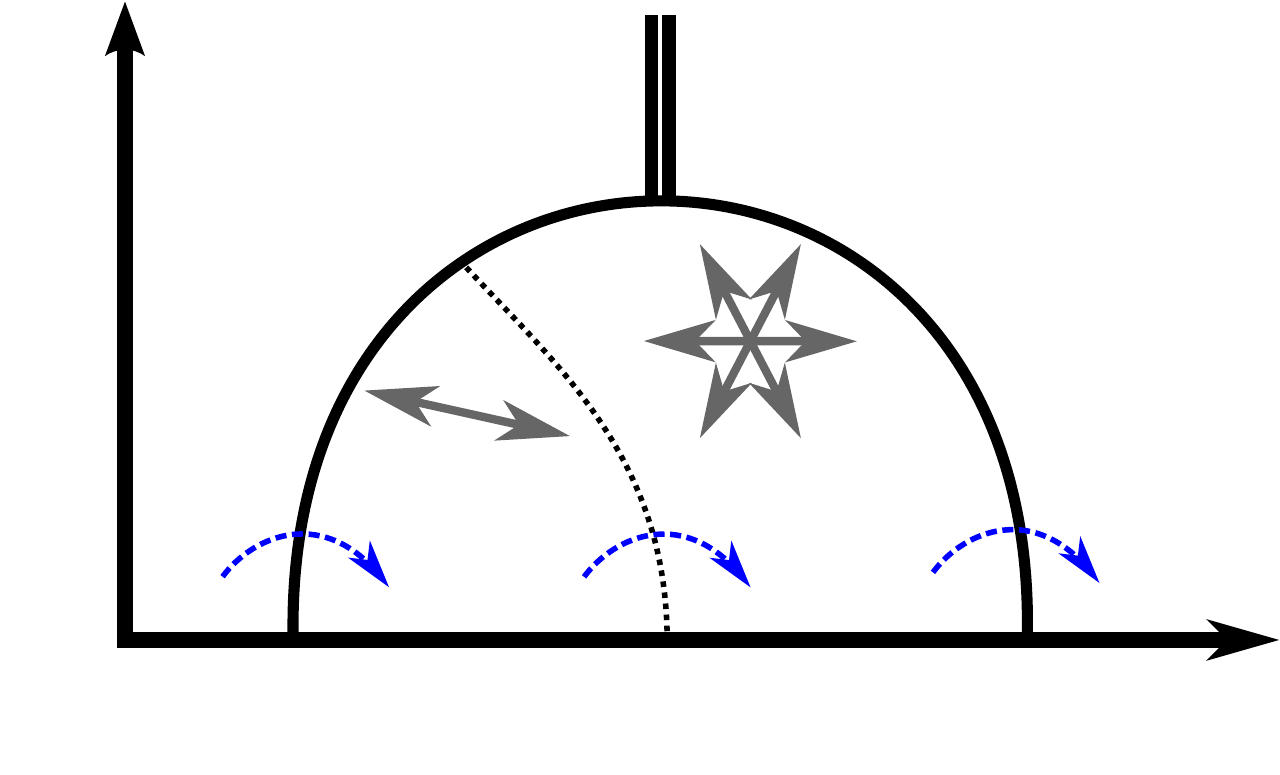}}%
    \put(0.50724124,0.01806734){\color[rgb]{0,0,0}\makebox(0,0)[lb]{\normalsize quantum disorder}}%
    \put(0.06599246,0.07){\color[rgb]{0,0,0}\rotatebox{90}{\makebox(0,0)[lb]{\normalsize inverse rotational stiffness}}}%
    \put(0.14208687,0.38795623){\color[rgb]{0,0,0}\makebox(0,0)[lb]{crystal}}%
    \put(0.82550977,0.38918617){\color[rgb]{0,0,0}\makebox(0,0)[lb]{superfluid}}%
    \put(0.30073679,0.21){\color[rgb]{0,0,0}\makebox(0,0)[lb]{smectic}}%
    \put(0.55,0.21){\color[rgb]{0,0,0}\makebox(0,0)[lb]{nematic}}%
    \put(0.23997547,0.12){\color[rgb]{0,0,1}\makebox(0,0)[lb]{\scriptsize dislocations}}%
    \put(0.82243049,0.12){\color[rgb]{0,0,1}\makebox(0,0)[lb]{\scriptsize disclinations}}%
    \put(0.53,0.12){\color[rgb]{0,0,1}\makebox(0,0)[lb]{\scriptsize dislocations}}%
  \end{picture}%
\endgroup%
\hfill\null
 \caption{Phase diagram for spatial ordering of bosonic matter at zero temperature. Consecutive defect-proliferation transitions are indicated in blue. We have chosen the point group $D_6$ for simplicity, but any rotational order follows the same pattern. The horizontal axis denotes increasing quantum disorder, that perturbs the crystalline state. On the vertical axis is inverse rotational stiffness, or the inverse energy cost to form disclinations. If the rotational stiffness is very large, dislocations proliferate in the absence of disclinations. When only dislocations with a single Burgers orientation proliferate, one obtains a quantum smectic. If the condensate is a superposition of all Burgers orientations, a quantum nematic forms, which has no preferred orientation but is nevertheless rotationally rigid. Finally, disclinations can proliferate to lead to the perfectly spatially symmetric superfluid. If rotational stiffness is low, instead dislocations and disclinations proliferate at the same time, resulting in a first-order transition to the superfluid state.}\label{fig:phase diagram sketch}
\end{figure}

We are led to the following scenario of quantum melting, sketched in Fig.~\ref{fig:phase diagram sketch}. In the  following sections, we address this phase diagram within the continuum field theory of dual elasticity laying bare the associated long-distance Goldstone modes and the topological defects, whose symmetries classify the disordered states. The dual theory is formulated in Sec.~\ref{sec:Dual elasticity} and the defect condensates are described by flavored {\em dual stress superconductors}, which are introduced in Sec.~\ref{sec:Dynamics of disorder fields}. The melting from a solid to a liquid can either take place directly as a first-order phase transition, which is the consequence of dislocations and disclinations melting simultaneously~\cite{Kleinert83}, or in two separate second-order phase transitions: (i) first dislocations proliferate to melt the solid to a smectic or directly to the nematic, and (ii) disclinations melt the liquid crystal to a liquid~\cite{HalperinNelson78,NelsonHalperin79}. This is depicted in the lower part of Fig. \ref{fig:phase diagram sketch}, where dislocations in one particular Burgers orientation proliferate to restore symmetry along an axis, say along the $x$-axis, via a second-order phase transition of the KTNHY-type~\cite{HalperinNelson78,NelsonHalperin79,Young79}.  The difference between these two paths is dictated by the propensity of disclinations to appear in the crystal phase, denoted the inverse of the rotational stiffness $\ell$ of Eq.~\eqref{eq:isotropic solid second gradient energy}. Similarly, the smectic phase results when the dislocations have a non-isotropic Hamiltonian \cite{OstlundHalperin81} favoring a Burgers direction. The zero-temperature phase diagram follows the same outline, see Fig.~\ref{fig:phase diagram sketch}. The only difference is that at $T=0$, the liquid phase is now a superfluid phase, with a well-defined zero-sound mode and no dissipation. Here we are interested in the limit of large rotational stiffness, implying the consecutive second-order phase transitions.

In the quantum smectic, one orientation is liquid-like, while the other one remains solid-like~\cite{CvetkovicZaanen06b}, which is the topic of Sec.~\ref{sec:Quantum smectic}. After dislocations with Burgers orientation along one axis have proliferated, all lattice points that differ by lattice vectors purely in this orientation become equivalent: this is the demise of the crystal and the restoration of translational symmetry. Therefore, regardless of the discrete symmetry of the original crystal lattice, the remaining translational order is always perpendicular to the `liquid direction' in two dimensions, see Fig.~\ref{fig:lattice vectors}.

Due to the highly anisotropic properties of this quantum phase, we will first discuss the simpler case of full translational symmetry restoration. Upon further disordering the smectic, also dislocations in the remaining crystal directions can proliferate, again through a second-order phase transition. This is the quantum nematic (liquid crystal), treated in Sec.~\ref{sec:Quantum nematic}. Then all translational symmetry is restored, but disclinations are still massive and there is rotational rigidity and its associated Goldstone modes. Note that there is no preferred direction in the isotropic quantum nematic and no point group symmetry inherited from the underlying crystal orientations; the situation is akin to the continuum treatment of the classical hexatic phase~\cite{NelsonHalperin79}. Here we are faced with a potential cause for confusion in our nomenclature ``nematic" for the quantum liquid crystal phases. However, in Sec.~\ref{sec:Order parameters for 2+1-dimensional nematics} we argue for the correct description of the full the long-distance hydrodynamical order parameter description of phases with translational symmetry but finite point group symmetry. The overall picture is in fact largely independent on which rotational order with point group symmetry is realized, which further justifies bundling them all under the umbrella ``nematic''.
 
 \section{Order parameter theory of two-dimensional quantum nematics}\label{sec:Order parameters for 2+1-dimensional nematics}
 In this somewhat independent section, we focus on the finite rotational symmetries of the nematic state. This is appropriate in the framework of quantum melting, since the rotational symmetries evident in Fig. \ref{fig:lattice vectors} are not affected by the dislocations restoring the translations. The task is to develop a long-distance hydrodynamic order parameter theory describing the nematic phase(s) and their transitions to the isotropic liquid, incorporating the correct rotational symmetries. In the main part of this review, the melting process is described in terms of a dual gauge field description of 2+1-dimensional \emph{isotropic} continuum quantum elasticity. The latter theory ignores the discrete rotational symmetries by construction. 

In what follows, starting from neither a correlated crystal nor an isotropic quantum liquid, the nematic quantum liquid is described by identifying the correct orientational degrees of freedom that lead to the long-distance hydrodynamic order parameter. We then proceed to describe the possible phase transitions between the nematic phases and the isotropic liquid. In the nematic phase all translational symmetries are restored, and what remains of the crystalline order is an unbroken, residual discrete point group symmetry $\bar{P}\subset O(D)$. Since the resulting nematic phases and phase transitions are then associated with the rotational point group symmetries $\bar{P}$, the order parameter also encodes for the correct unbroken symmetries describing the ordered phases. Similarly, a pure symmetry group classification of nematic states, not necessarily descending from crystal melting, leads to the possibility of unbroken subgroups $\bar{P}$ of the rotation group $O(D)$ in the context of the Landau paradigm of phase transitions. In two spatial dimensions, the subgroup classification will be relatively straightforward since the rotational group is given by $O(2)$ and we will in the following only focus on the proper rotational part $SO(2)$, which is Abelian.

The subgroups $P \subset SO(2)$ are finite and discrete. The order parameters and coarse-grained effective theories have to embody the discrete symmetry from the outset. We will formulate the nematic phases and their discrete symmetries using a gauge theory on a lattice following Refs.~\cite{LammertRoksharToner93, LammertRoksharToner95}. The gauge formulation also naturally leads to the incorporation of the topological defects, i.e. the disclinations, into the theory. We can then consider defects and melting from the nematic state(s) to the isotropic state in full generality in 2+1 dimensions. 

Before constructing the order parameter theory of 2+1-dimensional nematics, we first review the classical 2+0-dimensional nematic (hexatic) state, following Refs.~\cite{NelsonHalperin79, Nelson83}, emphasizing the role of the discrete symmetries. Further details regarding this section can be found in Ref.~\cite{LiuEtAl15}.

\subsection{Finite point group symmetries and melting}

Classical nematic states are usually imagined as phases of matter where microscopic constituents or ``mesogens'' of different shapes break the rotational $SO(D)$ symmetries of $D$-dimensional isotropic space by aligning along a macroscopic direction. Such `molecular' phases have inherent symmetries related to the point group symmetries of the `molecules' forming the phase. Similarly, in the theory of crystal melting, the nematic phase occurs in between the crystal and isotropic liquid phases, and possesses (quasi-)long-range orientational order with the rotational symmetries being inherited from the crystal due to local crystalline correlations, even in the absence of the crystalline translational order. 

Let us start reviewing the order parameters for crystalline states of matter, hereafter fixing the space dimensions to $D=2$. For the crystal phase the order parameter is the density $\rho_{\mathbf{G}}(\mathbf{x})$~\cite{Nelson83}, and long-range order is determined by the density correlation function $g(\mathbf{x},\mathbf{x}')$  (for a thermal phase transition),
\begin{align}
\rho_{\mathbf{G}}(\mathbf{x}) &= \te^{\ti \mathbf{G}\cdot (\mathbf{x} + \mathbf{u}(\mathbf{x}))}, \\
g(\mathbf{x},\mathbf{x}') &= \corr{\rho_{\mathbf{G}}(\mathbf{x}) \rho^*_{\mathbf{G}}(\mathbf{x}')} \propto \begin{cases} r^{-\eta_{\mathbf{G}}(T)},& T < T_\mathrm{c_1} \\ \te^{-r/\xi(T)}, & T >T_\mathrm{c_1}\end{cases},\nonumber
\end{align}
where the displacement vector $\mathbf{u}(\mathbf{x})$ is the Goldstone fluctuation (phonon) and the relevant wave vectors $\mathbf{G}$ are reciprocal lattice vectors \cite{AshcroftMermin76}. These comprise a discrete set of vectors instead of a continuum. The message is that in two dimensions the density shows power law correlations with (temperature-dependent) exponents $\eta_{\mathbf{G}}(T)$ at the reciprocal lattice wavevectors $\mathbf{G}$ in the ordered phase. This already hints at the discrete nature of proper order parameters for crystalline ordering and related phases.

In the case of thermal melting of classical crystals, the finite point group symmetries of the crystalline and nematic phases have been discussed in relation to the melting transition of the two-dimensional triangular lattice \cite{HalperinNelson78, NelsonHalperin79, Young79}. Driven by defect-mediated melting, there is a BKT-type transition from the crystal phase to a $C_6$-symmetric nematic phase, coined {\em hexatic} in Refs.~\cite{HalperinNelson78, NelsonHalperin79}. However, the ramifications of the finite point group are minimal due to the fact that the long-distance elastic properties of the triangular lattice are identical to the isotropic case, which was part of the original motivation to study them. Nevertheless, due to the absence of true long-range order in two dimensions and the topological properties of the defects, these phases and melting transitions have many interesting properties in their own right.

With regards to the defining properties of the nematic/hexatic phase, the Goldstone modes or low-energy fluctuations are constrained by the rotational part of the crystal symmetry. Physical fluctuations of the orientational order parameter are invariant under the rotational point group symmetry. In the context of continuum elasticity, the orientational order parameter and its fluctuations are simply the local crystal orientation in terms of the rotation $\theta(\mathbf{x}) = \frac{1}{2}\epsilon^{ij}\partial_{i}u^j$ representing an angle with respect to the chosen axes. Note that in this section we use the symbol $\theta$ to denote the orientational order parameter, in contrast with the local rotation field $\omega = \epsilon_{ab}\omega^{ab}$ of Eq.~\eqref{eq:strain Ehrenfest constraint}. Due to the short-range crystalline order, the order parameter $\theta(\mathbf{x})$ exists in the nematic, although the displacement field $\mathbf{u}(\mathbf{x})$ is not well defined. The hexatic $C_6$ order parameter is then~\cite{Nelson83} (for a thermal phase transition)
\begin{align}
\psi_{6}(\mathbf{x}) &= \te^{\ti 6\theta(\mathbf{x})}, \\
g_{6}(\mathbf{x},\mathbf{x}') &= \corr{\psi_6(\mathbf{x})\psi_6(\mathbf{x}')} \propto \begin{cases} r^{-\eta_6(T)},& T_{c_1} < T < T_{c_2}, \\ \te^{-r/\xi_6(T)},& T>T_{c_2}. \end{cases}\nonumber
\end{align}
Here $T_{c_1}$ is the critical temperature for the crystal-to-hexatic phase transition and $T_{c_2}$ is the critical temperature for the hexatic-to-liquid transition. The order parameter $\psi_6$ carries manifest $C_6$ symmetry $\theta \to \theta + 2\pi/6$. We note that in the crystal phase the same order parameter $\psi_6$ approaches a constant, instead of the algebraic decay. This is related to the confinement of rotational Goldstone modes, see Sec.~\ref{subsec:torque stress nematic}. Moreover, in the two-dimensional hexatic phase, the six-fold periodicity of the orientational degree order parameter $\theta$ does not affect the nature of the phase transition: it remains in the $XY$-universality class. The details including the exponents $\eta_6(T)$ and correlation lengths $\xi_6(T)$ are affected however \cite{Nelson83}.

\subsection{Classical crystal melting in two dimensions --- KTHNY theory}\label{subsec:classical melting}

Below, we will now quickly summarize the relevant aspects of the classical  $C_6$-melting transitions in two dimensions \cite{Nelson83} before we discuss the 2+1-dimensional quantum case.

The melting transition from the crystal to the nematic/hexatic occurs as an unbinding process of defect pairs (dislocations) and is similar to the BKT-transition of the two-dimensional $XY$-model, discussed in Sec.~\ref{sec:XY-duality}. In particular, there are  power-law correlations in the ordered phase with temperature-dependent exponents $\eta(T)$ and essential singularities in the correlation length $\xi(T)$. 

At the level of the Hamiltonian, the main difference of the crystal melting compared to the $XY$-model comes from the vectorial nature of the defect's Burgers charge carried by the dislocations. In terms of strains $u^{mn}(x)$ defined in Eq.~\eqref{eq:strain definition}, the reduced Hamiltonian is given by \cite{NelsonHalperin79} (see also Eq.~\eqref{eq:first gradient elastic energy}),
\begin{align}
\frac{\mathcal{H}}{k_\mathrm{B} T} &=  \int \td^2\mathbf{x}~\frac{1}{2} u^{mn} C_{mnkl} u^{kl} = \mathcal{H}_0 + \mathcal{H}_{\rm defect}, \label{eq:Hhexatic}\\ 
\mathcal{H}_0 &= \frac{1}{2} \int \td^2\mathbf{x}~ u^{mn}_\mathrm{smooth}\,C_{mnkl}\,u^{kl}_\mathrm{smooth}, \nonumber \\
\mathcal{H}_{\rm defect}[\{\mathbf{J}\}] &= \frac{-1}{8\pi} \sum_{\mathbf{x}\neq\mathbf{x}'} K_1(T)\mathbf{J}(\mathbf{x})\cdot \mathbf{J}(\mathbf{x}') \ln \left(\frac{\abs{\mathbf{x}-\mathbf{x}'}}{a}\right) - K_2(T)\frac{\mathbf{J}(\mathbf{x})\cdot (\mathbf{x}-\mathbf{x}') \mathbf{J}(\mathbf{x}')\cdot (\mathbf{x}-\mathbf{x}')}{\abs{\mathbf{x}-\mathbf{x}'}^2} \nonumber\\
&\phantom{mm}- E_\mathrm{c}(T)\sum_{\mathbf{x}} \mathbf{J}(\mathbf{x})^2 \label{eq:Hdefect}
\end{align} 
Here $a$ is the dislocation core size acting as a UV cutoff scale. The displacements are decomposed as $u^{a}(x) = u^{a}_\mathrm{smooth}(x) + u^{a}_\mathrm{sing}(x)$ in terms of the smooth and multivalued parts. This leads to a elastic Hamiltonian $\mathcal{H}_0$ and a defect Hamiltonian $\mathcal{H}_{\rm defect}$. The dislocation density $\mathbf{J}(\mathbf{x})$ entering the defect part is given by $\oint_{\mathbf{x}'} \td\mathbf{u} = \int \td^2 x\, \mathbf{J}(\mathbf{x})= \mathbf{B}(\mathbf{x}')$ with $\mathbf{B} = n\mathbf{e}_1 + m\mathbf{e}_2$ the Burgers vector, where $\{\mathbf{e}_1, \mathbf{e}_2\}$ are a lattice basis, cf. Eqs.~\eqref{eq:dislocation density}, \eqref{eq:D+0 dislocation density} and Fig~\ref{fig:lattice vectors}. There is the constraint of overall Burgers charge neutrality $\sum_{\mathbf{x}} \mathbf{J}(\mathbf{x}) = 0$. The smooth elastic part $\mathcal{H}_0$ gives rise to the crystal correlations as expected for isotropic bare elastic constants Eq.~\eqref{eq:elasticity tensor}, $C_{mnkl} = 2\mu_0 P^{(2)}_{mnkl} + 2\kappa_0 P^{(0)}_{mnkl}$. The Hamiltonian $\mathcal{H}_{\rm defect}$ represents the contribution from the elastic defects and has been written in terms of a lattice regularization. The non-zero defect densities weighted with $K_1$ and $K_2$ contribute to the renormalized values of the elastic constants and drive the phase transition from the crystal to the liquid crystal phases. 

The first term proportional to $K_1$ in $\mathcal{H}_{\rm defect}$ is the Coulomb-like logarithmic interaction between the vectorially charged dislocations with a dislocation core size of $a$. We will later see how this type of interactions arise from the interaction of the dislocations with the dual gauge fields. The second scalar term with coefficient $K_2$ is actually the divergent $V(x) \sim x^{2}$ potential between the non-zero disclination densities, see Eqs.~\eqref{eq:dislocation density from Frank vector} and \eqref{eq:D+0 disclination density},
\begin{align}
\Theta(\mathbf{x}) = \frac{1}{2\pi} \sum_{\mathbf{x}'} \frac{\mathbf{J}(\mathbf{x}')\cdot (\mathbf{x}-\mathbf{x}')}{\abs{\mathbf{x}-\mathbf{x}'}^2}
\end{align}
that arise due to the non-zero dislocation density $\mathbf{J}(\mathbf{x})$. 
That is, the scalar interaction is schematically of the form $K_{2}(T) \Theta(\mathbf{x}) \Theta(\mathbf{x}') V(|\mathbf{x}-\mathbf{x}'|)$ with a quadratically increasing kernel $V(|\mathbf{x}-\mathbf{x}'|)$
in the separation $\vert \mathbf{x}- \mathbf{x}' \vert$. The quadratic divergence implies the confinement of the disclinations at all temperatures $T$ where the Hamiltonian Eq. \eqref{eq:Hdefect} is valid. Since $\mathcal{H}_0$ is independent of rotations in first-order elasticity, the rotational degrees of freedom effectively drop out in the crystal phase. This  was discussed from the dual perspective in 2+1-dimensions in Ref.~\cite{BeekmanWuCvetkovicZaanen13} and will be discussed in Sec.~\ref{subsec:torque stress nematic}. Finally the term $E_\mathrm{c}$ is a dislocation core-energy. 

One of the main results of the KTHNY theory \cite{KosterlitzThouless73, HalperinNelson78, NelsonHalperin79, Young79} is the explicit renormalization group recursion relations for the couplings $K = K_1 = K_2$ and $y = \te^{ -E_\mathrm{c}(T)}$ and the elastic constants $\mu, \kappa$. These show the existence of a phase transition from the crystal to a liquid crystal via the unbinding of the dislocation pairs, $K\to 0, y \gg 0$. The liquid crystal phase has no translational order and a vanishing shear modulus $\mu$ but power law correlations in the orientational order parameter $\psi_6$, see below. 

As we mentioned, the elastic tensor $C_{mnkl}$ is isotropic for $C_6$ in two dimensions, so only two elastic constants $\mu$ and $\kappa$ enter and no difference is seen in the continuum limit to the isotropic case, at least at the level of the couplings of the Hamiltonian. Nematics with other point groups are more complex due to the additional elastic constants in $C_{mnkl}$. Ref.~\cite{OstlundHalperin81} considered these and furthermore a smectic phase was found at $T=0$ between the nematic and crystal due to anisotropies preferring certain types of dislocations. The unbinding of dislocation pairs is the thermal analogue of the quantum dislocation condensate in Sec.~\ref{sec:Dynamics of disorder fields}. Finally, two-dimensional crystals on commensurate or incommensurate substrates can also be considered~\cite{NelsonHalperin79}. Depending on the substrate lattice, these can feature true long-range orientational order $\corr{\psi_6} \sim \textrm{const.}$ and the transition from the liquid crystal to the liquid is then washed out due to the pinning with the substrate. 

In conclusion, the analysis of the Hamiltonian Eq.~\eqref{eq:Hhexatic} establishes the hexatic  liquid crystal, or nematic phase with symmetry $C_6$, in two dimensions via topological  dislocation melting from the crystal~\cite{KosterlitzThouless72, HalperinNelson78, NelsonHalperin79, Young79}. Numerical and experimental observations of this phase on various systems have been found in Refs.~\cite{RosenbaumEtAl83,KeimEtAl07,KapferKrauth15} and we should note that the transition can be first order, too, in some systems, see e.g. Ref. \cite{BernardKrauth11}. The reduced effective Hamiltonian for the hexatic liquid crystal, inherited from the crystal phase, is~\cite{NelsonHalperin79}
\begin{align}
\frac{\mathcal{H}_6}{k_\mathrm{B} T} &= \frac{K(T)}{2}\int \td^2\mathbf{x}\, (\partial_i \theta(\mathbf{x}))^2 \nonumber\\
&= \frac{1}{2} K(T) \int \td^2 \mathbf{x}\, (\partial_i\vartheta(\mathbf{x}))^2  -\frac{\pi K(T)}{36} \sum_{\mathbf{x}\neq\mathbf{x}'} \Theta'(\mathbf{x})\Theta'(\mathbf{x}')\ln\left(\frac{\abs{\mathbf{x}-\mathbf{x}'}}{a}\right)   + E_{\mathrm{c}'}(T) \sum_{\mathbf{x}} \Theta'^2(\mathbf{x}),  \label{eq:Hnematic}
\end{align}
where $\theta(\mathbf{x}) = \vartheta(\mathbf{x}) + \pi \Theta'(\mathbf{x})/3$ in terms of a smooth part $\vartheta(\mathbf{x})$ and a multivalued disclination part $\Theta'(\mathbf{x})\in \integers$. Here $\Theta'(\mathbf{x})$ is the integer-valued disclination density (as compared to the real-valued disclination density $\Theta(\mathbf{x})$ of Eq.~\eqref{eq:D+0 disclination density}), measured in terms of the elementary triangular $\pi/3$-disclination. Furthermore $E_{\mathrm{c}'}(T)$ is a phenomenological disclination core-energy. The Coulomb-like logarithmic interaction between disclinations is cut off by the disclination core size $a$. In the ordered phase at $T_{c_1} < T< T_{c_2}$, the system described by the Hamiltonian Eq.~\eqref{eq:Hnematic} features algebraic order in the orientational order parameter $\psi_6$.

\subsection{2+1-dimensional quantum nematics and point group symmetries}

In the 2+1-dimensional quantum case, we can in principle expect more quantum nematic states at very low temperatures. In this section we assume the existence of the quantum nematic states via dislocation melting but refrain from analyzing their existence or stability via the phase transition from the crystal phase. The feasibility of these quantum nematic states has been discussed in Refs.~\cite{BruunNelson14, LechnerBuchnerZoller14, WuBlockBruun15}, mostly regarding the quantum version of the hexatic. For the transition to the nematic from the crystal phase, one can use the finiteness of the shear rigidity $\mu$ as an order parameter defining the crystal.  We will show in Sec.~\ref{sec:Quantum nematic} that $1/\mu$ diverges at the transition to the nematic in terms of the dual gauge formulation of isotropic continuum elasticity. In the dual formulation, this role is played by $\Omega^2/\mu_0$ where $\Omega^2$ is the density of the dislocation condensate and $\mu_0$ is the `bare' shear modulus. The introduction of lattice symmetries will introduce additional elastic constants $C_{mnkl}$, and all the shear components vanish at the transition to the nematic. 

The starting point is then similar to postulating the quantum equivalent of the classical Hamiltonian in Eq.~\eqref{eq:Hnematic}. The question regarding the fate of the finite point group symmetries in quantum nematic phases was addressed in Ref.~\cite{LiuEtAl15}, focusing on the properties at $T=0$. Considering all possible crystal lattices in two spatial dimensions, in general one expects several different nematic phases. The resulting phases in terms of mere symmetry group analysis are summarized in Table~\ref{table:nematic phases}. These follow from the consideration that the dislocation melting effectively disorders all translational correlations while it leaves the global rotational correlations intact. In order to analyze the rotational symmetry group $P \subset SO(2)$, we must briefly recollect the full space group classification of crystal.

In two dimensions, the crystal carries a finite space or \emph{wallpaper} group $\bar{W} = T \rtimes \bar{P}$, where $T$ is the lattice translation group and $\bar{P}\subset O(2)$. In two dimensions there are 17 wallpaper groups; while in three dimensions there are 230 space groups. The translational part $T$ of the group $W$ generates the {\em Bravais lattice} related to $W$ with non-translational symmetries $\bar{P}$. In two and three dimensions there are 5 and 14 Bravais lattices, respectively. Thus the symmetry group of the nematic is given by a subgroup $\bar{P} \simeq \bar{W}/T \subset O(2)$ after the dislocation melting. If we focus on the strictly rotational part $P\subset SO(2)$ of $\bar{P}$ pertaining to the ordered nematic, the conjectured dislocation melting from two-dimensional crystals with space groups $W = T \rtimes P$ leads to the nematic phases with symmetries $C_N$ for $N=1,2,3,4$ and $6$, with the trivial group $C_1$ indicating no point group symmetry, see Table~\ref{table:nematic phases}.  

\setlength{\tabcolsep}{5pt}
\begin{table*}
\label{table:2Dnematics}
\begin{center}
\begin{tabular}{ c c c c }
\toprule 
Bravais-lattice structure ($\bar{P}$) & Bravais rotations  ($P$) & Space group ($\bar{P}$) & $C_{N}$ nematic phase\\
\hline
\multirow{5}{*}{Hexagonal (D6)}&\multirow{5}{*}{$C_{6}$ }&p6mm $(D_{6})$ &\multirow{2}{*}{$C_6$ nematic}\\
& &p6 $(C_{6})$ &\\ \cline{3-4}
& &p31m $(D_{3})$ &\multirow{3}{*}{$C_{3}$ nematic}\\
& &p3m1 $(D_{3})$ &\\
& &p3 $(C_{3})$ &\\
\hline
\multirow{3}{*}{Square (D4)}&\multirow{3}{*}{$C_{4}$ }&p4mm $(D_{4})$ &\multirow{3}{*}{$C_4$ nematic}\\
& &p4gm $(D_{4})$ &\\
& &p4 $(C_{4})$ &\\
\hline
\multirow{5}{*}{Rectangular (D2)}&\multirow{5}{*}{$C_{2}$ }&p2mm $(D_{2})$ &\multirow{3}{*}{$C_2$ nematic}\\
& & p2gm $(D_{2})$ &\\
& & p2gg $(D_{2})$ &\\ \cline{3-4}
& & pm $(D_{1})$ &\multirow{2}{*}{$C_{1}$ nematic}\\
& & pg  $(D_{1})$ &\\
\hline
\multirow{2}{*}{Rhombic (D2)}&\multirow{2}{*}{$C_{2}$ }&c2mm $(D_{2})$ &$C_2$ nematic\\ \cline{3-4}
& & cm $(D_{1})$ & $C_{1}$ nematic\\
\hline
\multirow{2}{*}{Oblique (C2)}&\multirow{2}{*}{$C_{2}$ }&p2 $(C_{2})$&$C_{2}$ nematic\\
\cline{3-4}
& & p1 $(C_{1})$&$C_{1}$ nematic\\
\bottomrule
\end{tabular}
\caption{Two dimensional nematic phases or $p$-atics which arise as descendants of crystals by dislocation melting. The first column shows the five Bravais lattice structures, with their corresponding point groups $\bar{P} \subset O(2)$. The second column displays the relevant $C_{N}$ group describing the rotational order associated with this Bravais lattice. The actual nematic phase is then obtained by considering the full space group and its associated point group, which may break the rotational symmetry to a smaller $C_{N}$ subgroup, as presented in the last two columns. Reproduced from Ref. \cite{LiuEtAl15}.}\label{table:nematic phases}
\end{center}
\end{table*}

In order to relate this to the results presented later in this review, let us reproduce here the imaginary time action Eq.~\eqref{eq:rotational Higgs term} of the nematic from Sec.~\ref{subsec:torque stress nematic},
\begin{align}
\mathcal{S}_{\rm nematic}[\omega] &= \frac{\Omega^2}{2 c_\tT\mu} \int \td\tau \td^2\mathbf{x}~ \frac{1}{2}(\partial_{\mu} \omega)^2 \equiv J' \int  \td\tau \td^2\mathbf{x}~ \frac{1}{2}(\partial_{\mu} \omega)^2.
\end{align}
which is obtained  within the dual gauge formulation of isotropic elasticity via a dislocation condensate in the limit of the Higgs mass $\Omega\to\infty$. Here the nematic orientational degree of freedom $\omega(\mathbf{x},\tau)$ represents local rotational fluctuations with rigidity $J'>0$. This action clearly has the expected 2+1-dimensional $XY$ form. Since the action is derived from isotropic continuum elasticity, different from the crystal with discrete point group symmetries, the theory is only valid for long-distance fluctuations $\omega = \omega_\mathrm{smooth} + \omega_\mathrm{sing}$ representing small distortion angles and infinitesimal disclination densities and lacking the additional role played by the $C_1$ periodicity $\omega \sim \omega + 2\pi$ (or its $C_N$ generalizations).  It is immediately clear that more realistic results for \emph{any} nematic are obtained by considering the compact angular variable $\theta(\mathbf{x},\tau) \equiv \theta(\mathbf{x},\tau) +2\pi$, with the lattice regulated action 
\begin{align}
\mathcal{S}_{XY}[\{\theta_i\}] =  -J \sum_{\corr{ij}\in \Lambda_{\tau}} \cos (\theta_i -\theta_j) \label{eq:C1nematic} =  -\frac{J}{2} \sum_{\corr{ij}\in \Lambda_{\tau}} z_i^* z_j + \textrm{c.c.},
\end{align}
where $\{\theta_i\} \in [0,2\pi)$ and $z_{i} = \te^{\ti \theta_i}$ are rotor variables on the sites $i$, with $\corr{ij}$ denoting nearest neighbor sites, of an auxiliary space-time lattice $\Lambda_{\tau}$. We take this to be  cubic and isotropic $J = J_{x,y}= J_{\tau} $ along the Euclidean time-direction $\tau \sim \beta\hbar$ for convenience with $J>0$ being the nematic coupling. The variables $\{\theta_i\}$ can now be taken to have manifest $2\pi$-periodicity due to the cosine form of the interaction. The model also implicitly features the long distance `spin-waves' $\delta\theta \sim \delta\omega$ and the corresponding rotational defects (disclinations) as $2\pi$-vortices, as is well known \cite{JoseEtAl77, Savit78}. The role of the defects becomes clear in the $2\pi$-periodic Gaussian or Villain model approximation of Eq.~\eqref{eq:C1nematic}.  We have thus arrived at a lattice regulated model of the $C_1$ quantum nematic in Euclidean time. 

The properties of the model Eq.~\eqref{eq:C1nematic} are as follows: there is a critical value $J_c = J_{c, 3D} \simeq 0.45$ of the $XY$ model that determines the  long-range ordering in the orientational degree of freedom $\{\theta_i\}$ in the sense that, in the three dimensional Euclidean formulation,
\begin{align}
\lim_{\abs{i-j}\to \infty} \corr{\te^{\ti(\theta_i-\theta_j)}} \propto \begin{cases} \textrm{const.}, & J> J_c \\ \te^{-\abs{i-j}/\xi}, & J<J_c. \end{cases}
\end{align}
We clearly identify the ordered phase as the nematic and the disordered phase as the isotropic liquid. A notable fact is that although the transition is tuned with respect to the coupling $J$, at the critical point $J_c$ the defect or vortex loops also unbind and proliferate~\cite{KajantieEtAl2000}. 

\subsection{Gauge theory and nematics}\label{subsec:Nematics gauge formulation}

Comparing the $XY$-model of the $C_1$ nematic above and the hexatic Hamiltonian Eq. \eqref{eq:Hnematic}, these are similar except for the order parameter $\psi_6$. Indeed, this is the traditional approach to the subject where one considers the physical order parameter and an invariant action for them. An example is the two-dimensional director order parameter $Q^{ab} = n^a n^b - \tfrac{1}{2}\delta^{ab}$, where the vector $\mathbf{n} = n^a$ determines the director via the identification $\mathbf{n} \equiv -\mathbf{n}$. One can then consider fluctuations of the order parameter by constructing effective actions that are invariant under this symmetry.  In the two-dimensional case, the natural (complex) order parameters are $\psi_N = \te^{\ti N \theta}$. From these one can of course construct traceless real tensor order parameters in terms of the vector $\mathbf{n} = (\cos \theta, \sin \theta)$~\cite{ParkLubensky96, LiuEtAl15} but this is not as convenient in two dimensions as the complex representation. 

We can conclude that in two dimensions, the nematic order parameters have the expected complex $XY$-rotor form and their excitations are of the type spin waves + vortices, and so are expected to feature transitions in the $XY$ universality class for the nematic-isotropic transition. However, one could expect that there are differences in the melting process due to the existence of $C_N$ disclinations not just the `vanilla' $2\pi$-vortices present already in the $XY$-model. The situation is reminiscent of Ref.~\cite{OstlundHalperin81}, where the effect of anisotropic lattices was considered and it was found that certain combinations of dislocation pairs could be more favorable to unbind, leading to different phase transitions.

There is a natural generalization of the order parameter approach that still retains simple representations of the degrees of freedom including the appropriate disclinations. This is achieved by demanding the correct order parameter symmetry for the degrees of freedom right from the outset. For the $C_N$ point groups this means that the nematic $O(2)$ rotor $\mathbf{n}_{i}\sim \te^{\ti \theta_i}$ should be identified as
\begin{align}
\theta_i \equiv \theta_i + 2\pi k/N, \quad k=1,2,\dots, N, \textrm{ for } i\in\Lambda_{\tau}. \label{eq:gaugesymm}
\end{align}
under the $C_N$ symmetry. This is a \emph{local, discrete} symmetry enforcing the point group symmetries on the rotor $\theta_i$. On the other hand, the order parameter $\psi_N(i) = \te^{\ti N \theta_i}$ is invariant under the transformations and well defined, as it should. In contrast to global symmetries that imply degeneracies of states, such local symmetries imply that two states related by the local symmetries are \emph{physically the same} or \emph{indistinguishable}. In effect, the rotor ceases to be a physical variable due to the local symmetry, similarly to the uniaxial nematic vector $\mathbf{n}\equiv -\mathbf{n}$ giving rise to the director. 

Such local symmetries are ubiquitous in physics and are usually referred to as \emph{gauge} symmetries. They do not, however, represent symmetries in the traditional sense but instead redundancies in the degrees of freedom and in this sense can be introduced whenever convenient~\cite{HenneauxTeitelboim92}. All physical observables are gauge invariant and the disappearance of the gauge degrees of freedom can be made explicit by imposing a suitable gauge fix. 

The gauge formulation was first applied in the context of (three-dimensional) nematics in Refs.~\cite{LammertRoksharToner93, LammertRoksharToner95} as a $\integers_2$-gauge theory of the nematic director $\mathbf{n}$. Following the same logic, the effective theories of the order parameters for 2+1-dimensional nematics with $C_N$ point group symmetries can be formulated using a gauge theory description~\cite{LiuEtAl15}. This follows from the $XY$-model Eq.~\eqref{eq:C1nematic}, since the principles of lattice gauge theories~\cite{KogutSusskind75, Kogut79} allow us to immediately write down a generalization of Eq.~\eqref{eq:C1nematic} to a theory of $C_N$ gauged rotors in the sense of Eq. \eqref{eq:gaugesymm} as
\begin{align}
\mathcal{S}_{N}[\{\theta_i\}, \{ U_{ij} \}] &= \mathcal{S}^U_{XY}[\{\theta_i\}, \{U_{ij}\}] + \mathcal{S}_{\rm gauge}[\{U_{ij}\}], \label{eq:CNNematic}
\end{align}
where
\begin{align}
\mathcal{S}_{XY}^U[\{\theta_i\}, \{U_{ij}\}] 
&= -J \sum_{\corr{ij}\in\Lambda_{\tau}} \cos (\theta_{i}-\theta_{j} - a_{ij})  = -\frac{J}{2} \sum_{\corr{ij}\in \Lambda_{\tau}} z^*_i U_{ij} z_j  + \textrm{ c.c. ,} \\ 
\mathcal{S}_{\rm gauge}[\{U_{ij}\}] 
&= -K \sum_{\Box \in\Lambda_{\tau}} \cos (\sum_{\corr{ij}\in\Box} a_{ij})  = -\frac{1}{2}K \sum_{\Box\in\Lambda_{\tau}} \prod_{\corr{ij} \in \doo\Box} U_{ij} + \textrm{ c.c.} \ .
\end{align}
Here the invariance of $\mathcal{S}_N$ under the local symmetry Eq.~\eqref{eq:gaugesymm} necessitates the introduction of the $C_N$ gauge field $U_{ij} = \te^{-\ti a_{ij}}\in C_N$ acting on the rotors $z_i = \te^{\ti \theta_i}$, where $a_{ij} =\frac{2\pi}{N}k_{ij}$ and $k_{ij}=0,1,\dots, N-1$ are $\integers_N$ valued fields on the links $\corr{ij}$ of the lattice. Consistency requires $U_{ji} = U^*_{ij}$ and all sums and products over the plaquettes $\Box \in \Lambda_{\tau}$ are taken in a counterclockwise manner along their boundary $\doo\Box$. 
The local gauge transformation at site $i$ now reads
\begin{align}
\theta_{i} &\to \theta_i + \frac{2\pi k_i}{N}, \nonumber\\ 
a_{ij} &\to a_{ij} + \frac{2\pi (k_i - k_j)}{N}, 
\end{align}
where the $k_i$ are $\integers_N$-valued fields; this local transformation is a symmetry of $\mathcal{S}_{N}$. By the isomorphism $C_N \simeq \integers_N$, the theory with action $\mathcal{S}_{N}$ was referred to as $O(2)/\integers_N$ gauge theory in Ref.~\cite{LiuEtAl15}.

Let us now briefly describe the interpretation of the $O(2)/\integers_N$ gauge theory in the context of nematics~\cite{LammertRoksharToner95, LiuEtAl15}. The meaning of the term $\mathcal{S}^U_{XY}$ is as usual: it favors the ordering of the orientational degrees of freedom but now modulo the gauge symmetries. On the other hand, the term $\mathcal{S}_{\rm gauge}$ represents the simplest gauge invariant term for the gauge fields $\{U_{ij}\}$, and has been included in the action with a coupling $K>0$. This gives independent dynamics to the gauge fields and it is therefore not determined by just symmetry considerations.

\begin{figure}
\begin{center}
\parbox[t][][t]{6cm}{
\includegraphics[height=4.8cm]{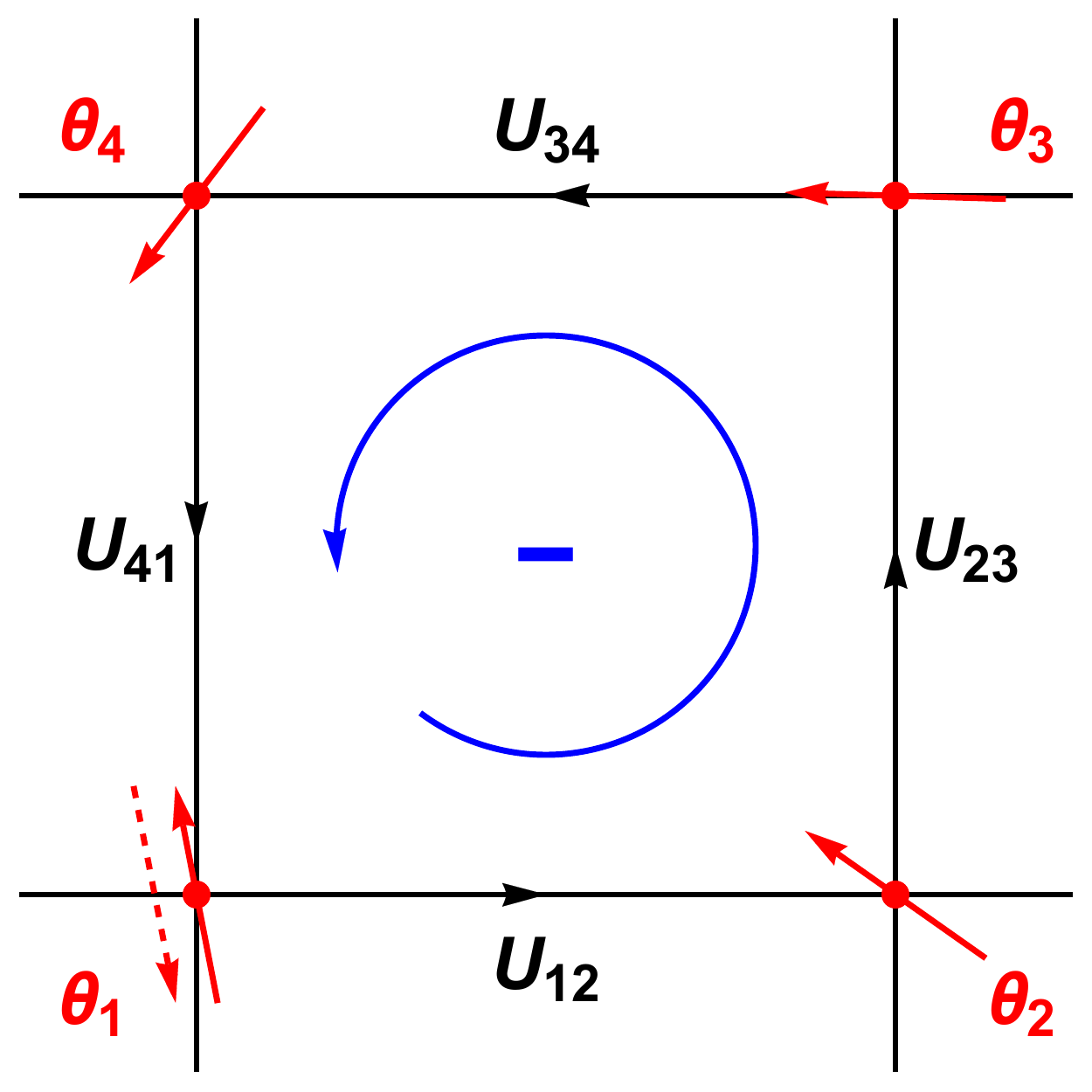}
\caption{The lattice model has rotors $\{\theta_i\}$ (red) defined on sites and gauge fields $\{U_{ij}\}$ on the links (black arrows). In the simplest case of $O(2)/\integers_2$, a disclination is given by a $\te^{-\ti \pi}=-1$ gauge flux over a plaquette and leads to a $\pi$-rotation ambiguity (dashed) at the site $1$ when encircling clockwise.}
\label{fig:plaquette}
}
\hspace{1em}
\parbox[t][][t]{9cm}{
\includegraphics[height=4.8cm]{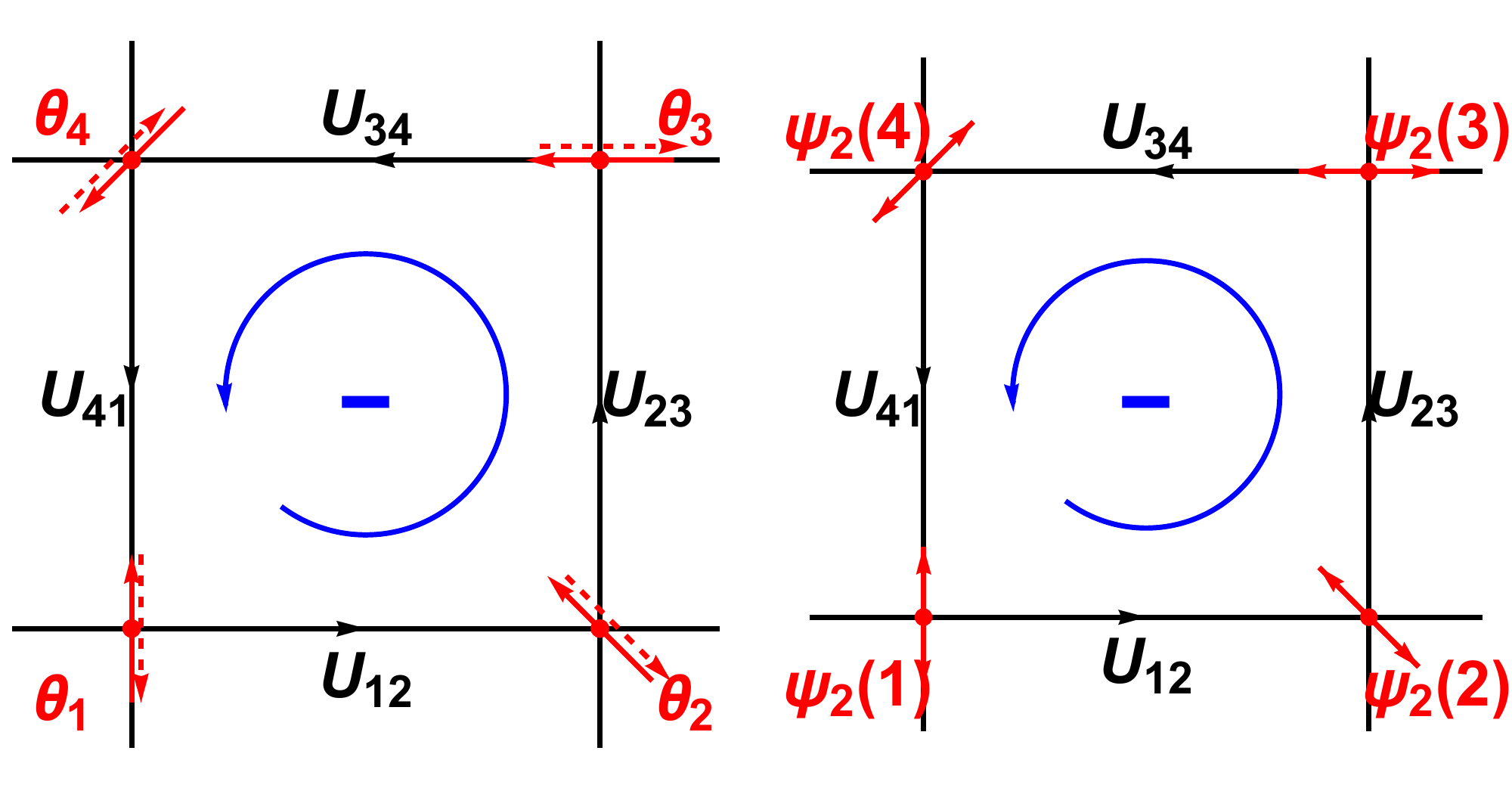}
\caption{Left: All $\integers_2$-gauge transformations (red and red dashed arrows $\theta_i$ and links $U_{ij}$) of the $C_2$ defect configuration in Fig. \ref{fig:plaquette}. All possible combinations of the rotors and gauge links are physically equivalent states. Right: The order parameter $\psi_{2}$ is unique and  single valued around the $C_2$ disclination, although cannot be continuously extended through the core region (center of the plaquette). }
\label{fig:OPplaquette}
}
\end{center}
\end{figure}

The interpretation of $\mathcal{S}_{\rm gauge}$ and the gauge fields $\{U_{ij}\}$ for nematics is as follows: by the Volterra process (Sec.~\ref{subsec:Volterra processes}), we know that the disclinations are labeled by the elements of the rotational symmetry group $C_N$. Upon encircling the disclination core, the local crystal axes rotate under parallel transport but otherwise the lattice has (nearly) perfect local order. This continues to be true for the nematic with the local (crystal) axis orientation but no translational crystalline order, see Fig.~\ref{fig:disclination}.  Since the gauge fields are labeled by the elements of $\integers_N$ and the order parameter fields are charged under it, we expect them to be related to the defects. More precisely, the gauge invariant field strength is related to the disclination charge density and the contributions of $\mathcal{S}_{\rm gauge}$ are then related to the action cost of disclinations \cite{LiuEtAl15}. These gauge fields are reminiscent of the {\em defect gauge fields} introduced by Kleinert~\cite{Kleinert89b,Kleinert08}; however, those take continuous values and are employed to describe infinitesimal contributions from topological defects in a continuum, while the $\{a_{ij}\}$ are $\integers_N$-valued and serve to encode the different, but $C_N$-equivalent local lattice orientations.

For the nematic described by Eq.~\eqref{eq:CNNematic} we can see this explicitly, as depicted in Fig. \ref{fig:plaquette}. We imagine a perfectly ordered nematic with $\theta_i = \theta_0 \ \forall i$ in a suitable gauge fix and trivial gauge fields $\{U_{ij}\} \sim w_iw^*_j$, where $U_{ij} = 1$ up to gauge transformations  $w_i \in C_N$. Then we put a non-zero gauge flux $\Delta \theta(\Box) =-\sum_{\corr{ij}\in \doo\Box} a_{ij}$ around the plaquette $\Box$. The rotor fields are now deformed from the aligned configuration in the sense that it is impossible to satisfy $\mathcal{S}_{XY}(\Box)=-4J$. Since the elastic part is $-J\sum_{\corr{ij}\in \doo\Box}\cos (\theta_i-\theta_j + a_{ij})$, the gauge fields rotate the rotor field $\{ \theta_i\}_{i\in\Box}$ by an amount $U_{ij}U_{jk}U_{kl}U_{li} = \te^{-\im\Delta \theta(\Box)}$ around the plaquette $\Box$, see Fig.~\ref{fig:plaquette}. We conclude that the presence of a non-trivial plaquette implies that there is a disclination in the rotor field $\theta_i$, and a consistent single-valued orientation of the rotor field (i.e. the local crystal axes) is impossible to define in the region surrounding the frustrated plaquette. The term $\mathcal{S}_{\rm gauge}$ assigns the action cost $-K \cos \Delta \theta(\Box)$ to the gauge field configuration $\{U_{ij}\}$ independent of the elastic contribution from $\mathcal{S}_{XY}$, which include the elastic distortion cost accompanying the disclination. A large positive $K$ acts as a defect suppression term analogous to the two-dimensional dislocation and disclination core-energies $E_b$ and $E_c$. It is linked to the rotational stiffness expressed by Eq.~\eqref{eq:isotropic solid second gradient energy}. We also note that it is most favorable to create disclination--antidisclination pairs, which are topologically neutral, due to the frustration introduced by the non-trivial gauge links $\{U_{ij}\}$.

Meanwhile, irrespective of the gauge field configuration $\{U_{ij}\}$, it is possible to define a consistent configuration of the gauge invariant variable $\psi_N(i) = \te^{\ti N \theta_i}$, since the ambiguity along the loop is an element of $C_N$. This is depicted in Fig.~\ref{fig:OPplaquette}. Therefore the physical order parameter $\psi_N$ is well-defined and winds to itself around the plaquette or defect. As is appropriate for the defect, however, the extension of $\psi_N$ is undefined at the core region (center of the plaquette). 

The physical interpretation of the $\integers_N$-gauge fluxes as $C_N$-disclinations is tied to the presence of the rotor fields, as is fitting for a defect in a nematic state. In the gauge theory formulation, however,  they have a life of their own and a physical status regardless of the charged matter fields $\{\theta_i\}$. This brings into question their role in the physics of the model, apart from fulfilling the required gauge symmetries.  In particular, the  field strength of the gauge fields $\{U_{ij}\}$ associated with the fluxes is gauge invariant and therefore always physical and is controlled by the coupling $K$. In the following, we will see how the presence of an independent weight $K$ affects the phase diagram of 2+1-dimensional nematics.

\subsection{Phase diagram of the $O(2)/\integers_N$ model}\label{subsec:Nematics phase diagram}

Having defined the gauge model of nematics encoding for the correct degrees of freedom and symmetries, we now want to analyze its universal properties and possible phases and transitions.

From the perspective of the original orientational degrees of freedom $\{\theta_i\}$, the most natural limit is $K\to \infty$, where the gauge fluxes are very expensive and the gauge fields are `frozen'. More precisely in this limit, assuming a topologically trivial lattice, $U_{ij} = w_iw^*_j$ for some $w_i\in C_N$ and we can choose a gauge where $U_{ij} = 1$ for all $\corr{ij}\in\Lambda_{\tau}$, indicating that the gauge fields are frozen. In this limit the action becomes
\begin{align}
\mathcal{S}_{N}[K\to\infty] = -J\sum_{\corr{ij}\in\Lambda_{\tau}} \cos(\theta_i + \frac{2\pi m_i}{N} - \theta_j-\frac{2\pi m_j}{N}),
\end{align} 
where $w_i = \te^{2\pi\ti m_i/N}$. We can now compute the partition function $\mathcal{Z}_N = \sum_{\{U_{ij}\}\in \integers_N} \int \mathcal{D}\theta\, \te^{-\mathcal{S}_N}$ as
\begin{align}
\mathcal{Z}_N[K\to\infty] = \sum_{\{m_i\}\in \integers_N} \int \prod_{i\in \Lambda_{\tau}} \td\theta_i~ \te^{-\mathcal{S}_N[\{\theta_i\}, \{m_i\}]} =  N^{\rm sites} \int \prod_{i\in\Lambda_{\tau}} \td\theta'_i~ \te^{-\mathcal{S}_{XY}[\{\theta'_i\}]} = N^{\rm sites} \mathcal{Z}_{XY}[J], 
\end{align}
where we have performed the change of variables $\theta_i' = \theta_i + 2\pi m_i/N$ and trivially summed over the gauge fields $\set{m_i}$. The conclusion is that the partition function in the limit $K\to\infty$ is exactly given by a multiple (the gauge group volume) of the $XY$ model at coupling $J$, from which all critical properties follow. However, the $XY$ variable $\theta_i$ cannot order in the usual sense of spontaneous symmetry breaking, since $\corr{\theta_i} \equiv 0$ due to the gauge symmetries (Elitzur's theorem~\cite{Elitzur75}), and moreover all physical quantities in the theory should be gauge invariant. The physical order parameter field $\psi_N(i) = \te^{\ti N \theta_i}$ is a composite, gauge-invariant quantity and follows a transition with some exponents and properties differing from the usual $XY$ model, dubbed the $XY^*$ universality class (for the $\integers_2$ case) in Refs. \cite{SenthilFisher00, SedgewickScalapinoSugar02, PodolskyDemler05,IsakovMelkoHastings12, MrossSenthil12}.

In the opposite limit $K\to 0$, the fluxes (disclinations) cost no action and will proliferate if the nematic stiffness $J$ is not too large, i.e. the gauge fields $\{U_{ij}\}$ are free to fluctuate~\cite{ZaanenBeekman12}. In this case we can conveniently sum over them in the Villain or $2\pi$-periodic Gaussian approximation of the $XY$-model, expected to be in the same universality class \cite{Villain75, JankeKleinert86}. The model is defined as the quadratic approximation of the cosine that retains the $2\pi$-periodicity by summing over auxiliary integer-valued $\{l_{i,\mu}\}$ variables as
\begin{align}
\mathcal{S}^{\rm Villain}_N = 
\sum_{(i,\mu) \in \Lambda_{\tau}}  \sum_{\{l_{i,\mu}\}\in\integers}  \frac{-J_\mathrm{V}}{2}(\triangle_{\mu} \theta_i - \frac{2\pi}{N} k_{i,\mu} + 2\pi l_{i,\mu})^2,
\end{align}
where the shorthand $(i,\mu)$ denotes the links from site $i$ in the directions $\mu = x,y,\tau$, $\triangle_{\mu}f(i) \equiv f(i+\mu)-f(i)$ and the Villain coupling $J_\mathrm{V}(J)$ is an analytic function of the original coupling $J$ \cite{Villain75,JankeKleinert86}. The sum over the gauge fields can now just be combined with $\{l_{i,\mu}\}$ to a sum over new integers $\{s_{i,\mu}\}$ in the partition function
\begin{align}
\mathcal{Z}^{\rm Villain}_N = \int \prod_{i\in\Lambda_{\tau}}\td\theta_i   \sum_{\{s_{i,\mu}\}\in\integers}  \prod_{(i,\mu)\in\Lambda_{\tau}} \te^{-\frac{J_\mathrm{V}}{N^2}(\triangle_{\mu}N\theta_i + 2\pi s_{i,\mu})^2},
\end{align}
which is the Villain approximation of the cosine model
\begin{align}
\mathcal{S}_{N,\rm eff}[K\to 0] = -J' \sum_{\corr{ij}} \cos N(\theta_i -\theta_j), \label{eq:NVillain}
\end{align}
with $J'\simeq J_\mathrm{V}(J)/N^2$ in the Villain approximation. Since the Villain model has a critical point $J_{\mathrm{V},\mathrm{c}} = 0.33$ in three dimensions, we find that the vanishing cost of arbitrary defects in the limit $K=0$ shifts the disordering transition to larger $J$ as a function of $N$. Another salient feature is that after summing over the gauge fields, only the gauge invariant combination $\psi_N(i) = \te^{\ti N\theta_i}$ appears in the model, as is to be expected.

\begin{figure}
\begin{center}
\includegraphics[width=0.5\textwidth]{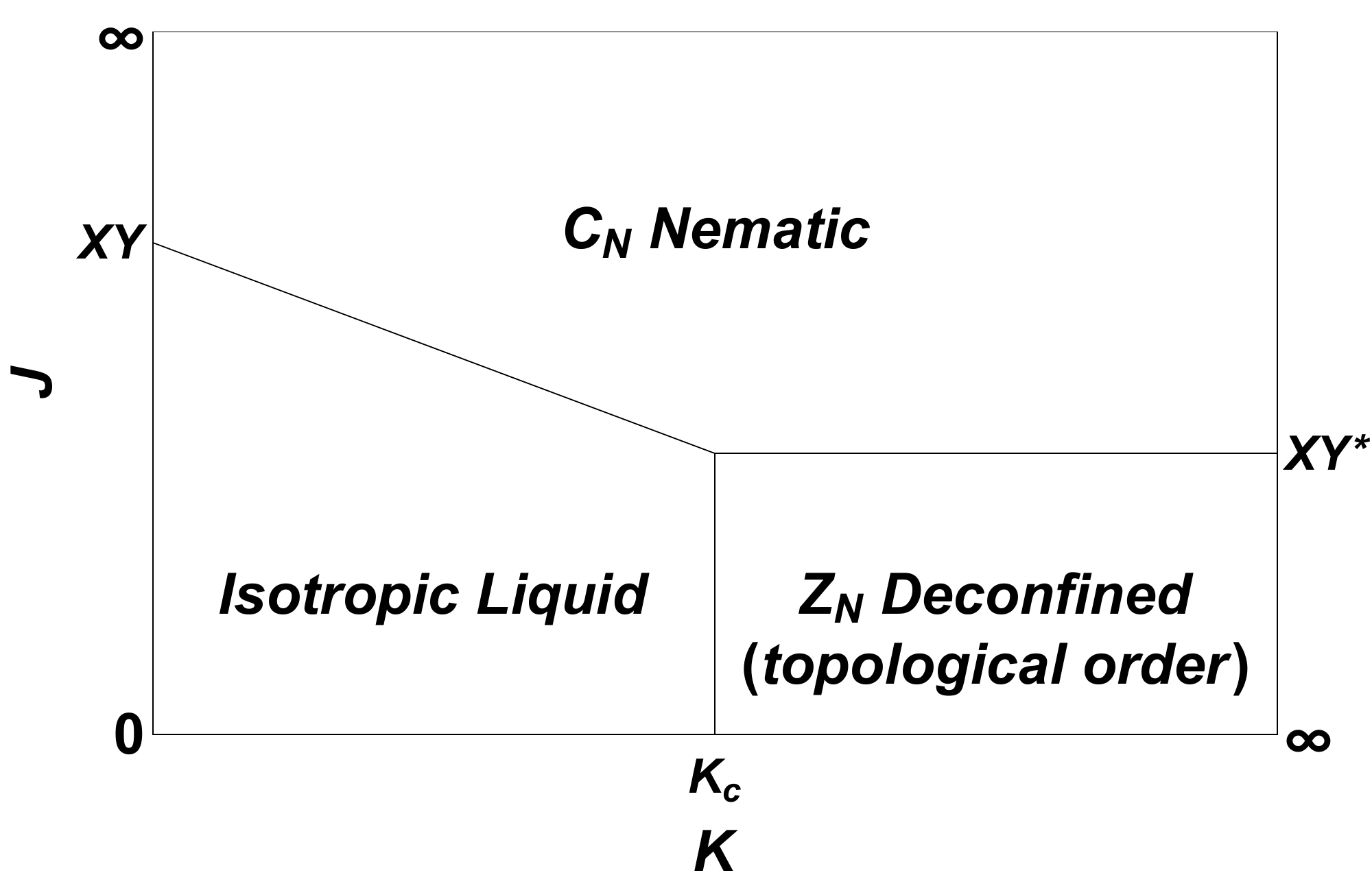}
\caption{The schematic phase diagram of the $O(2)/\integers_N$ theory.}
\label{fig:ZNphase}
\end{center}
\end{figure}

\begin{figure}
\begin{center}
\parbox[t][][t]{7cm}{
\includegraphics[width=7cm]{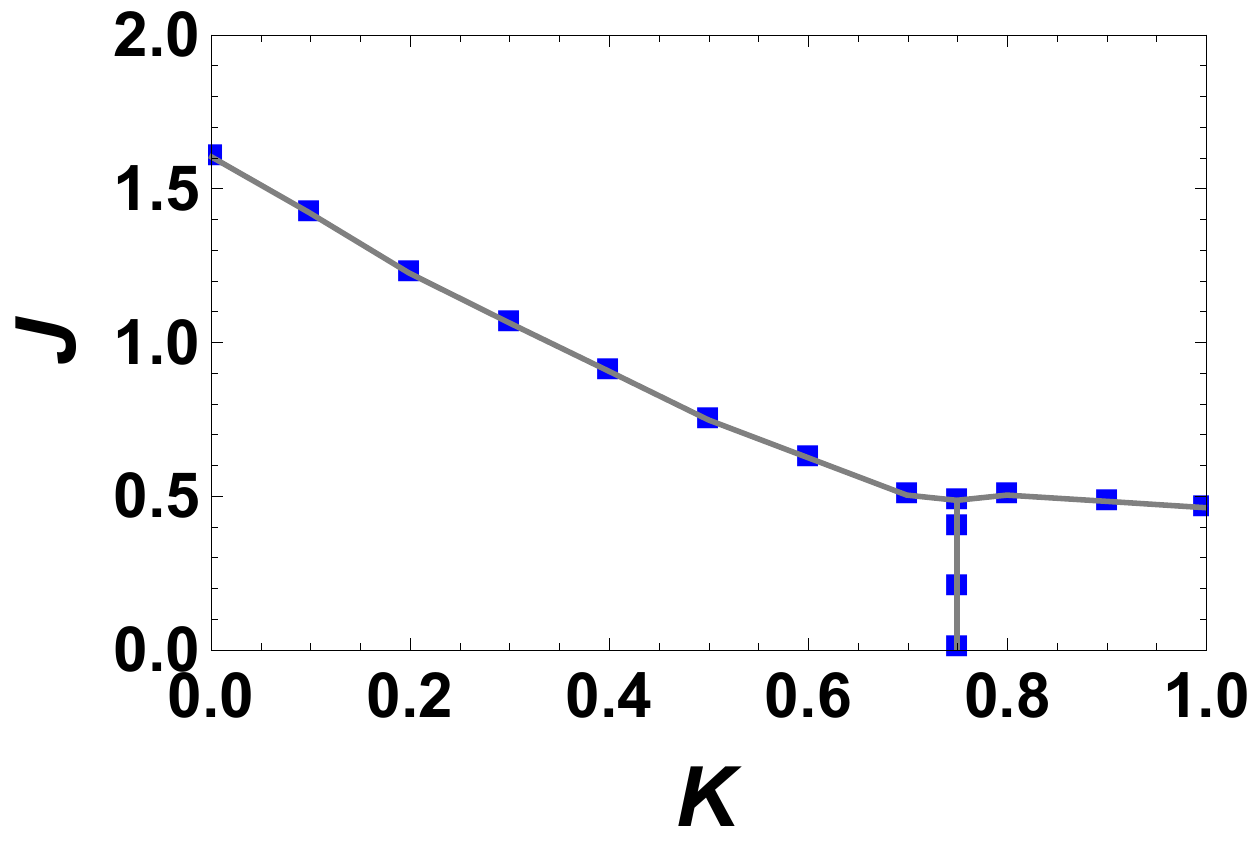}
\caption{The phase diagram of the $O(2)/\integers_2$ theory as determined by Monte-Carlo calculations, reproduced from Ref.~\cite{LiuEtAl15}.}
\label{fig:Z2phase}
}
\hspace{1em}
\parbox[t][][t]{7cm}{
\includegraphics[width=7cm]{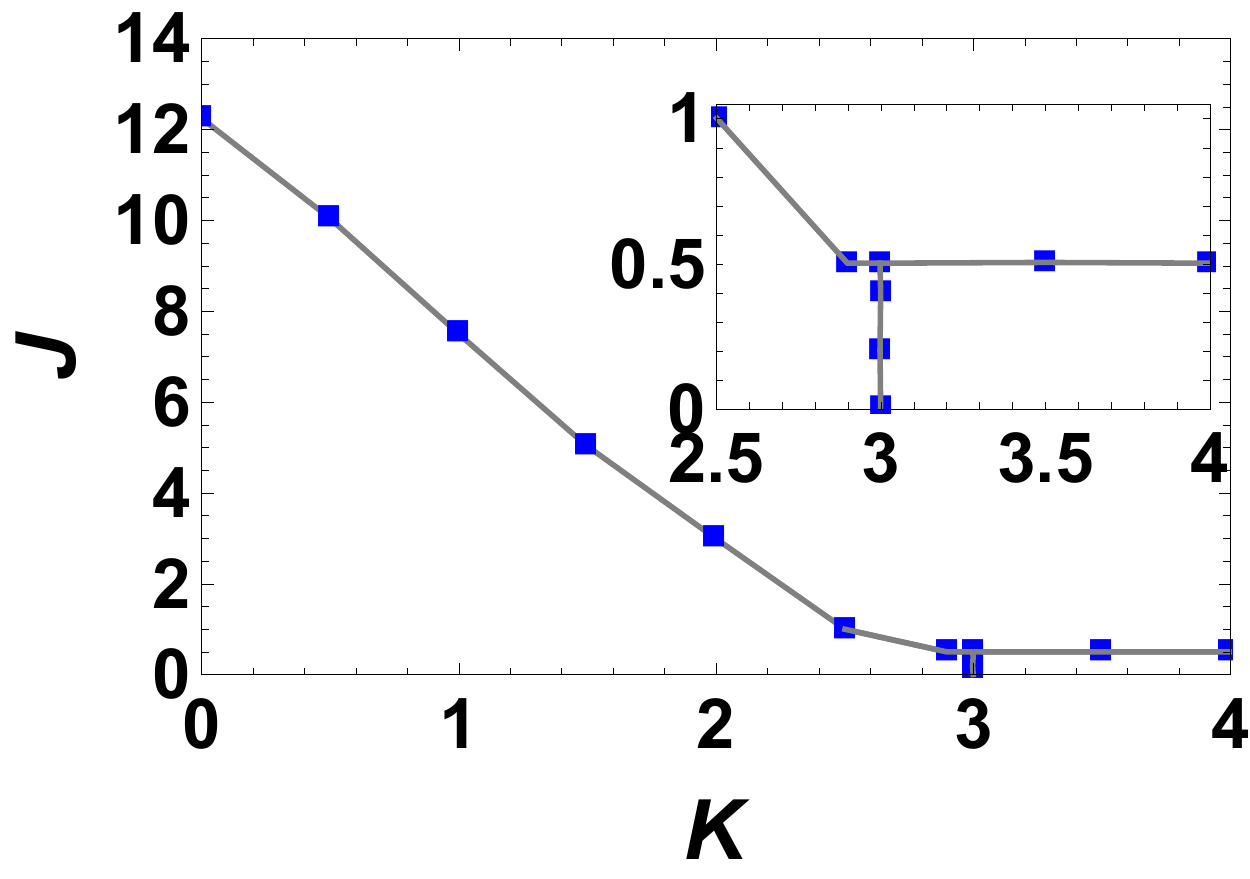}
\caption{The phase diagram of the $O(2)/\integers_6$ theory as determined by Monte-Carlo calculations, reproduced from Ref.~\cite{LiuEtAl15}. Note the larger $J_\mathrm{c}$ at $K=0$ and the shrinking of the deconfined region to larger $K$ as compared to $O(2)/\integers_2$.}
\label{fig:Z6phase}
}
\end{center}
\end{figure}

We are now in the position to discuss the various limiting behaviors in the $(K,J)$ plane that reveals the topology of the phase diagram. 
\begin{itemize}
\item[$J \ll 1$] In the limit $J\to 0$ the theory $\mathcal{S}_N$ is just pure $\integers_N$-gauge theory in three dimensions. There is a confinement--deconfinement phase transition at $K=K_\mathrm{c}$~\cite{HornWeinsteinYankielowicz79, BhanotCreutz80, BorisenkoEtAl14} in the pure gauge theory. The confined phase at $K<K_\mathrm{c}$ features condensed gauge fluxes and only $\integers_N$ charge-neutral fields can have non-zero expectation values. In the deconfined regime at $K>K_\mathrm{c}$, the gauge fields have a gap and the $\integers_N$ gauge fluxes remain well defined excitations over the vacuum. At finite but small $J$, we expect that the behavior of the gauge fields remain similar to those corresponding to the pure gauge theory phases. This phase is actually topological~\cite{Wen04, LammertRoksharToner95, GregorEtAl11, LiuEtAl15} given that this is the deconfined phase of $\integers_N$ gauge theory. 
\item[$J\gg 1$] In this limit, the rotors are ordered and similarly also the gauge fields are gapped due to the elastic terms proportional to $J$, irrespective of the value of $K$. There is therefore only one phase at large enough $J$ where the rotors are ordered (the $\integers_N$ Higgs phase~\cite{FradkinShenker79}). Gauge fluxes can however appear as tightly bound disclination--antidisclination pairs with Coulomb (logarithmic) interactions. 
\item[$K\ll 1$] Here we have the expected $XY$-transition of the gauge invariant and $\integers_N$-confined order parameter $\psi_N(i)= \te^{\ti N \theta_i}$, as described by Eq.~\eqref{eq:NVillain}. 
\item[$K\gg 1$] In this limit the $\integers_N$ fluxes are completely suppressed and an $XY^*$ transition from the nematic to the deconfined phase takes place as a function of $J$. For large but finite $K$ the gauge fluxes are well-defined and either show Coulomb (logarithmic) interactions in the nematic, or appear as free excitations over the vacuum in the topological phase.
\end{itemize}
We also expect that the critical value $J_\mathrm{c}(K)$ shifts upwards as the value of $K$ is lowered below $K_c$ due to the cheap $\integers_N$-fluxes that disorder the rotors. For $K>K_\mathrm{c}$ on the other hand, the $\integers_N$-disclinations are effectively always gapped and the disordering of the rotors is due to spin-waves and $2\pi$-vortices, both controlled by the coupling $J$.  The full phase diagram in the $(K,J)$-plane can then be completed and leads to the schematic phase diagram shown in Fig.~\ref{fig:ZNphase}, which was also verified by Monte-Carlo simulations for lattices of sizes $N_{\rm sites}=12^3$ and is shown in Figs.~\ref{fig:Z2phase} and~\ref{fig:Z6phase}~\cite{LiuEtAl15}. 

%%%%%%%%%%%%%%%%%%%%%%%%%%
%%%%%%%%%%%%%%%%%%%%%%%%%

The surprise is the appearance of the topological phase at small $J$ and large $K$. This phase has disordered rotors but deconfined $\integers_N$-gauge fields. An order parameter that discerns it from the isotropic liquid phase with confined gauge fields is given by the so-called Fredenhagen--Marcu (FM) order parameter~\cite{FredenhagenMarcu86,GregorEtAl11, LiuEtAl15}. It is defined as
\begin{align}
\corr{R(C_{L_{n,m}})} = \frac{\langle z_{n}^* (\prod_{\corr{ij}\in C_L} U_{ij})z_{m}\rangle}{\sqrt{\corr{\prod_{\corr{ij}\in C_{2L}} U_{ij}}}}.
\end{align}
where $z_i = \te^{\ti \theta_i}$, $C_{L_{n,m}}$ is a curve of length $L\sim \abs{n-m}$ between the sites $n,m$ and $C_{2L}$ is a closed loop formed by closing the curve  $C_L$. The idea of the non-local FM order parameter $R(C_L)$ is the following. In gauge theories with matter, the Wilson loop ceases to be a good order parameter and always has a perimeter law decay $\te^{-L}$. The form of the FM order parameter is given by the gauge string $C_L$ connecting the rotors at $n,m$ in the numerator, divided by the square root of a suitable Wilson loop $C_{2L}$ formed from $C_{L}$ in the denominator, so that a non-vanishing limit can be obtained. In the deconfined phase with disordered matter fields, the string of rotors connected by the gauge links decays faster than $\te^{-L}$. We conclude that
\begin{align}
\lim_{L\to \infty} \corr{R(C_L)} = \begin{cases} 0,& \integers_N \textrm{ deconfined phase}, \\ \textrm{const.} \neq 0,& \textrm{ otherwise}.   \end{cases}
\end{align}
It is noteworthy that
\begin{align}
\corr{\psi^*_N(n)\psi_N(m)} = \corr{z_n^{*N} z^{N}_m } \sim \corr{R(C_{L_{n,m}})^N},
\end{align}
from which we conclude that the transition to the topological $\integers_N$-deconfined phase is always accompanied by the $XY^*$-transition of the rotors when the gauge fields are deconfined and stays second order but with associated topological order in the gauge fields as measured by the FM loop order parameter. 

\subsection{Dual formulation and fractionalized charges}

To make connection to the duality methods used in this review and to gain further intuition, we can perform a duality transform on the rotors of the $O(2)/\integers_N$ model. The transformation is strictly valid in the large-$J$ limit and proceeds as for the $XY$-model (see e.g.~\cite{Savit78} and Sec.~ \ref{sec:XY-duality}), with the result \cite{LiuEtAl15}
\begin{align}
\widetilde{\mathcal{S}}_N &= -\frac{1}{8\pi^2 J}\sum_{\widetilde{\Box}\in\widetilde{\Lambda}_{\tau}} \widetilde{a}_{\widetilde{\Box}}^2 -\ti \sum_{\langle \widetilde{i},\widetilde{j} \rangle\in\widetilde{\Lambda}_{\tau}} \widetilde{a}_{\widetilde{i}\widetilde{j}}\left( J^\mathrm{V}_{\Box} + \frac{a_{\Box}}{2\pi} \right) - K\sum_{\Box} \cos (a_{\Box}).
\label{eq:XYZNdual}
\end{align}
Here $\{\widetilde{a}_{\widetilde{i}\widetilde{j}}\}$ is a non-compact $U(1)$-gauge field dual to the the rotors $\{\theta_{i}\}$ and is defined on the links $\langle \widetilde{i},\widetilde{j} \rangle$ of the dual lattice $\widetilde{\Lambda}_{\tau} \simeq (\integers+\tfrac{1}{2})^3$ obtained from the cubic lattice $\integers^3$ by a shift by half a lattice translation. The sites $\widetilde{i}$, links $\langle \widetilde{i},\widetilde{j} \rangle$ and plaquettes $\widetilde{\Box}$ of the dual lattice are canonically related to the cubes, plaquettes and links of the original lattice, respectively. We have used a shorthand for the gauge fluxes over plaquettes $\Box$ and $\widetilde{\Box}$ as $f_{\Box} \equiv \sum_{\corr{ij}\in\Box} f_{ij}$. Similarly, $J^\mathrm{V}_{\Box}$ denotes the winding number of the $2\pi$-vortices over $\Box$. An action cost (core energy) of the form $\sim K' \sum_{\Box\in \Lambda_{\tau}} (J^{\mathrm{V}}_{\Box})^2$ has been neglected for these. In the partition function, we are instructed to sum over the gauge fields $\widetilde{a}_{\widetilde{i}\widetilde{j}}$ (in a gauge fix), and the integer defect configurations $\{J^\mathrm{V}_{\Box}\}\in \integers$ and $\{a_{\Box}\} \in \integers_{N}$.

The actions in Eqs.~\eqref{eq:CNNematic} and \eqref{eq:XYZNdual} can be seen as the lattice formulations of the generalizations of Eq.~\eqref{eq:Hnematic} to 2+1 dimensions including the explicit point group symmetries. The continuum action, corresponding to the non-compact gauge field (dual to the spin waves) and the defects, arises from coarse graining the microscopic lattice degrees of freedom. We see that the gauge interactions in Eq.~\eqref{eq:XYZNdual} give rise to the Coulomb (logarithmic) interactions between the defects $\{J^\mathrm{V}_{\Box}\}$, whereas the $\{\widetilde{a}_{\widetilde{i}\widetilde{j}}\}$ gauge fields also contain the smooth spin-wave fluctuations when summing over the defect-free configurations. 

The difference compared to the continuum treatment is, however, the explicit presence of the $\integers_N$-vortices and their relation to the $2\pi$-vortices. Of course, these additional fields are due to the inclusion of the finite point group symmetries. The $2\pi$-vortices $J^\mathrm{V}_{\Box}$ and the $\integers_N$ defects are charged under the dual gauge field, as in the standard $XY$-duality. However, the $\integers_N$ defects carry only a $1/N$-charge compared to the $2\pi$-vortices and we conclude that the fractionalized charge represents the $\integers_N$-disclinations. The action is invariant under the dual gauge transformations
\begin{align}
\widetilde{a}_{\widetilde{i},\mu} \to \widetilde{a}_{\widetilde{i},\mu} + \triangle_{\mu}\lambda_{\widetilde{i}}, \quad \sum_{\mu} \triangle_{\mu}\left( J^\mathrm{V}_{\widetilde{i},\mu} + \frac{a_{\widetilde{i},\mu}}{2\pi} \right) =0.
\end{align}
where $\{\lambda_{\widetilde{i}}\}\in \real$ is an arbitrary function on the dual lattice (remember that the dual links $(\widetilde{i},\mu)$ uniquely label plaquettes $\Box $ and vice versa). We see that the dual gauge symmetry, or the conservation of the rotor current, leads to the conservation of the charged defect currents in units of $1/N$. It is still possible for the $\integers_N$ current to source the $2\pi$-vortices and $a_{\Box}$ is only conserved mod $2\pi$. Similar considerations of fractionalized line-charges in 3+1-dimensions appeared in Ref.~\cite{GeraedtsMotrunich14}.

Phases with disordered rotors $\{\theta_i\}$ are characterized by Higgs condensates of the vortices $J_{\Box}$ and $a_{\Box}/2\pi$ that break the dual gauge symmetry. It is however possible that only the $2\pi$-vortices proliferate leaving the $\integers_{N}$-vortices above the vacuum (and this happens at $J = J_{\mathrm{c},\mathrm{3D}}$). This is the topological $\integers_N$-deconfined phase at large $K$ and small $J$. If we naively sum over the integers $J_{\Box}^{\rm V}$ in Eq.~\eqref{eq:XYZNdual}, we get the constraint $\widetilde{a}_{\widetilde{i}\widetilde{j}} \in 2\pi \integers$ signaling the condensation of $N$-tuples of the $\integers_N$ vortices, and arrive at a large gap $\sim 1/4J$ for the dual gauge fields (rotors) and the pure $\integers_N$-gauge theory in its deconfined phase. This is the ``Villain analogue'' of the Higgs transition. Finally, if $K<K_\mathrm{c}$, the $\integers_N$-gauge fields always condense as the rotor fields disorder.

\subsection{$C_N$-nematic transitions}

The above gauge model is the simplest model incorporating the $C_N$-nematics and leads to three phases, with the deconfined $\integers_N$ phase as surprise due to the possibility of independent defect dynamics. We now consider slightly more general theories with anisotropies and additional order parameters in order to drive transitions between the different $C_N$-nematics and the $\integers_N$-deconfined phases.

Just as for the nematic and smectic phases formed from dislocation condensates with certain (an)isotropies (see Sec.~\ref{subsec:Preview to defect-mediated melting} and later Secs. \ref{sec:Quantum nematic} and \ref{sec:Quantum smectic}) we can imagine cases where it is favorable to condense only certain types of disclinations. This is achieved by considering generalizations of $\mathcal{S}_{\rm gauge}$ by including anisotropic disclination weights $K\to \{K_{m}\}_{m\in\integers_N}$ of the form
\begin{align}
\mathcal{S}'_{\rm gauge} = -\sum_{\Box\in \Lambda_{\tau}} \sum_{m\in\integers_N} K_m\delta_{m,k_{\Box}} \cos (2\pi k_{\Box}/N),
\end{align}
 where have expanded the coefficients in terms of the $\integers_N$ character decomposition of $k_{\Box}$ in the fluxes $a_{\Box} = 2\pi k_{\Box}/N$. In particular for the cases $\integers_N = \integers_{M} \times \integers_{N/M}$, we can consider the decomposition
 \begin{align}
a_{ij} = \frac{2\pi}{N}k_{ij} = \frac{2\pi}{N}(\frac{N}{M}l_{ij} + n_{ij}), 
\end{align}
where $l_{ij} \in \{ 0,1,\dots,M-1 \} = \integers_{M}$ and $n_{ij} \in \{ 0,1,\dots, N/M-1\} = \integers_{N/M}$. We can now envisage different disclination weights for the subgroups as
 \begin{align}
 \mathcal{S}'_{\rm gauge} = \sum_{\Box\in\Lambda_{\tau}} -K_{N/M} \delta_{k_{\Box},\integers_{N/M}} \cos (2\pi k_{\Box}/N) - K_M\delta_{k_{\Box},\integers_{M}} \cos (2\pi l_{\Box}/M).
\end{align}
By tuning the coefficients $K_M$ and $K_{N/M}$ we can in principle condense different subsets of the original fluxes, provided that the elastic term controlled by $J$ is small enough. We can see that this scenario is feasible from the phase diagrams in Figs.~\ref{fig:Z2phase} and~\ref{fig:Z6phase}, since the defect proliferation occurs with larger $J$ and $K$ as a function of $N$.  Any flux condensate will however always completely disorder the $O(2)$ rotor fields. Nevertheless, the conclusion is that we can drive transitions between the different $\integers_N \to \integers_M$ deconfined phases by tuning $K_{N/M}$, which then have the role of describing anisotropies in the disclination weights, analogous to Ref.~\cite{OstlundHalperin81}.
 
An explicit way to break the nematic symmetry $C_N \to C_{M} $ where $C_N= C_M \times C_{N/M}$ is by introducing additional Higgs fields of the form
 \begin{align}
 \mathcal{S}_{\rm Higgs} = -J_M\sum_{\corr{ij}\in\Lambda_{\tau}} \sigma_{i}^* U_{ij}^{M}\sigma_j + \textrm{ c.c.},
 \end{align}
where $\sigma_i \in C_{N/M}\simeq \integers_{N/M}$ is a discrete field, carrying charge $M$ under the original $C_N$ group. Summing over the field $\{\sigma_i\} \in \integers_{N/M}$ produces terms $\sim U^N_{ij}$ that are trivial in the $C_N$ theory. For $J_M$ large enough or by explicit symmetry breaking, the auxiliary Higgs field condenses $\corr{\sigma_i} \neq 0$, and the resulting theory is different. We can pick a gauge where $\sigma_i = \te^{2\pi\ti M m_i/ N} = 1 \ \forall i\in \Lambda_{\tau}$, where $m_i\in\{0,1,\dots,N/M-1\} = \integers_{N/M}$ and define a new gauge field $a'_{ij} = \frac{2\pi}{N}k'_{ij} = \frac{2\pi}{N} (k_{ij} + m_i  - m_{j})$, to get
\begin{align}
\mathcal{S}_{\rm Higgs} = - (\frac{N}{M})^{\rm sites} J_M \sum_{\corr{ij}\in\Lambda_{\tau}} \cos (2\pi M k'_{ij}/N), \label{eq:CMHiggs}
\end{align}
which forces $k'_{ij} \in \integers_M$, or $n'_{ij}\equiv 0$, for all $\corr{ij}\in \Lambda_{\tau}$ for $J_M \gg K, J$. In the gauge term of the action this leads to
\begin{align}
\mathcal{S}_{\textrm{gauge}, N} &= \sum_{\Box\in\Lambda_{\tau}} - K  \delta_{k'_{\Box},\integers_{N/M}} \cos (2\pi k'_{\Box}/N)  - K\delta_{k'_{\Box},\integers_{M}} \cos (2\pi l'_{\Box}/M)  \label{eq:CMgauge}\\
&\to  \sum_{\Box\in\Lambda_{\tau}} -K \cos (2\pi l'_{\Box}/M) \equiv \mathcal{S}_{\textrm{gauge}, M}. \nonumber
\end{align}
We see that the the resulting theory is exactly the $O(2)/\integers_M$ theory describing the $C_M$ nematic. From the results of Ref.~\cite{FradkinShenker79}, we know in general that the Higgs phase and the confined phase of $\integers_{N/M}/\integers_{N/M}$ gauge theory are in fact in the same phase, with a phase transition to the deconfined phase of $\integers_{N/M}$ gauge fields described by a $\integers_{N/M}$ clock-spin model. We do not expect that the Higgs model Eq.~\eqref{eq:CMHiggs} with slightly modified gauge terms Eq.~\eqref{eq:CMgauge} has different universal behavior. For the $C_N$ point groups this then leads to Ising transitions between $C_6 \to C_3$ or $C_4\to C_2$ nematics, and to a (vector) $\integers_3$-Potts transitions between $C_6 \to C_2$ nematics. Of course, there should also be a transition $C_N \to C_1$ to a nematic with no symmetries which is in the (vector) Potts universality classes Eq.~\eqref{eq:CMHiggs}. 

\subsection{Concluding remarks}

This section has discussed the finite rotational symmetries of the quantum nematics in two spatial dimensions. We classified the possible symmetries of nematics using the two-dimensional rotational symmetries inherited from the crystalline wallpaper group. We then described the  long-distance hydrodynamic order parameters of these quantum nematic states, using a gauge-theoretical framework.

We have seen also in this section that compared to the standard Abelian-Higgs duality with (a non-compact) $U(1)$ group (Sec.~\ref{sec:XY-duality}), the presence of the finite $C_N$-point group symmetries allows for a completely new phase: the topological deconfined $\integers_N$-phase, where the gauge fluxes have become the only entities discernible from the ground state. This is possible when the cost of $2\pi$-vortices is lower than that of the $\integers_N$ fluxes. It has been argued that such a phase has in principle a microscopic identification in the context of fluctuating electronic stripes~\cite{ZaanenNussinov02, ZaanenNussinov03, PodolskyDemler05, MrossSenthil12}, where also a spin nematic can be realized. However, dealing with the `molecular' microscopy of the quantum liquid crystals discussed here, such an identification remains obscure. 
 
 \section{Dual elasticity}\label{sec:Dual elasticity}
 We will now commence the development of the main topic of this review: the quantum analogue of dislocation-mediated melting (Sec.~\ref{subsec:classical melting}) within the continuum field theory of elasticity. We will do so by employing dual gauge fields (Sec.~\ref{sec:XY-duality}) that encode the phonon degrees of freedom (Sec.~\ref{sec:Field-theoretic elasticity}) mediating interactions between dislocations with Burgers vectors (Sec.~\ref{sec:Topological defects in solids}). We will ignore the $C_N$-periodicity of the real nematic order parameter discussed in Sec.~\ref{sec:Order parameters for 2+1-dimensional nematics}, focusing on the `spherical cow' isotropic limit. The discreteness of the Frank scalars can be restored at the very end of this development.

In Sec.~\ref{sec:XY-duality}, we have learned from the vortex--boson-duality that dual operators are uniquely suited to handle disordering (melting) phase transitions. Smoothness of the original operators dictates a conservation law for the dual operators (the canonical momenta), which leads to the definition of a dual gauge field. These gauge fields correspond to the Goldstone modes in the ordered state, and they communicate the perturbations in the medium caused by defect sources. When these defects proliferate, the dual gauge fields become gapped by the Anderson--Higgs mechanism, signaling the demise of order. In elasticity, this pattern moves largely along the same lines, but is complicated by the larger amount of fields since displacements $u^a$ are vectors whereas the superfluid phase field $\varphi$ is a scalar, and the strain components are not independent of each other. Furthermore, we must deal with the Ehrenfest constraint prohibiting local rotations and the glide constraint which will enforce compressional rigidity even in the disordered states. 

\subsection{Stress gauge fields}\label{subsec:Stress gauge fields}
The first part of the duality construction, which was pioneered by Kleinert~\cite{Kleinert89b}, reproduces the familiar stress--strain relations. In the duality context, the strain variables $\partial_m u^a$ are the original fields from which the dual stress fields $\sigma_m^a$ are derived. The stresses are the canonical momenta conjugate to the generalized velocities that the strains represent. Here we expand this description to the quantum, dynamical case with imaginary-time derivatives $\partial_\tau u^a$ taken into account~\cite{ZaanenNussinovMukhin04}. The standard textbook spatial elasticity tensor is now expanded to a spatio-temporal variant $C_{\mu\nu ab}$ as in Eq.~\eqref{eq:elasticity relativistic Lagrangian}. Unlike in the vortex--boson-duality of Sec.~\ref{sec:XY-duality}, the resulting theory is absolutely not Lorentz-invariant due to the absence of timelike displacements $u^\tau \equiv 0$. The dualization procedure is nevertheless rather straightforward thanks to the Abelian symmetries.

The Lagrangian in terms of strains is Eq.~\eqref{eq:elasticity relativistic Lagrangian}:
\begin{equation}\label{eq:elasticity relativistic Lagrangian repeat}
\mathcal{L}_0 = \frac{1}{2} \partial_\mu u^{a} C_{\mu \nu a b} \partial_\nu u^b, 
\end{equation}
and we remind the reader that this is quadratic in terms of the small strains $\partial_\mu u^a$ (we neglect non-linear strains). The stress tensor familiar from elasticity textbooks~\cite{LandauLifshitz86, Kleinert89b} is Eq.~\eqref{eq:classical stress tensor},
\begin{equation}
 \underline{\sigma}_{ma} = \frac{\delta \mathcal{E}_{\rm cl}}{\delta u^{ma}},
\end{equation}
where $\mathcal{E}_{\rm cl}$ is the static energy of a crystal, basically the non-kinetic part of the action and the strain $u^{ma}$ is defined in Eq.~\eqref{eq:strain definition}. This stress tensor is symmetric in its indices by definition. Here we consider dynamics as well, in imaginary time, and the stress tensor that we employ in the duality was briefly defined in Eq.~\eqref{eq:stress} as the canonical momentum analogous to Eq.~\eqref{eq:superfluid canonical momentum}:
\begin{equation}\label{eq:stress definition}
 \sigma_\mu^a = -\ti \frac{ \delta \mathcal{S}}{\delta (\partial_\mu u^a)} = -\ti  C_{\mu\nu ab} \partial_\nu u^b.
\end{equation}
The factor $-\ti$ stems from the imaginary-time formalism, and is the same as in Eq.~\eqref{eq:superfluid canonical momentum}. We need to choose a velocity for the temporal derivative, and it is most natural to choose $c_\tT$ from Eq.~\eqref{eq:transverse phonon velocity}, so that 
\begin{equation}\label{eq:imaginary time shear velocity rescaling}
\partial_\mu = (\partial_\ft , \partial_m) =  (\frac{1}{c_\tT} \partial_\tau, \partial_m). 
\end{equation}
Because there are no displacements in the timelike direction $u^\ft = 0$, the indices $\mu,a$ of $\sigma^a_\mu$ now have an inequivalent status, and we find it useful to keep them apart as lower and upper indices respectively. In the absence of local rotations $\omega^{ma}$ in the action $\mathcal{S}[\doo_{\mu}u^a]$, this stress tensor is still symmetric in its {\em spatial} indices: $\sigma^a_m = \sigma^m_a$. 

Next, we will take the stresses as the principal variables, so that we need to invert Eq.~\eqref{eq:stress definition}, and express the strains $\partial_\nu u^b$ in terms of stress $\sigma^a_\mu$. However, the tensor $C_{\mu\nu ab}$ is in general not invertible. This can be solved by making the Lagrangian block-diagonal and inverting only the non-singular square submatrices. Let us consider the isotropic solid of Eq.~\eqref{eq:elasticity tensor}. The antisymmetric strains are absent, leading to several zero-eigenvalues of the elasticity tensor. However, if we restrict ourselves to symmetric strains, we can invert those parts. The decomposed relations explicitly read,
\begin{align}
  P^{(0)}_{mnab} \sigma_n^b &=  -\ti D \kappa P^{(0)}_{mnab} \partial_n u^b, \label {eq:P0 stress} \\
  P^{(1)}_{mnab} \sigma_n^b &= 0, \label {eq:P1 stress} \\
  P^{(2)}_{mnab} \sigma_n^b &=  -\ti 2 \mu P^{(2)}_{mnab} \partial_n u^b, \label {eq:P2 stress} \\
  \sigma^a_\ft &= -\ti \mu \partial_\ft u^a.
\end{align}
Here Eq.~\eqref{eq:P1 stress} implies that the stress tensor must be symmetric in its spatial indices. This result, which holds in equilibrium in the absence of external torques, can for quantum solids be traced back to the Ehrenfest theorem~\cite{Ehrenfest27} which states that the net force on a particle is proportional to the negative gradient of the potential, see e.g. Ref.~\cite{NielsenMartin85}. Therefore we call this constraint Eq.~\eqref{eq:P1 stress} the {\em Ehrenfest constraint}. In the present duality derivation, it is a direct consequence of the absence of local rotations in first-gradient elasticity, Eq.~\eqref{eq:strain Ehrenfest constraint}. In the context of the Dirac program of categorizing constraints~\cite{Dirac01,HenneauxTeitelboim92}, it is a so-called {\em primary constraint} as an overcomplete set of momenta. This is opposed to a {\em secondary constraint}, which is a dynamical consequence of equations of motion, as for instance a conservation law. 

These relations are readily inverted. We formally define
\begin{equation}
 C^{-1}_{\mu\nu ab} = \frac{1}{\mu} \delta_{\mu\ft} \delta_{\nu\ft}\delta_{ab} + \frac{1}{D\kappa} P^{(0)}_{\mu\nu ab} + \frac{1}{2\mu} P^{(2)}_{\mu\nu ab}.\label{eq:isotropic solid inverse stress tensor}
\end{equation}
Using the fact that projectors are idempotent for the same spin ${P^{(a)}}^2 = P^{(a)}$ and orthogonal for different spin $P^{(a)}P^{(b)}=0,\ a \neq b$, one may verify that the Hamiltonian can be written as,
\begin{align}
  \mathcal{H} &= -\ti \sigma_\mu^{a} \partial_\mu u^a + \mathcal{L}_0 \nonumber \\
  &= -\ti \sigma_\ft^a \partial_\ft u^a -  \ti\sigma_m^{a} \left (
  P^{(0)}_{mnab} + P^{(1)}_{mnab} + P^{(2)}_{mnab} \right ) \partial_n u^b
 + \tfrac 1 2 \partial_m u^{a} \left \lbrack P^{(0)}_{mnab} D \kappa P^{(0)}_{nkbc}
  + P^{(2)}_{mn ab} 2 \mu P^{(2)}_{nkbc} \right \rbrack \partial_k u^c \nonumber \\
  & \phantom{mmm} + \tfrac 1 2 \mu (\partial_\ft u^a)^2 \nonumber\\
  &= \tfrac 1 2 \sigma_\mu^{a} C^{-1}_{\mu \nu a b} \sigma_\nu^b. \label{eq:dual elasticity Hamiltonian}
\end{align}
Here we used $P^2 = P$ for each of the projectors and the definition of the stresses. As we did for the vortex--boson-duality, we transform this Hamiltonian back into a Lagrangian, where now the $\sigma^a_\mu$ are the principal variables,
\begin{align}
  \mathcal{L}_\mathrm{dual}[\sigma^a_\mu] &= \mathcal{H} + \ti \sigma_\mu^{a} \partial_\mu u^a \nonumber\\
  &=  \tfrac 1 2 \sigma_\mu^{a} C^{-1}_{\mu \nu a b} \sigma_\nu^b +  \ti \sigma_\mu^{a} \partial_\mu u^a  \label{eq:dual elasticity Lagrangian}
\end{align}
Again, we could have arrived here directly by performing a Hubbard--Stratonovich transformation. For  isotropic solids we can substitute the elastic tensor Eq.~\eqref{eq:isotropic solid inverse stress tensor}, leading to 
\begin{align}
 \mathcal{Z} &= \int \mathcal{D} \sigma^a_\mu \mathcal{D} u^b \ \te^{-\mathcal{S}_\mathrm{dual}}, \\
 \mathcal{S}_\mathrm{dual} &= \int_0^\beta \td \tau \int \td^D {\mathbf{x}} \ \mathcal{L}_\mathrm{dual}, \\
\mathcal{L}_\mathrm{dual} &=  \frac{1}{2\mu} (\sigma^a_\ft)^2 + \frac{1}{8\mu} \big[ \sigma^{a}_m \sigma^a_m + \sigma^{a}_m \sigma^m_a - \frac{2\nu}{1+\nu} \sigma^{a}_a \sigma^b_b\big] + \ti \sigma_\mu^{a} \partial_\mu u^a.
\end{align}
Next we separate the displacement field into smooth and singular parts $u^a(x) = u^a_\mathrm{smooth}(x) + u^a_\mathrm{sing}(x)$. On the smooth part we perform integration by parts, and then integrate it out as a Lagrange multiplier for the (secondary) constraint
\begin{equation}
\partial_\mu \sigma^a_\mu =0. \label{eq:stress conservation}
\end{equation}
This relation represents the well-known fact that stress is locally conserved in a solid. We will now restrict ourselves to 2+1 dimensions. 

For two spatial dimensions, Eq.~\eqref{eq:stress conservation} can be explicitly enforced by expressing the stress as the curl of a dual gauge field called the {\em stress gauge field},
\begin{equation}
 \sigma^a_\mu (x) = \epsilon_{\mu\nu\lambda} \partial_\nu b^a_\lambda (x).\label{eq:stress gauge field definition}
\end{equation}
The field $b^a_\lambda$ has six components. Here $b^a_\mu = (b^a_\ft , b^a_m) = (\frac{1}{c_\tT} b^a_\tau, b^a_m)$ and there are two independent gauge transformations which leave the stress tensor invariant,
\begin{equation}\label{eq:stress gauge transformation}
 b^a_\lambda (x) \to b^a_\lambda (x) + \partial_\lambda \varepsilon^a (x),
\end{equation}
where $\varepsilon^x$ and $\varepsilon^y$ are two arbitrary scalar fields. The number of physical degrees of freedom is then reduced to four. Enforcing the Ehrenfest constraint Eq.~\eqref{eq:P1 stress}
\begin{equation}\label{eq:dual gauge field Ehrenfest constraint}
 \epsilon_{\ft ma }\sigma^a_{m} =  \partial_\ft b^n_n - \partial_m b^m_\ft =0,
\end{equation}
reduces this further to three degrees of freedom, which correspond to the three symmetric strains $u^{ab}$ of Eq.~\eqref{eq:strain definition}. Substituting the definition Eq.~\eqref{eq:stress gauge field definition} into the Lagrangian Eq.~\eqref{eq:dual elasticity Lagrangian} we find
\begin{align}
 \mathcal{L}_\mathrm{dual} &= \tfrac 1 2 ( \epsilon_{\mu\kappa\lambda} \partial_\kappa b^{a}_\lambda ) C^{-1}_{\mu \nu a b} (\epsilon_{\mu\rho\sigma} \partial_\rho b^b_\sigma)   + \ti (\epsilon_{\mu\kappa\lambda} \partial_\kappa b^{a}_\lambda ) \partial_\mu u^a \nonumber \\
 &= \tfrac 1 2 ( \epsilon_{\mu\kappa\lambda} \partial_\kappa b^{a}_\lambda ) C^{-1}_{\mu \nu a b} (\epsilon_{\mu\rho\sigma} \partial_\rho b^b_\sigma)   + \ti b^{a}_\lambda J^a_\lambda, \label{eq:stress gauge field Coulomb action}
\end{align}
where in the second line we have substituted the dislocation current $J^a_\lambda = \epsilon_{\lambda\mu\nu} \partial_\mu \partial_\nu u^a_\mathrm{sing}$ from Eq.~\eqref{eq:dislocation density multivalued displacement}, after performing partial integration. From this action we see that dislocations $J^a_\lambda $ interact in the solid by exchanging gauge fields $b^a_\lambda$. For this reason the fields $b^a_\lambda$ are interchangeably referred to as {\em stress photons} in this review. The dynamic term for the gauge fields is slightly more complicated than in the vortex--boson-duality Eq.~\eqref{eq:XY Coulomb action}, yet we will refer to the isotropic solid as the Coulomb phase in terms of these dual gauge fields. See Sec.~\ref{subsec:Isotropic solid dual Lagrangian} for explicit expressions.

\subsection{Physical content of the stress tensor}\label{subsec:Physical content of the stress tensor}
Using the angular momentum projectors Eqs.~\eqref{eq:spin-0 projector}--\eqref{eq:spin-2 projector}, we can give a direct physical interpretation of the different components of the stress tensor. This follows from the Clebsch--Gordan composition of $2\otimes2$-dimensional representations of the 2-dimensional rotation group~\cite{Cvetkovic06}. The real representations of a rotation over an angle $\alpha$ are 
\begin{equation}
 R^n : \alpha \to \begin{pmatrix}\cos n\alpha & -\sin n \alpha \\ \sin n\alpha & \cos n\alpha \end{pmatrix}.
\end{equation}
With a slight abuse of language, we will call these the spin-$n$ representations. Over a complex field, the irreducible representations are $\rho^n:\alpha \mapsto \te^{\ti n \alpha}$. Then $R^n$ is in fact a reducible representation if considered over the complex field, as $R^n \simeq \rho^n  \oplus \rho^{-n}$. In the spatial part of the stress tensor $\sigma^a_n$ both indices transform under rotations as $R^1$. The decomposition is
\begin{equation}
 R^1 \otimes R^1 \simeq R^0 \oplus R^2 \simeq \rho^{0+} \oplus \rho^{0-} \oplus R^2.
\end{equation}
Here $\rho^{0+}$ and $\rho^{0-}$ have signs of $+1$ respectively $-1$ under spatial inversion. The change under spatial inversion stems from the corresponding strains $\sigma^a_m \propto \partial_m u^a$. The eigenvectors of these representations when acting on the stress tensor are
\begin{align}
 \rho^{0+} &: \sigma^x_x + \sigma^y_y \propto P^{(0)} \cdot \sigma^a_m ,\label{eq:compression stress}\\
 \rho^{0-} &: \sigma^x_y -\sigma^y_x \propto P^{(1)} \cdot \sigma^a_m \label{eq:rotation stress}\\
 R^2 &: \sigma^x_x - \sigma^y_y {\text{ and }} \sigma^x_y + \sigma^y_x \propto P^{(2)} \cdot \sigma^a_m .\label{eq:shear stress}
\end{align}
Not surprisingly, we see that these representations correspond precisely with the previous decomposition in terms of the spin-projectors. The component $\sigma^a_a = \sigma^x_x + \sigma^y_y$ is called {\em compression}: it is the response to isostatic pressure. It corresponds to stress due to volume changes. The component $\epsilon_{ab} \sigma^b_a = \sigma^x_y -\sigma^y_x$ is called {\em rotation stress}. It is absent in a solid to lowest order (Ehrenfest constraint Eq.~\eqref{eq:P1 stress}), but see Sec.~\ref{subsec:Second-gradient elasticity}. The spin-2 components are called {\em shear stress}, the response in one direction when strained in the perpendicular direction. When expressed in coordinates in longitudinal ($\tL$) and transverse ($\tT$) relative to the momentum $\mathbf{q}$, we name these components: $\sigma^\tL_\tL - \sigma^\tT_\tT$ {\em electric shear} and $\sigma^\tL_\tT + \sigma^\tT_\tL$ {\em magnetic shear}. The reason for this is that the electric shear is part of the longitudinal sector of the Lagrangian and it is associated with the longitudinal phonon, while the magnetic shear is part of the transverse sector of the Lagrangian and is related to the transverse phonon, see Sec.~\ref{subsec:Isotropic solid dual Lagrangian}.

These arguments can be extended to 2+1 dimensions including the temporal components $\sigma^a_\ft$~\cite{Cvetkovic06}. These two components transform under $R^1 \otimes 1$ and therefore do not `mix' with the other components in the decomposition, forming a vector (doublet) under the spatial rotations.

\subsection{Constraints in the path integral}\label{subsec:stress constraints}
The action Eq.~\eqref{eq:stress gauge field Coulomb action} must be amended by the constraints in the system: the Ehrenfest constraint Eq.~\eqref{eq:dual gauge field Ehrenfest constraint}, the glide constraint Eq.~\eqref{eq:2D glide constraint} and gauge fixing. The Ehrenfest constraint can be implemented by a Lagrange multiplier, or equivalently, by a delta function in the measure of the path integral. Similarly we can add a gauge fixing term to remove the gauge volume due to the unphysical degrees of freedom Eq.~\eqref{eq:stress gauge transformation} in the path integral. The path integral then reads,
\begin{equation}
  \mathcal{Z}_\mathrm{dual} = \int \mathcal{D} b_\mu^a ~ \mbox{``} \mathcal{D} J_\mu^a \mbox {''}\!  \prod_{a = x, y} \!\!
  \mathcal{F}^a \lbrack b_\mu^a \rbrack  ~ \delta (\partial_m b_\ft^m - \partial_\ft b^n_n) \te^{-\mathcal{S}_\mathrm{dual}}. \label{eq:dual stress constrained path integral}
\end{equation}
Here $F^a[b^a_\mu]$ is a gauge fixing functional to remove the gauge degrees of freedom. For instance, for the generalized Lorenz gauge this would be $F^a[b^a_\mu] = \delta(\partial_\mu b^a_\mu)$. Furthermore, the path integral over the dislocation sources $J^a_\mu$ is not rigorously defined: at this stage we aim to describe the response of the solid to the disturbance by a small number of isolated dislocations,  or rather use the dislocation as external sources to probe the elasticity response of the solid. A full grand-canonical treatment of dislocations will be the topic of Sec.~\ref{sec:Dynamics of disorder fields}.

The glide constraint Eq.~\eqref{eq:2D glide constraint} is a constraint on dislocations, not on the gauge fields. Since in the solid dislocations are treated as external sources, it is possible to violate the glide constraint as an experimentalist may inject interstitials to engage climb motion. The constraint per se applies only to internal, spontaneously formed dislocations. To enforce the constraint in the bulk of the solid, a Lagrange multiplier term can be imposed as
\begin{equation}
 \mathcal{L}_\mathrm{glide\ constraint} = \ti \lambda (\epsilon_{\ft m a} J^a_m).
\end{equation}
This form will be revisited in Section \ref{sec:Dynamics of disorder fields}.

To understand the physical consequences of the glide constraint, consider the minimal coupling term  $\ti b^a_\lambda J^a_\lambda$ in Eq.~\eqref{eq:stress gauge field Coulomb action}. The terms relevant for the glide constraint can be rewritten as
\begin{align}
 b^x_y J^x_y + b^y_x J^y_x = \frac{1}{2}(b^x_y - b^y_x)(J^x_y -J^y_x)  +\frac{1}{2}(b^x_y + b^y_x)(J^x_y + J^y_x).
\end{align}
The first term on the right-hand side contains the glide motion $\epsilon_{am} J^a_m = J^x_y -J^y_x$ which must be set to zero. Therefore, the stress component $b^x_y - b^y_x$ has no sources in the bulk of a solid. Via a gauge transformation $\varepsilon^x = -b^\ft_x$ and $\varepsilon^y = -b^\ft_y$  in Eq.~\eqref{eq:stress gauge transformation} the time derivative of this is equivalent to 
\begin{equation}
 \partial_\ft (b^x_y - b^y_x) \to \partial_\ft (b^x_y - b^y_x) - \partial_y b^x_\ft + \partial_x b^y_\ft = -(\sigma^x_x + \sigma^y_y).\label{eq:compression stress gauge field}
\end{equation}
Here we use the definition of the stress gauge field Eq.~\eqref{eq:stress gauge field definition}. Therefore, it is precisely the compressional stress Eq.~\eqref{eq:compression stress} that has no dislocation sources in the bulk of the solid. In other words: dislocations in the solid only cause shear stress, not volume changes. In Sec.~\ref{subsec:Kinematic constraints} we had already shown that the glide constraint originates in the conservation of particle number. In the analysis of the quantum liquid crystal phases in Secs.~\ref{sec:Quantum nematic} and \ref{sec:Quantum smectic} we will see that this has the consequence that the compressional Goldstone mode is protected from the dual Higgs mechanism and remains massless even after dislocation proliferation.

\subsection{Second-gradient elasticity}\label{subsec:Second-gradient elasticity}
It is useful to consider higher-order terms, not only for completeness, but also to get a better handle on the Ehrenfest constraint for dual fields. Furthermore, we cannot include disclinations if the stress fields are included only up to lowest order. We depart from second-gradient curvature $\sim (\partial_m \omega)^2$, Eq.~\eqref{eq:isotropic solid second gradient energy}. The higher-order compression is neglected by taking $\ell' \equiv 0$, because it does not add any new physics other than a refinement of the longitudinal propagator in Eq.~\eqref{eq:longitudinal strain propagator}. Since the rotations $\omega = \partial_x u^y - \partial_y u^x$ are gradients of displacements field and the dynamics of displacements was already incorporated in Eq.~\eqref{eq:elastivity kinetic Lagrangian}, at this stage there is no separate kinetic term for rotations. This amounts to an extra non-dynamical field $\omega \equiv \omega_\tau = \frac{1}{2}\epsilon_{\tau ab} \partial_a u^b$ and an extra term in the Lagrangian Eq.~\eqref{eq:elasticity relativistic Lagrangian repeat},
\begin{equation}
  \mathcal{L}^{(2)} = \tfrac 1 2 4 \mu \ell^2 (\partial_m \omega)^2, \label{eq:second gradient rotation term}
\end{equation}
However, in the quantum nematic phase the displacement fields have lost their physical meaning, and then the rotation field acquires its own kinetic term originating in the dislocation condensate, see Sec.~\ref{subsec:torque stress nematic}.

This introduces a new canonical momentum, the so-called {\em torque stress} $\tau^\ell_m$,
\begin{equation}\label{eq:2nd gradient torque stress}
 \tau^\ell_m = -\ti \frac{\partial \mathcal{S}}{\partial (\partial_m \omega)} =  -4 \ti \mu \ell^2 \partial_m \omega.
\end{equation}
Torque stress is the canonical angular momentum conjugate to rotation strains, just as linear stress $\sigma^a_\mu$ is the canonical linear momentum conjugate to displacement strains~\cite{Kleinert89b}. Here the label $\ell$ indicates that it is the torque stress descendant from the second-gradient elasticity term. Later we shall also find a relation between shear stress and torque stress, and this label will discriminate between these different origins. The dual Lagrangian Eq.\eqref{eq:dual elasticity Lagrangian} has to be extended by 
\begin{equation}
  \mathcal{L}^{(2)}_\mathrm{dual} =   \tfrac 1 2 \frac{1}{4 \mu \ell^2} (\tau^\ell_m)^2 + \ti \tau^\ell_m \partial_m \omega.\label{eq:2nd gradient torque stress Lagrangian}
\end{equation}
Substituting $\omega = \tfrac{1}{2}\epsilon_{ab} \partial_a u^b$ back, and integrating out the smooth part of the displacement $u^a$ as a Lagrange multiplier,
\begin{align}
 \mathcal{L}_\mathrm{int} 
 &= \ti \sigma^{a}_\mu \partial_\mu u^a + \ti \tau^\ell_n \partial_n \omega = -\ti (\partial_\mu \sigma^{a}) u^a +  \ti \tfrac{1}{2}(\epsilon_{\ft \mu a} \partial_\mu \partial_n \tau^\ell_n) u^a \nonumber \\
 &= - \ti u^{a} \partial_\mu (\sigma^{a}_\mu - \tfrac{1}{2} \epsilon_{\ft \mu a} \partial_n \tau^\ell_n),
\end{align}
we find a modified form of the stress conservation Eq.~\eqref{eq:stress conservation},
\begin{equation}
\partial_\mu ( \sigma^a_\mu - \frac{1}{2} \epsilon_{\ft \mu a} \partial_n \tau^\ell_n)=0.
\end{equation}
The second term is non-zero only for the antisymmetric part of $\sigma^a_m$, and only the equation for those components is modified. We see that the Ehrenfest constraint Eq.~\eqref{eq:P1 stress} is modified to \cite{LandauLifshitz86}
\begin{equation}\label{eq:second-gradient Ehrenfest constraint}
 \epsilon_{ma}\sigma^a_m= \partial_n \tau^\ell_n.
\end{equation}

When the rotational stiffness length $\ell$ is finite, the Ehrenfest constraint is softened and antisymmetric stress can exist, but only when it is sourced by torque stress. This creates the possibility to invert the elasticity tensor $C_{\mu\nu ab}$, which was up to now only partially accomplished in Eq.~\eqref{eq:isotropic solid inverse stress tensor}. We mention briefly two other ways to achieve this inversion. First, to the first-order Lagrangian \eqref{eq:dual elasticity Lagrangian}, one can add an Ehrenfest-violating first-gradient local rotation term $\sim \eta (\epsilon_{ab} \partial_a u^ b)^2$ with small parameter $\eta$, which is sent to zero at the end of the calculation. Second, one could introduce an external source term $\mathcal{J} \epsilon_{ma} \partial_m u^a$ and include Lagrange multiplier field that enforces the Ehrenfest constraint with an external source. After the duality transformation one can integrate out this field to find the correlation function. The latter method was used in Ref.~\cite{Cvetkovic06}.

\subsection{Torque stress gauge field}\label{subsec:Torque stress gauge field}
We wish to elaborate on torque stress a bit more, since this will be greatly important in determining the rotational degrees of freedom in the quantum liquid crystals. Even in the absence of second-gradient elasticity it is possible to define the torque stress tensor, again because translations and rotations are not independent. The Ehrenfest constraint then turns into a dynamical constraint. This technique was developed in Ref.~\cite{BeekmanWuCvetkovicZaanen13}.

Consider again the displacement and rotation fields $u^a$ and $\omega$ on the strain side of the duality. We will consider them as independent variables only now, and enforce the relation between them as a constraint $\tfrac{1}{2}\epsilon_{nb} \partial_n u^b -\omega = 0$. Split the stress tensor $\sigma^a_m$ in a symmetric $\bar{\sigma}^a_m$ and antisymmetric part $\breve{\sigma}^a_m$: $\sigma^a_m = \bar{\sigma}^a_m +\breve{\sigma}^a_m$, and recognize that in first-gradient elasticity, the antisymmetric component of the stress tensor $\breve{\sigma}^a_m \equiv \sigma^a_m - \sigma^m_a$ is absent. Only $\bar{\sigma}^a_m$ features quadratically in Eq.~\eqref{eq:dual elasticity Lagrangian}. We are for this reason at liberty to use the linearly appearing antisymmetric component as a Lagrange multiplier for the constraint and we add $\epsilon_{ma} \breve{\sigma}^a_m (\omega - \tfrac{1}{2}\epsilon_{nb} \partial_n u^b)$ to the Lagrangian. Then,
\begin{align}
 \mathcal{L}_\mathrm{constr.} &= \ti \sigma^a_\ft \partial_\ft u^a + \ti \bar{\sigma}^a_m \partial_m u^a + \ti\epsilon_{ma} \breve{\sigma}^a_m (\tfrac{1}{2}\epsilon_{nb} \partial_n u^b - \omega) \nonumber\\
 &=  \ti \sigma^a_\mu \partial_\mu u^a - \ti\epsilon_{ma} \sigma^a_m \omega.
\end{align}
Integrating out the smooth part of $u^a$ leads again to the stress conservation law $\partial_\mu \sigma^a_\mu = 0$, while $\omega$ enforces the Ehrenfest constraint $\epsilon_{ma} \sigma^a_m =0$. However, we may take the stress gauge field $b^a_\lambda$ from Eq.~\eqref{eq:stress gauge field definition}, and formally define the torque stress via,
\begin{equation}\label{eq:torque stress definition}
 \tau_\mu \equiv \epsilon_{ba} \epsilon_{b\mu\lambda} b^a_\lambda,
\end{equation}
such that $\tau_{\ft} = -b^a_a$ and $\tau_a = b^a_{\ft}$. Note the different origin of this torque stress, which is applicable even in linear elasticity, and of the one originating from second-gradient elasticity in Eq.~\eqref{eq:2nd gradient torque stress}. Substituting Eq.~\eqref{eq:torque stress definition} into the Ehrenfest constraint $\epsilon_{ma} \sigma^a_m =0$ it follows that this constraint corresponds to
\begin{equation}
 \partial_\mu \tau_\mu = 0.\label{eq:dynamical Ehrenfest constraint}
\end{equation}
This turns the strict absence of antisymmetric stress into a {\em dynamical} constraint, that ensures that torque stress is conserved in any solid lacking a rotational elastic response. In this way we have accomplished to turn the primary constraint into a secondary constraint, in Dirac's terminology~\cite{Dirac01,HenneauxTeitelboim92}. Note that even though the constraint can be reduced to just eliminating the component $\tau_0$ along the space-time 3-momentum of the field (see Sec.~\ref{subsec:Conventions and notation}), all three components $\tau_\mu$ in Eq.~\eqref{eq:torque stress definition}  have a physical meaning. The temporal part $\tau_{\ft}$ corresponds to intrinsic `spin' angular momentum density (i.e. angular momentum not due to the linear momenta $\sigma_{\ft}^a$) and $\doo_a \tau_a$ is the intrinsic rotational moment or couple-stress density \cite{LandauLifshitz86, Toupin62, Toupin64}. Note the difference between this equation with temporal components and Eq.~\eqref{eq:second-gradient Ehrenfest constraint} of second-gradient elasticity with only spatial components; the consequence is that when the right-hand side of Eq.~\eqref{eq:second-gradient Ehrenfest constraint} is expressed in terms of the stress gauge field $b^a_\lambda$, these components do not match with those of Eq.~\eqref{eq:torque stress definition}. One should be careful with distinguishing between $\tau^\ell_m$ and $\tau_\mu$.

The constraint Eq.~\eqref{eq:dynamical Ehrenfest constraint} can be explicitly enforced by expressing the torque stress as the curl of yet another gauge field, the {\em torque stress gauge field} $h_\lambda$:
\begin{equation}\label{eq:torque stress gauge field definition}
 \tau_\mu = \epsilon_{\mu\nu\lambda} \partial_\nu h_\lambda.
\end{equation}
The transverse component $h_\tT$ correspond to the rotational Goldstone mode (in a suitable gauge fix), while the temporal component $h_{\ft}$ corresponds to static forces between disclination and dislocation sources. The great benefit of the definition Eq.~\eqref{eq:torque stress definition} is that the torque stress gauge field $h_\lambda$ couples directly to disclination sources $\Theta_\lambda$ of Eq.~\eqref{eq:2D disclination density}, just as the stress gauge field $b_\lambda$ couples to the dislocation density according to Eq.~\eqref{eq:stress gauge field Coulomb action}. This can be seen from
\begin{align}
 -\ti\epsilon_{ma} \sigma^a_m \omega_\mathrm{sing} &= -\ti(\partial_\mu \tau_\mu) \omega_\mathrm{sing} =  \ti\tau_\mu \partial_\mu \omega_\mathrm{sing} 
 = \ti(\epsilon_{\mu \nu \lambda} \partial_\nu h_\lambda)\partial_\mu \omega_\mathrm{sing} 
 = - \ti h_\lambda \epsilon_{\lambda \mu\nu} \partial_\nu \partial_\mu \omega_\mathrm{sing} \nonumber\\
 &= \ti h_\lambda \Theta_\lambda. \label{eq:disclination torque stress coupling}
\end{align}

Because in two dimensions there is only one plane of rotation, both the torque stress and the torque stress gauge field do not carry any upper indices. The number of indices follows those of the Frank tensor  and in particular in three dimensions we would have a vector index $c$ that is orthogonal to the plane of rotation $ab$.  We will come back to the torque stress and rotational gauge fields in Sec.~\ref{subsec:torque stress nematic}.

\subsection{Dual Lagrangian of the isotropic solid}\label{subsec:Isotropic solid dual Lagrangian}
In the remainder we will need an explicit expression for the Lagrangian Eq.~\eqref{eq:stress gauge field Coulomb action} of the isotropic solid.
This is accomplished by inserting the values of the inverse elastic tensor $C^{-1}_{\mu \nu a b}$ of the isotropic solid, Eq.~\eqref{eq:isotropic solid inverse stress tensor}. Interestingly, if we go to Fourier space and $(\ft,\tL,\tT)$-coordinates (see Sec.~\ref{subsec:Conventions and notation} and \ref{sec:Fourier space coordinate systems}) the Lagrangian becomes block diagonal with two blocks $\mathcal{L} = \mathcal{L}^\tL + \mathcal{L}^\tT$ that we shall call longitudinal $\mathcal{L}^\tL$ and transverse $\mathcal{L}^\tT$, as they contain the longitudinal and transverse phonon, respectively. Including the second-gradient rotation terms by substituting $\tau^\ell_\tL = -\frac{1}{q} \epsilon_{ma} \sigma_m^a = -\frac{1}{q}(\ti \tfrac{\omega_n}{c_\tT} b^\tT_\tT + \ti \tfrac{\omega_n}{c_\tT} b^\tL_\tL -qb^\tL_\ft)$ from Eq.~\eqref{eq:second-gradient Ehrenfest constraint} into Eq.~\eqref{eq:2nd gradient torque stress Lagrangian}, we find
\begin{align}
\mathcal{L}^\tL &= \tfrac{1}{2} \frac{1}{4\mu}\frac{2}{(1+\nu)} 
\begin{pmatrix} b^{\tT\dagger}_\ft \\ b^{\tT\dagger}_\tL \\ b^{\tL \dagger}_\tT \end{pmatrix}^\tT
\begin{pmatrix}
 q^2 & -\ti \tfrac{1}{c_\tT} \omega_n q & -\ti \nu \tfrac{1}{c_\tT}  \omega_n q \\
\ti \tfrac{1}{c_\tT} \omega_n q & \tfrac{1}{c_\tT^2 }\omega_n^2 & \nu\tfrac{1}{c_\tT^2} \omega_n^2 \\
\ti \nu \tfrac{1}{c_\tT} \omega_n q & \nu \tfrac{1}{c_\tT^2} \omega_n^2 & \tfrac{1}{c_\tT^2}\omega_n^2 + 2(1 + \nu) q^2 
\end{pmatrix}
\begin{pmatrix} b^{\tT}_\ft \\ b^{\tT}_\tL \\ b^{\tL}_\tT \end{pmatrix}
+ \ti b^{\tT\dagger}_\ft J^{\tT}_\ft  + \ti b^{\tT\dagger}_\tL  J^{\tT}_\tL + \ti b^{\tL \dagger}_\tT J^{\tL}_\tT ,
\label{eq:longitudinal Lagrangian stress gauge field}\\
\mathcal{L}^\tT &= \tfrac{1}{2} \frac{1}{4\mu} 
\begin{pmatrix} b^{\tL\dagger}_\ft \\ b^{\tL\dagger}_\tL \\ b^{\tT \dagger}_\tT \end{pmatrix}^\tT
\begin{pmatrix}
 q^2 (1 + \tfrac{1}{\ell^2 q^2})  & -\ti \tfrac{1}{c_\tT}\omega_n q(1 + \tfrac{1}{\ell^2 q^2})  & \ti \tfrac{1}{c_\tT} \omega_n q(1 - \tfrac{1}{\ell^2 q^2})  \\
\ti \tfrac{1}{c_\tT} \omega_n q (1 + \tfrac{1}{\ell^2 q^2})  & \tfrac{1}{c_\tT^2}\omega_n^2(1 + \tfrac{1}{\ell^2 q^2})  & - \tfrac{1}{c_\tT^2}\omega_n^2 (1 - \tfrac{1}{\ell^2 q^2}) \\
-\ti \tfrac{1}{c_\tT} \omega_n  q(1 - \tfrac{1}{\ell^2 q^2}) & - \tfrac{1}{c_\tT^2 }\omega_n^2(1 - \tfrac{1}{\ell^2 q^2}) & \tfrac{1}{c_\tT^2}\omega_n^2(1 + \tfrac{1}{\ell^2 q^2})  + 4 q^2 
\end{pmatrix}
\begin{pmatrix} b^{\tL}_\ft \\ b^{\tL}_\tL \\ b^{\tT}_\tT \end{pmatrix} \nonumber\\
&\phantom{mmm}
+ \ti b^{\tL\dagger}_\ft J^{\tL}_\ft  + \ti b^{\tL\dagger}_\tL  J^{\tL}_\tL + \ti b^{\tT \dagger}_\tT J^{\tT}_\tT .\label{eq:transverse Lagrangian stress gauge field}
\end{align}
Recall that the gauge fields and dislocation sources are real fields, and the $\dagger$-symbol merely denotes $b^{a\dagger}_\lambda(p) = b^a_\lambda(-p)$ in Fourier space. This also implies $\int_p \ti b^{a\dagger}_\lambda(p) J^a_\lambda(p) = \int_p \ti J^{a\dagger}_\lambda(p) b^a_\lambda(p)$ which can be seen by transforming $p \to -p$ for the 
whole integral. Note that longitudinal and transverse directions in the Burgers index $a$ get mixed in these two sectors.

The expressions Eqs. \eqref{eq:longitudinal Lagrangian stress gauge field}, \eqref{eq:transverse Lagrangian stress gauge field} do not yet take into account any gauge fixing or constraints. Each of them contains one gauge degree of freedom due to the gauge transformations $b^a_\lambda \to b^a_\lambda +\partial_\lambda \varepsilon^a$. This can be handled in two ways. One can either perform an explicit gauge fix, such as imposing the Coulomb gauge fix $\partial_m b^a_m = -qb^a_\tL =0$ removing the components $b^a_\tL$ from the Lagrangian altogether. The other option is to perform a coordinate transformation to the $(0,+1,-1)$-system and substitute $pb^a_{+1} = \ti \frac{\omega_n}{c_\tT} b^a_\tL - q b^a_\ft$, see \ref{sec:Fourier space coordinate systems}. The gauge degree of freedom corresponds then precisely to the $0$-direction in Fourier space, and is therefore completely absent from the Lagrangian. This is equivalent to choosing the explicit Lorenz gauge fix $\partial_\mu b_\mu^a = 0$. In either case, the superfluous gauge degree of freedom has to be gauge fixed or integrated out at some stage of the calculation. The  result is that the expressions simplify to
\begin{align}
\mathcal{L}^\tL &= \tfrac{1}{2} \frac{1}{4\mu}\frac{2}{(1+\nu)} 
\begin{pmatrix} b^{\tT\dagger}_{+1} \\ b^{\tL \dagger}_\tT \end{pmatrix}^\tT
\begin{pmatrix}
 p^2 &  \ti \nu \tfrac{1}{c_\tT} \omega_n p \\
-\ti \nu \tfrac{1}{c_\tT} \omega_n p & \tfrac{1}{c_\tT^2} \omega_n^2 + 2(1 + \nu) q^2 
\end{pmatrix}
\begin{pmatrix} b^{\tT}_{+1} \\ b^{\tL}_\tT \end{pmatrix}
+ \ti b^{\tT\dagger}_{+1} J^{\tT}_{+1} + \ti b^{\tL \dagger}_\tT J^{\tL}_\tT ,
\label{eq:longitudinal Lagrangian stress gauge field compact}\\
\mathcal{L}^\tT &= \tfrac{1}{2} \frac{1}{4\mu} 
\begin{pmatrix} b^{\tL\dagger}_{+1} \\ b^{\tT \dagger}_\tT \end{pmatrix}^\tT
\begin{pmatrix}
 p^2 (1 + \tfrac{1}{\ell^2 q^2})  & -\ti \tfrac{1}{c_\tT} \omega_n p (1 - \tfrac{1}{\ell^2 q^2})  \\
\ti \tfrac{1}{c_\tT} \omega_n p (1 - \tfrac{1}{\ell^2 q^2}) & \tfrac{1}{c_\tT^2}\omega_n^2(1 + \tfrac{1}{\ell^2 q^2})  + 4  q^2 
\end{pmatrix}
\begin{pmatrix} b^{\tL}_{+1} \\ b^{\tT}_\tT \end{pmatrix}
+ \ti b^{\tL\dagger}_{+1} J^{\tL}_{+1} + \ti b^{\tT \dagger}_\tT J^{\tT}_\tT .\label{eq:transverse Lagrangian stress gauge field compact}
\end{align}

\subsection{Correlation functions}\label{subsec:stress correlation functions}
To compare results on the original and dual sides, we should formulate correlation functions in terms of the dual stress:  we want to obtain relations of the kind  Eq.~\eqref{eq:superfluid Zaanen--Mukhin relation} for dual elasticity. The naive general form of these relations is,
\begin{equation}
 \langle \partial_\mu u^a \partial_\nu u^b \rangle = C^{-1}_{\mu\nu ab} - C^{-1}_{\mu\kappa ac}C^{-1}_{\nu \lambda bd} \langle \sigma^c_\kappa \sigma^d_\lambda \rangle.
\end{equation}
However, the elasticity tensor $C_{\mu\nu ab}$ is generally not invertible. We focus primarily on the longitudinal and transverse propagators Eqs. \eqref{eq:longitudinal strain propagator} and \eqref{eq:transverse strain propagator}. Following the recipe of Section \ref{subsec:Correlation functions}, we add external sources $\mathcal{K}$ and $\mathcal{J}$ to the strain Lagrangian Eq.~\eqref{eq:elasticity relativistic Lagrangian repeat}:
\begin{align}\label{eq:dual elasticity with sources}
 \mathcal{L} &= \frac{1}{2} \partial_\mu u^{a} C_{\mu \nu a b} \partial_\nu u^b + \mathcal{K} \partial_a u^a + \mathcal{J}  \epsilon_{ab} \partial_a u^b \nonumber\\
 &= \frac{1}{2} \partial_\mu u^{a} C_{\mu \nu a b} \partial_\nu u^b + \mathcal{K} \delta_{ma} P^{(0)}_{mnab} \partial_n u^b  + \mathcal{J}  \epsilon_{ma} P^{(1)}_{mnab} \partial_n u^b.
\end{align}
The stress tensor Eq.~\eqref{eq:stress definition} changes accordingly, and  we find the modified versions of Eqs \eqref{eq:P0 stress}--\eqref{eq:P2 stress} as,
\begin{align}
  P^{(0)}_{mnab} \sigma_n^b &=  -\ti D \kappa P^{(0)}_{mnab} \partial_n u^b -\ti \mathcal{K} \delta_{ma}, \label {eq:P0 stress with source} \\
  P^{(1)}_{mnab} \sigma_n^b &= -\ti \epsilon_{ma} \mathcal{J}, \label {eq:P1 stress with source} \\
  P^{(2)}_{mnab} \sigma_n^b &= -\ti 2 \mu P^{(2)}_{mnab} \partial_n u^b, \label {eq:P2 stress with source} \\
  \sigma^a_\ft &= -\ti \mu \partial_\ft u^a.
\end{align}
Due to the relation Eq.~\eqref{eq:P1 stress with source} this is still not invertible. For the longitudinal propagator this poses no problem, and we repeat the procedure of Section \ref{subsec:Correlation functions} to find,
\begin{align}
 \mathcal{H}[\mathcal{K}] &= -\ti \sigma^a_\mu \partial_\mu u^a + \frac{1}{2} \partial_\mu u^{a} C_{\mu \nu a b} \partial_\nu u^b + \mathcal{K} \partial_a u^a = \frac{1}{2} \sigma^a_\mu C^{-1}_{\mu\nu ab} \sigma^b_\nu - \frac{1}{2\kappa}  \mathcal{K}^2 + \ti \frac{1}{D\kappa} \mathcal{K} \sigma^a_a, \nonumber\\
 \mathcal{L}_\mathrm{dual} &= \frac{1}{2} \sigma^a_\mu C^{-1}_{\mu\nu ab} \sigma^b_\nu - \frac{1}{2\kappa}  \mathcal{K}^2 +  \ti \frac{1}{D\kappa}\mathcal{K} \sigma^a_a + \ti \sigma^a_\mu \partial_\mu u^a. 
\end{align}
The correlation function relation for the longitudinal propagator becomes,
\begin{align}\label{eq:longitudinal propagator stress tensor}
 G_\tL &= \langle \partial_a u^a \partial_b u^b \rangle 
 = \frac{1}{\mathcal{Z}[0]} \left. \frac{\delta}{\delta \mathcal{K}} \frac{\delta}{\delta \mathcal{K}} \mathcal{Z}[\mathcal{K}]\right\vert_{\mathcal{K} = 0} = \frac{1}{\kappa} - \frac{1}{(D\kappa)^2} \langle \sigma^a_a \sigma^b_b \rangle.
\end{align}
We can also express this in terms of stress gauge fields by using 
\begin{align}
 \sigma^a_a &= \partial_\ft(b_x^y - b_y^x) + \partial_y b^x_\ft - \partial_x b^y_\ft = -\ti \frac{\omega_n}{c_\tT} (b^\tT_\tL - b^\tL_\tT) + q b^\tT_\ft = \ti \frac{\omega_n}{c_\tT} b^\tL_\tT - p b^\tT_{+1}.
\end{align}
yielding the result in terms of stress photons
\begin{align}
 G_\tL &=  \frac{1}{\kappa} - \frac{1}{(D\kappa)^2} \Big[ p^2 \langle b^{\tT\dagger}_{+1} b^\tT_{+1} \rangle -\ti \frac{\omega_n}{c_\tT} p \langle b^{\tT\dagger}_{+1}  b^\tL_\tT \rangle + \ti \frac{\omega_n}{c_\tT} p \langle {b^\tL_\tT}^\dagger  b^\tT_{+1}\rangle + \frac{\omega_n^2}{c_\tT^2 }  \langle {b^\tL_\tT}^\dagger b^\tL_\tT \rangle \Big].\label{eq:longitudinal Zaanen-Mukhin stress gauge field}
\end{align}

For the transverse propagator, we must deal with the non-invertibility of Eq.~\eqref{eq:P1 stress with source}, using any one of the methods mentioned at the end of Section \ref{subsec:Second-gradient elasticity}. Since second-gradient elasticity has a well-established physical origin, it is arguably more elegant to use this rather than introducing unphysical fields and setting them to zero later. We therefore choose to incorporate the second-gradient elasticity term of Eq.~\eqref{eq:second gradient rotation term}, which gives rise to a new dual variable, the torque stress $\tau^\ell_m$. The transverse propagator is, cf. Eq.~\eqref{eq:transverse strain propagator},
\begin{equation}\label{eq:transverse propagator rotation field}
 G_\tT = \sum_{ab} \langle (\partial_a u^b - \partial_b u^a)^2 \rangle = 2 \sum_{ab} \langle \omega_{ab} \omega_{ab} \rangle = 4 \langle \omega \omega \rangle. 
\end{equation}
Recall that $\omega = \tfrac{1}{2} \epsilon_{ab} \omega_{ab} =  \tfrac{1}{2} \epsilon_{ab} \tfrac{1}{2} (\partial_a u^b - \partial_b u^a)$. We substitute this in the external source term of Eq.~\eqref{eq:dual elasticity with sources} added to Eq.~\eqref{eq:second gradient rotation term},
\begin{equation}
 \mathcal{L}^{(2)} = \tfrac 1 2 4 \mu \ell^2 (\partial_m \omega)^2 + 2 \omega \mathcal{J}.
\end{equation}
It follows that the last form in Eq.~\eqref{eq:transverse propagator rotation field} corresponds to $G_\tT = \frac{1}{Z[0]} \frac{\delta}{\delta \mathcal{J}} \frac{\delta}{\delta \mathcal{J}} \mathcal{Z}[\mathcal{J}]\rvert_{\mathcal{J} = 0}$. To obtain the torque stress we use $1 = \frac{\partial_m^2}{\partial_n^2} = \frac{\partial_m}{\partial_n^2} \partial_m$ to find
\begin{align}
 \tau^\ell_m &= -\ti 4 \mu \ell^2 \partial_m \omega + 2\ti \frac{\partial_m}{\partial_n^2}\mathcal{J},\\
 \partial_m \omega &= \frac{1}{4\mu \ell^2} \big( \ti \tau^\ell_m + 2  \frac{\partial_m}{\partial_n^2}\mathcal{J} \big).
\end{align}
Now we perform the dualization procedure as in Eq.~\eqref{eq:2nd gradient torque stress Lagrangian} but including the external source, to find,
\begin{align}
 \mathcal{H} &= - \ti \tau^\ell_m \partial_m \omega +  \tfrac 1 2 4 \mu \ell^2 (\partial_m \omega)^2 + 2 \omega \mathcal{J} 
 = \frac{1}{4 \mu \ell^2} \big( \tfrac{1}{2} (\tau^\ell_m)^2 - 2\ti \tau^\ell_m  \frac{\partial_m}{\partial_n^2}\mathcal{J} - 2  \frac{\partial_m}{\partial_n^2}\mathcal{J} \frac{\partial_m}{\partial_n^2}\mathcal{J} \big) \nonumber \\
 &= \frac{1}{4 \mu \ell^2} \big( \tfrac{1}{2} (\tau^\ell_m)^2 + 2\ti   \frac{\partial_m \tau^\ell_m}{\partial_n^2}\mathcal{J} + 2  \mathcal{J} \frac{1}{\partial_n^2}\mathcal{J} \big)
 = \frac{1}{4 \mu \ell^2} \big( \tfrac{1}{2} (\tau^\ell_m)^2 - 2\ti   \frac{(\partial_m \tau^\ell_m)}{q^2}\mathcal{J} - 2  \mathcal{J} \frac{1}{q^2}\mathcal{J} \big).
\end{align}
Here we used $\partial_n^2 = (\ti q_n)^2 = -q^2$ in the last step. The correlation function relation in terms of torque stress becomes
\begin{align}\label{eq:transverse propagator torque stress}
 G_\tT &= \frac{1}{\mathcal{Z}[0]} \left. \frac{\delta}{\delta \mathcal{J}} \frac{\delta}{\delta \mathcal{J}} \mathcal{Z}[\mathcal{J}]\right\vert_{\mathcal{J} = 0} 
 = \frac{1}{ \mu \ell^2 q^2} - \frac{1}{(2 \mu \ell^2 q^2)^2} \langle \partial_m \tau^\ell_m\  \partial_n \tau^\ell_n \rangle.
\end{align}
This is still expressed in terms of torque stress, pertaining to second-gradient fields. However, we can substitute the Ehrenfest constraint Eq.~\eqref{eq:second-gradient Ehrenfest constraint} to find the form which is also valid for linear elasticity in the limit $\ell \to 0$,
\begin{equation}\label{eq:transverse propagator stress tensor}
 G_\tT =  \frac{1}{ \mu \ell^2 q^2} - \frac{1}{(2 \mu \ell^2 q^2)^2} \langle \epsilon_{ma} \sigma_m^a \ \epsilon_{nb} \sigma_n^b  \rangle.
\end{equation}
This can also be expressed in terms of the stress gauge fields by using,
\begin{align}
 \epsilon_{ma} \sigma_m^a &= - \partial_\ft (b^x_x + b^y_y) + \partial_y b^y_\ft + \partial_x b^x_\ft 
 = -\ti \frac{\omega_n}{c_\tT} b^\tL_\tL - \ti \frac{\omega_n}{c_\tT} b^\tT_\tT + q b^\tL_\ft \nonumber\\
 &= - p b^\tL_{+1} - \ti \frac{\omega_n}{c_\tT} b^\tT_\tT,
\end{align}
yielding as the final result
\begin{align}
 G_\tT &=  \frac{1}{ \mu \ell^2 q^2} - \frac{1}{(2 \mu \ell^2 q^2)^2}\Big[  p^2 \langle  {b^\tL_{+1}}^\dagger b^\tL_{+1} \rangle + \ti \frac{\omega_n}{c_\tT} p \langle  {b^\tL_{+1}}^\dagger b^\tT_\tT \rangle - \ti \frac{\omega_n}{c_\tT} p \langle  {b^\tT_\tT}^\dagger b^\tL_{+1} \rangle  + \frac{\omega_n^2}{c_\tT^2}  \langle  {b^\tT_\tT}^\dagger b^\tT_\tT \rangle\Big].\label{eq:transverse Zaanen-Mukhin stress gauge field}
\end{align}
One should insert the correlation functions $\langle b^\dagger b  \rangle$ and set $\ell \to 0$ afterwards, to obtain results for first-gradient elasticity only. The terms with factors $\ell^2$ in the denominator will cancel, as we shall see later.

At this point one can wonder whether it is really necessary to use the dual gauge fields at all, since the longitudinal and transverse propagators appear to be complete and simpler when expressed in terms of the stress tensor $\sigma_\mu^a$. The reason for introducing gauge fields is twofold. First, the conservation of stress $\partial_\mu \sigma_\mu^a=0$ is automatically imposed, whereas otherwise one would have to keep track of an additional constraint, which at least makes the calculation of Eqs.~\eqref{eq:longitudinal propagator stress tensor} and \eqref{eq:transverse propagator stress tensor} not as easy as it may seem. Second, the true objective of this work is to describe the quantum smectic and nematic phases, which are Bose--Einstein condensates of dislocations. The dislocation sources couple not to the stress tensor (the `field strength') but instead to the gauge fields in Eq.~\eqref{eq:stress gauge field Coulomb action}. The dislocation condensate is the Higgs phase for these gauge fields, and mimics precisely the photon field in a superconductor. Just as the electrodynamics of superconductors follows directly from the Ginzburg--Landau action for the superconducting order parameter (Cooper pair condensate) with the minimally coupled photon field, so does the elasticity of quantum liquid crystal follow immediately from the action of the dislocation condensate with minimally coupled stress gauge fields.

For later use, let us also present the propagator relation for the chiral propagator Eq.~\eqref{eq:chiral strain propagator}, which follows from
\begin{align}
 G_{\tL\tT} &=  \langle \partial_a u^a  \omega_{bc} \rangle = \frac{1}{\mathcal{Z}[0]} \left. \frac{\delta}{\delta \mathcal{K}} \frac{\delta}{\delta \mathcal{J}} \mathcal{Z}[\mathcal{K},\mathcal{J}]\right\vert_{\mathcal{K} = \mathcal{J} =0} = \frac{1}{4\mu \kappa} \frac{1}{q^2 \ell^2} \langle \sigma^a_a \epsilon_{mb} \sigma_m^b \rangle \nonumber\\
 &=  \frac{1}{4\mu \kappa} \frac{1}{q^2 \ell^2} \Big[ - \frac{\omega_n^2}{c_\tT^2} \langle b_\tT^{\tL\dagger} b^\tT_\tT \rangle + p^2 \langle b_{+1}^{\tT\dagger} b^\tL_{+1} \rangle + \ti \frac{\omega_n p}{c_\tT} \langle b_\tT^{\tL\dagger} b^\tL_{+1} \rangle + \ti \frac{\omega_n p}{c_\tT}\langle b_{+1}^{\tT\dagger} b^\tT_\tT \rangle \Big], \label{eq:chiral propagator stress gauge field}
\end{align}
while $G_{\tT\tL} = G_{\tL\tT}^\dagger$.

\subsection{Correlation functions from stress gauge fields}
It is instructive to rederive the explicit expressions for the  propagators of the isotropic solid Eqs.~\eqref{eq:longitudinal strain propagator},\eqref{eq:transverse strain propagator} completely on the dual side. We can verify that the relations Eqs.~\eqref{eq:longitudinal Zaanen-Mukhin stress gauge field},\eqref{eq:transverse Zaanen-Mukhin stress gauge field} are in fact correct, and we will be able to practice such derivations that we need for the quantum liquid crystal phases with dislocation condensates of Secs.~\ref{sec:Quantum nematic} and \ref{sec:Quantum smectic}.

The recipe is as follows: take the partition function in terms of stress gauge fields Eq.~\eqref{eq:stress gauge field Coulomb action}; use the external dislocation sources $J^a_\lambda$ and integrate out all gauge fields $b^a_\lambda$ in the path integral, leaving only expression in terms of $J^a_\lambda$; take functional derivatives with respect to these sources and then set them to zero to obtain propagators for all the components separately; substitute these expressions into the relations Eqs.~\eqref{eq:longitudinal Zaanen-Mukhin stress gauge field} or \eqref{eq:transverse Zaanen-Mukhin stress gauge field}.

Since the coupling between the stress gauge fields and the dislocation sources is just $\ti b^{a\dagger}_\lambda J^a_\lambda = \ti J^{a\dagger}_\lambda b^a_\lambda$, and since the matrix in Eq.~\eqref{eq:longitudinal Lagrangian stress gauge field compact} in Fourier space is just algebraic, we can simply invert the matrices to immediately find
\begin{align}
 \mathcal{Z}^\tL[J] &= \int \mathcal{D}b\ \exp \Big( -\int \mathcal{L}^\tL \Big) = \exp\Big( -\int \tfrac{1}{2}\begin{pmatrix} J^{\tT\dagger}_{+1} & J^{\tL \dagger}_\tT \end{pmatrix}
 \mathcal{M}_\tL^{-1} 
  \begin{pmatrix} J^{\tT}_{+1} \\ J^{\tL}_\tT \end{pmatrix} \Big),\nonumber\\
 \mathcal{M}_\tL^{-1} &=  4\mu\frac{1}{4p^2(\frac{1}{c_\tL^2}\omega_n^2 + q^2)}
 \begin{pmatrix} \tfrac{1}{c_\tT^2} \omega_n^2 + 2(1+\nu) q^2 & -\ti \nu \tfrac{1}{c_\tT}\omega_n p \\
  \ti \nu\tfrac{1}{c_\tT} \omega_n p & p^2 
 \end{pmatrix}.\label{eq:longitudinal inverse matrix}
\end{align}
Being careful to use $\langle b^{\tT\dagger}_{+1}  b^\tL_\tT \rangle = \frac{\delta}{\delta \mathcal{J}^{\tL\dagger}_\tT} \frac{\delta}{\delta \mathcal{J}^\tT_{+1}} \mathcal{Z}$ etc., we find for Eq.~\eqref{eq:longitudinal Zaanen-Mukhin stress gauge field},
\begin{align}
 G_\tL &= \frac{1}{\kappa} - \frac{1}{(2\kappa)^2} 4\mu  \frac{1}{4(\tfrac{1}{c_\tL^2}\omega_n^2 + q^2)} \Big[ \tfrac{1}{c_\tT^2}\omega_n^2 + 2 (1+\nu) q^2  + 2\nu\omega_n^2 + \omega_n^2 \Big] \nonumber\\
 &= \frac{1}{\kappa} - \frac{1}{\kappa^2} \frac{\mu(1+\nu)}{2}\frac{\tfrac{1}{c_\tT^2}\omega_n^2 + q^2 }{\tfrac{1}{c_\tL^2}\omega_n^2 +  q^2}
 = \frac{1}{\kappa} - \frac{1}{\kappa} \frac{\omega_n^2 + c_\tT^2 q^2 }{\omega_n^2 + c_\tL^2 q^2}
  = \frac{c_\tL^2 - c_\tT^2}{\kappa} \frac{q^2}{\omega_n^2 + c_\tL^2 q^2}\nonumber\\
  &\to \frac{1}{\rho} \frac{-q^2}{\omega^2 -  c_\tL^2 q^2+ \ti \delta}. \label{eq:longitudinal propagator stress gauge fields}
\end{align}
This agrees with Eq.~\eqref{eq:longitudinal strain propagator}. We used $\mu \frac{1+\nu}{1-\nu} = \kappa$ from Eq.~\eqref{eq:compression modulus definition}; $c_\tL^2 = \frac{\kappa +\mu}{\rho}$ from Eq.~\eqref{eq:longitudinal phonon velocity}; $c_\tT^2 = \frac{\mu}{\rho}$ from Eq.~\eqref{eq:transverse phonon velocity}; and we can combine these to substitute $\frac{1-\nu}{2} c_\tL^2 = c_\tT^2$. We performed analytic continuation to real time $\omega_n \to \ti \omega - \delta$ in the last step.

The derivation for the transverse propagator is similar, with slight complications coming from the second-gradient terms. We can invert the matrix in Eq.~\eqref{eq:transverse Lagrangian stress gauge field compact} to find
\begin{align}
 \mathcal{Z}^\tT[J] &= \int \mathcal{D}b\ \exp \Big( -\int \mathcal{L}^\tT \Big)
 = \exp\Big( -\int \tfrac{1}{2}\begin{pmatrix} J^{\tL\dagger}_{+1} & J^{\tT \dagger}_\tT \end{pmatrix}
 \mathcal{M}_\tT^{-1} 
  \begin{pmatrix} J^{\tL}_{+1} \\ J^{\tT}_\tT \end{pmatrix} \Big),\nonumber\\
 \mathcal{M}_\tT^{-1} &=  4\mu \frac{\ell^2 q^2}{4p^2} \frac{1}{\tfrac{1}{c_\tT^2}\omega_n^2 + q^2 (1 + \ell^2q^2)} 
 \begin{pmatrix}  \big( \tfrac{1}{c_\tT^2 }\omega_n^2 (1 + \tfrac{1}{\ell^2 q^2}) + 4 q^2\big) & \ti \tfrac{1}{c_\tT} \omega_n p  (1 - \tfrac{1}{\ell^2 q^2}) \\
 -\ti \tfrac{1}{c_\tT} \omega_n p   (1 - \tfrac{1}{\ell^2 q^2}) & p^2(1 + \tfrac{1}{\ell^2 q^2})
 \end{pmatrix}.\label{eq:transverse inverse matrix}
\end{align}
We find for Eq.~\eqref{eq:transverse Zaanen-Mukhin stress gauge field},
\begin{align}\label{eq:transverse propagator stress gauge fields}
 G_\tT &= \frac{1}{\mu \ell^2 q^2} - \frac{1}{(2\mu\ell^2 q^2)^2} \frac{\mu \ell^2 q^2 }{\omega_n^2 + c_\tT^2 q^2 (1 + \ell^2q^2)}  \Big[ 2\omega_n^2 (1 + \tfrac{1}{\ell^2 q^2}) + 4 c_\tT^2 q^2 + 2 \omega_n^2 (1 - \tfrac{1}{\ell^2 q^2})\Big] \nonumber\\
 &= \frac{1}{\mu \ell^2 q^2} \Big[ 1 - \frac{\omega_n^2 + c_\tT^2 q^2 }{\omega_n^2 + c_\tT^2 q^2 (1 + \ell^2q^2)} \Big] 
 = \frac{1}{\mu \ell^2 q^2} \frac{ \ell^2 q^2 c_\tT^2 q^2}{\omega_n^2 + c_\tT^2 q^2 (1 + \ell^2q^2)} 
 = \frac{1}{\rho} \frac{  q^2}{\omega_n^2 + c_\tT^2 q^2 (1 + \ell^2q^2)} \nonumber\\
 &\to \frac{1}{\rho} \frac{ - q^2}{\omega^2 - c_\tT^2 q^2 (1 + \ell^2q^2) + \ti \delta}
\end{align}
This agrees with Eq.~\eqref{eq:transverse strain propagator}. We used $c_\tT^2 = \frac{\mu}{\rho}$ from Eq.~\eqref{eq:transverse phonon velocity} in the second-to-last step and performed analytic continuation to real time $\omega_n \to \ti \omega - \delta$ in the last step. As anticipated, this result is meaningful even in the limit of linear elasticity where $\ell \to 0$, even though the expression Eq.~\eqref{eq:transverse Zaanen-Mukhin stress gauge field} seems at first sight to be divergent in that limit.

We see here a quite convoluted way to reobtain the familiar propagators Eqs.~\eqref{eq:longitudinal strain propagator}, \eqref{eq:transverse strain propagator}. However, the relations Eqs.~\eqref{eq:longitudinal Zaanen-Mukhin stress gauge field}, \eqref{eq:transverse Zaanen-Mukhin stress gauge field} are valid in the liquid crystal phases as well. The calculations in this section set a template for those derivations.

\subsection{Collective modes of the isotropic solid}
Let us conclude this section by listing the inventory of static forces and dynamic modes in the isotropic solid. This is most readily done in the Coulomb gauge $\partial_l b^a_l = 0$ rather than the Lorenz gauge $\partial_\lambda b^a_\lambda =0$, because the temporal components of gauge field then correspond directly to static forces (cf. the scalar potential $V = A_t$ in electrodynamics). We can use the dislocation sources $J^a_\lambda$ directly to probe the modes carried by these gauge fields with no need to resort to the longitudinal and transverse propagators. Since we are interested in the perfect crystal, there are no dislocations in the bulk, and the sources $J^a_\lambda$ refer only external forces (strains) that interrogate the (stress) response. As the quantities presented here concern physical observable, we will present the final results in real time via analytic continuation $\omega_n \to \ti \omega - \delta$.

\begin{table}
\begin{center}
  \begin{tabular}{ccc}
  \toprule 
  component & character & name \\
  \hline
  $b^\tL_\tT$ & massless propagating & longitudinal phonon \\
  $b^\tT_\tT$ & massless propagating & transverse phonon \\
  $b^\tT_\ft$ & long-ranged static & dislocation  force \\
  $b^\tL_\ft$ & short-ranged static & rotational  force \\
  \bottomrule
 \end{tabular}
 \caption{Mode spectrum of the isotropic solid}\label{table:Mode content of the isotropic solid}
\end{center}
 \end{table}

We impose the Coulomb gauge fix $\partial_l b^a_l = -qb^a_\tL =0$, removing the $b^a_\tL$-components from the Lagrangians Eqs.~\eqref{eq:longitudinal Lagrangian stress gauge field}, \eqref{eq:transverse Lagrangian stress gauge field}. We can then integrate out the gauge fields, which amounts to inverting the matrices as we did in Eqs.~\eqref{eq:longitudinal inverse matrix}, \eqref{eq:transverse inverse matrix}. We will focus on the `diagonal' correlation functions, taking double derivatives of $\mathcal{Z}^{\tT,\tL}[J]$ with the corresponding currents. 

In the longitudinal sector we find
\begin{align}
 \langle b^{\tT\dagger}_\ft b^\tT_\ft \rangle &= \frac{4\kappa \mu}{\kappa + \mu} \frac{1}{q^2} \frac{\frac{1}{1-\nu^2}\omega^2 - c_\tL^2 q^2}{\omega^2 - c_\tL^2 q^2 + \ti \delta},\label{eq:isotropic solid instantaneous force}\\
 \langle b^{\tL\dagger}_\tT b^\tL_\tT \rangle &= \mu \frac{1}{\frac{-1}{c_\tL^2} \omega^2 - q^2 + \ti \delta}.
\end{align}
Clearly, the second equation is the longitudinal phonon propagating with velocity $c_\tL$. As expected, the transverse component of the dual gauge field corresponds to the massless, propagating mode. The first equation has a clear interpretation as well. Taking the high-energy limit $\omega \to \infty$, this propagator becomes proportional to $\frac{1}{q^2}$, and it is therefore an instantaneous or static force, and not a dynamic mode. In analogy with the photon gauge field in electrodynamics, we could call $b^\tL_\tT$ the longitudinal stress photon, and $b^\tT_\ft$ the Coulomb force between dislocation sources. Note that the Burgers index of the Coulomb force is transverse. In the static limit $\omega \to 0$, we find  $\langle b^{\tT\dagger}_\ft b^\tT_\ft \rangle = 4 \frac{\kappa\mu}{\kappa + \mu} \frac{1}{q^2}$. The Coulomb force disappears for either vanishing compression modulus $\kappa$ or vanishing shear modulus $\mu$. The first (compressionless) case cannot exist in nature; it would correspond to a hypothetical `compressionless' solid where the excess row of atoms (constituting a dislocation) can be stacked on top of another row at no energy cost. The second case is a medium without shear stress, i.e. a liquid or nematic liquid crystal, where dislocations cannot exist. 

For the transverse sector we find
\begin{align}
 \langle b^{\tL\dagger}_\ft b^\tL_\ft \rangle &= \mu\frac{1}{q^2} \frac{\omega^2 (1 + \ell^2 q^2) - 4 c_\tT^2 q^2 \ell^2 q^2}{\omega^2 - c_\tT^2 q^2(1+\ell^2 q^2) + \ti \delta},\\
 \langle b^{\tT\dagger}_\tT b^\tT_\tT \rangle &= \mu \frac{1(1 + \ell^2 q^2)}{\frac{1}{c_\tT^2} \omega^2 - q^2(1 + \ell^2 q^2) + \ti \delta}.
\end{align}
The second equation is readily identified as the transverse phonon propagating with velocity $c_\tT$. It persists even when $\ell = 0$, only considering linear elasticity. By taking the high-energy limit $\omega \to \infty$ the first equation becomes proportional to $\frac{1 + \ell^2 q^2}{q^2}$, which for $q \to 0$ goes as $\frac{1}{q^2}$; yet again this is not a propagating mode but a instantaneous force. More insightful is the static limit $\omega \to 0$, where 
\begin{equation}
 \langle b^{\tL\dagger}_\ft b^\tL_\ft \rangle (\omega \to 0) = 4\mu \frac{\ell^2}{1 + \ell^2 q^2} = 4\mu \frac{1}{q^2 + \frac{1}{\ell^2}}. \label{eq:solid rotational force}
\end{equation}
Here we see that this is also a Coulomb force, but short-ranged with a length scale set by $\ell$. For $\ell =0$,  this propagator vanishes completely, getting effectively removed by the Ehrenfest constraint. We call this therefore the rotational Coulomb force, since it only shows up when rotational (torque) stress is applied.

\begin{figure}
 \begin{center}
   \includegraphics[width=7.7cm]{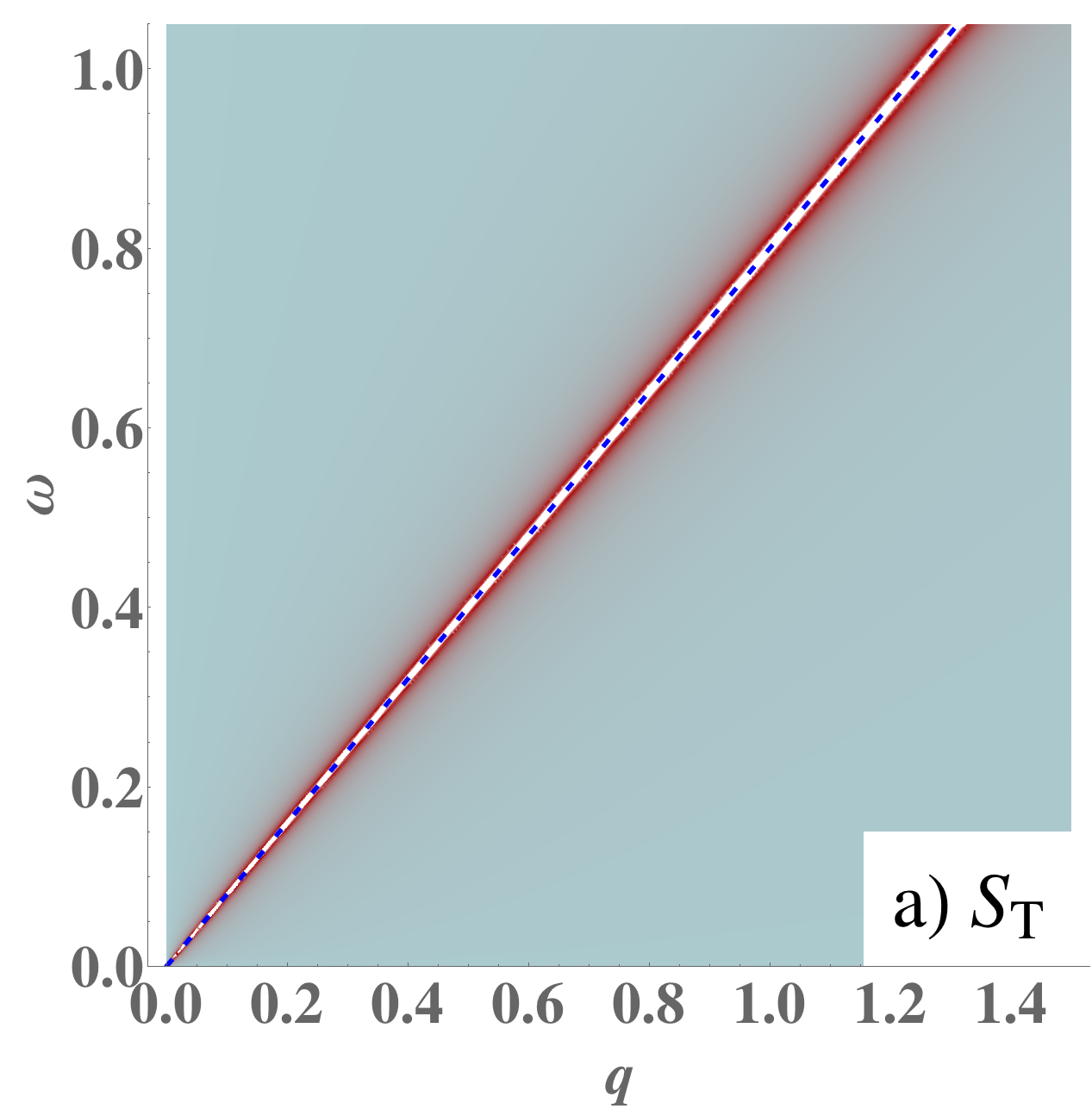}
 \hfill
\includegraphics[width=7.7cm]{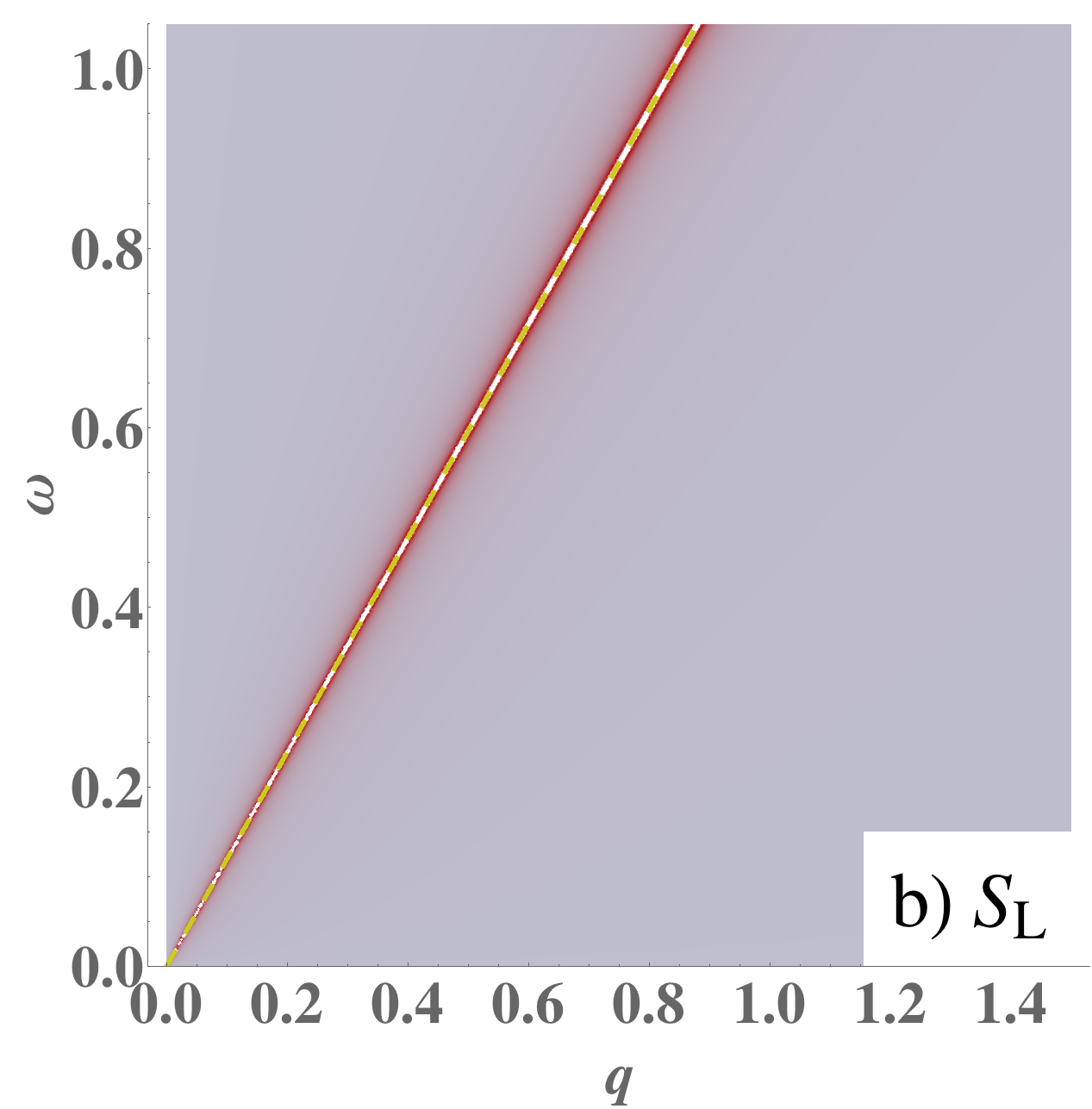}\\
{\centering
\includegraphics[height=0.7cm]{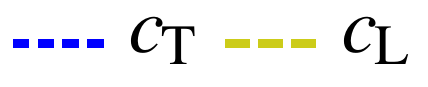}
}\\
 \caption{Spectral functions Eq.~\eqref{eq:spectral function definition} (left: transverse; right: longitudinal) of the isotropic solid in units of the inverse shear modulus $1/\mu \equiv 1$, with Poisson ratio $\nu = 0.1$. The width of the poles is artificial and denotes the relative pole strengths: these ideal poles are actually infinitely sharp.  Each sector has one propagating mode, the phonon, which is massless (zero energy as $q \to 0$) and has linear dispersion $\omega = cq$. The velocities are the transverse $c_\tT$ resp. longitudinal velocity $c_\tL$.}\label{fig:solid spectral functions}
 \end{center}
\end{figure}

This mode spectrum is summarized in Table~\ref{table:Mode content of the isotropic solid}. The dual variables capture both the dynamic and static interactions, even though they were derived from the strain action Eq.~\eqref{eq:elasticity relativistic Lagrangian repeat} which contained only the phonons. Here we note that conventional 2+0D ``elastostatics'' is obtained by going to the static limit $\omega \to 0$ and considering only densities and not currents, that is only $b^a_\ft$ and $J^a_\ft$ and not $b^a_m$ and $J^a_m$.

In the liquid crystal phases in Secs.~\ref{sec:Quantum nematic} and \ref{sec:Quantum smectic} we will analyze the spectral function $S(\omega , q)$:
\begin{equation}\label{eq:spectral function definition}
 S_{\tL,\tT} ( \omega , q ) =  \mathrm{Im}\;  G_{\tL,\tT} (  \ti \omega - \delta , q),
\end{equation}
using the analytic continuation of the frequency $-\ti \omega_n \to \omega + \ti \delta$ and $\delta \ll 1$. The poles of the spectral function are propagating modes. For comparison to the quantum liquid crystals responses later on, we have plotted the transverse and longitudinal spectral functions of the isotropic solid in Fig.~\ref{fig:solid spectral functions}. There is clearly one phonon in each sector, which is massless and has linear dispersion with the transverse and longitudinal velocity, respectively.

Now that we have identified the mode spectrum, we can return to another gauge fix: the Lorenz gauge where $\partial_\mu b^a_\mu = 0$ and let us examine the sum of the Lagrangians of Eqs.~\eqref{eq:longitudinal Lagrangian stress gauge field compact}, \eqref{eq:transverse Lagrangian stress gauge field compact}. Since $b^a_{-1} = b^a_\tT$, the components $b^\tL_{-1}, b^\tT_{-1}$ are to be identified as the longitudinal and transverse phonons. This implies that $b^\tL_{+1}, b^\tT_{+1}$ represent the (rotational, dislocation) Coulomb forces. In the quantum liquid crystal phases, the Lorenz gauge fix is more appropriate due to the nature of the minimal coupling to the dislocation condensate. With this identification, it becomes useful to compare the fate of the static forces resp. dynamic modes in the smectic and nematic phases.
 
 \section{Disorder field theory of dislocations: the dual stress superconductors}\label{sec:Dynamics of disorder fields}
 After setting the stage, we are now entering the core of this review. By describing the elastic stresses as gauge fields, up to this point we have been collecting all the pieces that are needed for the construction of the theories of the {\em dual stress superconductors} describing the physics of the maximally strongly-correlated quantum nematic and smectic crystals.  Now we will put them all together in a general disorder field theory framework, which will be further explored below for the specific cases of quantum nematic and smectic order. 

This disorder field theory is build upon the central notion of weak--strong (or Kramers--Wannier) duality that the 
physics of the {\em disordered} state can be regarded as an {\em ordered} state in terms of the disorder fields, coding for a system formed out of topological excitations of the original 
ordered state. Upon approaching the quantum melting transition from the ordered side, the number of closed loops formed from the (particle-like in two-dimensions) defect--antidefect pairs increases, while at the same time the loops 
grow in size. At the quantum critical point these loops in spacetime `blow out', becoming as large as the size of the system. The original order is destroyed since individual defects occur freely, while from the disordered viewpoint one obtains a `tangle' of free defect and antidefect worldlines. Such a tangle corresponds generally to a {\em relativistic superfluid}. However, 
when the topological excitations have long-range interactions that can be captured in terms of effective $U(1)$-gauge fields, as is the case for both vortices and dislocations due to Eqs.~\eqref{eq:XY Coulomb action} resp. \eqref{eq:stress gauge field Coulomb action}, one is dealing with a {\em charged}
superfluid, which is nothing else than the Abelian-Higgs problem describing the physics of the relativistic superconductor. 

This program has been carried out in full detail for the vortex--boson duality associated with the superfluid--Mott insulator system introduced in Sec.~\ref{sec:XY-duality}~\cite{Kleinert89a,Kleinert08,KiometzisKleinertSchakel95,NguyenSudbo99,HoveSudbo00,HoveMoSudbo00,CvetkovicZaanen06a}. These results employs only the Villain construction as a simplifying approximation. Such an explicit 
construction has not quite been accomplished for the more complex case of disorder fields associated with proliferating dislocations. However, the form of the dual disorder field 
theory describing the vortex condensate relies only on general phenomenological arguments. This same basic strategy then also applies to the construction of the theories of the dual stress superconductors. A crucial 
aspect in common with the vortex system is that one is dealing with {\em Abelian} global symmetries. In the vortex--boson problem this is the internal $U(1)$-symmetry, while in the crystal-to-liquid crystal melting problem we are coping with the Abelian `pure' translations since we regard the rotations to be broken, remaining frozen at the melting transition from the solid to the liquid crystal, recall Fig.~\ref{fig:phase diagram sketch}. An Abelian symmetry group has Abelian topological defects, so that the braiding of the defect worldlines is trivial. The tangle of proliferated defect worldlines then reduces to a featureless quantum fluid (a superfluid).

The generalization of the vortex duality to the liquid crystal context is nevertheless a bit of hairy affair. This was for the first time accomplished in Ref.~\cite{ZaanenNussinovMukhin04}; although the outcome was guessed right, the line of arguments was actually not quite correct. The general theory of nematic order parameters as reviewed in Sec.~\ref{sec:Order parameters for 2+1-dimensional nematics} was not available back then, leading to some confusion. Let us here present the case as it is now understood~\cite{Cvetkovic06,BeekmanWuCvetkovicZaanen13}.
 As a caution, the theory as it will unfold below might look overly general, but the reader should be aware
that all of it rests on the `low fugacity of defects' in the Villain approximation assumption. Kept implicit, the key assumption is that only the Goldstone bosons and defects are available as building material. But 
this requires that the defects are very dilute, which in turn implies that the length scale associated with the crystalline correlations in the  liquid are very large compared to the lattice constant/UV cut-off. 
In this way the ``maximally correlated liquid'' motive is hard-wired into the construction, rendering the theory actually to be tractable.  Let us first warm up by revisiting the $U(1)$ vortex--boson (or Abelian-Higgs) duality in 2+1D
that we already discussed at length in Sec.~\ref{sec:XY-duality}.

\begin{figure*}[t]
\hfill
 \subfigure[ vortex worldline]{
  \includegraphics[height=3cm]{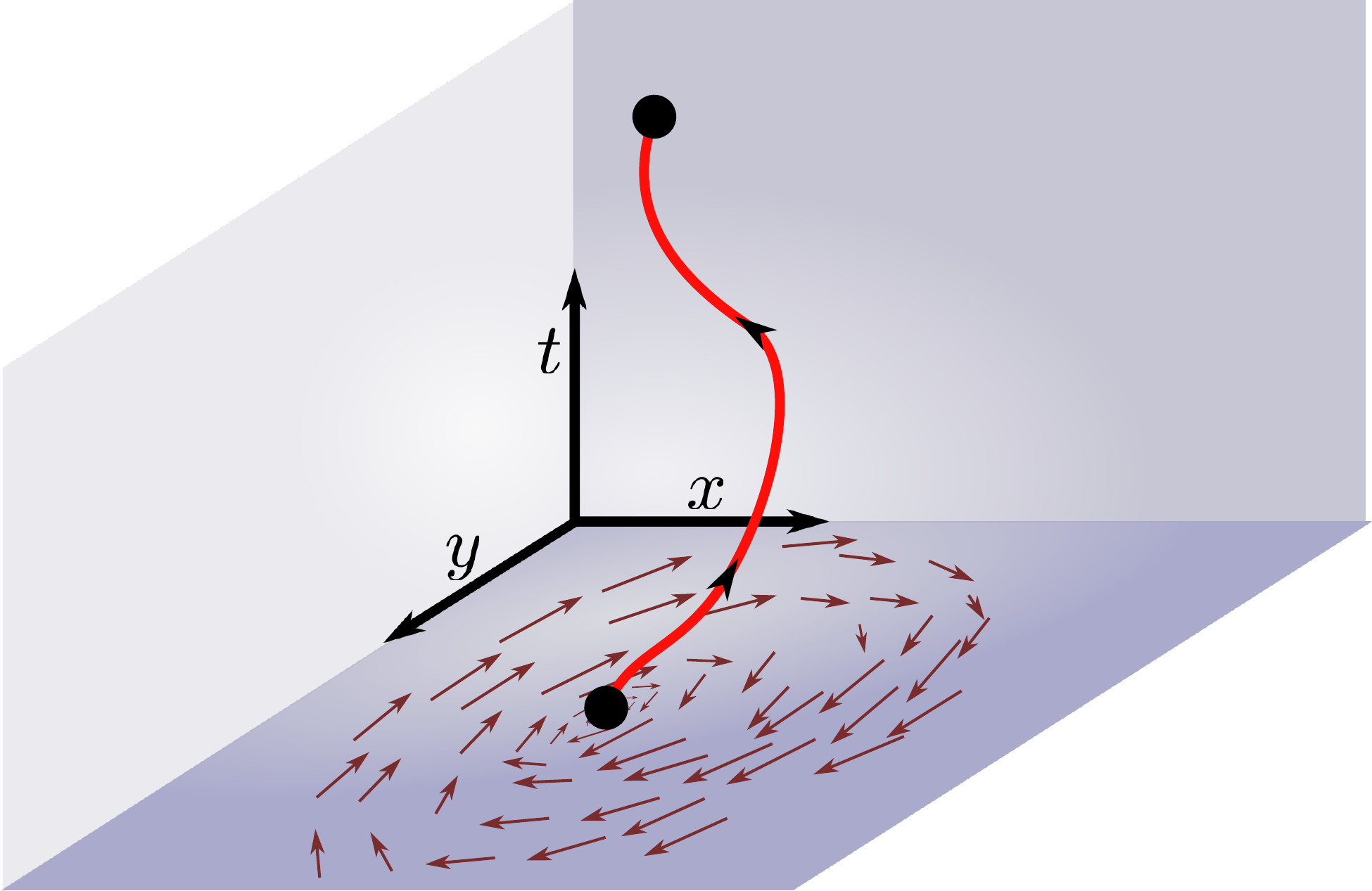}\label{fig:single vortex worldline}
 }
 \hfill
 \subfigure[ vortex Coulomb gas]{
  \includegraphics[height=3cm]{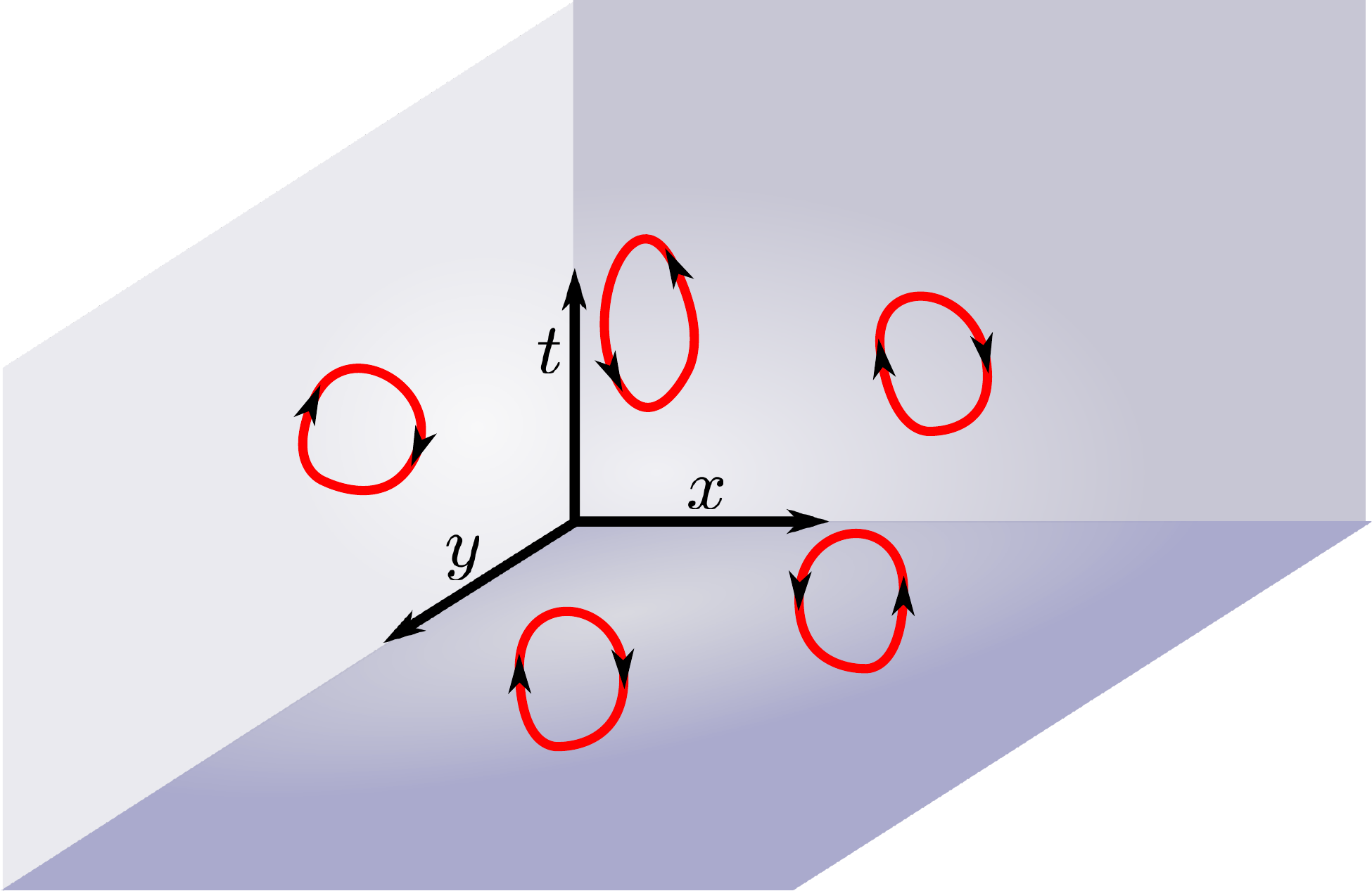}\label{fig:vortex Coulomb gas}
 }
 \hfill
 \subfigure[ vortex blowout]{
  \includegraphics[height=3cm]{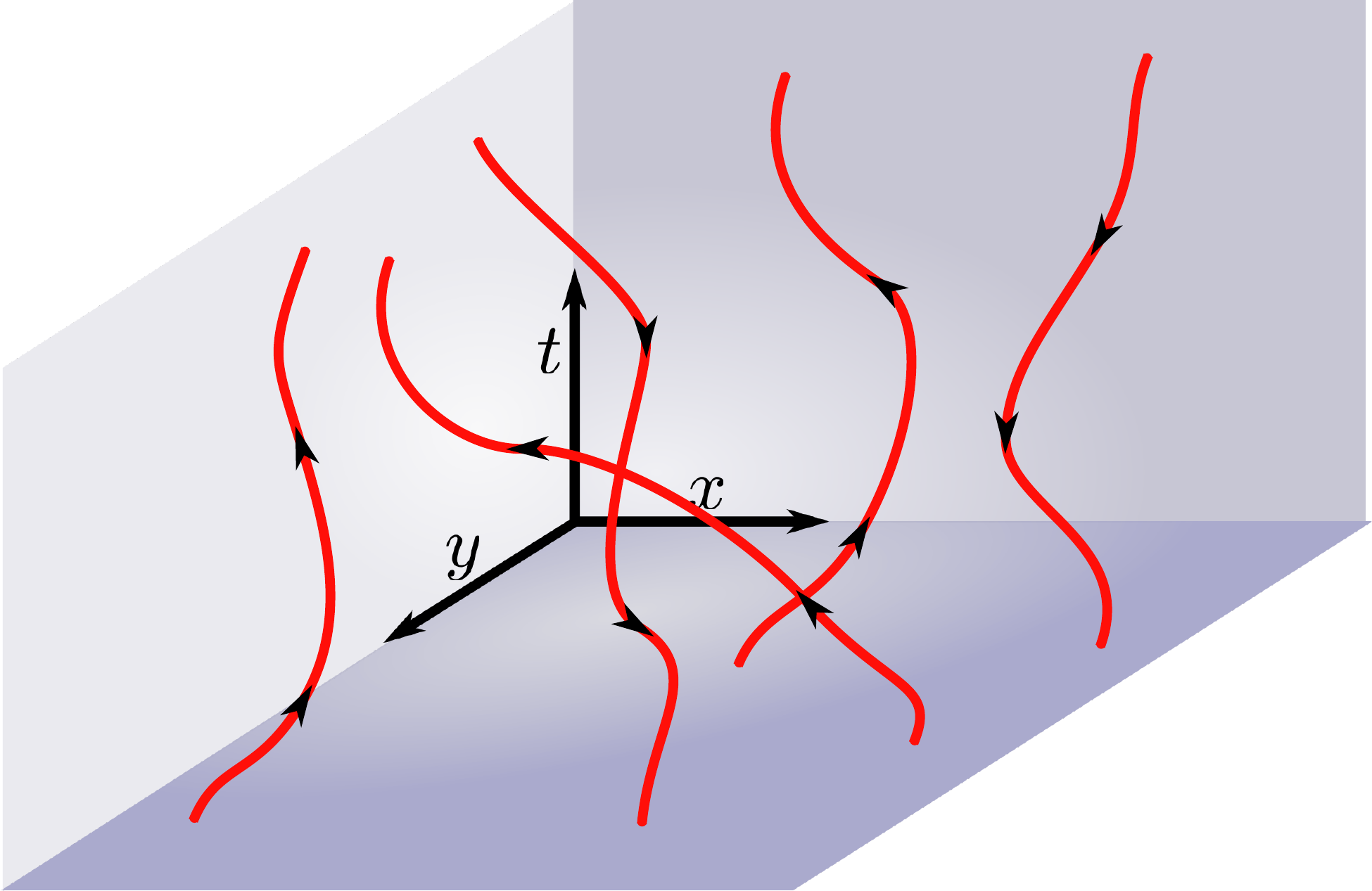}\label{fig:vortex blowout}
 }
 \hfill 
 \caption{Proliferation of vortices in 2+1 dimensions. \subref{fig:single vortex worldline} The worldline of a single $U(1)$-vortex in a superfluid. The configuration of the phase fields indicated by the in-plane arrows correspond to a singularity of the core. The evolution in time of this singularity is the vortex worldline. The orientation of the worldline corresponds to the vorticity of the vortex; an antivortex would have the opposite orientation. \subref{fig:vortex Coulomb gas} If the phase stiffness is large, vortices are confined. The combination of a vortex and an antivortex is topologically neutral, and the creation and subsequent annihilation of a vortex--antivortex pair can occur as a fluctuation, represented by a closed loop in spacetime. In terms of the dual gauge field, the superfluid is a Coulomb gas of vortex excitations. \subref{fig:vortex blowout} As the phase stiffness shrinks, it is easier to create vortex fluctuations. In dual language, the line tension of the vortex worldline becomes smaller. The size and occurrence of closed vortex worldlines increases. Upon the phase transition to the disordered phase, these loops have grown to the size of the whole system. The line tension is so low that vortices and antivortices can be created spontaneously; locally, there is no conservation of vortex (winding) number. The only difference between vortex worldlines in 2+1 dimensions and vortex lines in 3+0D dimensions is that the worldlines must still respect causality: they must always have a non-zero temporal component, whereas static vortex lines can lie in the $xy$-plane.}\label{fig:vortex proliferation}
\end{figure*}

\subsection{Vortex disorder field theory}

The action Eq.~\eqref{eq:XY Coulomb action} describes individual vortices $J^\mathrm{V}_\lambda(x) = 2\pi \delta_\lambda(L,x)$ that interact via dual gauge fields $b_\lambda(x)$, Fig.~\ref{fig:single vortex worldline}. The quantum fluctuations in the superfluid phase feature as a dilute gas of small vortex--antivortex loops in spacetime: see Fig.~\ref{fig:vortex Coulomb gas}. These have the same status as e.g. fermion loops in QED --- since these represent bound pairs of defects they just perturbatively dress the vacuum. Upon increasing the coupling constant $g$  these loops grow in size and number, and right at the quantum phase transition  matters turn non-perturbative: 
When the loops grow to become of the same size as the system, these unbind (the {\em loop blow-out}, see Fig.~\ref{fig:vortex blowout}) and on the disordered side of the phase transition 
one is dealing instead with a quantum liquid formed from freely occurring  vortices and anti-vortices.  Everything propagates with the same `velocity of light' (actually, the superfluid sound velocity) while 
the (anti)vortices have the same status as the (anti)particles of a relativistic field theory. Finally, the (anti)vortices obey a bosonic exchange statistics (because they descend from the symmetry breaking of an Abelian group) and this is in fact all one needs to know to construct the dual disorder field theory.  

Such a tangle of worldlines of relativistic bosons that interact via (compact) $U(1)$ gauge fields is very familiar: it is just a relativistic superconductor as described by Abelian-Higgs theory. It carries now a rigidity 
associated with the infinite winding of the worldlines around the imaginary time axis and this is captured by a collective field  $\Phi(x)$ describing the superconducting order parameter formed from (anti)-vortex matter. This 
is the literal disorder field.  By general principles, the effective field theory describing this vortex tangle has to be of the Abelian-Higgs form,
\begin{align}
 \mathcal{L}_\Phi  &=  \tfrac{1}{2 c_\mathrm{V}^2} \lvert (\partial_\tau - \ti b_\tau) \Phi\rvert^2 + \tfrac{1}{2} \lvert (\partial_m - \ti b_m) \Phi\rvert^2 
 + \tfrac{1}{2} \alpha \lvert \Phi \rvert^2 + \tfrac{1}{4} \beta\lvert \Phi \rvert^4 + \tfrac{1}{4} F_{\mu \nu} F_{\mu \nu} \nonumber \\
 &= \tfrac{1}{2} \lvert \tilde{D}_\mu \Phi\rvert^2 +\tfrac{1}{2} \alpha \lvert \Phi \rvert^2 + \tfrac{1}{4} \beta\lvert \Phi \rvert^4 + \tfrac{1}{4} F_{\mu \nu} F_{\mu \nu}.\label{eq:scalar disorder field theory}
\end{align}
 in  terms of the velocity-rescaled covariant derivative $\tilde{D}_\mu = \tilde{\partial}_\mu - \ti \tilde{b}_\mu$ (cf. Eq.~\eqref{eq:vortex condensate minimal coupling}), the disorder potential $V(\vert \Phi \vert) = \tfrac{1}{2} \alpha \lvert \Phi \rvert^2 + \tfrac{1}{4} \beta\lvert \Phi \rvert^4$ and the gauge field strength $F_{\mu \nu} = \partial_{\mu} b_{\nu} - \partial_{\nu} b_{\mu}$.  The mass of the 
 disorder field is now regulating the size of the vortex--antivortex loops: when $\alpha > 0$ these are pushed out of the vacuum (Fig.~\ref{fig:vortex Coulomb gas}) while a negative $\alpha$ signals that 
 they proliferate and condense (Fig.~\ref{fig:vortex blowout}). It has been demonstrated that the short-range repulsions between `vortex particles' turn into the quartic 
 self-interaction $\beta$ after coarse-graining~\cite{KiometzisKleinertSchakel95,Cvetkovic06}. We learned in Sec.~\ref{sec:XY-duality} that the long-range interactions between the vortices in 2+1D are precisely captured in terms of the $U(1)$-fields analogous to electromagnetism. Accordingly,  the disorder field $\Phi(x)$ is minimally coupled to these effective gauge fields via the covariant derivative, while the gauge fields themselves are governed by a
 Maxwell action, see Eq.~\eqref{eq:superfluid Higgs Lagrangian}. The further ramifications such as the identification of the dual superconductor as the Bose-Mott insulator were already discussed at length in Sec.~\ref{sec:XY-duality}.
 
 Once again, we emphasize that the dual condensate is fully relativistic, characterized  by a single `velocity of light' set by the sound velocity $c_\mathrm{ph}$ of the superfluid. However, it will be convenient to 
 introduce a separate velocity governing the vortex condensate $c_\mathrm{V}$ for the sole purpose of identifying the origin of the various excitations in the dual condensates. 

\subsection{Dislocations are different}

Let us now turn to the task of formulating the disorder field theory associated with disordering the crystal into quantum liquid crystals. Once again, this should have its basic features 
in common with vortex duality, rooted in the Abelian nature of the `pure' translations. This ensures that the exchange statistics of dislocations is bosonic.  Another crucial similarity is that dislocations are particle-like excitations in 2+1D, just as vortices are.
To restore the translational invariance characteristic of a liquid, dislocations have to proliferate, and since these are bosons they will condense. Last but not least, since dislocations have long-range interactions governed by electromagnetism-like stress gauge fields, they will form some sort of {\em charged} condensate: the {\em dual stress superconductor}.

Here the similarities end and compared to the vortex condensate we are facing a variety of complications, rendering the stress superconductor to be a much richer affair. First of all, instead of the simple scalar topological charge of the vortex (winding number $N$), dislocations carry the Burgers vector charge $B^a$ as discussed in Sec.~\ref{sec:Topological defects in solids}.  These Burgers vectors in turn are lattice translations also constrained by the point group symmetry that is broken both in the crystal and the liquid crystals, as discussed in section Sec.~\ref{sec:Order parameters for 2+1-dimensional nematics}. 
The rule is that the Burgers vector is a lattice vector, as was indicated in Fig.~\ref{fig:lattice vectors}. The dislocations in turn lead to shear stress in the material and therefore source the stress gauge fields that we discussed extensively in Sec.~\ref{sec:Dual elasticity}. We found that the dislocation currents that followed directly from the dualization Eq.~\eqref{eq:stress gauge field Coulomb action} couple to the dual gauge fields with the same flavor labels  according to the sourcing term $\sim \ti  b^a_{\mu} J^a_{\mu}$. This translates to stress gauge fields $b^a_{\mu}$ coming in two `Burgers flavors' that we have parametrized by either $a=x,y$- or 
$a=\tL,\tT$-coordinates.  Moreover, by carefully taking the constraints into account for these flavored gauge fields, we verified the existence of two propagating phonons associated with the two orthogonal directions $\tL, \tT$ along the momentum, fundamental for two space dimensions. We called $a$ casually the ``Burgers vector direction'', but dealing with the construction of the dislocation condensate we have now to pay extra attention. For instance, in the hexatic with its six-fold axis one can discern six possible directions of the Burgers vector Fig.~\ref{fig:triangular lattice vectors} with quantized Burgers vectors $B^a \in \{\integers \vec{e}_1 + \integers \vec{e}_2\}$. This leads to the quantization constraint $\int \de^2x J_{\ft}^{a} \in \{\integers \vec{e}_1 + \integers \vec{e}_2\}$ in the lattice basis. How can we reconcile this with the arbitrary dislocation densities in the $x,y$ directions of 2D space in the (non-compact dual gauge) continuum disorder field theory?

There is a constraint to impose on the dislocation condensate in the disorder theory that we mentioned in Sec.~\ref{subsec:Preview to defect-mediated melting}. In Sec.~\ref{subsec:Interdependence between dislocations and disclinations} we reviewed the interdependence between dislocations and disclinations, 
and we learned that a disclination is actually identical to an infinite number of dislocations with their Burgers vectors pointing in the same direction. The other side of the same coin is that a \emph{local} uncompensated dislocation (a ``net Burgers charge") corresponds to a disclination--antidisclination pair. However, to maintain 
rotational symmetry breaking, disclinations have to be ``kept out of the vacuum". This means that the quantum liquid crystals have to be characterized by a vanishing \emph{local} ``Burgers-vector magnetization" in addition to that the net Burgers vector must vanish. There is a universal way of accommodating this requirement: for all wallpaper groups of Table.~\ref{table:nematic phases}, the allowed Burgers vectors should occur in anti-parallel pairs, and one has to impose that in the condensate these 
precisely opposite Burgers vectors are equally populated on the \emph{microscopic} scale, so that they compensate each other. This ``\emph{local} Burgers neutrality'' is the motive in this topological language behind the director nature/tensor nature  of the order parameter in Sec.~\ref{subsec:Nematics gauge formulation}.

In order to avoid confusion with the magnetic field $\mathbf{B} = B^z$, here and below we shall denote the Burgers vector by $\mathbf{n}= (n^x,n^y)$. As we discussed, the directions of the Burgers vectors $\mathbf{n}$ are set directly by the point group symmetry of the crystal. However, in the liquid crystal one has to impose that $\mathbf{n}$ and $-\mathbf{n}$ are \emph{locally}
indistinguishable because of the topological condition of \emph{local} Burgers neutrality.  This is in turn a very basic example of the general mechanism by which gauge theories governing the collective physics
emerge in condensed matter physics (in this specific context, see also Refs.~\cite{ZaanenNussinov02,ZaanenNussinov03,ZaanenBeekman12}). This is just 
the condition encoded explicitly in the gauge-theory language of Secs.~\ref{subsec:Nematics gauge formulation} and \ref{subsec:Nematics phase diagram}  by the $\mathbb{Z}_N$-gauge fields. In fact, there we found that all possible Burgers vector directions should become gauge equivalent when dealing with the nematic order. In the continuum formulation at long-distance hydrodynamic level, its actually enough to just impose the Burgers neutrality condition, while the quantization of the Burgers charges can be added by hand in the end, as is usual in compact gauge theories. In addition, the Burgers neutrality condition following from the crystal topology is less stringent with an important consequence: it leaves room for the occurrence of the {\em quantum smectic} as will become clear soon. 

We are not done yet because we also learned in Sec.~\ref{subsec:stress constraints} that extra constraints  have to be imposed which are not standard in field theory. In the first place, we have to handle the Ehrenfest 
constraint Eq.~\eqref{eq:dual gauge field Ehrenfest constraint}, that directly relates to the fact that rotational motions and the associated torque stress should become alive in the quantum liquid crystal in the form of
a rotational Goldstone boson.  This is an elegant affair, that will be exposed in its full glory using the dynamical Ehrenfest constraint of Eq.~\eqref{eq:dynamical Ehrenfest constraint} in Sec.~\ref{subsec:torque stress nematic}. An unfamiliar but crucial ingredient 
in the construction of the disorder field theory is associated with the glide constraint discussed in Secs.~\ref{subsec:Kinematic constraints} and \ref{subsec:stress constraints}. This appears at 
first sight as a highly unusual  kinematic  constraint telling that the dislocation can only move in the direction of the Burgers vector. However, it is derived from particle number conservation or equivalently the condition that dislocations do not ``occupy volume'', and we will see that as a  natural consequence the dislocation condensate does not couple to compressional stress. 

Finally, there is yet one other property that the dislocation condensate has in common with the vortex condensate: it is `relativistic',  not only because it is formed out of opposite charges, but also because its characteristic propagation velocity should be related to the transverse sound velocity  $c_\tT$ of the crystal. As first realized by Friedel in the 1960s dealing with screw dislocations in three dimensions~\cite{Friedel64}, the inertial mass associated with glide motions of the dislocations is coincident with the mass of the atomic constituents. For the same reasons as in the vortex condensate, 
this just implies that there is only one velocity scale and this is the transverse phonon velocity since the dislocation condensate is associated with the shear modulus of the background elasticity. However, different from the vortex condensate, the glide constraint restricts the motions of the individual dislocations and this should imprint on the collective velocity characterizing the condensate. We expect that this velocity is parametrically different from the phonon velocity (by a factor of $\sqrt{2}$ or so). Resting on mere phenomenological arguments, this factor cannot be established and we leave it therefore as a free parameter in the remainder. We leave the determination of this factor in a microscopic calculation as a future challenge. Moreover, in order to decipher the role of the dislocate condensate in the dual stress superconductor it is convenient to keep the condensate velocity $c_\td$ explicit, while it should be put equal (or nearly equal) to the shear velocity to read off the physical outcomes. Using the same rescaling as for the vortex condensate Eq.~\eqref{eq:scalar disorder field theory} we indicate  the components that are rescaled to the velocity $c_\td$ with a tilde 
$\tilde{\phantom{a}}$: $\tilde{\partial}_\mu = ( \frac{1}{c_\td} \partial_\tau, \partial_m)$ and $\tilde{b}^a_\mu = ( \frac{1}{c_\td} b^a_\tau, b^a_m)$. For completeness, let us reproduce here the results for the isotropic solid where 
$\mathcal{L}_\mathrm{stress}$ contains both longitudinal and transversal sectors (Eqs.~\eqref{eq:longitudinal Lagrangian stress gauge field compact} and 
\eqref{eq:transverse Lagrangian stress gauge field compact}), rewritten in terms of the rescaled tilde fields $\tilde{b}^a_\lambda$,

\begin{align}
\mathcal{L}_\mathrm{solid} &=  \frac{1}{8\mu} 
\begin{pmatrix} \tilde{b}^{\tT\dagger}_{+1} \\ \tilde{b}^{\tL \dagger}_{-1} \\ \tilde{b}^{\tL\dagger}_{+1} \\ \tilde{b}^{\tT \dagger}_{-1}\end{pmatrix}^\tT  \!\!
\begin{pmatrix}
\tfrac{2}{1 + \nu} \tfrac{c_\td^2}{c_\tT^2} \tilde{p}^2 & \tfrac{ 2\ti \nu}{1+\nu} \tfrac{1}{c_\tT} \omega_n \tfrac{c_\td}{c_\tT} \tilde{p} &  \!\!\!\! 0 & 0\\
-\tfrac{ 2\ti \nu}{1+\nu}  \tfrac{1}{c_\tT} \omega_n\tfrac{c_\td}{c_\tT} \tilde{p} & \tfrac{2}{1+\nu} \tfrac{1}{c_\tT^2} \omega_n^2 + 4 q^2  & \!\! \!\!0 & 0\\
0 & 0 & \!\! \!\! \!\!\tfrac{c_\td^2}{c_\tT^2} \tilde{p}^2 (1 + \tfrac{1}{\ell^2 q^2})  & -\ti \tfrac{1}{c_\tT} \omega_n\tfrac{c_\td}{c_\tT} \tilde{p} (1 - \tfrac{1}{\ell^2 q^2})  \\
0 & 0 &  \!\! \!\! \!\!\ti \tfrac{1}{c_\tT} \omega_n \tfrac{c_\td}{c_\tT} \tilde{p} (1 - \tfrac{1}{\ell^2 q^2}) & \tfrac{1}{c_\tT^2}\omega_n^2(1 + \tfrac{1}{\ell^2 q^2})  + 4  q^2 
\end{pmatrix}
\begin{pmatrix}\tilde{b}^{\tT}_{+1} \\ \tilde{b}^{\tL}_{-1} \\ \tilde{b}^{\tL}_{+1} \\ \tilde{b}^{\tT}_{-1} \end{pmatrix}.\label{eq:crystal Lagrangian smectic coordinates full}
\end{align}
Here $\tilde{p} = \sqrt{ \tfrac{1}{c_\td^2} \omega_n^2 + q^2 }$ and  have imposed the the Lorenz gauge fix $\tilde{p}\tilde{b}^a_0 = 0$ (see below), which will turn out to be particularly convenient.
We can now express the stress photon propagators of Sec.~\ref{subsec:stress correlation functions} in terms of these tilde-fields $\tilde{b}^a_\mu$ in the Lorentz gauge fix:
\begin{align}
 G_\tL &=  \frac{1}{\kappa} - \frac{1}{(D\kappa)^2} \frac{c_\td^2}{c_\tT^2} \Big[ \tilde{p}^2 \langle \tilde{b}^{\tT\dagger}_{+1} \tilde{b}^\tT_{+1} \rangle -\ti \frac{\omega_n}{c_\td} \tilde{p} \langle \tilde{b}^{\tT\dagger}_{+1}  \tilde{b}^\tL_{-1} \rangle  + \ti \frac{\omega_n}{c_\td}  \tilde{p} \langle \tilde{b}^{\tL\dagger}_{-1} \tilde{b}^\tT_{+1}\rangle + \frac{\omega_n^2}{c_\td^2 }  \langle \tilde{b}^{\tL\dagger}_{-1} \tilde{b}^\tL_{-1} \rangle \Big]\label{eq:longitudinal propagator smectic gauge fix}, \\
G_\tT &=  \frac{1}{ \mu \ell^2 q^2} - \frac{1}{(2 \mu \ell^2 q^2)^2}\frac{c_\td^2}{c_\tT^2}\Big[  \tilde{p}^2 \langle \tilde{b}^{\tL\dagger}_{+1} \tilde{b}^\tL_{+1} \rangle  + \ti \frac{\omega_n}{c_\td} \tilde{p} \langle  \tilde{b}^{\tL\dagger}_{+1} \tilde{b}^\tT_{-1} \rangle - \ti \frac{\omega_n}{c_\td} \tilde{p} \langle  \tilde{b}^{\tT\dagger}_{-1} \tilde{b}^\tL_{+1} \rangle  + \frac{\omega_n^2}{c_\td^2}  \langle  \tilde{b}^{\tT\dagger}_{-1} \tilde{b}^\tT_{-1} \rangle\Big]\label{eq:transverse propagator smectic gauge fix},\\
G_{\tL\tT} &=  \frac{1}{4\mu \kappa} \frac{1}{q^2 \ell^2} \frac{c_\td^2}{c_\tT^2} \Big[ - \frac{\omega_n^2}{c_\td^2} \langle \tilde{b}_{-1}^{\tL\dagger} \tilde{b}^\tT_{-1} \rangle + \tilde{p}^2 \langle \tilde{b}_{+1}^{\tT\dagger} \tilde{b}^\tL_{+1} \rangle  + \ti \frac{\omega_n \tilde{p}}{c_\td} \langle \tilde{b}_{-1}^{\tL\dagger} \tilde{b}^\tL_{+1} \rangle + \ti \frac{\omega_n \tilde{p}}{c_\td}\langle \tilde{b}_{+1}^{\tT\dagger} \tilde{b}^\tT_{-1} \rangle \Big] \label{eq:chiral propagator smectic gauge fix}.
\end{align}

\subsection{Dislocation disorder field theory}\label{subsec:Dislocation disorder field theory}

In principle one would like to explicitly construct the disorder field theory departing from the Villain construction for isolated dislocations with the Burgers charge quantization condition of the associated lattice, to then evaluate numerically the full quantum partition sum describing the 
tangle of dislocation lines, mimicking results in vortex--boson duality~\cite{NguyenSudbo99,HoveSudbo00,HoveMoSudbo00}. This has not been accomplished yet but the structure of the 
effective disorder field theory in the continuum can be deduced via a phenomenological procedure just resting on general symmetry arguments we outlined above. 

The first conundrum  we have to overcome  was introduced in the previous section: the mismatch between the values 
of the Burgers vectors  as set by the lattice point group symmetry versus the outcome of the continuum stress--strain duality that only the number of orthogonal directions in
space should matter. Departing from a hexagonal crystal as associated with isotropic elasticity, one identifies a total of six distinguishable elementary vectors and including the `Burgers-neutrality' rule this would amount to three 
distinguishable Burgers-neutral combinations, Fig.~\ref{fig:triangular lattice vectors}. However, this does not appear to match the outcome of the dualization, in which there are two unquantized disorder fields  $J^{x,y}_{\mu}$ associated with the  two orthogonal $(x,y)$ directions of two dimensional space, Eq.~\eqref{eq:dislocation density multivalued displacement}.  However, this problem is easily addressed. As soon as a dislocation condensate with Burgers vectors in, say, the $x$-direction forms, all points which differ from each other in the $x$-direction have become equivalent, and therefore $x$ must be along on of the lattice vectors. Consequently, the remaining perfect translational order can exist only in the orthogonal direction of the (now deformed) lattice, here the $y$-direction (later we identify this with the quantum smectic). Any further dislocation condensation can only destroy this remaining order, leading to the quantum nematic liquid crystal.
 
 In this way, it was implicitly assumed in the stress--strain duality construction that the Burgers vector chosen from the `rotational cross' of  Fig.~\ref{fig:triangular lattice vectors} was actually aligned with the disordering direction. The conclusion is that there are only  {\em two independent disorder fields} in 2+1D formed by the condensed dislocations, associated with the two space directions in which translational symmetry can be restored. As expressed by e.g. the $\mathbb{Z}_6$-gauge invariance of the quantum hexatic, the distinction between the six directions in 
 the cross of Fig.~\ref{fig:triangular lattice vectors} has become immaterial anyhow in the quantum nematic. One can therefore supplement this in the continuum with a minimal $C_4$-cross in the $(x,y)$-directions, indicated in red in the figure, more as if one were to depart from a square lattice (that would also have quantized Burgers charges). Nonetheless, this continuum picture remains correct in terms of the net `disordering capacity' of the full dislocation condensate where all Burgers directions are populated equally. The quantization of Burgers charges can in principle be added by hand at the end and it remains to verify the predictions of the continuum theory with results from explicit lattice realizations incorporating the discrete Burgers charges for the disorder fields.

As compared to the vortex condensate characterized by a single $U(1)$-condensate field, the above implies that in $D$ space dimensions we have to cope with $D$ independent disorder fields
taking care of the restoration of translational symmetry in the $D$ orthogonal space directions. We therefore need in two dimensions 
{\em two complex scalar Higgs fields} $\Phi^a = | \Phi^a | \te^{\ti\phi^a}$ with $a = x,y$ to represent 
the dislocation condensate.  Given that these two fields are to taken to be independent, can this freedom have physical ramifications? Remarkably, this expresses the room in this topological formulation for the 
occurrence of the extra vestigial {\em quantum smectic} phases, occurring in between the solid and the nematic, see Fig~\ref{fig:phase diagram sketch}. 

To keep the disclinations out of the vacuum, the topological requirement is that the dislocations proliferate with their Burgers vectors
in opposite directions. This requirement can be satisfied not only by dislocations in all lattice directions, but also by dislocations in a single direction~\cite{ZaanenNussinovMukhin04,CvetkovicZaanen06b}. This will turn the system into a quantum liquid exclusively in the $x$-direction, while the translational order persists in the precisely orthogonal $y$-direction. This implies that only the $\Phi^x$ disorder field condenses while the $\Phi^y$ field stays massive.  The conclusion is that the description of
this (quantum) smectic order is just a natural part of the weak--strong duality, demonstrating that there are indeed $D$  linearly independent disorder fields are at work. This independence of dislocations was somewhat understood in the 1980s in the classical statistical physics literature following the KTHNY-tradition~\cite{OstlundHalperin81}, but is becomes very clear in the dual framework presented here. 

In the two-dimensional quantum realm, this symmetry structure was elucidated by Mathy and Bais~\cite{MathyBais07,BaisMathy06}. They put defect condensation by a disorder parameter, analogous to symmetry breaking by an order parameter, on firm mathematical footing  via the language of {\em quantum doubles} or {\em Hopf algebras}~\cite{MathyBais07,BaisMathy06}. Whereas in symmetry breaking one considers residual symmetry subgroups, defect condensation leads to sub-Hopf algebras. And just as the order parameter is a particular vector in order parameter space on which the group acts, so is the disorder parameter a vector in the space on which the Hopf algebra acts. Since we can take superpositions of vectors, in principle any condensate could be considered, in particular a symmetric combination of basis vectors, see Fig.~\ref{fig:lattice vectors}.

\begin{figure}
\begin{center}
  \includegraphics[width=8.5cm]{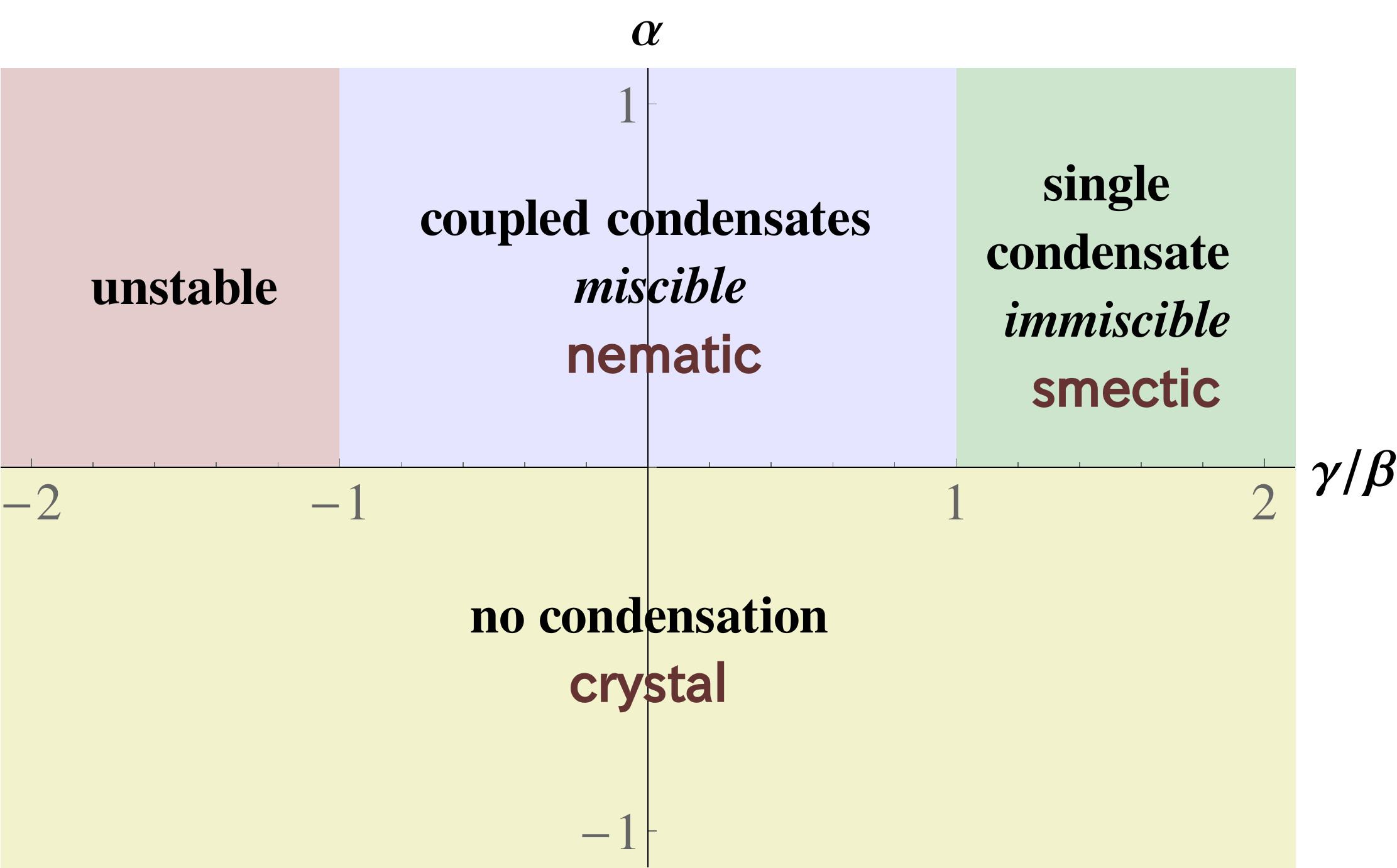}
 \caption{Phase diagram of the two-condensate model Eq.~\eqref{eq:two-condensate potential} for the simplified case 
 $\alpha_1 = \alpha_2 \equiv \alpha$ and $\beta_1 = \beta_2 \equiv \beta >0$, as function of $\alpha$ (arbitrary units) and $\gamma/\beta$. The corresponding phases in quantum elasticity are denoted in sans-serif font. If $\alpha >0$ no condensation takes place (the crystal), while for $\alpha<0$ some condensate (liquid crystal) will form. Its nature depends on the ratio $\gamma/\beta$. If $\gamma > \beta$ the repulsion between the two fields is so strong that only one will condense while the other stays massive. This corresponds to a dislocation condensate along one axis only: the quantum smectic. If $-\beta < \gamma < \beta$, both fields condense and automatically acquire the same expectation value. This corresponds to the quantum smectic. The case $\gamma < -\beta$ is unphysical.}\label{fig:binary Ginzburg-Landau phase diagram}
\end{center}
 \end{figure}

We have now sufficient information to write down the general form of the amplitude part of the Lagrangian describing the dual stress superconductors just using standard symmetry arguments, up to fourth order in the amplitude. It is just the potential of two condensates which couple through local density--density interactions, as in e.g. two-component Bose--Einstein condensates~\cite{KasamatsuEtAl05},
\begin{align}
 \mathcal{L}_{|\Phi |}= V(| \Phi^x |,| \Phi^y |) =  
 \tfrac{\alpha_x}{2} | \Phi^x |^2 + \tfrac{\alpha_y}{2} | \Phi^y |^2 + \tfrac{\beta_x}{4}| \Phi^x |^4 
 + \tfrac{\beta_y}{4} | \Phi^y |^4 + \tfrac{\gamma}{2}  | \Phi^x |^2 | \Phi^y |^2. \label{eq:two-condensate potential}
\end{align}
Here $\gamma$ describes the nature of interaction between the two condensates ($\gamma <0$ : attraction; $\gamma >0$ : repulsion). If we assume the $C_4$-symmetry of the lattice, there is {\em a priori} no difference between $x$- and $y$-directions, leading to $\alpha_x = \alpha_y \equiv \alpha$ and $\beta_x = \beta_y \equiv \beta$. The phase diagram of the $C_4$-symmetric model is sketched in Fig.~\ref{fig:binary Ginzburg-Landau phase diagram}. If $\alpha <0$ and $\beta>0$, a condensate forms as is familiar from basic Ginzburg--Landau theory. The nature of the condensation depends on the value of the inter-condensate coupling $\gamma$, which can be easily found by energy minimization of Eq.~\eqref{eq:two-condensate potential}. If $-\beta < \gamma < \beta$, attraction between the condensates dominates and we are in the {\em miscible} regime. Both condensate fields acquire the same expectation value  $| \Phi^x |  = | \Phi^y | =  | \Phi |$, this is the isotropic quantum nematic. When $\gamma > \beta $, the repulsion between condensates dominates and we are in the {\em immiscible} regime. Only one out of $\Phi^x$, $\Phi^y$ will condense (``phase separation''). This is the quantum smectic. In this simplified, isotropic case, which particular condensate form is a spontaneous choice. Therefore, even the simple theory Eq.~\eqref{eq:two-condensate potential} shows that one can go from a solid to a smectic to a nematic  depending on the ratio $\gamma/\beta$, which at this stage of development are phenomenological parameters. The case $\gamma < - \beta$ does not lead to physical solutions.

How to account further for the large family of nematic states characterized by the different point groups, as discussed in Sec.~\ref{sec:Order parameters for 2+1-dimensional nematics} and Table.~\ref{table:nematic phases}?  We just noticed that in principle the $\Phi^{x,y}$ condensates can acquire different strengths in the case of less symmetric point groups. The nematic will also reflect the corresponding rotational symmetry through 
the extra anisotropy in the rotational Goldstone modes. This is in accordance with the fact that for any wallpaper group different from the hexatic the elasticity itself becomes anisotropic. 
This requires one or two extra moduli besides the shear and compression modulus~\cite{Kleinert89b}. This has the effect that the phonons and thereby the stress photons become anisotropic as well.  Although not yet enumerated in detail, it is anticipated that the anisotropy in the dislocation condensates will properly reproduce the anisotropies in the mode spectra of these nematics. In particular, leads to different values for $\alpha_x$ and $\alpha_y$, and for $\beta_x$ and $\beta_y$. In the remainder we will ignore these cases because they do not give rise to anything essentially new: they just interpolate between the smectic and the isotropic nematic. 

Even after these lengthy arguments, at the present stage it might appear as rather obscure how the nematic states as constructed via the disorder condensates could `know' that they break rotational symmetry to a finite subgroup. We have actually incorporated the rotational symmetry breaking in a rather 
undynamical fashion by just imposing by hand that the Burgers vector have to lie along the lattice directions but otherwise live in the continuum. As highlighted in Sec.~\ref{subsec:torque stress nematic}, the isotropic Goldstone modes associated with the spontaneous rotational breaking are nevertheless correctly recovered. These are actually encoded in a seemingly implicit way 
already in the elasticity theory describing the crystal. As we will see, these are literally {\em confined} in the solid while they deconfine in the quantum liquid crystal phases. The disclinations have the same fate, becoming finite energy topological excitations in the nematics. Surely, their Frank scalars are determined by the elements of the point group. Again, since these involve finite quantized discontinuities of the nematic order parameter, the quantization of the Frank scalars can be restored by hand in the dual stress superconductors, leading to the theories discussed in Sec.~\ref{sec:Order parameters for 2+1-dimensional nematics} when regulated on a lattice (in the limit of $\lvert \Phi \rvert \to \infty$). 

Finally, we have to incorporate the condensate phase fluctuations and the dual stress gauge fields that mediate long-range interactions between the dislocations. Here we can rest on a general principle: 
the coupling between gauge fields and an isolated charged particle $ \ti A_{\mu} J_{\mu}$ turns into the minimal gauge coupling of the condensate field formed from these particles, 
enumerated by the covariant derivative $\partial_{\mu} - \ti A_\mu$ acting on the condensate field. Dealing with the stress photons we depart from the $a = x, y$ flavored  terms sourcing 
the stress gauge fields  $\ti b^a_{\mu} J^a_{\mu}$ suggesting that the disorder field theory should contain the covariant derivative terms  $\sim | (\tilde{\partial}_\mu - \ti \tilde{b}^a_\mu)\Phi^a|^2 $ for $a =x$ and/or $a=y$.
This intuition is correct, except that there is one more ingredient that has to be dealt with: the {\em glide constraint}. Glide motion is associated with the dynamics of isolated dislocations and it is therefore 
obvious that it has to enter via the kinetic (gradient) terms in the disorder field theory.  

This is accomplished as follows \cite{ZaanenNussinovMukhin04,CvetkovicNussinovZaanen06,Cvetkovic06}. The glide constraint Eq.~\eqref{eq:2D glide constraint} can be imposed by a Lagrange multiplier field $\lambda(x)$. A term  is added to the original Lagrangian,
\begin{equation}
 \mathcal{L}_\mathrm{glide} =  \ti  \lambda \epsilon_{\ft \mu a} J^a_\mu.
\end{equation}
Then at any stage of the calculation, integrating out $\lambda$ will impose the glide constraint. The  source term involving the dislocation current becomes ~\cite{ZaanenNussinovMukhin04},
\begin{equation}
  \mathcal{L}_\mathrm{source} = \ti \left( b^a_{\mu}  + \lambda \epsilon_{\ft \mu a} \right) J^a_\mu.
 \label{glideconstraintsingle}
\end{equation}
The recipe is now to replace $b^a_\mu$ with $b^a_{\mu}  + \lambda \epsilon_{\ft \mu a}$ when coupling to dislocations, which works equally well for single dislocations as for dislocation condensates.

The glide principle amounts to the statement that dislocations can move ballistically in the direction of their Burgers vectors.  We are again facing the conundrum that the Burgers vector
directions are quantized rather than just having two independent orthogonal directions in the continuum space where the disorder fields do their work. However, as we showed in Sec.~\ref{subsec:Kinematic constraints}, in the long-distance hydrodynamic theory the glide principle is just equivalent to the fact that dislocations do not occupy volume. Therefore, individual dislocations only communicate with shear stress and not with compressional stress. This in turn implies that the disorder fields formed from the dislocations {\em should not carry charge associated with the compressional stress gauge fields}. The single Lagrange multiplier field $\lambda(x)$ will protect the scalar ``spin-0"-component of the stress gauge fields $b^a_\lambda$ from obtaining a mass through the Anderson--Higgs mechanism. Impressively, the constraint in Eq.~\eqref{glideconstraintsingle} captures this regardless of the details of the dislocation condensate, and applies equally well to smectics, isotropic and anisotropic nematics. Concluding, the single dislocation source  term including the glide constraint  Eq.~\eqref{glideconstraintsingle} is just promoted to a covariant derivative ``dressed with the glide constraint" acting on the $\Phi^{x,y}$  disorder fields. This implies that we are dealing with gradient terms in the effective action, 
\begin{equation}
\mathcal{L}_\mathrm{kin} =  \frac{1}{2} \sum_{a=x,y} \lvert \left( \tilde{\partial}_\mu - \ti \tilde{b}^a_{\mu} - \ti \lambda \epsilon_{\ft \mu a} \right) \Phi^a \rvert^2
\label{higgsdualsstress}
\end{equation}
Again, note the tilde fields, referring to the rescaling in terms of the condensate velocity $c_\td$. Again, we do not know whether $c_\td$ differs from $c_\tT$ by factors of order 1. 

We have now completed the formulation of the disorder field theory describing the dual stress superconductors. In full it is given by the Lagrangian,
\begin{equation}
\mathcal{L}_\mathrm{stress\ SC} =  \mathcal{L}_\mathrm{kin} +  \mathcal{L}_\mathrm{| \Phi |}  + \mathcal{L}_\mathrm{stress}
\label{dualstressSC}
\end{equation}
where the minimal coupling term Eq.~\eqref{higgsdualsstress} takes care of the interactions between the stress gauge fields and the dual disorder parameters,
the condensation of the disorder fields is governed by the potential $ \mathcal{L}_\mathrm{| \Phi |} $,  Eq.~\eqref{eq:two-condensate potential}, and  $\mathcal{L}_\mathrm{stress}$, describing the stress gauge fields of the crystal, given by Eq.~\eqref{eq:crystal Lagrangian smectic coordinates full}.

The foundational work is hereby completed and in the next three sections we will just evaluate the theory on its physical consequences, finding out that it is a remarkably powerful affair. Before we turn to these computations, let us end this section with introducing some  convenient technicalities.  

In the remainder we will focus on the long-wavelength physics well inside the stable  quantum liquid crystals phases. We can regard the amplitude fluctuations to be frozen out such 
that  $\langle \lvert \Phi^a \rvert \rangle$ is constant, and only the phase fields $\phi^a$ remain as dynamical degrees of freedom in this London limit. We introduce the Higgs mass $\Omega^a$
associated with the dual stress superconductor via

\begin{equation}\label{eq:dislocation Higgs mass}
( \Omega^a)^2 = \mu c_\tT^2\lvert \Phi^a\rvert^2, \qquad a = x,y,
\end{equation}
having units of energy (since $\hbar \equiv 1$). In the isotropic nematic $\Omega^x = \Omega^y = \Omega$ while in the smectic $\Omega^x = \Omega \neq 0$ while $\Omega^y = 0$.  The phase action 
follows immediately from Eq. (\ref{higgsdualsstress}) as,

\begin{equation}
\mathcal{L}_\mathrm{London} =  \tfrac{1}{2} \sum_{a=x,y}   (\Omega^a)^2  \left( \tilde{\partial}_\mu \phi^a -  \tilde{b}^a_{\mu} -  \lambda \epsilon_{\ft \mu a} \right)^2 + \mathcal{L}_\mathrm{stress}
\label{phasedualsstress}
\end{equation}

Apart from the $x,y$-flavors we are of course dealing here with the usual Abelian-Higgs/relativistic superconductor problem.  For future reference let us introduce here the standard gauge fixes. 
For every Burgers flavor there is one gauge degree of freedom that becomes physical via the Anderson--Higgs mechanism. When dealing with the nematic, the unitary gauge is convenient, where both condensate phase fields vanish $\phi^x = \phi^y \equiv 0$, 
leading to
\begin{equation}
 \mathcal{L}_\mathrm{unitary} =    \sum_{a=x,y}   \frac{\Omega^2}{2 c_\tT^2 \mu} \lvert  \tilde{b}^a_\mu + \lambda \epsilon_{\ft \mu a} \rvert^2 + \mathcal{L}_\mathrm{stress}.
 \label{eq:dislocation unitary gauge fix}
\end{equation}
such that all stress gauge field components $b^a_\lambda$ correspond to physical degrees of freedom after imposing the Ehrenfest constraint. Another convenient  gauge fix is the Lorenz gauge,  
\begin{equation}
 \tfrac{1}{c_\td^2} \partial_\tau b^a_\tau + \partial_m b^a_m = \tilde{\partial}_\mu \tilde{b}^a_\mu = 0.
 \label{eq:dislocation Lorenz gauge fix}
\end{equation}
In this gauge we have to pay some extra care, keeping track of the condensate velocity $c_d$. Rescale according to 
$\tilde{b}^a_\mu \equiv (\tilde{b}_\ft , \tilde{b}_m) \equiv (\frac{1}{c_\td} b^a_\tau , b_m)$, $\tilde{p}_\mu = (\tfrac{1}{c_\td} \omega_n,q_m)$ and furthermore (see \ref{sec:Fourier space coordinate systems}),
\begin{align}
\tilde{b}^a_0 &= -\frac{\ti \omega_n}{c_\td \tilde{p}} \tilde{b}^a_\ft + \frac{q}{\tilde{p}} \tilde{b}^a_\tL,\label{eq:dislocation helical 0}\\
 \tilde{b}^a_{+1} &= -\frac{q}{\tilde{p}} \tilde{b}^a_\ft +\frac{\ti \omega_n}{c_\td \tilde{p}}  \tilde{b}^a_\tL,\label{eq:dislocation helical +1}\\
 \tilde{b}^a_{-1} &= \tilde{b}^a_\tT =b^a_\tT,\label{eq:dislocation helical -1}
\end{align}
such that indeed $\tilde{p}_\mu \tilde{b}^a_\mu = \ti \tilde{p} \tilde{b}^a_0 = \tfrac{1}{c_\td^2} \partial_\tau b^a_\tau + \partial_m b^a_m = 0$. 
The relation with the `ordinary' $b$-fields is $b^a_\ft = \frac{c_\td}{c_\tT} \tilde{b}^a_\ft$ and $b^a_{+1} = \frac{c_\td}{c_\tT} \frac{\tilde{p}}{p} \tilde{b}^a_{+1}$. 
In the Lorenz gauge fix Eq.~\eqref{eq:dislocation Lorenz gauge fix}, the condensate phase degrees of freedom $\phi^a$ decouple while the longitudinal component of the stress gauge fields is projected out via a factor $\delta_{\mu\nu} - \frac{\tilde{p}_\mu \tilde{p}_\nu}{\tilde{p}^2}$. The Higgs term in this gauge takes the form

\begin{equation}
 \mathcal{L}_\mathrm{Higgs, Lorenz\ gf} =\sum_{a=x,y} \frac{(\Omega^a)^2}{2 c_\tT^2 \mu}
  \Big[ \lvert \tilde{\partial}_\mu\phi^a \rvert^2 +  
  (\tilde{b}^{a\dagger}_\mu + \lambda^\dagger \epsilon_{\tau \mu a})  (\delta_{\mu\nu} - \frac{\tilde{p}_\mu \tilde{p}_\nu}{\tilde{p}^2} ) (\tilde{b}^a_\nu + \lambda \epsilon_{\tau \nu a})\Big].
 \label{eq:dislocation Higgs term} 
\end{equation}

This is the form we shall use most of the time in the calculations that follow in the next two sections. Perhaps confusingly, in this Lorenz gauge it appears that the now physical phase modes of the condensate $\phi^a$ appear to be completely decoupled from the stress gauge fields $b^a_{\mu}$ while in the computations of the propagators only the latter are sourced externally. 
However, we will find that the condensate degrees of freedom leave their mark on the dynamical response.
 
 \section{Quantum nematic}\label{sec:Quantum nematic}
 We are now facing the in principle straightforward task of exploring the physical consequences of the theory of the dual stress superconductor as formulated 
in the previous section. Symmetry is always a simplifying circumstance and obviously the quantum nematic is more symmetric and therefore simpler than the 
quantum smectic. We therefore first focus on the nematic states, that at the same time reveal the most striking consequences of the disorder field formulation. 
Specifically we will consider the isotropic nematic characterized by the condensed disorder fields having equal amplitudes $| \Phi^x | = |\Phi^y| \neq 0$. 
The focus is on the physics deep in the quantum nematic state where the full theory Eq.(\ref{dualstressSC}) reduces to its London/Josephson form 
Eq. (\ref{phasedualsstress}) since the amplitude fluctuations of the dislocation condensate can be regarded as frozen out. 

\subsection{Stress propagators in the quantum nematic}\label{subsec:Correlation functions quantum nematic}

For later convenience, we rescale the Higgs mass term $\Omega^2$ from Eq.~\eqref{eq:dislocation Higgs mass} by a factor of $2$; this could be conceived as `normalizing' the Higgs mass as the average over the two identical components $\Omega^x$ and $\Omega^y$. Taking the London limit and the Lorenz gauge fix, the Higgs term Eq.~\eqref{eq:dislocation Higgs term} that we shall use in this section is
\begin{align}\label{eq:nematic Higgs term}
 \mathcal{L}_\mathrm{Higgs} = \sum_{a=x,y}
  \frac{\Omega^2}{4 c_\tT^2 \mu} (\tilde{b}^{a\dagger}_\mu + \lambda^\dagger \epsilon_{\tau \mu a})  (\delta_{\mu\nu} - \tfrac{\tilde{p}_\mu \tilde{p}_\nu}{\tilde{p}^2}) (\tilde{b}^a_\nu + \lambda \epsilon_{\tau \nu a}).
\end{align}
We now have to deal with the Lagrange multiplier field $\lambda$ enforcing the glide constraint in the path integral. This is accomplished as follows. In the Lorenz gauge only `transverse'
stress gauge field components (orthogonal to spacetime momentum $\tilde{p}_\mu$) enter and this is conveniently accommodated to in the helical spacetime basis $\tilde{b}_{\mu}^a \rightarrow \tilde{b}_{\pm 1}^a$, Eqs.~\eqref{eq:dislocation helical 0}--\eqref{eq:dislocation helical -1}. 
First expand the Higgs term into these components. This further simplifies by going over to the $\tL,\tT$ basis for the Burgers index $a$, see \ref{sec:Fourier space coordinate systems}. One notices that the $\lambda$ field can be straightforwardly 
integrated out, which leads to (the sum over $a$ is now implicit)
\begin{align}
 \mathcal{L}_\mathrm{Higgs} &= \tfrac{1}{4} \frac{\Omega^2}  {c_\tT^2 \mu} \Big[   \lvert \tilde{b}^a_{+1} \rvert^2 + \lvert \tilde{b}^a_{-1} \rvert^2   + \lambda^\dagger \lambda (2 - \frac{q^2}{\tilde{p}^2})  + \lambda^\dagger ( b^\tL_{-1} +  \ti \frac{\omega}{c_\td \tilde{p}} \tilde{b}^\tT_{+1}) + \textrm{h.c.}  \Big] \nonumber\\
 &= \tfrac{1}{4}  \frac{\Omega^2}{c_\tT^2 \mu} \Big[   \lvert \tilde{b}^a_{+1} \rvert^2 + \lvert \tilde{b}^a_{-1} \rvert^2  - 
 \frac{\lvert \tilde{p}\tilde{b}^\tL_{-1} +  \ti \frac{\omega_n}{c_\td} \tilde{b}^\tT_{+1} \rvert^2 }{ \frac{\omega^2}{c_\td^2} + \tilde{p}^2} \Big] 
 = \tfrac{1}{4}\frac{\Omega^2}{c_\tT^2 \mu} \Big[ \frac{\lvert \ti \frac{\omega_n}{c_\td} \tilde{b}^\tL_{-1} + \tilde{p} \tilde{b}^\tT_{+1}\rvert^2}
{ \frac{1}{c_\td^2} \omega_n^2 + \tilde{p}^2 } +  \lvert \tilde{b}^\tL_{+1} \rvert^2 + \lvert \tilde{b}^\tT_{-1} \rvert^2 \Big].
\label{eq:quantum nematic Higgs Lagrangian}
\end{align}

With these simple operations we have accomplished quite a feat: the meaning of this result is that the stress gauge field component that is propagating purely {\em compressional} forces Eq.~\eqref{eq:compression stress gauge field} is decoupled from the dislocation condensate! As we already announced, this is rooted in the principle (encoded by the glide constraint) that the dislocations do not occupy volume. The other way around, the 
difference between a solid and liquid is that both carry pressure but the solid is in addition characterized  by a reactive response to shear stresses. Our quantum nematic does not break translations
but it is a liquid that should be capable of propagating sound. This is indeed the case: we will soon find out that at energies less than the Higgs mass $\Omega$ of the dislocation condensate, it carries a purely compressional massless sound mode, that crosses over into the stress-photon version of the longitudinal phonon at energies larger than $\Omega$. We are at zero temperature and this sound mode is clearly not the hydrodynamical sound of a classical liquid. Below it will become clear that it is actually the zero-sound (phase) mode of the superfluid. At the same time, Eq.~\eqref{eq:quantum nematic Higgs Lagrangian} also implies that the stress photons that propagate shear forces acquire a Higgs mass: the beauty of dual stress superconductor is that it
takes care that shear forces can only propagate over a finite distance in the liquid.

How does one infer these important physical consequences from the equations? By imposing the Lorenz gauge fix in the form of the helical $\pm 1$ spacetime indices we are dealing with 
the physical stress fields $\tilde{b}^\tL_{+1}, \tilde{b}^\tT_{+1}, \tilde{b}^\tL_{-1}, \tilde{b}^\tT_{-1}$. Their action in the solid  is given by  Eq.~\eqref{eq:crystal Lagrangian smectic coordinates full}, and one immediately reads off that the $\tilde{b}^\tL_{+1}$ and $\tilde{b}^\tT_{-1}$ stress photons are associated with the transverse phonons. According to the last line of 
Eq.~\eqref{eq:quantum nematic Higgs Lagrangian} these just acquire the Higgs mass: these photons in the transverse sector mediate shear stress, and one reads off immediately that ``shear stress acquires a Higgs mass'' in the quantum nematic.

The longitudinal phonon is associated with the other two components $\tilde{b}^\tT_{+1}$ and $\tilde{b}^\tL_{-1}$. The glide constraint has taken care that these occur only in one particular combination in the Higgs term. The orthogonal combination formed from these two fields is absent from the Higgs term, which is the one that ``does not carry charge under dislocation condensation''. We will soon find out that for energies much less than $\Omega$ this mode propagates with the real sound velocity set by the compression modulus only, $c_\tK^2 = \kappa/\rho$.

The bottom line is this: the longitudinal phonon of the solid automatically causes shear deformations when it occurs at finite wave vectors, such that its velocity also involves the shear modulus, $c_\tL^2 = (\kappa + \mu)/\rho$. 
 In the liquid this shear component has to be removed at long distances and this is precisely what the first term in our Higgs action Eq.~\eqref{eq:quantum nematic Higgs Lagrangian} accomplishes.

In order to get a clearer view on the physics of the quantum nematic let us now compute the propagators of the stress photons $\tilde{b}^a_\mu$.  For the quantum nematic this is a rather straightforward computation. 
We combine the Higgs term Eq.~\eqref{eq:quantum nematic Higgs Lagrangian} with the isotropic crystal result Eq.~\eqref{eq:crystal Lagrangian smectic coordinates full}. Then we calculate the stress photon propagators Eqs.~\eqref{eq:longitudinal propagator smectic gauge fix}--\eqref{eq:chiral propagator smectic gauge fix}. The chiral propagator vanishes since there are cross terms between longitudinal and transverse sectors in neither Eq.~\eqref{eq:quantum nematic Higgs Lagrangian} nor  Eq.~\eqref{eq:crystal Lagrangian smectic coordinates full}.  For the longitudinal and transverse propagators we obtain the important results, 
 \begin{align}
  G_\tL &= \frac{1}{\mu} \frac{c_\tT^2 q^2 ( \omega_n^2 + c_\tR^2 q^2 + \Omega^2)}{(\omega_n^2 + c_\tL^2 q^2)(\omega_n^2 + c_\tR^2 q^2) + \Omega^2 (\omega_n^2 + c_\tK^2 q^2)},
 \label{eq:nematic longitudinal propagator}\\
  G_\tT &= \frac{1}{\mu}\frac{c_\tT^2 q^2 (\omega_n^2 + c_\td^2 q^2) + \Omega^2 ( \omega_n^2 + 2 c_\tT^2 q^2 + c_\tR^2 q^2 + \Omega^2 )}{(\omega_n^2 + c_\td^2 q^2)(\omega_n^2 + c_\tT^2 q^2(1+\ell^2 q^2)) + \Omega^2( \omega_n^2 (1 + \ell^2 q^2) + c_\tR^2 q^2 (1+\ell^2 q^2 ) + \ell^2 q^2 (2 c_\tT^2 q^2 + \Omega^2 ))}.
 \label{eq:nematic transverse propagator} 
 \end{align}
We use $c_\tL = \sqrt{\frac{2}{1 - \nu}} c_\tT = \sqrt{(\kappa+\mu)/\rho}$, and introduce the compressional velocity $c_\tK = \sqrt{\frac{1+\nu}{1-\nu}} c_\tT = \sqrt{\kappa/\rho}$ 
and rotational velocity $c_\tR = \tfrac{1}{\sqrt{2}} c_\td$. The different velocities in the problem are summarized in Table~\ref{table:velocities}. 

The transverse propagator even takes into account the corrections coming from second-order elasticity. For the time being we can ignore these by  taking  $\ell \to 0$, and Eq.~\eqref{eq:transverse propagator smectic gauge fix} simplifies to,

\begin{align}
G_\tT &= \frac{1}{\mu}\frac{c_\tT^2 q^2 (\omega_n^2 + c_\td^2 q^2) + \Omega^2 ( \omega_n^2 + 2 c_\tT^2 q^2 + c_\tR^2 q^2 + \Omega^2 )}{(\omega_n^2 + c_\td^2 q^2)(\omega_n^2 + c_\tT^2 q^2) + \Omega^2( \omega_n^2  + c_\tR^2 q^2)}.\label{eq:transverse propagator nematic linear elasticity}
 \end{align}

\subsection{Collective modes of the quantum nematic}\label{subsec:Mode content of the quantum nematic}
\begin{table}
\begin{center}
  \begin{tabular}{cccc}
  \toprule 
 symbol & definition & name & occurrence\\
  \hline
 $c_\tT$ & $\sqrt{\mu / \rho}$ & transverse velocity & solid phonon\\
 $c_\tL$ & $\sqrt{(\kappa + \mu) / \rho}$ & longitudinal velocity & solid phonon \\
  $c_\tK$ & $\sqrt{\kappa / \rho}$ & compressional velocity & nematic compression mode \\
 $c_\td$ & -- & dislocation velocity  & dislocation condensate sound mode\\
 $c_\tR$ & $c_\td/\sqrt{2}$ & rotation velocity & rotational Goldstone mode \\
  \bottomrule
 \end{tabular}
 \caption{Velocities in the quantum liquid crystal. The dislocation velocity $c_\td$ should  be equal or almost equal to the transverse velocity $c_\tT$. Other velocities used in this article are the phase velocity $c_\mathrm{ph}$ and vortex velocity $c_\mathrm{V}$ in the $XY$-model (Sec.~\ref{sec:XY-duality}) and the velocity of light $c_\mathrm{l}$ (Sec.~\ref{sec:Dual elasticity of charged media}). }\label{table:velocities}
\end{center}
\end{table}

As in the case of the solid, the longitudinal and transverse propagators Eqs.~\eqref{eq:nematic longitudinal propagator},\eqref{eq:transverse propagator nematic linear elasticity} enumerate precisely the collective properties of the quantum nematic state. We immediately infer that the response is isotropic in space, depending only on the magnitude and not on the direction of  $\mathbf{q}$ as it should since we wired this in by using Eq.~\eqref{eq:quantum nematic Higgs Lagrangian}. To discern the nature of the excitation spectrum  the spectral functions Eq.~\eqref{eq:spectral function definition} are of central interest and  in Fig.~\ref{fig:nematic spectral functions} we 
show some representative results. As expected from Eqs.~\eqref{eq:nematic longitudinal propagator},\eqref{eq:transverse propagator nematic linear elasticity} we can discern a pair of propagating modes in both the longitudinal and the 
transverse sector. In both sectors these divide into a massless and a massive mode. This is in stark contrast with the solid, where we found a single propagating massless mode in each sector, just corresponding 
to the transverse and longitudinal phonons, Fig.~\ref{fig:solid spectral functions}. What is the physical nature of this richer mode spectrum of the nematic? To identify the various modes it is now convenient to heed the various velocities in the problem, where especially the artificial dislocation condensate velocity $c_\td$ becomes quite revealing --- in fact, the main reason for the effort to keep track of it during the derivations. 
The various velocities are tabulated in Table~\ref{table:velocities}.

\subsubsection{Longitudinal sector}
\begin{figure}
 \begin{center}
\includegraphics[width=7.7cm]{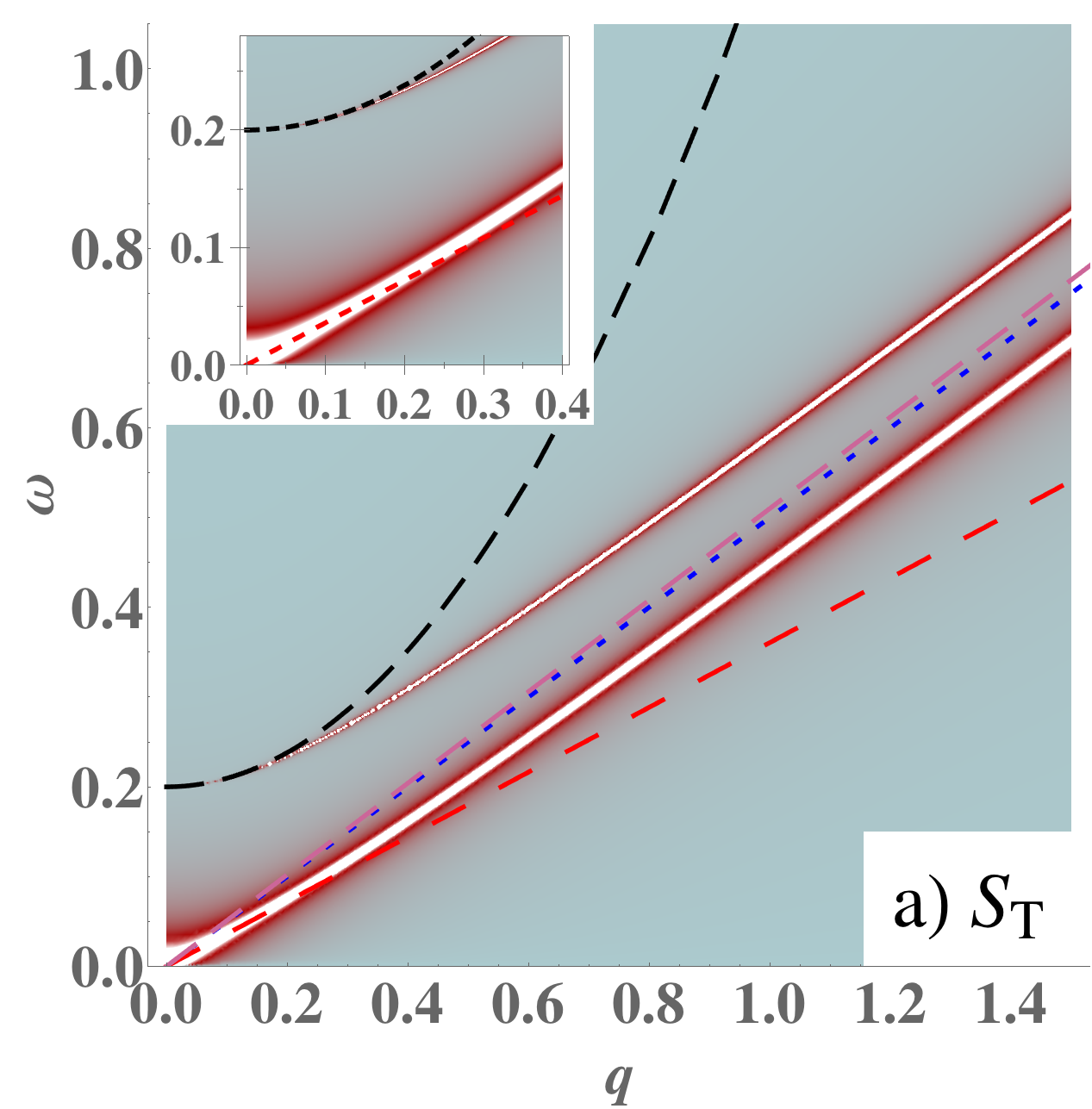}
 \hfill
\includegraphics[width=7.7cm]{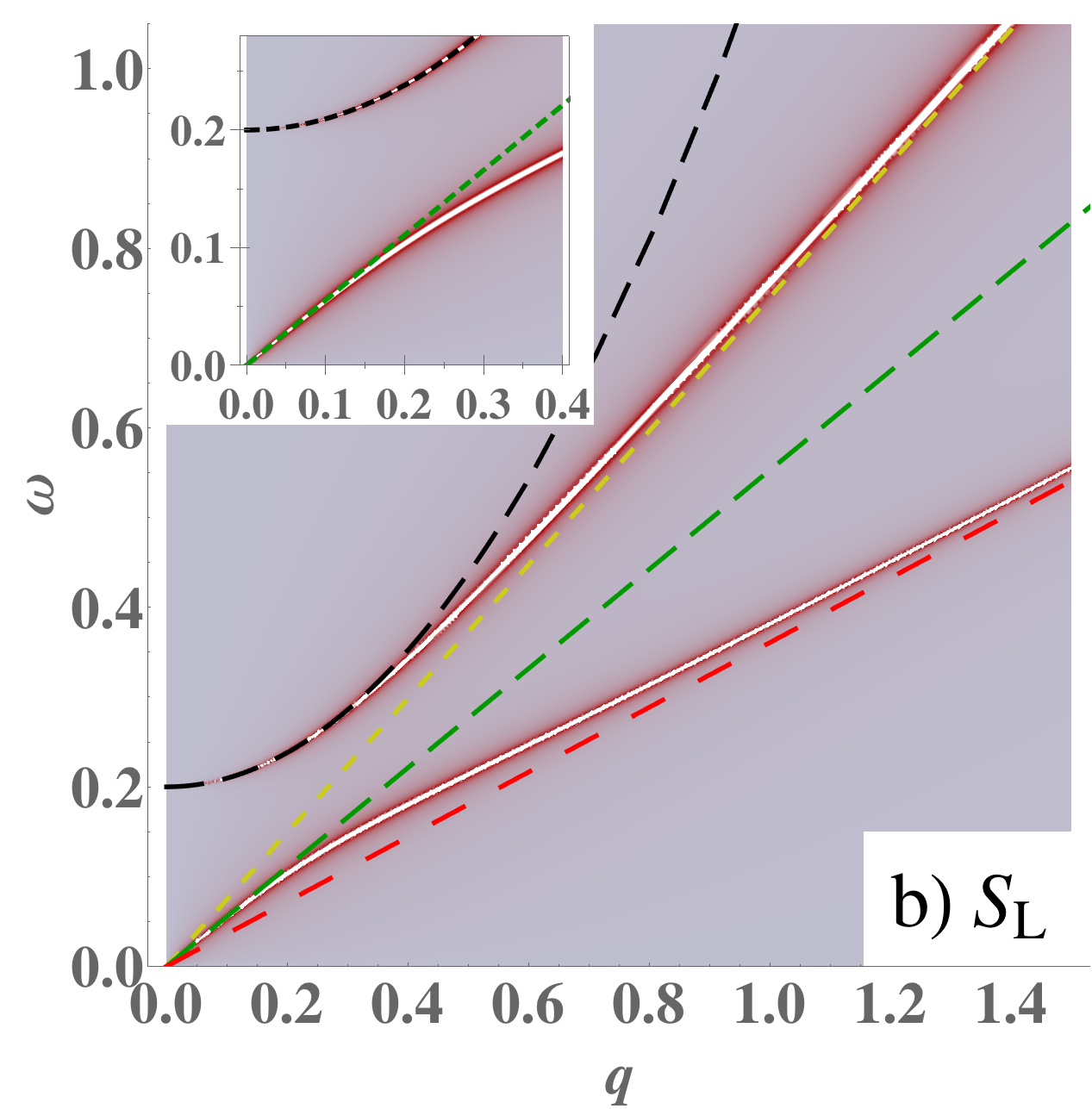}\\
 \includegraphics[width=7.7cm]{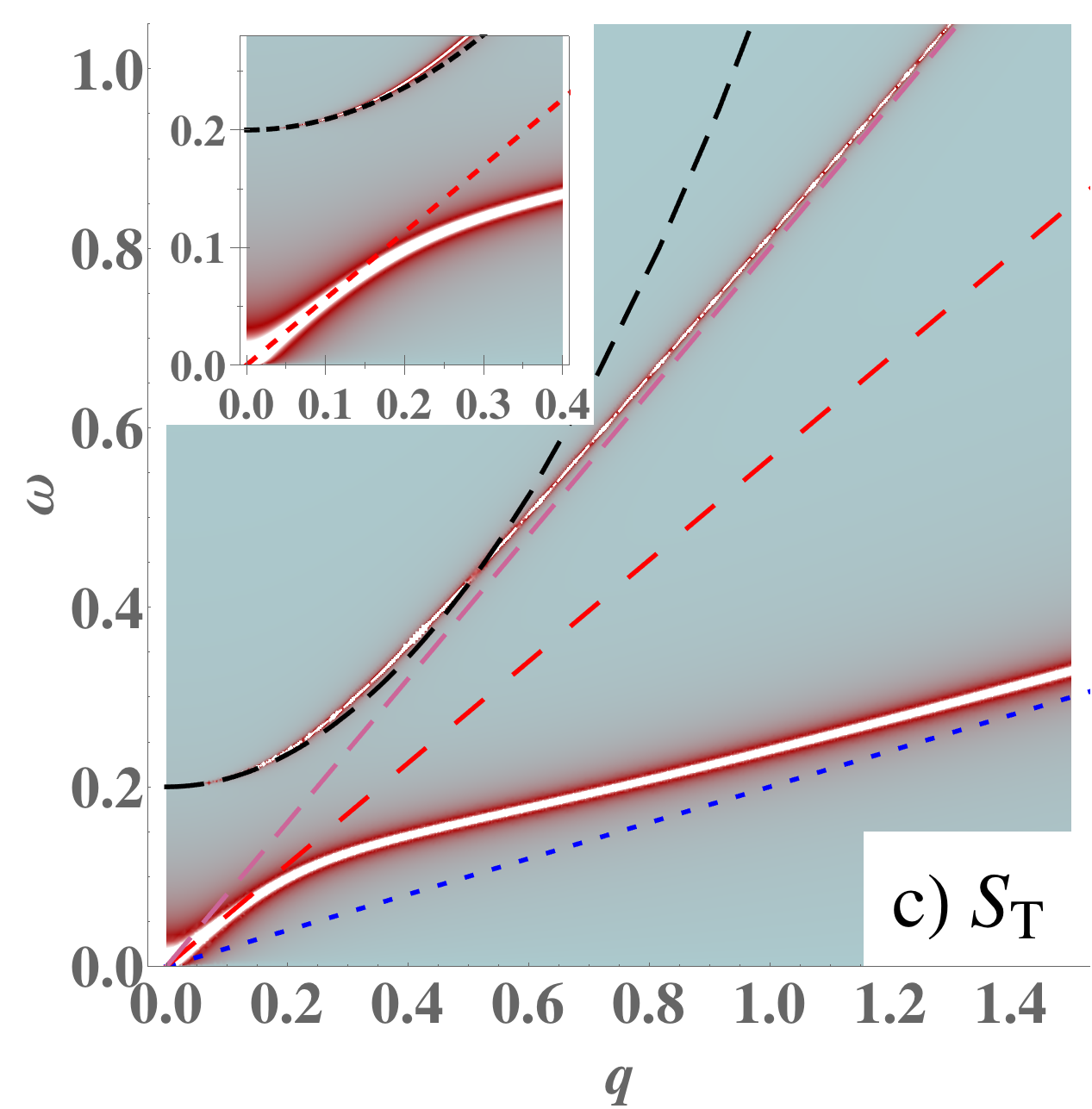}
\hfill
\includegraphics[width=7.7cm]{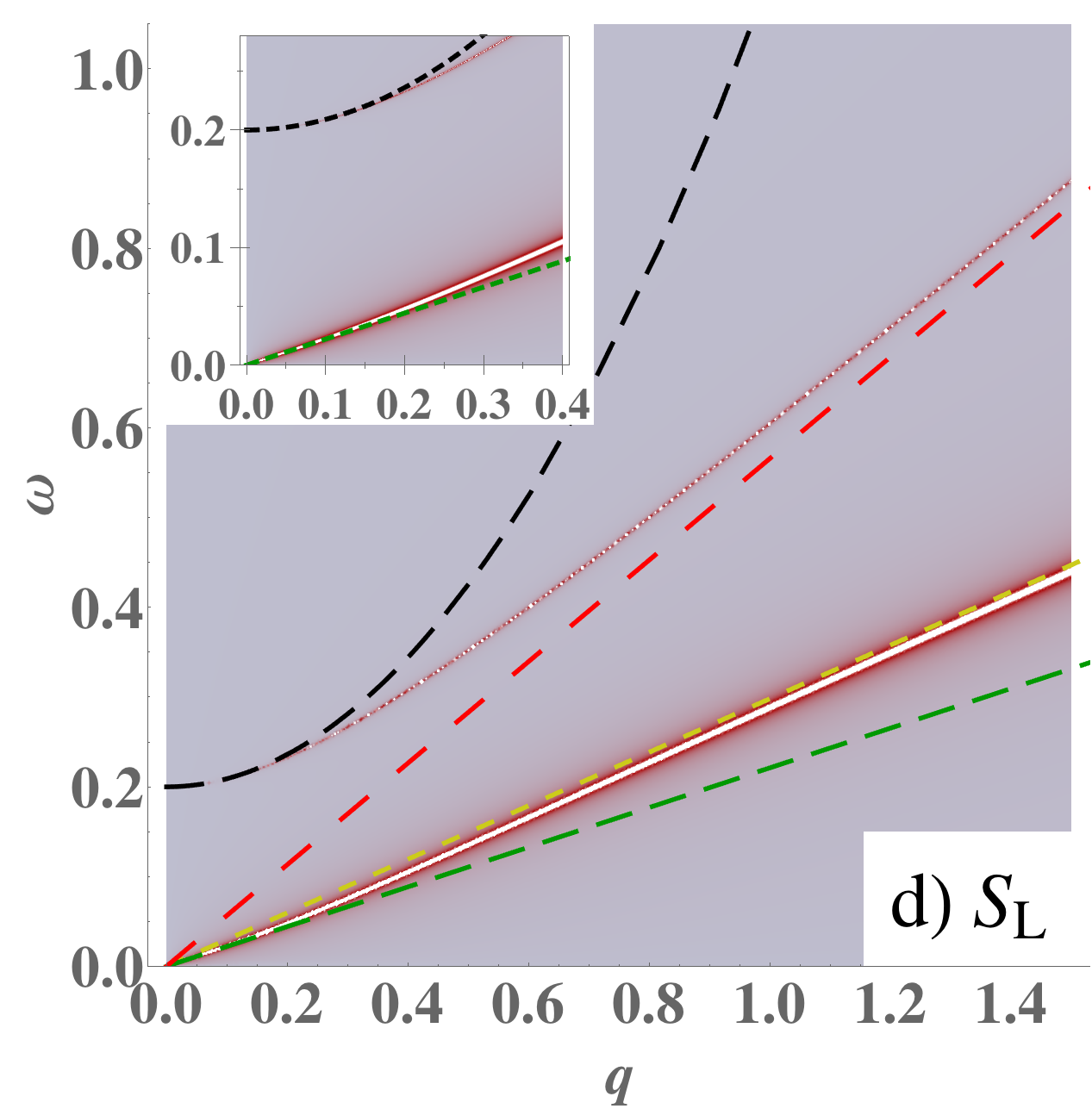}\\
{\centering
\includegraphics[height=0.7cm]{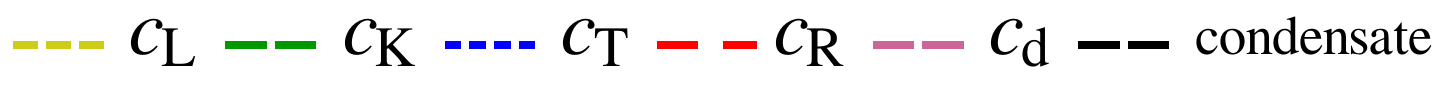}
}\\
 \caption{Spectral functions (left: transverse; right: longitudinal) of the nematic liquid crystal in units of 
 the inverse shear modulus $1/\mu \equiv 1$, with Poisson ratio $\nu = 0.1$ and dislocation Higgs mass $\Omega = 0.2$ (in arbitrary units). The top row shows a realistic situation with $c_\td \approx c_\tT$, whereas the bottom row artificially has $c_\td = 4 c_\tT$ which reduces mode couplings in order to trace the nature of the massive poles.  The width of the poles is artificial and denotes the relative pole strengths: these ideal poles are actually infinitely sharp. The insets show zoom-ins near the origin. In both the transverse and the longitudinal response we find a massive and a massless pole. The massive poles are identified as the collective modes of the dislocation condensate itself. They have a Higgs gap $\Omega$ and dispersion at low momenta following Eq.~\eqref{eq:nematic dislocation sound dispersion} while at high momenta  $q > \Omega/c_\td$ they approach linear dispersions with the condensate velocities $c_\td$ (transverse) resp.  $c_\tR = c_\td/\sqrt{2}$ (longitudinal). The massless pole in the longitudinal sector propagates at the pure compressional velocity $c_\tK$ at low momenta while it reduces to the longitudinal phonon with velocity $c_\tL$ at high momenta where the underlying crystal again becomes apparent. In the transverse response one  
 clearly identifies the rotational Goldstone mode with the rotational velocity $c_\tR$ at low momenta, while at high momenta it reduces to the transverse phonon with velocity $c_\tT$.}\label{fig:nematic spectral functions}
 \end{center}
\end{figure}

Let us first consider the longitudinal response.   From Fig.~\ref{fig:nematic spectral functions} we directly infer that for energies much less than the Higgs mass the massless mode propagates with the literal sound velocity set by only the compression
modulus $\kappa$ via $c_\tK = \sqrt{\kappa/\rho} = \sqrt{c_\tL^2 - c_\tT^2}$. In fact, by expanding Eq.~\eqref{eq:nematic longitudinal propagator} for $ \omega_n \ll \Omega$ it follows immediately, 

\begin{align}
 G_\tL (\Omega \to \infty) &= \frac{1}{\mu} \frac{c_\tT^2 q^2}{\omega_n^2 + c_\tK^2 q^2}.
\end{align}

This demonstrates explicitly the case we already discussed: in the liquid crystal translational symmetry is restored but at zero temperature it is still carrying a propagating sound mode. The shear component 
of the longitudinal mode has to be removed at low energy and this is accomplished by a mode coupling to the massive mode in the longitudinal spectrum. 
At energies larger than the dislocation Higgs mass $\Omega$, the characteristic length scales become smaller than the distance between dislocations and the medium restores its elastic, solid-like nature. 
At the Higgs mass the sound mode of the nematic therefore shows a crossover back to a longitudinal 
phonon velocity, as follows immediately from the longitudinal propagator Eq.~\eqref{eq:nematic longitudinal propagator} in the limit $\Omega \ll c_\tR q$. We still have to identify the nature of the massive mode in the longitudinal spectrum.  Compared to the solid we need an extra medium 
that can `vibrate': this is actually associated with the phase modes $\phi^a$ of the dislocation condensate that got shuffled into the stress sector using the Lorenz gauge fix! We now profit from the artificial condensate 
velocity as a diagnostic tool. We find that for small momenta $q$, the dispersion of the massive mode is given by (for real frequency $\omega$),
\begin{equation}\label{eq:nematic dislocation sound dispersion}
 \omega^2  = \Omega^2  +  (\tfrac{1}{2} c_\td^2 +  c_\tT^2) q^2 + \ldots
\end{equation}
This demonstrates that this mode is associated with the phase modes of the dislocation condensate since 
the velocity $c_\td$ enters. Similarly, the significance of this mode is that it `absorbs the shear component' of the longitudinal phonon and therefore the signature of shear through $c_\tT = \sqrt{\mu/\rho}$ is present even in the longitudinal response. At large energy it turns into a pure dislocation condensate mode 
propagating at a velocity $c_\tR = c_\td/\sqrt{2}$. 

In summary, we find out that the longitudinal phonon has changed at low energy into a pure (compressional) sound mode, of which by the mode coupling with the condensate mode
``the shear components have been removed''.  This mechanism is quite remarkable, tying together the requirements that (a) the liquid cannot support reactive shear 
responses, with (b) the liquid itself is descending from the solid in this dual language such that the longitudinal phonon turns into real sound through the action of the dual condensate, 
in turn rooted in the fact (c) that ``sound does not carry (stress) gauge charge in the dual stress superconductor''. As we repeatedly emphasized, this is in turn a ramification of the fact 
that dislocations ``do not have a volume'' which via the detour involving the glide constraint resurfaces in the final outcomes. 

With regard to macroscopic collective behavior, of course 
only the massless excitations matter and we have discovered that the quantum nematic as a {\em zero temperature} state of matter is characterized by a sound mode. Given that we
are dealing with zero temperature matter formed from bosons, what else can this be than a {\em superfluid}? We also have to demonstrate that this fluid is {\em irrotational}, i.e. the 
circulation can only occur in the form of massive quantized vortices. The easiest way to accomplish this is by considering the electrically charged system as we will do in Sec.~\ref{sec:Dual elasticity of charged media}. There we will find that the charge quantum nematic is at the same time a full-fledged superconductor characterized by expulsion of magnetic fields, i.e. the Meissner effect.

One can now
be concerned that at first sight this seems to violate the general principle that superfluidity is associated with the spontaneous breaking of global, internal $U(1)$-symmetry: eventually it has to invoke off-diagonal long-range 
order in terms of the constituent bosons. However, from a first-quantized path integral perspective this apparent paradox is directly  resolved. Off-diagonal long-range order means 
that infinite windings of the worldlines of the constituent bosons around the imaginary time axis acquire a finite probability in the superfluid. Now one should realize that the dislocation condensate `liberates' the 
constituent bosons from their crystal lattice sites so that they can move about freely: the system is a genuine liquid. Therefore the particle worldlines can braid arbitrarily and the dislocation condensate `stirs' the particle 
system into the infinite winding affair of the standard superfluid.  It is crucial in this regard that the particle condensate and the dislocation condensate perspectives actually reside 
`on the same side' of the duality --- shear stress is on the `opposite side'. As usual, in the dual language one can understand physics from a very different but yet equivalent perspective, 
summarized here by: {\em ``a bosonic crystal losing its shear rigidity by restoring translational symmetry automatically becomes a superfluid''}.

\subsubsection{Transverse sector}
Let us now turn to the transverse response.  We infer  a similar pattern as in the longitudinal sector but now the physics is very different. Yet again we find that at high energies the 
transverse phonon is recovered (the pole characterized by $c_\tT$) and a condensate phase mode propagating with $c_\td$. However, considering very small energies we find now a
massless pole,
 \begin{equation}\label{eq:transverse propagator quantum nematic deep Higgs}
 G_\tT (\Omega \to \infty) = \frac{1}{\mu} \frac{\Omega^2}{\omega_n^2 + c_\tR^2 q^2}.
\end{equation}

We interpret this massless mode as follows. The rotational symmetry is still broken. Although hidden in the crystal, this fact becomes manifest again in the quantum nematic crystal: this is just the Goldstone boson associated with the 
broken rotational symmetry! Remarkably, its stiffness is set by the dislocation Higgs mass as is manifest both from the fact that it propagates at the rotational velocity $c_\tR = c_\td/\sqrt{2}$,
as well as the prefactor $\Omega^2 / \mu$.  This rotational mode is rooted literally in `vibrating' the dislocation condensate. Below we will expose the precise `deconfinement' 
mechanism behind this phenomenon using the `dynamical Ehrenfest constraint' formalism which makes this explicit. 

The massive mode in the transverse sector is clearly identified as the transverse phonon 
which has acquired a Higgs mass, obviously because shear stress is `expelled from the liquid' very much in the same way as magnetic fields are expelled from a superconductor.   In fact,  
in analogy with the London penetration depth of the superconductor, the Higgs mass sets a characteristic length scale, the {\em dislocation penetration depth}, given by  $\lambda_\td = \Omega / c_\td$, which is equal or almost equal to the {\em shear penetration depth} $\lambda_\mathrm{s} = \Omega / c_\tT$ since $c_\td \approx c_\tT$. This is the length 
over which shear stresses can penetrate the liquid/liquid crystal. The dispersion relation at small momenta coincides with the one associated with the massive mode in the longitudinal sector Eq.~\eqref{eq:nematic dislocation sound dispersion}. There is a mode coupling between the
`pure' transverse stress photon and the condensate mode at work at these large length scales.
The presence of these propagating massive modes is actually tied into the strongly-correlated liquid assumption underlying the entire formalism. We departed from the assertion that the dislocations occur at a low fugacity: their typical separation is large compared to the lattice constant. The shear penetration depth 
coincides with this typical distance between dislocations. At larger length scales the dislocation condensate takes care that translational symmetry is restored such that shear rigidity cannot propagate. 
At shorter distances, however, the system discovers that it is still a crystal and therefore it supports propagating  shear stress.  The occurrence of such propagating `massive shear photons' can 
therefore be used as a diagnostic for the strength of the crystal correlations in the liquid. When the shear penetration depth becomes of the order of the lattice constant these massive modes get completely invisible. Even in the case of a reasonably strongly-interacting superfluid like $^4$He these modes have not be seen, which is no surprise 
because $\lambda_\td$ is at most a few times the lattice constant. One might wonder, however, whether the roton minimum is a leftover of propagating massive shear. 

Finally, we still have to address the nature of the instantaneous static stresses that are supported by the quantum nematic; since translational rigidity has perished, the instantaneous forces between individual dislocations must vanish as well. Looking more closely, the correlation functions of the instantaneous forces in the static limit $\omega \to 0$ are
\begin{align}
 \langle b^{\tT\dagger}_\ft b^\tT_\ft \rangle &= \frac{4\kappa \mu}{\kappa + \mu} \frac{1}{q^2 + (1+\nu) \Omega^2 /c_\td^2},\label{eq:nematic instantaneous force}\\
  \langle b^{\tL\dagger}_\ft b^\tL_\ft \rangle &= 4\mu \frac{1}{q^2 + 1/\ell^2  + 2 \Omega^2 /c_\td^2},
\end{align}
These should be compared with Eqs.~\eqref{eq:isotropic solid instantaneous force} and \eqref{eq:solid rotational force} of the solid. We see that in both cases, the range of the forces is cut off by the dislocation penetration depth $\lambda_\td = c_\td/\Omega$. 

The only long-range static rigidity is rotational, related to the rotational Goldstone boson mentioned above.  We will discuss this further below. The mode content of the quantum nematic is summarized in Table~\ref{table:Mode content of the quantum nematic}.   

\begin{table}
\begin{center}
 \begin{tabular}{cc}
  \toprule 
 name  & character \\
  \hline
 sound mode & longitudinal,  massless  \\
  rotational Goldstone mode & transverse, massless  \\
  dislocation condensate & longitudinal, massive \\
  shear mode & transverse, massive \\
 dislocation force & longitudinal, short-ranged\\ 
 dislocation force & transverse, short-ranged\\ 
  static torque & transverse, long-ranged  \\
  \bottomrule
 \end{tabular}
 \caption{Mode spectrum of the quantum nematic}\label{table:Mode content of the quantum nematic}
 \end{center}
\end{table}

There is one question remaining: for very high momenta $q \gg 1 / \lambda_\td$ we expect to recover the character of the underlying crystal. One way of looking at this is that the cores of the topological defects (the dislocations) should correspond to the `normal state' as is familiar from Abrikosov vortices in superconductors. The normal state of the dual stress superconductor is obviously the crystalline solid. Indeed we have noticed above that the longitudinal massless mode goes to $c_\tL q$ for high momenta and the transverse massless mode goes to $c_\tT q$, recovering the longitudinal and transverse phonons. But in the spectral functions of Fig.~\ref{fig:nematic spectral functions} we also see the two dislocation sounds modes with velocities $c_\tR$ resp. $c_\td$, that should be absent in the solid. The answer is that the spectral weight of these dislocation modes vanishes in the limit $q \lambda_\td \to \infty$, leaving only the phonons.

\subsection{Torque stress in nematics and the rotational Goldstone mode}\label{subsec:torque stress nematic}
Above we have seen the emergence of a massless mode in the transverse sector of the quantum nematic. 
Recall that the transverse propagator is in fact the rotation strain correlation function Eq.~\eqref{eq:transverse strain propagator}. 
The quantum nematic has full translational symmetry, but broken rotational symmetry---since even the `isotropic' dislocation condensate of 
Fig.~\ref{fig:lattice vectors} remembers the underlying crystal axes---and it is tempting to identify this mode as the Goldstone mode associated 
with this broken rotational symmetry. Here we will show that that is indeed the case. Then the important question is raised: 
what happens to this Goldstone mode in the ordered, solid phase, which also breaks rotational symmetry spontaneously? 
The duality construction provides a satisfactory and surprising answer: the rotational Goldstone mode is {\em confined} in the solid, 
in the sense that the interaction it mediates is stronger than logarithmic in separation. This goes hand-in-hand with the 
confinement of disclinations in the solid. The derivation presented here was carried out in Ref.~\cite{BeekmanWuCvetkovicZaanen13}. 
For complementary discussions on the interdependence between translational and rotational Goldstone modes, see for 
instance Refs.~\cite{LowManohar02,WatanabeMurayama13,BraunerWatanabe14,HayataHidaka14}.

It should be emphasized that this propagating rotational Goldstone boson is actually a
{\em unique} feature of the {\em superfluid} (or superconducting, see Sec.~\ref{sec:Dual elasticity of charged media}) nematic state. It is actually a well-known conundrum that in the thermal 
nematic state the Goldstone theorem appears to be violated in  the sense that despite the breaking of a global symmetry (the rotations) a 
{\em propagating} Goldstone mode is missing. The culprit is the hydrodynamical nature of the classical nematic: it breaks rotations, but is at the same time 
a normal fluid governed by the equations of hydrodynamics. Upon stirring the fluid, circulation in the hydrodynamical flow will appear which is subjected 
to viscous damping. The highly unusual feature that this viscous normal fluid circulation is coupled to the rotational  Goldstone `mode' and this 
coupling is ``relevant in the deep IR''~\cite{DeGennesProst95,ChaikinLubensky00}. The effect is that this Goldstone boson just gets completely overdamped. 
Similarly, dealing with a quantum nematic arising in a Fermi liquid by deforming the Fermi surface, one finds that the Goldstone boson 
is overdamped, in essence by Landau damping, but concurrently the perturbation expansion associated with the quasiparticle dressing blows up~\cite{OganesyanKivelsonFradkin01,FradkinEtAl10}. Remarkably, it appears that `fermionic nematics' have to be non-Fermi liquids! But here we are dealing with superfluid nematics 
and in superfluids ``circulation is massive'': rotational motions can only enter in the form of quantized vortices, and below the vortex creation energy scale the 
superfluid is irrotational. Therefore the rotational Goldstone boson is protected against the circulation in the fluid and can therefore propagate as an undamped mode.  

At the heart of the increased structure of Goldstone modes due to spatial symmetry breaking is the fact that translations and rotations are not independent. 
Recall that in the derivation of the Noether theorem, which constructs a conserved current for each global symmetry, one considers infinitesimal symmetry transformations. 
However, locally, an infinitesimal rotation amounts to just an infinitesimal translation~\cite{LowManohar02}. This is reflected in the definition of a local rotation deformation 
Eq.~\eqref{eq:strain Ehrenfest constraint}, $\omega = \partial_x u^y - \partial_y u^x$. Clearly, an infinitesimal rotation $\omega$ can be represented by an infinitesimal displacement $u^a$. 

Very recently, it has been cleared up that in such cases of interdependent symmetries (Noether currents), the correct spectrum of Goldstone modes needs to take account of so-called inverse 
Higgs constraints~\cite{NicolisEtAl13,WatanabeMurayama13,BraunerWatanabe14,HayataHidaka14}. This reasoning also explains why, for instance, 
we do not observe any Goldstone modes due to spontaneous breaking of symmetry under Galilean boosts (Galileons) in superfluids, or moreover those due to broken Lorentz boosts.
Next to these symmetry considerations, there is a neat kinematic explanation of what happens when both translations and rotations are spontaneously broken, which is the topic of the remainder of this section.
 Heuristically, if one attempts to excite a rotational Goldstone mode in a solid by exerting an external torque, instead one excites transverse, shear phonons. 

The rotational Goldstone mode is in fact a propagating excitation in the rotation field $\omega(x)$, which is dualized into the torque stress gauge field $h_\lambda(x)$ 
in Sec.~\eqref{subsec:Torque stress gauge field}. To accomplish this, we converted the Ehrenfest constraint $\epsilon_{ma} \sigma^a_m =0$ into a dynamical constraint, 
or conservation law, $\partial_\mu \tau_\mu =0$. This allowed us to investigate torque stress even in first-gradient elasticity of the solid phase. A part of the stress tensor 
$\sigma^a_\mu$ can be represented by the torque stress gauge field, by using consecutively Eqs.~\eqref{eq:stress gauge field definition}, \eqref{eq:torque stress definition} 
and \eqref{eq:torque stress gauge field definition}. In explicit form we find the exact relations:
\begin{align}
 \tau_\ft &= - b^\tL_\tL - b^\tT_\tT &&= -q h_\tT, \label{eq:temporal torque in gauge fields}\\
 \tau_\tL &= b^\tL_\ft &&= -\ti\tfrac{1}{c_\tT} \omega_n h_\tT, \label{eq:longitudinal torque in gauge fields}\\
 \tau_\tT &= b^\tT_\ft &&= \ti \tfrac{1}{c_\tT} \omega_n h_\tL - q h_\ft = p h_{+1}. \label{eq:transverse torque in gauge fields}
\end{align}
Upon taking the Coulomb gauge fix for the torque stress gauge field $\partial_m h_m = -q h_\tL =0$, we can interpret as usual the temporal component $h_\ft$ as the instantaneous, static force 
and the transverse component $h_\tT$ as the propagating excitation. Interestingly, the static force between sources of torque (disclinations) is carried by the same component as 
the static force between dislocations, via the last relation $b^\tT_\ft = -q h_\ft$. In the solid, the force between dislocations derives from $\langle b^{\tT\dagger}_\ft b^\tT_\ft \rangle \propto \frac{1}{q^2}$,
 amounting to a decay logarithmic in the separation between two sources. We now immediately find that the force between disclinations becomes,
\begin{equation}\label{eq:disclination confinement}
 \langle h_\ft^\dagger h_\ft \rangle = \frac{1}{q^2} \langle b^{\tT\dagger}_\ft b^\tT_\ft \rangle \propto \frac{1}{q^4},
\end{equation}
which implies that the energy cost of a disclination--anti-disclination pair at separation $r$ grows like $r^2$, see for instance Ref.~\cite{SeungNelson88}. This is obviously a much stronger dependence
 than for dislocations, and using standard terminology this implies that disclinations are {\em confined} in the solid phase. This is not dissimilar to the confinement of 
 quarks due to the strong interaction which grows with $r$ according to quantum chromodynamics (QCD). From Eqs.~\eqref{eq:temporal torque in gauge fields}, \eqref{eq:longitudinal torque in gauge fields}, we see that also the propagating  component of the torque stress gauge field $h_\tT$ has this behavior. In fact, substituting these equations in the Lagrangian of the solid Eq.~\eqref{eq:transverse Lagrangian stress gauge field} using the stress Coulomb gauge fix $\partial_m b^\tL_m = -q b^\tL_\tL =0$ leads to the simple expression,
\begin{equation}
 \mathcal{L}^\tT =  \tfrac{1}{2} \frac{1}{\mu} q^2 (\tfrac{1}{c_\tT^2} \omega_n^2 + q^2) h_\tT^\dagger h_\tT.
\end{equation}
Note that terms coming from second-gradient elasticity with factors $\ell^2$ drop out. The additional factor of $q^2$ with respect to the phonon Lagrangian $\sim (\frac{1}{c_\tT^2}\omega_n^2 + q^2)  b^{\tT\dagger}_\tT b^\tT_\tT$ implies that it is energetically more costly to excite rotational Goldstone modes than phonons at any finite momentum. Perturbing a solid by an external torque source will therefore excite transverse phonons rather than rotational Goldstone modes. But  the solid in fact behaves like a completely `incompressible' medium in its response to torque: it is the way that crankshafts do their work, and confinement of torque is thereby actually fundamental in mechanical engineering. 

Let us  now zoom in on the torque stress gauge field in 
 the quantum nematic phase. The derivation of the Higgs term for the quantum smectic and nematic was carried out in a gauge fix that removed cross terms for the stress gauge fields, 
Eq.~\eqref{eq:dislocation Lorenz gauge fix}. Here it is more convenient to choose the unitary gauge fix, which eliminates the phase fields of the dislocation condensate 
Eq.~\eqref{eq:dislocation unitary gauge fix}. In this gauge fix, all six components of the stress gauge field $b^a_\mu$ initially acquire a Higgs contribution, whereas in the other case 
two of these were assigned to the dislocation phase modes. Obviously, the end results are gauge invariant, so this procedure can be used without any problems. 
One component in the longitudinal sector will be removed by the glide constraint, but since torque resides purely in the transverse sector we can leave that aside here.

Since we are only interested in the massless mode, we can take the limit $\Omega^2 \to \infty$ immediately, and consider only the Higgs term, dropping the contribution 
from the original isotropic solid Eq.~\eqref{eq:stress gauge field Coulomb action}. Going back from $\tilde{b}^a_\mu$ to $b^a_\mu$ fields, the Higgs term in the unitary gauge fix reads,
\begin{equation}
 \mathcal{L}_\mathrm{Higgs} = \frac{\Omega^2}{2 c_\tT^2 \mu} \Big[ \frac{c_\tT^2}{c_\td^2} \lvert b^a_\ft \rvert^2 + \lvert b^a_m \rvert^2 \Big].
\end{equation}
We substitute the torque stress tensor Eq.~\eqref{eq:torque stress definition} in real-space coordinates,
 \begin{align}\label{eq:torque stress gauge field components}
 \tau_\ft &= - b^x_x - b^y_y,& \tau_x &= b^x_\ft,& \tau_y &= b^y_\ft. 
\end{align}
This leads to
\begin{align}
 \mathcal{L}_\mathrm{Higgs} &= \frac{\Omega^2}{2 c_\tT^2 \mu} \Big[ \tfrac{c_\tT^2}{c_\td^2} \lvert \tau_x \rvert^2 + \tfrac{c_\tT^2}{c_\td^2}\lvert \tau_y \rvert^2  + \tfrac{1}{2} \lvert \tau_\ft \rvert^2 +  \tfrac{1}{2} \lvert b^x_x - b^y_y \rvert^2  + \tfrac{1}{2} \lvert b^x_y + b^y_x \rvert^2   + \tfrac{1}{2} \lvert b^x_y - b^y_x \rvert^2 \Big].  
\end{align}
Here $b^x_y - b^y_x$ is part of the compression component which will be eliminated from the Higgs term by the glide constraint. 
Focusing on the components that appear in the torque stress $\tau_\mu$, we substitute the torque stress gauge field $h_\mu$ from Eq.~\eqref{eq:torque stress gauge field definition}. 
Taking care of appropriate factors of $c_\tT / c_\td$, the first line reads,
\begin{align}
 \mathcal{L}_\mathrm{Higgs}^\mathrm{torque} &= \frac{\Omega^2}{2 c_\tT^2 \mu} \Big[ (  \tfrac{1}{c_\td^2} \omega_n^2  + \tfrac{1}{2} q^2 )\lvert h_\tT \rvert^2  + \tfrac{c_\tT^2}{c_\td^2} q^2 \lvert h_\ft\rvert^2 \Big] = \frac{\Omega^2}{4 c_\tT^2 \mu} \Big[ (  \tfrac{1}{c_\tR^2} \omega_n^2  +  q^2 )\lvert h_\tT \rvert^2  +  q^2 \lvert \tfrac{c_\tT}{c_\tR} h_\ft\rvert^2 \Big].\label{eq:torque sector Higgs term}
\end{align}
Here $c_\tR \equiv c_\td/\sqrt{2}$, cf. Eq.~\eqref{eq:transverse propagator quantum nematic deep Higgs}. For convenience, we have imposed the Coulomb gauge fix $ \partial_m h_m = -q h_\tL =  0$. We recognize this Lagrangian as simply the dual of a real scalar field as in Eq.~\eqref{eq:XY Coulomb action}. 
Therefore we can `dualize back' in which case we retrieve the action
\begin{align}\label{eq:rotational Higgs term}
 \mathcal{L}_\mathrm{Higgs}^\mathrm{rot} &= \tfrac{1}{2} \frac{2 c_\tT^2 \mu}{\Omega^2} (\partial^\tR_\mu \omega)^2,
\end{align}
where $\partial^\tR_\mu \equiv (\tfrac{1}{c_\tR}\partial_\tau, \partial_m)$. Here $\omega(x)$ is in fact the rotation field $\omega = \partial_x u^y - \partial_y u^x$ from Eq.~\eqref{eq:strain Ehrenfest constraint}. 
That this must be so can be seen from the fact that the multivalued part of this field are the rotational defects, disclinations, that source the fields $h_\mu$.

Let us now interpret the physical meaning of Eq.~\eqref{eq:torque sector Higgs term}. In Sec.~\ref{subsec:Torque stress gauge field} we interpreted $h_\ft$ as the static force between disclinations. 
Clearly, this is now of the `ordinary'  $1/q^2$ form, leading to interactions logarithmic in separation. This is of course completely in line with what we know from classical nematics, 
cf. Sec.~\ref{sec:Order parameters for 2+1-dimensional nematics}. In the nematic phase disclinations are {\em deconfined} and can occur as ordinary topological defects, 
on the same footing as vortices in superfluids and dislocations in solids. One way to look at this is that by increasing the separation between a disclination and an antidisclination in the solid 
additional dislocations are introduced given Eq.~\eqref{eq:dislocation density from Frank vector}. But in the nematic phase dislocations are completely condensed which implies that dislocations can be pulled out of the condensate `for free'. Therefore, there is no longer a `topological barrier' obstructing the creation of disclination--antidisclination pairs. This is surely reflected by the static force mediated by $h_\ft$.

Another interesting observation is the prefactor of Eq.~\eqref{eq:torque sector Higgs term} which is proportional to $\Omega^2$, see also Eq.~\eqref{eq:transverse propagator quantum nematic deep Higgs}. 
Recall from Eq.~\eqref{eq:dislocation Higgs mass} that this Higgs mass is proportional to the density of the dislocation condensate $\lvert \Psi \rvert^2$. If this density goes to zero---upon returning to the solid phase---the rotational Goldstone
 mode disappears. Therefore, the dislocation condensate can be said to form a medium which carries the rotational Goldstone mode. Adopting fully the dual viewpoint, one must accept that a condensate of 
 topological defects is a real form of matter.

Also note that the velocity of this mode is $c_\tR = c_\td/\sqrt{2}$, as we have already seen in Eq.~\eqref{eq:transverse propagator quantum nematic deep Higgs}. 
We can now trace back the origin of this velocity reduction from $c_\td$ to Eq.~\eqref{eq:torque stress gauge field components}: 
the spatial propagation of the rotational mode is `shared' between the $x$- and $y$-directions.
 
 \section{Quantum smectic}\label{sec:Quantum smectic}
 In the soft matter literature the smectic order appears as a vestigial order in between the crystal and the nematic order. It is typically viewed from a `molecular perspective'~\cite{DeGennesProst95,ChaikinLubensky00}:
one asserts that rod-like molecules orient themselves, like in the uniaxial nematic of Fig.~\ref{fig:classical melting nematic}, while the interactions are so anisotropic that these want to form liquid layers. These liquid layers subsequently 
stack in a periodic array in the direction perpendicular to the layers, see Fig.~\ref{fig:classical melting smectic}. The resulting medium has collective elasticity-like properties that are unique: it is behaving neither as a solid, nor as a liquid but instead as something that is `hanging in the middle'. The most striking consequence is the appearance of a quadratically dispersing `phonon': consider a periodic transverse displacement that is propagating along the liquid direction.
Since the layers are liquid (no translational symmetry breaking), the reactive response associated with strains $\sim \partial u$  vanishes. However, considering the next order in the gradient expansion $\sim \partial^2 u$ an elastic response is
recovered associated with the curvature interactions between the liquid layers~\cite{DeGennesProst95,ChaikinLubensky00}. Accordingly, one is dealing 
with an effective action of the form $ (\partial_t u)^2 + (\partial_m^2 u)^2$  describing this `liquid direction transverse phonon', called {\em undulation mode}, which is therefore characterized by a quadratic dispersion $\omega \propto q^2$.

Turning to the zero temperature realms, the idea of a quantum incarnation of  a smectic state was introduced by Fradkin, Kivelson, Emery and Lubensky in the specific microscopic setting  of 
the electron stripes in doped Mott insulators and the stripe phases formed in quantum Hall systems~\cite{FradkinKivelson99,EmeryFradkinKivelsonLubensky00}. 
This school of thought departs from the assertion that in the two-dimensional setting  of the copper-oxide planes, the electrons self-organize in Mott-insulating domains separated by `rivers of charge' (the stripes of the introduction)
which are to zeroth order considered to form an array of independent 1+1D Luttinger 
liquids. One can then demonstrate that for very specific intra- and inter-Luttinger-liquid couplings the interactions between the different Luttinger liquids in the array can become irrelevant in the IR. 
The effect is that a state is obtained in 2+1D which is in the scaling limit behaving like a 1+1D Luttinger liquid metal in the `liquid direction', while it is Mott-insulating in the perpendicular direction. 

This is based on a microscopic model associated with strong electron--electron interactions in the presence of a very strong  {\em periodic background potential}. Like in the soft-matter context, one 
would like to find out whether such quantum smectic states can be formed from bosonic matter living in the {\em Galilean continuum}.  In fact, in Sec.~\ref{sec:Dynamics of disorder fields} we already demonstrated that this arises as a natural part of the dual stress superconductor portfolio. As we argued, in $D$ space dimensions there 
are $D$ disorder fields required to represent the dislocation condensate.  Since these are symmetry-wise inequivalent, they do not have to condense simultaneously. 
Given the appropriate microscopic conditions (such as `rod-like bosons') it can therefore happen that the dislocations proliferate exclusively in  one of the high symmetry (``$x$'') directions of the crystal, leaving the
orthogonal (``$y$'') space direction unaffected. This dual perspective on smectic order is still obscure in the soft matter tradition. Based on these mathematical considerations it was put forward for the first time in  Ref.~\cite{ZaanenNussinovMukhin04}, where it was initially misidentified as an ``ordered nematic state''. This was  put in the proper context in Refs.~\cite{CvetkovicZaanen06b,Cvetkovic06}.

As we will show in detail in this section, these proper bosonic quantum smectics form an entertaining and quite rich stage. We learned in Sec.~\ref{sec:Quantum nematic} that the bosonic quantum nematics are at the same time 
genuine superfluids.  Accordingly, we will find that the quantum smectics are also behaving like superfluids in so far as their `liquid side' is concerned. Just as in the case of the classical smectics, they 
also have a `solid side', but eventually the solid and superfluid characteristics are `glued together' into an indivisible entity with unique characteristics that we will present in this section. We focus on the elastic properties.

Following precisely the template that we exposed for  the quantum nematic in Sec.~\ref{sec:Quantum nematic}, we will now compute the dual stress propagators and mode spectra for the quantum smectics. Since there is less
symmetry, the computations become more involved with the reward that there is also more going on.

\begin{figure}[tb]
\begin{center}
  \includegraphics[height=3cm]{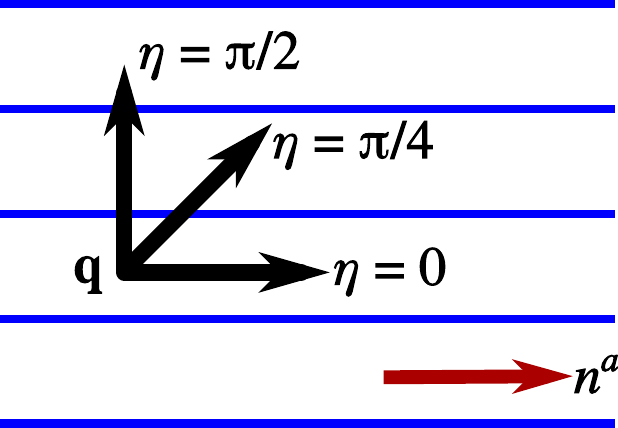}
 \caption{The smectic liquid crystal is a periodic stack of liquid layers (blue lines). Within the layers there is full translational symmetry and therefore it is referred to as the {\em liquid direction}. In the perpendicular direction translation symmetry is still broken and it is referred to as the {\em solid direction}.  Dislocations have proliferated with their Burgers vectors $n^a$ in the liquid direction. Dislocations with Burgers vectors in the solid direction are still allowed as stable topological defects. When probing the system with linear response, we define the angle $\eta$ between the momentum and the smectic layers via $\epsilon_{\ft m a } q_m n^a = q \sin \eta$.}\label{fig:smectic layers}
\end{center}
\end{figure}

\subsection{Lagrangian of the quantum smectic}

Let us depart from the general expression for the dual stress superconductor Eq.~\eqref{dualstressSC}. Without loss of generality, we consider the case where only the $\Phi^x$-disorder field 
condenses while the $\Phi^y$-field stays massive. We shall simply use $\Omega = \Omega^x$ to denote the Higgs mass Eq.~\eqref{eq:dislocation Higgs mass}. Dropping the pure phase field terms $\lvert \partial_\mu \phi^x \rvert^2$ because they are decoupled in the Lorenz gauge fix, the Higgs term in the Josephson limit Eq.~\eqref{eq:dislocation Higgs term} is explicitly,
 \begin{align}\label{eq:smectic Higgs term}
 \mathcal{L}_\mathrm{Higgs} =\frac{1}{2} \frac{\Omega^2}{ c_\tT^2 \mu} (\tilde{b}^{x\dagger}_\mu + \lambda^\dagger \epsilon_{\ft \mu x}) ( \delta_{\mu\nu} - \frac{\tilde{p}_\mu \tilde{p}_\nu}{\tilde{p}^2} ) (\tilde{b}^x_\nu + \lambda \epsilon_{\ft \nu x} ).
 \end{align}
We define an angle $\eta$ between the Burgers direction of the dislocation condensate and the momentum, see Fig.~\ref{fig:smectic layers}, via:
\begin{equation}
 \epsilon_{\ft m a } q_m n^a = q \sin \eta. \label{eq:eta definition}
\end{equation}
This implies $q_m n^m  =q \cos \eta$. For our choice $n^x = 1, n^y = 0$ we have $\sin \eta = -q_y /q$ and $\cos \eta = q_x/q$. Furthermore we define $\tilde{p}_\eta^2 \equiv \omega_n^2 / c_\td^2 + \cos^2 \eta\; q^2$. With this we expand, using $\tilde{p}_\mu \tilde{b}^x_\mu = \ti \tilde{p} \tilde{b}^x_0$,
\begin{align}
  \mathcal{L}_\mathrm{Higgs} &= \frac{1}{2} \frac{\Omega^2}{ c_\tT^2 \mu} \Big( 
  \lvert \tilde{b}^x_{+1} \rvert^2 + \lvert \tilde{b}^x_{-1} \rvert^2 + \big(1 - \frac{ \lvert\epsilon_{\ft \mu x} \tilde{p}_\mu\rvert^2 }{\tilde{p}^2}\big) \lambda^\dagger \lambda
  + \lambda^\dagger ( - \tilde{b}^x_y - \ti \sin \eta \tfrac{q}{\tilde{p}} \tilde{b}^x_0) + \text{h.c.}
  \Big) \nonumber\\ 
  &= \frac{1}{2}\frac{\Omega^2}{ c_\tT^2 \mu} \Big( 
  \lvert \tilde{b}^x_{+1} \rvert^2 + \lvert \tilde{b}^x_{-1} \rvert^2 +  \tfrac{\tilde{p}_\eta^2}{\tilde{p}^2} \lambda^\dagger \lambda 
 + \lambda^\dagger (\sin \eta \tfrac{\omega_n}{c_\td \tilde{p}} \tilde{b}^x_{+1} - \ti \cos \eta\; \tilde{b}^x_{-1} )  + \text{h.c.}  \Big).
\end{align}
Here we used the transformations in \ref{sec:Fourier space coordinate systems} to expand
\begin{align}
\tilde{b}^x_y &= \ti \tfrac{q_y}{\tilde{p}} \tilde{b}^x_0 + \tfrac{\omega_n q_y}{c_\td \tilde{p} q} \tilde{b}^x_{+1} + \ti \tfrac{q_x}{q} \tilde{b}^x_{-1} 
= - \ti \sin \eta \tfrac{q}{\tilde{p}}  \tilde{b}^x_0 - \sin \eta \tfrac{\omega_n }{c_\td \tilde{p}} \tilde{b}^x_{+1} + \ti \cos \eta\; \tilde{b}^x_{-1}.
\end{align}
Integrating out the Lagrange multiplier field $\lambda$ leads to
\begin{align}
  \mathcal{L}_\mathrm{Higgs} &= \frac{1}{2}\frac{\Omega^2}{ c_\tT^2 \mu} \Big( \lvert b^x_{+1} \rvert^2 + \lvert b^x_{-1} \rvert^2  - \frac{1}{\tilde{p}_\eta^2} \lvert \sin \eta \tfrac{\omega_n}{c_\td} \tilde{b}^x_{+1} - \ti \cos \eta\; \tilde{p}\; \tilde{b}^x_{-1}\rvert^2 \Big)\nonumber\\
  &= \frac{1}{2}\frac{\Omega^2}{ c_\tT^2 \mu} \frac{1}{\tilde{p}_\eta^2} \lvert \cos \eta\; \tilde{p}\; \tilde{b}^x_{+1} + \ti \sin \eta \tfrac{\omega_n}{c_\td} \tilde{b}^x_{-1} \rvert^2 = \frac{1}{2}\frac{\Omega^2}{ c_\tT^2 \mu} \frac{1}{\tilde{p}_\eta^2} \lvert - q_x \tilde{b}^x_\ft + \tfrac{\omega_n}{c_\td} \tilde{b}^x_x \rvert^2 
  = \frac{1}{2} \frac{\Omega^2}{c_\tT^2 \mu} \frac{c_\tT^2}{c_\td^2} \frac{1}{\tilde{p}_\eta^2} \lvert - q_x b^x_\ft + \tfrac{\omega_n}{c_\tT} b^x_x \rvert^2 \nonumber\\
  &= \frac{1}{2}\frac{\Omega^2}{ c_\td^2 \mu} \frac{1}{\tilde{p}_\eta^2} \lvert \sigma^x_y \rvert^2.
\end{align}
In the last line we used the definition of the stress gauge field Eq.~\eqref{eq:stress gauge field definition}. We started out in Eq.~\eqref{eq:smectic Higgs term} by giving a Higgs mass to two components $\tilde{b}^x_{+1},\tilde{b}^x_{-1}$, but by imposing the glide constraint we find out that only one particular linear combination of these remains, while the other combination stays massless. We saw this mechanism at work as well in the quantum nematic, where out of the initial four components, one remains massless. For completeness, we note that if the crystal axis along which the dislocations condense does not line up with the $x$-axis, we have  an arbitrary two-vector $n^a$, and the last equation reads:
\begin{equation}
   \mathcal{L}_\mathrm{Higgs} =\frac{1}{2} \frac{\Omega^2}{c_\td^2\mu} \frac{1}{\tilde{p}_\eta^2} \lvert  \epsilon_{ac} n^a n^b \sigma^b_c \rvert^2. \label{eq:smectic Higgs stress tensor}
\end{equation}
Writing out this term explicitly gives
 \begin{align}
  \mathcal{L}_\mathrm{Higgs} &=
   \frac{1}{2} \frac{\Omega^2}{c_\tT^2 \mu} 
  \begin{pmatrix} \tilde{b}^{\tT\dagger}_{+1} & \tilde{b}^{\tL \dagger}_\tT & \tilde{b}^{\tL\dagger}_{+1} & \tilde{b}^{\tT \dagger}_\tT\end{pmatrix}
  \mathcal{M}_\eta
  \begin{pmatrix}\tilde{b}^{\tT}_{+1} \\ \tilde{b}^{\tL}_\tT \\ \tilde{b}^{\tL}_{+1} \\ \tilde{b}^{\tT}_\tT \end{pmatrix}. \nonumber \\ 
  \mathcal{M}_\eta &= \frac{1}{\tilde{p}_\eta^2} 
  \begin{pmatrix}
   \tilde{p}^2 \cos^2 \eta \sin^2\eta & 
   \ti \frac{\omega_n}{c_\td}  \tilde{p} \cos^2 \eta \sin^2\eta &
   \tilde{p}^2 \cos^3 \eta \sin\eta &
   \ti \frac{\omega_n}{c_\td} \tilde{p} \cos \eta \sin^3\eta 
   \\
   -\ti \frac{\omega_n}{c_\td} \tilde{p} \cos^2 \eta \sin^2\eta &
   \frac{\omega_n^2}{c_\td^2} \cos^2 \eta \sin^2\eta &
   -\ti \frac{\omega_n}{c_\td} \tilde{p} \cos^3 \eta \sin\eta &
   \frac{\omega_n^2}{c_\td^2} \cos \eta \sin^3\eta 
   \\
   \tilde{p}^2 \cos^3 \eta \sin\eta &
   \ti \frac{\omega_n}{c_\td} \tilde{p} \cos^3 \eta \sin\eta &
   \tilde{p}^2 \cos^4 \eta &
   \ti \frac{\omega_n}{c_\td} \tilde{p} \cos^2 \eta \sin^2\eta &
   \\
   -\ti \frac{\omega_n}{c_\td} \tilde{p} \cos \eta \sin^3\eta &
   \frac{\omega_n^2}{c_\td^2} \cos \eta \sin^3\eta &
   -\ti \frac{\omega_n}{c_\td} \tilde{p} \cos^2 \eta \sin^2\eta &
   \frac{\omega_n^2}{c_\td^2} \sin^4\eta 
  \end{pmatrix}.\label{eq:quantum smectic Higgs term full}
  \end{align}

As we already learned in Sec.~\ref{sec:Quantum nematic}  the principle governing the longitudinal sector is the fact that the glide constraint decouples the compressional stress from the dislocation
condensate, which only gives a Higgs mass to the shear stress component. The character of this component depends on the angle $\eta$, referring to the orientation of the observer looking at the 
stress response relative to the liquid direction.  Zooming in on the liquid direction $\eta = 0$, it follows immediately from Eqs.~\eqref{eq:smectic Higgs stress tensor},\eqref{eq:quantum smectic Higgs term full} that only $b^\tL_{+1}$ and only $\sigma^\tL_\tT$ acquire a Higgs mass. 
In contrast, for momentum  along the solid direction  $\eta = \pi/2$, only
 $b^\tT_\tT$ and only $\sigma^\tT_\tL$ acquire  the Higgs mass. In these two cases the gapped component belongs to the `magnetic shear' $\sigma^\tL_\tT + \sigma^\tT_\tL$  (see Sec.~\ref{subsec:Physical content of the stress tensor}). Observing the system  halfway these directions ($\eta = \pi/4$) the 
stress component $\sigma^\tL_\tL - \sigma^\tT_\tT$ gets the Higgs  mass, corresponding to the `electric shear' of Sec.~\ref{subsec:Physical content of the stress tensor}. For angles $\eta$
in between these special values, linear combinations of these stress components acquire the mass. As we will see soon,  this will have important ramifications for the collective modes of the quantum smectic.

\subsection{Stress propagators of the quantum smectic}\label{subsec:smectic correlation functions}

The effects of smectic order are fully encoded in the Higgs action  Eq.~\eqref{eq:quantum smectic Higgs term full}  and together with the crystal Lagrangian 
Eq.~\eqref{eq:crystal Lagrangian smectic coordinates full} we have available all the information necessary to determine the mode spectrum. 
As we did for the quantum nematic, we shall compute the longitudinal and transverse propagators Eqs.~\eqref{eq:longitudinal Zaanen-Mukhin stress gauge field}, \eqref{eq:transverse Zaanen-Mukhin stress gauge field} for the quantum smectic. However, since we are now dealing with the 
lowered symmetry of the smectic we also have to pay attention to the chiral propagator Eq.~\eqref{eq:chiral propagator stress gauge field} since the longitudinal and transverse components
mix via the Higgs sector: Eq.~\eqref{eq:quantum smectic Higgs term full} is no longer block-diagonal for intermediate angles $0 < \eta < \pi/2$. Let us first seek some simplifications of these 
relatively complicated equations.  

Let us depart from the matrix of propagators, 
\begin{equation}
 G = \begin{pmatrix} G_\tL & G_{\tL\tT} \\ G_{\tT\tL} & G_\tT \end{pmatrix}.\label{eq:full propagator matrix}
\end{equation}
Defining the  `unperturbed' propagator matrix   $G^0 = \mathrm{diag}(G^0_\tL,G^0_\tT)$ in terms of the isotropic solid propagators Eqs.\eqref{eq:longitudinal strain propagator}, \eqref{eq:transverse strain propagator},
the effects of the condensate can be absorbed in an (anomalous) self-energy $\Pi$ according to, 
\begin{align}
 G &= \frac{1}{{G^0}^{-1} - \Pi},& \Rightarrow && \Pi = - (G^{-1} - {G^0}^{-1}).
\end{align}
This involves a matrix inversion of a $4 \times 4$-matrix, which becomes a quite lengthy affair when one also wants to keep track of the second-order elasticity corrections. This is easily enough accomplished
using a computer; ignoring the second-order corrections ($\ell  \to 0$) the expression stays manageable. Here we  derive the leading-order results in explicit form. 
The inverse of the diagonal matrix $G^0$ is of course trivial. Introducing the notation $(\alpha ,\beta)^{\otimes 2} = \begin{pmatrix} \alpha^2 & \alpha \beta \\
 \alpha \beta & \beta^2  \end{pmatrix}$, the self-energy matrix $\Pi$ becomes,
 
\begin{align}
 \Pi = \frac{\mu \Omega^2}{c_\tT^2 q^2} \frac{1}{c_\tT^2 q^2 ( \omega_n^2 + c_\td^2 q^2 \cos^2 \eta) + \Omega^2 (\omega_n^2 + 4 c_\tT^2 q^2 \cos^2\eta )} 
 \begin{pmatrix} -c_\tT^2 q^2 \sin 2\eta, & \omega_n^2 + 2 c_\tT^2 q^2 \cos^2 \eta \end{pmatrix}^{\otimes 2}
\end{align}
The stress-photon self-energy $\Pi$ is clearly  highly anisotropic, depending on angle $\eta$ between the direction of propagation of the photon relative to the `smectic frame'. 

The general expressions for the propagators are quite complicated. However, the essence of the physics is revealed by focusing on momenta which are parallel to the liquid ($\eta =0$) and
solid ($\eta = \pi/2$) directions~\cite{CvetkovicZaanen06b}. We shall see that also the $\eta = \pi/4$ case is quite informative. 
Although one has to cope with the full $4 \times 4$ propagator matrix  at intermediate angles $0 < \eta < \pi/2$, the outcomes just interpolate 
between these limits as can be easily checked by solving the system on the computer. A first simplification that one encounters along the highest  symmetry directions $\eta =0, \pi/2$
is that the chiral $G_{\tL\tT} $
propagator vanishes as is also the case in the nematic and the solid. This is just associated with the lowered symmetry in  the smectic that renders a decomposition in terms of purely transverse 
and longitudinal modes impossible at intermediate $\eta$ angles. For the momentum 
along the solid direction ($\eta = \pi/2$),
\begin{align}
 G_\tL &= \frac{1}{\mu} \frac{c_\tT^2 q^2}{\omega_n^2 + c_\tL^2 q^2},\label{eq:smectic eta 90 longitudinal propagator}\\
 G_\tT &= \frac{1}{\mu}\frac{ c_\tT^2 q^2 + \Omega^2 }{\omega_n^2 + (c_\tT^2 q^2 +\Omega^2)(1 + \ell^2 q^2)}  ,\label{eq:smectic eta 90 transverse propagator}\\
 G_{\tL\tT} &= 0.\label{eq:smectic eta 90 chiral propagator}
\end{align}
While for the momentum in the liquid direction ($\eta = 0$),
\begin{align}
 G_\tL &= \frac{1}{\mu} \frac{c_\tT^2 q^2}{\omega_n^2 + c_\tL^2 q^2},\label{eq:smectic eta 0 longitudinal propagator}\\
 G_\tT &= \frac{1}{\mu}\frac{ \omega_n^2 \Omega^2 + c_\tT^2 q^2 (\omega_n^2 + c_\td^2q^2 + 4 \Omega^2)}{(\omega_n^2 + c_\td^2 q^2)\big(\omega_n^2 + c_\tT^2 q^2 (1 + \ell^2 q^2)\big) + \omega_n^2 \Omega^2 (1+\ell^2 q^2) + 4 c_\tT^2 q^2 \Omega^2 \ell^2 q^2}  ,\label{eq:smectic eta 0 transverse propagator}\\
 G_{\tL\tT} &= 0.\label{eq:smectic eta 0 chiral propagator}
\end{align}
In these cases we can even keep track of the effects of the second-order elasticity corrections and the artificial condensate velocity $c_\td \approx c_T$, tied to the rotational velocity $c_\tR = \frac{1}{\sqrt{2}} c_\td$, introduced in 
Sec.~\ref{subsec:Correlation functions quantum nematic}. One anticipates that the momentum direction
precisely in between the solid and liquid directions $\eta = \pi/4$ is `fairly' symmetric. The full expressions get quite lengthy but when we ignore the second-order elasticity by putting $\ell \to 0$ we find,
 \begin{align}
  G_\tL &= \frac{1}{\mu} \frac{c_\tT^2 q^2 ( \omega_n^2 + c_\tR^2 q^2 + \Omega^2)}{(\omega_n^2 + c_\tL^2 q^2)(\omega_n^2 + c_\tR^2 q^2) + \Omega^2 (\omega_n^2 + c_\tK^2 q^2)},\label{eq:smectic eta 45 longitudinal propagator}\\
  G_\tT &= \frac{1}{\mu} \Big[ \frac{c_\tT^2 q^2}{\omega_n^2 + c_\tT^2 q^2} +  \Omega^2 \frac{\omega_n^2 +  c_\tL^2 q^2}{(\omega_n^2 + c_\tL^2 q^2)(\omega_n^2 + c_\tR^2 q^2 + \Omega^2) - c_\tT^2 q^2 \Omega^2}\Big],\label{eq:smectic eta 45 transverse propagator} \\
  G_{\tL\tT} &= - \frac{1}{\mu} \frac{c_\tT^2 q^2 \Omega^2}{(\omega_n^2 + c_\tL^2 q^2)(\omega_n^2 + c_\tR^2 q^2 + \Omega^2) - c_\tT^2 q^2 \Omega^2}.\label{eq:smectic eta 45 chiral propagator}
 \end{align}

In this direction the chiral propagator $G_{\tL\tT}$ is no longer vanishing. Notice also that along this direction the longitudinal propagator becomes exactly equal to that of the quantum 
nematic Eq.~\eqref{eq:nematic longitudinal propagator} and we will see below that this has a quite profound meaning. 

At stake is  that these equations reveal that the responses of the 
quantum smectic, even along the `liquid' and `solid' directions, are neither those of a superfluid, nor of a solid but instead that of a medium where these responses are truly intertwined. To shed light on this interesting affair we analyze the collective modes of the quantum smectic.

  \begin{figure}
   \begin{center}
 \includegraphics[width=7.7cm]{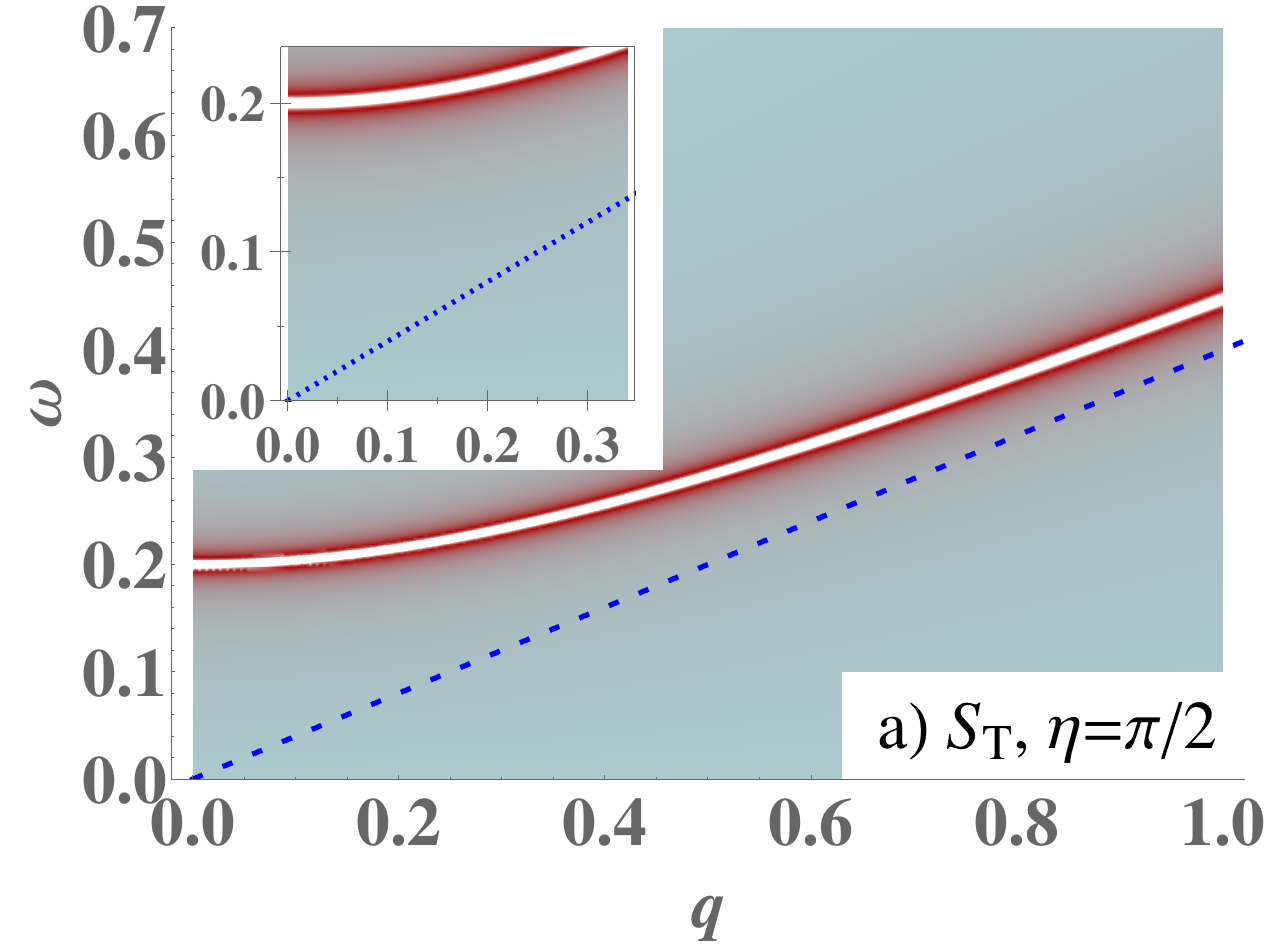}
 \includegraphics[width=7.7cm]{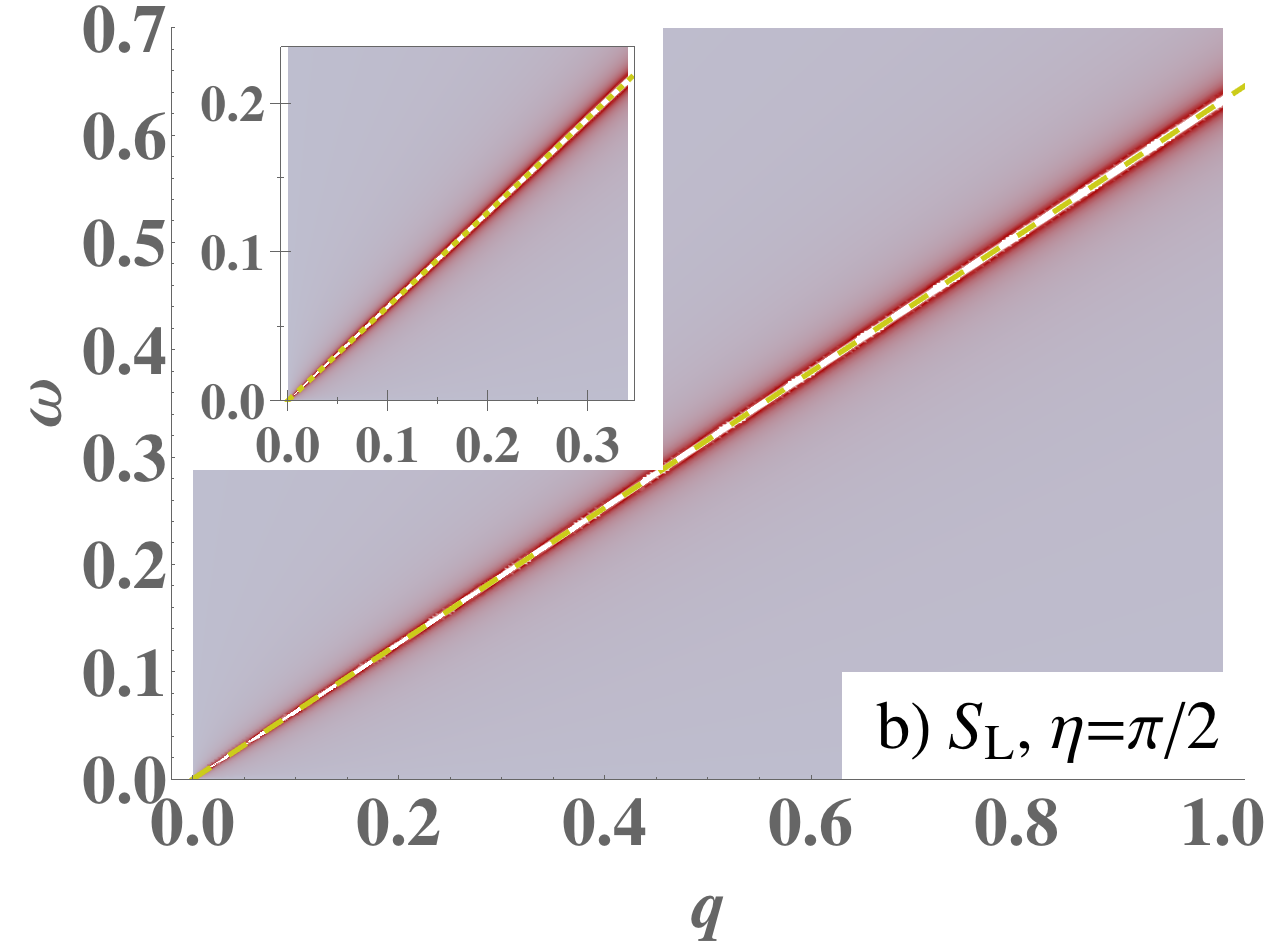}\\
 \includegraphics[width=7.7cm]{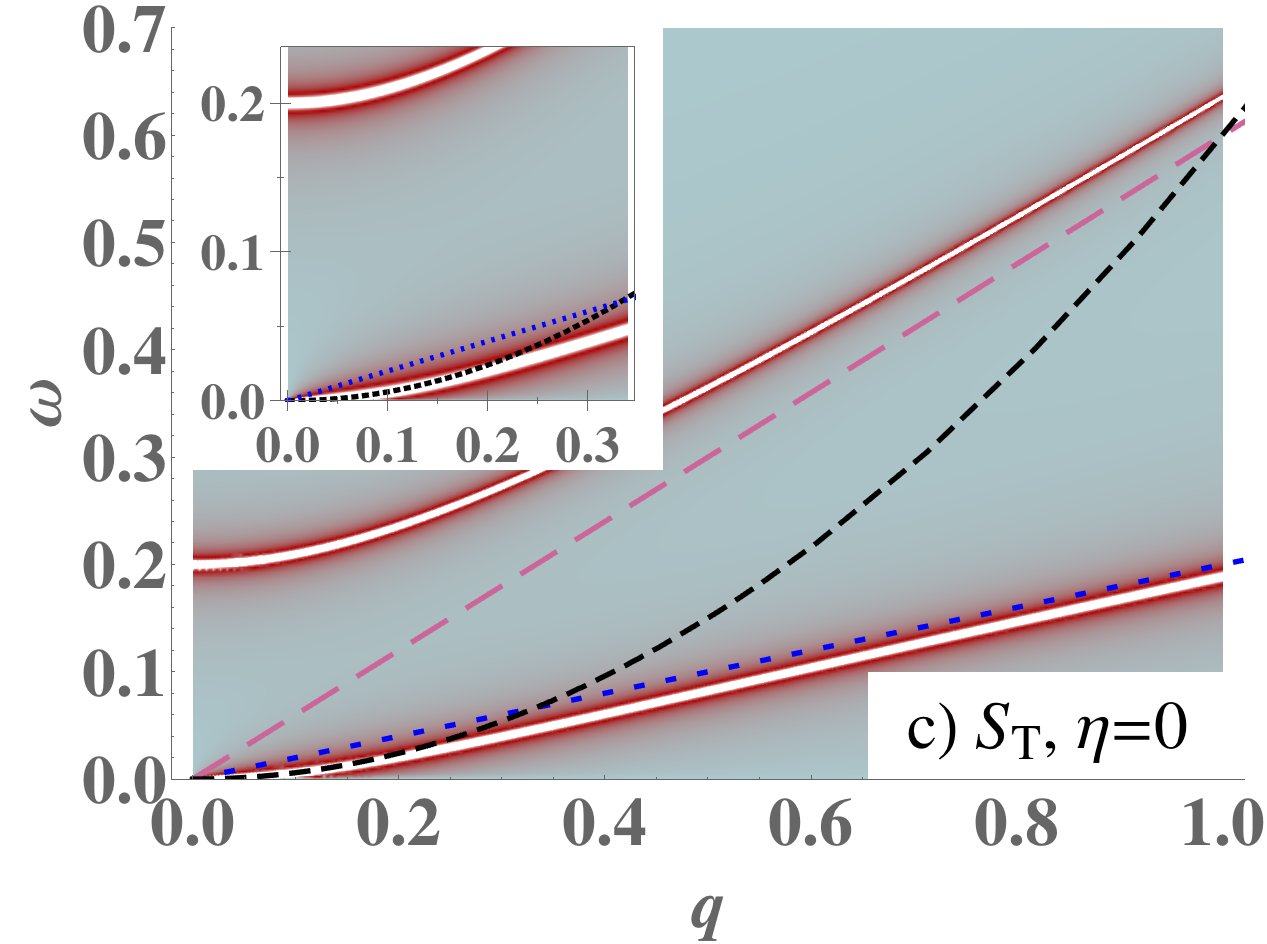}
 \includegraphics[width=7.7cm]{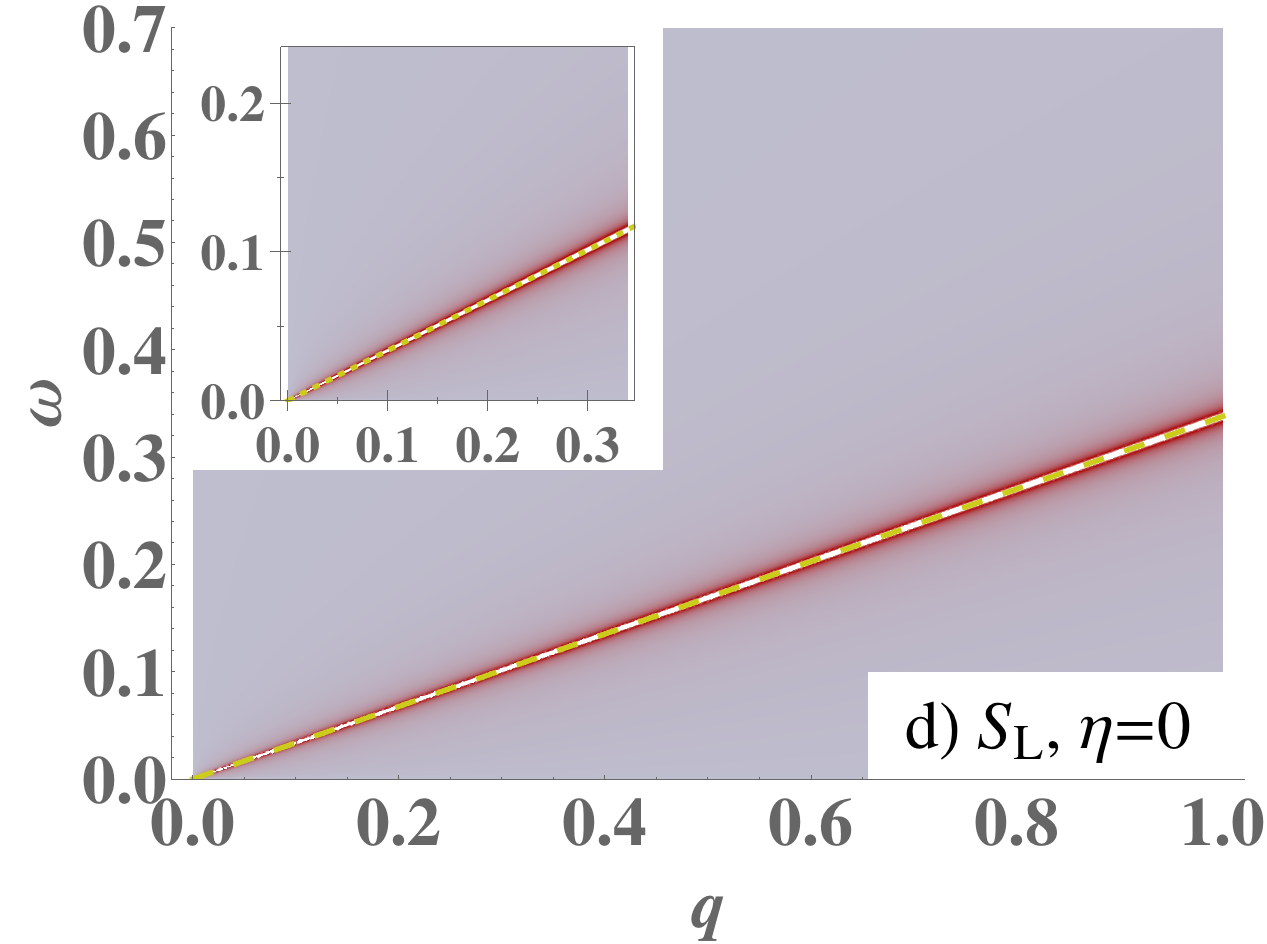}\\
  {\centering
\includegraphics[height=0.7cm]{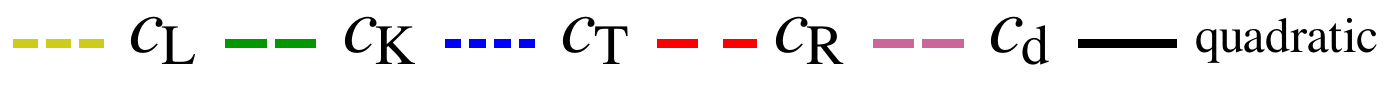}
}\\
  \caption{Spectral functions (left: transverse; right: longitudinal) of the smectic liquid crystal in units of the inverse shear modulus $1/\mu \equiv 1$, for the special angles $\eta = \pi/2,0$ between the momentum $\mathbf{q}$ and condensate Burgers vector $\mathbf{n}$. We have again artificially set $c_\td = 3 c_\tT$ for clarity only. The width of the poles is artificial and denotes the relative pole strengths: these ideal poles are actually infinitely sharp. The insets show zoom-ins near the origin. 
 a,b) $\eta = \pi/2$, momentum along the solid direction. Here we find the longitudinal phonon in the longitudinal sector, whereas the shear stress is completely gapped. c,d) $\eta = 0$, momentum along the liquid direction. In the longitudinal response one finds again a massless mode with the longitudinal phonon velocity $c_\tL$, implying that the shear of the stack of smectic layers still contributes. In the transverse sector we find a massless mode with quadratic dispersion at low momenta, the {\em undulation mode}, while at high momenta it reduces to the transverse phonon. It also features the massive dislocation condensate mode. }\label{fig:smectic spectral functions special angles}
 \end{center}
\end{figure}

\begin{figure} 
\begin{center}
\includegraphics[width=7.7cm]{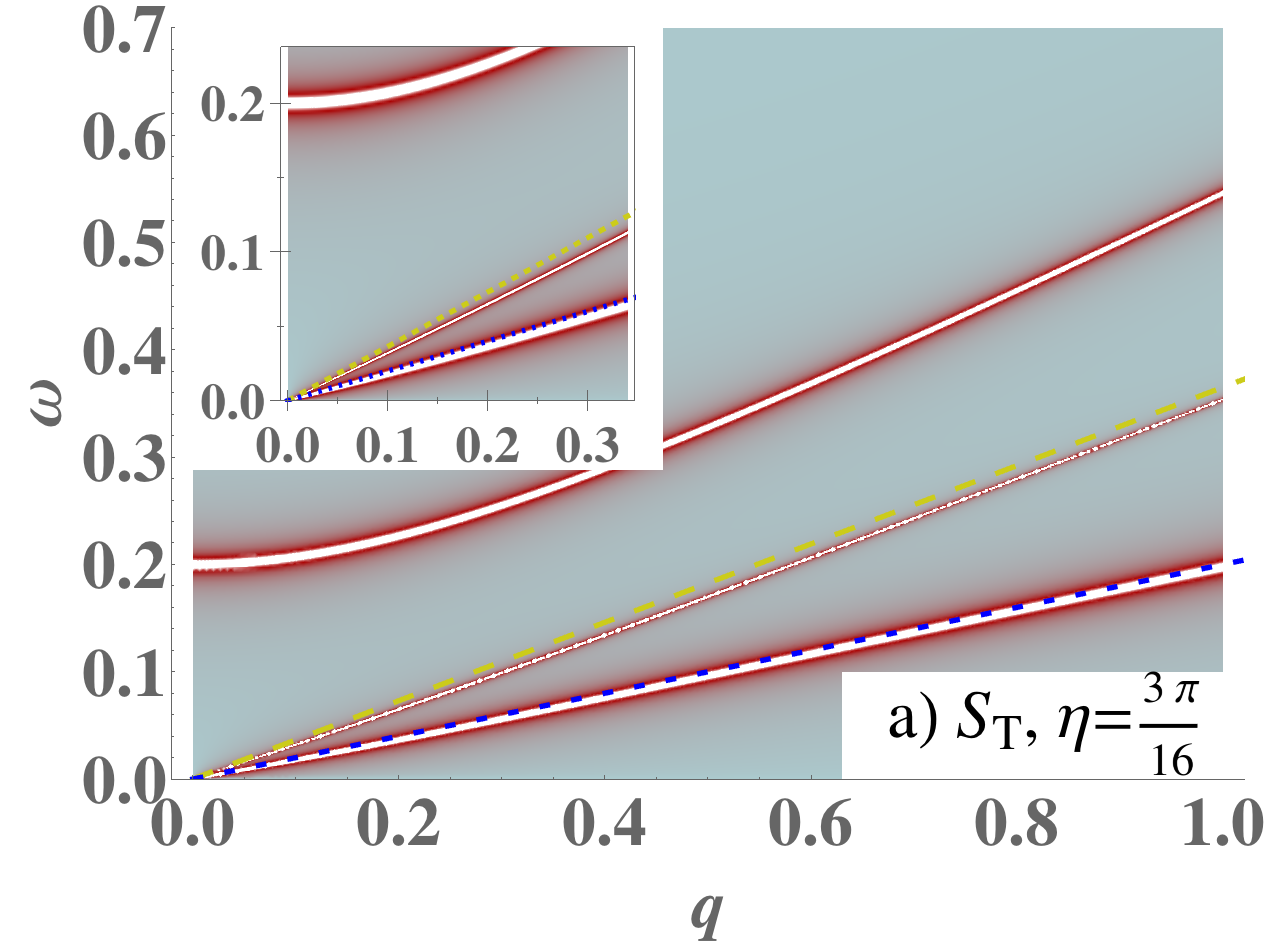}
  \includegraphics[width=7.7cm]{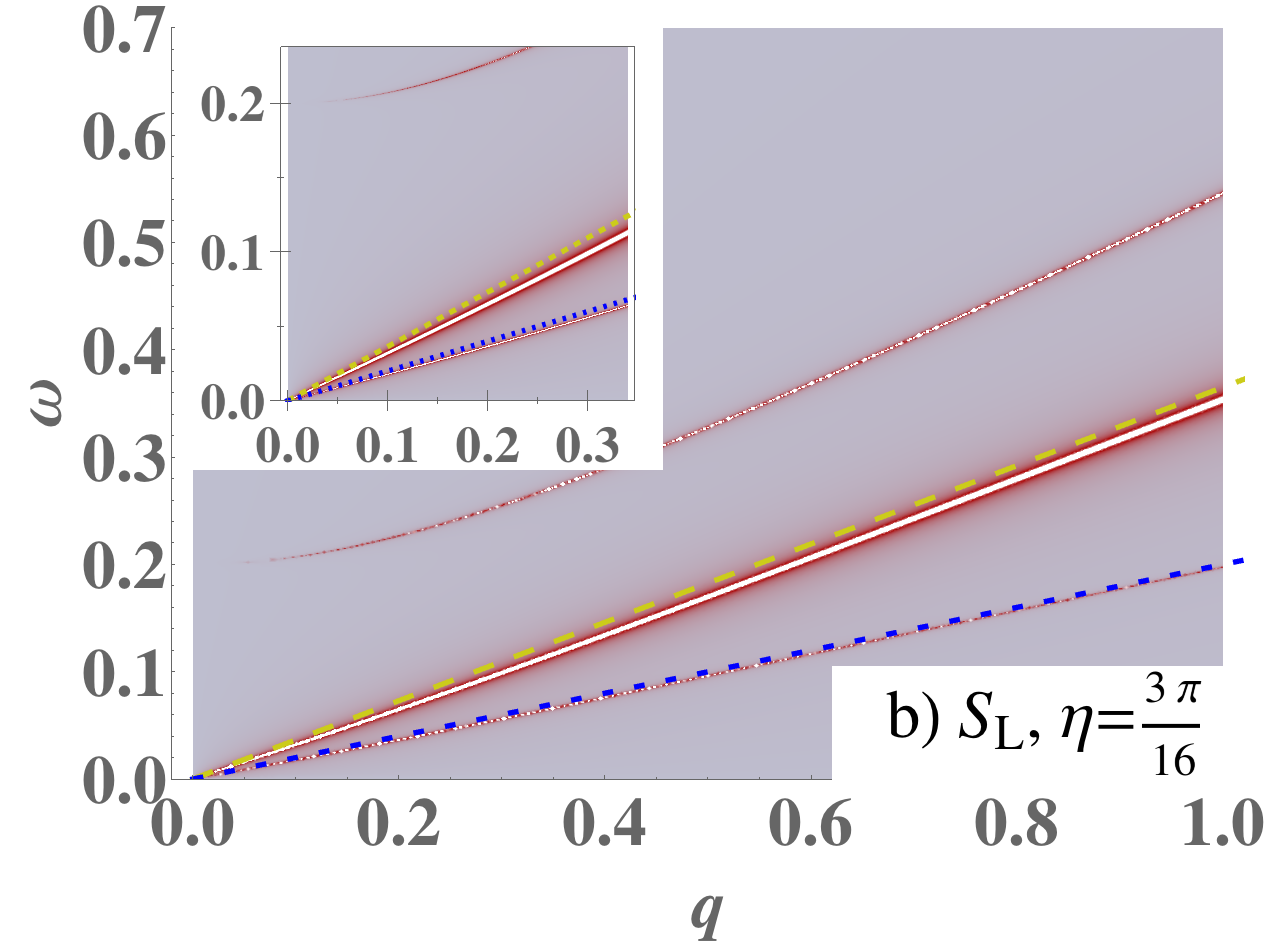}\\
  \includegraphics[width=7.7cm]{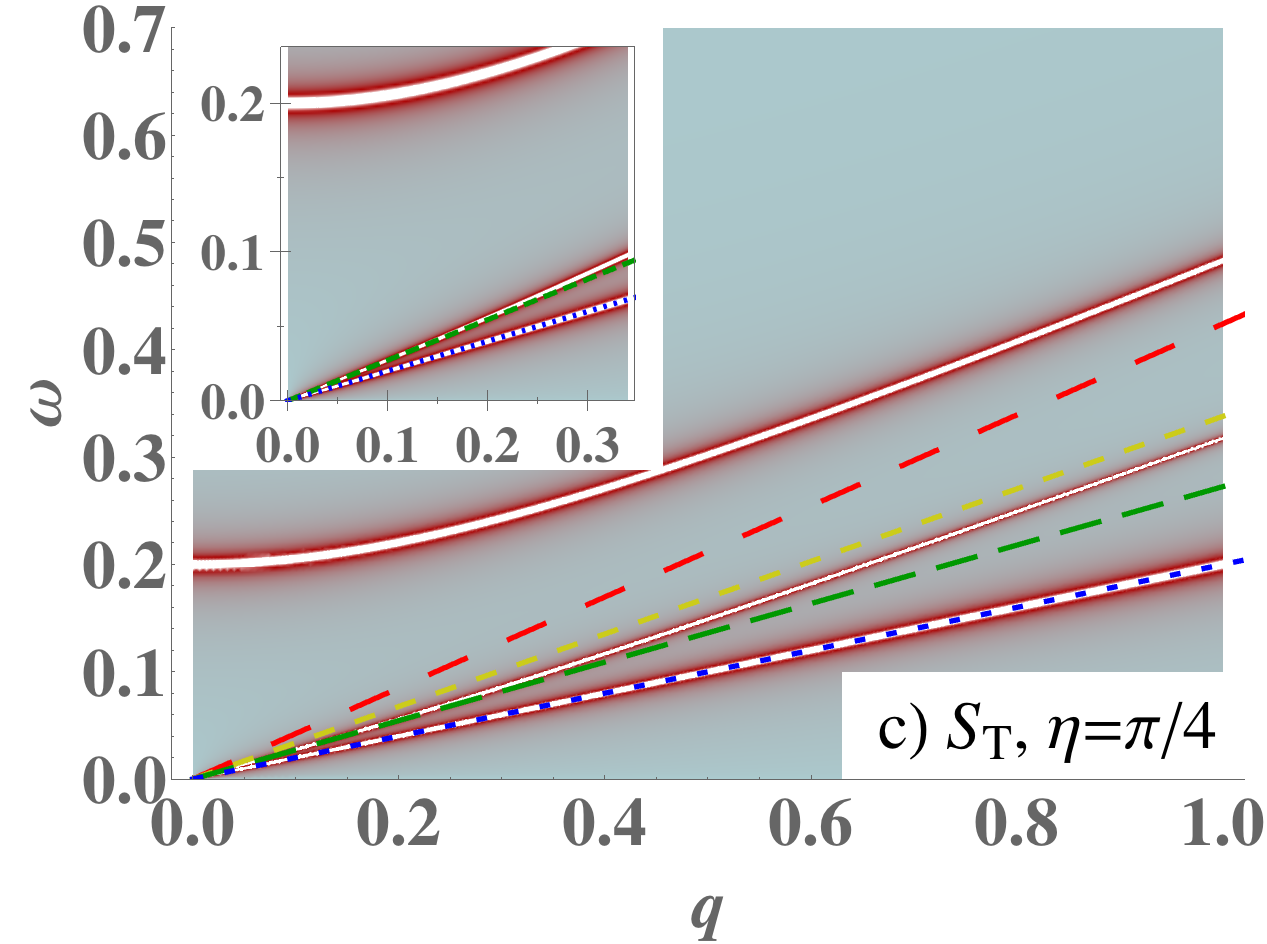}
 \includegraphics[width=7.7cm]{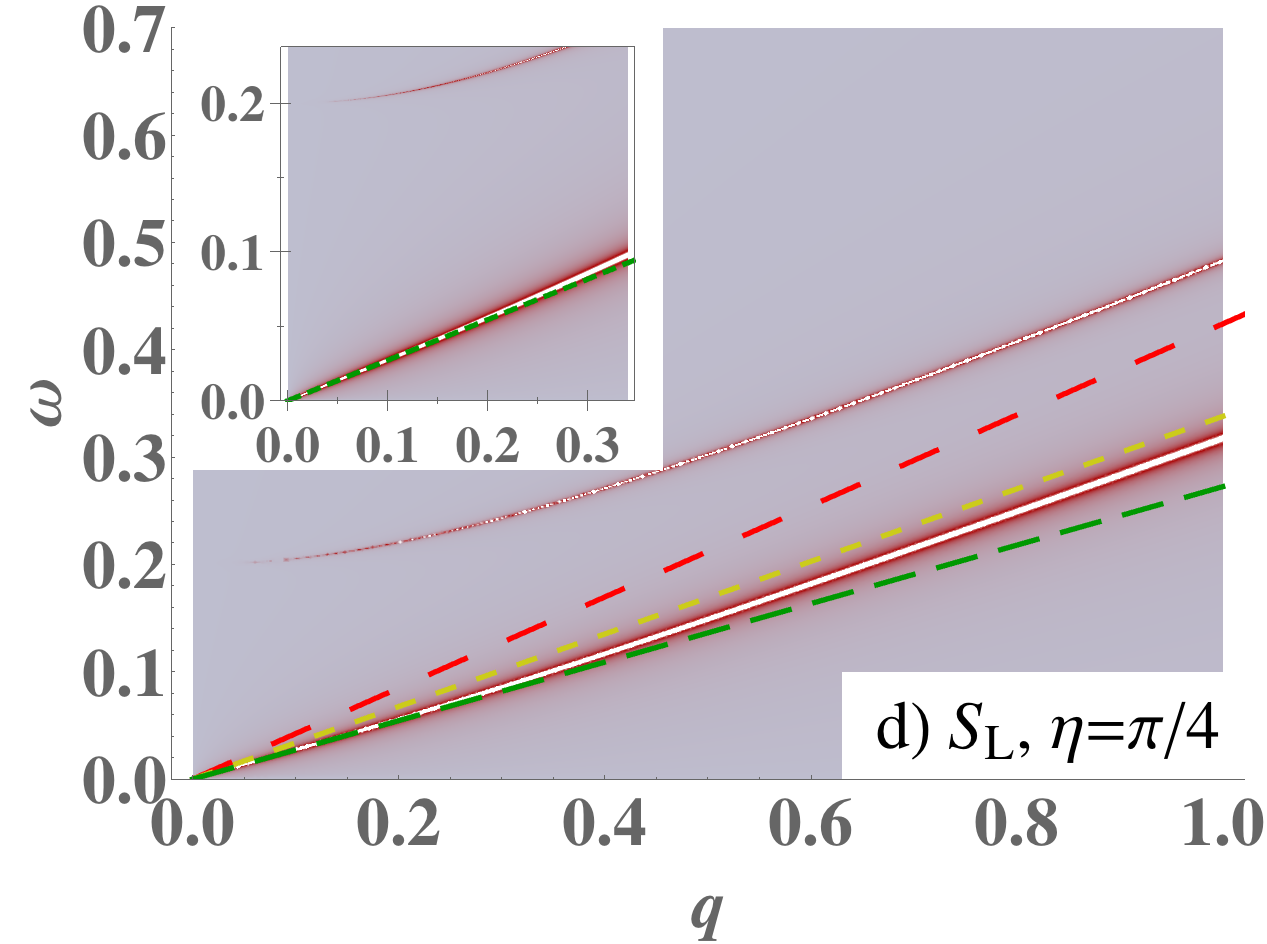}\\  
 {\centering
\includegraphics[height=0.7cm]{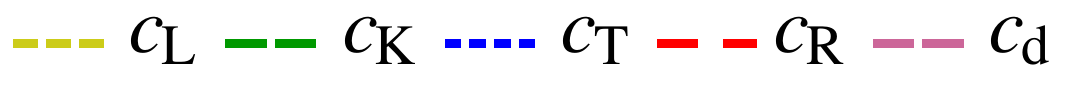}
}\\
  \caption{ Spectral functions  (left: transverse; right: longitudinal) of the smectic liquid crystal. a,b) Here $\eta=3\pi/16$ is taken as a representative angle. Both transverse and longitudinal sectors share the same three modes, the only difference is in the spectral weights. The massive mode originates in the dislocation condensate sound and carries zero spectral weight at $q=0$ in the longitudinal response. The two massless modes extrapolate to the longitudinal resp. transverse phonons at high momenta. At low momenta their velocities follow Eq.~\eqref{eq:smectic massless poles dispersion} and Fig.~\ref{fig:eta velocity}. c,d) $\eta = \pi/4$ `right in the middle' between solid and liquid directions. Here the massless modes obtain the `ideal' velocities $c_\tK$ (purely compressional) and $c_\tT$ (perfect shear). The transverse massless mode vanishes from the longitudinal sector.}\label{fig:smectic spectral functions general angle}
 \end{center}
 \end{figure}

\subsection{Collective modes of the quantum smectic}\label{subsec:Mode content of the quantum smectic}

To get a handle on the special nature of the modes of the quantum smectic let is first zoom in on the high-symmetry directions $\eta =0, \pi/2$~\cite{CvetkovicZaanen06b}. The response in the  solid direction  $\eta = \pi/2$ is quite simple, see Fig.~\ref{fig:smectic spectral functions special angles} a),b).  From Eq.~\eqref{eq:smectic eta 90 longitudinal propagator} one reads off directly that the longitudinal response is governed by a literal longitudinal phonon: there is just a single pole at $\omega = c_\tL q$. However, the transverse mode is now Higgsed: the pole 
is at $\omega = \sqrt {c_\tT^2 q^2 +\Omega^2}$ (ignoring the higher-order corrections) showing that the transverse acoustic phonon of the solid has turned into a `massive transverse stress photon', the simplest manifestation of the dual stress superconductor principle. This is according to the intuition that a smectic is a solid in one, and a liquid in the other direction.  However, this simplicity is deceptive. The quantum nematic 
is a true liquid and there we found that the Higgsing of the shear sector always involves the degrees of freedom of the dislocation condensate itself, signaled by factors of $c_\td$ which are now absent at $\eta = \pi/2$. That the longitudinal mode propagates with the {\em longitudinal} 
phonon velocity $c_\tL = \sqrt{(\kappa + \mu)/\rho}$ is even more striking.  Shear rigidity is impossible dealing with a strictly one-dimensional translational symmetry breaking, since shear requires at least two space directions and a phonon in a 
one-dimensional crystal is therefore always purely compressional with velocity  $c_\tK = \sqrt{\kappa/\rho}$. But the velocity of the smectic longitudinal mode is altered by the shear modulus $\mu$ and in this regard the quantum smectic is still behaving as a two-dimensional solid.    

Within the naive `solid$\times$liquid' intuition, what would one expect when the momentum is directed along the liquid direction? One would anticipate that the longitudinal mode should now be a 
pure compressional sound mode with velocity $c_\tK$, while in the transverse direction it appears that it should `shear the solid':  naively one expects a transverse phonon. However, for $\eta = 0$ this intuition is completely defeated, see Fig.~\ref{fig:smectic spectral functions special angles} c),d). It follows immediately 
from Eq.~\eqref{eq:smectic eta 0 longitudinal propagator} that the longitudinal response is identical to that of a solid, characterized by a single mode propagating with the longitudinal phonon velocity $c_\tL$! Giving 
it a bit of thought this makes sense. Just as for the standard longitudinal phonon, this mode is in first instance associated with compressional motions in the propagation direction and it does not matter whether
in this direction the system is solid or liquid. However, this compressional wave can relax by `expanding' in the transverse direction with the effect that the shear modulus enters its velocity: in the smectic at $\eta = 0$, this transverse direction is the `solid' direction where shear rigidity is still present. Remarkably, the massless longitudinal mode turns into a precise {\em  sound mode} when the propagation direction is precisely in the middle between the liquid and the solid directions: one reads off  from the $\eta = \pi/4$ propagator $G_\tL$ Eq.~\eqref{eq:smectic eta 45 transverse propagator} that it is characterized by a massless
mode propagating with the compressional velocity $c_\tK$, see Fig.~\ref{fig:smectic spectral functions special angles} f).  This is a first signal of  a quite profound phenomenon, as we will find out below.   

Considering the transverse response in the $\eta=0$ direction, one would perhaps expect from the `solid $\times$ liquid' intuition that this correspond to shearing motions in the solid direction,
giving rise to a simple transverse acoustic phonon. But this intuition is completely missing the point. One infers immediately  that the $G_\tT$  propagator in Eq.~\eqref{eq:smectic eta 0 transverse propagator}
has similarities with the transverse propagator of the quantum nematic Eq.~\eqref{eq:nematic transverse propagator}, although it is in detail yet different. Again the dislocation condensate mode enters with velocity $c_\td$ and upon expanding for long wavelengths ($q \ll \Omega / c_\td \equiv 1/\lambda_\td$) one finds a massive mode with real frequency dispersion relation,
\begin{equation}
\omega^2_1 =  \Omega^2 + (c^2_\td + c^2_\tT)q^2 + \cdots.
\label{eq:smectic massive mode eta 0}
\end{equation}
This is quite similar to the `massive shear photon' we found in the transverse response of the quantum nematic, involving a mixing with the condensate mode as 
signaled by the appearance of the $c_\td$ velocity. For the nematic we found that this goes hand in hand with a massless excitation corresponding with the torque mode/rotational 
Goldstone mode, that `acquired rigidity through the dislocation condensate'. In the quantum smectic something else happens, miraculously. Computing the other, massless pole for small momenta from   Eq.~\eqref{eq:smectic eta 0 transverse propagator}  we find the dispersion relation, 
\begin{equation}
 \omega^2_2  =  (\lambda_\td^2 + 4\ell^2) c^2_\tT q^4 + \cdots.
 \label{eq:smectic eta 0 transverse dispersion}
\end{equation}
Here we recall the definition of the dislocation penetration depth $\lambda_\td = c_\td/\Omega$, and keep the second 
order correction $\sim \ell^2$ to make manifest that this is rooted in a similar `rotational logic' as the emergent torque rigidity discussed in the previous section. The 
conclusion is that for a momentum oriented  precisely in the liquid direction $\eta = 0$ the quantum smectic is characterized by a {\em transverse massless mode with a quadratic 
dispersion relation} $\omega \propto q^2$! Upon going to large momenta $q \gg \Omega / c_\td$ one finds that these two modes turn into a transverse phonon and a pure condensate mode (take $\Omega \rightarrow 0$ in Eq.~\eqref{eq:smectic eta 0 transverse propagator}), linearly dispersing with $c_\tT$ and $c_\td$ respectively, in the same way as we found in the transversal response of the nematic. The difference is that both in the $\eta = 0$ and $\pi/2$ cases only a single longitudinal phonon is present regardless the momentum while the condensate mode is absent, see Fig.~\ref{fig:smectic spectral functions special angles}.

Yet again, the dual stress superconductor has accomplished a miracle which is similar to the `resurrection' of the rotational Goldstone boson in the quantum nematic. 
The massless quadratic mode has to be there, although for a universal reason that is specific to smectic order. In the context of  the `molecular perspective'  of 
the classical smectic corresponding with a stack of liquid layers this quadratic mode is canonical: it is well known as the {\em undulation mode}  (see for instance Refs.~\cite{DeGennesProst95,ChaikinLubensky00}). The elastic energy density  of the classical smectic is,
\begin{equation}
  e ({\bf x}) = \tfrac 1 2 \big( B (\nabla_\parallel u)^2 +
  K (\nabla_\perp^2 u )^2 \big). \label{eq:classical smectic energy}
\end{equation}
where $\perp$ refers to the liquid direction and $B$ and $K$ are elastic constants. The leading order gradient term $\sim (\nabla_\perp u )^2$ {\em has} to vanish because global rotations cause a finite modification of $\nabla_\perp u$ while $e({\bf x})$ must be invariant under such transformations. Therefore only the term $\sim  (\nabla_\perp^2 u )^2$ survives, explaining the appearance of the quadratic dispersion. We note that, as stated, the free energy Eq.~\eqref{eq:classical smectic energy} is invariant under global rotations of the smectic direction only to first order, whereas non-linear strains need to be included for full rotational invariance~\cite{ChaikinLubensky00}. The non-linear contributions to fluctuations can be important for the stability of the smectic phase (see Refs.~\cite{GrinsteinPelcovits82, GolubovicWang94, RadzihovskyViswanath09}) but Eq.~\eqref{eq:classical smectic energy} encodes for the general form of the dispersion of the harmonic, linear modes.

Remarkably, via an apparently 
completely different route, the `unidirectional' dislocation condensate has correctly reproduced this universal property of smectic order. Although ultimately caused by rotational symmetry breaking, just like the rotational Goldstone mode  of the quantum nematic, its physical origin is yet very different. The undulation mode is encoding for the fact that the smectic can still support a reactive response to 
shear stresses transverse to the liquid direction, although these are no longer of the usual elastic kind (leading-order gradients in the displacement) because  these cannot be supported by the 
liquid nature of the layers.  However, these liquid layers can support an extrinsic curvature-like reactive response involving  second-order derivatives. 

There is yet another, deeper way to understand how this works.  As we argued in Secs.~\ref{subsec:Dislocations and disclinations},\ref{subsec:Torque stress gauge field}, disclinations are primarily associated with {\em curvature}
which is confined in the solid.  However,  disclinations  deconfine in the dislocation condensate as we emphasized in the previous section and this in turn has the effect that in the dislocation condensate `curvature stress' (torque stress)  becomes deconfined and it can freely propagate as a reactive force. The quadratic mode Eq.~\eqref{eq:smectic eta 0 transverse dispersion} is from this general perspective just
the expression that the dislocation condensate turns the system into a quasi-liquid state that has the capacity to propagate torque stress, and torque stress is related to linear stress via an additional derivative Eq.~\eqref{eq:torque stress definition}.  

Let us now find out what happens at intermediate angles $0 < \eta < \pi/2$ --- the limiting cases $\eta = 0, \pi/2$ are qualitatively different and beforehand it is unclear how the results extrapolate to the general case. Since $\tL$ and $\tT$ no longer refer to a normal mode decomposition away from these limits, these sectors mix.  At $\eta = 0$ we found a total of three modes (1 longitudinal and 2 transverse modes) and this number does not change as function of $\eta$. Therefore,  there is a total of three modes and for  $\eta  \neq 0, \pi/2, \pi/4$ these will show up as  the poles of all propagators $G_\tL,  G_\tT$  or   $G_{\tL\tT}$ since these sectors are just mixed. In Fig.~\ref{fig:smectic spectral functions general angle} a,b) we plot the spectral function for a representative angle $\eta = 3\pi/16$ and for the special angle $\eta = \pi/4$.

\begin{figure}
\begin{center}
 \includegraphics[width=12.7cm]{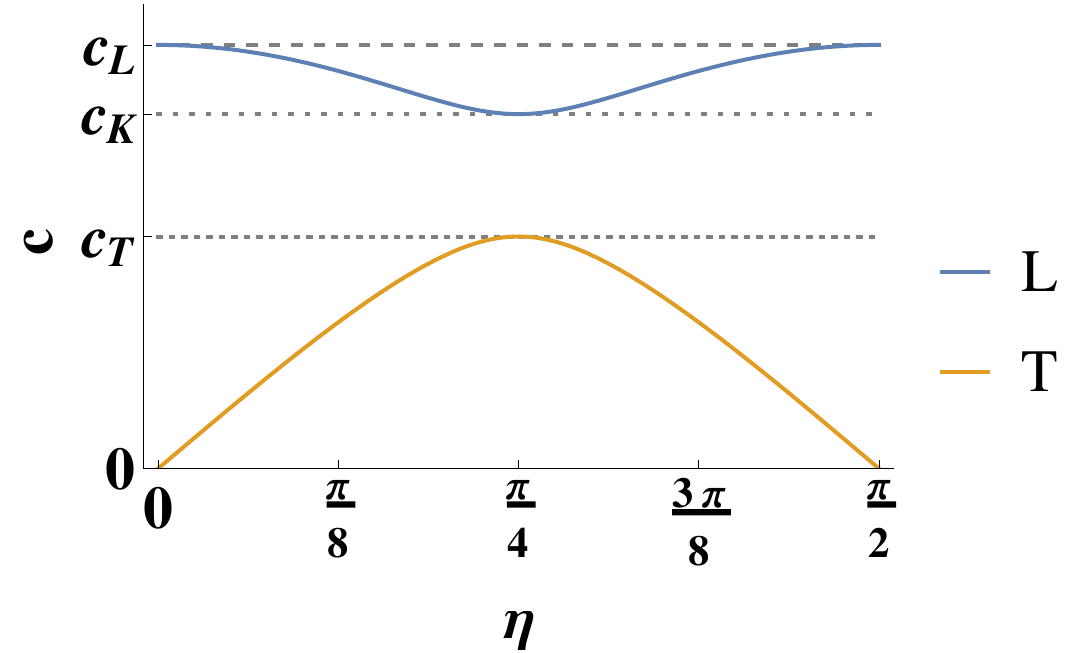}
 \caption{Velocities of the leading terms of the two massless poles of the quantum smectic Eq.~\eqref{eq:smectic massless poles dispersion} as a function of the angle $\eta$ between momentum and condensate Burgers vector, for the Poisson ratio $\nu = 0.4$. We can identify these as the longitudinal and transverse phonons. For $\eta \to \pi/2$ and $\eta \to 0$ the transverse velocity completely softens. At $\eta=\pi/2$ the transverse sound is simply gapped, while at $\eta = 0$ we find instead the quadratic undulation mode Eq.~\eqref{eq:smectic eta 0 transverse dispersion}. For $\eta =\pi/4$, the maximum velocity is reached, which is that of the transverse phonon $c_\tT$. In the longitudinal sector, the velocity varies between $c_\tL$ (same as the solid) and $c_\tK$ (same as the quantum nematic/superfluid).}\label{fig:eta velocity}
 \end{center}
\end{figure}

It is helpful to consider what these spectra look like at long wavelength. To lowest order in $q$ (and real frequency $\omega$) we find the dispersion relations
\begin{align}
 \omega_1 &= \Omega + \frac{ (c_\tT^2 + \cos^2 \eta \, c_\td^2 + \ell^2 \Omega^2) q^2}{2\Omega} + \mathcal{O}(q^4),\\
 \omega_{2,3} &= c_\tL q  \sqrt{  \tfrac{1}{2} \pm \tfrac{1}{2} \sqrt{ \cos^2 2\eta + \nu^2 \sin^2 2 \eta}} + \mathcal{O}(q^3).\label{eq:smectic massless poles dispersion}
\end{align}
The $\omega_1$ mode is obviously associated with the `massive shear photon' showing up in the transverse response at the end points ($\eta = 0,\pi/2$), which is just evolving smoothly with varying $\eta$, changing only its velocity from $\sqrt {c^2_d + c^2_T}$ at $\eta = 0$ to the pure shear velocity $c_T$ at $\eta = \pi/2$. Similarly, the $\omega_2$ mode (plus sign) is 
the continuation of the massless longitudinal mode that we found for $\eta=0,\pi/2$. Its velocity is changing smoothly from $c_\tL$ at $\eta = 0$ to $c_\tK$ at  $\eta = \pi/4$ back to $c_L$ at   $\eta =  \pi/2$ as we already anticipated in  the above
(see Fig.~\ref{fig:eta velocity}). As expected, these two modes can be smoothly interpolated between the two  $\eta$ end points, but the remaining the third mode $\omega_3$ (minus sign) is the most interesting one: for $\eta = 0$ this is the quadratic undulation mode, while it apparently disappears from the $\eta = \pi/2$ spectrum.  What is going on here? 

 The key to understanding is the `right in the middle' $\eta = \pi/4$ propagation direction. We already observed that the massless longitudinal mode turns into a pure compression sound mode at this angle. 
 However, according to  Eq.~\eqref{eq:smectic massless poles dispersion} a second mode has developed which is massless and linear, {\em propagating with precisely the transverse phonon velocity $c_\tT$}!  Surprisingly, at this magic angle the `merger of fluid and solid' reaches Platonic perfection! Upon applying density oscillations in the 
this  direction of the wave propagation the quantum smectic responds as if it were just a run-of-the-mill superfluid. However, upon application of transverse stress it responds as if 
nothing happened to the crystal, exhibiting an impeccable reactive response to shear forces! Notice that this was overlooked in the work ten years ago\cite{CvetkovicZaanen06b,Cvetkovic06}: we discovered it while writing this review. 

This is clearly rooted in an intricate balance of the remnant shear in the quantum smectic. We in fact have not managed to reconstruct this on basis of phenomenological arguments of the kind behind Eq.~\eqref{eq:classical smectic energy} --- it is a typical instance where it is better to rely on the remarkable powers of the duality formalism which computes 
these delicate matters for us. That such a `partitioning' in solid and liquid characteristics can occur at special angles must undoubtedly be rooted in symmetry. Such a phenomenon does not 
occur within the soft matter `liquid layer' perspective: after all, there is no room for magic angles of this kind just taking  Eq.~\eqref{eq:classical smectic energy}.  However, one should realize that these classical constructions are based on  the highly anisotropic `rod-like molecule' perspective. One wires in that the nematic state will be of the uniaxial 
$O(2)/\mathbb{Z}_2$ kind in the classification scheme of Sec.~\ref{sec:Order parameters for 2+1-dimensional nematics}.  The associated solid is of the extreme orthorhombic kind, while the vestigial smectic phase between the uniaxial nematic 
and this solid can be pictured in terms of the stack of liquid layers. The present construction  of the quantum smectic departs however from the maximally isotropic situation that implicitly refers to a hexagonal crystal and a hexatic state: the $O(2)/\mathbb{Z}_6$ class of Sec.~\ref{sec:Order parameters for 2+1-dimensional nematics}.  For the reasons spelled out above, a smectic state can still be realized that at first sight appears to have the same `liquid versus solid' direction appearance as the conventional `liquid layers'. However, it is actually much more symmetric and the effective isotropy in this situation opens up room for the `magic angle' phenomenon. 

One can also view this from the perspective of the present weak--strong duality. It has to be so that when one constructs the `smectic stress-superconductor' departing from the {\em anisotropic} elasticity theory describing the orthorhombic crystal, the propagating shear-like mode will disappear from the spectrum, to be supplemented by the undulation mode also at intermediate angles as implied by Eq.~\eqref{eq:classical smectic energy}. Fundamentally, the trouble 
is that a general symmetry classification scheme for smectic order parameters is still lacking. The classical soft matter literature~\cite{DeGennesProst95,ChaikinLubensky00}
suffers from a `uniaxial tunnel vision'.  In the context of nematic order, this was already recognized in the 1970s by the introduction of the hexatic order, which has been brought to completion (at least in two space dimensions) by the recent developments outlined in Sec.~\ref{sec:Order parameters for 2+1-dimensional nematics}.  In order to keep the
mathematics simple, we just described in this section a smectic characterized by maximal spatial isotropy: it is the `smectic precursor to the hexatic order'. It behaves very differently from the `uniaxial smectic' of conventional soft matter physics, which is no wonder because symmetry-wise they are at opposite sides of the spectrum. Presently, the full order parameter theory of superfluid smectic order yielding symmetry-based predictions for the spectrum of Goldstone modes is an open problem that remains as a challenge. 

We have not yet pinned down precisely what happens to the `transverse-like' massless mode $\omega_3$ at arbitrary angles $\eta$. To recall, we found at $\eta = 0$ the quadratic undulation mode, it disappears completely at $\eta = \pi/2$, but it reappears as a full-fledged transverse phonon at $\eta = \pi/4$. How to smoothly interpolate between these different cases? The answer can be read off from Eq.~\eqref{eq:smectic massless poles dispersion}:
for any $\eta \neq 0, \pi/2$ this transverse-like mode acquires a {\em finite} velocity $c^{*}_\tT(\eta)$ that depends on the observation angle $\eta$, see Figs.~\ref{fig:eta velocity} and \ref{fig:smectic spectral functions general angle}. In fact, expanding the dispersion relation of the transverse-like pole for both small $q$ and small $\eta$, and $\ell =0$, the lowest order terms read
\begin{align}
 \omega_3 = c_\tT q \sqrt{\frac{1 - \sqrt{\cos^2 2 \eta + \nu^2 \sin^2 2\eta}}{1 - \nu} + \frac{c_\td^2}{\Omega^2}q^2}+\cdots
\end{align}
For small but finite $\eta$ this gives the linear dispersion Eq.~\eqref{eq:smectic massless poles dispersion} to lowest order in $q$, but for $\eta = 0$ it gives the quadratic dispersion Eq.~\eqref{eq:smectic eta 0 transverse dispersion} of the undulation mode. For $\eta >0$ the transverse oscillation picks up the remnant shear rigidity of the isotropic smectic that reaches its maximum at $\eta = \pi/4$. Increasing $\eta$ further
this velocity goes down again to vanish again precisely at $\eta = \pi/2$. But according to Eq.~\ref{eq:smectic eta 90 transverse propagator}) and Fig.~\ref{fig:smectic spectral functions special angles}a)
this mode has completely disappeared from the spectrum! The reason for this can be deduced from the propagators, by zooming in on  the limit $\eta \rightarrow \pi/2$. The would-be transverse pole is canceled by an identical term in the numerator exactly at $\eta = \pi/2$.  For $\eta < \pi/2$ this cancellation  is incomplete, such that a massless pole emerges with diminishing velocity as $\eta \rightarrow \pi/2$. The physical interpretation is in fact obvious. The transverse-like velocity $c^*_\tT = \sqrt { \mu^* (\eta) / \rho}$ (cf. Eq.~\eqref{eq:transverse phonon velocity}) where $\rho$ is the mass density 
which of course is constant, while $\mu^*(\eta)$ is an effective shear modulus associated with the transverse mode propagating  at the angle $\eta$. This effective modulus
{\em vanishes} precisely at $\pi/2$: the system {\em can no longer propagate shear stresses} and therefore there is no longer a corresponding Goldstone boson! Different from the liquid direction $\eta =0$ also the curvature rigidity that is responsible for the undulation mode does not act when the system is transversely oscillates in the solid direction and therefore there is no longer a propagating massless mode at all!

Finally, we can investigate the three poles at high momenta $q \gg 1 / \lambda_\td$. Here we find that the three modes acquire a linear dispersion with velocities $c_\tL$, $c_\tT$ and $\cos \eta\; c_\td$ respectively. When $\cos \eta\; c_\td < c_\tT$, there is a mode coupling between the massive and the massless mode, such that the massive mode extrapolates continuously to the transverse phonon, which amongst others accounts for the behavior we encountered for $\eta = \pi/2$. As was the case for the quantum nematic, the spectral weight of the pole with velocity $\cos \eta\; c_\td$ vanishes in the limit $q \lambda_\td \to \infty$, such that we recover the response of the crystalline solid with longitudinal and transverse phonons only.

We have reached the end of this exhibition of the amazingly rich landscape formed by the elastic properties of the {\em isotropic quantum smectics}. But this not yet the end of the whole story; besides its solid-like properties highlighted in this section this quantum smectic is also a superfluid of a quite unusual kind. A convenient way to address the nature of this superfluidity is to couple in electromagnetic charge to find out how it behaves as a superconductor when it is interrogated by external electromagnetic fields. This is the subject of the next two sections.      
 
 \section{Elasticity and electromagnetism: the formalism}\label{sec:Dual elasticity of charged media}
 Up to this point we have focused on electrically neutral matter. To complete the story, in this section we will discuss {\em charged} matter, coupling elasticity to electromagnetic fields. In part this is done for practical reasons: extracting information from the quantum elasticity that is in principle experimentally available, as to be discussed in Sec.~\ref{sec:Electromagnetic observables}. For quantum elasticity, we need forms of matter that are highly quantal, capable of avoiding full crystallization at zero temperature.  In earthly laboratories the only form of neutral matter that is available is formed out of atoms. There is hope that the cold-atom community might eventually get so much control that they can access the regime required to form the kind of quantum liquid crystals that are the subject of this review. On the other hand, the questions that formed the initial motivation for the subject of this review arose in the field of strongly-interacting electrons in solids. As we highlighted in the introduction, could it be that the superconductors in underdoped cuprates are characterized by very strong solid-like (``stripy'') correlations, which are also behind the nematic properties that appear to have been observed? If this is the case, how to measure this in the laboratory? In practice, the only way to interrogate electrons is by exploiting the fact that they interact with external electromagnetic fields. We have therefore to find out how the physics discussed up to this point imprints on the electromagnetic response. In fact, it will become clear that the `fluctuating solid order' will give rise to unambiguous fingerprints in specific electromagnetic responses, which are however not so easy to measure. 

There is also a good theoretical reason to consider the charged versions of quantum nematics and smectics. As we found out in the previous sections, the `liquid' aspect of the quantum liquid crystals we are dealing with refers to superfluidity. To interrogate the nature of this superfluidity it is just very convenient to couple in electromagnetic fields, turning them into superconductors, and study the electromagnetic responses which directly reflect the specific rigidities associated with their superfluid side. It will turn out that the marriage between our `generalized elasticity' and electromagnetism is a quite happy one. We will demonstrate that the whole portfolio of electromagnetic responses such as optical conductivities, dielectric functions, polariton spectra and the Meissner effect involve rather effortless computations resting on the ground work we have laid out 
in the previous sections.  

In this section we will just derive the equations governing the electromagnetic response of the solid and the quantum nematic/smectic, analyzing them in further detail in the next section. 
In fact, the only hurdle that amusingly takes some effort is to find out how to deal with the electromagnetism of charged `bosonic Wigner crystals', in the field theoretical formulation of elasticity we have used throughout this review. It should be emphasized that we consider strictly 2+1 dimensions where not only the matter but also the EM fields are confined to the plane. We avoid in this way the unnecessary complications
that one is facing dealing with the laboratory two-dimensional systems such as graphene where the 2+1D matter interacts with 3+1D EM fields. 

The study of charged elastic systems is of course a rather classic subject~\cite{Toupin63} but it appeared that in the early stages of the current research program the wheel had to be reinvented. Ref.~\cite{ZaanenNussinovMukhin04} introduced the formalism that incorporates electromagnetism with the stress-gauge field formulation of elasticity. We will review this here, climaxing in a highly economical --- and to a degree surprising --- description of the electrodynamics of Wigner crystals in the stress-photon description. The key is that the stress photons (dual to the phonons) are the natural partners of the electromagnetic photons and the theory becomes much more transparent than in the conventional formulations that keep the matter on the strain side of the 
stress--strain duality. We derive the effective action governing the electromagnetism by integrating out the stress photons, which is then rather effortlessly lifted to the description of the quantum nematic and quantum smectic states by just augmenting the crystal with the dislocation condensates.  

\subsection{Preliminaries: electromagnetic fields and crystals}

Let us first find out how to incorporate Maxwell theory in the theory of elasticity. This is straightforward but we have to pay some care  dealing with the non-relativistic nature of the crystal, as well as the theory 2+1D electromagnetism in the Euclidean signature.  

Let us depart from standard Maxwell theory in real time $t$, with its
the electric field $E_m$ and magnetic field $B$, 
\begin{align}\label{eq:electric magnetic field in potentials}
 E_m &= - \partial_m V - \partial_t A_m, & 
 B &= \epsilon_{mn} \partial_m A_n,
\end{align}
in terms of the scalar potential $V$ and vector potential $A_m$.  In two space dimensions the Maxwell action is,
\begin{align}
 \mathcal{S}_\mathrm{Maxw} &= \int \td^2 x\, \td t  \left[  \frac{1}{2 \mu_0} ( \tfrac{1}{c_l^2} E_m^2 - B^2) - V \rho_Q + A_m j_m \right] 
 = \int \td^2 x\, \td t  \left[ -\frac{1}{4 \mu_0} F_{\mu\nu} F^{\mu\nu} + A_\mu j^\mu  \right]  \label{eq:real time EM action},
\end{align}
where $\rho_{Q}$ is the charge density, $j_m$ the electric current density, $\mu_0$ is the magnetic constant and $c_l$ is the speed of light. 
We will also use the dielectric constant $\varepsilon_0 = 1/\mu_0 c_l^2$. Furthermore, we have defined the three-potential $A^\mu = (\frac{1}{c_l}V, A_m)$, 
the three-current $j^\mu = (c_l \rho , j_m)$ and the  Maxwell tensor $F_{\mu\nu} = \partial^l_\mu A_\nu - \partial^l_\nu A_\mu$, where the three-derivative is $\partial^l_\mu = (\frac{1}{c_l} \partial_t , \partial_m)$ .
We use the mostly-plus Minkowski metric $\eta_{\mu\nu} = \mathrm{diag}(-1,+1,+1)$ throughout. 

In two spatial dimensions, the units of the electromagnetic constants differ from 3D as $\mu_0 \equiv  \mu_{0}^{\mathrm{3D}}/[\textrm{length}]$ with units  $[\mu_0] =\frac{\mathrm{J}\,\mathrm{s}^2}{\mathrm{C}^2\,\mathrm{m}^2}$. The dielectric constant changes to $\varepsilon_0 \equiv [\textrm{length}] \varepsilon_0^{\mathrm{3D}}$ with units $[\varepsilon_0] = \frac{\mathrm{C}^2 }{\mathrm{J}}$. 

To avoid potential confusion, let us pay some care to the Wick rotation to imaginary time. 
 We use $\tau = \ti t$ such that $\partial_\tau = -\ti \partial_t$; all covariant tensors with temporal components undergo this coordinate transformation.  It is useful to keep track of this
 defining for instance $V^\imath = - \ti V$, $E^\imath_m = -\ti E_m$ and $\rho^\imath_Q = -\ti \rho_Q$ (see \ref{sec:Euclidean electromagnetism conventions}).
This results in the Euclidean Maxwell action $S_\mathrm{E}$, where we shall always use covariant (lower) indices referring to spacetime directions, while upper indices will be reserved for the 
Burgers flavors:
\begin{align}\label{eq:Euclidean Maxwell action}
 &\mathcal{S}_\mathrm{E,Maxw} = \int \td \tau \; \td^2 x\;  \left[  \frac{1}{4 \mu_0} F^\imath_{\mu\nu} F^\imath_{\mu\nu} - A^{\imath,l}_\mu j^{\imath,l}_\mu \right] 
 = \int \td \tau \; \td^2 x\; \left[ \frac{1}{2 \mu_0} ( \tfrac{1}{c_l^2} E_m^{\imath 2} + B^2) - V^\imath \rho^\imath_Q -A_m j_m \right].
\end{align}
The definition of the electric current follows immediately,
\begin{equation}\label{eq:electric current from action}
 j^{\imath}_\mu = (\rho^{\imath}_Q, j_m) = - \frac{\partial \mathcal{S}_\mathrm{E}}{\partial A_{\mu}}.
\end{equation}
A new, and crucially important ingredient is that we are now dealing with two velocity scales that are very different. Besides the light velocity $c_l$ introduced above for the Maxwell action, the material system 
is characterized by the phonon velocity that we take as before in units of the shear velocity  $c_\tT$. We will compute matters in terms of the phonon velocity such  that $\partial_\ft = \frac{1}{c_\tT} \partial_\tau$,
and this forces us to rescale the velocity factors in the electrodynamics in the following way (in contrast to the quantities with the label $^l$ above), 
\begin{align}
 A^\imath_\ft &= - \tfrac{1}{c_\tT} V^\imath ,& A^\imath_\mu &= (-\tfrac{1}{c_\tT} V^\imath , A_m) ,\\
 j^\imath_\ft & = - c_\tT \rho^\imath ,& j^\imath_\mu &= (-c_\tT \rho^\imath  , j_m) ,\\
E^\imath_m &= c_\tT ( \partial_m A^\imath_\ft - \partial_\ft A_m), \label{eq:imaginary time electric field}
\end{align}
and the  Maxwell action becomes,
\begin{align}
 \mathcal{S}_\mathrm{E,Maxw} 
&= \int \td \tau \; \td^2 x\; \frac{1}{2 \mu_0} \Big( \tfrac{c_\tT^2}{c_l^2} ( \partial_m A^\imath_\ft -\partial_\ft A_m)^2 + (\epsilon_{mn} \partial_m A_n)^2 \Big)  - A^\imath_\mu j^\imath_\mu.
\label{eq:Euclidean action EM fields}
\end{align}

Having set the various definitions, let us now spell out, following Ref.~\cite{ZaanenNussinovMukhin04}, how the electromagnetic fields couple to the strain fields of the elastic solid of Sec.~\ref{sec:Quantum elasticity}.
The standard  (real time) action  of a  system of $N$ non-relativistic microscopic particles of charge $e^*$ with trajectories $\mathbf{R}_{i}(t)$ in an external potential $A_\mu(x)$ without any retardation effects is~\cite{Jackson99}:
\begin{equation}
 \mathcal{S}_{\rm EM} =  \sum_{i}^N \int \td t\; \Big\{ - e^* V \big(\mathbf{R}_{i}(t)\big) + e^* \big(\partial_t \mathbf{R}_i(t)\big) \cdot \mathbf{A} \big( \mathbf{R}(t) \big) \Big\}.
 \label{eq:jacksonpart}
\end{equation}
In elasticity, we consider a macroscopic state of such particles forming a periodic lattice with fluctuations: $\mathbf{R}_{i}(t)\simeq \mathbf{R}(x,t) = \mathbf{R}_0(x) + \mathbf{u}(x,t)$, where $\mathbf{R}_0$ are 
the equilibrium positions that do not vary in time and $\mathbf{u}(x,t)$ are small displacements; as in Sec. \ref{sec:Quantum elasticity} 
we have taken the continuum limit with $x$ denoting the spatial coordinates. The continuum version follows immediately from Eq.~\eqref{eq:jacksonpart}
\begin{align}
 \mathcal{S}_{\rm EM} =  n e^*  \int \td t\; \td^D x\;\Big\{-V \big(\mathbf{R}(x,t)\big) 
 +\big(\partial_t \mathbf{u}(x,t)\big) \cdot \mathbf{A} \big( \mathbf{R}(x,t) \big)\Big\},
\end{align}
where $n=N/\mathcal{V}$ is the particle density and $\mathcal{V}$ the volume of the system. We now exploit the fact that the displacements $\mathbf{u}$ are small compared to the lattice and we can Taylor-expand $V(\mathbf{R}_0 + \mathbf{u}) = V(\mathbf{R}_0) + \mathbf{u} \cdot \nabla V (\mathbf{R}_0) + \mathcal{O}(\mathbf{u}^2)$, keeping only the  term linear in $\mathbf{u}$. After a partial integration,
\begin{equation}\label{eq:EM coupling real time}
 \mathcal{S}_{\rm EM} =  n e^*  \int \td t\; \td^D x\; \Big\{ V (\nabla \cdot \mathbf{u})  +  \big(\partial_t \mathbf{u} \big) \cdot \mathbf{A} \Big\}.
\end{equation}
Comparing this with the coupling term $\int - V \rho_Q + A_m j_m$ from Eq.~\eqref{eq:real time EM action},
we identify the deformation charge density and current as $\rho_Q = -n e^* \nabla \cdot \mathbf{u}$ and $\mathbf{j} = n e^* \partial_t \mathbf{u}$, according to expectation. 

Upon performing the Wick rotation to imaginary time,  $\mathcal{S} = \ti \mathcal{S}_\mathrm{E}$, $V = - \ti c_\tT A^\imath_\ft$, $\rho_Q = -\frac{\ti}{c_\tT} j^\imath_\ft$, $\partial_t = \ti c_\tT \partial_\ft$, the 
Euclidean action becomes 
\begin{align}\label{eq:basicEMstrains}
 \mathcal{S}_\mathrm{E,EM} &= -\ti n e^* c_\tT  \int \td \tau \; \td^D x\; \Big\{ -  A^\imath_\ft  (\nabla \cdot \mathbf{u})  +  \big(\partial_\ft \mathbf{u} \big) \cdot \mathbf{A} \Big\} 
 = -\int \td\tau \; \td^D x\; j^{\imath}_{\mu}A_{\mu},
\end{align}
with
\begin{align}\label{eq:deformation current}
 j^\imath_\ft &= -\ti n e^* c_\tT \nabla \cdot \mathbf{u},&
 \mathbf{j}  &= +\ti n e^* c_\tT \partial_\ft \mathbf{u}.
\end{align}
The coupling of elasticity to electromagnetism is completely captured by Eq.~\eqref{eq:basicEMstrains}, where we just rested on the physical EM law~\eqref{eq:jacksonpart} and the fact that the displacement fields are small (the governing principle of continuum elasticity). 

The total theory is defined by adding the coupling Eq.~\eqref{eq:basicEMstrains} to the Maxwell action Eq.~\eqref{eq:Euclidean action EM fields} and the strain action Eq.~\eqref{eq:elasticity relativistic Lagrangian} of elasticity. It will turn out to be convenient to define~\cite{ZaanenNussinovMukhin04}
\begin{equation}
 \mathcal{A}_\mu^a = n e^*c_\tT ( \delta_{\mu a} A^\imath_\ft - \delta_{\mu \ft} A^\imath_a ), \label{eq:curlyA}
\end{equation}
and the action Eq.~\eqref{eq:basicEMstrains} reduces to the form
\begin{equation}\label{eq:displacement EM coupling}
  \mathcal{S}_\mathrm{E,EM} =  \int \td \tau\; \td^D x\;  \ti  \mathcal{A}_\mu^a \partial_\mu u^a 
\end{equation}
explicitly featuring the strains $\doo_{\mu}u^a$. Finally, notice that there is a natural relation between the gauge invariance of the electromagnetic potentials and the glide principle. 
The conservation law for the deformation current $\partial_\mu j_\mu = 0$ is tied to the gauge invariance of the electromagnetic field $A_\mu \to A_\mu + \partial_\mu \varepsilon$. 
Using Eq.~\eqref{eq:deformation current} and the definition of the dislocation current Eq.~\eqref{eq:dislocation density multivalued displacement}
 $J^a_\mu = \epsilon_{\mu\nu\lambda} \partial_\nu \partial_\lambda u^a$, it follows immediately that $\partial_\mu j_\mu = 0$ is coincident with the glide constraint $\epsilon_{ba} J^a_b = 0$.
Therefore, gauge invariance of the electromagnetic field must imply the validity of the glide constraint. Given that the glide constraint is a consequence of the conservation of particle number, 
 it has to be reflected in the conservation of electric charge that is responsible for the $U(1)$ gauge symmetries of $A_\mu$.

\subsection{Stress--strain duality and electromagnetism}

The photons of electromagnetism encode of the capacity of the fundamental vacuum to propagate electromagnetic forces. In the previous sections we learned to appreciate that it is quite convenient to treat elastic media in a very similar way, in terms of the stress photons that reveal the capacity of the elastic medium to propagate elastic forces. To describe the charged crystal one could of course stick the more familiar strain side of the stress--strain duality, to depart from Eq.~\eqref{eq:deformation current} and compute the way that electromagnetism modifies the elasticity. However, since EM photons and stress photons are  close siblings a much more convenient and transparent formalism is obtained in the dual stress language.  To the best of our knowledge this formalism was discovered for the first time in Ref.~\cite{ZaanenNussinovMukhin04}. Even for the simple Wigner crystal it reveals surprising features, such as that the ``crystal is on the verge of becoming a superconductor''.  

The plan is easy: starting with the coupling Eq.~\eqref{eq:displacement EM coupling} between strains and electromagnetic potentials we have just to  `carry the photon field through the stress-strain duality transformation' following the steps in Sec.~\ref{subsec:Stress gauge fields}. Adding the coupling  Eq.~\eqref{eq:displacement EM coupling} to the elasticity action of the solid Eq.~\eqref{eq:elasticity relativistic Lagrangian} gives, 
 \begin{equation}\label{eq:charged strain Lagrangian}
 \mathcal{L} = \frac{1}{2} \partial_\mu u^a C_{\mu\nu ab} \partial_\nu u^b + \ti \mathcal{A}^a_\mu \partial_\mu u^a + \frac{1}{4\mu_0} F^\imath_{\mu\nu} F^\imath_{\mu\nu}.
\end{equation}
Here $\partial_\mu = (\frac{1}{c_\tT} \partial_\tau, \partial_m) = (\partial_\ft ,\partial_m)$. This leads to a modification in the canonical momenta and the stress tensor becomes,
\begin{align}
 \sigma^a_\ft &= -\ti \frac{\partial \mathcal{S}}{\partial (\partial_\ft u^a)} = -\ti \mu  \partial_\ft u^a +    \mathcal{A}^a_\ft, \label{eq:EM stress tensor temporal}\\
 \sigma^a_m &= -\ti \frac{\partial \mathcal{S}}{\partial (\partial_m u^a)} = -\ti C_{mnab} \partial_n u^b +  \mathcal{A}^a_m. \label{eq:EM stress tensor spatial}
\end{align}
This definition turns $\sigma^a_{\mu}$ into a gauge covariant object, while the gauge transformations of $\sigma^a_{\mu}$ follow from those of Eq.~\eqref{eq:curlyA}. 

Using the spin-projectors Eqs.~\eqref{eq:spin-0 projector}--\eqref{eq:spin-2 projector}, in addition to the temporal component $\mathcal{A}^a_\ft =  -n e^* c_\tT A^\imath_a$, we find that only the spin-0 components of strains are influenced by the electromagnetic field, 
\begin{align}
  P^{(0)}_{mnab} \mathcal{A}^b_n &= \mathcal{A}^a_m = n e^* c_\tT \delta_{ma} A^\imath_\ft , \\ 
  P^{(1)}_{mnab} \mathcal{A}^b_n &= 0, \\
  P^{(2)}_{mnab} \mathcal{A}^b_n &=  0.
\end{align}
Using Eq.~\eqref{eq:isotropic solid inverse stress tensor} from the inverse elastic tensor, the inverse relations are:
\begin{align}
 \partial_\ft u^a &= \ti \frac{1}{\mu} ( \sigma^a_\ft -  \mathcal{A}^a_\ft), \label{eq:charged strain stress temporal}\\
 \partial_m u^a &= \ti C^{-1}_{mnab} ( \sigma^b_n - \mathcal{A}^b_n).
\end{align}
We can now repeat the derivation of the dual Lagrangian from Sec.~\ref{subsec:Stress gauge fields} with the outcome,
\begin{align}
 \mathcal{L}_\mathrm{dual} &= \tfrac 1 2 \sigma_\mu^{a} C^{-1}_{\mu \nu a b} \sigma_\nu^b + \tfrac 1 2 \mathcal{A}_m^{a} C^{-1}_{mn a b} \mathcal{A}_n^b + \tfrac{1}{2}  \frac{1}{\mu}  (\mathcal{A}_\ft^a)^2  \nonumber\\
&\phantom{mm}
 - \sigma^a_m  C^{-1}_{mn a b} \mathcal{A}_n^b - \frac{1}{\mu}\sigma^a_\ft \mathcal{A}^a_\ft 
 + \ti \sigma_\mu^{a} \partial_\mu u^a 
 +  \frac{1}{4 \mu_0} F^\imath_{\mu\nu} F^\imath_{\mu\nu} + \mathcal{L}_\mathrm{constraints}.\label{eq:dual elasticity EM Lagrangian}
\end{align}
The glide and Ehrenfest constraints are left  here implicit and they are treated as in the charge-neutral case. 
This is a very elementary result that appears to be missing from the textbooks: it is just the action describing the charged crystal in terms of the stress tensors $\sigma^a_\mu$ and the electromagnetic field $A_\mu$. 

It has surprising features, foremost a Meissner-like term  $\mathcal{L}_{\rm Meissner}$  quadratic in photon fields that has emerged in  terms of the dressed potentials $\mathcal{A}^a_{\mu}$. This has the  explicit form 
\begin{align}\label{eq:Meissner in hiding}
 &\mathcal{L}_{\rm Meissner} = \tfrac 1 2 \mathcal{A}_m^{a} C^{-1}_{mn a b} \mathcal{A}_n^b + \tfrac{1}{2}  \frac{1}{\mu}  (\mathcal{A}_\ft^a)^2 = \tfrac{1}{2} \frac{1}{D \kappa} \mathcal{A}_m^{a} P^0_{mnab} \mathcal{A}^b_n + \tfrac{1}{2}  \frac{1}{\mu}  (\mathcal{A}_\ft^a)^2 \nonumber\\
 &\phantom{mmmm}  = \tfrac{1}{2} (n e^* c_\tT)^2 \big( \frac{1}{\mu} (A^\imath_a)^2 + \frac{1}{\kappa} (A^\imath_\ft)^2 \big) 
   = \tfrac{1}{2} \varepsilon_0\omega_\mathrm{p}^2 \big(  (A^\imath_a)^2 + \tfrac{c_\tT^2}{c_\mathrm{K}^2} (A^\imath_\ft)^2 \big).
\end{align}
Here we used $c_\tT^2 = \mu/\rho$ and $c_\mathrm{K}^2 = \kappa/\rho$, and defined the plasmon frequency via
\begin{equation}\label{eq:plasmon frequency}
\omega_\mathrm{p}^2 =  \frac{(n e^*)^2}{\rho\varepsilon_0} = \frac{n {e^*}^2}{ m^*\varepsilon_0}. 
\end{equation}
Here the mass density is given by $\rho = m^* n$ where $m^*$ is the mass of the charge carriers; recalling 
that in two dimensions $[\varepsilon_0] = \mathrm{C}^2 /\mathrm{J}$, the plasmon frequency $\omega_\mathrm{p}$ has units of $1/\mathrm{s}$ as expected, while the plasmon energy is  $\hbar \omega_\mathrm{p}$.

Since we are dealing with 2+1D charged matter coupled to 2+1D electromagnetism, we anticipate that a plasmon frequency $\omega_\mathrm{p}$ has to be around. According to textbook wisdom, regardless whether one is dealing with a classical plasma, a degenerate electron gas or a charged crystal which is not pinned, 
the  longitudinal oscillations should acquire the plasmon energy at small momentum. We will find that the longitudinal phonon of the neutral crystal acquires indeed a plasmon mass in Sec.~\ref{subsec:Electromagnetism of the isotropic Wigner crystal}. However, at first sight it appears that the term Eq.~\eqref{eq:Meissner in hiding} also `gives mass' to the transverse electromagnetic fields: if this would be the whole story the charged crystal should also be a superconductor! Of course this is not the case and we shall see that upon integrating out the massless shear photons/transverse phonons a counterterm is produced that precisely cancels this electromagnetic 
Higgs mass. We already emphasized the dual way of understanding the superfluidity of the quantum liquid crystals, in terms of the principle that a ``bosonic solid  that loses its shear rigidity turns automatically in a superfluid''. This becomes elegantly exposed in the presence of the EM fields: in our liquids the 
shear photons acquire a mass and upon integrating them out the counterterms no longer cancel the `Higgs mass lying in hide' of the crystal, with the outcome that the liquid crystals do expel magnetic fields. This is precisely the Meissner effect and it leaves no doubt that the quantum liquid crystals are {\em electromagnetic superconductors} as well. These have in turn peculiar superconducting properties that will be the subject of Sec.~\ref{subsec:Superconductivity and the electromagnetism of the quantum nematic}.    

As before the smooth part of $u^a$ in the Lagrangian Eq.~\eqref{eq:dual elasticity EM Lagrangian} can be integrated out, leaving the constraint $\partial_\mu \sigma^a_\mu =0$. The stress fields however now contain a contribution from the electromagnetic field via Eqs.~\eqref{eq:EM stress tensor temporal}--\eqref{eq:EM stress tensor spatial}. We introduce the symbol $\underline{\sigma}$ for the stress originating only in elastic and not electromagnetic degrees of freedom, 
\begin{align}
 \sigma^a_\ft &= \underline{\sigma}^a_\ft +  \mathcal{A}^a_\ft, & 
 \sigma^a_m &= \underline{\sigma}^a_m +  \mathcal{A}^a_m,
\label{eq:totalstressconservation}
\end{align}
Using  Eq.~\eqref{eq:imaginary time electric field} we find the conservation laws,
\begin{align}
 \partial_\mu \sigma^a_\mu &= 0\\
 \partial_\mu \underline{\sigma}^a_\mu &= - \partial_\ft \mathcal{A}^a_\ft - \partial_m \mathcal{A}^a_m 
 = n e^* c_\tT (\partial_\ft A_a - \partial_a A^\imath_\ft) = - n e^* E^\imath_a,
 \label{eq:partialsresssourcing}
 \end{align}
 Elastic stress is no longer conserved by itself for the simple reason that the electromagnetic fields exert forces on the system: Eq.~\eqref{eq:totalstressconservation} expresses the total stress that is conserved while the purely elastic stress  $\underline{\sigma}^a_\mu$ is sourced by the external EM field. Since the conservation law applies the total stress, it is this quantity that relates to the dual stress gauge field $b^a_\mu$ imposing stress conservation, 
  \begin{align} \label{eq:EMstressgauge}
  \sigma^a_\mu &= \epsilon_{\mu\nu\lambda} \partial_\nu b^a_\lambda, \\
 \underline{\sigma}^a_\mu &= \epsilon_{\mu\nu\lambda} \partial_\nu b^a_\lambda -  \mathcal{A}^a_\mu. 
 \end{align} 
These stress gauge fields $b^a_{\mu}$ are no longer EM gauge-invariant objects. We already observed that this relates to the glide constraint. Tracking the definitions it follows that the gauge transformation $A_{\mu} \to A_{\mu}+\partial_{\mu}\varepsilon$ is represented  as $b^a_{\mu} \to b^a_{\mu}+\epsilon_{\ft a \mu} \varepsilon$, which coincides with the form of the dual gauge field Lagrange multiplier enforcing the glide constraint.
 
 \subsection{Photons versus stress photons}
 
We want to know how external electromagnetic fields `pull' on the elastic media. 
Therefore we expand the interaction terms in Eq.~\eqref{eq:dual elasticity EM Lagrangian} in terms of the stress gauge fields Eq.~\eqref{eq:EMstressgauge} which impose the conservation of the total stress,
\begin{align}
 \mathcal{L}_\mathrm{int} &=
 - \sigma^a_m  C^{-1}_{mn a b} \mathcal{A}_n^b - \frac{1}{\mu} \sigma^a_\ft \mathcal{A}^a_\ft \nonumber\\
 &=  -\frac{n e^* c_\tT}{2\kappa} \sigma^a_a A^\imath_\ft + \frac{ne^* c_\tT}{\mu} \sigma^a_\ft A_a 
 = -\frac{n e^* c_\tT}{2\kappa} (-\ti \frac{\omega_n}{c_\tT} b^{\tL\dagger}_\tT - p b^{\tT\dagger}_{+1}) A^{\imath}_\ft + \frac{ne^* c_\tT}{\mu}  (-q b^{a\dagger}_\tT) A_a \nonumber\\
 &= \frac{n e^* c_\tT}{2\kappa} \ti \frac{\omega_n}{c_\tT} b^{\tL\dagger}_\tT A^{\imath}_\ft  + \frac{n e^* c_\tT}{2\kappa} p\, b^{\tT\dagger}_{+1} A^{\imath}_\ft 
  - \frac{ne^* c_\tT}{\mu} q\, b^{\tL\dagger}_\tT A_\tL - \frac{ne^* c_\tT}{\mu} q b^{\tT\dagger}_\tT A_\tT. \label{eq:EM-stress interaction}
 \end{align}
Here we transformed directly to momentum space in the third line, using $p b^a_{+1} = -q b^a_\ft + \ti \frac{\omega_n}{c_\tT} b^a_\tL$, where $p = \sqrt{\frac{1}{c_\tT^2} \omega_n^2 + q^2}$. 
As usual, the dagger symbol denotes $A^\dagger(p) = A(-p)$ . Notice once  again that the scalar potential $A^\imath_\ft$ couples to $\sigma^a_a$, which is precisely compression stress: 
a volume change leads to an increase or decrease of the charge density.

We use the above $\mathcal{L}_{\rm int}$, specializing as usual to the isotropic form of the elastic tensor $C^{-1}_{mn a b}$, as the defining relation for how external EM fields couple to dual stress fields. With the dual formulation of elasticity coupled to electromagnetism, we shall now compute the electromagnetic properties of elastic media in the crystal, nematic and smectic phases. 
 The electromagnetic response of the medium $\mathcal{L}^\mathrm{EM}_{\rm stress}$ consists of a `diamagnetic' contribution $\mathcal{L}_{\rm Meissner}$ from Eq.~\eqref{eq:Meissner in hiding} 
 and a `paramagnetic' contribution $\mathcal{L}_{\rm para}$ that follows from integrating out  the stress gauge fields, which we now shall calculate. 
 Surely also the bare Maxwell term must be added: 
 \begin{align}\label{eq:effective total EM Lagrangian}
\mathcal{L}^{\rm EM}_{\rm eff} = \mathcal{L}_{\rm Maxw}+ \mathcal{L}^\mathrm{EM}_{\rm stress} 
=\mathcal{L}_{\rm Maxw}+\mathcal{L}_{\rm Meissner} + \mathcal{L}_{\rm para}.
\end{align}

To determine  $\mathcal{L}_{\rm para}$ we have to integrate out the elastic stress fields. The general form of the stress part of the Lagrangian is 
$\mathcal{L} = \tfrac{1}{2} b^a_\mu (G^{-1})^{ab}_{\mu\nu} b^b_\nu$ and the dual gauge interaction \eqref{eq:EM-stress interaction} is of the form 
$\mathcal{L}_{\rm int} = b^{a\dagger}_\mu g^a_{\mu\nu} A^\imath_\nu + A^{\imath\dagger}_\mu g^{\dagger a}_{\mu\nu} b^a_\nu$, where $g^a_{\mu\nu}$
are momentum-dependent coefficients, with the Hermitian conjugate $(g^\dagger)^a_{\mu\nu} = (g^{a}_{\nu\mu})^*$. 
Define the matrix $G$ via $(G^{-1})^{ab}_{\mu\nu} G^{bc}_{\nu\kappa} = \delta_{ac} \delta_{\mu\kappa}$. We are now in the position to integrate out the stress gauge fields,
 finding out how these dress the EM contributions,
  \begin{align}\label{eq:stress contribution to EM fields}
  \mathcal{L}_{\rm para}  &=  \tfrac{1}{2} b^{a\dagger}_\mu (G^{-1})^{ab}_{\mu\nu} b^b_\nu + \tfrac{1}{2} b^{a\dagger}_\mu g^a_{\mu\nu} A^{\imath}_\nu + \tfrac{1}{2} A^{\imath\dagger}_\mu g^{\dagger a}_{\mu\nu} b^a_\nu \nonumber\\
  &= - \tfrac{1}{2} A^{\imath \dagger}_\mu g^{a\dagger}_{\mu\nu} G^{ab}_{\nu\kappa} g^b_{\kappa \lambda} A^\imath_\lambda
 \end{align}
  Collecting  the indices $a,\mu$ pertaining to the stress gauge field in a four-vector, and the index $\nu$ pertaining to the EM gauge field in a three-vector,
  the couplings $g^a_{\mu\nu}$ are,
  \begin{align}
  g^\tL_{\tT \ft} &=  \ti \frac{n e^*}{2\kappa}\omega_n, &
  g^\tT_{+1 \ft} &=  \frac{n e^* }{2\kappa}c_\tT p, \nonumber\\
  g^\tL_{\tT \tL} &=  -\frac{n e^*}{\mu} c_\tT q , &
  g^\tT_{\tT \tT} &=  -\frac{n e^*}{\mu} c_\tT q ,
 \end{align}
 while all other components vanish. The coupling terms can be written as,
 \begin{equation}\label{eq:stress-EM coupling matrix}
  b^{a\dagger}_\mu g^a_{\mu\nu} A^\imath_\nu = 
  \begin{pmatrix} b^\tT_{+1} \\ b^\tL_\tT \\ b^\tL_{+1} \\ b^\tT_\tT \end{pmatrix}^\dagger
  n e^* \begin{pmatrix}
  \frac{1}{2 \kappa} c_\tT p & 0 & 0 \\
   \ti \frac{1}{2 \kappa} \omega_n & -\frac{1}{\mu} c_\tT q & 0 \\
   0 & 0 & 0\\
   0 & 0 & -\frac{1}{\mu} c_\tT q
  \end{pmatrix}
  \begin{pmatrix} A^\imath_\ft \\ A_\tL \\ A_\tT \end{pmatrix}.
 \end{equation}

 The recipe for calculating the total effective electromagnetic response of an elastic state of matter is: (a) compute the inverse stress propagator  $G^{-1}$ in terms of the dual gauge fields, 
(b) invert the propagator matrix to find $G$, (c) insert it in Eq.~\eqref{eq:stress contribution to EM fields} to find the `paramagnetic' contribution $ \mathcal{L}_{\rm para}$,  and (d)
add this result to the `diamagnetic' contribution $\mathcal{L}_{\rm Meissner}$ from Eq.~\eqref{eq:Meissner in hiding}, that arises in the duality construction, and the bare Maxwell term to obtain 
the total Lagrangian $\mathcal{L}^{\rm EM}_{\rm eff}$ from Eq.~\eqref{eq:effective total EM Lagrangian}.

Since everything is quadratic in the stress fields, integrating them out is equivalent to substituting their on-shell equations of motion,
meaning  that there is much one can infer  directly regarding  $\mathcal{L}^{\rm EM}_{\rm eff} $. For example,  we noticed already that $A^\imath_\ft$ couples to compression stress, 
and therefore the term $\sim A^{\imath\dagger}_\ft A^\imath_\ft$ is proportional to the compression correlator $\langle \sigma^a_a \sigma^b_b \rangle$. In fact, 
we can see directly from Eq.~\eqref{eq:charged strain Lagrangian} that
\begin{equation}\label{eq:longitudinal propagator potential correlator}
 \frac{1}{\mathcal{Z}[0]} \frac{\delta}{\delta A^\imath_\ft}  \frac{\delta}{\delta A^{\imath\dagger}_\ft} \mathcal{Z} = (n e^* c_\tT)^2 \langle \partial_a u^a \partial_b u^b \rangle = (n e^* c_\tT)^2 G_\tL,
\end{equation}
where we have used the Coulomb gauge $A_{\tL} = 0$. Since in this gauge fix $E^\imath_\tL = c_\tT q A^\imath_{\ft}$, the longitudinal electric field directly measures compression response. We will see this confirmed below.

\subsection{The effective electromagnetic actions}

The ground work is completed and in this subsection we will compute the effective actions governing the electromagnetism of the charged solid, quantum nematic and quantum smectic. This is now a straightforward affair,  amounting to integrating out the stress-gauge fields $b^a_\mu$ as developed for the neutral case, relying on the coupling between the EM photons and stress photons as introduced in the previous section. The template is set by the computation for the solid, and in order to determine the effective actions for the quantum liquid crystals one just has to add the Higgs terms of the appropriate dual stress superconductors Eqs.~\eqref{eq:quantum nematic Higgs Lagrangian} resp. \eqref{eq:quantum smectic Higgs term full}.  This is just a technical affair --- what matters are the results Eqs.~\eqref{eq:solid EM Lagrangian}, \eqref{eq:nematic EM Lagrangian}, \eqref{eq:smectic EM-stress Lagrangian} for the effective electromagnetic theories in the Coulomb gauge for respectively the solid, nematic and smectic that will be analyzed in detail in the next subsections.  
  
\subsubsection{Charged isotropic crystal} 
 
 For the isotropic solid, or rather a Wigner crystal in the charged case, the inverse propagator is (ignoring second-gradient elasticity),  
 \begin{align}
  \mathcal{L}_\textrm{Xtal} &= \tfrac{1}{2} b^{a\dagger}_\mu (G_{\textrm{Xtal}}^{-1})^{ab}_{\mu\nu} b^b_\nu, \nonumber\\
  b^a_\mu &= \begin{pmatrix} b^\tT_{+1} & b^\tL_\tT & b^\tL_{+1} & b^\tT_\tT \end{pmatrix}^\tT, \nonumber\\
  G_{\textrm{Xtal}}^{-1} &= \frac{1}{8\mu}
  \begin{pmatrix}
   \tfrac{2}{1+\nu}p^2 & \ti \tfrac{2\nu}{1+\nu} \frac{\omega_n p }{c_\tT} & 0 & 0  \\
   - \ti \tfrac{2\nu}{1+\nu} \frac{\omega_n p }{c_\tT} & \tfrac{2}{1+\nu}\frac{\omega_n^2}{c_\tT^2} + 4 q^2 & 0 & 0 \\
   0 & 0 & p^2 & - \ti \frac{\omega_n p }{c_\tT} \\
   0 & 0 & \ti \frac{\omega_n p }{c_\tT} & \frac{\omega_n^2}{c_\tT^2} + 4q^2
  \end{pmatrix}.
 \end{align}
 The propagator is then
\begin{equation}
 G_{\textrm{Xtal}} = \frac{\mu}{p^2} 
 \begin{pmatrix} \frac{\omega_n^2/c_\tT^2 + 2 (1+\nu) q^2}{ \omega_n^2/c_\tL^2 + q^2} & - \ti \nu \frac{\omega_n p / c_\tT}{ \omega_n^2/c_\tL^2 + q^2} & 0 & 0 \\
  \ti \nu \frac{ \omega_n p / c_\tT}{ \omega_n^2/c_\tL^2 + q^2} & \frac{p^2}{ \omega_n^2/c_\tL^2 + q^2} & 0 & 0 \\
  0 & 0 & \frac{\omega_n^2}{c_\tT^2 p^2} & - \ti \frac{\omega_n}{c_\tT p} \\
  0 & 0 &   \ti \frac{\omega_n }{c_\tT p} & 1 
 \end{pmatrix}.
\end{equation}
Inserting this in Eq.~\eqref{eq:stress contribution to EM fields} and using Eq.~\eqref{eq:stress-EM coupling matrix} yields,
\begin{equation}
 \mathcal{L}^\mathrm{para}_\mathrm{Xtal} = 
 -\tfrac{1}{2} \varepsilon_0\omega_\mathrm{p}^2   \begin{pmatrix} A^\imath_\ft \\ A^\imath_\tL \\ A^\imath_\tT \end{pmatrix}^\dagger 
 \begin{pmatrix}
  \tfrac{c_\tT^2}{c_\tK^2} \frac{\omega^2 + c_\tT^2 q^2}{\omega_n^2 + c_\tL^2 q^2} &  \ti \frac{\omega_n c_\tT q}{\omega_n^2 + c_\tL^2 q^2} & 0 \\
   -\ti \frac{\omega_n c_\tT q}{\omega_n^2 + c_\tL^2 q^2} & \frac{c_\tL^2 q^2}{\omega_n^2 + c_\tL^2 q^2} & 0 \\
   0 & 0 &  \frac{c_\tT^2 q^2}{\omega_n^2 + c_\tT^2 q^2}
 \end{pmatrix}
  \begin{pmatrix} A^\imath_\ft \\ A^\imath_\tL \\ A^\imath_\tT \end{pmatrix},
\end{equation}
where we have used the definition of the plasmon frequency Eq.~\eqref{eq:plasmon frequency}, and $\frac{1-\nu}{1+\nu} = \frac{c_\tT^2}{c_\tK^2}$.

Taking the EM Coulomb gauge $\partial_m A^\imath_m = - q A^\imath_\tL = 0$, we find for the `paramagnetic'  contribution to the EM action,
\begin{align}
 \mathcal{L}^\mathrm{para}_\mathrm{Xtal} &= -\tfrac{1}{2} \varepsilon_0 \omega_\mathrm{p}^2   \Big( \frac{c_\tT^2}{c_\tK^2} \frac{\omega_n^2 + c_\tT^2 q^2}{\omega^2 + c_\tL^2 q^2} \lvert A^\imath_\ft \rvert^2 +   \frac{c_\tT^2 q^2}{\omega_n^2 + c_\tT^2 q^2} \lvert A^\imath_\tT \rvert^2 \Big).
\end{align}
This must be added to the `diamagnetic' contribution Eq.~\eqref{eq:Meissner in hiding} yielding the total stress contribution to the EM Lagrangian,
\begin{align}
 \mathcal{L}^\mathrm{EM}_{\rm Xtal}  
  &= \tfrac{1}{2}\varepsilon_0 \omega_\mathrm{p}^2 \Big(  \frac{c_\tT^2 q^2}{\omega_n^2 + c_\tL^2 q^2} \lvert A^\imath_\ft \rvert^2 + \frac{\omega_n^2 }{\omega_n^2 + c_\tT^2 q^2}  \lvert A^\imath_\tT\rvert^2 \Big).
\end{align}
This agrees with what one would obtain integrating out directly the  
displacement fields $u^a$ and not invoking the duality mapping. Using $\varepsilon_0 \omega_\mathrm{p}^2 = (n^2 e^{*2})/\rho$ and comparing with Eq.~\eqref{eq:longitudinal strain propagator} 
we see that the coefficient of $A^{\imath \dagger}_\ft A^\imath_\ft$ is $\tfrac{1}{2} (n e^* c_\tT)^2 G_\tL$, as expected from Eq.~\eqref{eq:longitudinal propagator potential correlator}.

To compute the full EM response, we must add the Maxwell contribution Eq.~\eqref{eq:Euclidean action EM fields} and this yields the effective action which is the point of departure for the further
analysis of the electromagnetism of the charged crystals: 
\begin{align}
 \mathcal{L}^{\rm EM}_\mathrm{eff, Xtal} &=  \tfrac{1}{2} \varepsilon_0  \Big( c_\tT^2 q^2 + \omega_\mathrm{p}^2   \frac{c_\tT^2 q^2}{\omega_n^2 + c_\tL^2 q^2} \Big) \lvert A^\imath_\ft \rvert^2 
 + \tfrac{1}{2}\varepsilon_0 \Big(  \omega_n^2 +  c_l^2 q^2 +  \omega_\mathrm{p}^2\frac{\omega_n^2 }{\omega_n^2 + c_\tT^2 q^2}  \Big) \lvert A^\imath_\tT\rvert^2.\label{eq:solid EM Lagrangian}
\end{align}
 
\subsubsection{Charged quantum nematic}

Let us now turn to the quantum nematic -- this follows the same pattern as for the charged solid except that we have now to account for the effects of the dislocation condensate. Given that this dislocation condensate plays its part, we first need to reintroduce the tilde-coordinates which associated with the condensate velocity $c_d$: $\tilde{p} = \sqrt{\frac{1}{c_\td^2} \omega_n^2 + q^2}$, see Eq.~\eqref{eq:crystal Lagrangian smectic coordinates full},
\begin{align}
 b^a_\ft &= \frac{c_\td}{c_\tT} \tilde{b}^a_\ft, & b^a_{+1} &= \frac{c_\td}{c_\tT} \frac{\tilde{p}}{p} \tilde{b}^a_{+1},&  b^a_\tT &= \tilde{b}^a_\tT. 
\end{align}
The matrix Eq.~\eqref{eq:stress-EM coupling matrix}  of photon-stress photon couplings becomes
 \begin{equation}\label{eq:tilde stress-EM coupling matrix}
  \tilde{b}^{a\dagger}_\mu \tilde{g}^a_{\mu\nu} A^\imath_\nu = 
  \begin{pmatrix} \tilde{b}^\tT_{+1} \\ \tilde{b}^\tL_\tT \\ \tilde{b}^\tL_{+1} \\ \tilde{b}^\tT_\tT \end{pmatrix}^\dagger
  n e^* \begin{pmatrix}
   \frac{1}{2 \kappa} c_\td \tilde{p}& 0 & 0 \\
  \frac{\ti}{2 \kappa} \omega_n & - \frac{1}{\mu} c_\tT q & 0 \\
   0 & 0 & 0\\
   0 & 0 & -\frac{1}{\mu}c_\tT q
  \end{pmatrix}
  \begin{pmatrix} A^\imath_\ft \\ A^\imath_\tL \\ A^\imath_\tT \end{pmatrix},
 \end{equation}
 while the paramagnetic contribution to the effective EM action becomes, 
  \begin{align}\label{eq:tilde stress contribution to EM fields}
  \mathcal{L}^\mathrm{para}_\mathrm{nem}  &=  \frac{1}{2} \tilde{b}^{a\dagger}_\mu (\tilde{G}_{\textrm{nem}}^{-1})^{ab}_{\mu\nu} \tilde{b}^b_\nu + \tfrac{1}{2} \tilde{b}^{a\dagger}_\mu \tilde{g}^a_{\mu\nu} A^{\imath}_\nu + \tfrac{1}{2} A^{\imath\dagger}_\mu \tilde{g}^{\dagger a}_{\mu\nu} \tilde{b}^a_\nu \nonumber\\
  &= - \tfrac{1}{2} A^{\imath \dagger}_\mu \tilde{g}^{a\dagger}_{\mu\nu} (\tilde{G}_{\textrm{nem}})^{ab}_{\nu\kappa} \tilde{g}^b_{\kappa \lambda} A^\imath_\lambda.
 \end{align}
 To evaluate this contribution, we need the specifics of the stress gauge fields of the neutral quantum nematic encoded in the inverse propagator $\tilde{G}^{-1}_\textrm{nem}$. This was discussed in Sec.~\ref{sec:Quantum nematic}:
 \begin{align}\label{eq:nematic inverse Green function}
  \tilde{G}_{\textrm{nem}}^{-1} = \tfrac{1}{2} \frac{1}{4\mu} \Big[ & 
  \begin{pmatrix}
   \tfrac{2}{1+\nu}\frac{c_\td^2}{c_\tT^2}\tilde{p}^2 & \ti \tfrac{2\nu}{1+\nu} \frac{ \omega_n c_\td\tilde{p} }{c_\tT^2} & 0 & 0  \\
   - \ti \tfrac{2\nu}{1+\nu} \frac{\omega_n c_\td \tilde{p} }{c_\tT^2} & \tfrac{2}{1+\nu}\frac{\omega_n^2}{c_\tT^2} + 4 q^2 & 0 & 0 \\
   0 & 0 & \frac{c_\td^2}{c_\tT^2} \tilde{p}^2 & - \ti \frac{c_\td \omega_n \tilde{p} }{c_\tT^2} \\
   0 & 0 & \ti \frac{c_\td \omega_n \tilde{p} }{c_\tT^2} & \frac{\omega_n^2}{c_\tT^2} + 4q^2
  \end{pmatrix} 
+ \frac{2\Omega^2}{c_\tT^2} \begin{pmatrix}
   \frac{c_\td^2 \tilde{p}^2}{ \omega_n^2 + c_\td^2 p^2} & \ti \frac{ \omega_n c_\td\tilde{p} }{\omega_n^2 + c_\td^2 p^2} & 0 & 0  \\
   - \ti \frac{\omega_n c_\td \tilde{p} }{ \omega_n^2 + c_\td^2 p^2}  & \frac{\omega_n^2}{ \omega_n^2 + c_\td^2 p^2}  & 0 & 0 \\
   0 & 0 & 1 &0 \\
   0 & 0 & 0 & 1
  \end{pmatrix}
  \Big]
 \end{align}
 which has to be  inverted to obtain the propagator matrix $\tilde{G}_{\textrm{nem}}$. The components entering the electromagnetic effective action are Eqs.~\eqref{eq:nematic longitudinal propagator}, \eqref{eq:nematic transverse propagator} in Sec.~\ref{subsec:Correlation functions quantum nematic},
\begin{align}\label{eq:nematic Green function phonon components}
(\tilde{G}_{\textrm{nem}})^{\tL\tL}_{\tT\tT} &= \frac{\mu}{q^2} \frac{ c_\tL^2 q^2 ( \omega_n^2 + c_\tR^2 q^2) + \Omega^2 c_\tK^2 q^2  }{ (\omega_n^2  + c_\tL^2 q^2)(\omega_n^2  + c_\tR^2 q^2) + \Omega^2(\omega_n^2  + c_\tK^2 q^2)},\nonumber\\
(\tilde{G}_{\textrm{nem}})^{\tT\tT}_{\tT\tT} &= \frac{\mu}{q^2} \frac{ c_\tT^2 q^2 ( \omega_n^2 + c_\td^2 q^2)  }{ (\omega_n^2  + c_\tT^2 q^2)(\omega_n^2  + c_\td^2 q^2) + \Omega^2(\omega_n^2  + c_\tR^2 q^2)},
\end{align}
while the cross-terms are zero. The full matrix $(\tilde{G}_{\textrm{nem}})^{ab}_{\mu\nu}$ is substituted in Eq.~\eqref{eq:tilde stress contribution to EM fields} and adding the `diamagnetic' term
Eq.~\eqref{eq:Meissner in hiding} we find,
\begin{align}
 \mathcal{L}^\mathrm{EM}_{\rm nem} &= \tfrac{1}{2} \varepsilon_0\omega_\mathrm{p}^2 
 \begin{pmatrix} A^\imath_\ft \\ A^\imath_\tL \\ A^\imath_\tT \end{pmatrix}^\dagger
\begin{pmatrix}
   c_\tT^2 q^2 g_b^{-1} & -\ti  \omega_n c_\tT q g_b^{-1} & 0 \\
   \ti  \omega_n c_\tT q g_b^{-1} & \omega_n^2 g_b^{-1} & 0 \\
   0 & 0 & \frac{ \omega_n^2 (\omega_n^2 + c_\td^2 q^2) + \Omega^2 ( \omega_n^2 + c_\tR^2 q^2)}{(\omega_n^2 + c_\tT^2 q^2)(\omega_n^2 + c_\td^2 q^2) + \Omega^2(\omega_n^2 + c_\tR^2 q^2)}
 \end{pmatrix}
 \begin{pmatrix} A^\imath_\ft \\ A^\imath_\tL \\ A^\imath_\tT \end{pmatrix}, \\
 g_b^{-1} &= \frac{(\omega_n^2 + c_\tR^2 q^2 + \Omega^2)}{(\omega_n^2  + c_\tL^2 q^2)(\omega_n^2  + c_\tR^2 q^2) + \Omega^2(\omega_n^2  + c_\tK^2 q^2)}.
\end{align}
We add this to the the Maxwell term Eq.~\eqref{eq:Euclidean action EM fields} and upon taking the electromagnetic Coulomb gauge $q A_\tL = 0$, we obtain the final result, the effective EM Lagrangian of the nematic:
\begin{align}\label{eq:nematic EM Lagrangian}
  \mathcal{L}^{\rm EM}_\mathrm{eff, nem}
 =& \phantom{+} \tfrac{1}{2} \varepsilon_0  \Big(c_\tT^2q^2 +  \omega_\mathrm{p}^2 \frac{c_\tT^2 q^2(\omega_n^2 + c_\tR^2 q^2 + \Omega^2)}{(\omega_n^2  + c_\tL^2 q^2)(\omega_n^2  + c_\tR^2 q^2) + \Omega^2(\omega_n^2  + c_\tK^2 q^2)} \Big) \lvert A^\imath_\ft \rvert^2 \nonumber\\
 & + \tfrac{1}{2} \varepsilon_0  \Big(\omega_n^2 + c_l^2 q^2 + \omega_\mathrm{p}^2  \frac{ \omega_n^2 (\omega_n^2 + c_\td^2 q^2) + \Omega^2 ( \omega_n^2 + c_\tR^2 q^2)}{(\omega_n^2 + c_\tT^2 q^2)(\omega_n^2 + c_\td^2 q^2) + \Omega^2(\omega_n^2 + c_\tR^2 q^2)} \Big)\lvert A^\imath_\tT\rvert^2 .
\end{align}
As a check, this expression reduces to the one we obtained for the solid Eq.~\eqref{eq:solid EM Lagrangian}  upon taking  the limit $\Omega \to 0$, as it should.

\subsubsection{Charged quantum smectic}

The derivation for the quantum smectic just follows the pattern of the quantum nematic, substituting instead of the disorder field Higgs term in Eq.~\eqref{eq:nematic inverse Green function} the smectic version given by Eq.~\eqref{eq:quantum smectic Higgs term full}. While the computations are straightforward, the expressions are long and tedious and we just state the end result here. Recall that the smectic is parametrized by the angle $\eta$ between the liquid direction and the momentum or propagation direction of the stress photons (Fig.~\ref{fig:smectic layers}), which are of course the same as  the momentum directions of the electromagnetic photons that probe the system. 
For arbitrary angle $\eta$ the stress contribution (that is added to the Maxwell action) to the electromagnetic response is, in the electromagnetic Coulomb gauge $\partial_m A_m = -q A_\tL =0$, given by
\begin{align}\label{eq:smectic EM-stress Lagrangian}
  \mathcal{L}^\mathrm{EM}_{\rm smec} &= \tfrac{1}{2} \frac{\omega_\mathrm{p}^2}{\mathcal{D}} 
\Big[
  c_\tT^2 q^2 \big\{ c_\td^2 p_\eta^2 (\omega_n^2 + c_\tT^2 q^2) + \Omega^2 (\omega_n^2 + c_\tT^2 q^2 \sin^2 2 \eta) \big\} \lvert A^\imath_\ft\rvert^2 \nonumber\\
&\phantom{mmmm} +  \omega_n^2 \big\{ c_\td^2 p_\eta^2 (\omega_n^2 + c_\tL^2 q^2) + \Omega^2 (\omega_n^2 + c_\tL^2 q^2 - c_\tT^2 q^2 \sin^2 2 \eta ) \big\}  \lvert A^\imath_\tT \rvert^2 \\
&\phantom{mmmm}  + \ti \omega_n c_\tT^3 q^3 \Omega^2 \cos 2 \eta \sin 2 \eta ( A^{\imath \dagger}_\tT A^\imath_\ft - A^{\imath \dagger}_\ft A^\imath_\tT)  \Big], \nonumber  \\
  \mathcal{D} &= c_\td^2 p_\eta^2 ( \omega_n^2 + c_\tT^2 q^2)(\omega_n^2+ c_\tL^2 q^2) + \Omega^2(\omega_n^4 + \omega_n^2 c_\tL^2 q^2 +  c_\tK^2 q^2 c_\tT^2 q^2 \sin^2 2\eta)
\end{align}
Here we have used $p_\eta^2 = \frac{1}{c_\td^2} \omega_n^2 + \cos^2 \eta \, q^2$. Notice that for angles that are not multiples of $\eta = \pi/4$ there are now cross terms between $A_\ft$ and $A_\tT$. The final answer in terms of the effective EM action of the smectic is then given by adding the Maxwell term Eq.~\eqref{eq:Euclidean action EM fields},
\begin{align}
\mathcal{L}_{\rm eff, smec}^{\rm EM} = \mathcal{L}_{\rm Maxw} + \mathcal{L}^\mathrm{EM}_{\rm smec} \label{eq:smecticEMfull}
\end{align}
and will be analyzed further in the next section.
 
 \section{Electromagnetic observables of the quantum nematic and smectic}\label{sec:Electromagnetic observables}
 
The effective electromagnetic actions for the solid \eqref{eq:solid EM Lagrangian}, quantum nematic ~\eqref{eq:nematic EM Lagrangian}  and quantum smectic \eqref{eq:smectic EM-stress Lagrangian}
encode in full generality all the information of these phases that can be retrieved using electromagnetic means. As we already argued, this is the only access experimentalists have to interrogate these systems. It is now just a matter of extracting these observables using  standard electromagnetic linear response theory.  The reader might already anticipate that a lot more is going on in these `near-solids' than in the usual `gaseous' Drude metals. For  the neutral systems we learned about the existence of a plethora of propagating elastic modes. Just by glancing at the effective EM actions of the previous section, it is obvious that these have left fingerprints on the EM response that are measurable, at least in principle. The massless, long-wavelength modes are universal, governed by symmetry principles and they therefore do not convey direct information regarding the strength of the solid-like correlations in the liquid-crystalline phases. This information is instead encoded by the massive modes that propagate at finite momenta when the solid-like correlations are strong enough. One way to obtain finite-frequency information is by interrogating the system with light. Photons probe the transverse response and the effective actions indicate that there is much going on in principle. Here we encounter a frustrating practical limitation of electromagnetic experiments in condensed matter: it is just the habit of non-relativistic matter that the velocity of light $c_l$ is very large compared to the material velocities ($c_\tT$, $c_\tL$, $c_\td$, $\ldots$) and therefore one can access only the small-momentum limit, $q \to 0$, in the laboratory. Due to universal, in essence hydrodynamical, reasons our ``massive stress photons'' lose their spectral weights in this limit and there is just no way to discriminate between the ultimate gaseous and solid-like quantum liquids using only information extracted at small momenta. 

This is perhaps the most important take-home message: it could well be that some correlated electron systems are quite solid-like but, for the practical reasons we just explained, this information is hidden from the 
experimentalist's view. There is however a way to circumvent this: one can also probe the system using electron beams. These measure the dynamical, longitudinal density response through the electron loss functions and one can probe larger momenta, easily up to the reciprocal lattice factors. As we will show below, the massive stress photons do leave characteristic fingerprints on the longitudinal responses and these are therefore the spectroscopy of choice to address the physics.  Dealing with real systems (say the cuprates) this requires machinery with an energy resolution of the order of meV and momenta of the order of inverse nanometers. Such machines can be built and it appears that these are on the verge of being operational~\cite{VigEtAl15}. We will further discuss this is in Sec.~\ref{subsec:Superconductivity and the electromagnetism of the quantum nematic}. 

Besides these practical issues there is also the fundamental question: {\em How does the electric charge alter the nature of the solid, quantum nematic and smectic?} The highlight is the simple demonstration that the effective action of the nematic Eq.~\eqref{eq:nematic EM Lagrangian} contains the Meissner effect: this proves that it is a superconductor. However,  there is more entertainment around such as our finding  that the genuine Wigner crystal is characterized by a massless `polariton' with a quadratic dispersion (Sec~\ref{subsec:Electromagnetism of the isotropic Wigner crystal}), while the electromagnetism of the quantum smectic will turn out to be a dizzying affair. We start out with a short review of general electromagnetic response theory. 

\subsection{Electromagnetic linear responses}

Given an effective electromagnetic action, it is easy to compute the response of the medium to electromagnetic fields by performing the Wick rotation $\omega_n \to \ti \omega - \delta $ to real time where $\delta$ is an infinitesimal factor that is only important when taking imaginary parts of quantities. This factor $\delta$ will be almost always suppressed to keep the notation compact. In this section we will express all results in real time and real frequency. Let us shortly recall this, using the standard 
Drude metal as a reference to textbook wisdoms~\cite{Mahan93, Coleman15}. As a matter of principle one has to distinguish the longitudinal and transverse responses, 
a division that we already wired in as much as possible in our effective actions Eqs.~ \eqref{eq:solid EM Lagrangian}, \ref{eq:nematic EM Lagrangian}, \eqref{eq:smectic EM-stress Lagrangian}. From Eq.~\eqref{eq:electric magnetic field in potentials} and using \ref{sec:Fourier space coordinate systems} and the definition $A_\ft = -\frac{1}{c_\tT}V$, the longitudinal and transverse components of the EM fields are given by
\begin{align}
E_{\tL}(\omega,q) &= c_{\tT}q A_{\ft}(\omega,q) -\ti  \omega A_{\tL}(\omega,q), \\ 
E_{\tT}(\omega,q) &= -\ti \omega A_{\tT}(\omega,q),\\
B(\omega,q) &= -q A_{\tT}(\omega,q).
\end{align}
The magnetic field $B(\omega,q)$ and propagating photons are purely transverse; $A_\tT$ is gauge invariant, whereas, under gauge transformations $\varepsilon$, the longitudinal component transforms $A_L\to A_L + q \varepsilon$ and is not physical. Therefore everything that is measured by propagating light is in the transverse sector, while the static electric screening and the electron-loss-density spectra are exclusively longitudinal and couple to the charge density. 

Let us first consider the {\em transverse response}. The natural point of departure is the photon propagator,
\begin{align}
 \langle A^{\dagger}_\tT(\omega, q) A_\tT(-\omega,-q) \rangle = \frac{1}{\varepsilon_0}\frac{1}{\omega^2 - c_l^2 q^2 -\Pi_\tT (\omega, q)} 
\label{photonpropdef}
\end{align}
where the transverse photon self-energy $\Pi_\tT$ (polarization propagator) contains all the information regarding the charged medium. One notices that one can read off this self-energy directly 
from our effective actions  Eqs.~\eqref{eq:solid EM Lagrangian}, \eqref{eq:nematic EM Lagrangian},\eqref{eq:smectic EM-stress Lagrangian}, since the factors 
multiplying $ | A_\tT(\omega,q)|^2$ are just the inverse of this propagator after Wick rotation to real frequency.

The photon propagator can also be written in an equivalent form, 
\begin{equation}
\langle A_{\tT}^{\dagger}(\omega,q) A_{\tT}(-\omega,-q) \rangle  =  \frac{1}{\hat{\varepsilon}_{\tT\tT} (\omega,q) \omega^2 - q^2/\hat{\mu}(\omega,q)} 
 \label{dielmagnper}
 \end{equation}
defining the transverse dielectric function $\hat{\varepsilon}_{\tT \tT} (\omega,q)$ and magnetic permeability $\hat{\mu}(\omega,q)$ of the medium. We will use the approximation $\hat{\mu}(\omega,q) = \mu_0(1+\mathcal{O}(c^2_\tT/c_l^2))$. For instance, frequency dependent  refractive
index $N(\omega)$ is given by the usual formula $N(\omega)^2 = c_l^2 \lim_{q \to 0} \hat{\varepsilon}_{TT} (\omega,q) \hat{\mu}(\omega,q)$. From now on we always take $\hat{\mu}(\omega,q) = \mu_0$. The poles of the photon propagator define the transverse modes of the combined matter--light system while the pole strengths define the EM spectral weight. The dispersion of the transverse modes is determined by solutions of 
\begin{equation} 
\omega^2 = \frac{\varepsilon_0}{\hat{\varepsilon}_{\tT\tT}(\omega,q)} c_l^2 q^2.
\label{transversedisp}
\end{equation}
In semiconductors these will reveal the `polaritons' or coupled photon-exciton modes. We emphasize yet again that in general only the $q \to 0$ limit is experimentally accessible because of the mismatch between the light and material velocities. We will see that the charged elastic medium has here a surprise in store. 

The quantity that is typically first extracted from experiment is the optical conductivity. In the \ref{sec:Dual Kubo formula}, its Kubo formula is carefully rederived when one is dealing with the stress gauge fields. The quantity that is plotted by the experimentalist is the real part $\hat{\sigma}_1 (\omega,q)$ of the momentum- and frequency-dependent optical conductivity $\hat{\sigma}_{\tT \tT} (\omega,q) = \hat{\sigma}_{\tT,1} (\omega,q) + i \hat{\sigma}_{\tT, 2} (\omega,q)$, which is related to the transverse dielectric function $\hat{\varepsilon}_{\tT \tT}(\omega,q)$ and to the photon self-energy $\Pi_{\tT} (\omega,q)$ as
\begin{align} 
\hat{\sigma}_{\tT \tT} (\omega,q) &= -\ti \omega \left(\hat{\varepsilon}_{\tT \tT}(\omega,q) -\varepsilon_0\right) \\
&= \varepsilon_0\frac{\ti}{\omega} \Pi_{\tT} (\omega,q).
\label{opticalcond}
\end{align}
The conductivity tensor $\hat{\sigma}_{ab}$ is not to be confused with the stress tensor $\sigma^a_\mu$.
We have identified the neutral nematic phase, or the dual stress superconductor,  as a true superfluid of the constituent bosons that becomes a superconductor in the charged case. To reveal the superconducting state emerging from  the unpinned Wigner crystal with perfect conductivity, we analyze the behavior of the magnetic fields in the phases. Given that $B = -q A_\tT$, the interest is in the the equation of motion for $A_\tT$ following from the effective electromagnetic action. In any conductor, the transverse fields $E_\tT$ and $B$ decay exponentially at high frequencies $\omega$ due to the skin effect: the oscillating EM field induces a current which in turn induces a magnetic field opposing the external field~\cite{LandauLifshitz84}. In some sense, the Meissner effect is the extreme version of this that even holds at $\omega = 0$. We can capture both these effects by a slight generalization of textbook electromagnetism. The response of a medium ($x\ge 0$) to an outside ($x<0$) plane wave of frequency $\omega$, momentum $q= \omega/c_l$ and amplitude $A_0$ is the general solution (one-dimensional for simplicity)
\begin{equation}
 A_\tT(x,t) = A_0 \te^{- \kappa(\omega,q) \frac{\omega}{c_l}x} \te^{\ti n(\omega,q) \frac{\omega}{c_l}x - \ti \omega t}.
\end{equation}
Here $N(\omega,q) = n(\omega,q) + \ti \kappa(\omega,q)$ is the complex and momentum-dependent refractive index. This quantity is determined from Eqs.~\eqref{dielmagnper}--\eqref{transversedisp}. In our case $\hat{\mu}(\omega,q) \approx \mu_0$, and we shall use as a definition $N(\omega,q) = \sqrt{\hat{\varepsilon}_{\tT\tT}(\omega,q)/\varepsilon_0}$. Clearly, if the imaginary part of the refractive index $\kappa$ is positive, the field decays exponentially with distance with length scale
\begin{equation}
 \lambda (\omega,q) = \frac{c_l}{\omega \kappa(\omega,q)}. \label{eq:penetration depth}
\end{equation}
This defines the penetration depth $\lambda$, also called skin depth in the case of the skin effect at finite $\omega$. It is directly related to the photon propagator via $N = \sqrt{\hat{\varepsilon}_{\tT\tT}/\varepsilon_0}$ and Eq.~\eqref{opticalcond} as
\begin{equation}\label{eq:penetration depth from self-energy}
 \lambda(\omega,q) = \frac{c_l}{\omega \mathrm{Im} \sqrt{1 - \frac{1}{\omega^2} \Pi_\tT}} = \frac{c_l}{\mathrm{Im} \sqrt{\omega^2 - \Pi_\tT}}.
\end{equation}
We will find surprising behavior of the penetration depth for both the quantum nematic and the quantum smectic.

Let us now turn to the {\em  longitudinal} response. This can only be measured through `material' probes such as electron beams, or by looking at the screening of 
static electrical monopole sources in the system. The longitudinal dielectric function $\hat{\varepsilon}_{\tL \tL}  (\omega,q)$ is defined through the longitudinal electrical field propagator,
\begin{align}
\langle E_{\tL}(\omega,q) E_{\tL}(-\omega,-q)  \rangle &= c_{\tT}^2 q^2 \langle A^{\dagger}_{\ft}(\omega,q) A_{\ft}(-\omega,-q) \rangle  = \frac{1} {\hat{\varepsilon}_{\tL \tL}(\omega,q)},
\label{longdielecdef}
\end{align}
where $c_\tT A_\ft = -V$ is the scalar potential and we have taken the Coulomb gauge fix $-q A_\tL = 0$. As before, the longitudinal dielectric function can be simply read off directly from  the effective actions  Eq. (\ref{eq:solid EM Lagrangian},\ref{eq:nematic EM Lagrangian},\ref{eq:smectic EM-stress Lagrangian}).

In the static limit, this determines the Debye electrical screening length $l_\mathrm{D} = q^{-1}_\mathrm{D}$ via $\hat{\varepsilon}_{\tL \tL}  (\omega = 0 , q_\mathrm{D}) = 0$.
In fact, using {\em electron energy-loss spectroscopy} (EELS), this longitudinal dielectric function can be accessed over a large kinematical range which is of primary physical relevance. The quantity measured in transmission spectroscopy is the transmissive energy loss function $f_\tT(\omega,q)$, defined as
\begin{equation}\label{eq:energy loss definition}
f_\tT(\omega ,q ) = - \mathrm{Im} \frac{\varepsilon_0}{\hat{\varepsilon}_{\tL\tL} (\omega, q)},
\end{equation}
while in reflection mode used in low-energy high-resolution (HR) EELS, the the quantity measured is the reflective energy loss function $f_\tR(\omega,q)$, 
s\begin{equation}\label{eq:relective energy loss definition}
f_\tR(\omega ,q ) = - \mathrm{Im} \frac{1}{1+\hat{\varepsilon}_{\tL\tL} (\omega, q)/\varepsilon_0}.
\end{equation}
We will focus in the remainder on the transmissive loss function Eq.~\eqref{eq:energy loss definition}, since these both are determined essentially just by $\hat{\varepsilon}_{\tL \tL}$. This should hold also for the longitudinal charge response, that can be extracted from the novel {\em resonant inelastic X-ray scattering} (RIXS) experiments. Finally, notice that in the limit $q \to 0$ the momentum space directions $\tL,\tT$ become degenerate, yielding the identity $\hat{\varepsilon}_{\tL \tL} (\omega, q \rightarrow 0) = \hat{\varepsilon}_{\tT \tT} (\omega , q \rightarrow 0)$ satisfied by our results, see e.g.~\cite{Pines99} (note that when also $\omega\to 0$, the order of limits is important in the presence of long-range order). 

\subsection{The Drude model and the charged viscous fluid}

We are done with our summary of  the standard repertoire of electromagnetic response functions.  Before we turn to the EM response for our generalized elasticity, let us first fill this in with the expectations for the `gaseous' limit. Instead of the usual RPA treatment of the electron gas, a more informative reference frame is given by a classical single-component plasma. This just corresponds to a charged version of a dissipative hydrodynamical fluid; its electromagnetism has been studied in full detail only recently~\cite{ForcellaEtAl14} and we present here the simplified cartoon version. 
 
\begin{figure}
\parbox[t][][t]{.48\textwidth}{
\includegraphics[width=.48\textwidth]{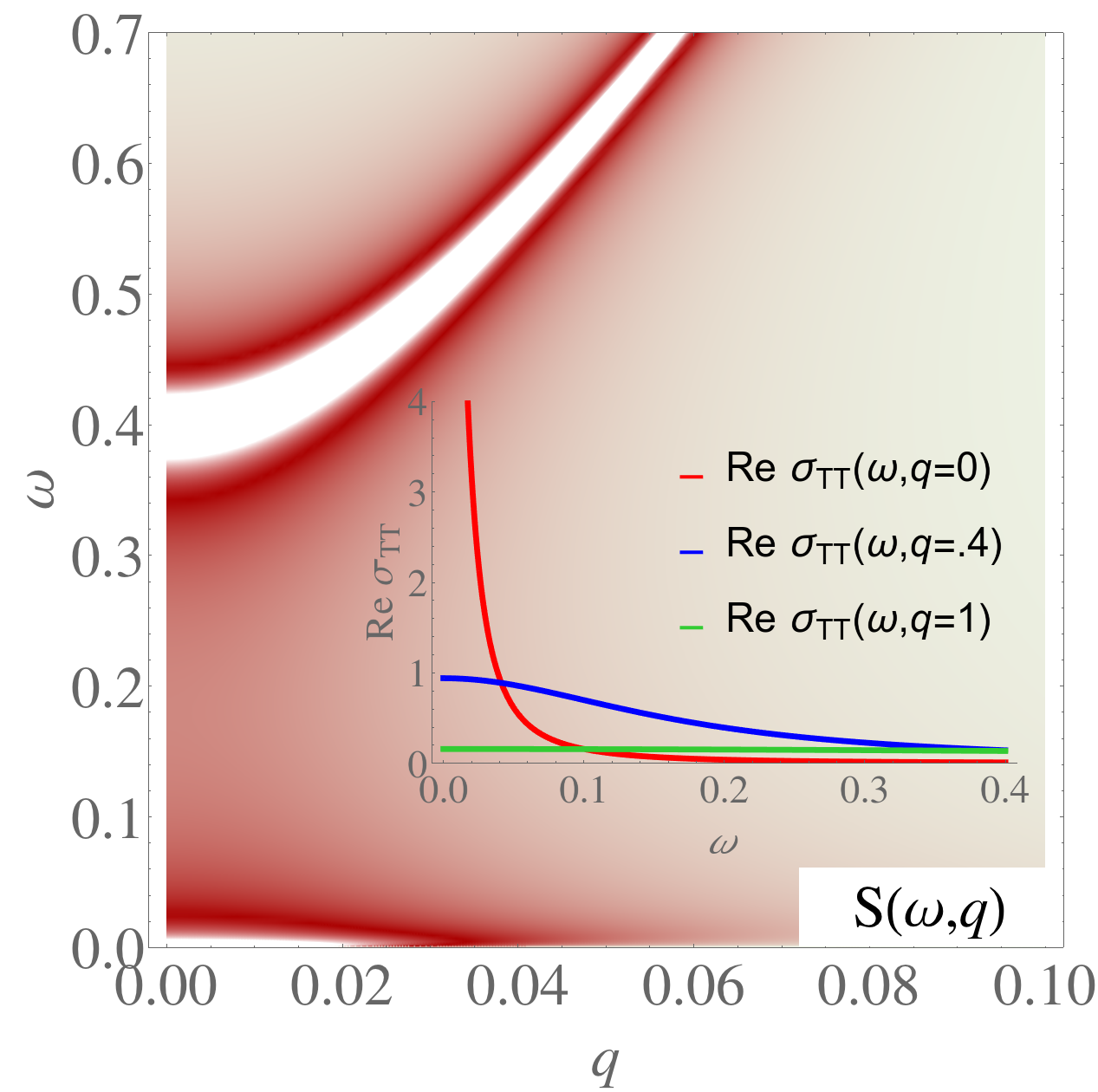}
\caption{The spectral function Eq.~\eqref{photonpropdef} using the self-energy Eq.~\eqref{transpropfluid} of the viscous charged fluid for $D_\tT =0.5$, $\tau=100$, $\omega_\mathrm{p}=1$, $c_l=10$. A transverse plasma-polariton with mass $\omega_\mathrm{p}$ is a propagating mode. The zero-momentum DC Drude peak is almost not visible. The broadening is due to the Drude scattering $~1/\tau$ and to the viscous diffusion $\sim D_\tT q^2$. Inset: the diffusive broadening of the Drude peak as function of the momentum $q$.}\label{Fig:viscous}
}
\hspace{.02\textwidth}
\parbox[t][][t]{.48\textwidth}{
\includegraphics[width=0.48\textwidth]{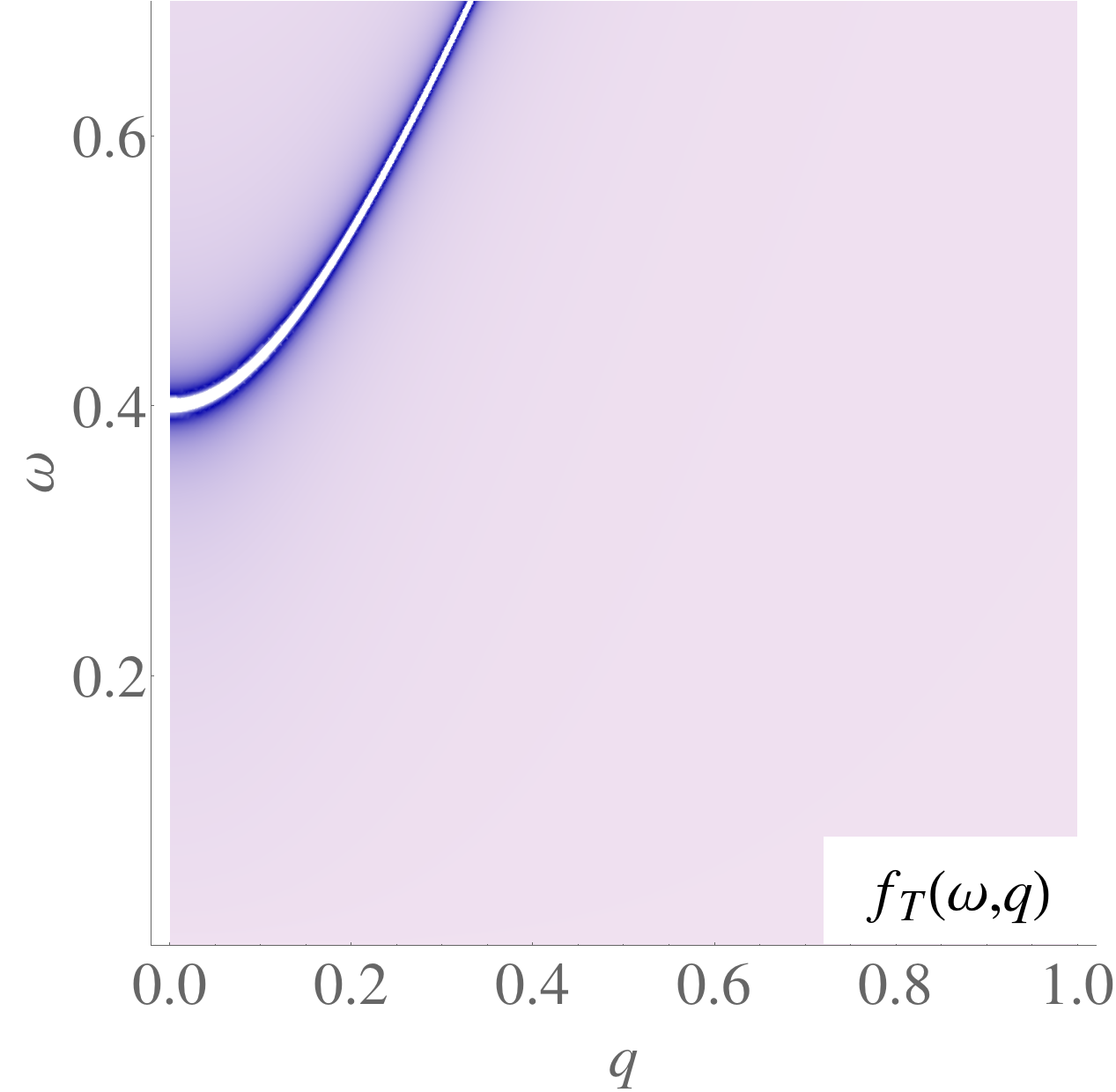}
\caption{The EELS spectral function $f_\tT(\omega,q)$ for the charged viscous fluid for $c_\tK = c_\tL \simeq 1.73$ corresponding to $\nu=0.5$ and $c_\tT=1$, and $\omega_\mathrm{p}=1$. We see a propagating mode with velocity $c_\tK$ and the plasmon mass $\omega_\mathrm{p}$. The pole strength is given by $Z(q)=\omega_\mathrm{p}/2\sqrt{\omega_\mathrm{p}^2+c_\tK^2 q^2}$.}
\label{fig:viscous energy loss}
}
\end{figure}

The transverse response of a classical fluid is dissipative; this is the regime of hydrodynamics. It follows from general hydrodynamical principles that the transverse photon self-energy have the form 
\begin{equation}
\Pi_{\tT} (\omega, q) =  \omega_\mathrm{p}^2 \frac{\omega} {\omega + \ti ( D_{\tT}  q^2 + 1 / \tau) }.
\label{transpropfluid}
\end{equation}
where, as before, $\omega_\mathrm{p}$ is the plasmon frequency. This self-energy implies that translation symmetry is broken, or equivalently that the total momentum is no longer conserved, as parametrized by a momentum relaxation time $\tau$. This parameter is equivalent to the phenomenological {\em scattering time} in Drude theory~\cite{AshcroftMermin76}. In addition, the fluid is viscous, and this is parametrized in  terms of the transverse momentum diffusion constant $D_{\tT}$,
which is set by the  shear viscosity $\eta_{\rm shear}$ and mass density $\rho$ of the fluid,
\begin{align}
D_\tT = \frac{\eta_{\rm shear}}{\rho}.
\label{diffcon}
\end{align}
Obviously the transverse self-energy is characterized by a purely dissipative pole. In the small momentum limit, assuming $\tau$ is finite, this reduces to the familiar 
Drude results
\begin{align}
\hat{\varepsilon}_{\tT \tT} (q \rightarrow 0, \omega) & = \varepsilon_0\big( 1 - \frac{\omega_\mathrm{p}^2}{\omega^2} \frac {1} {1 + i/ ( \omega \tau)} \big), \nonumber \\
\hat{\sigma}_{\tT \tT} (q \rightarrow 0, \omega) & = \varepsilon_0\frac{\omega^2_\mathrm{p} \tau}{ 1 - i \omega \tau }.
\label{drudeEM}
\end{align}
For later reference, we show in Fig.~\ref{Fig:viscous} the photon spectral function as a function of the momentum and frequency, together with Drude peak of the real part of the optical conductivity. 

Turning to the longitudinal response, the fluid is now characterized by a propagating sound mode which is hydrodynamically protected in the Galilean continuum. Let us
for simplicity ignore the damping $\propto q^2$ and momentum relaxation. It follows immediately that
\begin{equation}
\hat{\varepsilon}_{\tL\tL} (\omega,q) = \varepsilon_0 (1  - \frac{\omega^2_\mathrm{p}}{\omega^2 - c_\tK^2 q^2} ),
\label{fluidsound}
\end{equation}
where $c_\tK$ is the sound velocity. The Debye screening length must be $l_\mathrm{D} \approx c_\tK / \omega_\mathrm{p}$ and we find for the EELS spectrum Eq.~\eqref{eq:energy loss definition},
\begin{equation}
 f_\tT(\omega ,q ) = - \mathrm{Im} \frac{\omega^2 - c_\tK^2 q^2}{\omega^2 -c_\tK^2 q^2 - \omega_\mathrm{p}^2}.
 \label{eq:viscous EELSspectrum}
\end{equation}
This is plotted in Fig.~\ref{fig:viscous energy loss}, and one immediately discerns the  canonical plasmon characterized by a dispersion $\omega = \sqrt{\omega_\mathrm{p}^2 + c^2_\tK q^2}$ and pole strength $Z(q)=\frac{\omega_\mathrm{p}}{2\sqrt{\omega_\mathrm{p}^2+c_\tK^2 q^2}}$.  

\subsection{Electromagnetism of the isotropic Wigner crystal}\label{subsec:Electromagnetism of the isotropic Wigner crystal}

Let us now turn to the electromagnetic response of charged version of isotropic quantum elasticity. This in principle entails the bosonic Wigner crystal and one would expect 
it to be textbook material. However, we do find a surprise in the form of a `polariton-like' massless mode that appears at very long wavelengths.  We did not manage find any reference to this phenomenon in the literature. One should keep in mind that we are considering a charged crystal living in a perfectly homogeneous isotropic background that, at the same time, carries the compensating positive charge to avoid the Coulomb catastrophe. This is an unphysical limit: any laboratory Wigner crystal will be subjected to a translational symmetry breaking of the background charges that will immediately pin the crystal. However, as matter of principle it is entertaining to contemplate the question of what happens in the impeccable Euclidean continuum. In addition,  the 
charged elastic medium  is another useful template to enhance the intuition of how things work in the quantum liquid crystals. As usual, at high momenta/short length scale one will recover the `solid' behavior in the liquids.

It is a famous observation that in the continuum limit any charged object will behave like an ideal metal. Let us start with a simple notion: the longitudinal 
dielectric function. It follows immediately from the effective action Eq.~\eqref{eq:solid EM Lagrangian} that the longitudinal dielectric function of the charged isotropic medium is 
\begin{equation}
\hat{\varepsilon}_{\tL \tL} (\omega,q) = \varepsilon_0 (1  - \frac{\omega^2_\mathrm{p}}{\omega^2 - c^2_\tL q^2}).
\label{eq:epslongXtal}
\end{equation}
The Debye screening length can be read off to be $l_\mathrm{D} \sim c_{\tL}/\omega_\mathrm{p}$: the Wigner crystal also behaves like an ideal metal, in the regard that a charge is screened perfectly. Compared to the fluid of the previous section, the only difference is that the longitudinal phonon velocity $c_\tL$ enters instead 
of the velocity of sound $c_\tK$. 

The loss function of the  Wigner crystal follows immediately from Eqs.~\eqref{eq:energy loss definition},\eqref{eq:epslongXtal}, 
\begin{equation}
 f_\tT(\omega ,q ) = -\mathrm{Im} \frac{\omega^2 - c_\tL^2 q^2}{\omega^2 -c_\tL^2 q^2 - \omega_\mathrm{p}^2}.
 \label{crystaleels}
\end{equation}
This reveals yet again that the crystal carries the same plasmon as the `ordinary' liquid Eq.~\eqref{eq:viscous EELSspectrum} with the only difference that it now propagates with the {\em longitudinal phonon velocity}. In hindsight this is not all surprising realizing that we are dealing with a 2D charged solid interacting with 2D electromagnetic fields. After all, one can explain the plasmon frequency invoking charged plates that freely oscillate relative to each other. 

\begin{figure}
\parbox[t][][t]{.48\textwidth}{
\includegraphics[width=0.48\textwidth]{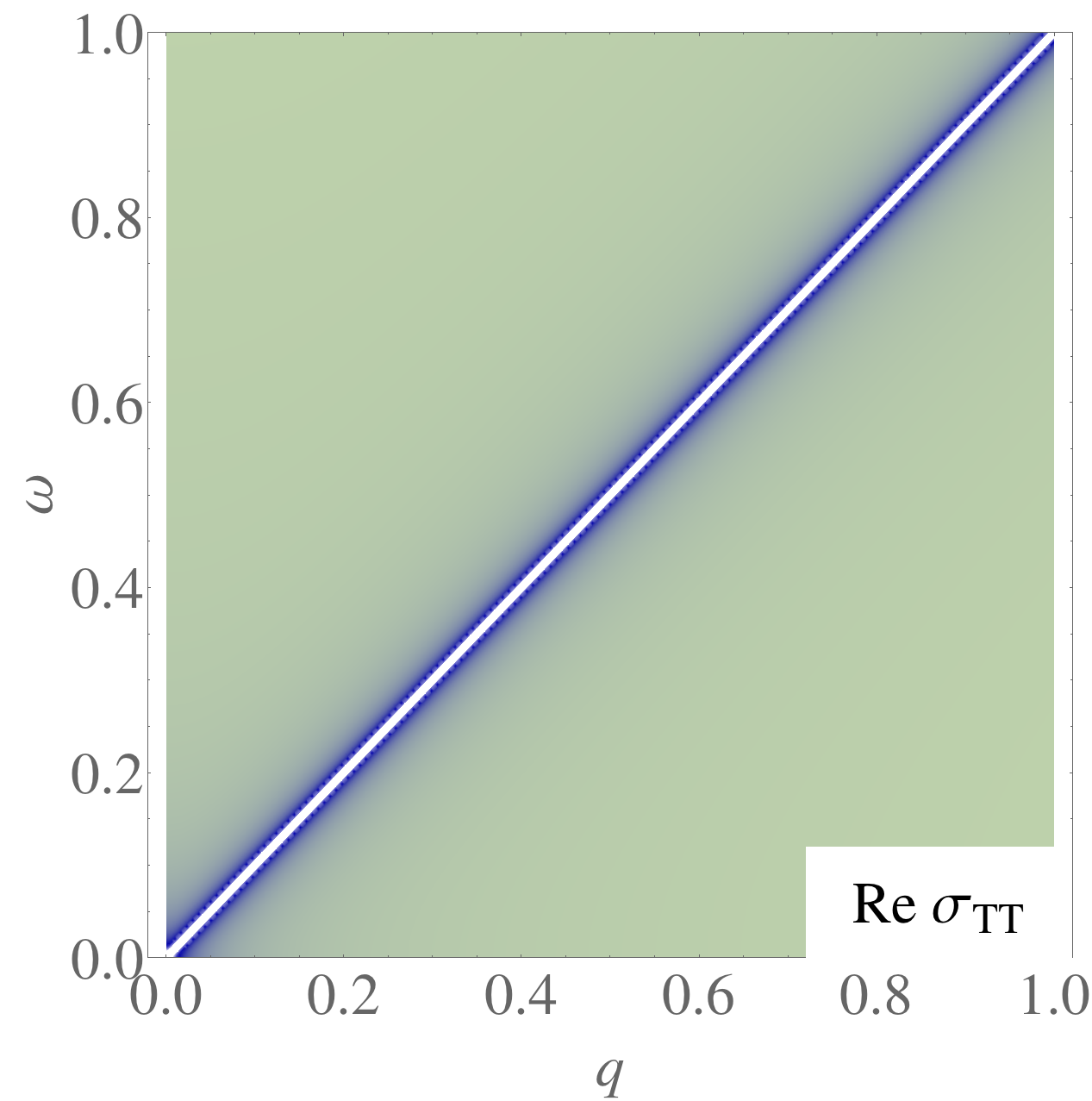}
\caption{The real part of the transverse optical conductivity $\mathrm{Re }~\hat{\sigma}_{\tT\tT}(\omega,q)$ for the Wigner crystal for $c_\tT =1$ and $\omega_\mathrm{p} =1$, showing a propagating pole with constant pole strength corresponding to the transverse phonon.}
\label{Fig:sigmaTTWigner}
}
\hspace{.02\textwidth}
\parbox[t][][t]{.48\textwidth}{
\includegraphics[width=0.48\textwidth]{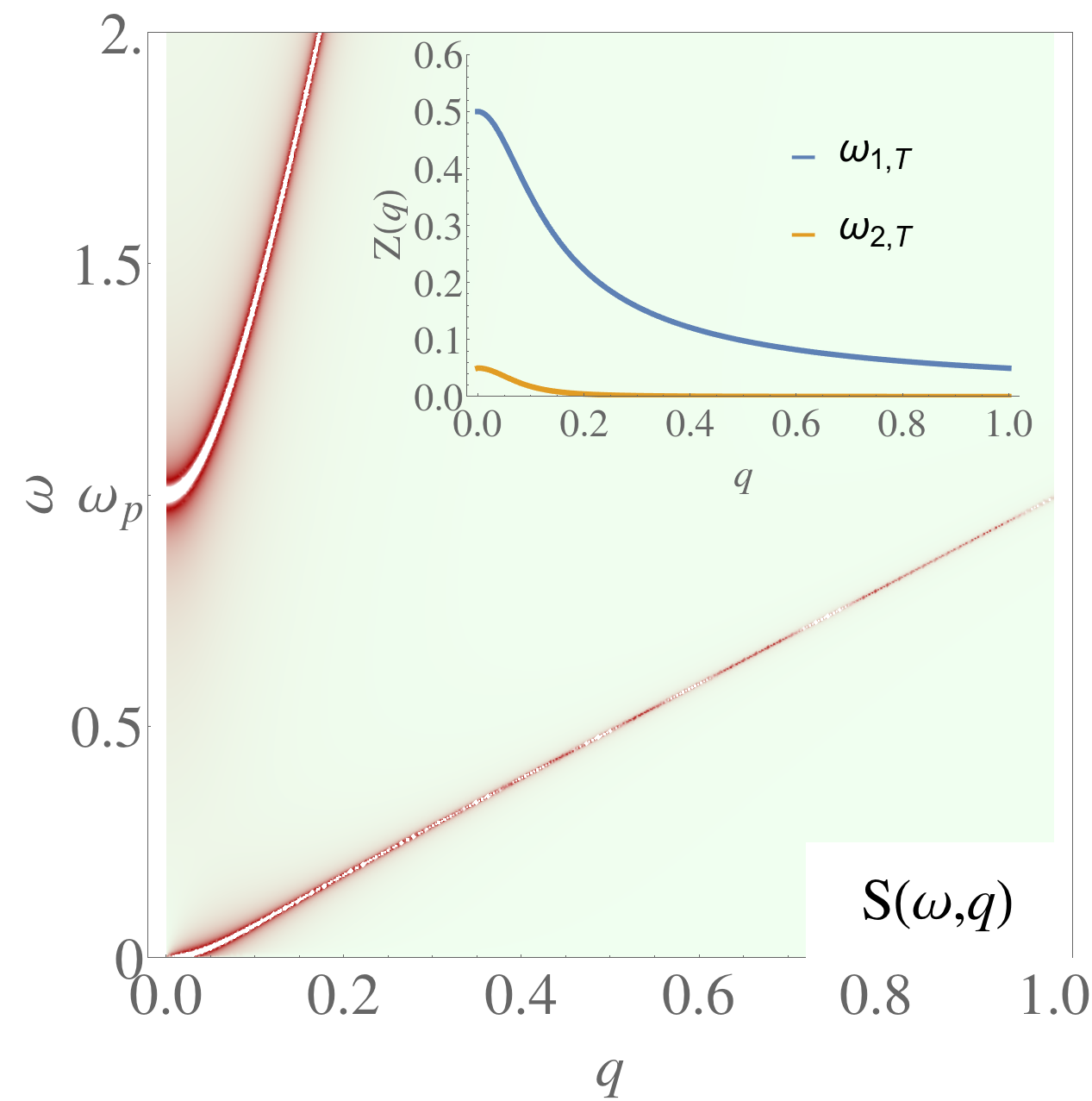}
\caption{The transverse spectral function of the Wigner crystal from Eq.~\eqref{eq:Wigner photon propagator} in units set by $\omega_\mathrm{p} = c_\tT =1$. For clarity of the figure we take $c_l = 10 c_\tT$ which is much smaller than reality. The poles are infinitely sharp and the width denotes the relative pole strength. We identify the usual photonic plasma-polariton with only slightly modified velocity $\sqrt{c_l^2 + c_\tT^2} \approx c_l$ and a much weaker transverse phonon. The disappearance of the phonon near $q\to 0$ is an artifact due to the numerical resolution.  Inset: pole strengths in units of $1/\varepsilon_0$. The phonon is weaker by a factor of $c_\tT/c_l$ than the polariton at $q=0$. Both pole strengths decrease as $1/q^2$ at low momenta.}
\label{Fig:WignerSpectral}
}
\end{figure}

Let us now turn to the transverse response. One reads off the photon self-energy from Eq.~\eqref{eq:solid EM Lagrangian} for the charged solid, 
\begin{equation} 
\Pi_\tT (\omega,q) = \omega^2_\mathrm{p} \frac{\omega^2}{\omega^2 - c^2_\tT q^2},
\label{photonsesolid}
\end{equation}
which leads to the transverse conductivity,
\begin{align} 
\hat{\sigma}_{\tT \tT} (\omega, q)  &=  \varepsilon_0 \omega_\mathrm{p}^2 \frac{ \ti \omega}{\omega^2 - c_\tT^2 q^2}.
 \label{eq:Wigner crystal conductivity}
\end{align}
At zero momentum $\hat{\sigma}_{\tT\tT} (\omega, 0) =  \varepsilon_0 \omega_\mathrm{p}^2 ( \ti /\omega )$: the pole $\sim \frac{1}{\omega}$ has weight $\varepsilon_0 \omega_\mathrm{p}^2$, and it is related to a Drude  delta-function peak at zero frequency in  $\hat{\sigma}_1(\omega, 0)$, indicative of a perfect conductor.  This is yet again according to the expectations. However, in stark contrast  with the dissipative fluid result in Fig. \ref{Fig:viscous} with broad Drude peaks; here the finite momentum conductivity Eq.~\eqref{eq:Wigner crystal conductivity} is characterized by a sharp propagating pole with a strength $\varepsilon_0\omega_\mathrm{p}^2/2$ that is independent of momentum, see Fig. \ref{Fig:sigmaTTWigner}. This is of course the transverse phonon that does carry electromagnetic spectral weight.

However, hell breaks loose when the transverse modes of the {\em combined} photon--stress photon system are considered. For this purpose we zoom in on the full photon propagator $\langle A^{\dagger}_\tT(\omega,q)A_\tT(-\omega,-q)\rangle$ and the spectral function $\mathrm{Im}~\langle A^{\dagger}_\tT(\omega,q) A_\tT(0,0)\rangle$. The poles or the propagating modes are the solutions of Eq.~\eqref{transversedisp}. Using the exact expression,
\begin{align}
\hat{\varepsilon}_{\tT \tT}(\omega,q) = \varepsilon_0(  1 - \frac{\omega^2_\mathrm{p}}{\omega^2 - c^2_\tT q^2}),
\label{xtalTdiel}
\end{align}
the photon propagator is,
\begin{equation}
 \langle A^{\dagger}_\tT A_\tT\rangle = \frac{1}{\varepsilon_0}\frac{\omega^2 - c_\tT^2 q^2}{(\omega^2 - c_\tT^2q^2)(\omega^2 - c_l^2 q^2) - \omega_\mathrm{p}^2 \omega^2}.\label{eq:Wigner photon propagator}
\end{equation}
There are two pairs of solutions associated with the mode-coupled photon resp. transverse stress photon, 
\begin{align}
\omega^2_{\tT} (q) &= \frac{\omega^2_\mathrm{p} + (c_l^2 + c_\tT^2)q^2}{2} 
 \pm \sqrt{ (\omega^2_\mathrm{p}+ (c_l^2+c^2_\tT)q^2)^2 - 4c^2_\tT c^2_l q^4}.
\label{Xtalpolaritons}
\end{align}
Here the Galilean velocity $\sqrt{c^2_l+c^2_\tT} \simeq c_l$ is an artifact of the non-relativistic theory of elasticity coupled to EM fields. We see two modes that are composed of the transverse photon and phonon. Due to $c_\tT \ll c_l$ and the finite plasmon gap $\omega_\mathrm{p}$, the modes couple very little in general. At high momenta $c_l q \gg \omega_\mathrm{p}$, we simply recover the photon and phonon dispersion.  However, at very large wavelength a surprise is happening. The characteristic momentum turns out to be set by $\omega_\mathrm{p}/c_l = 1/\lambda_\tL$, the quantity that is associated with the London penetration 
depth. Expanding Eq. (\ref{Xtalpolaritons}) for small momenta $c_ l k \ll \omega_\mathrm{p} $ yields
\begin{align}
\omega_{\tT,1} (q) & = \sqrt{ (\omega^2_\mathrm{p} + (c_l^2+c^2_\tT)q^2} \simeq \sqrt{\omega_\mathrm{p}^2+c_l^2 q^2} + \mathcal{O}(q^2),\nonumber \\
\omega_{\tT,2} (q) & = \frac{c_\tT c_l q^2}{\sqrt{ (\omega^2_\mathrm{p} + (c_l^2+c^2_\tT)q^2)}} \simeq \frac{c_l c_\tT}{\omega_\mathrm{p}} q^2 + \mathcal{O}(q^4).
\label{Xtallongwave}
\end{align}
The $\omega_{\tT,1}$ mode is familiar: it is just the plasma-polariton with a mass $\omega_\mathrm{p}$. However, there is also a massless mode inherited from the transverse phonon with a quadratic dispersion and a very small `mass-coefficient' $\sim 2 c_l c_\tT/\omega_\mathrm{p}$! In Fig.~\ref{Fig:WignerSpectral} we show the full photon spectral function and the spectral weights. Both pole strengths decrease as $\sim q^2$ from the zero-momentum values $1/2\omega_\mathrm{p}$ and $c_\tT/2c_l\omega_\mathrm{p}$, for the photon and phonon, respectively. Note that taking $q = 0$ first, the photon propagator obtains only the plasma-polariton. This is a singular limit.

\subsection{Superconductivity and electromagnetism of the quantum nematic}\label{subsec:Superconductivity and the electromagnetism of the quantum nematic}

This might well be the most important part of this long theoretical story with regard to the  potential relevance to experiments in high-$T_\mathrm{c}$ superconductors. It is of course 
obvious that at large momenta one will recover the `maximally-correlated' electromagnetic features of the Wigner crystal also in the nematic.  We just learned that the gross qualitative differences reside at rather large momenta in transverse probes revealing the reactive responses towards shear stresses in the medium, instead of the dissipative responses of the weakly-interacting liquid. Unfortunately this kinematical regime is not experimentally accessible because of the smallness of $c_\tT/c_l$ ratio. But what is 
happening at scales small compared to the shear Higgs mass $\Omega$ and the inverse dislocation penetration depth $1/\lambda_\mathrm{d}$, where the system has turned into a fluid? Is there any unconventional feature to be discerned at these small momenta, which is perhaps even in reach of optical experiments? 

The longitudinal response is a different story since the small-momentum regime is in principle accessible by EELS and RIXS. As we will review below, already in 2007 a {\em smoking gun} prediction was put forward in the form of an extra mode visible in the loss spectrum~\cite{CvetkovicNussinovMukhinZaanen08}, which has been on the benchmark list of the machine builders ever since. Before we turn to these matters, let us first highlight perhaps the most stunning outcome of this whole endeavor: the proof that the quantum nematic is also a superconductor. 

As we already discussed, from the transverse photon propagator one can directly deduce the Meissner effect, using Eq.~\eqref{eq:penetration depth from self-energy}. The photon self-energy follows
immediately from the effective action for the nematic Eq.~\eqref{eq:nematic EM Lagrangian},
\begin{equation}
\Pi_{\tT} (\omega,q) =  \omega^2_\mathrm{p} \frac{\omega^2 (\omega^2 - c^2_\td q^2) - \Omega^2 (\omega^2 - c^2_\tR q^2)}{ (\omega^2 -c^2_\tT q^2) (\omega^2 - c^2_\td q^2) - \Omega^2 ( \omega^2 - c^2_\tR q^2)}.
\label{eq:nematicphotSE}
\end{equation}
The Meissner effect is the existence of a finite penetration depth in the static limit $\omega \to 0$. Taking this limit and using the definition $c_\tR = c_\td / \sqrt{2}$, we find
\begin{align}\label{eq:penetration depth nematic}
\lambda(\omega = 0,q) = \lambda_\tL \sqrt{1 + \lambda_\mathrm{s}^2q^2}.
\end{align}
Here $\lambda_\tL = c_l/ \omega_\mathrm{p}$ is the usual London penetration depth while $\lambda_\mathrm{s} = \sqrt{2} c_\tT / \Omega$ is the {\em shear penetration  depth}. Since $c_\tT \approx c_\td$ up to factors of order 1, this is related to the dislocation penetration depth $\lambda_\td = c_\td /\Omega$ of Sec.~\ref{sec:Quantum nematic}. Obviously, in the Wigner crystal the photon self-energy vanishes at zero frequency and a static magnetic field can freely penetrate the medium. However, at the moment the Higgs mass $\Omega$ and the associated length scale $\lambda_\mathrm{s}$ of the dual shear superconductor become finite the {\em magnetic field is expelled}. In fact, for a uniform ($q=0$) magnetic field the penetration depth becomes $\lambda_\tL$ as if one is dealing with a meat-and-potato BCS superconductor! This is the hard evidence we have in the
offering for the `theorem': {\em upon losing its shear rigidity a charged bosonic Wigner crystal at zero temperature will become a superconductor}.     

\begin{figure}
 \begin{center}
  \includegraphics[width=.7\textwidth]{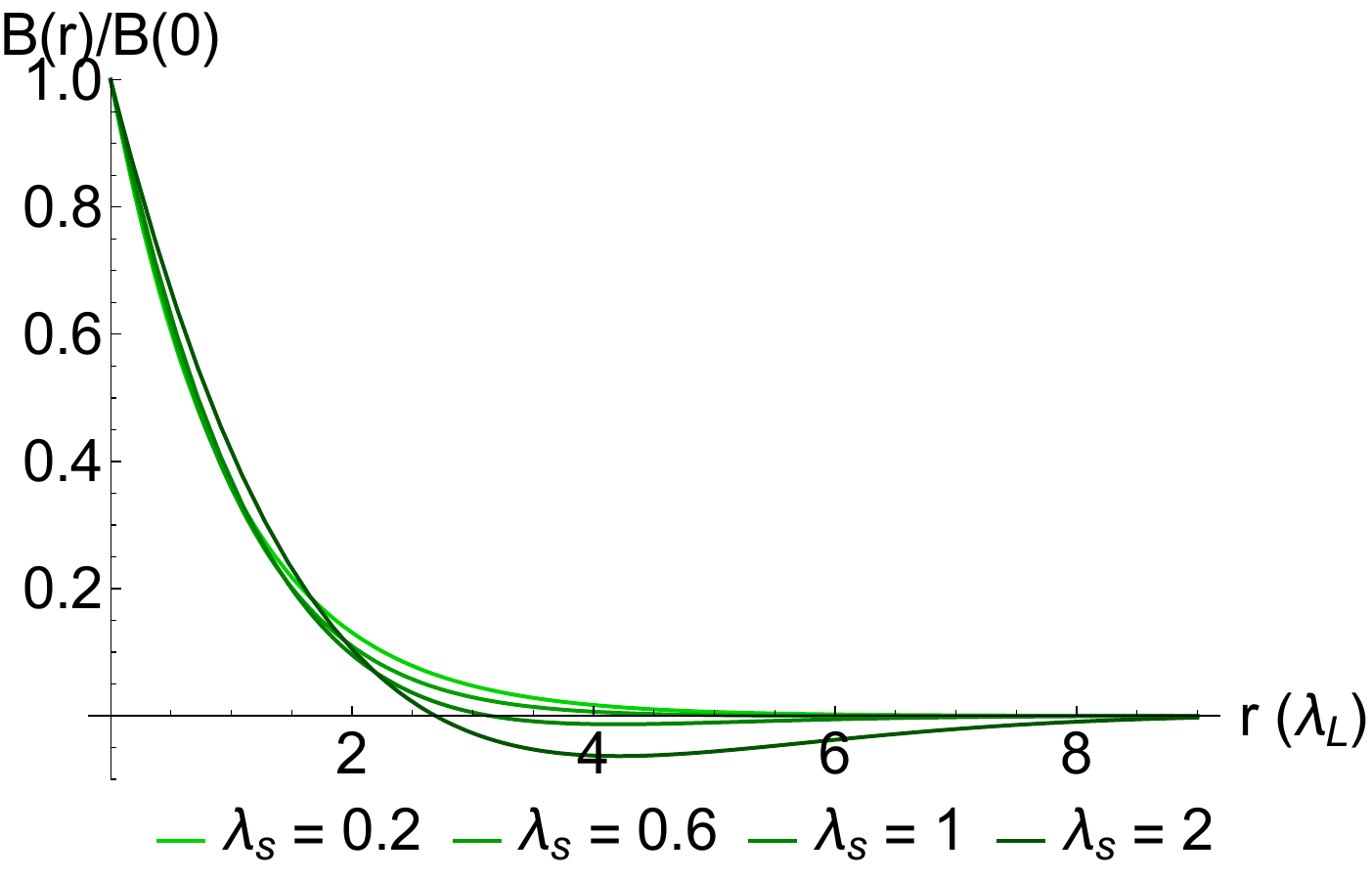}
 \caption{Profile of the DC magnetic field $B$ in the quantum nematic as a function of distance $r$ within the medium in units of the London penetration depth $\lambda_\tL \equiv 1$, normalized with respect to $B(0)$, following from Eq.~\eqref{eq:penetration depth nematic}. 
 For small shear penetration depth $\lambda_\mathrm{s} \ll \lambda_\mathrm{L}$, the screening follows the ordinary exponential decay. However when $\lambda_\mathrm{s}$ is of the order of $\lambda_\mathrm{L}$ or greater, the magnetic field shows damped oscillatory screening, where the sign of the field is even reversed at certain distances. }\label{fig:nematic Meissner screening}
 \end{center}
\end{figure}

Is the response to static magnetic fields of our maximally-correlated superconductor completely conventional? In fact, there is a second scale in the problem: the shear penetration depth $\lambda_\mathrm{s}$. This length scale has a very different status from the usual coherence length (``pair size''), since it signals that at this length scale the system rediscovers that there 
 is a reactive response to shear forces. This has a very unusual consequence for the penetration of spatially modulated ($q>0$) magnetic fields. Surely, when the length scale of this modulation is 
 large compared to $\lambda_\mathrm{s}$, one finds the ubiquitous London response. However, when it becomes of order and smaller than $\lambda_\mathrm{s}$, according to
 Eq.~\eqref{eq:penetration depth nematic} a momentum dependence in the magnetic penetration depth develops.  The meaning of this becomes clear by computing the $r$-dependence of the magnetic field strength $B(r)$, penetrating the superconductor close to its boundary with the vacuum as shown in 
 Fig.~\ref{fig:nematic Meissner screening}. When $\lambda_\tL \gg \lambda_\mathrm{s}$ one finds the usual exponential decay. However, when $\lambda_\tL \lesssim  \lambda_\mathrm{s}$ one finds 
 a surprising {\em overscreening} of the magnetic field~\cite{ZaanenNussinovMukhin04}.  However, in the systems of physical relevance the shear penetration  will be typically quite small compared to the London penetration depth. The 
 take-home message is that with regard to the response to static magnetic fields our maximally-correlated nematic superconductor will behave just as an extreme type-II conventional superconductor. As we have encountered more often, the unusual solid-like features are perfectly hidden for the experimentalist's eye. 
 
Let us now turn to the frequency dependence of the transverse response. We first inspect the easier optical conductivity before turning to the problem of the spectrum of 
transverse modes. The optical conductivity follows directly from Eqs.~\eqref{opticalcond},\eqref{eq:nematicphotSE}, 
\begin{align}
 \hat{\sigma}_{\tT \tT} (\omega,q) = \frac{\ti \varepsilon_0\omega_\mathrm{p}^2}{\omega} 
 \frac{\omega^2 ( \omega^2 - c_\td^2 q^2) - \Omega^2 (\omega^2 - c_\tR^2 q^2)}{(\omega^2 - c_\tT^2 q^2)(\omega^2 - c_\td^2 q^2 ) - \Omega^2 (\omega^2 - c_\tR^2 q^2)},
 \label{eq:optconnematic}
\end{align}
where $c_\tR = c_\td/\sqrt{2}$ and $c_\td$ is of order $c_\tT$; the precise proportionality factor is presently not known. As expected, one immediately infers that in the $q \to 0$ limit this yields just the perfect conductor in the from of the delta function peak in $\hat{\sigma}_1(0,\omega)$ at zero frequency. At momenta much larger than $1 /\lambda_\mathrm{s}$, the response will become similar to the Wigner crystal: in this regime one can send $\Omega \rightarrow 0$ and recovers the transverse-phonon optical conductivity. 

What happens at frequencies and momenta in the fluid regime, up to scales of order $\Omega$ and $1 /\lambda_\mathrm{s}$? We can first look at a simplified case, where we artificially set $c_\td = c_\tR$:
\begin{equation}
 \hat{\sigma}_{\tT \tT} (\omega,q) = \varepsilon_0\omega_\mathrm{p}^2 \frac{\ti}{\omega} \frac{ \omega^2 - \Omega^2} { \omega^2 - c_\tT^2 q^2 - \Omega^2}.
 \label{eq:optconnematiceasy}
\end{equation}
In this case, there is a positive frequency pole and a zero frequency Drude peak with pole strengths that follow 
\begin{align}
 \hat{\sigma}_{\tT,1} (\omega,q) &\sim \frac{\pi \omega_\mathrm{p}^2}{2}  \frac{c^2_\tT q^2} {\Omega^2 + c_\tT^2 q^2} \delta (\omega - \sqrt{\Omega^2 + c^2_\tT q^2 }) + \frac{\pi \omega_\mathrm{p}^2}{2}\frac{\Omega^2}{c_\tT^2 q^2+\Omega^2}\delta(\omega).
 \label{eq:optconnematicpole}
\end{align}
The pole in the first line reduces to the transverse phonon for large momenta $c_\tT^2 q^2 \gg \Omega$, but obtains a Higgs gap for small momenta $c_\tT^2 q^2 \ll \Omega$, indicating the loss of shear rigidity. Its pole strength decreases as $q^2$ for small momenta $c_\tT^2 q^2 \ll \Omega$, and becomes momentum-independent for large momenta $c_\tT^2 q^2 \gg \Omega$. Apparently for small $q$, all the spectral weight is transferred to the Drude peak.

\begin{figure}
\parbox[t][][t]{.48\textwidth}{
\includegraphics[width=0.48\textwidth]{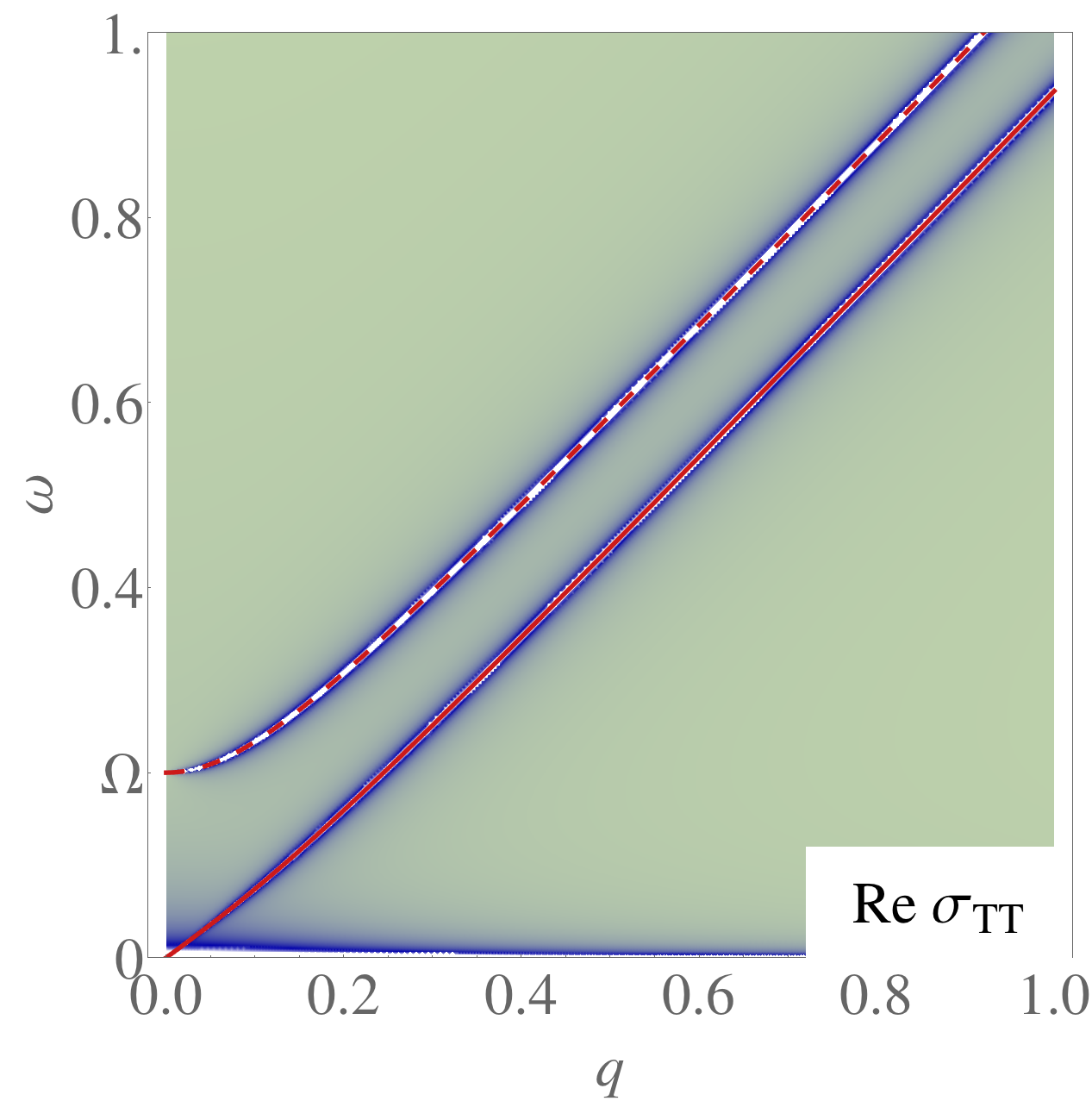}
\caption{The real part of the transverse optical conductivity $\hat{\sigma}_{\tT \tT}(\omega,q)$ of the nematic with $c_\tT = c_\td=1$, $c_\tR = c_\td/\sqrt{2}$, $\Omega=0.2$, showing the transverse phonon that has obtained a Higgs gap from loss of shear rigidity, a low-energy massless mode with velocity $c_\tR$ that we identify as the rotational Goldstone mode, and the zero-frequency Drude peak. The red dashed and solid lines are Eqs.~\eqref{eq:nematic conductivity plasmon}, \eqref{eq:nematic conductivity rotational Goldstone} respectively.}
\label{Fig:sigmaNematicFull}
}
\hspace{.02\textwidth}
\parbox[t][][t]{.48\textwidth}{
\includegraphics[width=0.48\textwidth]{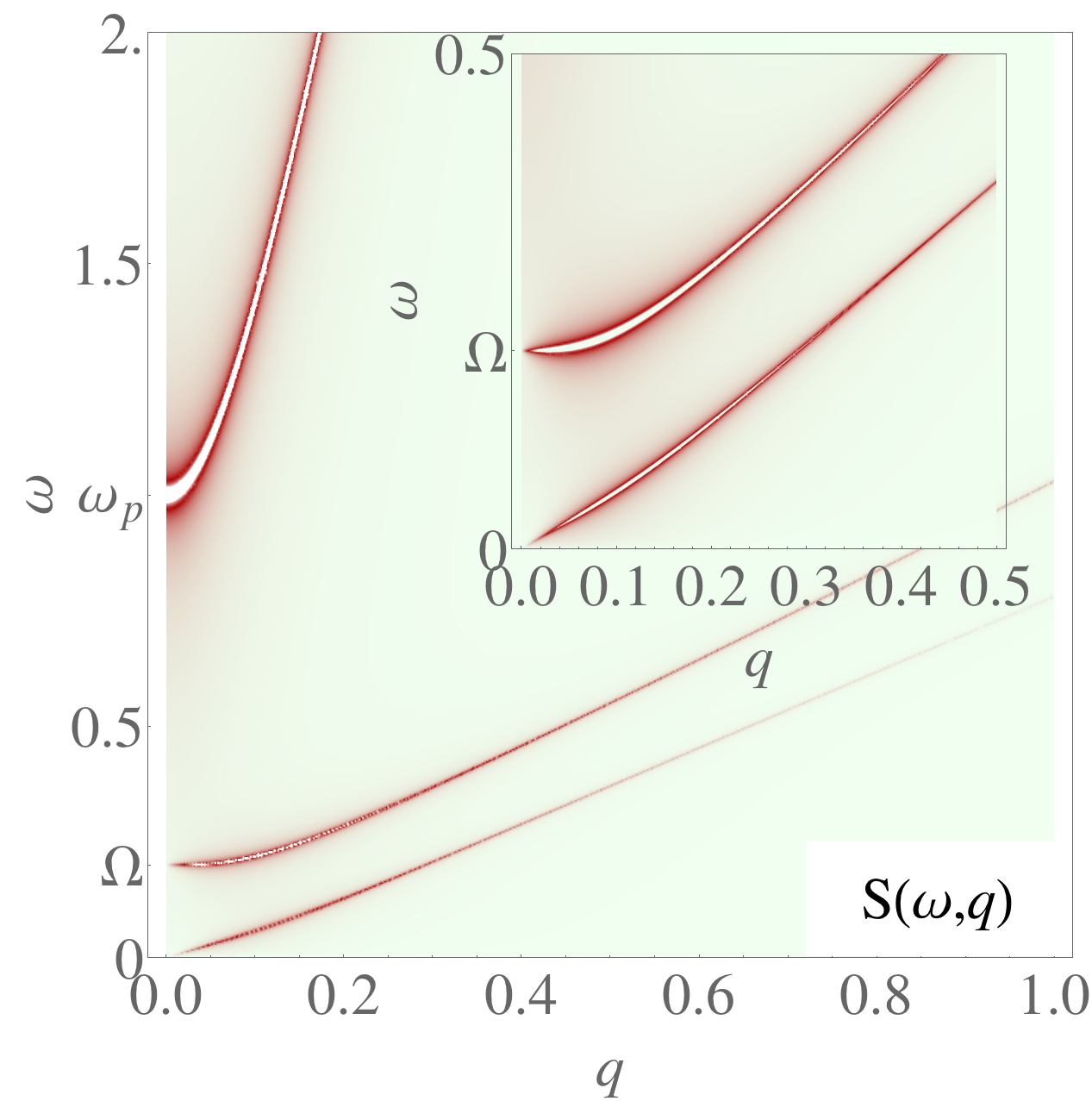}
\caption{The transverse photon spectral function of the nematic for $c_l=10, c_\tT=1, c_\td=0.8, c_\tR = c_\td/\sqrt{2}, \omega_\mathrm{p} =1,\Omega=0.2$ with artificial broadening proportional to the pole strengths.
Most of the spectral weight is concentrated on the transverse photon Eq. \eqref{eq:nematic transverse photon} with mass $\omega_\mathrm{p}$, with the massive transverse phonon and rotational Goldstone appearing at finite $q$ but weak spectral weight. Inset: zoom on the much weaker transverse phonon and the rotational Goldstone mode, Eqs. \eqref{eq:nematic transverse phonon} and \eqref{eq:nematic polariton massless}.}
\label{Fig:Nematic tranverse spectral}
}
\end{figure}

In the physically relevant case $c_\td \approx c_\tT$ and $c_\tR = c_\td/\sqrt{2}$, next to the zero-frequency Drude peak we find two propagating modes with dispersion relations,
\begin{align}
\omega_{2}^2(q) &= c_\tT^2 q^2 + \frac{\Omega^2}{2}\left(1+\sqrt{1+\frac{4(c_\tT^2-c_\tR^2)q^2}{\Omega^2}}\right) \label{eq:nematic conductivity plasmon}\\
\omega_{3}^2(q) &=  c_\tT^2 q^2 + \frac{\Omega^2}{2}\left(1-\sqrt{1+\frac{4(c_\tT^2-c_\tR^2)q^2}{\Omega^2}}\right).\label{eq:nematic conductivity rotational Goldstone}
\end{align}
The first equation contains the massive shear phonon. However, the second equation shows the emergence of an additional massless mode with velocity $c_\tR$ at low momenta. This mode we identify as the rotational Goldstone mode, see below. The three poles show up clearly in the plot of the real part of the optical conductivity  Fig.~\ref{Fig:sigmaNematicFull}.

Let us now turn to the full transverse spectrum of coupled photon--stress photon modes and the way they show up in the photon spectral function. The full transverse propagator is given by Eq.~\eqref{photonpropdef} with self-energy Eq.~\eqref{eq:nematicphotSE}, and its spectral function is depicted in Fig.~\ref{Fig:Nematic tranverse spectral}. We observe the expected transverse photon or plasma-polariton with gap $\omega_\mathrm{p}$ and two weaker propagating modes at finite momenta.  To solve for the dispersions of the poles $\langle A_{\tT}^{\dagger}(\omega,q) A_{\tT}(0,0)\rangle^{-1} = 0$, we take the limit $\Omega \ll \omega_\mathrm{p}$ and expand around $q = 0$. There are three pairs of solutions, namely
\begin{align}
 \omega^2_1 &= \omega_\mathrm{p}^2 + c_l^2 q^2 + c_\tT^2 q^2 \big(1 + \frac{\Omega^2}{\omega_\mathrm{p}^2} + \frac{\Omega^4}{\omega_\mathrm{p}^4} \big)  + \mathcal{O}(q^4), \label{eq:nematic transverse photon}\\
 \omega^2_2 &= \Omega^2 + c_\tR^2 q^2 - c^2_\tT q^2( \frac{\Omega^2}{\omega_\mathrm{p}^2}+\frac{\Omega^4}{\omega_\mathrm{p}^4}) + \mathcal{O}(q^4), \label{eq:nematic transverse phonon}\\
 \omega^2_3 &=  c_\tR^2 q^2(1 - \frac{c_\tR^2 q^2}{\Omega^2}) + \mathcal{O}(q^6).\label{eq:nematic polariton massless}
\end{align}
The plasma-polariton in the first pair of solutions and has the same form as in the Wigner crystal Eq.~\eqref{Xtallongwave}, but it has corrections due to the Higgs mass $\Omega$. This is the result from mode coupling to the transverse phonon, identified now with the dispersion on the second line with $c_\td \sim c_\tT$, where the $\mathcal{O}(q^4)$-transverse phonon dispersion is dominated by quadratic terms from the condensate, in addition to the Higgs mass. At high momenta, this mode indeed obtains the linear dispersion of the transverse phonon, although all spectral weight is shifted to the polariton. Both the transverse phonon and the remaining third mode now involve the effects of the dislocation condensate, as can be seen from the appearance of the rotational velocities $c_\tR$. A massless propagating mode Eq.~\eqref{eq:nematic polariton massless} emerges and is related to the massless elastic mode of the quantum nematic: the rotational Goldstone mode. This is how the emergence of the additional Goldstone mode shows in the EM spectrum; meanwhile the transverse phonon just gets gapped in the nematic, as required for liquid-like correlations.

Finally we analyze the longitudinal response of the quantum nematic.  From the scalar potential contribution of Eq. (\ref{eq:nematic EM Lagrangian}), we read off the 
longitudinal dielectric function of the quantum nematic, 
\begin{equation}
\hat{\varepsilon}_{\tL \tL} (\omega,q) = 1 - \frac{ \omega^2_\mathrm{p}( \omega^2 -c^2_\tR q^2 -  \Omega^2)}{(\omega^2 -c^2_\tL q^2)(\omega^2 -c^2_\tR q^2) - \Omega^2 ( \omega^2 -c^2_\tK q^2)}.
\label{eq:longdielnematic}
\end{equation}
In the static limit this simplifies to
\begin{equation}
\hat{\varepsilon}_{\tL \tL} (q, \omega=0) = 1 + \omega^2_\mathrm{p} \frac{ c^2_\tR q^2 +  \Omega^2}{c^2_\tL c^2_\tR q^4 + \Omega^2 c^2_\tK q^2}.
\label{eq:longdielnematicstatic}
\end{equation}
At momenta large compared to $\Omega/c_\tL$, one recovers the screening properties of the Wigner crystal.  However, the dual shear superconductor exerts control at small momenta. Then the terms containing $\Omega^2$ dominate and it follows that $\hat{\varepsilon}_{\tL\tL} (q, \omega=0) \simeq 1 + \omega^2_\mathrm{p} / (c^2_\tK q^2)$ The Debye screening 
length is now governed by the velocity of sound $c_\tK$, as it should be in a true fluid! 

\begin{figure}
\parbox[t][][t]{.48\textwidth}{
\includegraphics[width=.48\textwidth]{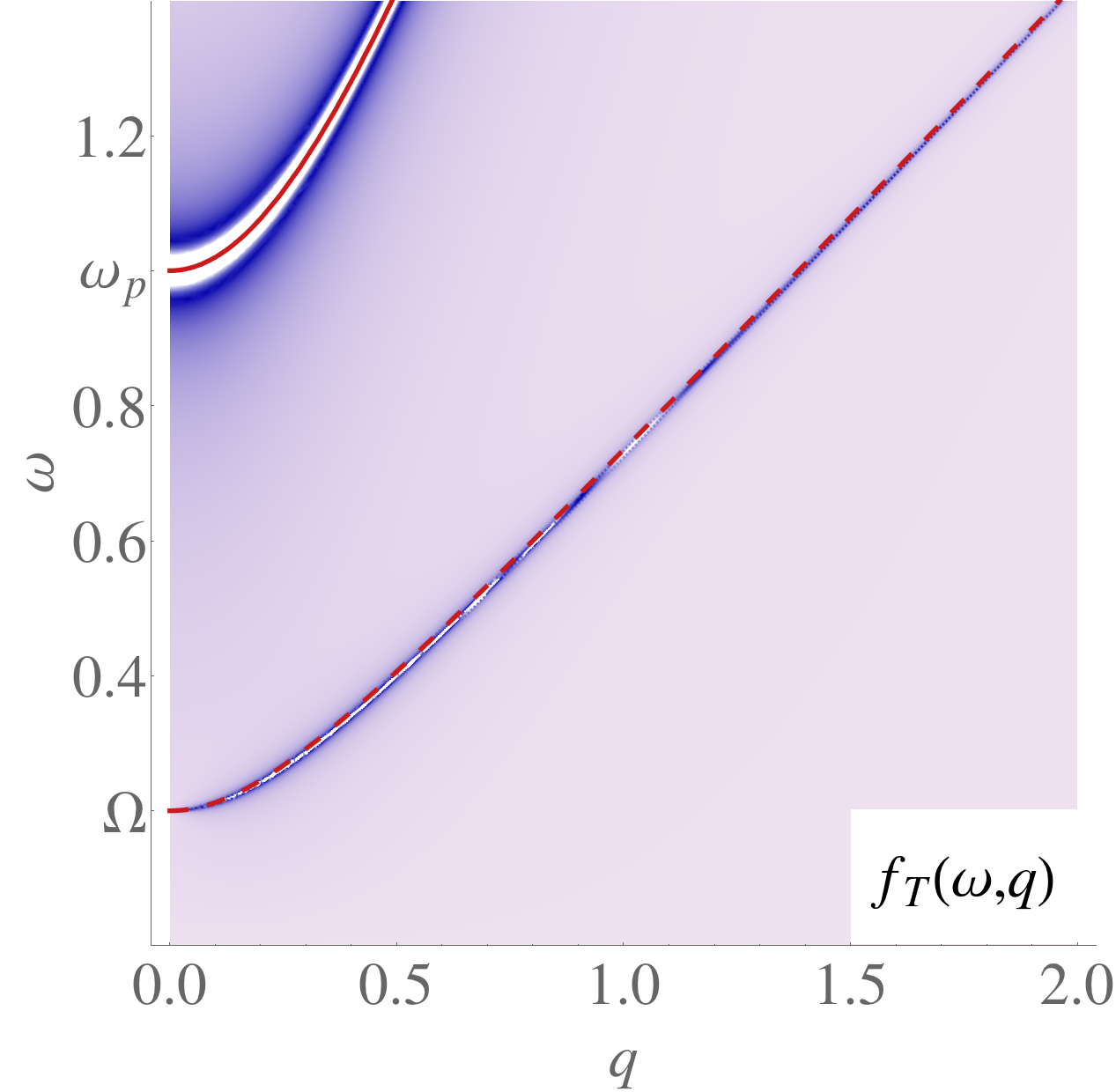}
\caption{The transmissive EELS spectrum $f_\tT(\omega,q)$ for the nematic with $c_\tT= c_\td=1$, $\nu=0.5$, $c_\tR = c_\td/\sqrt{2}$,$\omega_\mathrm{p}=1$, $\Omega=0.2$. The poles are infinitely sharp and the width of the features indicates the spectral weight. There is the longitudinal massive plasmon mode and a massive mode with gap $\Omega$ with much weaker spectral weight, see Fig.~\ref{fig:EELS intensity}. The red dashed and solid lines are the approximate dispersions Eqs.~\eqref{eq:nematic loss condensate dispersion}, \eqref{eq:nematic loss plasmon dispersion} respectively.}
\label{Fig:LongSpec}
 }
\hspace{.02\textwidth}
\parbox[t][][t]{.48\textwidth}{
 \includegraphics[width=.48\textwidth]{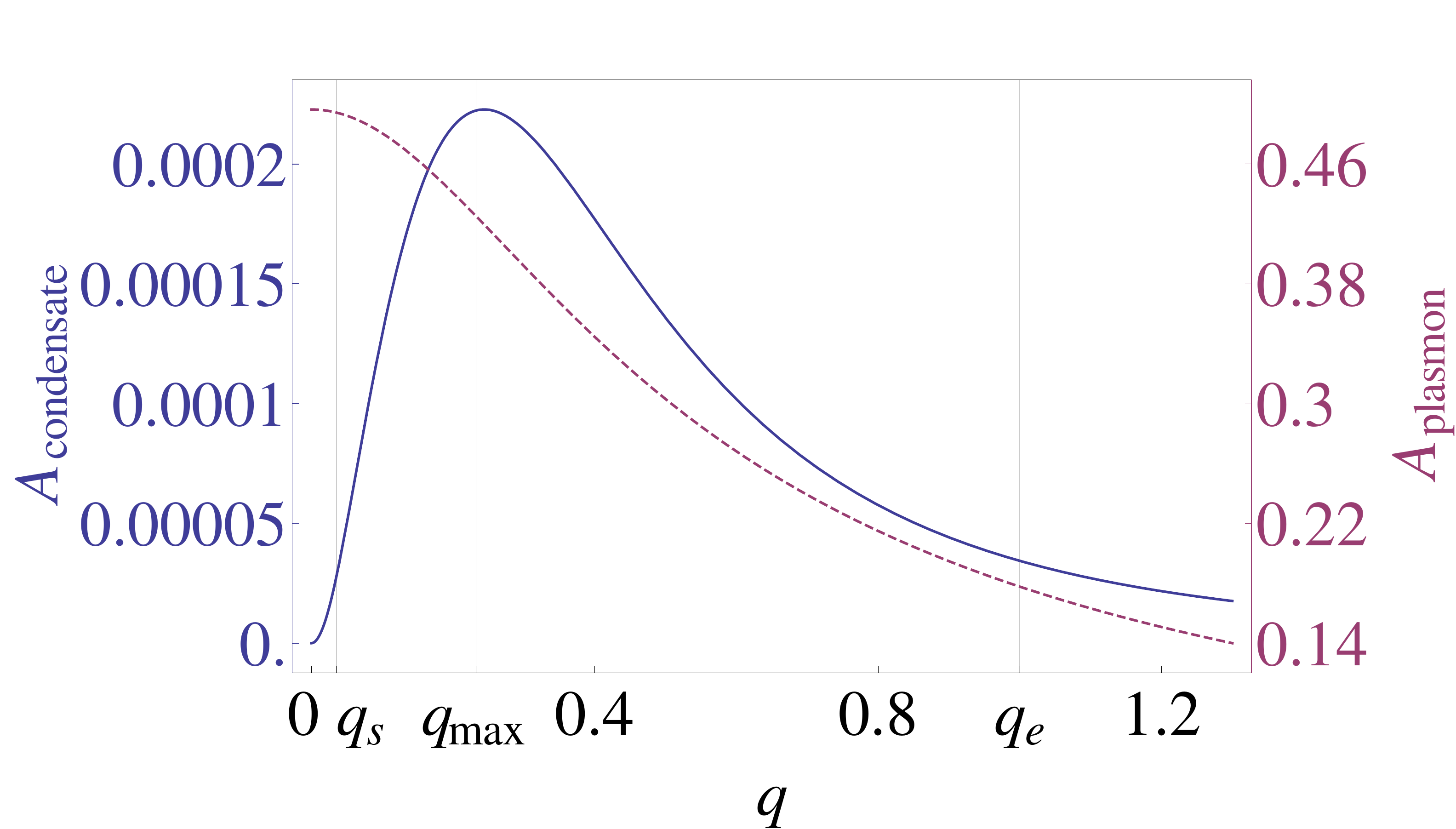}
 \caption{Intensity $Z(q)$ of the two poles, plasmon and massive shear, of the transmissive loss function $f_\tT(\omega,q)$ of the quantum nematic in units of the inverse dielectric constant $1/\varepsilon_0$, for the values $\omega_\mathrm{p} =1\mathrm{eV}$, $\Omega = 0.05 \omega_\mathrm{p}$, $c_\tT = c_\td \equiv 10^6 \mathrm{m}/\mathrm{s}$ and $\nu = 0.5$. Here we assume that the phonon velocity of a Wigner crystal is of electronic origin and therefore of the order of the Fermi velocity $v_\mathrm{F} \approx 10^6\mathrm{m}/\mathrm{s} $. The purple dashed line is the plasmon mode with axis on the right, and the solid blue line is the massive shear mode with axis on the left. Notice the difference in scales: the plasmon is about 1000 times stronger. The intensity of the shear mode has a strong non-monotonic behavior, with a maximum intensity near $q_\mathrm{max} \approx \sqrt{\frac{2 - 2\nu}{9 + 3\nu}} \frac{\omega_\mathrm{p}}{c_\tT}$. Also indicated are the inverse shear penetration depth $q_\mathrm{s} = 1/\lambda_\mathrm{s} = \Omega/\sqrt{2} c_\tT$ and the inverse Debye length $q_\mathrm{e} = 1/\lambda_\mathrm{e} = \omega_\mathrm{p}/c_\tT$. 
  }\label{fig:EELS intensity}
  }
\end{figure}

The loss function of the quantum nematic follows  from Eqs.~\eqref{eq:energy loss definition}, \eqref{eq:longdielnematic}, 
\begin{align}
f_\tT(\omega ,q ) = - \frac{1}{\varepsilon_0}\mathrm{Im} \frac{(\omega^2 - c_\tL^2 q^2)(\omega^2-c_\tR^2 q^2) - \Omega^2 (\omega^2 - c_\tK^2 q^2)}{(\omega^2 -c_\tR^2 q^2)(\omega^2-c_\tL^2 q^2- \omega_\mathrm{p}^2 )-\Omega^2 (\omega^2 -c_\tK^2 q^2 - \omega_{\rm p}^2)}.
\label{EELSnematic} 
\end{align}
This has two propagating poles at 
\begin{align}
\omega(q)_{\tL1, \tL 2}^2 &= \frac{1}{2}\bigg( (c_\tL^2+c_\tR^2) q^2+\omega_\mathrm{p}^2+\Omega^2 \nonumber\\
& \phantom{mmmm} \pm \sqrt{\left((c_\tL^2+c_\tR^2)q^2+\omega_\mathrm{p}^2+\Omega ^2\right)^2-4 \left(\Omega ^2 \left(c_\tK^2 q^2+\omega_\mathrm{p}^2\right)+c_\tL^2 c_\tR^2q^4+c_\tR^2 q^2 \omega_\mathrm{p}^2\right)} \bigg) \label{eq:nematiclongdisp}.
\end{align}
The loss function $f_{\tT}(\omega,q)$ is shown in Fig.~\ref{Fig:LongSpec}.
 At small momenta $c_\tT q, \Omega \ll \omega_\mathrm{p}$, the poles look like
\begin{align}
\omega^2_{\tL,1}(q) &= c_\tR^2 q^2 + \Omega^2 + \mathcal{O}(q^3, \Omega^2/\omega_\mathrm{p}^2), \label{eq:nematic loss condensate dispersion}\\
\omega^2_{\tL,2}(q) &= c_\tL^2 q^2 + \omega_\mathrm{p}^2 + \mathcal{O}(q^3, \Omega^2/\omega_\mathrm{p}^2),\label{eq:nematic loss plasmon dispersion}
\end{align}
whereas in the opposite, liquid-like limit $\Omega \gg \omega_\mathrm{p}$, we have
\begin{align}
\omega^2_{\tL,1}(q) &= (c_\tT^2+c_\tR^2) q^2 + \Omega^2 + \mathcal{O}(q^3, \omega_\mathrm{p}^2/\Omega^2),\\
\omega^2_{\tL,2}(q) &= c_\tK^2 q^2 + \omega_\mathrm{p}^2 + \mathcal{O}(q^3,\omega_\mathrm{p}^2/\Omega^2)
\end{align}
In the Wigner crystal we found a single longitudinal excitation, corresponding to the longitudinal phonon that had acquired a plasmon gap, Eq.~\eqref{crystaleels}. 
In the fluid regime $1/\lambda_\tL \ll q \ll 1/ \lambda_\mathrm{s}$ of the quantum nematic, the plasmon propagates with the sound velocity $c_\tK$ instead, just as in the generic liquid, Eq.~\eqref{eq:viscous EELSspectrum}! This of course is expected but the mechanism at work is quite intriguing. We already encountered it in Sec.~\ref{subsec:Mode content of the quantum nematic}. The difference between a sound mode
and the longitudinal phonon is that the latter automatically involves shear deformations with the effect that its velocity contains the shear modulus via $c_\tL^2 = c_\tK^2 + c_\tT^2$. This implies that only in the $q=0$ limit can purely compressional stresses be applied.  In the fluid regime, the shear component has to somehow be `removed' from the longitudinal oscillation. The theory is Gaussian and the only way that this can be accomplished is by a mode coupling. One reads off from Eq.~\eqref{EELSnematic} that this second mode is a condensate mode since it propagates with the velocity $c_\tR$. What happens is that for $\Omega \gg \omega_\mathrm{p}$, the transverse part $\sim c_\tT^2 q^2$ has been transferred into the condensate mode by the coupling $\big(c_\tL^2 - c_\tK^2 (\frac{\Omega^2}{\omega_\mathrm{p}^2}) \big) q^2$.   

Accordingly, the always-`shearish' condensate mode has spectral weight only at finite momenta. This mechanism has universal consequence for the spectral weight distribution as function of momentum in the EELS spectrum: this is the smoking gun of the quantum nematic, making it possible to prove or disprove whether this is of relevance to experimental systems. The condensate mode in isolation does not carry electric charge: as we repeatedly emphasized the dislocation condensate is electrically neutral. However, due to the mode coupling related to the removal of the shear of the plasmon it acquires in return a finite electromagnetic weight that is growing as function of momentum. The plasmon completely dominates the loss function at small momenta. However, for increasing momentum the lower condensate momentum becomes better visible, while its spectral weight again decreases for quite large momenta $q \gg 1 /\lambda_\mathrm{d}$.  The spectral weight  of both poles is plotted in Fig.~\ref{fig:EELS intensity} for the parameters that might have bearing on the ``fluctuating stripes'' of the high-$T_\mathrm{c}$ superconductors~\cite{CvetkovicNussinovMukhinZaanen08}: $\Omega = 0.05 \omega_\mathrm{p}$, $c_\tT = c_\td = 10^6 \mathrm{m}/\mathrm{s}$ and $\nu = 0.5$. Notice the difference between the scales: the condensate mode is indeed much weaker. For small momenta the weight of the condensate (``massive shear photon'') mode increases $\sim q^2$. However, upon exceeding the inverse shear penetration depth the mode coupling becomes again weaker because the plasmon turns into a longitudinal phonon
again. In fact, one finds a maximum in the pole strength of the condensate mode $ Z(q_\mathrm{max})  \approx \frac{3}{16} \sqrt{\frac{3-3\nu}{3+\nu}} \frac{\Omega^2}{\omega_\mathrm{p}}$
at a momentum $ q_\mathrm{max} \approx \sqrt{\frac{2 - 2\nu}{9 + 3\nu}} \frac{\omega_\mathrm{p}}{c_\tT}$~\cite{CvetkovicNussinovMukhinZaanen08}.

There are  considerable complications in trying to detect the ``massive shear photon'' in the laboratory. Believing that the energy scales seen in the spin fluctuations measured by neutron scattering
have any bearing on maximal charge correlations one would estimate $\Omega \approx 40$ meV, while $\lambda_\mathrm{s} \sim 10$ nm. The `phonon' velocity should be of a typical electronic kind
$\sim 1$ eV \AA, while the plasmon energy is established to be of order $1$ eV. This does imply that the EELS spectral weight of the shear photon should be quite small. In addition, this kinematical 
regime is littered with other excitations, especially the normal phonons of the real crystal. In order to stand a chance to detect this feature a  very high (meV) resolution appears to be required. 
High-energy (transmissive) EELS and RIXS have still a long way to go, and it remains to be seen whether low energy electron loss is capable of picking up such genuine electronic features~\cite{VigEtAl15}.

\subsection{Electromagnetism of the charged quantum smectic}\label{subsec:Electromagnetism of the charged quantum smectic}

As we learned in Sec.~\ref{sec:Quantum smectic}, the neutral  quantum smectic is behaving like an intricate and counterintuitive mixture of `intertwined' solid and fluid characteristics. It is entertaining to find out how these characteristics get further amplified in the charged quantum smectic. 

For convenience, let us first shortly recollect the conventions introduced for the neutral case. The angle $\eta$ parametrizes the direction of propagation relative to the smectic liquid ($\eta =0$) and
solid ($\eta = \pi/2$) directions. The decomposition into longitudinal and transverse responses is only possible at the special angles $\eta =0, \pi/4,\pi/2$. Accordingly, only at 
these angles one can separate the electromagnetic response in terms of transverse and longitudinal response functions: as Eq.~\eqref{eq:smectic EM-stress Lagrangian} shows,
one finds non-diagonal terms in the self-energy matrix associated with the transverse photon $A_\tT$ and the longitudinal scalar potential $A_\ft$. This follows simply from the spatial anisotropy of the smectic, and the propagator matrices do not decompose into longitudinal and transverse sectors, except at the special angles.

The expression in Eq.~\eqref{eq:smectic EM-stress Lagrangian} is not very illuminating and we start by analyzing the special angles, where Eq.~\eqref{eq:smectic EM-stress Lagrangian} gives the following effective actions:
 {\paragraph{Solid direction $\eta = \pi/2$}
 \begin{align}
   \mathcal{L}^\mathrm{EM}_\mathrm{smec}
 &= \tfrac{1}{2} \varepsilon_0 \omega_\mathrm{p}^2 \Big(  \frac{c_\tT^2 q^2}{\omega_n^2 + c_\tL^2 q^2} \lvert A_\ft \rvert^2 +  \frac{\omega_n^2 + \Omega^2}{\omega_n^2 + c_\tT^2 q^2 + \Omega^2}  \lvert A_\tT\rvert^2 \Big).
\label{eq:smecticEMsol}
\end{align}
 The longitudinal response is equal to that of the Wigner crystal, while the transverse response simply picks up the Higgs mass. 

 \paragraph{Liquid direction $\eta = 0$}
 \begin{align}
   \mathcal{L}^\mathrm{EM}_\mathrm{smec}
 &= \tfrac{1}{2} \varepsilon_0\omega_\mathrm{p}^2 \Big(  \frac{c_\tT^2 q^2}{\omega_n^2 + c_\tL^2 q^2} \lvert A_\ft \rvert^2  + \frac{\omega_n^2 (\omega_n^2 + c_\td^2 q^2 + \Omega^2)}{(\omega_n^2 + c_\tT^2 q^2)(\omega_n^2 + c_\td^2 q^2) + \omega_n^2 \Omega^2}  \lvert A_\tT\rvert^2 \Big).
\label{eq:smecticEMliq}
\end{align} 
 The longitudinal response is independent of $\Omega$ and still equal to the Wigner crystal but the transverse response is a bit more complicated with the condensate velocity $c_\td$ entering. 
 
 \paragraph{Intermediate $\eta = \pi/4$}
  \begin{align}
  \mathcal{L}^\mathrm{EM}_\mathrm{smec}
 &= \tfrac{1}{2} \varepsilon_0\omega_\mathrm{p}^2 \Big(  \frac{c_\tT^2 q^2(\omega_n^2 + c_\tR^2 q^2 + \Omega^2)}{(\omega_n^2 + c_\tL^2 q^2)(\omega_n^2 +c_\tR^2 q^2) + \Omega^2(\omega_n^2 + c_\tK^2 q^2)} \lvert A_\ft \rvert^2 + \frac{\omega_n^2 }{\omega_n^2 + c_\tT^2 q^2}  \lvert A_\tT\rvert^2 \Big).
 \label{eq:smecticEMmid}
\end{align} 
Interestingly and coincidentally, the transverse sector is identical to the isotropic solid; the propagating photons do not pick up the smectic condensate. The longitudinal response is however affected by the condensate, and is identical to the quantum nematic. This is not a surprise: already in Sec.~\ref{subsec:smectic correlation functions} we had noticed that the longitudinal response of the $\pi/4$ smectic is identical to that of the quantum nematic.

 \paragraph{other angles}
As it turns out the angles $\eta = 0, \pi/4 , \pi/2$ are very special and not representative of other interrogation directions. In particular, at those angles the off-diagonal terms in the electromagnetic Lagrangian Eq.~\eqref{eq:smectic EM-stress Lagrangian} vanish, while at other angles they do not. This leads e.g. to the possibility of anomalous Hall conductivity (a non-zero $\sigma_{xy}$ in the absence of an external magnetic field). }

Let us first zoom in on the {\em longitudinal} response along the special directions which is easy to comprehend. In Sec.~\ref{sec:Quantum smectic}, we learned that both along the liquid   ($\eta =0$) and solid ($\eta = \pi/2$)  directions the smectic just exhibits a solid response in the form of
a longitudinal phonon, while along the $\eta = \pi/4$ directions this is turned into the sound mode as if one would be dealing with a pure liquid. From Eqs.~\eqref{eq:smecticEMsol}, \eqref{eq:smecticEMliq}
we read off the longitudinal dielectric function $\varepsilon_{\tL \tL}(\omega,q) = \varepsilon_0 \big( 1 - \omega^2_\mathrm{p}/(\omega^2 - c_\tL^2 q^2) \big)$, identical to the result for the Wigner crystal Eq.~\eqref{eq:epslongXtal}.
More interestingly, what happens at the special angle $\eta = \pi/4$? For the neutral case we found the genuine sound mode propagating with velocity $c_\tK$. 
Inspecting the temporal part in Eq.~\eqref{eq:smecticEMmid}, we find a longitudinal dielectric function identical to that of the quantum nematic Eq.~\eqref{eq:longdielnematic}, with the same ramifications for the electron-loss spectrum that we just discussed at length! 

Naively one would perhaps anticipate that apparently the smectic wants to behave like a crystal in the transverse sector along the $\eta = 0$ or in the longitudinal sector along $\eta = \pi/2$, while it should behave like a superconductor in the intermediate direction $\eta = \pi/4$. But we already learned from the neutral smectic that this is not at all the case. Superconductivity is associated with the transverse response and  
we found there the undulation $\omega \propto q^2$ mode in the ($\eta =0$) liquid direction, while in the $\pi/4$-direction the transverse phonon with velocity $c_\tT$ got completely resurrected. The 
solid direction ($\eta = \pi/2$) appeared to have a special status exhibiting just the massive shear phonon, different for the nematic not accompanied by any condensate mode. According
to Eq.~\eqref{eq:smectic EM-stress Lagrangian}, the transverse photon propagator becomes particularly simple for the $\eta = \pi/4$ case. One reads off from Eq.~\eqref{eq:smecticEMmid} that it is actually identical to the transverse photon propagator of the Wigner crystal Eq.~\eqref{photonsesolid}! As for the crystal, for this angle the transverse mode spectrum is characterized by the plasma-polariton and the transverse phonon that turns into the remarkable `polariton-phonon' with quadratic dispersion at very small $q \ll 1/ \lambda_\tL$. 

This is as far as the relationship between the charged quantum smectic properties and the now familiar properties of the charged Wigner crystal and quantum nematic extends. Everything else
is unique to the quantum smectic. Let us now focus on the {\em transverse} response in the liquid ($\eta =0$) direction. In the neutral case we learned that precisely for this direction we 
recovered the famous undulation mode with a quadratic dispersion. What happens in the charged system? The photon propagator follows from Eq.~\eqref{eq:smecticEMliq},
\begin{align}
\langle A^{\dagger}_{\tT}(\omega,q) A _{\tT}(-\omega,-q)\rangle_{\eta=0} = 
\frac{1}{\varepsilon_0}\frac {1}{\omega^2 - c^2_l q^2 -  \omega^2_\mathrm{p}\frac{\omega^2 ( \omega^2 - c^2_\td q^2 - \Omega^2)}{(\omega^2 - c^2_\tT q^2) (\omega^2 -c^2_\td q^2) - \omega^2 \Omega^2}}.
\label{eq:eta0photprop}
\end{align}
Upon closer inspection it becomes clear that this describes quite different physics than any of the other cases we have considered. The interest is in the spectrum of transverse modes and 
their pole strengths in the photon propagator at small momenta. For $c_l q, \Omega \ll \omega_\mathrm{p}$, 
\begin{align}
\omega^2_{1,\eta=0}(q) &= \omega_\mathrm{p}^2+ (c_l^2+c_\tT^2)q^2 + \mathcal{O}(q^4, \Omega^2/\omega_\mathrm{p}^2), \label{eq:smectic liquid photon}\\
\omega^2_{2,\eta=0}(q) &= \Omega^2+ c_\td^2 q^2 + \mathcal{O}(q^4, \Omega^2/\omega_\mathrm{p}^2), \label{eq:smectic liquid phonon}\\
\omega^2_{3,\eta=0}(q) &= \frac{c_l^2 c_\tT^2 c_\td^2}{\omega_\mathrm{p}^2\Omega^2}q^6  + \mathcal{O}(q^8). \label{eq:smectic liquid undulation}
\end{align} 
We find, as before, the plasma-polariton with velocity $\sqrt{c_l^2 + c_\tT^2} \approx c_l$ and a condensate-like mode with velocity $c_\td \approx c_\tT$ characterized by the masses $\omega_\mathrm{p}$ and $\Omega$ respectively.
However, there is also a gapless mode characterized by a {\em cubic} dispersion $\omega_{3,\eta=0}(q) \sim \tfrac{c_l c_\tT^2}{\omega_\mathrm{p}^2\Omega^2} q^3$! In hindsight this actually makes sense. We already mentioned that the transverse phonon of the Wigner crystal turns at distances larger than $\lambda_\tL$ into a quadratic mode, indicating that the 
influence of the EM screening is to `add an extra derivative' to the dispersion. In the liquid direction of the smectic, we started out with the transverse undulation mode which has a quadratic dispersion in the neutral case. The EM screening will have yet again the same effect, causing the undulation mode to acquire a cubic dispersion. Given that it has the status of a Goldstone mode excitation dressed with EM fields, this has to the best of our knowledge an unique status as record holding power-of-momentum dependence!

\begin{figure}
\begin{center}
\includegraphics[width=0.6\textwidth]{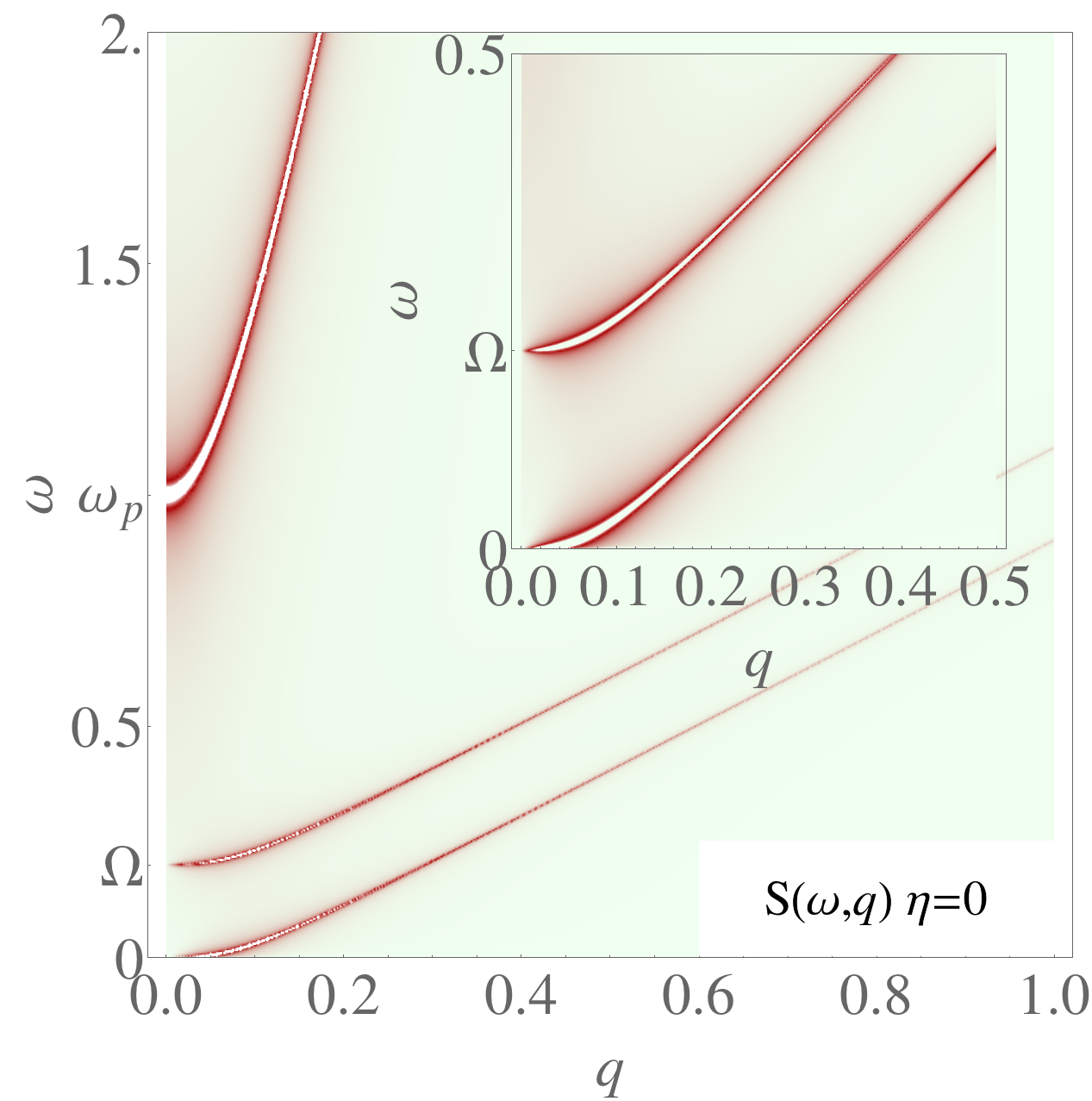}
\caption{The transverse photon spectral function of the smectic along the liquid direction $\eta=0$ for $c_l=10$, $c_\tT = c_\td = 1$, $\omega_\mathrm{p} =1$, $\Omega=0.2$. The poles are infinitely sharp and the width indicates the spectral weight. We identify the strong plasma-polariton, and the weaker massive condensate and massless undulation modes. Inset: zoom to the origin, where the two elastic modes have non-zero spectral weights.}
\label{fig:SmecticLiquidSpectral}
\end{center}
\end{figure}

Similar to the case of  the neutral smectic in Sec.~\ref{subsec:Mode content of the quantum smectic}, the undulation mode turns into the linear condensate mode $\omega \propto c_\td q$ and the transverse phonon for $q \gg 1/\lambda_s$. The transverse photon spectral function for $\eta=0$ is shown in Fig.~\ref{fig:SmecticLiquidSpectral}. Again, the spectral density for the elastic modes around $q=0$ is zero and vanishes quickly for finite $q$.

\begin{figure}
\begin{center}
\includegraphics[width=7.7cm]{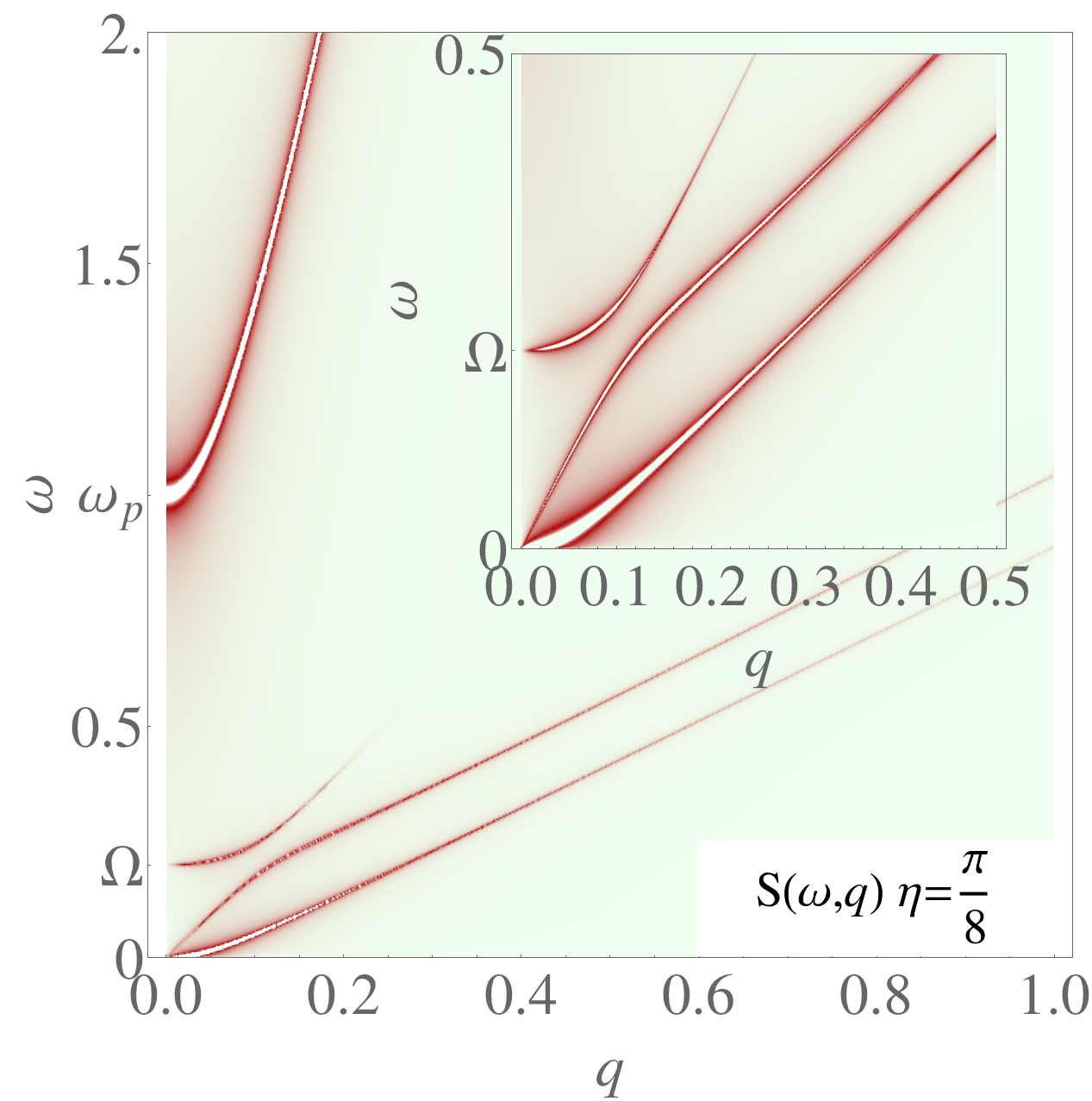}
  \includegraphics[width=7.7cm]{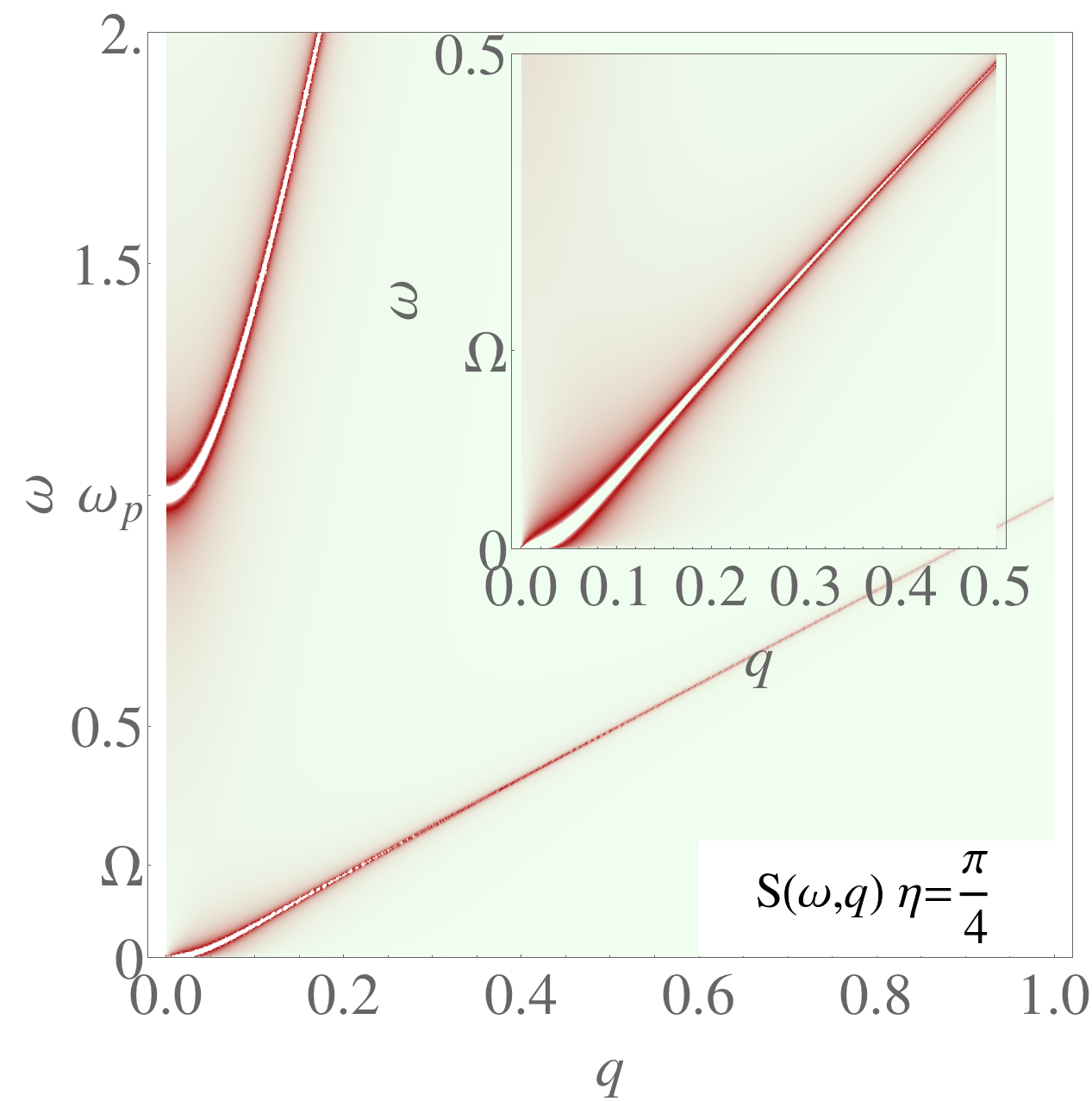}\\
  \includegraphics[width=7.7cm]{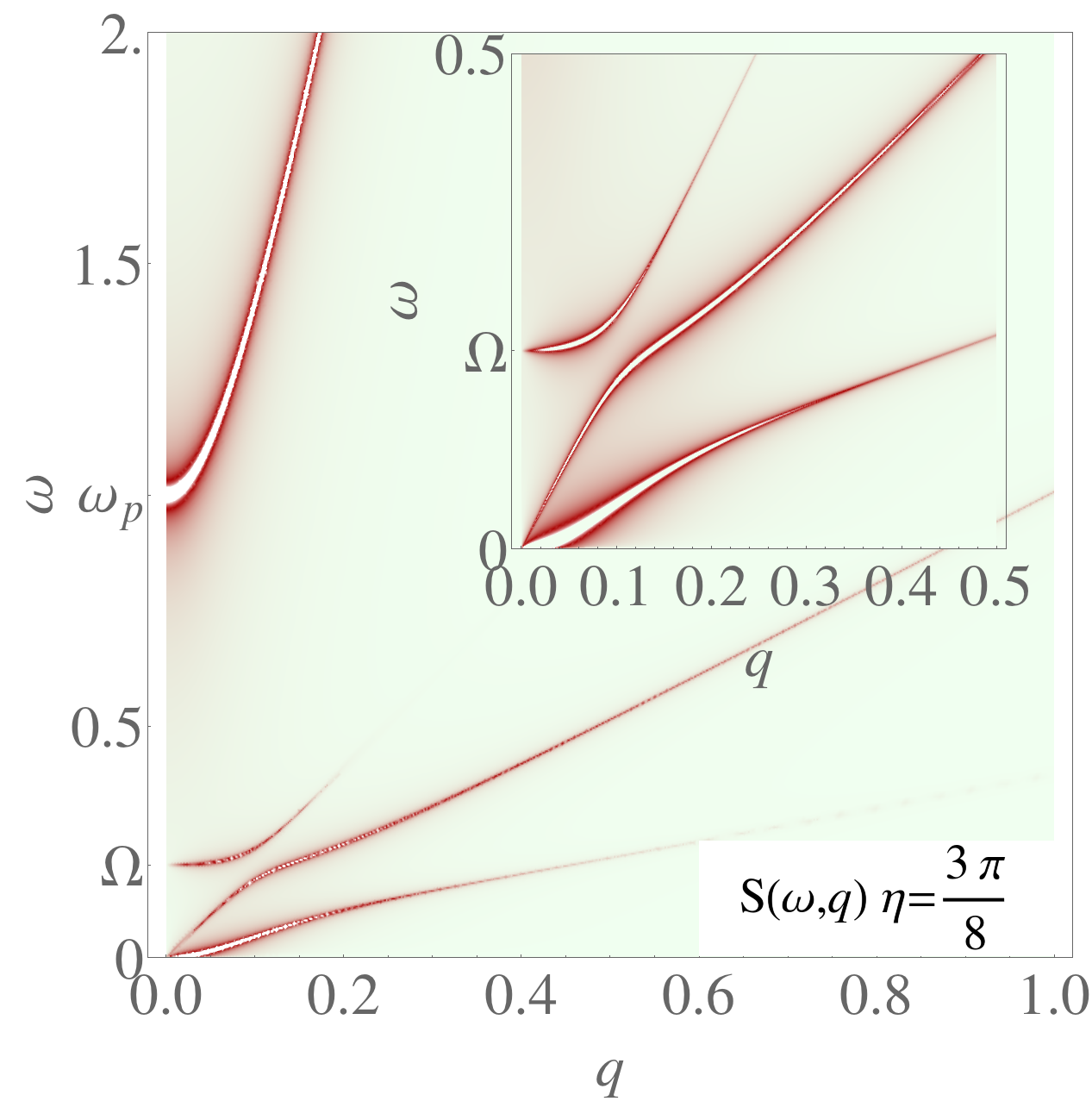}
 \includegraphics[width=7.7cm]{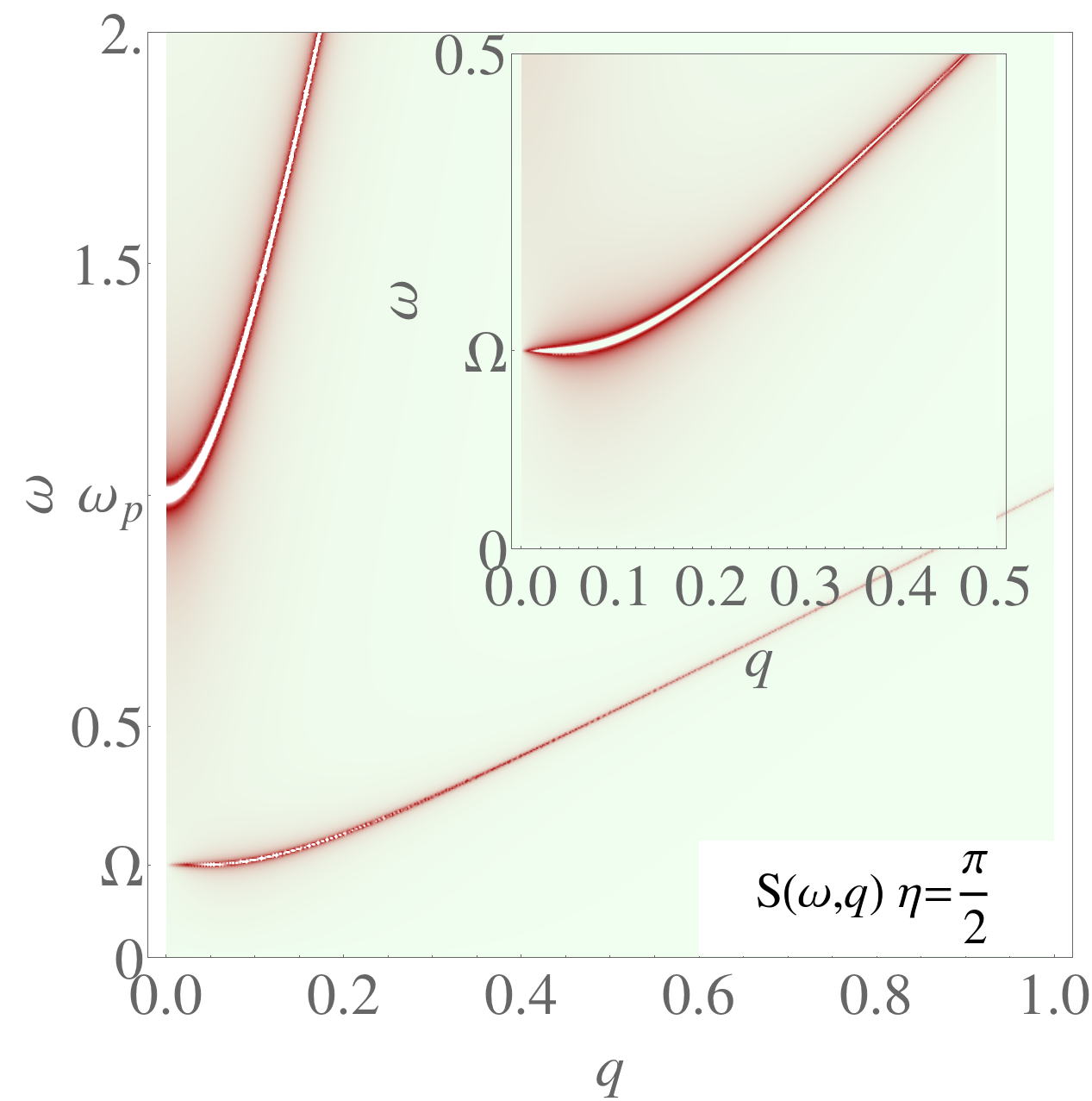}
\caption{The transverse photon spectral function for $\eta = \pi/8, \pi/4, 3\pi/8, \pi/2$ respectively. The parameters were chosen as $c_l=10$, $c_\tT=c_\td=1$, $\nu=0.5$, $\omega_\mathrm{p}=1$, $\Omega=0.2$. The broadening of the dispersions is artificial and is proportional to the spectral weight. For general angle we observe four poles: the plasma-polariton with gap $\omega_\mathrm{p}$, the condensate mode with gap $\Omega$ and two massless modes which we assign to the longitudinal (compression) and transverse sound modes. The velocities of these two modes at low momenta vary similar to Fig.~\ref{fig:eta velocity}, although the transverse mode is also coupled to the polariton. For the special angle $\eta = \pi/2$, we observe only the polariton and the condensate mode, as expected from the neutral case. However, at $\eta = \pi/4$ the response of the Wigner crystal is recovered.}
\label{fig:SmecticTransverseSpectral}
\end{center}
\end{figure}

Now let us turn to the transverse response in the solid ($\eta =\pi/2$) direction. In the neutral case, we found in this direction a `minimalistic Higgsed shear response' in  the form of
a single massive shear photon, with no sign of condensate modes taking part. Again the interest is in the transverse photon propagator, which follows from Eq.~\eqref{eq:smecticEMsol} as
\begin{align}
\langle A^{\dagger}_T A_T \rangle_{\eta=\pi/2} = \frac {1}{\omega^2-c_l^2 q^2 -\omega^2_\mathrm{p}\frac{\omega^2-\Omega^2}{\omega^2-c_\tT^2 q^2 -\Omega^2}}.
\label{eq:etasolidphotprop}
\end{align}       
This is actually coincident with the `minimal superconductor' Eq. \eqref{eq:optconnematiceasy} that arises when $c_\tT=c_\td=c_\tR$,  and we discussed this for the charged nematic. We learn therefore directly that precisely in  this direction the smectic turns into the simplest incarnation of a superconducting state that can be imagined in the present setting! In this direction there will be a fully developed Meissner effect. However, what happens when we deviate by any amount from $\eta = \pi/2$? We have already seen in the neutral case that this angle was strangely singular. We found that for all $\eta \neq 0, \pi/2$ we were dealing with a genuine transverse phonon, be it one with a velocity that vanishes both in both limits $\eta \rightarrow 0, \pi/2$, see Fig.~\ref{fig:eta velocity}. Let us therefore inspect what happens to the transverse response at generic angles. Since the longitudinal and transverse sectors mix, the full photon propagator becomes a complicated expression, but the spectrum of transverse modes is again of primary interest and the transverse propagator $\langle A^{\dagger}_\tT(\omega,q) A_\tT(-\omega,-q) \rangle$ follows simply from the self-energy $\Pi_\tT(\omega,q;\eta)$ which is the $\tT\tT$ entry of Eq.~\eqref{eq:smectic EM-stress Lagrangian}. In Fig.~\ref{fig:SmecticTransverseSpectral} we show the spectral function for several angles. 

The general picture is clear: for all angles there is the plasma-polariton that carries most of the spectral weight. At finite momenta, there are elastic modes that quickly vanish for large momenta. These include the massive condensate mode, and up to two massless modes: the longitudinal/sound mode and the transverse phonon mode. The spectral weight of the elastic modes interpolates as follows. For $\eta=0$, we have a condensate mode with mass $\Omega$ and a massless cubic mode Eq.~\eqref{eq:smectic liquid undulation}. As the angle grows, the longitudinal sound mode emerges, while also spectral weight is transferred from the massive condensate mode to the massless modes. Half-way at $\eta=\pi/4$, both the condensate mode and the sound mode vanish and there is only a massless transverse phonon carrying all the spectral weight. This is precisely the response of the Wigner crystal in Fig.~\ref{Fig:WignerSpectral}! As the angle approaches the solid direction $\eta \to \pi/2$, there are re-emerging massive and sound modes to which spectral weight is transferred back from the massless transverse mode. At $\eta=\pi/2$, all the spectral weight is carried by the massive shear mode with mass $\Omega$, as both massless modes disappear.

Summarizing we identify the photonic plasmon at $\omega_\mathrm{p}$ and the massive shear photon (condensate mode) at $\Omega$ and two massless modes with changing spectral weight as a function of $\eta$ that interpolates between the limiting cases related to the solid and liquid behaviors of the smectic. For small momenta they are 
\begin{align}
\omega_1^2(q;\eta)&=\omega_\mathrm{p}^2 + c_l^2 q^2+ \frac{c_\tT^2 \omega_\mathrm{p}^2 - 
   c_\tT^2 \Omega^2 \sin^2
     2 \eta}{\omega_\mathrm{p}^2 - \Omega^2} q^2 + \mathcal{O}(q^4)\\
\omega_2^2(q;\eta)&= \Omega^2+\cos^2 \eta c_\td q^2+ \frac{\sin ^2 2 \eta \omega _p^2-\Omega ^2}{\omega _p^2-\Omega ^2}c_\tT^2 q^2+\mathcal{O}(q^4) \\
\omega_3^2(q;\eta)&= (c_\tL^2-\sin^2 2\eta c_\tT^2) q^2 + \mathcal{O}(q^4)\\
\omega_4^2(q;\eta)&= \bigg(\tfrac{c_\tK^2+c_l^2}{\omega_\mathrm{p}^2} - (\tfrac{1}{\Omega^2}+\tfrac{1}{\omega_\mathrm{p}^2})(c_\tL^2- \sin^2 2\eta c_\tT^2)  +\tfrac{ c_\td^2 \cos^2 \eta}{\Omega^2})\bigg)\sin^2 2 \eta c_\tT^2  q^4 + \mathcal{O}(q^6).
\end{align}
For the limiting cases of the special angles, we recover the results discussed above with at most a single massless mode within the accuracies stated. The bottom line is that we have just re-identified the surprise we already encountered in the case of the neutral smectic: for all angles except $\eta = \pi/2$,
the transverse response of the smectic shows that it behaves like a solid as signaled by the presence of the velocity-renormalized but massless transverse phonon. Remarkably, at the intermediate angles and at small momenta, this massless transverse phonon mixes with the longitudinal phonon with velocities $c_\tL$ and $c_\tK$ but this is what we intuitively expect: when the transverse modes are in between the liquid layers they can mix with the longitudinal modes that always exist in the solid and liquid. We emphasize that in the special cases of $\eta=0, \pi/4 ,\pi/2$ the spectral weight of this mixed mode vanishes.

With regards to the superconducting identity of the dislocation condensate, we were facing the conundrum in the neutral case that the superfluid density `spikes' precisely at $\eta = \pi/2$. But now we are in a better shape to address what is going on since in the charged case we can apply the coupling to external fields, and the electromagnetic response has to be susceptible to a reasonable physical interpretation. Let us therefore zoom in on the behavior of the frequency- and momentum-dependent magnetic penetration depth $\lambda(\omega,q)$, 
as defined in Eq.~\eqref{eq:penetration depth from self-energy}.

\begin{figure}
\begin{center}
\includegraphics[width=.5\textwidth]{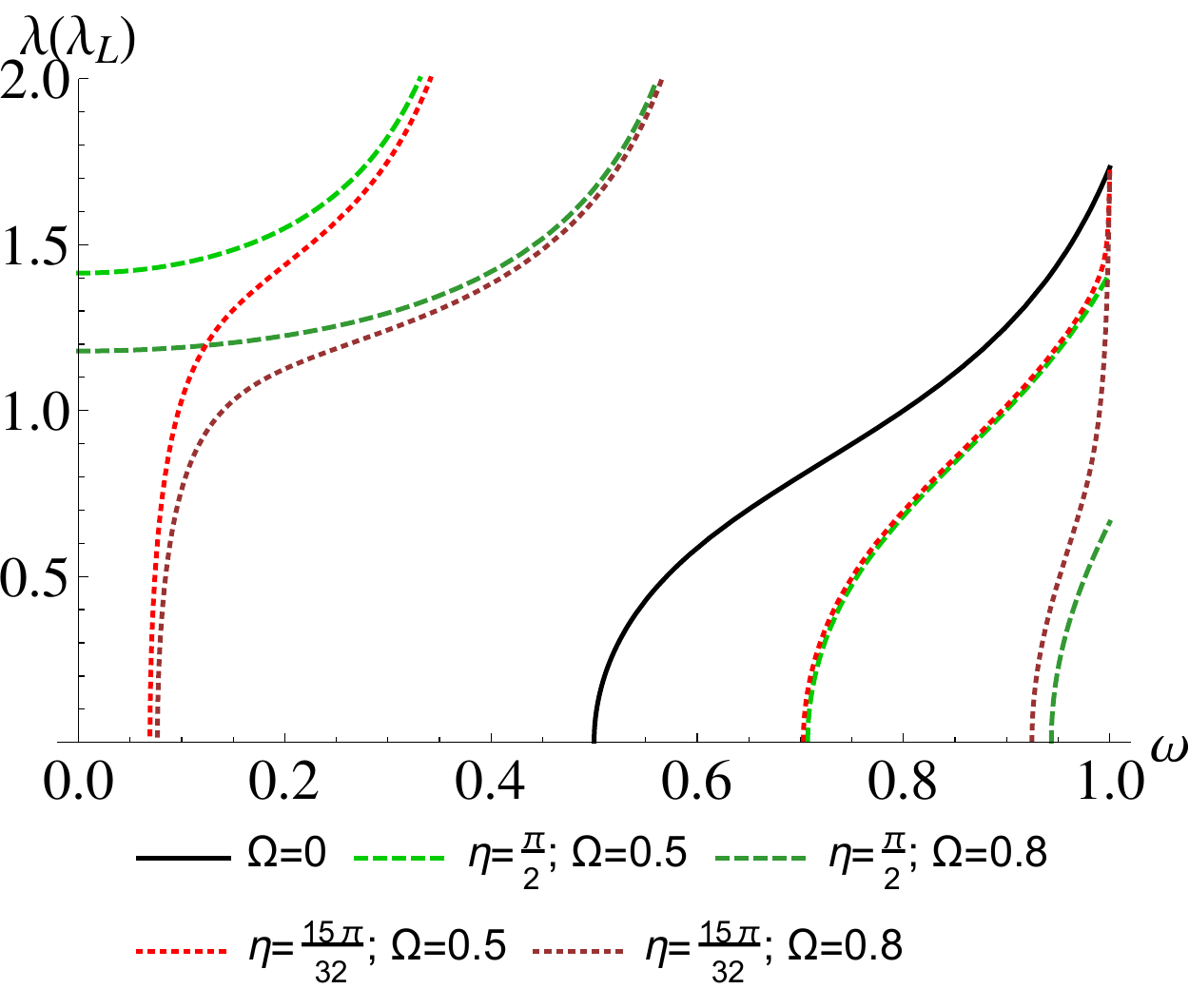}
\caption{Frequency dependence of the penetration depth Eq.~\eqref{eq:smectic penetration depth} in units of the London penetration depth $\lambda_\tL$ at momentum $q = 0.5$, for the parameters $c_\tT = c_\td = 1$, $\nu = 0.5$, for angles $\eta=\pi/2$ and $\eta = 15\pi/32$ and for varying Higgs mass $\Omega$. The black solid line is the case of no condensate $\Omega=0$ i.e. the Wigner crystal. The skin effect sets in above a certain momentum-dependent energy gap. The dashed green lines are for $\eta=\pi/2$ and we find a true Meissner effect up to $\omega = 0$. The weaker the condensate (lower $\Omega$), the longer the momentum-dependent penetration depth Eq.~\eqref{eq:smectic shear penetration depth}. The dotted red lines are for the angle $\eta = 15\pi/32$. There is no Meissner effect in the static limit $\omega \to 0$, but at finite frequencies the penetration depth quickly follows the behavior of $\eta =\pi/2$ and the energy gap is small. In other words, the skin effect at finite momentum is much stronger in the quantum smectic than in the Wigner crystal.}
\label{fig:skineffect}
\end{center}
\end{figure}

Let us first consider the general expression for this dynamical penetration depth for arbitrary angle $\eta$ as follows directly from Eqs.~\eqref{eq:penetration depth from self-energy},  \eqref{eq:smectic EM-stress Lagrangian}.
 \begin{equation}\label{eq:smectic penetration depth}
\frac{\lambda(\omega,q;\eta)}{c_l} = \left( \mathrm{Im} \sqrt{\omega^2 - \omega_\mathrm{p}^2\frac{\omega^2 \big((\omega^2 - \cos^2 \eta c_\td^2 q^2)(\omega^2 - c_\tL^2 q^2) - \Omega^2 (\omega^2 - c_\tL^2 q^2 + c_\tT^2 q^2 \sin^2 2 \eta ) \big)}{ (\omega^2 - \cos^2 \eta c^2_\td q^2) (\omega^2 - c_\tT^2 q^2)(\omega^2 - c_\tL^2 q^2) - \Omega^2(\omega^4 - \omega^2 c_\tL^2 q^2 +  c_\tK^2 c_\tT^2 q^4 \sin^2 2\eta)}} \right)^{-1}
\end{equation}
where we used the definition $p_{\eta}(\omega,q) = -\frac{1}{c_\td^2} \omega^2 + \cos^2 \eta q^2$. One infers an overall factor of $\omega$ in the denominator, having as a consequence that  $\lambda \rightarrow \infty$ in  the static limit  $\omega \to 0$, and the penetration depth diverges. 
However, by first taking the limit  $\eta \rightarrow \pi/2$  the above expression reduces to
\begin{equation}\label{eq:smectic contribution eta pi/2}
\lambda(\omega,q;\eta = \tfrac{\pi}{2}) = c_l \mathrm{Im} \sqrt {\frac{\omega^2 - c_\tT q^2 - \Omega^2}{
\omega^4 -\omega^2 (c_\tT^2 q^2 + \Omega^2 + \omega_\mathrm{p}^2) + \Omega^2 \omega_\mathrm{p}^2}}.
 \end{equation}
In the static limit, this results in a penetration depth,
\begin{equation}
\lambda(\omega =0,q;\eta=\pi/2) = \lambda_\tL\sqrt{ 1 + \tfrac{1}{2}\lambda^2_{\rm s} q^2 }, \label{eq:smectic shear penetration depth}
\end{equation}
which is identical to the expression for the quantum nematic with $\lambda_\mathrm{s} = \sqrt{2}c_\tT/\Omega$ except for the factor of $\tfrac{1}{2}$. Apparently the order of limits $\eta \to \pi/2, \omega \to 0$ is
crucial.  The clue is however that this is only singular precisely at zero frequency. The screening of magnetic fields becomes a smooth affair at any finite frequency and momenta, which usually is called the skin-effect \cite{LandauLifshitz84}, this 
can be neatly studied by looking at the finite-frequency behavior of $\lambda(\omega,q)$. This is shown in Fig. \ref{fig:skineffect} for the angles $\eta=\pi/2, 15\pi/32$. We conclude that the superconducting condensate is really measured only at $\eta=\pi/2$, whereas the angular dependence at finite frequencies is very weak. 

 \section{Conclusions}\label{sec:Conclusions}
 The reader who is still with us has shown resilience -- we hope that he/she shares with us a sense of wonder regarding the scenery we have exposed in this story. What has happened? We departed from a topic that perhaps at first sight seemed antiquated: the theory of elasticity, the quantum melting of solids into superfluids, subject matter that was supposed to be brought under complete control some fifty years ago. Arguably, the quantum {\em liquid crystal} in its bosonic incarnation adds some novelty but at least dealing with the quantum nematics, we could have as well written a short note pointing out that the continuum version should be characterized by a propagating rotational Goldstone mode since circulation is quantized in the superfluid. 

Instead, we arrived at a mathematical theory of this part of everyday reality where everything familiar seems turned upside down. It is the {\em Alice in Wonderland} version of ``Condensed Matter Physics 101''. Upon going through the mirror, phonons turn into photons, neutral superfluids are actually stress superconductors characterized by massive shear photons, the Meissner effect is lying in hide in the Wigner crystal and  torque stresses deconfine like gluons upon entering the liquid-crystal phase. There is a lot more, but we already listed this full portfolio in the introduction, Sec.~\ref{subsec:The warped dual view on quantum liquid crystals}. 

Surely, the reader has already realized a long while ago that the difference is that we are just starting out from a limit that is opposite to the one that figures in the textbooks. The conventional way to deal with (quantum) liquids is to begin with in essence kinetic gas theory. As point of departure the weakly-interacting system of microscopic constituents is chosen and subsequently one works in the effects of correlations using perturbative means. Here we have exposed the {\em dual perspective}, insisting the liquid is locally as solid-like as permitted by the laws of physics: from the limit of the strongest interactions imaginable. The miracle is that it is not at all that hard to describe this in minute detail resting on a surprisingly elegant mathematical theory.   

The basic notions are of course dating back to the Kosterlitz--Thouless--Nelson--Halperin--Young theory of classical topological melting as developed in the 1970s. However, to make this work in the 2+1D quantum case a more powerful mathematical formalism was required, and this was supplied by Kleinert in the 1980s. Kleinert was however focused on the 3D classical case and our discovery has been that its remarkable powers become
to full fruition when twisted into the 2+1D quantum realms. The reason is simply that a lot more is going on, and the stress gauge fields are exquisitely suited to deal with the zero-temperature realms. 

Is it good for anything in real nature? It is far from obvious whether the microscopic circumstances in systems that do occur in nature will be ever of the kind that this maximally correlated circumstances can be realized. However, it is the usual credo that reality is always in the middle and in order to figure what it is about, one better know the limiting cases. Although presently this is a strictly theoretical affair, we of course hope that it will enlighten the experimentalists to have a look in the unusual corners suggested by the dual view. A typical case in point are the ``fluctuating stripes'' that we already discussed in the introduction, forming the initial motivation for this work. Based on the somewhat indirect information coming from the spin fluctuations, it is literally asserted that the underdoped cuprate superconductors are characterized by stripe (crystalline-type) correlations that extends over many lattice constants.  We conclude directly that the shear penetration depth is large, which in turn implies that there should be at least remnants of our massive stress photons to be discerned in the electromagnetic responses of Sec.~\ref{subsec:Superconductivity and the electromagnetism of the quantum nematic}.

However, despite the fact that these massive stress photons leave behind big fingerprints in the dynamical response functions, these are just not accessible using standard spectroscopic techniques. As we discussed  at length in Sec.~\ref{sec:Electromagnetic observables}, this is tied in the practical circumstance that the velocity of light is very large compared to the material velocities with the effect that the experimentalists can only access the $q \rightarrow 0$ limit of the electromagnetic response functions, where by principle the massive stress photons lose their electromagnetic spectral weight. Plain vanilla light does not work, but there are other ways of measuring these response functions giving in principle access to the relevant kinematical regime involving energies in the range of $1-100$ meVs and momenta of order of inverse nanometers. The longitudinal electron energy-loss function is in fact directly measured by electron energy-loss spectroscopy (EELS), and 
Figs.~\ref{Fig:LongSpec}, \ref{fig:EELS intensity} show in great detail where to expect the massive shear photon (see also Ref.~\cite{CvetkovicZaanen06b}). The trouble is yet again on the experimental side: the high-energy transmission EELS technique would be exquisitely suited to observe these features but unfortunately the technical 
development came to a standstill in the 1970s with the effect that the energy resolution is measured in hundreds of meVs, way beyond what is needed to resolve the shear photon. A next contender is Resonant Inelastic X-ray Scattering (RIXS); yet again the energy resolution of the existing beam lines falls short by an order of magnitude while perhaps even a bigger problem is that although this technique is supposed to measure the loss function as
well, up to now the evidence that this is actually the case is still missing. Last but not least, one can attempt to use a low-energy, reflection mode EELS spectrometer. Only very recently proof of principle was delivered that such a machine can pick up bosonic excitations of the strongly-correlated electron system~\cite{VigEtAl15}, and perhaps this approach might in some near future deliver conclusive information. The verdict is that because of the shortcomings of the existing experimental machinery, it cannot be decided whether fluctuating stripes exist or whether the interpretations are flawed. The benefit of the dual description is that it delivers an unambiguous prediction for a signature of `serious' fluctuating order that in principle is accessible by experiment. 

\subsection{Open problems}

The reader might have the impression that the theory we have presented is fairly complete, while only some details remain to be settled. In fact, there are quite a number  of challenging problems becoming visible given the state of the art as presented in this review. Let us finalize this review with a list of the open questions, roughly in  order of increasing complexity and profundity. 

\subsubsection{The dual stress superconductor by first principle}

In Sec.~\ref{sec:Dynamics of disorder fields} we explained the central wheel around which everything else revolves: the construction of the dual dislocation  condensate. The sections that follow can be viewed as a confirmation that this framework has to be correct. This construction is however just resting on a phenomenological approach. In essence it takes for granted that dislocation quantum melting is governed by the same general weak--strong duality mechanism as vortex--boson duality, as validated by the Abelian nature of pure translations. There is just a unique way to incorporate the additional features rendering the dual stress condensate to be richer than the vortex condensate such as the rotational structure and the glide principle. This completely determines the structure of the effective dual theory. Even the number of free parameters is kept to nearly the absolute minimum. Assuming that the mass density, shear and compression moduli of the `background crystal' are known, the ideal theory would only have the dual Higgs mass (or shear penetration depth) as free parameter. We get very close: we know that the condensate velocity $c_\td$ is parametrically related to the phonon velocity $c_\tT$. However, as we discussed in Secs.~\ref{sec:Dynamics of disorder fields}, \ref{sec:Dual elasticity of charged media}, and \ref{sec:Electromagnetic observables}, we cannot be sure whether this involves a $\mathcal{O}(1)$ factor in between.

It is an open challenge to construct a `microscopic' theory of the dislocation quantum melting. This has been accomplished successfully for (multi-component) vortices in the Abelian-Higgs duality in 2+1 and 3 dimensions~\cite{Peskin78, Savit78, DasguptaHalperin81, KiometzisKleinertSchakel95, HerbutTessanovic96, Herbut97, NguyenSudbo99, HoveSudbo00, MoHoveSudbo02, SmisethSmorgravSudbo04, SmisethEtAl05, SmorgravEtAl05}, as well as for dislocations in two-dimensional melting \cite{Kleinert83, Saito82a, Saito82b, Strandburg88, JankeKleinert90}, and it appears that this can be rather straightforwardly generalized to dislocations in 2+1-dimensions, just involving an extensive numerical Monte Carlo effort. The strategy is to exploit the Villain construction with finite, quantized Burgers charges to derive the dual gauge theory. The ``dislocation-loop blow-out" of the topological defects can be explicitly constructed in the framework of lattice gauge theory, where the stress superconductor is just a flavored version of the vortex condensate. Surely extra complications such as the glide principle can be readily incorporated and it would be very interesting whether this will indeed reproduce the universal long-wavelength description that we have highlighted in this review.  

\subsubsection{Duality squared is one}

First of all, since duality is supposed to work both ways, one should be able to dualize the disorder parameter  of the quantum nematics (and/or smectics) once more, to return to the original ordered state by a condensation of the topological defects that are singularities in the disorder field. For instance, one should be able to go from the nematic to the solid by a proliferation of dual topological defects. What are these defects? Nominally, the  disorder fields of the nematic are coincident with the superfluid. Their topological excitations are in turn 
 the familiar vortices, and we learned in Sec.~\ref{sec:XY-duality}, dealing with the vortex--boson duality that the condensate of vortices forms a dual superconductor which is coincident with the boson-Mott insulator~\cite{BeekmanSadriZaanen11, BeekmanZaanen12}. But the Mott insulator does not break translations in the way of the crystal since translational symmetry is already broken by the background lattice.  The moral is that the superfluid by itself does not carry sufficient information to reconstruct the crystal dual. 
 
 Obviously one has to focus on the dislocation condensates since these are the disorder fields that themselves have been `formed' departing from the crystal. It has to be so that this follows the template of the vortex--boson duality in backward gear, that can be viewed as the transition from a charged superconductor to its metallic Coulomb phase. The topological excitations are now the fluxoids of the superconductor with their short-range interactions and it is well understood that when these proliferate and condense one gets back to the Coulomb phase/superfluid with its massless Goldstone mode. In case of the present stress duality this becomes technically a much more involved affair given the many extra modes that have to be tracked, that eventually should resurrect the phonons of the solid. 
 
 This backward dualization might also have some interesting ramifications for the interpretation of experiments that presently attract much attention: the field-induced charge density waves in the YBCO and BISCO cuprate superconductors~\cite{HoffmanEtAl02, LangEtAl02, ChangEtAl12, WuEtAl13, CominEtAl14, CominEtAl15, KeimerEtAl15}. A closely related subject is the field-induced magnetism in the LSCO superconductors that were discovered 
 in the 1990s, often interpreted as as a signature of freezing of the charge stripes~\cite{LakeEtAl02}.  Departing from an underdoped cuprate which appears to be nominally just a superconductor at a low temperature, the effect of an applied magnetic field is to stabilize an Abrikosov lattice of fluxoids. It is now observed that in the core of these vortices a charge density modulation appears that might lock into a long-range charge density order when the field exceeds a critical value~\cite{WuEtAl11, WuEtAl13}.
 
 This can be rationalized on basis of simple Ginzburg--Landau theory involving just competing charge and superconducting order \cite{SachdevLaPlaca13, FradkinKivelsonTranquada15, Lee14, HaywardEtAl14}. However, our dual stress superconductor perspective suggests a quite natural and highly elegant alternative explanation. As we just argued, the superconducting fluxoid is falling short as primary topological defect since it has not the capacity to encode for the resurrection of the crystalline order. However, given that the superconductivity and the dual stress condensate `live on the same side' of the duality it has to be the case that the superconducting fluxes, stabilized as usual by the magnetic field, are coincident with a composite of vortices in the dual stress superconductor. The dislocations are expelled from the core of such a topological defect with the effect that the crystalline order {\em has to} re-emerge in the core of the superconducting fluxoid! It is a no-brainer: upon destroying the superconductivity our system can only rediscover its crystalline heritage! The question arises whether any observable consequences can be deduced from this alternative perspective. We have just not attempted to work this out in any detail and we present it here as an interesting and quite timely challenge.  
 
Finally, we should note that the disorder theory of crystal melting can become considerably more complex when considered in a crystal formed in a topological state of matter with intrinsic couplings to strains and curvature, i.e. the lattice defects themselves~\cite{ChoEtAl15}. These results emerge from the concept of \emph{geometric response} of two-dimensional topological states, see e.g.~\cite{CanLaskinWiegmann15} and references therein, and in addition to Ref.~\cite{ChoEtAl15} would certainly be an interesting avenue to apply the theory developed here. At the moment these geometric considerations would rest on an uncontrolled assumption about the coexistence of the (liquid) crystalline correlations and the topological state --- but there is work to be done even without the assumption of the extra topological correlations, as we discuss in Item 7. of this list.

\subsubsection{Anisotropy and  the quantum smectic}

With regard to the symmetry characteristics we have been focused on the quite literal `spherical-cow' limit of generalized elasticity. Symmetry is a simplifying circumstance and for this reason we considered the maximally isotropic versions of solid, quantum nematic and smectic order. For solids, it is well understood how to deal with the complications associated with lower-dimensional space groups: this just introduces extra elastic moduli and these are classified and available in tabulated form~\cite{LandauLifshitz86, Kleinert89b}. Although we have not studied this in detail, it should also be straightforward to generalize the theory to less symmetric quantum nematics. In Sec.~\ref{sec:Order parameters for 2+1-dimensional nematics} we presented the systematic order parameter theory for nematics in two space dimensions. Departing from the stress fields of the anisotropic crystals  it appears to be straightforward to arrive at a full dynamical description of the anisotropic stress superconductors. 

Another issue is the {\em quantum smectic}: although we did not emphasize it, to quite a degree the general order parameter theory is still quite a mysterious affair. This is especially the case at zero temperature. We highlighted the very surprising interplay of superfluid and solid characteristics, noting that these orders are genuinely intertwined in the quantum smectic. It seems obvious that the behavior we found is tied into the isotropic limit --- we argued that this would work in a quite different dealing with the  quantum generalization of the conventional `liquid-layer' smectics of soft matter that are the smectic partners of the uniaxial nematics. To find their long-wavelength physics one does need the heavy machinery of the full duality which also processes the information associated with the massive modes. Embarrassingly, the Ginzburg--Landau--Wilson theory that prescribes what to expect at long wavelengths merely on basis of symmetry principles, is still lacking. We perceive it as a confusing affair. On the one hand, the rotational symmetry of the crystal and nematic phases that border the smectic on opposite sides is still remembered --- we dealt implicitly with a hexagonal crystal and a hexatic quantum nematic. However, at face value the rotational symmetry appears to be broken to a $C_2$-symmetry associated with the liquid and solid direction, followed by the rather intricate way that the system behaves depending on the transverse and longitudinal fluctuations in particular momentum directions.  Embarrassingly, we have not found out how to formulate such a general theory and we present this as a challenge to the readership.

\subsubsection{Generalization to 3+1D dimensions}

A major limitation of the work reported in this review is that is exclusively focused on two space dimensions. This is mostly for technical reasons: 2+1D is just much easier than 3+1D. Adding one dimension has the effect  that hell breaks loose. But this also has its benefits: there is much more going on in the three space dimensions of our universe. This becomes already manifest dealing with the bare-bone order parameter theory as discussed in Sec.~\ref{sec:Order parameters for 2+1-dimensional nematics}. As we learned, the principle is that 
the generalized nematic states are classified by the point groups, and in 2D this is just all about discrete Abelian symmetries. However, in 3+1D the point groups are nearly always {\em non-Abelian}. Nematic liquid crystals have been extensively studied in 3D but for the practical reason that this only works well for rod-like molecules this is nearly exclusively dealing with uniaxial nematics. It is just a coincidence that these are singular in the regard that these are still governed by an Abelian $\integers_2$ symmetry related to the simple $D_{\infty h}$ point group. As it turns out, not much is known regarding the order parameter theories associated with the generic non-Abelian point groups: these are surprisingly complicated, involving order parameter tensors of a high rank. In this context, the gauge-theoretical formulation of Sec.~\ref{sec:Order parameters for 2+1-dimensional nematics} comes to full fruition: one just generalizes the $O(2) / \mathbb{Z}_N$ theory to $O(3) / P$ where $P$ is any of the 3D point groups and by taking the limit of strong gauge coupling it turns into a generating functional for the order parameter theories of the generalized nematics. We are presently exploring this fascinating landscape~\cite{LiuEtAl15b, NissinenEtAl16, LiuEtAl16}.

Yet another matter is to formulate the full elastic duality in 3+1D. Here one meets questions of deep and general principle. In 2+1D the dislocations are `particles' and in the construction of the dual stress superconductor we relied heavily on the machinery of second quantization: the {\em system} of bosonic particles can be described in terms of the disorder field theory describing the evolution of the condensate order parameter. The difficulty is that in 3+1D dislocations become {\em strings or line defects} on the quantum level (see Sec.~\ref{subsec:Dislocations and disclinations}). It is still possible to formulate a stress gauge theory, describing how these dislocation-strings source long-range interactions, now involving {\em two-form} gauge fields. To parametrize a worldsheet we need two space-time indices, and as such a 3+1D dislocation is given 
by $J^a_{\mu\nu}$, where $a$ denotes the Burgers vector. Accordingly, the stress gauge fields that couple to these dislocations are $b^a_{\mu\nu}$, the so-called two-form, or Kalb--Ramond gauge fields. Dealing with the solid this works fine as a quite efficient framework to compute how stresses are sourced and propagate in the 3+1D elastic medium.
However, it is just not known how to generalize second quantization to deal with the ``tangle" (better, ``foam") formed by such strings that have proliferated into the liquid (crystal). This is a very basic incarnation of the infamous hurdle called {\em string field theory}. One could then argue that the duality principle should be universal, with the consequence that the 3+1D quantum liquid crystal can be still viewed as some kind of dual stress superconductor fueled by a stringy dislocation condensate. It is instructive to take a step back 
to vortex--boson duality as in Sec.~\ref{sec:XY-duality}. In the 3+1D superfluid one has vortex worldsheets and two-form gauge fields, which in the high-energy literature are referred to as Nielsen--Olesen strings and Kalb--Ramond gauge fields respectively. In the context of condensed matter systems (as opposed to critical, coreless strings), the condensation of such vortex strings was considered in Ref.~\cite{BeekmanSadriZaanen11}. The bottom line is that it is possible to describe the effects of the string condensate  in terms of a Josephson-limit Higgs term deduced on phenomenological grounds. In an upcoming publication we will show how to  generalize this the `flavored version'  associated with the dislocation condensation~\cite{WuEtAl16}.

\subsubsection{Perturbing from the strong correlation limit: interstitials}

It is insightful in physics to know the limiting cases, but it is even more useful to perturb away from the limits to find out what is happening in the middle. This is perhaps the most pressing open question we are facing.  The  maximally-correlated fluid limit on which we have focused rests on the assumption that the typical distance between the dislocation is large compared to the lattice constant: the low-fugacity requirement. The physics involved in the perturbation theory is clear: one has to restore the constituent bosons which are the building blocks of the kinetic gas theory of weak coupling as degrees of freedom. Departing from the low-fugacity limit it is clear when these appear in a strong-coupling perturbative sense. In the solid they correspond to interstitial/vacancy excitations; the loose atoms in the lattice. These have a clear identity with respect to the topology of the strongly-coupled crystals: interstitials are equal to bound pairs of dislocation--antidislocation pairs that are {\em separated by a lattice constant}.  One can picture a perturbative procedure where one systematically `dresses' the strong-coupling limit with dislocation-loop corrections, but we have not developed this in any detail. 

Such a perturbation theory could be quite useful to address a number of physics phenomena which are beyond the scope of the present theory. First, one can imagine that these massive excitations could overdamp the remnant massive modes left over from the crystal for finite Higgs mass $\Omega$, making their observation more challenging. Another example is the supersolid~\cite{KimChan04, NussinovEtAl07, BalatskyEtAl07, Prokofev07, BoninsegniProkofev12}, how does this fit into the greater scheme? In the two-dimensional Galilean continuum, supersolids will {\em not} be encountered. The supersolid should be simply viewed as Bose gas of interstitials peacefully coexisting with the crystal. The reason that this can happen is of course that topologically interstitials can be regarded as bound pairs of neutral dislocation--antidislocation pairs that do not destroy the crystal when they proliferate. However, this also provides the reason why supersolids do not exist in the 2D continuum: both the dislocations and interstitials are particle-like and the core energies of the latter are higher than those of individual dislocations and antidislocations that generically repel each other. In three dimensions this is a different affair since dislocations are strings, while the interstitials are still point particles which can carry a much larger zero-point (kinetic) energy~\cite{ProkofevSvistunov05}. However, in the presence of a background lattice these rules change: when the commensuration energy becomes large, the long-range deformation fields associated with the dislocations become very large and eventually the dislocations will get tightly bound, stabilizing the dislocation gas and the supersolid.  Accordingly, supersolids are an ubiquitous theme in the main-stream condensed matter literature dealing with (boson-)Hubbard models, since these models can be viewed as dealing with the limit of strong periodic background potentials~\cite{LiuFisher73, FisherEtAl89, BatrouniEtAl95, WesselTroyer05, MelkoEtAl05, HeidarianDamle05, SenguptaEtAl05}. Obviously one would like to understand better how this balance between dislocation quantum melting and supersolid formation develops as function of the potential strength, degree of commensuration and so forth, but for this purpose one needs the {\em interstitial perturbation theory}. In this regard, we notice that the `stripy' electronic charge order as seen by scanning tunneling spectroscopy (STS) in cuprate superconductors appears to be riddled with dislocations~\cite{MesarosEtAl11}. In fact, the data seem to show smooth, continuum-like long-range deformation fields associated with the dislocations. This is quite confusing since this stripy charge order is supposed to be rooted itself in the proximity to the Mott insulator which is in turn the hallmark of strong lattice commensuration. As we stressed early on, this is perhaps the best experimental indication that is currently available to take the idea of dislocation melting serious in this context.
 
Another question that arises is: what to think about the well-documented case of superfluid $^4$He? In a way this is a quite strongly-correlated liquid. On the microscopic scale it is basically a dense Van der Waals liquid that is kept fluid by the quantum zero-point motions~\cite{Ceperley95, ClarkCeperley06}. Historically it is also the instance where the Bogoliubov weak-coupling treatment of the Bose gas fails, among the highlights of quantum Monte Carlo calculations~\cite{Ceperley95}. The hallmark of the strong-correlation effects is the roton minimum, as seen directly by inelastic neutron scattering. Its essence was explained by Feynman's single-mode approximation linking it to the wiggles in the liquid structure factor. This in turn can be viewed as the signal of solid-like correlations extending over a distance of order of the lattice constant in the liquid. The crystallization transition in $^4$He is actually strongly first order, and the shear penetration depth (in our language) in the liquid is therefore short. Viewed from the dual perspective the question that immediately arises is: {\em 
 is there any relation between the roton and the massive shear photons of Sec.~\ref{sec:Quantum nematic}?} One would expect that in the presence of interstitials the propagating shear modes should be destroyed at small momenta while perhaps a remnant can survive at the roton momentum (inverse lattice constant). Conversely, from this perspective one expects that the roton should light up in the dynamical shear response. In principle this information can be extracted from quantum Monte Carlo calculations but to the best of our knowledge 
this has never been tried.  

\subsubsection{Translational order in the background: lattice pinning}

The most plausible instance of the quantum liquid crystals we consider seem to be the electronic kind of Secs.~\ref{sec:Dual elasticity of charged media}--\ref{sec:Electromagnetic observables}, which always live on an underlying ionic lattice, that invariably provides a periodic potential to some extent, and as such breaks spatial symmetry explicitly, not spontaneously. The hope is that the `pinning' of the electronic ordering to this background potential is weak enough that the liquid crystal features remain apparent. It is still useful to consider which effects a (mild) explicit symmetry breaking has on the phenomenology in this work. 

In general, departing from a spontaneously broken continuous symmetry, upon applying a field that breaks this symmetry explicitly the Goldstone mode will acquire a gap proportional to the strength of this external field. 
This is not different from the mass of the pion in particle physics associated with the almost-spontaneous breaking of chiral symmetry, or of the gap forming in the magnon spectrum in a ferromagnet in an external magnetic field. Dealing with the Goldstone modes of the smectic, nematic and even the elastic state this should of course be the same affair. We have already discussed the ramifications for experimental systems in the introduction. In the cuprates and pnictides, one departs from a tetragonal square lattice with a fourfold rotational symmetry that gets lowered in the electronic nematic state to a twofold symmetry, corresponding to  an orthorhombic lattice, and accordingly the symmetry of the order parameter is of the Ising kind. Surely, in the present context the anisotropy gap should not be too large. When it gets of order of other scales in  the system (e.g., the scale where pairs break up and so forth) the propagating excitations that are at the central entities in the field theory become irrelevant. 

In fact, from a microscopic perspective, strong lattice potentials should eventually be {\em detrimental} for the formation of quantum liquid crystals in general, and especially for the ``maximally correlated'' ones described by duality. The reason is well known: for dislocations to be of relevance it should be possible to be accompanied by smooth strain deformations that are described by the massless stress photons of our formalism. When the lattice pinning becomes strong this is no longer possible with the effect that dislocations become energetically very costly. At the same time, the interstitial defects (loose particles) come down in energy relative to the dislocations. This situation is addressed just above.

Turning to the empirical context of the cuprates, lattice pinning should be a major concern. After all, the stripes themselves are supposedly formed {\em because} of the strong Umklapp scattering. As explained already in the seminal paper Ref.~\cite{ZaanenGunnarsson89}, the stripe phenomenon on doped Mott insulators is best understood as a commensuration phenomenon. The Mott insulator is most generally viewed as an electron density wave commensurate with the ionic potential. Upon doping, the mismatch in periodicities between the ionic crystal and the electronic density wave is stored in  a periodic array of {\em discommensurations}: the charge stripes. To repeat ourselves: the fact that the stripe patterns as seen directly by scanning tunneling spectroscopy appear to be littered by static dislocations, accompanied by smooth strain fields~\cite{MesarosEtAl11}, is perhaps the best validation of possible dislocation quantum melting in the cuprates. The `hard rules' for commensuration of stripes are however tied to a conventional (Hartree--Fock) mean-field description. The latest numerical results indicate however that this somehow works out differently: there is a lack of preferred charge density on these stripes~\cite{ZhengEtAl16}.  It would be quite interesting to find out what these numerical big guns have to say about stripe dislocations.

Finally, in the case of the nematic on an $n$-periodic substrate, things are much simpler: the lattice pinning introduces the external field $ \propto h_n \cos(2\pi \theta n)$ for the nematic order parameter $\theta$ (for the $\mathbb{Z}_N$ generalization, $n\to k$, where $k$ is the smallest integer so that $N k/n$ is integer). For $n > n_c \equiv 3.41$, these couplings are irrelevant, and for $n < n_c$, the XY-like universality class of the nematic isotropic transition switches to the Ising or chiral Potts universality class~\cite{Balents96}. Of course, in the former case, to lowest order the lattice pinning introduces a mass $\propto h_n$ for the rotational Goldstone $\delta \theta = \omega$.

\subsubsection{Translational disorder in the background: glassy stress superconductivity}

 A closely related theme is, what happens when the dual stress superconductor lives in a background characterized by a disorder potential? The central theme of metallurgy is of course that dislocations pin easily to lattice defects, rendering the metal to be more brittle or malleable. The same principle of course applies to our quantum dislocations and the question arises: what to expect when the dual stress superconductor is exposed a to a background with a disorder potential of increasing  strength?  The quantum particles subjected to the disorder potential are now the delocalized dislocations forming the condensate and one anticipates these will tend to bind to the deepest minima in the potential landscape. Thereby they will localize and the effect is that a piece of solid emerges surrounding the static dislocations, while at the same time the superfluid density in this region is suppressed. To the observer with eyes focused on the density modulations this will appear as a glass-like substance,  although strangely structured since the mediators of the translational disorder (dislocations) appear to be at the same time the agents that {\em cause} the crystalline features in the first place. 
 
It would be interesting to explore this further, yet again looking for unambiguous observational consequences. Could it be that the rather mysterious disorder observed in the stripy electronic patterns observed by scanning tunneling spectroscopy (STS) on cuprate superconductors are in fact of this kind~\cite{FischerEtAl07, CarlsonEtAl15}? At the very least, as we mentioned, these STS patterns in cuprate clearly show the abundant presence of dislocations~\cite{MesarosEtAl11}.   

\subsubsection{Relations to gravity}

Let us finish this review with a subject matter that is perhaps most appealing to the imagination of the physicist: the multifaceted relations between field theoretical elasticity and gravitational physics~\cite{Kondo52, Kroner59, BilbyBulloughSmith55, McCreaHehlMielke90, KatanaevVolovich92}. 
In more than one regard the theory of general relativity (GR) is around the corner. By and large we have ignored  this aspect: it is complementary to the material we have discussed here and it deserves a thorough treatise by itself. To quite a degree Kleinert was in first instance fascinated by this aspect in his development of the fundamentals of the formalism~\cite{Kleinert89b,Kleinert08}. The idea underlying this pursuit dates back in a way to Einstein himself: the notion that the `fabric of spacetime' is metaphorically not unlike an elastic medium. But it is crucial regards yet completely different. Are there ways to make this metaphor more literal? 

The next step has also a long history, it seems to be dating back to Landau: the theory of elasticity can be geometrized and the resulting structure has several critical aspects in common with GR. This departs from 
the notion that a crystal defines an {\em internal geometry}. The crystal lattice is given the status of a coordinate system that is explored by an internal observer that measures distances by hopping from site to site.
The metric relevant to such an observer has the form $\td s^2 = \delta_{ab} + u^{ab}$ focusing on space dimensions where $u^{ab} = \tfrac{1}{2} (\partial_a u^b + \partial_b u^a )$, our familiar strain fields. One directly infers why spacetime is not a crystal: the diffeomorphisms (coordinate transformations) have turned into the physical elastic strain fields. The crystal is a manifold with a preferred frame! Despite this disaster, this crystal geometry still knows about the geometrical properties of {\em torsion} and {\em curvature}: the dislocation and disclinations precisely encapsulate these properties, be it in quantized form. A final aspect is the role of time. Real crystals are formed from matter at a finite density and this caused the bad breaking of Lorentz invariance that has been so important in this review. In order to get anywhere with gravity, this has to be restored. This can be done by hand by insisting that the point of departure is the {\em world crystal}: an elastic medium that is isotropic in (Euclidean) spacetime. One should be very aware that this is a strange object, very different from a normal crystal: since it is immaterial (zero density), it does not carry sound when the `liquid' has formed. Neither does the corresponding quantum liquid crystal carry rotational modes. 

Let us first zoom in on a less esoteric matter. Crystals and liquid crystals can be sourced by curvature in the background geometry. This is actually a flourishing subject in soft matter, dealing with matter living on curved two-dimensional surfaces embedded in three external dimensions~\cite{BowickGiomi09, KoningVitelli14}. What happens when one tries to cover the surface of a sphere with a solid or nematic layer? In a flat background, the crystal curvature is confined: this is encapsulated by the confined disclinations. However, the background curvature coerces the crystal to curve, which causes a periodic array of disclinations to be formed since Frank scalars are quantized. A famous point in case is the football (or buckyball) which can be viewed as a lattice of pentagons (disclinations) in a hexagonal lattice background. In the language of stress gauge fields this can be elegantly addressed~\cite{Kleinert89b,Kleinert08}. In space directions the stress tensor is a symmetric tensor and for this reason it can be parametrized in terms of double-curl gauge fields $h_{ab}$ as $\sigma_{a a'} = \varepsilon_{abc} \varepsilon_{a'b'c'} 
\partial_b \partial_b' h_{c c'}$ where $h$ itself is a symmetric rank-2 tensor field. But this is precisely coincident with the form of the Einstein tensor $G_{ij}$ in linearized gravity, where $h_{ab}$ is identified
with the graviton! 

Ignoring the dislocations, one now finds that $h_{ab}$ is sourced by the disclination current according to $h_{ab}
\Theta_{ab}$: the Belinfante form to enter stress energy in linearized gravity.  It is a simple matter to demonstrate that the background curvature enters as $h_{ab} ( \Theta_{ab} - G_{ab} )$ where $G_{ab}$
is the Einstein tensor of the background geometry.  The bad `net' curvature $\Theta_{ab}-G_{ab}$ can now be expelled from the confining vacuum by matching the background curvature with the disclination
currents so that their sum disappears. In two space dimensions this becomes an especially simple affair since only scalar curvature enters, which is easily matched with the Frank scalars of the disclinations.
      
At the next stage, it is amusing to ask the question what  happens when the world crystal is melted into a relativistic {\em world nematic}~\cite{KleinertZaanen04}. Because the stress tensors are now symmetric in  relativistic spacetime this becomes a very simple exercise that works out regardless the number of dimensions. Schematically the action of the crystal is  $S \propto \int \sigma_{\mu \nu} \sigma^{\mu \nu}$. Upon proliferating the dislocations
isotropically one finds in addition a Higgs term, in shorthand $\propto \sigma_{\mu \nu} \sigma^{\mu \nu} / \partial^2$. Identifying $\sigma_{\mu\nu}$ with the linearized Einstein tensor and expanding in the graviton field $h_{\alpha \beta}$, the Higgs term is just the same as the Fierz--Pauli action of linearized gravity! In the relativistic nematic curvature is deconfined as the proper elastic rigidity, and in 3+1D one will find a pair of massless spin-2 modes, that now mediate interactions between the stress-energy sources: the disclinations with their quantized curvature that can be identified with cosmic strings in the 3+1D setting. It can actually be precisely argued that this is rooted in  the restoration of general covariance by the dislocation condensate~\cite{ZaanenBeekman12}. However,  it fails badly with regard to describing GR in its full glory. The rotations/Lorentz boosts are still in the fixed frame and only infinitesimal Einstein translations are restored. Therefore this relativistic nematic is the vacuum of linearized gravity,  but there is nothing more: the theory cannot be bootstrapped in its full non-linear glory. We notice that Kleinert himself has taken a qualitatively different way is his pursuit of the world crystal spacetime by postulating a {\em floppy world crystal} space-time which has little in common with earthly crystals and has thereby the freedom to overcome the difficulties we just noticed~\cite{Kleinert89b,Kleinert08}.

Finally, field-theoretical elasticity might be quite useful in yet a very different gravitational context: the description of matter by translating it into a gravitational {\em holographic dual} using the AdS/CFT correspondence. Only very recently a proposal has appeared for the gravitational dual of elasticity in terms of the ``Stueckelberg star''~\cite{AlberteEtAl16}. This is in a boundary language which is closely related to the language we have been using. There is yet another long standing, and deep mystery in holographic duality related to the description of liquid crystals related to the deep problem that this has to be associated with spin-2 modes in the bulk that are already used up by the stress energy in the boundary~\cite{LangleyVanacorePhillips15}. It might well be the stress duality of this review can be translated to the gravitational bulk, where it should reveal profound structure.

\section*{Acknowledgements}
We thank Dirk van der Marel for useful discussions concerning Sec.~\ref{sec:Electromagnetic observables}. A.J.B. is partially supported by the MEXT-Supported Program for the Strategic Research Foundation at Private Universities ``Topological Science'' (Grant No. S1511006) and by the Foreign Postdoctoral Researcher program at RIKEN. This work was supported by the Netherlands foundation for Fundamental Research of Matter (FOM). The work of Z.N. has been supported from the NSF grant DMR 1411229.

\appendix
\section{Fourier space coordinate systems}\label{sec:Fourier space coordinate systems}
Throughout the text we often employ coordinate systems for the spacetime indices relative to the momentum of the fields. Here we give explicit coordinate transformations between all these systems.

The point of departure is the real space imaginary time coordinate system $(\ft,x,y)$, where $\ft = \frac{1}{c} \tau$ and $c$ a velocity, usually the shear velocity $c_\tT$. The second system $(\ft,\tL,\tT)$ rotates the spatial coordinates into parallel and orthogonal to the spatial momentum $\mathbf{q} = (q_x,q_y)$ (and $q = \sqrt{q_x^2 + q_y^2}$), in other words, the longitudinal and transverse directions $\tL$ resp. $\tT$. In the third system $(0,+1,-1)$ the temporal and longitudinal components are mixed so that the $0$-component is parallel to the spacetime momentum $p_\mu = (\frac{1}{c}\omega_n,\mathbf{q})$ (and $p = \sqrt{\frac{1}{c^2}\omega_n^2 + \mathbf{q}^2}$), where $c$ is the velocity of the medium (e.g. the transverse phonon velocity $c_\tT$); the $+1$-component is also in the $\ft$--$\tL$ plane but orthogonal to $0$, and $-1$ is parallel to the spatial-transverse direction $\tT$. Please note that there is a sign difference between our $-1$-component and that of Ref.~\cite{ZaanenNussinovMukhin04,Cvetkovic06}. Also note that the temporal components have been rescaled $A_\ft = \frac{1}{c} A_\tau$ to have the same dimension as the other field components (cf. Sec.~\ref{subsec:Conventions and notation}).

\begin{figure}
\begin{center}
 \begingroup%
  \makeatletter%
  \providecommand\rotatebox[2]{#2}%
  \ifx\svgwidth\undefined%
    \setlength{\unitlength}{8cm}%
    \ifx\svgscale\undefined%
      \relax%
    \else%
      \setlength{\unitlength}{\unitlength * \real{\svgscale}}%
    \fi%
  \else%
    \setlength{\unitlength}{\svgwidth}%
  \fi%
  \global\let\svgwidth\undefined%
  \global\let\svgscale\undefined%
  \makeatother%
  \begin{picture}(1,0.73534958)%
    \put(0,0){\includegraphics[width=\unitlength]{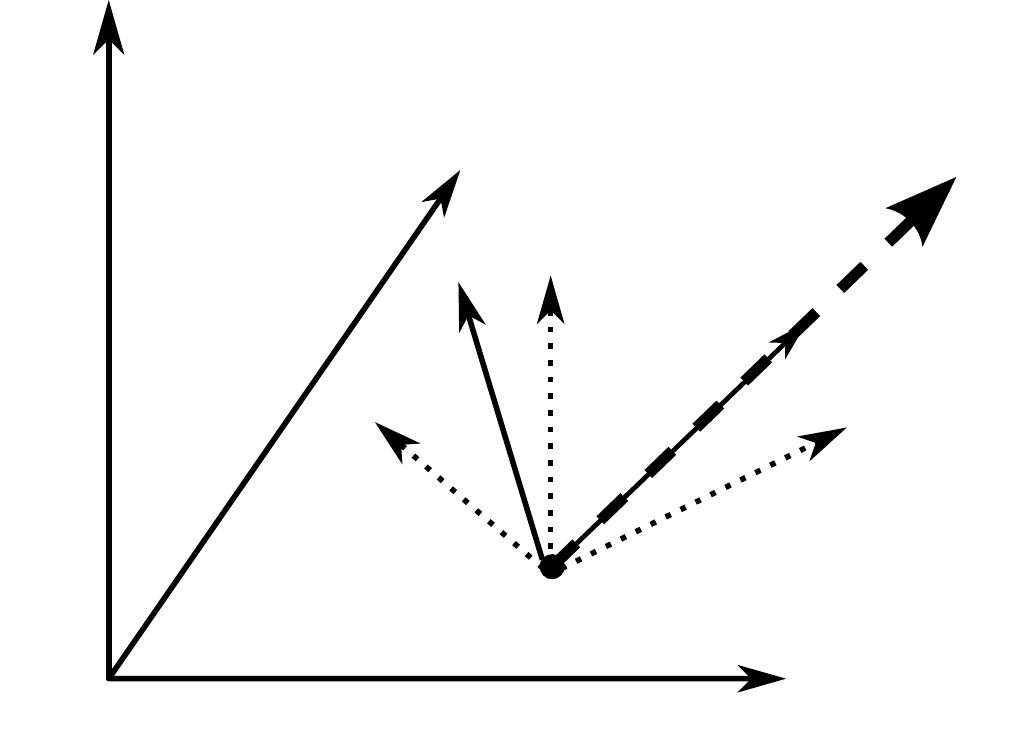}}%
    \put(0.65,0.00){\color[rgb]{0,0,0}\makebox(0,0)[lb]{$x$}}%
    \put(0.32,0.50){\color[rgb]{0,0,0}\makebox(0,0)[lb]{$y$}}%
    \put(0.00,0.61){\color[rgb]{0,0,0}\makebox(0,0)[lb]{$\ft$}}%
    \put(0.70,0.39){\color[rgb]{0,0,0}\makebox(0,0)[lb]{\smash{0}}}%
    \put(0.94,0.57){\color[rgb]{0,0,0}\makebox(0,0)[lb]{\smash{$p_\mu$}}}%
    \put(0.75,0.21){\color[rgb]{0,0,0}\makebox(0,0)[lb]{$\tL$}}%
    \put(0.32,0.20){\color[rgb]{0,0,0}\makebox(0,0)[lb]{$\tT/-1$}}%
    \put(0.55,0.38){\color[rgb]{0,0,0}\makebox(0,0)[lb]{$\ft$}}%
    \put(0.40,0.36){\color[rgb]{0,0,0}\makebox(0,0)[lb]{$+1$}}%
  \end{picture}%
\endgroup%
\caption{Coordinate systems relative to momentum of the fields. Pictorial explanation of the $(\tau,\tL,\tT)$ and $(0,+1,-1)$ systems.}
\end{center}
\end{figure}
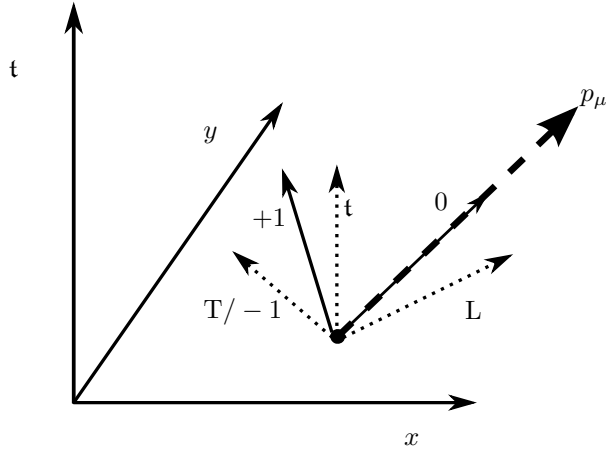

A general coordinate transformation between fields $A_\mu$ and $A'_\alpha$ is defined by $A_\mu = e_\mu^\alpha A'_\alpha$. For instance $A_\tL \sim \sum_{\alpha = \ft,x,y} e_\tL^\alpha A_\alpha$ where $ e_\tL^\alpha = (0,q_x/q,q_y/q)$. However, since these coordinate systems are defined in Fourier space only, we should insist on the condition for any field
\begin{equation}
 A(-p) = A^\dagger(p)
\end{equation}
so that the basis vectors are real-valued. Therefore we are led to the following relation for $\mu \in (\tau,x,y)$
\begin{align}\label{eq:coordinates factor i}
A_\mu &= e_\mu^\ft A_\ft + \sum_{E = \tL,\tT} \ti e_\mu^E A_E \nonumber\\
&=  \ti e_\mu^0 A_0 +  e_\mu^{+1}A_{+1} +  \ti e_\mu^{-1} A_{-1}.
\end{align}
Below these factors of $\ti$ are already incorporated. Note that because of these relations $q_\tL = -\ti q$. Also we have $\partial_m A_m =  - q A_\tL$ while $\partial_m A_m^\dagger = q A_\tL^\dagger$ etc.

The explicit coordinate transformations are:

{\bf(i)} $A_{\ft,x,y} \leftrightarrow A_{\ft,L,T}$ 
\begin{align}
 \begin{pmatrix} 
  A_\ft \\ A_x \\ A_y 
 \end{pmatrix}
&=
\begin{pmatrix}
 1 & 0 & 0 \\
 0 & \ti \frac{q_x}{q} & -\ti \frac{q_y}{q} \\
 0 & \ti \frac{q_y}{q} & \ti \frac{q_x}{q} \\
\end{pmatrix}
\begin{pmatrix}
 A_\ft \\ A_\tL \\ A_\tT 
\end{pmatrix},\\
\begin{pmatrix}
 A_\ft \\ A_\tL \\ A_\tT 
\end{pmatrix}
&=
\begin{pmatrix}
 1 & 0 & 0 \\
 0 & -\ti \frac{q_x}{q} & -\ti \frac{q_y}{q} \\
 0 & \ti \frac{q_y}{q} & -\ti \frac{q_x}{q} \\
\end{pmatrix}
\begin{pmatrix} 
  A_\ft \\ A_x \\ A_y 
 \end{pmatrix}.
\end{align}

{\bf(ii)} $A_{\ft,x,y} \leftrightarrow A_{0,+1,-1}$ 
\begin{align}
 \begin{pmatrix} 
  A_\ft \\ A_x \\ A_y 
 \end{pmatrix}
&=
\begin{pmatrix}
 \ti  \frac{\omega_n}{cp} & -\frac{q}{p} & 0 \\
 \ti \frac{q_x}{p} & \frac{\omega_n q_x}{cpq} & -\ti \frac{q_y}{q} \\
 \ti \frac{q_y}{p} & \frac{\omega_n q_y}{cpq} &  \ti \frac{q_x}{q}
\end{pmatrix}
\begin{pmatrix}
 A_0 \\ A_{+1} \\ A_{-1}
\end{pmatrix},\\
\begin{pmatrix}
 A_0 \\ A_{+1} \\ A_{-1} 
\end{pmatrix}
&=
\begin{pmatrix}
 -\ti \frac{\omega_n}{cp} & -\ti \frac{q_x}{p} & -\ti \frac{q_y}{p}\\
 -\frac{q}{p} & \frac{\omega_n q_x}{cp q} &  \frac{\omega_n q_y}{cp q} \\
 0 & \ti \frac{q_y}{q} & -\ti \frac{q_x}{q} \\
\end{pmatrix}
\begin{pmatrix} 
  A_\ft \\ A_x \\ A_y 
 \end{pmatrix}.
\end{align}

{\bf(iii)} $A_{\ft,\tL,\tT} \leftrightarrow A_{0,+1,-1}$ 
\begin{align}
 \begin{pmatrix} 
  A_\ft \\ A_\tL \\ A_\tT 
 \end{pmatrix}
&=
\begin{pmatrix}
 \ti  \frac{\omega_n}{cp} & -\frac{ q}{p} & 0 \\
 \frac{q}{p} & -\ti \frac{\omega_n}{cp} & 0 \\
 0 & 0 &  1
\end{pmatrix}
\begin{pmatrix}
 A_0 \\ A_{+1} \\ A_{-1}
\end{pmatrix},  \label{eq:transf tLT 0+1-1}\\
\begin{pmatrix}
 A_0 \\ A_{+1} \\ A_{-1} 
\end{pmatrix}
&=
\begin{pmatrix}
 -\ti \frac{\omega_n}{cp} &  \frac{q}{p} & 0\\
 -\frac{q}{p} & \ti \frac{\omega_n }{cp } & 0 \\
 0 & 0 & 1 \\
\end{pmatrix}
\begin{pmatrix} 
  A_\ft \\ A_\tL \\ A_\tT 
 \end{pmatrix}.
\end{align}

Some useful relations are (for any fields $A_\mu$ and $B_\nu$),
\begin{align}
A^\dagger_x B_x + A^\dagger_y B_y &= A^\dagger_\tL B_\tL + A^\dagger_\tT B_\tT.\\
\int \td^3 x\  A_\mu(x) B_\nu(x) &= \int \frac{\td^3 p}{(2\pi)^3}A_\mu(-p) B_\nu(p) \equiv \int A^\dagger_\mu B_\nu.\\
\int \td^3 x\ \epsilon_{\mu\kappa\lambda}\partial_\kappa A_\lambda \epsilon_{\mu\rho\sigma} \partial_\rho B_\sigma 
&= \int \frac{\td^3 p}{(2\pi)^3} \epsilon_{\mu\kappa\lambda}\epsilon_{\mu\rho\sigma} p_\kappa p_\rho A^\dagger_\lambda B_\sigma \nonumber\\
&= \int_p A^\dagger_\mu (p^2 \delta_{\mu\nu} - p_\mu p_\nu) B_\nu = \int_p \Big( A^\dagger_{+1} B_{+1} + A^\dagger_{-1} B_{-1} \Big). \label{eq:curl-curl contraction tLT}
\end{align}.

\section{Euclidean electromagnetism conventions}\label{sec:Euclidean electromagnetism conventions}

The metric is ``mostly plus" $\eta_{\mu\nu} = \mathrm{diag}(-1,+1,+1)$. In imaginary time $\tau = \ti t$, all covariant tensors undergo transformations. In particular, we define $V^\imath =-\ti V$ and $E_{m}^\imath = -\ti E_m$ and $\rho^\imath_Q = -\im \rho_Q$. This leads to the following three-vectors,
\begin{align}
 \partial^l_\mu &= ( \ti\tfrac{1}{c_l} \partial_\tau , \partial_m) & \partial^{l,\mu} &= (-\ti\tfrac{1}{c_l}\partial_\tau , \partial_m),\\
 A_\mu &= (-\ti\tfrac{1}{c_l} V^\imath , A_m) , & A^\mu &= (\ti\frac{1}{c_l} V^\imath , A_m) , \\
 A^{\imath,l}_\mu &= (-\tfrac{1}{c_l} V^\imath , A_m) , & A^{\imath,l\ \mu} &= (\frac{1}{c_l} V^\imath , A_m) , \\
 j_\mu &= ( -\ti c_l \rho^\imath_Q , j_m), & j^\mu &= ( \ti c_l \rho^\imath_Q , j_m),\\
 j^{\imath,l}_\mu &= ( - c_l\rho^\imath_Q , j_m), & j^{\imath,l\ \mu} &= ( c_l\rho^\imath_Q , j_m).
\end{align}
Note that $A^\imath_m = A_m$. We also define $F^\imath_{\mu\nu} = (F^\imath_{\ft n} , F_{mn}) \equiv (-\ti F_{\imath, \ft n} , F_{mn})$ such that
\begin{align}
 F^\imath_{\ft n } &= - F^{\imath, \ft n} 
 = \tfrac{1}{c_l}(\partial_\tau A_n + \partial_n V^\imath) = - \tfrac{1}{c_l} E^\imath_n.
 \end{align}
 The Euclidean Maxwell action $S_\mathrm{E}$, where only covariant (lower) indices are used, is:
\begin{align}\label{eq:Euclidean Maxwell action repeat}
 S_\mathrm{Maxw} &= \ti S_\mathrm{E,Maxw} = \ti \int \td \tau \; \td^2 x\;  \frac{1}{4 \mu_0} F^\imath_{\mu\nu} F^\imath_{\mu\nu} - A^{\imath,l}_\mu j^{\imath,l}_\mu \nonumber\\
 &= \ti \int \td \tau \; \td^2 x\; \frac{1}{2 \mu_0} ( \tfrac{1}{c_l^2} E_m^{\imath 2} + B^2) 
- V^\imath \rho^\imath_Q -A_m j_m .
\end{align}

The imaginary time deformation current is:
\begin{align}
 j^\imath_\ft &= -\ti n e^* c_\tT \nabla \cdot \mathbf{u},&
 \mathbf{j}  &= +\ti n e^* c_\tT \partial_\ft \mathbf{u}.
\end{align}
Substituting back the various imaginary time quantities, we correctly retrieve the real time deformation charge current $j^{\mu}$.

\section{Dual Kubo formula}\label{sec:Dual Kubo formula}

In a spirit similar to Sec.~\ref{subsec:stress correlation functions}, we can also express the conductivity directly in terms of the stress propagator $G$ from Eq.~\eqref{eq:stress contribution to EM fields}. In fact, since the actions are quadratic, we can always replace fields with their equations of motion, which is equivalent to integrating them out.

The conductivity tensor $\hat{\sigma}_{ab}$ (not to be confused with the stress tensor $\sigma^a_\mu$) is defined by the relation in real time,
\begin{equation}\label{eq:conductivity definition}
 j_a = \hat{\sigma}_{ab} E_b = \hat{\sigma}_{ab} ( - \partial_b V - \partial_t A_b) \to \hat{\sigma}_{ab} \omega_n A_b. 
\end{equation}
In the last step, we adopted the radiation gauge fix $V \equiv 0$ as we will do throughout this part of the appendix only. If $\hat{\sigma}_{ab}$ is diagonal and constant then this is just Ohm's law, but in general the conductivity tensor is a function of momentum and energy. Here $j_a$ is the electric current, defined by $j_a = -\partial S_\mathrm{E} / \partial A_a$ in Eq.~\eqref{eq:electric current from action}. For the charged crystal, it can be found from Eq.~\eqref{eq:deformation current}
\begin{equation}
 j_a = -\frac{\partial S_\mathrm{E} } {\partial A_a} = \ti n e^* c_\tT \partial_\ft u^a. 
\end{equation}
For dual variables, we use the inversion Eq.~\eqref{eq:charged strain stress temporal} or directly Eq.~\eqref{eq:dual elasticity EM Lagrangian} to find,
\begin{align}\label{eq:current dual gauge field}
 j_a &= - \frac{n e^*}{\mu} c_\tT \sigma^a_\ft - \varepsilon_0 \omega_p^2 A_a 
 =  \frac{n e^*}{\mu} c_\tT q\, b^a_\tT - \varepsilon_0 \omega_p^2 A^\imath_a. 
\end{align}
Here we used $\sigma^a_\ft = (\partial_x b^a_y - \partial_y b^a_x) \to - q b^a_\tT$. The first term is the so-called paramagnetic current $j^\mathrm{p}_a$ while the second term is the diamagnetic current, although these are not gauge-invariant notions. To get this into a form like Eq.~\eqref{eq:conductivity definition}, we need to substitute $b^a_\tT$ for $A_\mu$. This can be done through the equation of motion, obtained from varying the first line of Eq.~\eqref{eq:stress contribution to EM fields} with respect to $b^{a\dagger}_\mu$:
\begin{equation}
 (G^{-1})^{ab}_{\mu\nu} b^b_\nu + g^a_{\mu\nu} A^\imath_\nu = 0.
\end{equation}
Using $G G^{-1} =1$ as in Eq.~\eqref{eq:stress contribution to EM fields} this can be inverted to
\begin{equation}
 b^a_\mu = - G^{ab}_{\mu \nu} g^b_{\nu\lambda} A^\imath_\lambda.
\end{equation}
In the radiation gauge fix, from Eq.~\eqref{eq:stress-EM coupling matrix} the only non-zero components of $g^b_{\nu\lambda}$ are $g^\tL_{\tT \tL}$ and $g^\tT_{\tT\tT}$, both of which have value $-n e^* c_\tT q/\mu$. Inspecting Eq.~\eqref{eq:current dual gauge field}, we are only interested in the case $\mu = \tT$. Then we find,
\begin{equation}
 b^a_\tT = \frac{n e^* c_\tT}{\mu} q G^{ab}_{\tT \tT} A^\imath_b.
\end{equation}
Substituting in Eq.~\eqref{eq:current dual gauge field} we finally find,
\begin{align}
j_a &= \frac{(n e^* c_\tT)^2}{\mu} \frac{1}{\mu} q^2 G^{ab}_{\tT \tT} A^\imath_b - \varepsilon_0 \omega_p^2 A^\imath_a 
= - \frac{\varepsilon_0\omega_p^2}{\omega_n} ( \delta_{ab} -  \frac{1}{\mu} q^2 G^{ab}_{\tT \tT} ) \omega_n A^\imath_b.
\end{align}
Using the definition Eq.~\eqref{eq:conductivity definition} we find the conductivity tensor to be
\begin{equation}\label{eq:dual Kubo formula}
 \hat{\sigma}_{ab} =  -\frac{\varepsilon_0\omega_p^2}{\omega_n} ( \delta_{ab} -  \frac{1}{\mu} q^2 G^{ab}_{\tT \tT} )
\end{equation}
Keep in mind that $\mu$ is the shear modulus and not the magnetic permeability. Note that we do not need to additionally worry about the original Meissner contribution Eq.~\eqref{eq:Meissner in hiding}, this is already implicitly incorporated through Eq.~\eqref{eq:current dual gauge field}. Also note that in the tilde-coordinate system with respect to the dislocation velocity, $G^{ab}_{\tT\tT} = \tilde{G}^{ab}_{\tT\tT}$ since $b^a_\tT = \tilde{b}^a_\tT$. Recall that $G^{ab}_{\mu\nu} = \langle b^{a\dagger}_\mu b^b_\nu \rangle$. Realizing that the paramagnetic current is $j^\mathrm{p}_a = - \frac{n e^*}{\mu} c_\tT q\, b^a_\tT$, from Eq.~\eqref{eq:current dual gauge field}, we see that this is equivalent to
\begin{equation}
 \hat{\sigma}_{ab} =  -\varepsilon_0  \frac{\omega_p^2}{\omega_n} \delta_{ab} + \frac{1}{\omega_n}\langle j^{\mathrm{p}\dagger}_a j^\mathrm{p}_b \rangle .
\end{equation}
If we transform back to real time and real frequencies via $-\ti \omega_n = \omega + \ti \delta$, we retrieve the usual form of the Kubo formula
\begin{equation}
 \hat{\sigma}_{ab} (\omega,\mathbf{q}) =\frac{\ti}{\omega} \varepsilon_0 \omega_p^2 \delta_{ab}  + \frac{1}{\omega} \langle \big[ j^\mathrm{p}_a ( \omega, \mathbf{q}) , j^\mathrm{p}_b (-\omega ,-\mathbf{q}) \big]\rangle.
\end{equation}
Here the additional factor of $\ti$ in front of the current Green's function comes from its Fourier transform prescription 
\begin{equation}
\int \td \tau\; \te^{- \ti \omega_n \tau} \langle j_a(\omega_n) j_b (-\omega_n) \rangle  \to \ti \int \td t\; \te^{\ti \omega t} \langle j_a(\omega) j_b (-\omega) \rangle 
\end{equation}
and the commutator comes from time ordering in the retarded Green's function. See for instance Refs.~\cite{Mahan93,Coleman15} (but note they use a different convention $\ti \omega_n \to \omega +\ti \delta$).

Alternatively, if we have already integrated out the stress gauge fields and obtained the stress Lagrangian in the form
\begin{equation}\label{eq:stress Lagrangian EM fields}
 \mathcal{L}_\mathrm{stress} = \tfrac{1}{2} A_\mu \mathcal{G}^{-1}_{\mu\nu} A_\nu,
\end{equation}
where the electromagnetic Green's function $\mathcal{G}_{\mu\nu}$ should be distinguished from stress Green's function $G^{ab}_{
\mu\nu}$, then the conductivity can be found directly from Eq.~\eqref{eq:conductivity definition} and the definition of the current $J_a = -\partial S_\mathrm{E} / \partial A_a$, to be
\begin{equation}\label{eq:conductivity from EM propagator}
 \hat{\sigma}_{ab} = -\frac{1}{\omega_n} \mathcal{G}^{-1}_{ab}  \qquad \text{(radiation gauge)}.
\end{equation}
If we have used the (photon field) Coulomb gauge fix instead of the radiation gauge fix, then the conductivities are
\begin{align}
 \hat{\sigma}_{\tL\tL} &= -\frac{1}{\omega_n} \frac{\omega_n^2}{c_l^2 q^2} \mathcal{G}^{-1}_{\ft\ft} = -\frac{\omega_n}{c_l^2 q^2} \mathcal{G}^{-1}_{\ft\ft}, \  \text{(Coulomb gauge)}\nonumber\\
 \hat{\sigma}_{\tL\tT} &= -\frac{1}{\omega_n} \frac{-\ti \omega_n}{c_l q} \mathcal{G}^{-1}_{\ft\tT} = -\frac{-\ti}{c_l q} \mathcal{G}^{-1}_{\ft\tT},\nonumber\\
 \hat{\sigma}_{\tT\tL} &=- \frac{1}{\omega_n} \frac{\ti \omega_n}{c_l q} \mathcal{G}^{-1}_{\tT\ft} = -\frac{\ti}{c_l q} \mathcal{G}^{-1}_{\tT\ft},\nonumber\\
 \hat{\sigma}_{\tT\tT} &= -\frac{1}{\omega_n} \mathcal{G}^{-1}_{\tT\tT}.\label{eq:Coulomb gauge conductivity from EM propagator}
 \end{align}

Using Eq.~\eqref{eq:dual Kubo formula}, we can find the conductivity directly from the dual stress propagator for the uncharged crystal. The dielectric function is defined as
\begin{equation}\label{eq:dielectric function repeat}
 \hat{\varepsilon}_{ab} = \varepsilon_0 \delta_{ab} + \ti \frac{\hat{\sigma}_{ab}}{\omega} \to \varepsilon_0 \delta_{ab} -\frac{\hat{\sigma}_{ab}}{\omega_n} .
\end{equation}
This relation is found from the material Maxwell equations. Substituting Eq.~\eqref{eq:dual Kubo formula} we can derive the dielectric function directly from the dual stress propagator:
\begin{equation}\label{eq:dielectric dual Kubo formula}
 \hat{\varepsilon}_{ab} = \varepsilon_0\Big( (1 + \frac{\omega_p^2}{\omega_n^2} )\delta_{ab} - \frac{\omega_p^2}{\omega_n^2} q^2 \frac{1}{\mu} G^{ab}_{\tT \tT}\Big).  
\end{equation}

The DC conductivity is defined as taking first the low-momentum and then the low-frequency limit
\begin{equation}\label{eq:DC conductivity}
 \hat{\sigma}_\mathrm{DC} = \lim_{\omega \to 0 } \lim_{q \to 0} \hat{\sigma} (\omega,q).
\end{equation}

\bibliographystyle{model1a-num-names}
\bibliography{qlc_references}

\begin{thebibliography}{216}
\expandafter\ifx\csname natexlab\endcsname\relax\def\natexlab#1{#1}\fi
\providecommand{\bibinfo}[2]{#2}
\ifx\xfnm\relax \def\xfnm[#1]{\unskip,\space#1}\fi
%Type = Article
\bibitem[{Ando et~al.(2002)Ando, Segawa, Komiya, and Lavrov}]{AndoEtAl02}
\bibinfo{author}{Y.~Ando}, \bibinfo{author}{K.~Segawa},
  \bibinfo{author}{S.~Komiya}, \bibinfo{author}{A.~N. Lavrov},
  \bibinfo{journal}{Phys. Rev. Lett.} \bibinfo{volume}{88}
  (\bibinfo{year}{2002}) \bibinfo{pages}{137005}.
%Type = Article
\bibitem[{Vojta(2009)}]{Vojta09}
\bibinfo{author}{M.~Vojta}, \bibinfo{journal}{Adv. Phys.} \bibinfo{volume}{58}
  (\bibinfo{year}{2009}) \bibinfo{pages}{699--820}.
%Type = Article
\bibitem[{Oganesyan et~al.(2001)Oganesyan, Kivelson, and
  Fradkin}]{OganesyanKivelsonFradkin01}
\bibinfo{author}{V.~Oganesyan}, \bibinfo{author}{S.~A. Kivelson},
  \bibinfo{author}{E.~Fradkin}, \bibinfo{journal}{Phys. Rev. B}
  \bibinfo{volume}{64} (\bibinfo{year}{2001}) \bibinfo{pages}{195109}.
%Type = Article
\bibitem[{Borzi et~al.(2007)Borzi, Grigera, Farrell, Perry, Lister, Lee,
  Tennant, Maeno, and Mackenzie}]{BorziEtAl07}
\bibinfo{author}{R.~A. Borzi}, \bibinfo{author}{S.~A. Grigera},
  \bibinfo{author}{J.~Farrell}, \bibinfo{author}{R.~S. Perry},
  \bibinfo{author}{S.~J.~S. Lister}, \bibinfo{author}{S.~L. Lee},
  \bibinfo{author}{D.~A. Tennant}, \bibinfo{author}{Y.~Maeno},
  \bibinfo{author}{A.~P. Mackenzie}, \bibinfo{journal}{Science}
  \bibinfo{volume}{315} (\bibinfo{year}{2007}) \bibinfo{pages}{214--217}.
%Type = Article
\bibitem[{Fradkin et~al.(2010)Fradkin, Kivelson, Lawler, Eisenstein, and
  Mackenzie}]{FradkinEtAl10}
\bibinfo{author}{E.~Fradkin}, \bibinfo{author}{S.~A. Kivelson},
  \bibinfo{author}{M.~J. Lawler}, \bibinfo{author}{J.~P. Eisenstein},
  \bibinfo{author}{A.~P. Mackenzie}, \bibinfo{journal}{Ann. Rev. Cond. Mat.
  Phys.} \bibinfo{volume}{1} (\bibinfo{year}{2010}) \bibinfo{pages}{153--178}.
%Type = Incollection
\bibitem[{Fradkin(2012)}]{Fradkin12}
\bibinfo{author}{E.~Fradkin}, in: \bibinfo{booktitle}{Modern Theories of
  Many-Particle Systems in Condensed Matter Physics}, volume
  \bibinfo{volume}{843} of \textit{\bibinfo{series}{Lecture Notes in Physics}},
  \bibinfo{publisher}{Springer Berlin Heidelberg}, \bibinfo{year}{2012}, pp.
  \bibinfo{pages}{53--116}.
%Type = Article
\bibitem[{Chu et~al.(2012)Chu, Kuo, Analytis, and Fisher}]{ChuEtAl12}
\bibinfo{author}{J.-H. Chu}, \bibinfo{author}{H.-H. Kuo},
  \bibinfo{author}{J.~G. Analytis}, \bibinfo{author}{I.~R. Fisher},
  \bibinfo{journal}{Science} \bibinfo{volume}{337} (\bibinfo{year}{2012})
  \bibinfo{pages}{710--712}.
%Type = Article
\bibitem[{Fernandes et~al.(2014)Fernandes, Chubukov, and
  Schmalian}]{FernandesChubukovSchmalian14}
\bibinfo{author}{R.~M. Fernandes}, \bibinfo{author}{A.~V. Chubukov},
  \bibinfo{author}{J.~Schmalian}, \bibinfo{journal}{Nat. Phys.}
  \bibinfo{volume}{10} (\bibinfo{year}{2014}) \bibinfo{pages}{97--104}.
%Type = Article
\bibitem[{Kivelson et~al.(1998)Kivelson, Fradkin, and
  Emery}]{KivelsonFradkinEmery98}
\bibinfo{author}{S.~Kivelson}, \bibinfo{author}{E.~Fradkin},
  \bibinfo{author}{V.~Emery}, \bibinfo{journal}{Nature} \bibinfo{volume}{393}
  (\bibinfo{year}{1998}) \bibinfo{pages}{550--553}.
%Type = Book
\bibitem[{Kleinert(1989)}]{Kleinert89b}
\bibinfo{author}{H.~Kleinert}, \bibinfo{title}{Gauge Fields in Condensed
  Matter, Vol.II Stress and Defects}, \bibinfo{publisher}{World Scientific},
  \bibinfo{address}{Singapore}, \bibinfo{year}{1989}.
%Type = Book
\bibitem[{Kleinert(2008)}]{Kleinert08}
\bibinfo{author}{H.~Kleinert}, \bibinfo{title}{Mulivalued Fields in Condensed
  Matter, Electromagnetism, and Gravitation}, \bibinfo{publisher}{World
  Scientific}, \bibinfo{address}{Singapore}, \bibinfo{year}{2008}.
%Type = Article
\bibitem[{Zaanen et~al.(2004)Zaanen, Nussinov, and
  Mukhin}]{ZaanenNussinovMukhin04}
\bibinfo{author}{J.~Zaanen}, \bibinfo{author}{Z.~Nussinov},
  \bibinfo{author}{S.~Mukhin}, \bibinfo{journal}{Ann. Phys.}
  \bibinfo{volume}{310} (\bibinfo{year}{2004}) \bibinfo{pages}{181--260}.
%Type = Phdthesis
\bibitem[{Cvetkovic(2006)}]{Cvetkovic06}
\bibinfo{author}{V.~Cvetkovic}, \bibinfo{title}{Quantum Liquid Crystals}, Ph.D.
  thesis, Leiden University, \bibinfo{year}{2006}.
%Type = Article
\bibitem[{Kleinert and Zaanen(2004)}]{KleinertZaanen04}
\bibinfo{author}{H.~Kleinert}, \bibinfo{author}{J.~Zaanen},
  \bibinfo{journal}{Phys. Lett. A} \bibinfo{volume}{324} (\bibinfo{year}{2004})
  \bibinfo{pages}{361--365}.
%Type = Article
\bibitem[{Cvetkovic et~al.(2006)Cvetkovic, Nussinov, and
  Zaanen}]{CvetkovicNussinovZaanen06}
\bibinfo{author}{V.~Cvetkovic}, \bibinfo{author}{Z.~Nussinov},
  \bibinfo{author}{J.~Zaanen}, \bibinfo{journal}{Phil. Mag.}
  \bibinfo{volume}{86} (\bibinfo{year}{2006}) \bibinfo{pages}{2995--3020}.
%Type = Article
\bibitem[{Cvetkovic and Zaanen(2006{\natexlab{a}})}]{CvetkovicZaanen06a}
\bibinfo{author}{V.~Cvetkovic}, \bibinfo{author}{J.~Zaanen},
  \bibinfo{journal}{Phys. Rev. B} \bibinfo{volume}{74}
  (\bibinfo{year}{2006}{\natexlab{a}}) \bibinfo{pages}{134504}.
%Type = Article
\bibitem[{Cvetkovic and Zaanen(2006{\natexlab{b}})}]{CvetkovicZaanen06b}
\bibinfo{author}{V.~Cvetkovic}, \bibinfo{author}{J.~Zaanen},
  \bibinfo{journal}{Phys. Rev. Lett.} \bibinfo{volume}{97}
  (\bibinfo{year}{2006}{\natexlab{b}}) \bibinfo{pages}{045701}.
%Type = Article
\bibitem[{Cvetkovic et~al.(2008)Cvetkovic, Nussinov, Mukhin, and
  Zaanen}]{CvetkovicNussinovMukhinZaanen08}
\bibinfo{author}{V.~Cvetkovic}, \bibinfo{author}{Z.~Nussinov},
  \bibinfo{author}{S.~Mukhin}, \bibinfo{author}{J.~Zaanen},
  \bibinfo{journal}{Euro. Phys. Lett.} \bibinfo{volume}{81}
  (\bibinfo{year}{2008}) \bibinfo{pages}{27001}.
%Type = Article
\bibitem[{Zaanen and Beekman(2012)}]{ZaanenBeekman12}
\bibinfo{author}{J.~Zaanen}, \bibinfo{author}{A.~Beekman},
  \bibinfo{journal}{Ann. Phys. (N.Y.)} \bibinfo{volume}{327}
  (\bibinfo{year}{2012}) \bibinfo{pages}{1146--1161}.
%Type = Article
\bibitem[{Beekman et~al.(2013)Beekman, Wu, Cvetkovic, and
  Zaanen}]{BeekmanWuCvetkovicZaanen13}
\bibinfo{author}{A.~J. Beekman}, \bibinfo{author}{K.~Wu},
  \bibinfo{author}{V.~Cvetkovic}, \bibinfo{author}{J.~Zaanen},
  \bibinfo{journal}{Phys. Rev. B} \bibinfo{volume}{88} (\bibinfo{year}{2013})
  \bibinfo{pages}{024121}.
%Type = Article
\bibitem[{Liu et~al.(2015)Liu, Nissinen, Nussinov, Slager, Wu, and
  Zaanen}]{LiuEtAl15}
\bibinfo{author}{K.~Liu}, \bibinfo{author}{J.~Nissinen},
  \bibinfo{author}{Z.~Nussinov}, \bibinfo{author}{R.-J. Slager},
  \bibinfo{author}{K.~Wu}, \bibinfo{author}{J.~Zaanen}, \bibinfo{journal}{Phys.
  Rev. B} \bibinfo{volume}{91} (\bibinfo{year}{2015}) \bibinfo{pages}{075103}.
%Type = Article
\bibitem[{Kivelson et~al.(2003)Kivelson, Bindloss, Fradkin, Oganesyan,
  Tranquada, Kapitulnik, and Howald}]{KivelsonEtAl03}
\bibinfo{author}{S.~A. Kivelson}, \bibinfo{author}{I.~P. Bindloss},
  \bibinfo{author}{E.~Fradkin}, \bibinfo{author}{V.~Oganesyan},
  \bibinfo{author}{J.~M. Tranquada}, \bibinfo{author}{A.~Kapitulnik},
  \bibinfo{author}{C.~Howald}, \bibinfo{journal}{Rev. Mod. Phys.}
  \bibinfo{volume}{75} (\bibinfo{year}{2003}) \bibinfo{pages}{1201--1241}.
%Type = Article
\bibitem[{Vig et~al.(2015)Vig, Kogar, Mishra, Venema, Rak, Husain, Gu, Fradkin,
  Norman, and Abbamonte}]{VigEtAl15}
\bibinfo{author}{S.~Vig}, \bibinfo{author}{A.~Kogar},
  \bibinfo{author}{V.~Mishra}, \bibinfo{author}{L.~Venema},
  \bibinfo{author}{M.~S. Rak}, \bibinfo{author}{A.~A. Husain},
  \bibinfo{author}{G.~D. Gu}, \bibinfo{author}{E.~Fradkin},
  \bibinfo{author}{M.~R. Norman}, \bibinfo{author}{P.~Abbamonte},
  \bibinfo{journal}{arXiv:1509.04230}  (\bibinfo{year}{2015}).
%Type = Article
\bibitem[{Tranquada et~al.(1995)Tranquada, Sternlieb, Axe, Nakamura, and
  Uchida}]{TranquadaEtAl95}
\bibinfo{author}{J.~Tranquada}, \bibinfo{author}{B.~Sternlieb},
  \bibinfo{author}{J.~Axe}, \bibinfo{author}{Y.~Nakamura},
  \bibinfo{author}{S.~Uchida}, \bibinfo{journal}{Nature} \bibinfo{volume}{375}
  (\bibinfo{year}{1995}) \bibinfo{pages}{561--563}.
%Type = Article
\bibitem[{Hanaguri et~al.(2004)Hanaguri, Lupien, Kohsaka, Lee, Azuma, Takano,
  Takagi, and Davis}]{HanaguriEtAl04}
\bibinfo{author}{T.~Hanaguri}, \bibinfo{author}{C.~Lupien},
  \bibinfo{author}{Y.~Kohsaka}, \bibinfo{author}{D.-H. Lee},
  \bibinfo{author}{M.~Azuma}, \bibinfo{author}{M.~Takano},
  \bibinfo{author}{H.~Takagi}, \bibinfo{author}{J.~Davis},
  \bibinfo{journal}{Nature} \bibinfo{volume}{430} (\bibinfo{year}{2004})
  \bibinfo{pages}{1001--1005}.
%Type = Article
\bibitem[{Comin et~al.(2015)Comin, Sutarto, He, da~Silva~Neto, Chauviere,
  Frano, Liang, Hardy, Bonn, Yoshida, Eisaki, Achkar, Hawthorn, Keimer,
  Sawatzky, and Damascelli}]{CominEtAl15}
\bibinfo{author}{R.~Comin}, \bibinfo{author}{R.~Sutarto},
  \bibinfo{author}{F.~He}, \bibinfo{author}{E.~da~Silva~Neto},
  \bibinfo{author}{L.~Chauviere}, \bibinfo{author}{A.~Frano},
  \bibinfo{author}{R.~Liang}, \bibinfo{author}{W.~Hardy},
  \bibinfo{author}{D.~Bonn}, \bibinfo{author}{Y.~Yoshida},
  \bibinfo{author}{H.~Eisaki}, \bibinfo{author}{A.~Achkar},
  \bibinfo{author}{D.~Hawthorn}, \bibinfo{author}{B.~Keimer},
  \bibinfo{author}{G.~Sawatzky}, \bibinfo{author}{A.~Damascelli},
  \bibinfo{journal}{Nature Mat.} \bibinfo{volume}{14} (\bibinfo{year}{2015})
  \bibinfo{pages}{796--800}.
%Type = Article
\bibitem[{da~Silva~Neto et~al.(2014)da~Silva~Neto, Aynajian, Frano, Comin,
  Schierle, Weschke, Gyenis, Wen, Schneeloch, Xu et~al.}]{DaSilvaNetoEtAl14}
\bibinfo{author}{E.~H. da~Silva~Neto}, \bibinfo{author}{P.~Aynajian},
  \bibinfo{author}{A.~Frano}, \bibinfo{author}{R.~Comin},
  \bibinfo{author}{E.~Schierle}, \bibinfo{author}{E.~Weschke},
  \bibinfo{author}{A.~Gyenis}, \bibinfo{author}{J.~Wen},
  \bibinfo{author}{J.~Schneeloch}, \bibinfo{author}{Z.~Xu}, et~al.,
  \bibinfo{journal}{Science} \bibinfo{volume}{343} (\bibinfo{year}{2014})
  \bibinfo{pages}{393--396}.
%Type = Article
\bibitem[{Comin and Damascelli(2016)}]{CominDamascelli16}
\bibinfo{author}{R.~Comin}, \bibinfo{author}{A.~Damascelli},
  \bibinfo{journal}{Ann. Rev. Cond. Mat. Phys.} \bibinfo{volume}{7}
  (\bibinfo{year}{2016}) \bibinfo{pages}{369--405}.
%Type = Article
\bibitem[{Keimer et~al.(2015)Keimer, Kivelson, Norman, Uchida, and
  Zaanen}]{KeimerEtAl15}
\bibinfo{author}{B.~Keimer}, \bibinfo{author}{S.~Kivelson},
  \bibinfo{author}{M.~Norman}, \bibinfo{author}{S.~Uchida},
  \bibinfo{author}{J.~Zaanen}, \bibinfo{journal}{Nature} \bibinfo{volume}{518}
  (\bibinfo{year}{2015}) \bibinfo{pages}{179--186}.
%Type = Article
\bibitem[{Zaanen and Gunnarsson(1989)}]{ZaanenGunnarsson89}
\bibinfo{author}{J.~Zaanen}, \bibinfo{author}{O.~Gunnarsson},
  \bibinfo{journal}{Phys. Rev. B} \bibinfo{volume}{40} (\bibinfo{year}{1989})
  \bibinfo{pages}{7391--7394}.
%Type = Article
\bibitem[{Mesaros et~al.(2016)Mesaros, Fujita, Edkins, Hamidian, Eisaki,
  Uchida, Davis, Lawler, and Kim}]{MesarosEtAl16}
\bibinfo{author}{A.~Mesaros}, \bibinfo{author}{K.~Fujita},
  \bibinfo{author}{S.~D. Edkins}, \bibinfo{author}{M.~H. Hamidian},
  \bibinfo{author}{H.~Eisaki}, \bibinfo{author}{S.-i. Uchida},
  \bibinfo{author}{J.~C.~S. Davis}, \bibinfo{author}{M.~J. Lawler},
  \bibinfo{author}{E.-A. Kim}, \bibinfo{journal}{PNAS} \bibinfo{volume}{113}
  (\bibinfo{year}{2016}) \bibinfo{pages}{12661--12666}.
%Type = Article
\bibitem[{Zheng et~al.(2016)Zheng, Chung, Corboz, Ehlers, Qin, Noack, Shi,
  White, Zhang, and Chan}]{ZhengEtAl16}
\bibinfo{author}{B.-X. Zheng}, \bibinfo{author}{C.-M. Chung},
  \bibinfo{author}{P.~Corboz}, \bibinfo{author}{G.~Ehlers},
  \bibinfo{author}{M.-P. Qin}, \bibinfo{author}{R.~M. Noack},
  \bibinfo{author}{H.~Shi}, \bibinfo{author}{S.~R. White},
  \bibinfo{author}{S.~Zhang}, \bibinfo{author}{G.~K. Chan},
  \bibinfo{journal}{arXiv:1701.00054}  (\bibinfo{year}{2016}).
%Type = Article
\bibitem[{Zaanen et~al.(2001)Zaanen, Osman, Kruis, Nussinov, and
  Tworzydlo}]{ZaanenEtAl01}
\bibinfo{author}{J.~Zaanen}, \bibinfo{author}{O.~Osman},
  \bibinfo{author}{H.~Kruis}, \bibinfo{author}{Z.~Nussinov},
  \bibinfo{author}{J.~Tworzydlo}, \bibinfo{journal}{Phil. Mag. B}
  \bibinfo{volume}{81} (\bibinfo{year}{2001}) \bibinfo{pages}{1485--1531}.
%Type = Article
\bibitem[{Zaanen et~al.(1996)Zaanen, Horbach, and van
  Saarloos}]{ZaanenHorbachSaarloos96}
\bibinfo{author}{J.~Zaanen}, \bibinfo{author}{M.~L. Horbach},
  \bibinfo{author}{W.~van Saarloos}, \bibinfo{journal}{Phys. Rev. B}
  \bibinfo{volume}{53} (\bibinfo{year}{1996}) \bibinfo{pages}{8671--8680}.
%Type = Article
\bibitem[{Tranquada et~al.(2004)Tranquada, Woo, Perring, Goka, Gu, Xu, Fujita,
  and Yamada}]{TranquadaEtAl04}
\bibinfo{author}{J.~Tranquada}, \bibinfo{author}{H.~Woo},
  \bibinfo{author}{T.~Perring}, \bibinfo{author}{H.~Goka},
  \bibinfo{author}{G.~Gu}, \bibinfo{author}{G.~Xu},
  \bibinfo{author}{M.~Fujita}, \bibinfo{author}{K.~Yamada},
  \bibinfo{journal}{Nature} \bibinfo{volume}{429} (\bibinfo{year}{2004})
  \bibinfo{pages}{534--538}.
%Type = Article
\bibitem[{Huang et~al.(2016)Huang, Mendl, Liu, Johnston, Jiang, Moritz, and
  Devereaux}]{HuangEtAl16}
\bibinfo{author}{E.~W. Huang}, \bibinfo{author}{C.~B. Mendl},
  \bibinfo{author}{S.~Liu}, \bibinfo{author}{S.~Johnston},
  \bibinfo{author}{H.-C. Jiang}, \bibinfo{author}{B.~Moritz},
  \bibinfo{author}{T.~P. Devereaux}, \bibinfo{journal}{arXiv:1612.05211}
  (\bibinfo{year}{2016}).
%Type = Book
\bibitem[{de~Gennes and Prost(1995)}]{DeGennesProst95}
\bibinfo{author}{P.~de~Gennes}, \bibinfo{author}{J.~Prost}, \bibinfo{title}{The
  Physics of Liquid Crystals}, International Series of Monographs on Physics,
  \bibinfo{publisher}{Clarendon Press}, \bibinfo{year}{1995}.
%Type = Article
\bibitem[{Hinkov et~al.(2008)Hinkov, Haug, Fauqu\'e, Bourges, Sidis, Ivanov,
  Bernhard, Lin, and Keimer}]{HinkovEtAl08}
\bibinfo{author}{V.~Hinkov}, \bibinfo{author}{D.~Haug},
  \bibinfo{author}{B.~Fauqu\'e}, \bibinfo{author}{P.~Bourges},
  \bibinfo{author}{Y.~Sidis}, \bibinfo{author}{A.~Ivanov},
  \bibinfo{author}{C.~Bernhard}, \bibinfo{author}{C.~T. Lin},
  \bibinfo{author}{B.~Keimer}, \bibinfo{journal}{Science} \bibinfo{volume}{319}
  (\bibinfo{year}{2008}) \bibinfo{pages}{597--600}.
%Type = Article
\bibitem[{Daou et~al.(2010)Daou, Chang, LeBoeuf, Cyr-Choiniere, Laliberte,
  Doiron-Leyraud, Ramshaw, Liang, Bonn, Hardy, and Taillefer}]{DaouEtAl10}
\bibinfo{author}{R.~Daou}, \bibinfo{author}{J.~Chang},
  \bibinfo{author}{D.~LeBoeuf}, \bibinfo{author}{O.~Cyr-Choiniere},
  \bibinfo{author}{F.~Laliberte}, \bibinfo{author}{N.~Doiron-Leyraud},
  \bibinfo{author}{B.~J. Ramshaw}, \bibinfo{author}{R.~Liang},
  \bibinfo{author}{D.~A. Bonn}, \bibinfo{author}{W.~N. Hardy},
  \bibinfo{author}{L.~Taillefer}, \bibinfo{journal}{Nature}
  \bibinfo{volume}{463} (\bibinfo{year}{2010}) \bibinfo{pages}{519--522}.
%Type = Article
\bibitem[{Howald et~al.(2003)Howald, Eisaki, Kaneko, and
  Kapitulnik}]{HowaldEtAl03}
\bibinfo{author}{C.~Howald}, \bibinfo{author}{H.~Eisaki},
  \bibinfo{author}{N.~Kaneko}, \bibinfo{author}{A.~Kapitulnik},
  \bibinfo{journal}{PNAS} \bibinfo{volume}{100} (\bibinfo{year}{2003})
  \bibinfo{pages}{9705--9709}.
%Type = Article
\bibitem[{Kohsaka et~al.(2007)Kohsaka, Taylor, Fujita, Schmidt, Lupien,
  Hanaguri, Azuma, Takano, Eisaki, Takagi, Uchida, and Davis}]{KohsakaEtAl07}
\bibinfo{author}{Y.~Kohsaka}, \bibinfo{author}{C.~Taylor},
  \bibinfo{author}{K.~Fujita}, \bibinfo{author}{A.~Schmidt},
  \bibinfo{author}{C.~Lupien}, \bibinfo{author}{T.~Hanaguri},
  \bibinfo{author}{M.~Azuma}, \bibinfo{author}{M.~Takano},
  \bibinfo{author}{H.~Eisaki}, \bibinfo{author}{H.~Takagi},
  \bibinfo{author}{S.~Uchida}, \bibinfo{author}{J.~C. Davis},
  \bibinfo{journal}{Science} \bibinfo{volume}{315} (\bibinfo{year}{2007})
  \bibinfo{pages}{1380--1385}.
%Type = Article
\bibitem[{Kosterlitz and Thouless(1972)}]{KosterlitzThouless72}
\bibinfo{author}{J.~M. Kosterlitz}, \bibinfo{author}{D.~J. Thouless},
  \bibinfo{journal}{J. Phys. C} \bibinfo{volume}{5} (\bibinfo{year}{1972})
  \bibinfo{pages}{L124}.
%Type = Article
\bibitem[{Kosterlitz and Thouless(1973)}]{KosterlitzThouless73}
\bibinfo{author}{J.~M. Kosterlitz}, \bibinfo{author}{D.~J. Thouless},
  \bibinfo{journal}{J. Phys. C} \bibinfo{volume}{6} (\bibinfo{year}{1973})
  \bibinfo{pages}{1181}.
%Type = Article
\bibitem[{Halperin and Nelson(1978)}]{HalperinNelson78}
\bibinfo{author}{B.~Halperin}, \bibinfo{author}{D.~Nelson},
  \bibinfo{journal}{Phys. Rev. Lett.} \bibinfo{volume}{41}
  (\bibinfo{year}{1978}) \bibinfo{pages}{121--124}.
%Type = Article
\bibitem[{Nelson and Halperin(1979)}]{NelsonHalperin79}
\bibinfo{author}{D.~Nelson}, \bibinfo{author}{B.~Halperin},
  \bibinfo{journal}{Phys. Rev. B} \bibinfo{volume}{19} (\bibinfo{year}{1979})
  \bibinfo{pages}{2457--2484}.
%Type = Article
\bibitem[{Young(1979)}]{Young79}
\bibinfo{author}{A.~Young}, \bibinfo{journal}{Phys. Rev. B}
  \bibinfo{volume}{19} (\bibinfo{year}{1979}) \bibinfo{pages}{1855--1866}.
%Type = Article
\bibitem[{Chuang et~al.(2010)Chuang, Allan, Lee, Xie, Ni, Bud'ko, Boebinger,
  Canfield, and Davis}]{ChuangEtAl10}
\bibinfo{author}{T.-M. Chuang}, \bibinfo{author}{M.~P. Allan},
  \bibinfo{author}{J.~Lee}, \bibinfo{author}{Y.~Xie}, \bibinfo{author}{N.~Ni},
  \bibinfo{author}{S.~L. Bud'ko}, \bibinfo{author}{G.~S. Boebinger},
  \bibinfo{author}{P.~C. Canfield}, \bibinfo{author}{J.~C. Davis},
  \bibinfo{journal}{Science} \bibinfo{volume}{327} (\bibinfo{year}{2010})
  \bibinfo{pages}{181--184}.
%Type = Article
\bibitem[{Mesaros et~al.(2011)Mesaros, Fujita, Eisaki, Uchida, Davis, Sachdev,
  Zaanen, Lawler, and Kim}]{MesarosEtAl11}
\bibinfo{author}{A.~Mesaros}, \bibinfo{author}{K.~Fujita},
  \bibinfo{author}{H.~Eisaki}, \bibinfo{author}{S.~Uchida},
  \bibinfo{author}{J.~C. Davis}, \bibinfo{author}{S.~Sachdev},
  \bibinfo{author}{J.~Zaanen}, \bibinfo{author}{M.~J. Lawler},
  \bibinfo{author}{E.-A. Kim}, \bibinfo{journal}{Science} \bibinfo{volume}{333}
  (\bibinfo{year}{2011}) \bibinfo{pages}{426--430}.
%Type = Article
\bibitem[{Kramers and Wannier(1941)}]{KramersWannier41}
\bibinfo{author}{H.~A. Kramers}, \bibinfo{author}{G.~H. Wannier},
  \bibinfo{journal}{Phys. Rev.} \bibinfo{volume}{60} (\bibinfo{year}{1941})
  \bibinfo{pages}{252--262}.
%Type = Article
\bibitem[{Berezinskii(1970)}]{Berezinskii70}
\bibinfo{author}{V.~L. Berezinskii}, \bibinfo{journal}{Sov. J. Exp. Theor.
  Phys.} \bibinfo{volume}{32} (\bibinfo{year}{1970}) \bibinfo{pages}{493}.
%Type = Book
\bibitem[{Zaanen et~al.(2015)Zaanen, Liu, Sun, and Schalm}]{ZaanenEtAl15}
\bibinfo{author}{J.~Zaanen}, \bibinfo{author}{Y.~Liu}, \bibinfo{author}{Y.-W.
  Sun}, \bibinfo{author}{K.~Schalm}, \bibinfo{title}{Holographic Duality in
  Condensed Matter Physics}, \bibinfo{publisher}{Cambridge University Press},
  \bibinfo{year}{2015}.
%Type = Article
\bibitem[{Burgers(1939)}]{Burgers39}
\bibinfo{author}{J.~Burgers}, \bibinfo{journal}{Proc. Kon. Ned. Akad.
  Wetensch.} \bibinfo{volume}{42} (\bibinfo{year}{1939}) \bibinfo{pages}{293}.
%Type = Article
\bibitem[{Frank(1958)}]{Frank58}
\bibinfo{author}{F.~Frank}, \bibinfo{journal}{Disc. Farad. Soc.}
  \bibinfo{volume}{25} (\bibinfo{year}{1958}) \bibinfo{pages}{19}.
%Type = Article
\bibitem[{Park and Lubensky(1996)}]{ParkLubensky96}
\bibinfo{author}{J.-M. Park}, \bibinfo{author}{T.~C. Lubensky},
  \bibinfo{journal}{Phys. Rev. E} \bibinfo{volume}{53} (\bibinfo{year}{1996})
  \bibinfo{pages}{2648--2664}.
%Type = Article
\bibitem[{Ostlund and Halperin(1981)}]{OstlundHalperin81}
\bibinfo{author}{S.~Ostlund}, \bibinfo{author}{B.~I. Halperin},
  \bibinfo{journal}{Phys. Rev. B} \bibinfo{volume}{23} (\bibinfo{year}{1981})
  \bibinfo{pages}{335--358}.
%Type = Article
\bibitem[{Marchetti and Nelson(1990)}]{MarchettiNelson90}
\bibinfo{author}{M.~C. Marchetti}, \bibinfo{author}{D.~R. Nelson},
  \bibinfo{journal}{Phys. Rev. B} \bibinfo{volume}{41} (\bibinfo{year}{1990})
  \bibinfo{pages}{1910--1920}.
%Type = Book
\bibitem[{Chaikin and Lubensky(2000)}]{ChaikinLubensky00}
\bibinfo{author}{P.~Chaikin}, \bibinfo{author}{T.~Lubensky},
  \bibinfo{title}{Principles of Condensed Matter Physics},
  \bibinfo{publisher}{Cambridge University Press}, \bibinfo{year}{2000}.
%Type = Book
\bibitem[{Sachdev(2011)}]{Sachdev99}
\bibinfo{author}{S.~Sachdev}, \bibinfo{title}{Quantum phase transitions},
  \bibinfo{publisher}{Cambridge University Press}, \bibinfo{edition}{2nd}
  edition, \bibinfo{year}{2011}.
%Type = Article
\bibitem[{Kr\"uger et~al.(2009)Kr\"uger, Kumar, Zaanen, and van~den
  Brink}]{KruegerEtAl09}
\bibinfo{author}{F.~Kr\"uger}, \bibinfo{author}{S.~Kumar},
  \bibinfo{author}{J.~Zaanen}, \bibinfo{author}{J.~van~den Brink},
  \bibinfo{journal}{Phys. Rev. B} \bibinfo{volume}{79} (\bibinfo{year}{2009})
  \bibinfo{pages}{054504}.
%Type = Article
\bibitem[{Hertz(1976)}]{Hertz76}
\bibinfo{author}{J.~A. Hertz}, \bibinfo{journal}{Phys. Rev. B}
  \bibinfo{volume}{14} (\bibinfo{year}{1976}) \bibinfo{pages}{1165--1184}.
%Type = Article
\bibitem[{Millis(1993)}]{Millis93}
\bibinfo{author}{A.~J. Millis}, \bibinfo{journal}{Phys. Rev. B}
  \bibinfo{volume}{48} (\bibinfo{year}{1993}) \bibinfo{pages}{7183--7196}.
%Type = Article
\bibitem[{Metlitski and Sachdev(2010)}]{MetlitskiSachdev10}
\bibinfo{author}{M.~A. Metlitski}, \bibinfo{author}{S.~Sachdev},
  \bibinfo{journal}{Phys. Rev.B} \bibinfo{volume}{82} (\bibinfo{year}{2010})
  \bibinfo{pages}{075127}.
%Type = Article
\bibitem[{Schattner et~al.(2016)Schattner, Lederer, Kivelson, and
  Berg}]{SchattnerEtAl16}
\bibinfo{author}{Y.~Schattner}, \bibinfo{author}{S.~Lederer},
  \bibinfo{author}{S.~A. Kivelson}, \bibinfo{author}{E.~Berg},
  \bibinfo{journal}{Phys. Rev. X} \bibinfo{volume}{6} (\bibinfo{year}{2016})
  \bibinfo{pages}{031028}.
%Type = Article
\bibitem[{Lederer et~al.(2016)Lederer, Schattner, Berg, and
  Kivelson}]{LedererEtAl16}
\bibinfo{author}{S.~Lederer}, \bibinfo{author}{Y.~Schattner},
  \bibinfo{author}{E.~Berg}, \bibinfo{author}{S.~A. Kivelson},
  \bibinfo{journal}{arXiv:1612.01542}  (\bibinfo{year}{2016}).
%Type = Article
\bibitem[{Emery et~al.(2000)Emery, Fradkin, Kivelson, and
  Lubensky}]{EmeryFradkinKivelsonLubensky00}
\bibinfo{author}{V.~J. Emery}, \bibinfo{author}{E.~Fradkin},
  \bibinfo{author}{S.~A. Kivelson}, \bibinfo{author}{T.~C. Lubensky},
  \bibinfo{journal}{Phys. Rev. Lett.} \bibinfo{volume}{85}
  (\bibinfo{year}{2000}) \bibinfo{pages}{2160--2163}.
%Type = Article
\bibitem[{Koulakov et~al.(1996)Koulakov, Fogler, and
  Shklovskii}]{FoglerKoulakovShklovskii96}
\bibinfo{author}{A.~A. Koulakov}, \bibinfo{author}{M.~M. Fogler},
  \bibinfo{author}{B.~I. Shklovskii}, \bibinfo{journal}{Phys. Rev. Lett.}
  \bibinfo{volume}{76} (\bibinfo{year}{1996}) \bibinfo{pages}{499--502}.
%Type = Article
\bibitem[{Moessner and Chalker(1996)}]{MoessnerChalker96}
\bibinfo{author}{R.~Moessner}, \bibinfo{author}{J.~T. Chalker},
  \bibinfo{journal}{Phys. Rev. B} \bibinfo{volume}{54} (\bibinfo{year}{1996})
  \bibinfo{pages}{5006--5015}.
%Type = Article
\bibitem[{Fradkin and Kivelson(1999)}]{FradkinKivelson99}
\bibinfo{author}{E.~Fradkin}, \bibinfo{author}{S.~A. Kivelson},
  \bibinfo{journal}{Phys. Rev. B} \bibinfo{volume}{59} (\bibinfo{year}{1999})
  \bibinfo{pages}{8065--8072}.
%Type = Article
\bibitem[{Lilly et~al.(1999)Lilly, Cooper, Eisenstein, Pfeiffer, and
  West}]{LillyEtAl99}
\bibinfo{author}{M.~P. Lilly}, \bibinfo{author}{K.~B. Cooper},
  \bibinfo{author}{J.~P. Eisenstein}, \bibinfo{author}{L.~N. Pfeiffer},
  \bibinfo{author}{K.~W. West}, \bibinfo{journal}{Phys. Rev. Lett.}
  \bibinfo{volume}{82} (\bibinfo{year}{1999}) \bibinfo{pages}{394--397}.
%Type = Article
\bibitem[{Xia et~al.(2011)Xia, Eisenstein, Pfeiffer, and West}]{XiaEtAl11}
\bibinfo{author}{J.~Xia}, \bibinfo{author}{J.~Eisenstein},
  \bibinfo{author}{L.~N. Pfeiffer}, \bibinfo{author}{K.~W. West},
  \bibinfo{journal}{Nature Phys.} \bibinfo{volume}{7} (\bibinfo{year}{2011})
  \bibinfo{pages}{845--848}.
%Type = Article
\bibitem[{Balents(1996)}]{Balents96}
\bibinfo{author}{L.~Balents}, \bibinfo{journal}{Euro. Phys. Lett.}
  \bibinfo{volume}{33} (\bibinfo{year}{1996}) \bibinfo{pages}{291}.
%Type = Article
\bibitem[{Mulligan et~al.(2010)Mulligan, Nayak, and
  Kachru}]{MulliganNayakKachru10}
\bibinfo{author}{M.~Mulligan}, \bibinfo{author}{C.~Nayak},
  \bibinfo{author}{S.~Kachru}, \bibinfo{journal}{Phys. Rev. B}
  \bibinfo{volume}{82} (\bibinfo{year}{2010}) \bibinfo{pages}{085102}.
%Type = Article
\bibitem[{Mulligan et~al.(2011)Mulligan, Nayak, and
  Kachru}]{MulliganNayakKachru11}
\bibinfo{author}{M.~Mulligan}, \bibinfo{author}{C.~Nayak},
  \bibinfo{author}{S.~Kachru}, \bibinfo{journal}{Phys. Rev. B}
  \bibinfo{volume}{84} (\bibinfo{year}{2011}) \bibinfo{pages}{195124}.
%Type = Article
\bibitem[{Maciejko et~al.(2013)Maciejko, Hsu, Kivelson, Park, and
  Sondhi}]{MaciejkoEtAl13}
\bibinfo{author}{J.~Maciejko}, \bibinfo{author}{B.~Hsu}, \bibinfo{author}{S.~A.
  Kivelson}, \bibinfo{author}{Y.~Park}, \bibinfo{author}{S.~L. Sondhi},
  \bibinfo{journal}{Phys. Rev. B} \bibinfo{volume}{88} (\bibinfo{year}{2013})
  \bibinfo{pages}{125137}.
%Type = Article
\bibitem[{You et~al.(2014)You, Cho, and Fradkin}]{YouChoFradkin14}
\bibinfo{author}{Y.~You}, \bibinfo{author}{G.~Y. Cho},
  \bibinfo{author}{E.~Fradkin}, \bibinfo{journal}{Phys. Rev. X}
  \bibinfo{volume}{4} (\bibinfo{year}{2014}) \bibinfo{pages}{041050}.
%Type = Article
\bibitem[{Radzihovsky and Dorsey(2002)}]{RadzihovskyDorsey02}
\bibinfo{author}{L.~Radzihovsky}, \bibinfo{author}{A.~T. Dorsey},
  \bibinfo{journal}{Phys. Rev. Lett.} \bibinfo{volume}{88}
  (\bibinfo{year}{2002}) \bibinfo{pages}{216802}.
%Type = Article
\bibitem[{Cho et~al.(2015)Cho, Parrikar, You, Leigh, and Hughes}]{ChoEtAl15}
\bibinfo{author}{G.~Y. Cho}, \bibinfo{author}{O.~Parrikar},
  \bibinfo{author}{Y.~You}, \bibinfo{author}{R.~G. Leigh},
  \bibinfo{author}{T.~L. Hughes}, \bibinfo{journal}{Phys. Rev. B}
  \bibinfo{volume}{91} (\bibinfo{year}{2015}) \bibinfo{pages}{035122}.
%Type = Article
\bibitem[{{Z. Nussinov} and {J. Zaanen}(2002)}]{ZaanenNussinov02}
\bibinfo{author}{{Z. Nussinov}}, \bibinfo{author}{{J. Zaanen}},
  \bibinfo{journal}{J. Phys. IV France} \bibinfo{volume}{12}
  (\bibinfo{year}{2002}) \bibinfo{pages}{245--250}.
%Type = Article
\bibitem[{Zhang et~al.(2002)Zhang, Demler, and Sachdev}]{ZhangDemlerSachdev02}
\bibinfo{author}{Y.~Zhang}, \bibinfo{author}{E.~Demler},
  \bibinfo{author}{S.~Sachdev}, \bibinfo{journal}{Phys. Rev. B}
  \bibinfo{volume}{66} (\bibinfo{year}{2002}) \bibinfo{pages}{094501}.
%Type = Article
\bibitem[{Fradkin et~al.(2015)Fradkin, Kivelson, and
  Tranquada}]{FradkinKivelsonTranquada15}
\bibinfo{author}{E.~Fradkin}, \bibinfo{author}{S.~A. Kivelson},
  \bibinfo{author}{J.~M. Tranquada}, \bibinfo{journal}{Rev. Mod. Phys.}
  \bibinfo{volume}{87} (\bibinfo{year}{2015}) \bibinfo{pages}{457}.
%Type = Article
\bibitem[{Hamidian et~al.(2016)Hamidian, Edkins, Joo, Kostin, Eisaki, Uchida,
  Lawler, Kim, Mackenzie, Fujita et~al.}]{HamidianEtAl16}
\bibinfo{author}{M.~Hamidian}, \bibinfo{author}{S.~Edkins},
  \bibinfo{author}{S.~H. Joo}, \bibinfo{author}{A.~Kostin},
  \bibinfo{author}{H.~Eisaki}, \bibinfo{author}{S.~Uchida},
  \bibinfo{author}{M.~Lawler}, \bibinfo{author}{E.-A. Kim},
  \bibinfo{author}{A.~Mackenzie}, \bibinfo{author}{K.~Fujita}, et~al.,
  \bibinfo{journal}{Nature} \bibinfo{volume}{532} (\bibinfo{year}{2016})
  \bibinfo{pages}{343--347}.
%Type = Article
\bibitem[{Berg et~al.(2009)Berg, Fradkin, and Kivelson}]{BergFradkinKivelson09}
\bibinfo{author}{E.~Berg}, \bibinfo{author}{E.~Fradkin}, \bibinfo{author}{S.~A.
  Kivelson}, \bibinfo{journal}{Nature Phys.} \bibinfo{volume}{5}
  (\bibinfo{year}{2009}) \bibinfo{pages}{830--833}.
%Type = Article
\bibitem[{Mross and Senthil(2012)}]{MrossSenthil12}
\bibinfo{author}{D.~F. Mross}, \bibinfo{author}{T.~Senthil},
  \bibinfo{journal}{Phys. Rev. B} \bibinfo{volume}{86} (\bibinfo{year}{2012})
  \bibinfo{pages}{115138}.
%Type = Article
\bibitem[{Mross and Senthil(2015)}]{MrossSenthil15}
\bibinfo{author}{D.~F. Mross}, \bibinfo{author}{T.~Senthil},
  \bibinfo{journal}{Phys. Rev. X} \bibinfo{volume}{5} (\bibinfo{year}{2015})
  \bibinfo{pages}{031008}.
%Type = Article
\bibitem[{Fulde and Ferrell(1964)}]{FuldeFerrel64}
\bibinfo{author}{P.~Fulde}, \bibinfo{author}{R.~A. Ferrell},
  \bibinfo{journal}{Phys. Rev.} \bibinfo{volume}{135} (\bibinfo{year}{1964})
  \bibinfo{pages}{A550}.
%Type = Article
\bibitem[{Larkin and Ovchinnikov(1965)}]{LarkinOvchinnokov65}
\bibinfo{author}{A.~Larkin}, \bibinfo{author}{I.~Ovchinnikov},
  \bibinfo{journal}{Sov. Phys. JETP} \bibinfo{volume}{20}
  (\bibinfo{year}{1965}) \bibinfo{pages}{762--769}.
%Type = Article
\bibitem[{Radzihovsky and Vishwanath(2009)}]{RadzihovskyViswanath09}
\bibinfo{author}{L.~Radzihovsky}, \bibinfo{author}{A.~Vishwanath},
  \bibinfo{journal}{Phys. Rev. Lett.} \bibinfo{volume}{103}
  (\bibinfo{year}{2009}) \bibinfo{pages}{010404}.
%Type = Article
\bibitem[{Fisher et~al.(1989)Fisher, Weichman, Grinstein, and
  Fisher}]{FisherEtAl89}
\bibinfo{author}{M.~Fisher}, \bibinfo{author}{P.~Weichman},
  \bibinfo{author}{G.~Grinstein}, \bibinfo{author}{D.~Fisher},
  \bibinfo{journal}{Phys. Rev. B} \bibinfo{volume}{40} (\bibinfo{year}{1989})
  \bibinfo{pages}{546--570}.
%Type = Article
\bibitem[{Fisher and Lee(1989)}]{FisherLee89}
\bibinfo{author}{M.~Fisher}, \bibinfo{author}{D.~Lee}, \bibinfo{journal}{Phys.
  Rev. B} \bibinfo{volume}{39} (\bibinfo{year}{1989})
  \bibinfo{pages}{2756--2759}.
%Type = Article
\bibitem[{Lee and Fisher(1991)}]{LeeFisher91}
\bibinfo{author}{D.~Lee}, \bibinfo{author}{M.~Fisher}, \bibinfo{journal}{Int.
  J. Mod. Phys. B} \bibinfo{volume}{5} (\bibinfo{year}{1991})
  \bibinfo{pages}{2675}.
%Type = Book
\bibitem[{Kleinert(1989)}]{Kleinert89a}
\bibinfo{author}{H.~Kleinert}, \bibinfo{title}{Gauge Fields in Condensed
  Matter, Vol.I Superflow and Vortex Lines}, \bibinfo{publisher}{World
  Scientific}, \bibinfo{address}{Singapore}, \bibinfo{year}{1989}.
%Type = Article
\bibitem[{Nguyen and Sudb\o{}(1999)}]{NguyenSudbo99}
\bibinfo{author}{A.~K. Nguyen}, \bibinfo{author}{A.~Sudb\o{}},
  \bibinfo{journal}{Phys. Rev. B} \bibinfo{volume}{60} (\bibinfo{year}{1999})
  \bibinfo{pages}{15307--15331}.
%Type = Article
\bibitem[{Hove and Sudb\o{}(2000)}]{HoveSudbo00}
\bibinfo{author}{J.~Hove}, \bibinfo{author}{A.~Sudb\o{}},
  \bibinfo{journal}{Phys. Rev. Lett.} \bibinfo{volume}{84}
  (\bibinfo{year}{2000}) \bibinfo{pages}{3426--3429}.
%Type = Article
\bibitem[{Herbut and Te\v{s}anovi\'{c}(1996)}]{HerbutTessanovic96}
\bibinfo{author}{I.~Herbut}, \bibinfo{author}{Z.~Te\v{s}anovi\'{c}},
  \bibinfo{journal}{Phys. Rev. Lett.} \bibinfo{volume}{76}
  (\bibinfo{year}{1996}) \bibinfo{pages}{4588--4591}.
%Type = Article
\bibitem[{Beekman et~al.(2011)Beekman, Sadri, and
  Zaanen}]{BeekmanSadriZaanen11}
\bibinfo{author}{A.~Beekman}, \bibinfo{author}{D.~Sadri},
  \bibinfo{author}{J.~Zaanen}, \bibinfo{journal}{New J. Phys.}
  \bibinfo{volume}{13} (\bibinfo{year}{2011}) \bibinfo{pages}{033004}.
%Type = Book
\bibitem[{Zee(2010)}]{zee_book}
\bibinfo{author}{A.~Zee}, \bibinfo{title}{Quantum Field Theory in a Nutshell},
  \bibinfo{publisher}{Princeton University Press},
  \bibinfo{address}{Princeton}, \bibinfo{year}{2010}.
%Type = Book
\bibitem[{Landau and Lifshitz(1986)}]{LandauLifshitz86}
\bibinfo{author}{L.~D. Landau}, \bibinfo{author}{E.~Lifshitz},
  \bibinfo{title}{Theory of Elasticity}, volume~\bibinfo{volume}{7} of
  \textit{\bibinfo{series}{Course of Theoretical Physics}},
  \bibinfo{publisher}{Elsevier New York}, \bibinfo{year}{1986}.
%Type = Article
\bibitem[{Low and Manohar(2002)}]{LowManohar02}
\bibinfo{author}{I.~Low}, \bibinfo{author}{A.~V. Manohar},
  \bibinfo{journal}{Phys. Rev. Lett.} \bibinfo{volume}{88}
  (\bibinfo{year}{2002}) \bibinfo{pages}{101602}.
%Type = Article
\bibitem[{Watanabe and Murayama(2013)}]{WatanabeMurayama13}
\bibinfo{author}{H.~Watanabe}, \bibinfo{author}{H.~Murayama},
  \bibinfo{journal}{Phys. Rev. Lett.} \bibinfo{volume}{110}
  (\bibinfo{year}{2013}) \bibinfo{pages}{181601}.
%Type = Article
\bibitem[{Yang et~al.(2004)Yang, Z.-M., Li, Shi, Xie, and Yang}]{YangEtAl04}
\bibinfo{author}{W.~Yang}, \bibinfo{author}{Z.-M.}, \bibinfo{author}{Li},
  \bibinfo{author}{W.~Shi}, \bibinfo{author}{B.-H. Xie}, \bibinfo{author}{M.-B.
  Yang}, \bibinfo{journal}{J. Mater. Sci.} \bibinfo{volume}{39}
  (\bibinfo{year}{2004}) \bibinfo{pages}{3269}.
%Type = Article
\bibitem[{Kleiner et~al.(1964)Kleiner, Roth, and Autler}]{KleinerRothAutler64}
\bibinfo{author}{W.~H. Kleiner}, \bibinfo{author}{L.~M. Roth},
  \bibinfo{author}{S.~H. Autler}, \bibinfo{journal}{Phys. Rev.}
  \bibinfo{volume}{133} (\bibinfo{year}{1964}) \bibinfo{pages}{A1226--A1227}.
%Type = Article
\bibitem[{Kittinger et~al.(1981)Kittinger, Tich\'y, and
  Bertagnolli}]{KittingerTichyBertagnolli81}
\bibinfo{author}{E.~Kittinger}, \bibinfo{author}{J.~Tich\'y},
  \bibinfo{author}{E.~Bertagnolli}, \bibinfo{journal}{Phys. Rev. Lett.}
  \bibinfo{volume}{47} (\bibinfo{year}{1981}) \bibinfo{pages}{712--714}.
%Type = Article
\bibitem[{Lakes(1987)}]{Lakes87}
\bibinfo{author}{R.~Lakes}, \bibinfo{journal}{Science} \bibinfo{volume}{235}
  (\bibinfo{year}{1987}) \bibinfo{pages}{1038--1040}.
%Type = Article
\bibitem[{Ahsan et~al.(2007)Ahsan, Rudnick, and
  Bruinsma}]{AhsanRudnickBruinsma07}
\bibinfo{author}{A.~Ahsan}, \bibinfo{author}{J.~Rudnick},
  \bibinfo{author}{R.~Bruinsma}, \bibinfo{journal}{Phys. Rev. E}
  \bibinfo{volume}{76} (\bibinfo{year}{2007}) \bibinfo{pages}{061910}.
%Type = Article
\bibitem[{Mermin(1979)}]{Mermin79}
\bibinfo{author}{N.~Mermin}, \bibinfo{journal}{Rev. Mod. Phys.}
  \bibinfo{volume}{51} (\bibinfo{year}{1979}) \bibinfo{pages}{591--648}.
%Type = Book
\bibitem[{Friedel(1964)}]{Friedel64}
\bibinfo{author}{J.~Friedel}, \bibinfo{title}{Dislocations},
  \bibinfo{publisher}{Pergamon}, \bibinfo{year}{1964}.
%Type = Article
\bibitem[{Kleman and Friedel(2008)}]{KlemanFriedel08}
\bibinfo{author}{M.~Kleman}, \bibinfo{author}{J.~Friedel},
  \bibinfo{journal}{Rev. Mod. Phys.} \bibinfo{volume}{80}
  (\bibinfo{year}{2008}) \bibinfo{pages}{61--115}.
%Type = Article
\bibitem[{Kim and Chan(2004)}]{KimChan04}
\bibinfo{author}{E.~Kim}, \bibinfo{author}{M.~Chan}, \bibinfo{journal}{Nature}
  \bibinfo{volume}{427} (\bibinfo{year}{2004}) \bibinfo{pages}{225--227}.
%Type = Article
\bibitem[{Nussinov et~al.(2007)Nussinov, Balatsky, Graf, and
  Trugman}]{NussinovEtAl07}
\bibinfo{author}{Z.~Nussinov}, \bibinfo{author}{A.~V. Balatsky},
  \bibinfo{author}{M.~J. Graf}, \bibinfo{author}{S.~A. Trugman},
  \bibinfo{journal}{Phys. Rev. B} \bibinfo{volume}{76} (\bibinfo{year}{2007})
  \bibinfo{pages}{014530}.
%Type = Article
\bibitem[{Balatsky et~al.(2007)Balatsky, Graf, Nussinov, and
  Trugman}]{BalatskyEtAl07}
\bibinfo{author}{A.~V. Balatsky}, \bibinfo{author}{M.~J. Graf},
  \bibinfo{author}{Z.~Nussinov}, \bibinfo{author}{S.~A. Trugman},
  \bibinfo{journal}{Phys. Rev. B} \bibinfo{volume}{75} (\bibinfo{year}{2007})
  \bibinfo{pages}{094201}.
%Type = Article
\bibitem[{Prokof'ev(2007)}]{Prokofev07}
\bibinfo{author}{N.~Prokof'ev}, \bibinfo{journal}{Adv. Phys.}
  \bibinfo{volume}{56} (\bibinfo{year}{2007}) \bibinfo{pages}{381--402}.
%Type = Article
\bibitem[{Boninsegni and Prokof'ev(2012)}]{BoninsegniProkofev12}
\bibinfo{author}{M.~Boninsegni}, \bibinfo{author}{N.~V. Prokof'ev},
  \bibinfo{journal}{Rev. Mod. Phys.} \bibinfo{volume}{84}
  (\bibinfo{year}{2012}) \bibinfo{pages}{759--776}.
%Type = Article
\bibitem[{Bais and Mathy(2006)}]{BaisMathy06}
\bibinfo{author}{F.~A. Bais}, \bibinfo{author}{C.~J.~M. Mathy},
  \bibinfo{journal}{Phys. Rev. B} \bibinfo{volume}{73} (\bibinfo{year}{2006})
  \bibinfo{pages}{224120}.
%Type = Article
\bibitem[{Mathy and Bais(2007)}]{MathyBais07}
\bibinfo{author}{C.~Mathy}, \bibinfo{author}{F.~Bais}, \bibinfo{journal}{Ann.
  Phys. (N.Y.)} \bibinfo{volume}{322} (\bibinfo{year}{2007})
  \bibinfo{pages}{709--735}.
%Type = Article
\bibitem[{Kleinert(1983)}]{Kleinert83}
\bibinfo{author}{H.~Kleinert}, \bibinfo{journal}{Phys. Lett. A}
  \bibinfo{volume}{95} (\bibinfo{year}{1983}) \bibinfo{pages}{381 -- 384}.
%Type = Article
\bibitem[{Lammert et~al.(1993)Lammert, Rokhsar, and
  Toner}]{LammertRoksharToner93}
\bibinfo{author}{P.~E. Lammert}, \bibinfo{author}{D.~S. Rokhsar},
  \bibinfo{author}{J.~Toner}, \bibinfo{journal}{Phys. Rev. Lett.}
  \bibinfo{volume}{70} (\bibinfo{year}{1993}) \bibinfo{pages}{1650--1653}.
%Type = Article
\bibitem[{Lammert et~al.(1995)Lammert, Rokhsar, and
  Toner}]{LammertRoksharToner95}
\bibinfo{author}{P.~E. Lammert}, \bibinfo{author}{D.~S. Rokhsar},
  \bibinfo{author}{J.~Toner}, \bibinfo{journal}{Phys. Rev. E}
  \bibinfo{volume}{52} (\bibinfo{year}{1995}) \bibinfo{pages}{1778--1800}.
%Type = Inbook
\bibitem[{Nelson(1983)}]{Nelson83}
\bibinfo{author}{D.~R. Nelson}, \bibinfo{title}{Defect-mediated phase
  transitions}, volume~\bibinfo{volume}{7} of \textit{\bibinfo{series}{Phase
  transitions and critical phenomena}}, \bibinfo{publisher}{Academic Press New
  York}, pp. \bibinfo{pages}{2--165}.
%Type = Book
\bibitem[{Ashcroft and Mermin(1976)}]{AshcroftMermin76}
\bibinfo{author}{N.~W. Ashcroft}, \bibinfo{author}{N.~D. Mermin},
  \bibinfo{title}{Solid {S}tate {P}hysics}, \bibinfo{publisher}{Saunders,
  Philadelphia}, \bibinfo{year}{1976}.
%Type = Article
\bibitem[{Rosenbaum et~al.(1983)Rosenbaum, Nagler, Horn, and
  Clarke}]{RosenbaumEtAl83}
\bibinfo{author}{T.~F. Rosenbaum}, \bibinfo{author}{S.~E. Nagler},
  \bibinfo{author}{P.~M. Horn}, \bibinfo{author}{R.~Clarke},
  \bibinfo{journal}{Phys. Rev. Lett.} \bibinfo{volume}{50}
  (\bibinfo{year}{1983}) \bibinfo{pages}{1791--1794}.
%Type = Article
\bibitem[{Keim et~al.(2007)Keim, Maret, and von Gr\"unberg}]{KeimEtAl07}
\bibinfo{author}{P.~Keim}, \bibinfo{author}{G.~Maret}, \bibinfo{author}{H.~H.
  von Gr\"unberg}, \bibinfo{journal}{Phys. Rev. E} \bibinfo{volume}{75}
  (\bibinfo{year}{2007}) \bibinfo{pages}{031402}.
%Type = Article
\bibitem[{Kapfer and Krauth(2015)}]{KapferKrauth15}
\bibinfo{author}{S.~C. Kapfer}, \bibinfo{author}{W.~Krauth},
  \bibinfo{journal}{Phys. Rev. Lett.} \bibinfo{volume}{114}
  (\bibinfo{year}{2015}) \bibinfo{pages}{035702}.
%Type = Article
\bibitem[{Bernard and Krauth(2011)}]{BernardKrauth11}
\bibinfo{author}{E.~P. Bernard}, \bibinfo{author}{W.~Krauth},
  \bibinfo{journal}{Phys. Rev. Lett.} \bibinfo{volume}{107}
  (\bibinfo{year}{2011}) \bibinfo{pages}{155704}.
%Type = Article
\bibitem[{Bruun and Nelson(2014)}]{BruunNelson14}
\bibinfo{author}{G.~M. Bruun}, \bibinfo{author}{D.~R. Nelson},
  \bibinfo{journal}{Phys. Rev. B} \bibinfo{volume}{89} (\bibinfo{year}{2014})
  \bibinfo{pages}{094112}.
%Type = Article
\bibitem[{Lechner et~al.(2014)Lechner, B\"uchler, and
  Zoller}]{LechnerBuchnerZoller14}
\bibinfo{author}{W.~Lechner}, \bibinfo{author}{H.-P. B\"uchler},
  \bibinfo{author}{P.~Zoller}, \bibinfo{journal}{Phys. Rev. Lett.}
  \bibinfo{volume}{112} (\bibinfo{year}{2014}) \bibinfo{pages}{255301}.
%Type = Article
\bibitem[{{Wu} et~al.(2016){Wu}, {Block}, and {Bruun}}]{WuBlockBruun15}
\bibinfo{author}{Z.~{Wu}}, \bibinfo{author}{J.~K. {Block}},
  \bibinfo{author}{G.~M. {Bruun}}, \bibinfo{journal}{Sci. Rep.}
  \bibinfo{volume}{6} (\bibinfo{year}{2016}) \bibinfo{pages}{19038}.
%Type = Article
\bibitem[{Jos\'e et~al.(1977)Jos\'e, Kadanoff, Kirkpatrick, and
  Nelson}]{JoseEtAl77}
\bibinfo{author}{J.~V. Jos\'e}, \bibinfo{author}{L.~P. Kadanoff},
  \bibinfo{author}{S.~Kirkpatrick}, \bibinfo{author}{D.~R. Nelson},
  \bibinfo{journal}{Phys. Rev. B} \bibinfo{volume}{16} (\bibinfo{year}{1977})
  \bibinfo{pages}{1217--1241}.
%Type = Article
\bibitem[{Savit(1978)}]{Savit78}
\bibinfo{author}{R.~Savit}, \bibinfo{journal}{Phys. Rev. B}
  \bibinfo{volume}{17} (\bibinfo{year}{1978}) \bibinfo{pages}{1340--1350}.
%Type = Article
\bibitem[{{Kajantie} et~al.(2000){Kajantie}, {Laine}, {Neuhaus}, {Rajantie},
  and {Rummukainen}}]{KajantieEtAl2000}
\bibinfo{author}{K.~{Kajantie}}, \bibinfo{author}{M.~{Laine}},
  \bibinfo{author}{T.~{Neuhaus}}, \bibinfo{author}{A.~{Rajantie}},
  \bibinfo{author}{K.~{Rummukainen}}, \bibinfo{journal}{Phys. Lett. B}
  \bibinfo{volume}{482} (\bibinfo{year}{2000}) \bibinfo{pages}{114--122}.
%Type = Book
\bibitem[{Henneaux and Teitelboim(1992)}]{HenneauxTeitelboim92}
\bibinfo{author}{M.~Henneaux}, \bibinfo{author}{C.~Teitelboim},
  \bibinfo{title}{Quantization of Gauge Systems}, \bibinfo{publisher}{Princeton
  University Press}, \bibinfo{address}{Princeton, NJ}, \bibinfo{year}{1992}.
%Type = Article
\bibitem[{Kogut and Susskind(1975)}]{KogutSusskind75}
\bibinfo{author}{J.~Kogut}, \bibinfo{author}{L.~Susskind},
  \bibinfo{journal}{Phys. Rev. D} \bibinfo{volume}{11} (\bibinfo{year}{1975})
  \bibinfo{pages}{395--408}.
%Type = Article
\bibitem[{Kogut(1979)}]{Kogut79}
\bibinfo{author}{J.~B. Kogut}, \bibinfo{journal}{Rev. Mod. Phys.}
  \bibinfo{volume}{51} (\bibinfo{year}{1979}) \bibinfo{pages}{659--713}.
%Type = Article
\bibitem[{Elitzur(1975)}]{Elitzur75}
\bibinfo{author}{S.~Elitzur}, \bibinfo{journal}{Phys. Rev. D}
  \bibinfo{volume}{12} (\bibinfo{year}{1975}) \bibinfo{pages}{3978--3982}.
%Type = Article
\bibitem[{Senthil and Fisher(2000)}]{SenthilFisher00}
\bibinfo{author}{T.~Senthil}, \bibinfo{author}{M.~P.~A. Fisher},
  \bibinfo{journal}{Phys. Rev. B} \bibinfo{volume}{62} (\bibinfo{year}{2000})
  \bibinfo{pages}{7850--7881}.
%Type = Article
\bibitem[{Sedgewick et~al.(2002)Sedgewick, Scalapino, and
  Sugar}]{SedgewickScalapinoSugar02}
\bibinfo{author}{R.~D. Sedgewick}, \bibinfo{author}{D.~J. Scalapino},
  \bibinfo{author}{R.~L. Sugar}, \bibinfo{journal}{Phys. Rev. B}
  \bibinfo{volume}{65} (\bibinfo{year}{2002}) \bibinfo{pages}{054508}.
%Type = Article
\bibitem[{Podolsky and Demler(2005)}]{PodolskyDemler05}
\bibinfo{author}{D.~Podolsky}, \bibinfo{author}{E.~Demler},
  \bibinfo{journal}{New J. Phys.} \bibinfo{volume}{7} (\bibinfo{year}{2005})
  \bibinfo{pages}{59}.
%Type = Article
\bibitem[{Isakov et~al.(2012)Isakov, Melko, and
  Hastings}]{IsakovMelkoHastings12}
\bibinfo{author}{S.~V. Isakov}, \bibinfo{author}{R.~G. Melko},
  \bibinfo{author}{M.~B. Hastings}, \bibinfo{journal}{Science}
  \bibinfo{volume}{335} (\bibinfo{year}{2012}) \bibinfo{pages}{193--195}.
%Type = Article
\bibitem[{Villain(1975)}]{Villain75}
\bibinfo{author}{J.~Villain}, \bibinfo{journal}{J. Phys. France}
  \bibinfo{volume}{36} (\bibinfo{year}{1975}) \bibinfo{pages}{581--590}.
%Type = Article
\bibitem[{Janke and Kleinert(1986)}]{JankeKleinert86}
\bibinfo{author}{W.~Janke}, \bibinfo{author}{H.~Kleinert},
  \bibinfo{journal}{Nucl. Phys. B} \bibinfo{volume}{270} (\bibinfo{year}{1986})
  \bibinfo{pages}{135 -- 153}.
%Type = Article
\bibitem[{Horn et~al.(1979)Horn, Weinstein, and
  Yankielowicz}]{HornWeinsteinYankielowicz79}
\bibinfo{author}{D.~Horn}, \bibinfo{author}{M.~Weinstein},
  \bibinfo{author}{S.~Yankielowicz}, \bibinfo{journal}{Phys. Rev. D}
  \bibinfo{volume}{19} (\bibinfo{year}{1979}) \bibinfo{pages}{3715--3731}.
%Type = Article
\bibitem[{Bhanot and Creutz(1980)}]{BhanotCreutz80}
\bibinfo{author}{G.~Bhanot}, \bibinfo{author}{M.~Creutz},
  \bibinfo{journal}{Phys. Rev. D} \bibinfo{volume}{21} (\bibinfo{year}{1980})
  \bibinfo{pages}{2892--2902}.
%Type = Article
\bibitem[{Borisenko et~al.(2014)Borisenko, Chelnokov, Cortese, Gravina, Papa,
  and Surzhikov}]{BorisenkoEtAl14}
\bibinfo{author}{O.~Borisenko}, \bibinfo{author}{V.~Chelnokov},
  \bibinfo{author}{G.~Cortese}, \bibinfo{author}{M.~Gravina},
  \bibinfo{author}{A.~Papa}, \bibinfo{author}{I.~Surzhikov},
  \bibinfo{journal}{Nucl. Phys. B} \bibinfo{volume}{879} (\bibinfo{year}{2014})
  \bibinfo{pages}{80 -- 97}.
%Type = Book
\bibitem[{Wen(2004)}]{Wen04}
\bibinfo{author}{X.~Wen}, \bibinfo{title}{Quantum Field Theory of Many-Body
  Systems: From the Origin of Sound to an Origin of Light and Electrons},
  Oxford Graduate Texts, \bibinfo{publisher}{OUP Oxford}, \bibinfo{year}{2004}.
%Type = Article
\bibitem[{Gregor et~al.(2011)Gregor, Huse, Moessner, and Sondhi}]{GregorEtAl11}
\bibinfo{author}{K.~Gregor}, \bibinfo{author}{D.~A. Huse},
  \bibinfo{author}{R.~Moessner}, \bibinfo{author}{S.~L. Sondhi},
  \bibinfo{journal}{New J. Phys.} \bibinfo{volume}{13} (\bibinfo{year}{2011})
  \bibinfo{pages}{025009}.
%Type = Article
\bibitem[{Fradkin and Shenker(1979)}]{FradkinShenker79}
\bibinfo{author}{E.~Fradkin}, \bibinfo{author}{S.~H. Shenker},
  \bibinfo{journal}{Phys. Rev. D} \bibinfo{volume}{19} (\bibinfo{year}{1979})
  \bibinfo{pages}{3682--3697}.
%Type = Article
\bibitem[{Fredenhagen and Marcu(1986)}]{FredenhagenMarcu86}
\bibinfo{author}{K.~Fredenhagen}, \bibinfo{author}{M.~Marcu},
  \bibinfo{journal}{Phys. Rev. Lett.} \bibinfo{volume}{56}
  (\bibinfo{year}{1986}) \bibinfo{pages}{223--224}.
%Type = Article
\bibitem[{Geraedts and Motrunich(2014)}]{GeraedtsMotrunich14}
\bibinfo{author}{S.~D. Geraedts}, \bibinfo{author}{O.~I. Motrunich},
  \bibinfo{journal}{Phys. Rev. B} \bibinfo{volume}{90} (\bibinfo{year}{2014})
  \bibinfo{pages}{214505}.
%Type = Article
\bibitem[{Zaanen and Nussinov(2003)}]{ZaanenNussinov03}
\bibinfo{author}{J.~Zaanen}, \bibinfo{author}{Z.~Nussinov},
  \bibinfo{journal}{Phys. Stat. Sol. B} \bibinfo{volume}{236}
  (\bibinfo{year}{2003}) \bibinfo{pages}{332--339}.
%Type = Article
\bibitem[{Ehrenfest(1927)}]{Ehrenfest27}
\bibinfo{author}{P.~Ehrenfest}, \bibinfo{journal}{Zeit. Phys.}
  \bibinfo{volume}{45} (\bibinfo{year}{1927}) \bibinfo{pages}{455--457}.
%Type = Article
\bibitem[{Nielsen and Martin(1985)}]{NielsenMartin85}
\bibinfo{author}{O.~H. Nielsen}, \bibinfo{author}{R.~M. Martin},
  \bibinfo{journal}{Phys. Rev. B} \bibinfo{volume}{32} (\bibinfo{year}{1985})
  \bibinfo{pages}{3780--3791}.
%Type = Book
\bibitem[{Dirac(2001)}]{Dirac01}
\bibinfo{author}{P.~Dirac}, \bibinfo{title}{Lectures on Quantum Mechanics},
  Dover Books on Physics, \bibinfo{publisher}{Dover Publications},
  \bibinfo{year}{2001}.
%Type = Article
\bibitem[{Toupin(1962)}]{Toupin62}
\bibinfo{author}{R.~A. Toupin}, \bibinfo{journal}{Arch. Rat. Mech. Anal.}
  \bibinfo{volume}{11} (\bibinfo{year}{1962}) \bibinfo{pages}{385--414}.
%Type = Article
\bibitem[{Toupin(1964)}]{Toupin64}
\bibinfo{author}{R.~A. Toupin}, \bibinfo{journal}{Arch. Rat. Mech. Anal.}
  \bibinfo{volume}{17} (\bibinfo{year}{1964}) \bibinfo{pages}{85--112}.
%Type = Article
\bibitem[{Kiometzis et~al.(1995)Kiometzis, Kleinert, and
  Schakel}]{KiometzisKleinertSchakel95}
\bibinfo{author}{M.~Kiometzis}, \bibinfo{author}{H.~Kleinert},
  \bibinfo{author}{A.~M.~J. Schakel}, \bibinfo{journal}{Fortschr. Phys.}
  \bibinfo{volume}{43} (\bibinfo{year}{1995}) \bibinfo{pages}{697--732}.
%Type = Article
\bibitem[{Hove et~al.(2000)Hove, Mo, and Sudb\o{}}]{HoveMoSudbo00}
\bibinfo{author}{J.~Hove}, \bibinfo{author}{S.~Mo},
  \bibinfo{author}{A.~Sudb\o{}}, \bibinfo{journal}{Phys. Rev. Lett.}
  \bibinfo{volume}{85} (\bibinfo{year}{2000}) \bibinfo{pages}{2368--2371}.
%Type = Article
\bibitem[{Kasamatsu et~al.(2005)Kasamatsu, Tsubota, and Ueda}]{KasamatsuEtAl05}
\bibinfo{author}{K.~Kasamatsu}, \bibinfo{author}{M.~Tsubota},
  \bibinfo{author}{M.~Ueda}, \bibinfo{journal}{Int. J. Mod. Phys. B}
  \bibinfo{volume}{19} (\bibinfo{year}{2005}) \bibinfo{pages}{1835--1904}.
%Type = Article
\bibitem[{Brauner and Watanabe(2014)}]{BraunerWatanabe14}
\bibinfo{author}{T.~Brauner}, \bibinfo{author}{H.~Watanabe},
  \bibinfo{journal}{Phys. Rev. D} \bibinfo{volume}{89} (\bibinfo{year}{2014})
  \bibinfo{pages}{085004}.
%Type = Article
\bibitem[{Hayata and Hidaka(2014)}]{HayataHidaka14}
\bibinfo{author}{T.~Hayata}, \bibinfo{author}{Y.~Hidaka},
  \bibinfo{journal}{Phys. Lett. B} \bibinfo{volume}{735} (\bibinfo{year}{2014})
  \bibinfo{pages}{195--199}.
%Type = Article
\bibitem[{Nicolis et~al.(2013)Nicolis, Penco, Piazza, and
  Rosen}]{NicolisEtAl13}
\bibinfo{author}{A.~Nicolis}, \bibinfo{author}{R.~Penco},
  \bibinfo{author}{F.~Piazza}, \bibinfo{author}{R.~A. Rosen},
  \bibinfo{journal}{JHEP} \bibinfo{volume}{2013} (\bibinfo{year}{2013})
  \bibinfo{pages}{55}.
%Type = Article
\bibitem[{Seung and Nelson(1988)}]{SeungNelson88}
\bibinfo{author}{H.~S. Seung}, \bibinfo{author}{D.~R. Nelson},
  \bibinfo{journal}{Phys. Rev. A} \bibinfo{volume}{38} (\bibinfo{year}{1988})
  \bibinfo{pages}{1005--1018}.
%Type = Article
\bibitem[{Grinstein and Pelcovits(1982)}]{GrinsteinPelcovits82}
\bibinfo{author}{G.~Grinstein}, \bibinfo{author}{R.~A. Pelcovits},
  \bibinfo{journal}{Phys. Rev. A} \bibinfo{volume}{26} (\bibinfo{year}{1982})
  \bibinfo{pages}{915--925}.
%Type = Article
\bibitem[{Golubovi\ifmmode~\acute{c}\else \'{c}\fi{} and
  Wang(1994)}]{GolubovicWang94}
\bibinfo{author}{L.~Golubovi\ifmmode~\acute{c}\else \'{c}\fi{}},
  \bibinfo{author}{Z.-G. Wang}, \bibinfo{journal}{Phys. Rev. E}
  \bibinfo{volume}{49} (\bibinfo{year}{1994}) \bibinfo{pages}{2567--2578}.
%Type = Article
\bibitem[{Toupin(1963)}]{Toupin63}
\bibinfo{author}{R.~Toupin}, \bibinfo{journal}{Int. J. Eng. Sci.}
  \bibinfo{volume}{1} (\bibinfo{year}{1963}) \bibinfo{pages}{101--126}.
%Type = Book
\bibitem[{Jackson(1999)}]{Jackson99}
\bibinfo{author}{J.~D. Jackson}, \bibinfo{title}{Classical electrodynamics},
  \bibinfo{publisher}{Wiley}, \bibinfo{address}{New York, {NY}},
  \bibinfo{edition}{3rd ed.} edition, \bibinfo{year}{1999}.
%Type = Book
\bibitem[{Mahan(1993)}]{Mahan93}
\bibinfo{author}{G.~D. Mahan}, \bibinfo{title}{{Many-Particle Physics}},
  \bibinfo{publisher}{Plenum}, \bibinfo{address}{New York, N.Y.},
  \bibinfo{edition}{2nd} edition, \bibinfo{year}{1993}.
%Type = Book
\bibitem[{Coleman(2015)}]{Coleman15}
\bibinfo{author}{P.~Coleman}, \bibinfo{title}{{Introduction to Many-Body
  Physics}}, \bibinfo{publisher}{Cambridge University Press},
  \bibinfo{year}{2015}.
%Type = Book
\bibitem[{Landau et~al.(1984)Landau, Pitaevskii, and
  Lifshitz}]{LandauLifshitz84}
\bibinfo{author}{L.~D. Landau}, \bibinfo{author}{L.~Pitaevskii},
  \bibinfo{author}{E.~Lifshitz}, \bibinfo{title}{Electrodynamics of continuous
  media}, volume~\bibinfo{volume}{8}, \bibinfo{publisher}{Elsevier},
  \bibinfo{year}{1984}.
%Type = Book
\bibitem[{Pines(1999)}]{Pines99}
\bibinfo{author}{D.~Pines}, \bibinfo{title}{Elementary Excitations in Solids:
  Lectures on Protons, Electrons, and Plasmons}, Advanced book classics,
  \bibinfo{publisher}{Advanced Book Program, Perseus Books},
  \bibinfo{year}{1999}.
%Type = Article
\bibitem[{Forcella et~al.(2014)Forcella, Zaanen, Valentinis, and Van
  Der~Marel}]{ForcellaEtAl14}
\bibinfo{author}{D.~Forcella}, \bibinfo{author}{J.~Zaanen},
  \bibinfo{author}{D.~Valentinis}, \bibinfo{author}{D.~Van Der~Marel},
  \bibinfo{journal}{Phys. Rev. B} \bibinfo{volume}{90} (\bibinfo{year}{2014})
  \bibinfo{pages}{035143}.
%Type = Article
\bibitem[{Peskin(1978)}]{Peskin78}
\bibinfo{author}{M.~E. Peskin}, \bibinfo{journal}{Ann. Phys. (N.Y.)}
  \bibinfo{volume}{113} (\bibinfo{year}{1978}) \bibinfo{pages}{122--152}.
%Type = Article
\bibitem[{Dasgupta and Halperin(1981)}]{DasguptaHalperin81}
\bibinfo{author}{C.~Dasgupta}, \bibinfo{author}{B.~Halperin},
  \bibinfo{journal}{Phys. Rev. Lett.} \bibinfo{volume}{47}
  (\bibinfo{year}{1981}) \bibinfo{pages}{1556}.
%Type = Article
\bibitem[{Herbut(1997)}]{Herbut97}
\bibinfo{author}{I.~F. Herbut}, \bibinfo{journal}{J. Phys. A}
  \bibinfo{volume}{30} (\bibinfo{year}{1997}) \bibinfo{pages}{423}.
%Type = Article
\bibitem[{Mo et~al.(2002)Mo, Hove, and Sudb\o{}}]{MoHoveSudbo02}
\bibinfo{author}{S.~Mo}, \bibinfo{author}{J.~Hove},
  \bibinfo{author}{A.~Sudb\o{}}, \bibinfo{journal}{Phys. Rev. B}
  \bibinfo{volume}{65} (\bibinfo{year}{2002}) \bibinfo{pages}{104501}.
%Type = Article
\bibitem[{Smiseth et~al.(2004)Smiseth, Sm\o{}rgrav, and
  Sudb\o{}}]{SmisethSmorgravSudbo04}
\bibinfo{author}{J.~Smiseth}, \bibinfo{author}{E.~Sm\o{}rgrav},
  \bibinfo{author}{A.~Sudb\o{}}, \bibinfo{journal}{Phys. Rev. Lett.}
  \bibinfo{volume}{93} (\bibinfo{year}{2004}) \bibinfo{pages}{077002}.
%Type = Article
\bibitem[{Smiseth et~al.(2005)Smiseth, Sm\o{}rgrav, Babaev, and
  Sudb\o{}}]{SmisethEtAl05}
\bibinfo{author}{J.~Smiseth}, \bibinfo{author}{E.~Sm\o{}rgrav},
  \bibinfo{author}{E.~Babaev}, \bibinfo{author}{A.~Sudb\o{}},
  \bibinfo{journal}{Phys. Rev. B} \bibinfo{volume}{71} (\bibinfo{year}{2005})
  \bibinfo{pages}{214509}.
%Type = Article
\bibitem[{Sm\o{}rgrav et~al.(2005)Sm\o{}rgrav, Smiseth, Babaev, and
  Sudb\o{}}]{SmorgravEtAl05}
\bibinfo{author}{E.~Sm\o{}rgrav}, \bibinfo{author}{J.~Smiseth},
  \bibinfo{author}{E.~Babaev}, \bibinfo{author}{A.~Sudb\o{}},
  \bibinfo{journal}{Phys. Rev. Lett.} \bibinfo{volume}{94}
  (\bibinfo{year}{2005}) \bibinfo{pages}{096401}.
%Type = Article
\bibitem[{Saito(1982{\natexlab{a}})}]{Saito82a}
\bibinfo{author}{Y.~Saito}, \bibinfo{journal}{Phys. Rev. Lett.}
  \bibinfo{volume}{48} (\bibinfo{year}{1982}{\natexlab{a}})
  \bibinfo{pages}{1114--1117}.
%Type = Article
\bibitem[{Saito(1982{\natexlab{b}})}]{Saito82b}
\bibinfo{author}{Y.~Saito}, \bibinfo{journal}{Phys. Rev. B}
  \bibinfo{volume}{26} (\bibinfo{year}{1982}{\natexlab{b}})
  \bibinfo{pages}{6239--6253}.
%Type = Article
\bibitem[{Strandburg(1988)}]{Strandburg88}
\bibinfo{author}{K.~J. Strandburg}, \bibinfo{journal}{Rev. Mod. Phys.}
  \bibinfo{volume}{60} (\bibinfo{year}{1988}) \bibinfo{pages}{161}.
%Type = Article
\bibitem[{Janke and Kleinert(1990)}]{JankeKleinert90}
\bibinfo{author}{W.~Janke}, \bibinfo{author}{H.~Kleinert},
  \bibinfo{journal}{Phys. Rev. B} \bibinfo{volume}{41} (\bibinfo{year}{1990})
  \bibinfo{pages}{6848}.
%Type = Article
\bibitem[{Beekman and Zaanen(2012)}]{BeekmanZaanen12}
\bibinfo{author}{A.~J. Beekman}, \bibinfo{author}{J.~Zaanen},
  \bibinfo{journal}{Phys. Rev. B} \bibinfo{volume}{86} (\bibinfo{year}{2012})
  \bibinfo{pages}{125129}.
%Type = Article
\bibitem[{Hoffman et~al.(2002)Hoffman, Hudson, Lang, Madhavan, Eisaki, Uchida,
  and Davis}]{HoffmanEtAl02}
\bibinfo{author}{J.~Hoffman}, \bibinfo{author}{E.~Hudson},
  \bibinfo{author}{K.~Lang}, \bibinfo{author}{V.~Madhavan},
  \bibinfo{author}{H.~Eisaki}, \bibinfo{author}{S.~Uchida},
  \bibinfo{author}{J.~Davis}, \bibinfo{journal}{Science} \bibinfo{volume}{295}
  (\bibinfo{year}{2002}) \bibinfo{pages}{466--469}.
%Type = Article
\bibitem[{Lang et~al.(2002)Lang, Madhavan, Hoffman, Hudson, Eisaki, Uchida, and
  Davis}]{LangEtAl02}
\bibinfo{author}{K.~Lang}, \bibinfo{author}{V.~Madhavan},
  \bibinfo{author}{J.~Hoffman}, \bibinfo{author}{E.~Hudson},
  \bibinfo{author}{H.~Eisaki}, \bibinfo{author}{S.~Uchida},
  \bibinfo{author}{J.~Davis}, \bibinfo{journal}{Nature} \bibinfo{volume}{415}
  (\bibinfo{year}{2002}) \bibinfo{pages}{412--416}.
%Type = Article
\bibitem[{{Chang} et~al.(2012){Chang}, {Blackburn}, {Holmes}, {Christensen},
  {Larsen}, {Mesot}, {Liang}, {Bonn}, {Hardy}, {Watenphul}, {Zimmermann},
  {Forgan}, and {Hayden}}]{ChangEtAl12}
\bibinfo{author}{J.~{Chang}}, \bibinfo{author}{E.~{Blackburn}},
  \bibinfo{author}{A.~T. {Holmes}}, \bibinfo{author}{N.~B. {Christensen}},
  \bibinfo{author}{J.~{Larsen}}, \bibinfo{author}{J.~{Mesot}},
  \bibinfo{author}{R.~{Liang}}, \bibinfo{author}{D.~A. {Bonn}},
  \bibinfo{author}{W.~N. {Hardy}}, \bibinfo{author}{A.~{Watenphul}},
  \bibinfo{author}{M.~V. {Zimmermann}}, \bibinfo{author}{E.~M. {Forgan}},
  \bibinfo{author}{S.~M. {Hayden}}, \bibinfo{journal}{Nature Phys.}
  \bibinfo{volume}{8} (\bibinfo{year}{2012}) \bibinfo{pages}{871--876}.
%Type = Article
\bibitem[{{Wu} et~al.(2013){Wu}, {Mayaffre}, {Kr{\"a}mer}, {Horvati{\'c}},
  {Berthier}, {Kuhns}, {Reyes}, {Liang}, {Hardy}, {Bonn}, and
  {Julien}}]{WuEtAl13}
\bibinfo{author}{T.~{Wu}}, \bibinfo{author}{H.~{Mayaffre}},
  \bibinfo{author}{S.~{Kr{\"a}mer}}, \bibinfo{author}{M.~{Horvati{\'c}}},
  \bibinfo{author}{C.~{Berthier}}, \bibinfo{author}{P.~L. {Kuhns}},
  \bibinfo{author}{A.~P. {Reyes}}, \bibinfo{author}{R.~{Liang}},
  \bibinfo{author}{W.~N. {Hardy}}, \bibinfo{author}{D.~A. {Bonn}},
  \bibinfo{author}{M.-H. {Julien}}, \bibinfo{journal}{Nature Comm.}
  \bibinfo{volume}{4} (\bibinfo{year}{2013}) \bibinfo{pages}{2113}.
%Type = Article
\bibitem[{{Comin} et~al.(2014){Comin}, {Frano}, {Yee}, {Yoshida}, {Eisaki},
  {Schierle}, {Weschke}, {Sutarto}, {He}, {Soumyanarayanan}, {He}, {Le Tacon},
  {Elfimov}, {Hoffman}, {Sawatzky}, {Keimer}, and {Damascelli}}]{CominEtAl14}
\bibinfo{author}{R.~{Comin}}, \bibinfo{author}{A.~{Frano}},
  \bibinfo{author}{M.~M. {Yee}}, \bibinfo{author}{Y.~{Yoshida}},
  \bibinfo{author}{H.~{Eisaki}}, \bibinfo{author}{E.~{Schierle}},
  \bibinfo{author}{E.~{Weschke}}, \bibinfo{author}{R.~{Sutarto}},
  \bibinfo{author}{F.~{He}}, \bibinfo{author}{A.~{Soumyanarayanan}},
  \bibinfo{author}{Y.~{He}}, \bibinfo{author}{M.~{Le Tacon}},
  \bibinfo{author}{I.~S. {Elfimov}}, \bibinfo{author}{J.~E. {Hoffman}},
  \bibinfo{author}{G.~A. {Sawatzky}}, \bibinfo{author}{B.~{Keimer}},
  \bibinfo{author}{A.~{Damascelli}}, \bibinfo{journal}{Science}
  \bibinfo{volume}{343} (\bibinfo{year}{2014}) \bibinfo{pages}{390--392}.
%Type = Article
\bibitem[{Lake et~al.(2002)Lake, R{\o}nnow, Christensen, Aeppli, Lefmann,
  McMorrow, Vorderwisch, Smeibidl, Mangkorntong, Sasagawa et~al.}]{LakeEtAl02}
\bibinfo{author}{B.~Lake}, \bibinfo{author}{H.~R{\o}nnow},
  \bibinfo{author}{N.~Christensen}, \bibinfo{author}{G.~Aeppli},
  \bibinfo{author}{K.~Lefmann}, \bibinfo{author}{D.~McMorrow},
  \bibinfo{author}{P.~Vorderwisch}, \bibinfo{author}{P.~Smeibidl},
  \bibinfo{author}{N.~Mangkorntong}, \bibinfo{author}{T.~Sasagawa}, et~al.,
  \bibinfo{journal}{Nature} \bibinfo{volume}{415} (\bibinfo{year}{2002})
  \bibinfo{pages}{299--302}.
%Type = Article
\bibitem[{Wu et~al.(2011)Wu, Mayaffre, Kr{\"a}mer, Horvati{\'c}, Berthier,
  Hardy, Liang, Bonn, and Julien}]{WuEtAl11}
\bibinfo{author}{T.~Wu}, \bibinfo{author}{H.~Mayaffre},
  \bibinfo{author}{S.~Kr{\"a}mer}, \bibinfo{author}{M.~Horvati{\'c}},
  \bibinfo{author}{C.~Berthier}, \bibinfo{author}{W.~Hardy},
  \bibinfo{author}{R.~Liang}, \bibinfo{author}{D.~Bonn}, \bibinfo{author}{M.-H.
  Julien}, \bibinfo{journal}{Nature} \bibinfo{volume}{477}
  (\bibinfo{year}{2011}) \bibinfo{pages}{191--194}.
%Type = Article
\bibitem[{Sachdev and La~Placa(2013)}]{SachdevLaPlaca13}
\bibinfo{author}{S.~Sachdev}, \bibinfo{author}{R.~La~Placa},
  \bibinfo{journal}{Phys. Rev. Lett.} \bibinfo{volume}{111}
  (\bibinfo{year}{2013}) \bibinfo{pages}{027202}.
%Type = Article
\bibitem[{Lee(2014)}]{Lee14}
\bibinfo{author}{P.~A. Lee}, \bibinfo{journal}{Phys. Rev. X}
  \bibinfo{volume}{4} (\bibinfo{year}{2014}) \bibinfo{pages}{031017}.
%Type = Article
\bibitem[{Hayward et~al.(2014)Hayward, Hawthorn, Melko, and
  Sachdev}]{HaywardEtAl14}
\bibinfo{author}{L.~E. Hayward}, \bibinfo{author}{D.~G. Hawthorn},
  \bibinfo{author}{R.~G. Melko}, \bibinfo{author}{S.~Sachdev},
  \bibinfo{journal}{Science} \bibinfo{volume}{343} (\bibinfo{year}{2014})
  \bibinfo{pages}{1336--1339}.
%Type = Article
\bibitem[{{Can} et~al.(2015){Can}, {Laskin}, and
  {Wiegmann}}]{CanLaskinWiegmann15}
\bibinfo{author}{T.~{Can}}, \bibinfo{author}{M.~{Laskin}},
  \bibinfo{author}{P.~B. {Wiegmann}}, \bibinfo{journal}{Ann. Phys. (N.Y.)}
  \bibinfo{volume}{362} (\bibinfo{year}{2015}) \bibinfo{pages}{752--794}.
%Type = Article
\bibitem[{Liu et~al.(2016)Liu, Nissinen, Slager, Wu, and Zaanen}]{LiuEtAl15b}
\bibinfo{author}{K.~Liu}, \bibinfo{author}{J.~Nissinen}, \bibinfo{author}{R.-J.
  Slager}, \bibinfo{author}{K.~Wu}, \bibinfo{author}{J.~Zaanen},
  \bibinfo{journal}{Phys. Rev. X} \bibinfo{volume}{6} (\bibinfo{year}{2016})
  \bibinfo{pages}{041025}.
%Type = Article
\bibitem[{Nissinen et~al.(2016)Nissinen, Liu, Slager, Wu, and
  Zaanen}]{NissinenEtAl16}
\bibinfo{author}{J.~Nissinen}, \bibinfo{author}{K.~Liu}, \bibinfo{author}{R.-J.
  Slager}, \bibinfo{author}{K.~Wu}, \bibinfo{author}{J.~Zaanen},
  \bibinfo{journal}{Phys. Rev. E} \bibinfo{volume}{94} (\bibinfo{year}{2016})
  \bibinfo{pages}{022701}.
%Type = Article
\bibitem[{Liu et~al.(2017)Liu, Nissinen, de~Boer, Slager, and
  Zaanen}]{LiuEtAl16}
\bibinfo{author}{K.~Liu}, \bibinfo{author}{J.~Nissinen},
  \bibinfo{author}{J.~de~Boer}, \bibinfo{author}{R.-J. Slager},
  \bibinfo{author}{J.~Zaanen}, \bibinfo{journal}{Phys. Rev. E}
  \bibinfo{volume}{95} (\bibinfo{year}{2017}) \bibinfo{pages}{022704}.
%Type = Unpublished
\bibitem[{Beekman and {\em et al.}(2017)}]{WuEtAl16}
\bibinfo{author}{A.~Beekman}, \bibinfo{author}{{\em et al.}},
  \bibinfo{title}{Dual gauge field theory of quantum liquid crystals in three
  dimensions}, \bibinfo{year}{2017}. \bibinfo{note}{In preparation}.
%Type = Article
\bibitem[{Prokof'ev and Svistunov(2005)}]{ProkofevSvistunov05}
\bibinfo{author}{N.~Prokof'ev}, \bibinfo{author}{B.~Svistunov},
  \bibinfo{journal}{Phys. Rev. Lett.} \bibinfo{volume}{94}
  (\bibinfo{year}{2005}) \bibinfo{pages}{155302}.
%Type = Article
\bibitem[{Liu and Fisher(1973)}]{LiuFisher73}
\bibinfo{author}{K.-S. Liu}, \bibinfo{author}{M.~E. Fisher},
  \bibinfo{journal}{J. Low. Temp. Phys.} \bibinfo{volume}{10}
  (\bibinfo{year}{1973}) \bibinfo{pages}{655--683}.
%Type = Article
\bibitem[{Batrouni et~al.(1995)Batrouni, Scalettar, Zimanyi, and
  Kampf}]{BatrouniEtAl95}
\bibinfo{author}{G.~Batrouni}, \bibinfo{author}{R.~Scalettar},
  \bibinfo{author}{G.~Zimanyi}, \bibinfo{author}{A.~Kampf},
  \bibinfo{journal}{Phys. Rev. Lett.} \bibinfo{volume}{74}
  (\bibinfo{year}{1995}) \bibinfo{pages}{2527}.
%Type = Article
\bibitem[{Wessel and Troyer(2005)}]{WesselTroyer05}
\bibinfo{author}{S.~Wessel}, \bibinfo{author}{M.~Troyer},
  \bibinfo{journal}{Phys. Rev. Lett.} \bibinfo{volume}{95}
  (\bibinfo{year}{2005}) \bibinfo{pages}{127205}.
%Type = Article
\bibitem[{Melko et~al.(2005)Melko, Paramekanti, Burkov, Vishwanath, Sheng, and
  Balents}]{MelkoEtAl05}
\bibinfo{author}{R.~Melko}, \bibinfo{author}{A.~Paramekanti},
  \bibinfo{author}{A.~Burkov}, \bibinfo{author}{A.~Vishwanath},
  \bibinfo{author}{D.~Sheng}, \bibinfo{author}{L.~Balents},
  \bibinfo{journal}{Phys. Rev. Lett.} \bibinfo{volume}{95}
  (\bibinfo{year}{2005}) \bibinfo{pages}{127207}.
%Type = Article
\bibitem[{Heidarian and Damle(2005)}]{HeidarianDamle05}
\bibinfo{author}{D.~Heidarian}, \bibinfo{author}{K.~Damle},
  \bibinfo{journal}{Phys. Rev. Lett.} \bibinfo{volume}{95}
  (\bibinfo{year}{2005}) \bibinfo{pages}{127206}.
%Type = Article
\bibitem[{Sengupta et~al.(2005)Sengupta, Pryadko, Alet, Troyer, and
  Schmid}]{SenguptaEtAl05}
\bibinfo{author}{P.~Sengupta}, \bibinfo{author}{L.~P. Pryadko},
  \bibinfo{author}{F.~Alet}, \bibinfo{author}{M.~Troyer},
  \bibinfo{author}{G.~Schmid}, \bibinfo{journal}{Phys. Rev. Lett.}
  \bibinfo{volume}{94} (\bibinfo{year}{2005}) \bibinfo{pages}{207202}.
%Type = Article
\bibitem[{Ceperley(1995)}]{Ceperley95}
\bibinfo{author}{D.~M. Ceperley}, \bibinfo{journal}{Rev. Mod. Phys.}
  \bibinfo{volume}{67} (\bibinfo{year}{1995}) \bibinfo{pages}{279}.
%Type = Article
\bibitem[{Clark and Ceperley(2006)}]{ClarkCeperley06}
\bibinfo{author}{B.~K. Clark}, \bibinfo{author}{D.~M. Ceperley},
  \bibinfo{journal}{Phys. Rev. Lett.} \bibinfo{volume}{96}
  (\bibinfo{year}{2006}) \bibinfo{pages}{105302}.
%Type = Article
\bibitem[{Fischer et~al.(2007)Fischer, Kugler, Maggio-Aprile, Berthod, and
  Renner}]{FischerEtAl07}
\bibinfo{author}{O.~Fischer}, \bibinfo{author}{M.~Kugler},
  \bibinfo{author}{I.~Maggio-Aprile}, \bibinfo{author}{C.~Berthod},
  \bibinfo{author}{C.~Renner}, \bibinfo{journal}{Rev. Mod. Phys.}
  \bibinfo{volume}{79} (\bibinfo{year}{2007}) \bibinfo{pages}{353--419}.
%Type = Article
\bibitem[{Carlson et~al.(2015)Carlson, Liu, Phillabaum, and
  Dahmen}]{CarlsonEtAl15}
\bibinfo{author}{E.~Carlson}, \bibinfo{author}{S.~Liu},
  \bibinfo{author}{B.~Phillabaum}, \bibinfo{author}{K.~Dahmen},
  \bibinfo{journal}{J. Sup. Novel Magn.} \bibinfo{volume}{28}
  (\bibinfo{year}{2015}) \bibinfo{pages}{1237--1243}.
%Type = Inproceedings
\bibitem[{Kondo(1952)}]{Kondo52}
\bibinfo{author}{K.~Kondo}, in: \bibinfo{booktitle}{Proc. 2nd Japan Nat. Congr.
  Applied Mechanics}, volume~\bibinfo{volume}{2}, pp. \bibinfo{pages}{41--47}.
%Type = Article
\bibitem[{Kr{\"o}ner(1959)}]{Kroner59}
\bibinfo{author}{E.~Kr{\"o}ner}, \bibinfo{journal}{Arch. Rat. Mech. Anal.}
  \bibinfo{volume}{4} (\bibinfo{year}{1959}) \bibinfo{pages}{273--334}.
%Type = Inproceedings
\bibitem[{Bilby et~al.(1955)Bilby, Bullough, and Smith}]{BilbyBulloughSmith55}
\bibinfo{author}{B.~Bilby}, \bibinfo{author}{R.~Bullough},
  \bibinfo{author}{E.~Smith}, in: \bibinfo{booktitle}{Proc. Roy. Soc. Lond. A},
  volume \bibinfo{volume}{231}, \bibinfo{organization}{The Royal Society}, pp.
  \bibinfo{pages}{263--273}.
%Type = Article
\bibitem[{McCrea et~al.(1990)McCrea, Hehl, and Mielke}]{McCreaHehlMielke90}
\bibinfo{author}{J.~D. McCrea}, \bibinfo{author}{F.~W. Hehl},
  \bibinfo{author}{E.~W. Mielke}, \bibinfo{journal}{Int. J. Theor. Phys.}
  \bibinfo{volume}{29} (\bibinfo{year}{1990}) \bibinfo{pages}{1185--1206}.
%Type = Article
\bibitem[{Katanaev and Volovich(1992)}]{KatanaevVolovich92}
\bibinfo{author}{M.~Katanaev}, \bibinfo{author}{I.~Volovich},
  \bibinfo{journal}{Ann. Phys. (N.Y.)} \bibinfo{volume}{216}
  (\bibinfo{year}{1992}) \bibinfo{pages}{1--28}.
%Type = Article
\bibitem[{Bowick and Giomi(2009)}]{BowickGiomi09}
\bibinfo{author}{M.~J. Bowick}, \bibinfo{author}{L.~Giomi},
  \bibinfo{journal}{Adv. Phys.} \bibinfo{volume}{58} (\bibinfo{year}{2009})
  \bibinfo{pages}{449--563}.
%Type = Article
\bibitem[{Koning and Vitelli(2014)}]{KoningVitelli14}
\bibinfo{author}{V.~Koning}, \bibinfo{author}{V.~Vitelli},
  \bibinfo{journal}{arXiv:1401.4957}  (\bibinfo{year}{2014}).
%Type = Article
\bibitem[{Alberte et~al.(2016)Alberte, Baggioli, Khmelnitsky, and
  Pujol{\`a}s}]{AlberteEtAl16}
\bibinfo{author}{L.~Alberte}, \bibinfo{author}{M.~Baggioli},
  \bibinfo{author}{A.~Khmelnitsky}, \bibinfo{author}{O.~Pujol{\`a}s},
  \bibinfo{journal}{JHEP} \bibinfo{volume}{2016} (\bibinfo{year}{2016})
  \bibinfo{pages}{1--47}.
%Type = Article
\bibitem[{Langley et~al.(2015)Langley, Vanacore, and
  Phillips}]{LangleyVanacorePhillips15}
\bibinfo{author}{B.~W. Langley}, \bibinfo{author}{G.~Vanacore},
  \bibinfo{author}{P.~W. Phillips}, \bibinfo{journal}{JHEP}
  \bibinfo{volume}{2015} (\bibinfo{year}{2015}) \bibinfo{pages}{1--8}.

\end{thebibliography}

\end{document}